\documentclass[longlist]{aa}

\usepackage{natbib}
\usepackage{graphicx}
\usepackage[varg]{txfonts}
\usepackage{float}
\usepackage[caption = false]{subfig}
\usepackage{lscape}

\def\l{~$\lambda$}
\def\ll{~$\lambda\lambda$}

\def\hea{He\,{\sc i}}
\def\heb{He\,{\sc ii}}

\def\kms{km\,s$^{-1}$}

\def\s{$\sigma$}

\DeclareSymbolFont{matha}{OML}{txmi}{m}{it}
\DeclareMathSymbol{\varv}{\mathord}{matha}{118}
\def\drv{\Delta \varv_{\mathrm{rad}}}

\def\gtrsim{\mathrel{\hbox{\rlap{\hbox{\lower3pt\hbox{$\sim$}}}\hbox{\raise2pt\hbox{$>$}}}}}
\def\lesssim{\mathrel{\hbox{\rlap{\hbox{\lower3pt\hbox{$\sim$}}}\hbox{\raise2pt\hbox{$<$}}}}}

\defcitealias{SdMdK12}{S+2012}

\defcitealias{KK+2014}{K+2014}

\begin{document}

\title{The Tarantula Massive Binary Monitoring:}
\subtitle{I. Observational campaign and OB-type spectroscopic binaries}

\author{L.A. Almeida\inst{\ref{inst1},\ref{inst2}}
\and H. Sana\inst{\ref{inst3},\ref{inst4}}
\and W. Taylor\inst{\ref{inst5}}
\and R. Barb\'a\inst{\ref{inst6}}
\and A.Z. Bonanos\inst{\ref{inst7}}
\and P. Crowther\inst{\ref{inst8}}
\and A. Damineli\inst{\ref{inst1}}
\and A. de Koter\inst{\ref{inst9},\ref{inst3}}
\and S.E. de Mink\inst{\ref{inst9}}
\and C.J. Evans\inst{\ref{inst5}}
\and M. Gieles\inst{\ref{inst10}}
\and N.J. Grin\inst{\ref{inst12}}
\and V. H\'enault-Brunet\inst{\ref{inst11}}
\and N. Langer\inst{\ref{inst12}}
\and D. Lennon\inst{\ref{inst13}}
\and S. Lockwood\inst{\ref{inst4}}
\and J. Ma\'{i}z Apell\'aniz\inst{\ref{inst14}}
\and A.F.J. Moffat\inst{\ref{inst15}}
\and C. Neijssel\inst{\ref{inst9}}
\and C. Norman\inst{\ref{inst2}}
\and O.H. Ram\'{i}rez-Agudelo\inst{\ref{inst5}}
\and N.D. Richardson\inst{\ref{inst16}}
\and A. Schootemeijer\inst{\ref{inst12}}
\and T. Shenar\inst{\ref{inst17}}
\and I. Soszy{\'n}ski\inst{\ref{inst18}}
\and F. Tramper\inst{\ref{inst13}}
\and J.S. Vink\inst{\ref{inst19}}
}

\institute{
Instituto de Astronomia, Geof\'isica e Ci\^encias Atmosf\'ericas, Rua do Mat\~ao 1226, Cidade Universit\'aria S\~ao Paulo, SP, Brasil, 05508-090, \email{leonardodealmeida.andrade@gmail.com}\label{inst1}
\and
Department of Physics \& Astronomy, Johns Hopkins University, Bloomberg Center for Physics and Astronomy, Room 520, 3400 N Charles St\label{inst2}
\and Institute of Astrophysics, 
           KU Leuven, 
           Celestijnenlaan 200 D, 
           3001, Leuven, Belgium\label{inst3}
\and Space Telescope Science Institute, 3700 San Martin Drive, Baltimore, MD 21218, United States\label{inst4}
\and UK Astronomy Technology Centre, Royal Observatory Edinburgh, Blackford Hill, Edinburgh, EH9 3HJ, UK \label{inst5}
\and Departamento de F\'isica y Astronom\'ia, Universidad de La Serena, Av. Cisternas 1200 Norte, La Serena, Chile\label{inst6}
\and IAASARS, National Observatory of Athens, GR-15326 Penteli, Greece\label{inst7}
\and Dept of Physics and Astronomy, University of Sheffield, Hounsfield Road, Sheffield, S3 7RH, United Kingdom\label{inst8}
\and Anton Pannekoek Astronomical Institute, University of Amsterdam, 1090 GE Amsterdam, The Netherlands\label{inst9}
\and Department of Physics, University of Surrey, Guildford GU2 7XH, UK\label{inst10}
\and Department of Astrophysics/IMAPP, Radboud University, PO Box 9010, NL-6500 GL Nijmegen, the Netherlands\label{inst11}
\and Argelander-Institut für Astronomie, der Universit{\"a}t Bonn, Auf dem Hügel 71, 53121 Bonn, Germany\label{inst12}
\and European Space Astronomy Centre (ESA/ESAC), PO Box 78, 28691 Villanueva de la Ca\~nada, Madrid, Spain\label{inst13}
\and Centro de Astrobiolog\'{\i}a, CSIC-INTA, campus ESAC, camino bajo del castillo s/n, E-28\,692 Madrid, Spain\label{inst14}
\and Dépt de physique, Univ. de Montréal, C.P. 6128, Succ. C-V, Montréal, QC, H3C 3J7,  and Centre de Recherche en Astrophsyique de Québec, Canada\label{inst15}
\and Ritter Observatory, Department of Physics and Astronomy, The University of Toledo, Toledo, OH 43606-3390, USA\label{inst16}
\and Institut f\"ur Physik und Astronomie, Universit\"at Potsdam, Karl-Liebknecht-Str. 24/25, D-14476 Potsdam, Germany\label{inst17}
\and Warsaw University Observatory, Al. Ujazdowskie 4, 00-478 Warszawa, Poland\label{inst18}
\and Armagh Observatory, College Hill, Armagh, BT61 9DG, Northern Ireland, UK\label{inst19}
}

\newpage

\abstract
{Massive binaries play a crucial role in the Universe. Knowing the distributions of their orbital parameters is important for a wide range of topics from stellar feedback to binary evolution channels and from the distribution of supernova types to gravitational wave progenitors, yet no direct measurements exist outside the Milky Way.}
{The Tarantula Massive Binary Monitoring project was designed to help fill this gap by obtaining multi-epoch radial velocity (RV) monitoring of 102 massive binaries in the 30 Doradus region.}
{In this paper we analyze 32 FLAMES/GIRAFFE observations of 93 O- and 7 B-type binaries. We performed a Fourier analysis and obtained orbital solutions for 82 systems: 51 single-lined  (SB1) and 31 double-lined (SB2) spectroscopic binaries.}
{Overall, the binary fraction and orbital properties across the 30 Doradus region are found to be similar to existing Galactic samples. This indicates that within these domains environmental effects are of second order in shaping the properties of massive binary systems. A small difference is found in the distribution of orbital periods, which is slightly flatter (in log space) in 30 Doradus than in the Galaxy, although this may be compatible within error estimates and differences in the fitting methodology. Also, orbital periods in 30 Doradus can be as short as 1.1 d, somewhat shorter than seen in Galactic samples. Equal mass binaries ($q > 0.95$) in 30 Doradus are all found outside NGC 2070, the central association that surrounds R136a, the very young and massive cluster at 30 Doradus’s core. Most of the differences, albeit small, are compatible with expectations from binary evolution. One outstanding exception, however, is the fact that earlier spectral types (O2--O7) tend to have shorter orbital periods 
than later spectral types (O9.2--O9.7).}
{Our results  point to a relative universality of the incidence rate of massive binaries and their orbital properties in the metallicity range from solar ($Z_{\odot}$) to  about half solar. This provides the first direct constraints on massive binary properties in massive star-forming galaxies at the Universe's peak of star formation at redshifts $z \sim 1$ to 2 which are estimated to have $Z \sim 0.5 Z_{\odot}$.
}


\titlerunning{TMBM I: Observational campaign and spectroscopic binaries}
\authorrunning{Almeida et al.}

\maketitle

\section{Introduction}

Massive stars are one of the most important cosmic engines driving the evolution of galaxies throughout the history of the Universe \citep{BCP08}. Recent observational evidence \citep{Kobulnicky+2007, MHG09, SaE11, SdMdK12, KK2012, sota+2014} shows that over half of all massive stars belong to binary systems where the two stars are close enough to interact during their lifetime. This challenges the long-held view of the predominance of the single-star evolutionary channel. Establishing the multiplicity properties of large samples of massive stars, including the distribution of their orbital parameters, is thus crucial in order to understand and properly compute the evolution of these objects \citep{PJH92, LCY08, Eldridge+2008, dMCL09}, including the frequency of high-mass X-ray and double compact binaries \citep{Sadowski+2008, Belczynski2008}, Type Ib/c supernovae \citep{YWL10, Eldridge+2013}, short and long-duration gamma-ray bursts \citep{Podsiadlowski+2010}, and gravitational wave progenitors \citep{dMB15}
.

The Tarantula Nebula (30~Doradus or NGC 2070) in the Large Magellanic Cloud (LMC) is our closest view of a massive starburst region in the local Universe, and one of the largest concentrations of massive stars in the Local Group \citep{WaB97}. It is thus an ideal laboratory to investigate a number of important outstanding questions regarding the physics, evolution, and multiplicity of the most massive stars.  

Among recent large observing campaigns towards the 30 Doradus region, the VLT FLAMES-Tarantula survey \citep[VFTS;][]{ETHB11} was an ESO Large Programme (182.D-0222, PI: Evans) that obtained multi-epoch spectroscopy of over 800 massive OB and WR stars in the 30~Dor region, providing a nearly complete census of the massive star content of 30~Dor \citep[][]{ETHB11}. Using a few epochs spread over a one-year baseline, the VFTS observational strategy was designed to detect short- and intermediate-period binaries ($\sim$ 1\,yr). Dedicated analyses of the O- and B-type binary samples have been published by \citet{SdKdM13} and \citet{DDS15}. For the O-type stars, the VFTS identified 116 spectroscopic binary candidates within the observed sample of 360 O stars, based on the amplitude of their radial velocity (RV) variations. This is a population comparable in size to the number of Galactic O-type binaries with computed orbits \citep{SaE11, barba+2010}.

\begin{figure}
\centering
\includegraphics[width=\columnwidth]{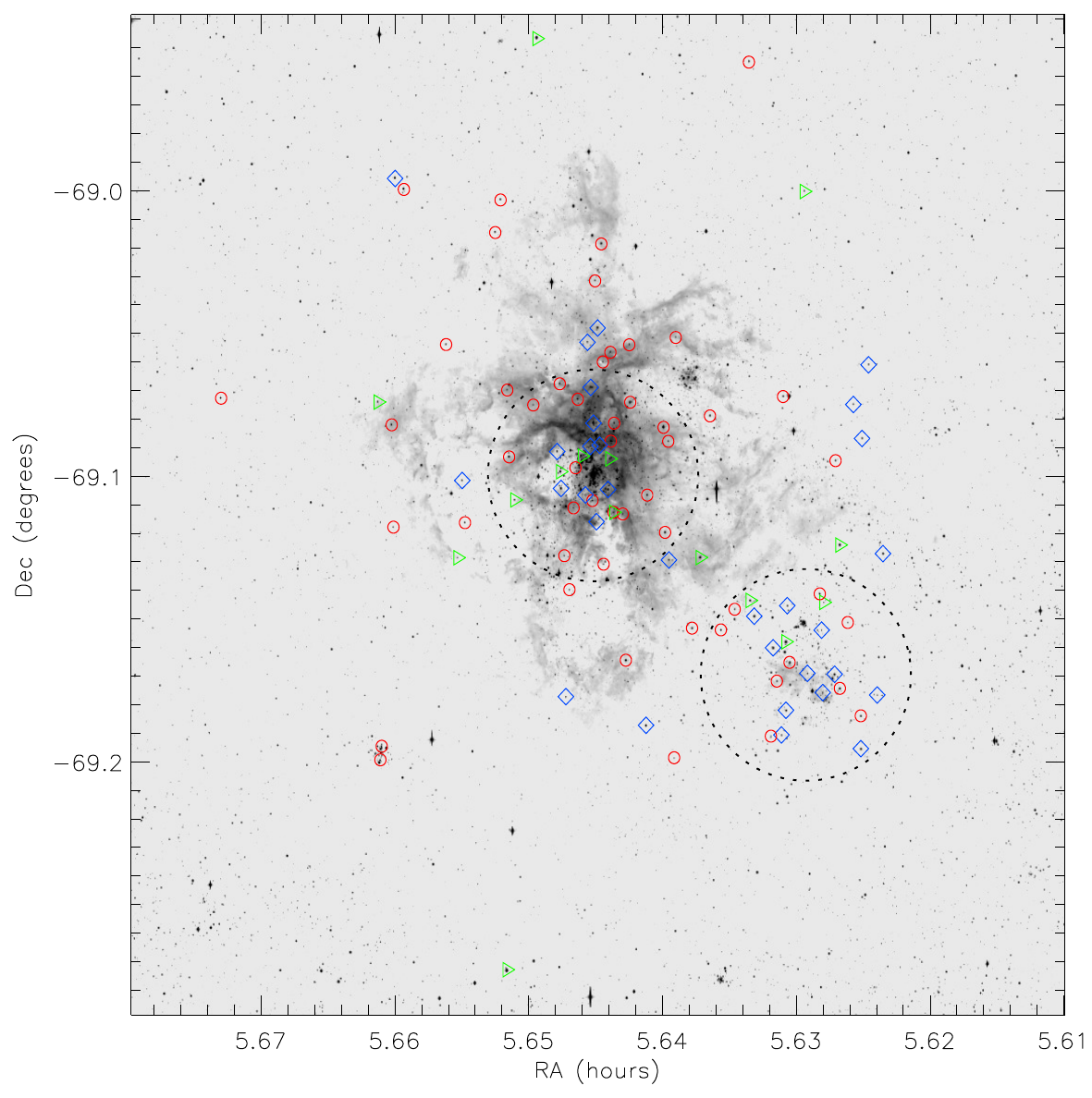}
\caption{Spatial distribution of TMBM O-type stars in the 30 Dor field. Red circles, blue diamonds, and green triangles are SB1 systems, SB2 systems, and stars with no periodicity, respectively. The dashed circles in the center and in the bottom right corner of the figure, which have 2.4$'$ of radius each, indicate the adopted regions for NGC~2070 and NGC~2060, respectively.
}
\label{f:alloc}
\end{figure}

The high incidence of short- and intermediate-period binaries in the Tarantula region confirms once more that binarity is central to the understanding of massive stars. The Tarantula Massive Binary Monitoring (hereafter TMBM) takes the next step. The project  aims to characterize the massive OB binaries that have been identified in the Tarantula region. Upon completion, the project will provide us with the unique opportunity to compare the multiplicity properties of the massive binaries in the Tarantula region with those from the Galaxy. This will allow us to investigate the impact of a dense and dynamically complex environment, as well as the possible role of metallicity in setting up the initial orbital parameters. Of particular interest is the shape of the period, eccentricity, and mass ratio distributions and whether the predominance of short-period systems is a general property of massive binaries. Quantitatively addressing this question is critical to properly predicting the end products of binary 
evolution, as we outlined earlier. 

At the same time, TMBM will obtain accurate orbital solutions of individual cornerstone systems. Detailed studies of these objects will provide unprecedented constraints on the nature of their components (including minimum mass estimates and, for eclipsing binaries, absolute values). They will be used as probes to test and calibrate our understanding of massive star evolution and of binary interaction. In this paper we focus on the overall multiplicity properties. In-depth studies of individual systems are deferred to future papers in this series.

Our observational strategy has been designed to measure the orbital properties of systems with orbital periods from about 1~d up to slightly over 1~yr, thus covering over two orders of magnitudes in the period distribution. Indeed, over 85\%\ of the binaries detected in VFTS show significant RV changes over time scales of one month, so that their periods are likely to be on the order of several months at most. This is also in line with properties of the Galactic massive binaries, whose population are largely dominated by systems with periods of less than a month \citep{SaE11, SdMdK12,KK+2014}.

This paper is organized as follows. Section~\ref{s:campaign} describes the observational campaign and data reduction. Sect.~\ref{s:orbits} provides the radial velocity measurements and orbital solution determination. Sect.~\ref{s:discuss} presents the observed distributions of orbital parameters and discusses our results. Our conclusions are summarized in Sect.~\ref{s:ccl}. 

\section{Observational campaign}\label{s:campaign}

\begin{table}
\centering
\caption{Target allocation ratio of the TMBM compared to the VFTS. Only O-type binaries are considered. As in \citet{SdKdM13}, membership to a cluster is defined as having an angular separation of less than 2.4$'$ with respect to the center of the considered association.}\label{t:alloc}
\begin{tabular}{c c c c}
\hline \hline
Category & TMBM & VFTS & Completion\\
\hline
\multicolumn{4}{c}{Entire region}\\
\hline
SBc   & 80 & 116 & 69\% \\ 
RV var. & 13 &  36 & 36\%  \\
\hline
\multicolumn{4}{c}{NGC 2070}\\
\hline
SBc   & 28 &  51 & 55\% \\ 
RV var. &  3 &  15 & 20\%  \\
\hline
\multicolumn{4}{c}{NGC 2060}\\
\hline
SBc    & 18 &  28 & 64\% \\ 
RV var. &  5 &  13 & 38\%  \\
\hline
\multicolumn{4}{c}{Outside clusters}\\
\hline
SBc   & 34 &  37 & 92\% \\ 
RV var. &  5 &   8 & 63\%  \\
\hline
\end{tabular}
\end{table}

\subsection{Target selection}\label{ss:targets}
The TMBM has obtained spectroscopic monitoring of 102 massive stars in the Tarantula region. Ninety-three O-type binary candidates and seven later-type (super)giants were chosen from the list of stars observed by the VFTS (360 O stars and 52 B supergiants, respectively) following the criteria described below. In addition, two WNh systems, RMC~144 and RMC~145 \citep[also known as R144 an R145;][]{SMVS09, SvBT13}, were added to the VFTS target list in view of their astrophysical interest. The WNh stars will be analyzed in separate papers, beginning with R145 \citep{Shenar}.

Operated in its MEDUSA+UVES mode, the ESO/FLAMES instrument offers 132 and 6 fibers, linked to the FLAMES/GIRAFFE and UVES spectrographs, respectively, thus  allowing simultaneous observations of up to 138 different objects. We used the FLAMES FPOSS fiber configuration software to allocate fibers to our targets. The main limiting factor in the allocation was crowding and the physical size of the buttons carrying the fibers, both of which prevent observations of objects too close to one another. A second limitation was the collision of buttons with the fibers. As a consequence, only about two-thirds of the known binary population could be observed in a single-plate configuration.

The initial target list is based on the RV analysis of the O-type stars observed by the VFTS in the MEDUSA configuration \citep{SdKdM13}. In doing so, we prioritize the allocation as follows. We assign the highest priority to objects with significant and large RV variations ($\drv > 20$~\kms), labeled spectroscopic binaries in \citet{SdKdM13} and considered here as high-likelihood spectroscopic binary candidates (SBc). Eighty of these binary candidates, hence 69\%\ of the detected population, could be allocated fibers.

As second priority objects, we assigned fibers to the objects presenting significant but low-amplitude RV variations ($\drv \lesssim 20$~\kms). These objects, labeled RV variables (RV var.) in \citet{SdKdM13}, could either be spectroscopic binaries or objects displaying photospheric or wind activity. Thirteen such objects could be allocated fibers. Finally, a handful of the remaining fibers could be allocated to interesting B-type (6) and A-type (1) supergiants \citep{MEDE15}, mostly in the outer part of the FLAMES field of view. These objects are believed to be the descendants of O-type binaries. As the primary star evolves into a B supergiant, it expands and will eventually transfer material to the secondary star. At least one of these systems shows clear evidence of past or present binary interactions and has already been analyzed by \citet{Howat+2015}.

\begin{figure}
\centering
\includegraphics[angle=0,width=\columnwidth]{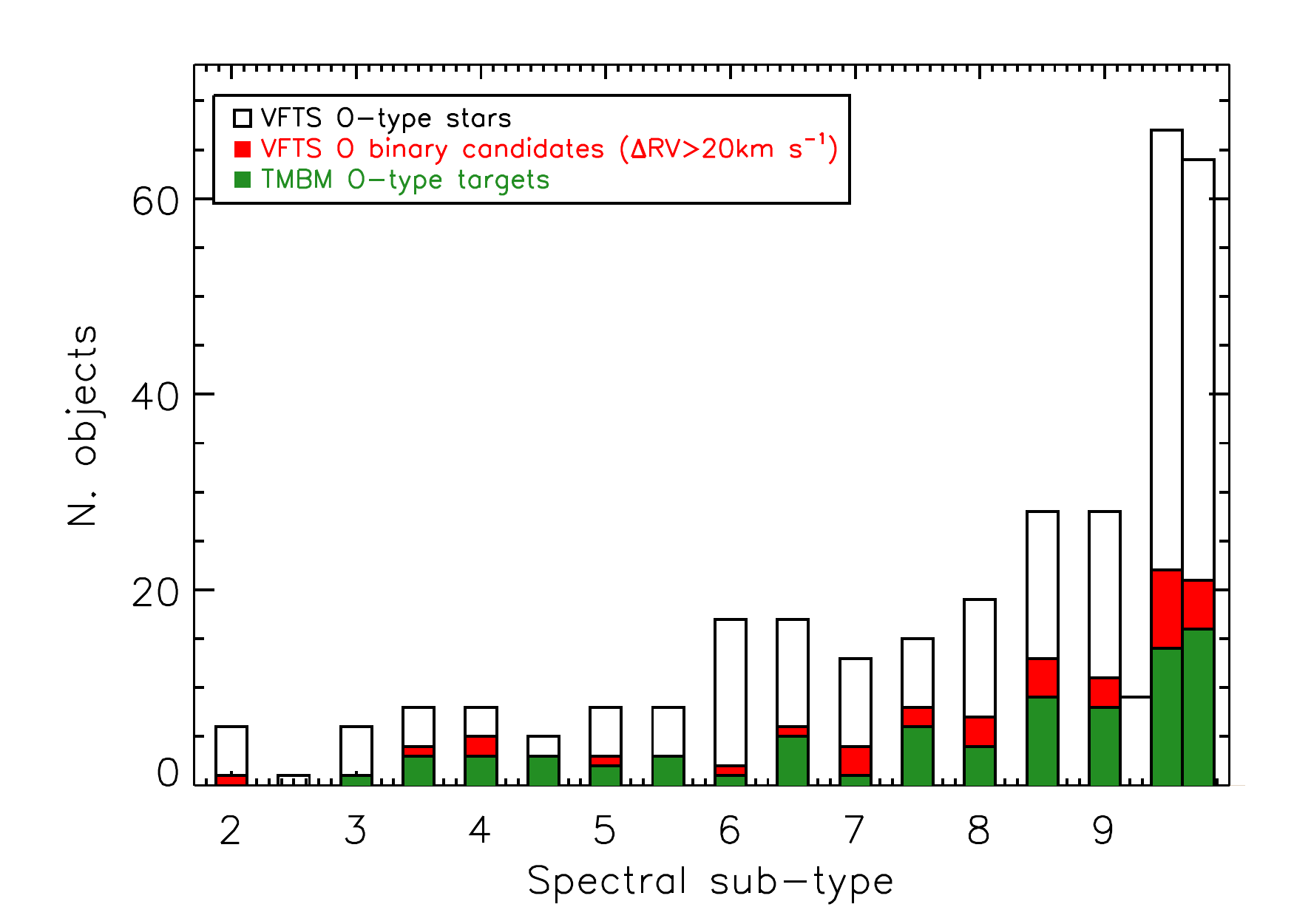}
\includegraphics[angle=0,width=\columnwidth]{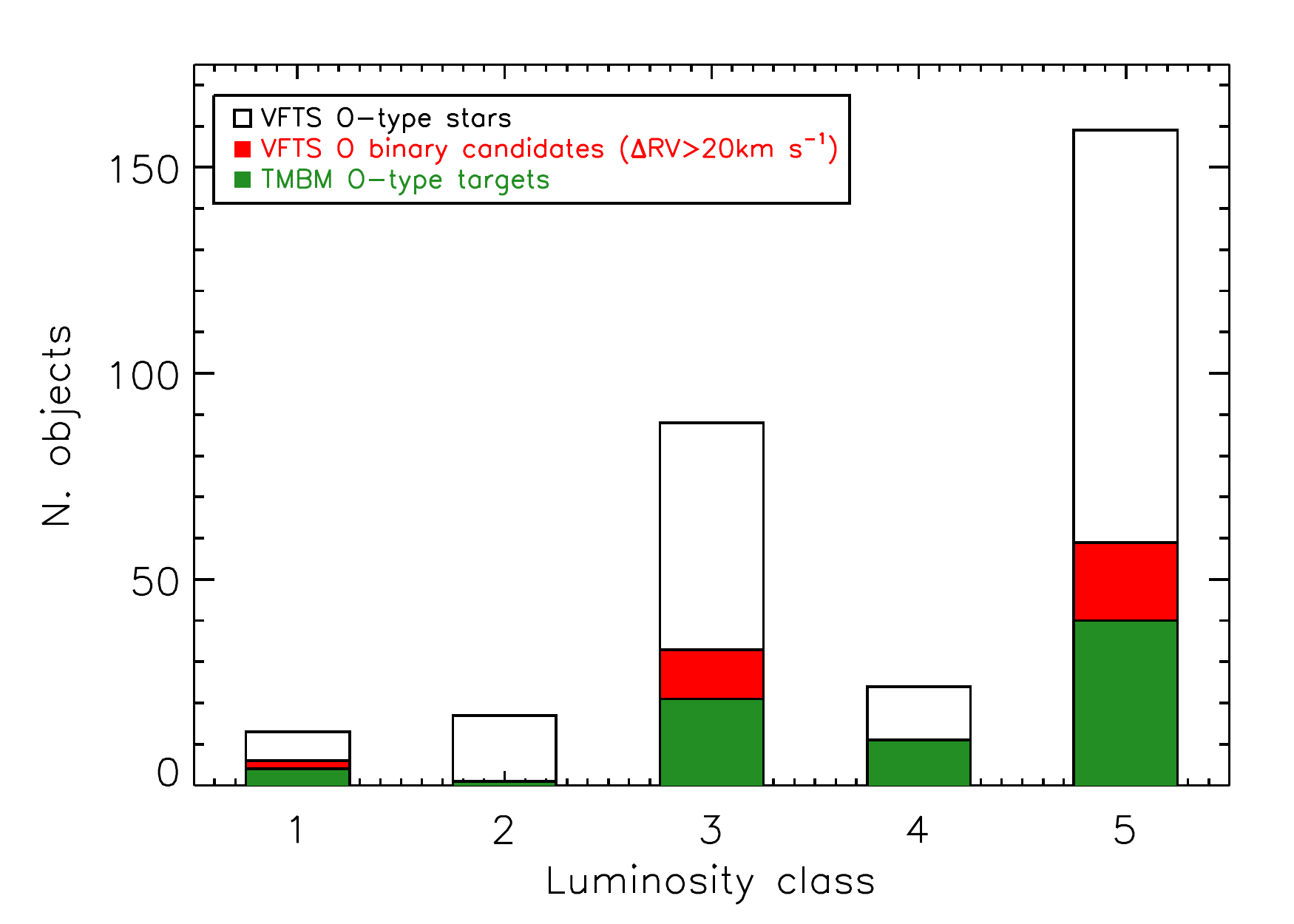}
\caption{Spectral types and luminosity classes of the O-type primary stars in TMBM compared to that of VFTS (see legend).}
\label{fig:spec_class}
\end{figure}

Table~\ref{t:alloc} provides the number of O-type targets and the fraction of the VFTS sample that has been monitored. Figure~\ref{f:alloc} shows the distribution of the targets in the field of view. Using spectral classification from \citet{WSSD14}, Fig.~\ref{fig:spec_class} compares the spectral type and luminosity class distributions for the primary components of our sample to that of the VFTS sample. It reveals that TMBM samples the entire O-type binary population in the VFTS with no particular selection effects. Finally, Table~\ref{tab:all} shows some basic information, e.g., spectral type(s), photometry, and previous line profile information (SB1/2, $\drv$), of all our targets.

\subsection{Observations}\label{ss:obs}
The bulk of the spectroscopic observations were obtained with the fiber-fed multi-object FLAMES/GIRAFFE spectrograph. We use the L427.2 (LR02) grating, which provides continuous coverage of the 3964--4567\AA\ wavelength range at a spectral resolving power $\lambda/\Delta \lambda$ of 6400. One object, R144, is significantly brighter that the rest of the targets  and was observed with UVES, the ESO high-resolution UV-optical spectrograph, in parallel to the FLAMES/GIRAFFE observations. The UVES data will be discussed in a separate paper.

We obtained 32 individual epochs of our 102 FLAMES targets: 18 epochs were spread from October 2012 to March 2013 (ESO period P90) and 14 additional epochs were acquired from October 2013 to March 2014 (P92). Because of issues with defective fibers, the P90 and P92 plate configurations that we used were not identical. In particular, two targets, VFTS~802 and VFTS~806, could not be allocated fibers in P92 and thus only have 18 epochs. Each epoch consisted of three 900~sec exposures taken back to back. The journal of the observations is available at the Centre de Donn\'ees astronomiques de Strasbourg\footnote{{\tt http://vizier.u-strasbg.fr/}} (CDS).

\longtab{
\small
\begin{landscape}
\begin{longtable}{lccccccccccc}
\caption{Photometric and spectral informations of the TMBM targets.\label{tab:all}}.\\
\hline\hline
\vspace*{-2mm}\\
ID & Spec. Type & V & B-V & J & $\sigma_\mathrm{\rm J}$ &  H & $\sigma_\mathrm{\rm H}$ &  K & $\sigma_\mathrm{\rm K}$ & $\drv$ & SB   \\
\vspace*{-2mm}\\
\hline
\endfirsthead
\caption{continued.}\\
\hline\hline
\vspace*{-2mm}\\
ID & Spec. Type & V & B-V & J & $\sigma_\mathrm{\rm J}$ &  H & $\sigma_\mathrm{\rm H}$ &  K & $\sigma_\mathrm{\rm K}$ & $\drv$ & SB  \\
\vspace*{-2mm}\\
\hline
\endhead
\hline
\endfoot
042    & O9.5 III((n))                           & 14.66  &  -0.12 & 14.75 & 0.02 & 14.77 & 0.01 & 14.82 & 0.02  &    30.1     &    SB1  \\
047    & O9 V + O9.5 V                           & 16.91  &   0.28 & 15.56 & 0.02 & 15.32 & 0.02 & 15.17 & 0.03  &    325.2    &  SB2   \\
055    & O8.5 V + O9.5 IV                        & 15.56  &   0.23 & 14.86 & 0.02 & 14.74 & 0.02 & 14.62 & 0.02  &    295.3    &  SB2   \\
061    & ON8.5 III: + O9.7: V:                   & 15.15  &   0.22 & 14.88 & 0.02 & 14.80 & 0.02 & 14.78 & 0.02  &    560.2    &  SB2   \\
063    & O5 III(n)(fc) + sec                     & 14.23  &   0.00 & 13.90 & 0.02 & 13.83 & 0.01 & 13.82 & 0.02  &    62.2     &  SB2   \\
064    & O7.5\,II(f)                             & 14.62  &   0.13 & 13.49 & 0.02 & 13.29 & 0.01 & 13.19 & 0.02  &    14.8     &  RV var.  \\  
066    & O9.5 III(n)                             & 15.54  &   0.10 & 15.20 & 0.02 & 15.14 & 0.02 & 15.09 & 0.02  &    46.8     &    SB1  \\
073    & O9.5\,III                               & 16.14  &   0.38 & 15.25 & 0.02 & 15.10 & 0.02 & 15.00 & 0.02  &    31.7     &    SB1  \\
086    & O9\,III((n))                            & 14.47  &   0.12 & 13.69 & 0.02 & 13.55 & 0.01 & 13.46 & 0.01  &    26.4     &    SB1  \\
087    &                     O9.7 Ib-II          & 13.58  &  -0.14 & 13.77 & 0.02 & 13.79 & 0.01 & 13.81 & 0.01  &     8.4     & RV var.    \\
093    & O9.2\,III-IV                            & 15.03  &   0.10 & 14.92 & 0.02 & 14.88 & 0.02 & 14.85 & 0.02  &     9.1     & RV var.    \\
094    & O3.5 Inf*p + sec?                       & 14.12  &   0.11 & 13.49 & 0.01 & 13.38 & 0.01 & 13.28 & 0.01  &    353.8    &  SB2   \\
113    &             O9.7 II or B0 IV ?          & 16.69  &   0.19 & 16.27 & 0.02 & 16.20 & 0.02 & 16.12 & 0.06  &    15.2     & RV var. \\
114    & O8.5 IV + sec                           & 15.97  &   0.17 & 15.33 & 0.01 & 15.21 & 0.01 & 15.09 & 0.02  &    181.1    &  SB2   \\
116    & O9.7: V: + B0: V:                       & 16.44  &   0.21 & 15.89 & 0.02 & 15.76 & 0.02 & 15.71 & 0.04  &     199.9   &   SB2  \\
120    & O9.5 IV:                                & 14.95  &   0.12 & 14.66 & 0.02 & 14.62 & 0.02 & 14.54 & 0.02  &    40.9     &  SB2   \\   
140    & O8.5 Vz                                 & 16.05  &   0.23 & 15.48 & 0.02 & 15.37 & 0.01 & 15.29 & 0.02  &    68.9     &    SB1  \\
148    &                 O9.7 II-III(n)          & 16.04  &  -0.04 & 16.39 & 0.02 & 16.42 & 0.03 & 16.27 & 0.07  &    54.0     &    SB1  \\
171    & O8\,II-III(f)                           & 14.06  &  -0.05 & 13.90 & 0.02 & 13.87 & 0.01 & 13.85 & 0.01  &    15.3     & RV var. \\
174    & O8 V + B0: V:                           & 15.50  &   0.25 & 15.07 & 0.02 & 14.97 & 0.02 & 14.96 & 0.03  &    320.3    & SB2    \\
176    & O6 V:((f)) + O9.5: V:                   & 14.78  &   0.03 & 14.50 & 0.01 & 14.49 & 0.01 & 14.45 & 0.02  &    59.58    & SB2    \\
178    &                       O9.7 Iab          & 12.91  &  -0.05 & 12.84 & 0.02 & 12.83 & 0.01 & 12.81 & 0.01  &    10.19    & RV var. \\
184    & O6.5 Vnz                                & 15.38  &  -0.09 & 15.29 & 0.02 & 15.28 & 0.02 & 15.28 & 0.03  &    4.47     & RV var. \\
187    & O9 IV: + B0: V:                         & 15.81  &   0.22 & 15.32 & 0.01 & 15.24 & 0.01 & 15.19 & 0.02  &   22.99     &    SB1  \\
191    & O9.5 V                                  & 15.74  &   0.12 & 15.49 & 0.01 & 15.49 & 0.01 & 15.45 & 0.03  &   9.32      & RV var.\\  
197    & O9 III                                  & 13.86  &  -0.06 & 13.84 & 0.02 & 13.86 & 0.02 & 13.85 & 0.01  &   36.18     &    SB1  \\
201    & O9.7 V + sec                            & 16.43  &   0.16 & 16.15 & 0.01 & 16.12 & 0.02 & 16.05 & 0.05  &    9.88     & SB2 / RV var.   \\     
217    & O4 V((fc)): + O5 V((fc)):               & 13.79  &  -0.11 & 13.86 & 0.02 & 13.87 & 0.01 & 13.87 & 0.01  &   139.77    & SB2    \\
223    &                        O9.5 IV          & 14.77  &  -0.05 & 14.94 & 0.02 & 14.93 & 0.01 & 14.93 & 0.02  &   5.07      & RV var.    \\
225    & B0.7-1III-II                            & 15.07  &  -0.01 & 15.12 & 0.02 & 15.19 & 0.02 & 15.16 & 0.04  &   --        &    --  \\
231    & O9.7 IV:(n) + sec                       & 16.18  &   0.11 & 15.89 & 0.02 & 15.84 & 0.02 & 15.76 & 0.05  &   143.7     & SB2    \\      
243    & O7 V(n)((f))                            & 15.26  &   0.21 & 14.73 & 0.02 & 14.62 & 0.01 & 14.55 & 0.02  &   128.5     &    SB1  \\      
256    & O7.5-8 V((n))z                          & 15.02  &  -0.10 & 15.20 & 0.02 & 15.22 & 0.02 & 15.24 & 0.03  &   42.9      &    SB1  \\
259    &                         O6 Iaf          & 13.65  &   0.21 & 12.85 & 0.02 & 12.73 & 0.01 & 12.61 & 0.01  &    13.9     & RV var.       \\ 
267    &                  O3 III-I(n)f*          & 13.49  &  -0.05 & 13.34 & 0.02 & 13.29 & 0.01 & 13.28 & 0.01  &   9.1       & RV var. \\
277    & O9 V                                    & 15.04  &   0.01 & 15.01 & 0.02 & 15.02 & 0.02 & 14.99 & 0.02  &   30.1      &    SB1  \\  
314    & O9.7 IV:(n) + sec                       & 16.06  &   0.03 & 16.08 & 0.03 & 16.07 & 0.02 & 16.07 & 0.06  &   244.7     &  SB2   \\ 
318    & O((n))p                                 & 16.56  &  -0.07 & 16.86 & 0.02 & 16.92 & 0.04 & 16.93 & 0.10  &   56.9      &    SB1  \\         
327    & O8.5 V(n) + sec                         & 15.33  &   0.01 & 15.15 & 0.02 & 15.14 & 0.02 & 15.12 & 0.03  &   154.4     &  SB2   \\
329    & O9.7 II-III(n)                          & 15.55  &   0.00 & 15.57 & 0.02 & 15.55 & 0.02 & 15.55 & 0.04  &    32.8     &    SB1  \\
332    & O9.2 II-III                             & 14.07  &   0.08 & 13.98 & 0.02 & 13.94 & 0.01 & 13.94 & 0.01  &    19.9     &    SB1  \\
333    & O8 II-III((f))                          & 12.49  &  -0.06 & 12.61 & 0.02 & 12.62 & 0.01 & 12.64 & 0.01  &    3.2      & RV var.    \\
350    & O8 V                                    & 14.95  &   0.12 & 14.46 & 0.03 & 14.42 & 0.02 & 14.38 & 0.03  &   115.9     &    SB1  \\
352    & O4.5 V(n)((fc)):z: + O5.5 V(n)((fc)):z: & 14.38  &  -0.10 & 14.32 & 0.01 & 14.31 & 0.01 & 14.31 & 0.01  &   375.8     &  SB2   \\
386    & O9 IV(n)                                & 14.75  &   0.20 & 14.67 & 0.04 & 14.83 & 0.05 & 14.67 & 0.10  &   35.2      &    SB1  \\
390    & O5-6 V(n)((fc))z                        & 15.49  &   0.14 & 15.13 & 0.01 & 14.92 & 0.02 & 14.89 & 0.04  &   112.4     &    SB1  \\
404    & O3.5 V(n)((fc))                         & 14.14  &   0.02 & 13.77 & 0.02 & 13.70 & 0.01 & 13.69 & 0.01  &    39.7     &    SB1  \\
409    & O4 V((f))z                              & 15.75  &   0.59 & 14.86 & 0.03 & 14.78 & 0.02 & 14.68 & 0.03  &    56.5     &    SB1  \\
429    & O7.5-8 V                                & 14.69  &  -0.11 & 14.68 & 0.02 & 14.66 & 0.02 & 14.71 & 0.02  &    174.2    &  SB2   \\
432    &                   O3.5 V((f))           & 15.65  &   0.25 & 14.90 & 0.03 & 14.76 & 0.02 & 14.71 & 0.05  &    49.5     &   SB2: \\
440    & O6-6.5 II(f)                            & 13.66  &   0.02 & 13.33 & 0.02 & 13.31 & 0.01 & 13.28 & 0.02  &      6.7    &   SB2: / RV var. \\
441    & O9.5 V                                  & 15.07  &  -0.07 & 15.09 & 0.03 & 15.10 & 0.03 & 15.05 & 0.04  &     105.0   &  SB2   \\
445    & O3-4 V:((fc)): +  O4-7 V: ((fc)         & 14.75  &   0.10 & 14.18 & 0.02 & 14.08 & 0.01 & 14.03 & 0.02  &    191.8    &  SB2   \\
450    & O9.7 III: + O7::                        & 13.60  &   0.20 & 13.08 & 0.02 & 12.91 & 0.01 & 12.89 & 0.03  &    417.9    &  SB2   \\
475    & O9.7 III                                & 16.43  &   0.37 & 15.38 & 0.02 & 15.23 & 0.02 & 15.13 & 0.03  &    124.4    &    SB1  \\
479    & O4-5 V((fc))z                           & 15.90  &   0.14 & 15.40 & 0.03 & 15.31 & 0.02 & 15.26 & 0.04  &    142.0    &    SB1  \\
481    & O8.5 III                                & 14.16  &  -0.04 & 14.09 & 0.01 & 14.06 & 0.01 & 14.08 & 0.03  &     73.1    &    SB1  \\
487    & O6.5: IV:((f)): + O6.5: IV:((f)):       & 16.88  &   0.12 &   $-$ &  $-$ & 15.69 & 0.09 & 15.37 & 0.12  &     272.5   &   SB2  \\
500    & O6.5 IV((fc)) + O6.5 V((fc))            & 14.19  &  -0.08 & 13.80 & 0.02 & 13.73 & 0.01 & 13.68 & 0.03  &    407.1    &  SB2   \\
508    & O9.5 V                                  & 15.98  &   0.17 & 15.57 & 0.02 & 15.33 & 0.04 & 15.48 & 0.05  &     39.9    &    SB1  \\
514    & O9.7 III                                & 15.84  &  -0.13 & 16.12 & 0.02 & 16.16 & 0.03 & 16.25 & 0.08  &     37.8    &    SB1  \\
526    & O8.5 I((n))fp                           & 14.92  &   0.54 & 13.57 & 0.02 & 13.31 & 0.01 & 13.15 & 0.01  &     27.6    &    SB1  \\
527    & O6.5 Iafc + O6 Iaf                      & 11.94  &   0.10 & 11.52 & 0.02 & 11.42 & 0.01 & 11.31 & 0.01  &     39.9    & SB2    \\
532    & O3 V(n)((f*))z+OB                       & 14.76  &   0.20 & 14.24 & 0.02 & 14.12 & 0.02 & 14.08 & 0.02  &     58.6    &    SB1  \\
538    & ON9 Ia: + O7.5: I:(f):                  & 13.99  &  -0.03 & 13.89 & 0.02 & 13.84 & 0.02 & 13.77 & 0.02  &    472.6    &  SB2   \\
543    & O9 IV + O9.7: V                         & 15.41  &   0.02 & 15.50 & 0.06 & 15.43 & 0.05 & 15.74 & 0.13  &    372.5    &  SB2   \\
555    & O9.5 Vz                                 & 15.88  &   0.08 & 15.51 & 0.02 & 15.37 & 0.02 & 15.29 & 0.04  &    38.9     &    SB1  \\
563    & O9.7 III: + B0: V:                      & 15.91  &   0.19 & 15.54 & 0.02 & 15.40 & 0.02 & 15.37 & 0.04  &     327.3   & SB2    \\
588    &                           O9.5          & 16.53  &   0.14 & 16.31 & 0.04 & 16.84 & 0.15 & 16.34 & 0.09  &    47.1     &    SB1  \\
603    & O4 III(fc)                              & 13.99  &   0.04 & 13.49 & 0.02 & 13.40 & 0.01 & 13.29 & 0.02  &    29.7     &    SB1  \\
613    & O8.5 Vz                                 & 15.78  &   0.16 & 15.38 & 0.02 & 15.38 & 0.02 & 15.19 & 0.03  &    20.4     &    SB1  \\
619    & O7-8 V(n)                               & 15.98  &   0.12 & 15.71 & 0.02 & 15.65 & 0.02 & 15.54 & 0.03  &    73.4     &    SB1  \\
631    & O9.7 III(n)                             & 16.00  &   0.14 & 15.76 & 0.02 & 15.66 & 0.02 & 15.66 & 0.04  &    293.7    &    SB1  \\
642    & O5 Vz: + O8 Vz:                         & 16.03  &   0.38 & 15.20 & 0.01 & 15.04 & 0.01 & 15.03 & 0.03  &    263.7    &  SB2   \\
645    & O9.5 V((n))                             & 16.29  &   0.16 & 15.91 & 0.02 & 15.80 & 0.02 & 15.82 & 0.05  &    71.0     &    SB1  \\
652    & B2 Ip + O9 III:                         & 13.88  &   0.20 & 13.40 & 0.02 & 13.28 & 0.01 & 13.22 & 0.01  &    --       &    --  \\
656    &               O7.5 III(n)((f))p         & 14.24  &   0.06 & 14.14 & 0.02 & 14.10 & 0.01 & 14.09 & 0.02  &    22.9     &    SB1  \\
657    & O7-8 II(f)                              & 15.45  &   0.44 & 14.19 & 0.02 & 13.96 & 0.02 & 13.73 & 0.02  &    92.7     &    SB1  \\
661    & O6.5 V(n) + O9.7: V:                    & 15.13  &   0.08 & 15.15 & 0.02 & 15.12 & 0.02 & 15.12 & 0.03  &    362.0    &  SB2   \\
696    &                 B0.7 Ib-Iab Nwk         & 12.73  &  -0.02 & 12.50 & 0.01 & 12.45 & 0.01 & 12.40 & 0.02  &    --       &    --  \\
702    & O8 V(n)                                 & 16.31  &   0.33 & 15.35 & 0.02 & 15.12 & 0.02 & 14.91 & 0.03  &    186.8    &    SB1  \\
728    &               O9.7 II-III((n))          & 15.61  &   0.07 & 15.50 & 0.02 & 15.45 & 0.02 & 15.42 & 0.04  &     28.4    &    SB1  \\
733    & O9.7p                                   & 14.28  &   0.12 & 13.90 & 0.02 & 13.82 & 0.01 & 13.74 & 0.02  &   135.6     &    SB1  \\
736    & O9.5 V                                  & 15.85  &   0.05 & 15.51 & 0.02 & 15.50 & 0.02 & 15.41 & 0.03  &    56.3     &    SB1  \\
739    &                          A0 Ip          & 12.26  &   0.31 & 11.13 & 0.02 & 10.93 & 0.03 & 10.85 & 0.02  &    --       &    --  \\
743    & O9.5 V((n))                             & 15.04  &  -0.17 & 15.33 & 0.01 & 15.39 & 0.02 & 15.41 & 0.04  &     40.4    &    SB1  \\
750    & O9.5 IV                                 & 15.43  &  -0.11 & 15.70 & 0.01 & 15.71 & 0.02 & 15.68 & 0.06  &      62.7   &    SB1  \\
764    &                   O9.7 Ia Nstr          & 12.26  &   0.09 & 12.34 & 0.01 & 12.32 & 0.01 & 12.32 & 0.01  &      13.7   & RV var.    \\
769    & O9.7 II-III                             & 15.83  &   0.08 & 15.37 & 0.02 & 15.30 & 0.02 & 15.31 & 0.03  &      79.3   &    SB1  \\
771    & O9.7 III:(n)                            & 15.66  &   0.16 & 15.17 & 0.02 & 15.08 & 0.02 & 15.04 & 0.03  &    104.9    &    SB1  \\
774    &            O7.5 IVp + O8.5: V:          & 16.89  &   0.49 & 15.48 & 0.02 & 15.28 & 0.02 & 15.17 & 0.03  &    133.5    & SB2    \\
779    & B1 II-Ib                                & 15.46  &   0.19 & 15.01 & 0.02 & 14.78 & 0.01 & 14.71 & 0.03  &    --       &    --  \\
802    & O7.5 Vz                                 & 14.14  &  -0.19 & 14.49 & 0.01 & 14.56 & 0.01 & 14.59 & 0.03  &    121.0    &  SB2   \\
806    & O5.5 V((fc)):z + O7 Vz:                 & 14.06  &  -0.17 & 14.35 & 0.01 & 14.40 & 0.01 & 14.44 & 0.03  &    306.4    &  SB2   \\
810    & O9.7 V + B1: V:                         & 16.36  &   0.07 & 16.24 & 0.02 & 16.17 & 0.02 & 16.26 & 0.07  &   235.1     &  SB2   \\                  
812    & O4-5 V((fc))                            & 14.81  &   0.05 & 14.46 & 0.02 & 14.40 & 0.01 & 14.36 & 0.02  &   88.1      &    SB1  \\                 
827    & B1.5 Ib                                 & 15.34  &   0.31 & 14.71 & 0.02 &   $-$ &  $-$ &   $-$ &  $-$  &    --       &    --  \\              
829    & B1.5-2 II                               & 15.13  &   0.41 & 14.31 & 0.02 & 14.17 & 0.01 & 14.09 & 0.02  &    --       &    --  \\                  
830    &                 O5-6 V(n)((f))          & 15.39  &  -0.03 & 15.29 & 0.01 & 15.31 & 0.02 & 15.34 & 0.03  &    24.0     &    SB1  \\ 
887    & O9.5 II-IIIn                            & 14.96  &  -0.06 & 15.09 & 0.01 & 15.08 & 0.01 & 15.15 & 0.02  &    59.1     &  SB2   \\   
\end{longtable}
\flushleft
\tablefoottext{Spectral types, optical and near-IR photometry are from \citet[][]{WSSD14}, \citet{ETHB11}, and \citet{kato+2007}. Binary classification in SB2 and $\Delta$ RV are from \citet{SdKdM13}}\\
\end{landscape}
}

\subsection{Data reduction}\label{ss:reduction}
The data were reduced using the ESO CPL FLAMES/GIRAFFE pipeline v.1.1.0 under the {\it esorex} environment. We consecutively applied the bias and dark subtraction, flat-field correction and wavelength calibration recipes. The spectra were extracted in SUM mode. To perform the sky-subtraction, we used the median of 14 sky-fibers that sample the field of view of our observations. Finally, the weighted mean  of the back-to-back exposures, together with a 3-\s\ clipping, was computed to remove cosmic rays. The obtained spectra of each target and each epoch were individually normalized by fitting a polynomial through the continuum region following the semi-automatic procedure described in \citet{SdKdM13}.

\subsection{Ancillary data}\label{ss:add_data}
While the TMBM driver is the FLAMES data, ancillary spectroscopic observations of three systems which were obtained with the VLT/X-shooter spectrograph. Similarly, $V$ and $I$ photometry of a large fraction of our targets are available through the OGLE-III and IV surveys \citep[][]{Udalski+2008,Udalski+2015} and additional photometry has been obtained at the Faulkes Telescope South. These data will be described in subsequent papers.

\begin{table}
  \centering
  \caption{Adopted rest wavelengths ($\lambda_0$).}
  \label{t:lines}
\begin{tabular}{lc}
\hline \hline 
Line & $\lambda_0$ (\AA)  \\
\hline
\vspace*{-2mm}\\
\hea+{\sc ii}\l4026 &  4026.072   \\
\heb\l4200      &  4199.832  \\ 
\hea\l4387      &  4387.929  \\
\hea\l4471      &  4471.480  \\
\heb\l4541      &  4541.591  \\
\hline\end{tabular}
\end{table}

\begin{figure}
\centering
\includegraphics[width=\columnwidth]{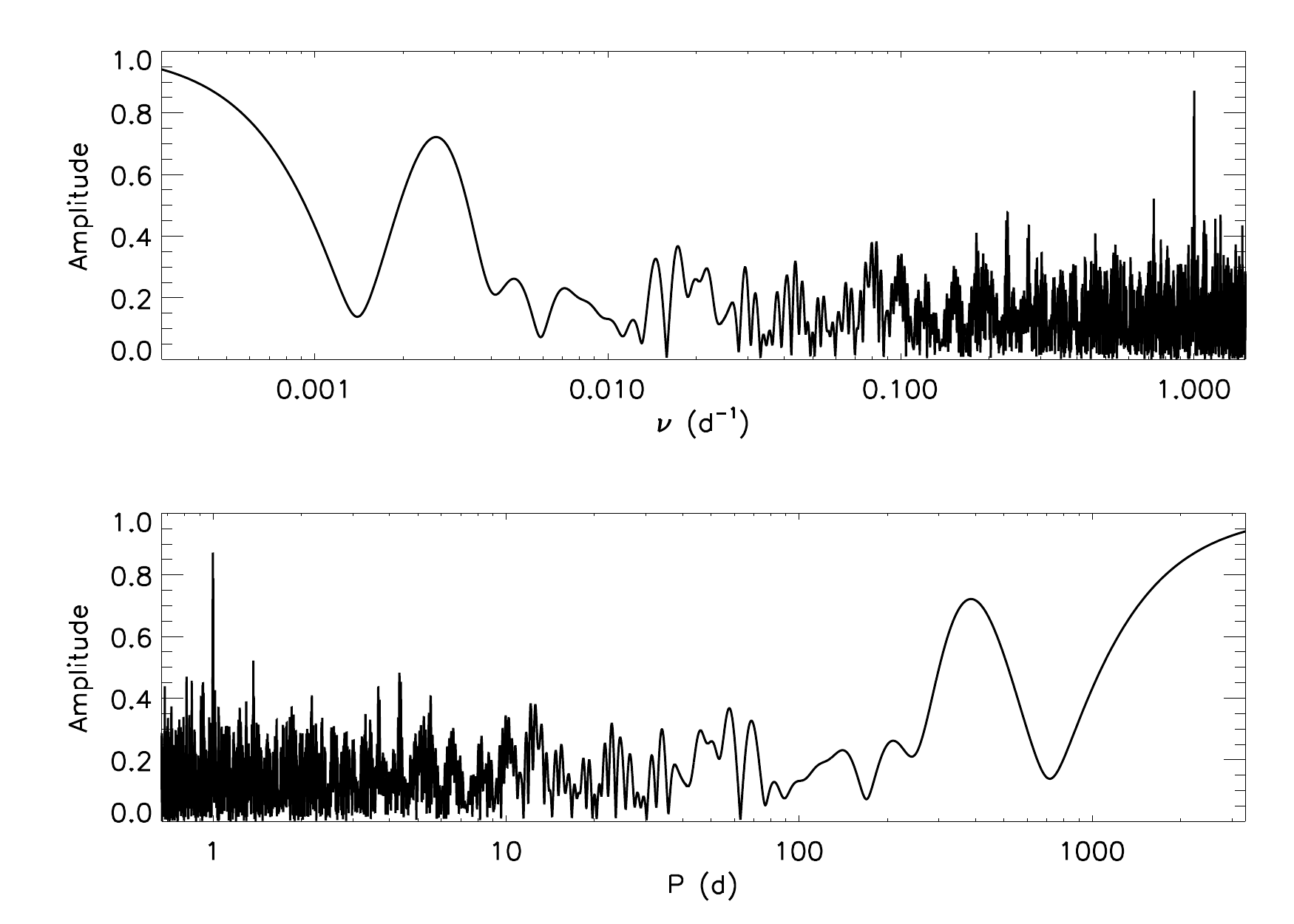}
\caption{Power spectral window of the TMBM campaign.}
\label{f:sw}
\end{figure}

\section{Orbital properties}\label{s:orbits}

\subsection{Radial velocities}\label{ss:rvs}
We measured the RVs of the OB stars in our target list by fitting Gaussian profiles to a selection of \hea\ and \heb\ absorption lines (see Table~\ref{t:lines}) present in the LR02 wavelength range. We used one or two Gaussian profiles per spectral line for single-lined (SB1) and double-lined (SB2) spectroscopic binaries, respectively.

Following  \citet{SdKdM13}, we simultaneously adjusted all the lines and all the epochs of a given target. We required that the Gaussian profile for a given line and binary component is identical throughout all the epochs and that all lines provide the same RV shift. In this process, we also included the LR02 data from the VFTS campaign, hence consistently re-deriving the RVs for the entire set of FLAMES/GIRAFFE LR02 data available for our objects.

Because the He\l4026 line is actually a blend of \hea\ and \heb, its effective wavelength varies with spectral subtypes \citep[see appendix in][]{SdKdM13}. For spectral subtypes O8 and earlier and for stars  where the \hea+\heb\l4026 blend was used for RV measurements, we also allowed for a possible zero-point shift between the rest frame of the fitted line profiles and the \heb\l4541 rest frame. A similar approach has been used in, e.g., \citet{TES11} and \citet{TSF16}. As a result, all our RV measurements are expressed in the \heb\l4541 reference frame, regardless of the spectral subtype. The table with the obtained RV measurements is available at the Centre de Donn\'ees astronomiques de Strasbourg. The RV measurements of all our targets present variability at a 3\s\ level, or more, confirming the validity of the selection criteria for our target list.

\begin{table}
\centering
\caption{TMBM objects for which no periodicity is found.}
\label{t:no_period}
\begin{tabular}{r r r}
\hline \hline
VFTS ID \# & Sp.\ Type & $\drv$ \\
        &           & (\kms) \\
\hline
    087 &                     O9.7 Ib-II &  11.9 \\ 
    113 &             O9.7 II or B0 IV ? &  28.0 \\ 
    148 &                 O9.7 II-III(n) &  50.9 \\ 
    178 &                       O9.7 Iab &  15.4 \\ 
    223 &                        O9.5 IV &   4.3 \\ 
    267 &                  O3 III-I(n)f* &  21.6 \\ 
    445 & O3-4 V:((fc)): + \hspace*{7mm} &  230 \\ 
        &                  O4-7 V: ((fc) &       \\ 
    588 &                           O9.5 &  57.7 \\ 
    656 &              O7.5 III(n)((f))p &  46.1 \\ 
    696 &                B0.7 Ib-Iab Nwk &  24.1 \\ 
    728 &               O9.7 II-III((n)) &  25.6 \\ 
    739 &                          A0 Ip &   6.1 \\ 
    774 &            O7.5 IVp + O8.5: V: & 333.2 \\ 
    830 &                 O5-6 V(n)((f)) &  50.3 \\ 
\hline
\end{tabular}
\end{table}

\begin{table}
\centering
\caption{TMBM objects for which a period is found but not an orbital solution.}
\label{t:no_orbit}
\begin{tabular}{r r r r}
\hline \hline
VFTS ID \# & Sp.\ Type & $\drv$ & $P$ \\
        &           & (\kms) & (d)\\
\hline 
    259 &                         O6 Iaf &  22.5 & 3.69  \\ 
    432 &                   O3.5 V((f))  &  91.3 & 4.87 \\ 
    526 &                   O8.5 I((n))fp&  40.6 & 10.98 \\     
    764 &                   O9.7 Ia Nstr &  27.0 & 1.22  \\
\hline
\end{tabular}
\end{table}

\begin{table*}
\centering
\caption{Orbital periods and eccentricities of systems that do not pass the Lucy-Sweeney test at the 5\%\ significance level ($e/\sigma_\mathrm{e} \leq 2.49$). The last two columns provide the 68.5 and 95\%\ confidence (upper) limits on the eccentricity.}
\label{t:LStest}
\begin{tabular}{r r r r r r r r}
\hline \hline
VFTS &          $P_\mathrm{orb}$    & $\sigma_\mathrm{P}$ & $e$ &  $\sigma_\mathrm{e}$  & $e / \sigma_\mathrm{e} $ & $< \epsilon_{68.3}$ & $< \epsilon_{95}$ \\
ID   &         (d)   & (d) \\
\hline
 066 &   1.141160  &  0.000005 &  0.008  & 0.005  & 1.600  & 0.009  & 0.015 \\
 140 &   1.611655 &   0.000007 &  0.002  & 0.002  & 1.000  & 0.003  & 0.005 \\
 217 &   1.855341 &   0.000002 &  0.001  & 0.001  & 1.000  & 0.001  & 0.002 \\
 243 &  10.402840 &   0.000140 &  0.016  & 0.008  & 2.000  & 0.021  & 0.027 \\
 318 &  14.004270 &   0.002910 &  0.083  & 0.044  & 1.886  & 0.105  & 0.145 \\
 352 &   1.124143 &   0.000002 &  0.012  & 0.005  & 2.400  & 0.016  & 0.019 \\
 500 &   2.875370 &   0.000004 &  0.001  & 0.001  & 1.000  & 0.001  & 0.002 \\
 538 &   4.159758 &   0.000023 &  0.001  & 0.001  & 1.000  & 0.001  & 0.002 \\
 543 &   1.383987 &   0.000003 &  0.016  & 0.007  & 2.286  & 0.021  & 0.026 \\
 563 &   1.217341 &   0.000007 &  0.015  & 0.010  & 1.500  & 0.017  & 0.029 \\
 619 &  14.504290 &   0.002580 &  0.085  & 0.040  & 2.125  & 0.111  & 0.142 \\
 631 &   5.374869 &   0.000184 &  0.007  & 0.005  & 1.400  & 0.008  & 0.014 \\
 642 &   1.726822 &   0.000011 &  0.008  & 0.005  & 1.600  & 0.009  & 0.015 \\
 702 &   1.981437 &   0.000022 &  0.008  & 0.006  & 1.333  & 0.009  & 0.016 \\
 733 &   5.922121 &   0.000068 &  0.002  & 0.001  & 2.000  & 0.003  & 0.003 \\
 743 &  14.947320 &   0.000890 &  0.012  & 0.008  & 1.500  & 0.013  & 0.023 \\
 769 &   2.365628 &   0.000018 &  0.023  & 0.013  & 1.769  & 0.028  & 0.041 \\
 \hline
\end{tabular}
\end{table*}

\subsection{Period search}\label{ss:periods}

The TMBM was designed to collect data covering time scales of days, weeks, and months in sufficient amount to allow us to derive the periodicity of most of the monitored targets. Figure~\ref{f:sw} displays the power spectrum window of the TMBM campaign. It shows that, besides the 1~d and 1.3~yr alias, the power is uniformly spread across the period range. The 1~d alias is the result of our observational and data handling strategy (no more than one observing point per night, stacking of the back-to-back exposures). The 1.3~yr alias is the actual duration of the TMBM campaign. The inclusion of the VFTS data further improves the performance at longer periods as, the present data set combined with the VFTS offers a time base of 1908 days ($\approx 5.2$~yr)

To search for periodicities in the RV time series, we computed the Lomb-Scargle periodogram \citep{Lomb1976, Scargle1982} using the IDL program MPRVFIT\footnote{http://www.vanderbilt.edu/AnS/physics/vida/mprvfit.htm} developed by \citet{DeLee2013}. To confirm the periodicities obtained by the Lomb-Scargle periodogram, we used a variant of the string-length method \citep[SL,][]{Dworetsky1983} and found similar results. The Lomb-Scargle periodograms for all objects are shown in Appendix~\ref{app:B}.

We only considered a periodogram peak to be significant if its associated false alarm probability (FAP) is less than 1\%\footnote{The periodograms of two SB2 systems, VFTS 404 and 810, do not show peaks above the 1\% FAP threshold, but they are still considered periodic. As discussed in Appendix~\ref{app:notes}, the RV measurements in these two SB2 systems are significantly affected by line blending, as is the strength of the peaks in the periodogram.}. Fourteen of our targets show no peak that passes this criterion and show no coherent RV curves. These objects are listed in Table~\ref{t:no_period} and include 7 of the 13 VFTS low-amplitude RV variable objects in our survey and two BA-type supergiants. Notes on these objects are presented in Appendix~\ref{app:notes}. The TMBM is thus able to find periodicity for over 90\%\ for the SBc and for over 50\%\ of the RV variable objects in our target list (see Sect.~\ref{ss:targets}), confirming the adequacy of the time-sampling strategy. 

For SB2 systems with almost identical components, the period search is complicated by possible misidentification of the components. In those cases, we also performed a period search on the absolute differences between the component RVs. This yields a peak in the periodogram corresponding to $P_\mathrm{orb}/2$, which allowed us to better discriminate between the binary components.

\subsection{Orbital solutions}\label{ss:orbits}

Best-fit orbital parameters for SB1 and SB2 systems, and their uncertainties, were obtained by fitting the orbital RV curve described by the equation given below to the time series of the RV measurements using a Markov chain Monte Carlo (MCMC) procedure:

\begin{equation}
RV_{j} = \gamma + K_{j} (\cos(\theta+\omega_j)+ e \cos \omega_j),
\end{equation}
with $j=1$ for the primary star and 2 for the secondary. In this equation, $K$ is the amplitude of the RV curve, $\theta$ is the true anomaly, $e$ is the eccentricity, $\omega$ is the periastron argument of the system's orbit ($\omega_2=\omega_1+\pi$), and $\gamma$ is the systemic velocity. The velocity amplitudes are related to the orbital parameters via 
\begin{equation}
K_{1,2} = \left(\frac{2\pi G}{P_\mathrm{orb}} \right)^{\frac{1}{3}}\frac{m_{2,1}\sin i}{m_{1,2}^{2/3}}\frac{1}{(1-e^2)^{1/2}},
\end{equation}
where $G$ is the gravitational constant, $P_\mathrm{orb}$ is the orbital period, $i$ is the orbital inclination, and $m_{1,2}$ are the masses of the components. We use the Lomb-Scargle periodogram to constrain the initial value for the  orbital period in our MCMC procedure.

The fitted and derived parameters of the SB1 systems, and their spectral classification following \citet[][]{WSSD14}, are listed in Table~\ref{tab:sb1}. For the SB2 systems, the fitted and derived parameters are given in Tables~\ref{tab:sb2} and~\ref{tab:sb2_2}, respectively. The parameter uncertainties given in these tables are the 68\% confidence intervals built from the  marginal MCMC posterior distributions.The best-fit RV curves for the SB1 and SB2 systems are shown in Figs.~\ref{sb1:orb_solution} and \ref{sb2:orb_solution1}. 

For the longer period systems (VFTS~064, 171, 332, 333, 440, 750; $P_\mathrm{orb}>1$~yr), the RV data often covers only one periastron passage. In such cases, establishing the correct periodicity is challenging. Their orbital solutions should be considered tentative at best. For 4 systems with determined periods we were not able to find a satisfactory solution (see Table~\ref{t:no_orbit}). Finally, we note that the best-fit orbital solutions of VFTS~171 ($P_\mathrm{orb}=677$~d), VFTS~332 ($P_\mathrm{orb}=1025$~d) and VFTS~802 ($P_\mathrm{orb}=183$~d) have converged to orbital periods that correspond to approximately half the value of the maximum peak in their Lomb-Scargle periodogram. We have checked that these solutions indeed provide a smaller $\chi^2$ and are thus formally better; however, additional data allowing us to double check these periods is desirable.

\subsubsection{Eccentricities}\label{sss:ecc}

To test the significance of the eccentricity values of the derived orbital solutions, we used the Lucy-Sweeney test \citep{LuS71,Luc13} and adopt a 5\%\ significance level to consider the eccentricity to be significant. Table~\ref{t:LStest} lists the systems that do not pass this test, i.e.\ the systems for which the eccentricity that we derived is not significantly different from zero. Table~\ref{t:LStest} also indicates the 68.5\%\ and 95\%\ confidence (upper) limits on the eccentricities computed following \citet{Luc13}. We performed the fitting procedure shown in Sec.~\ref{ss:orbits} again keeping $e=0$ and $\omega=90$\,deg for the systems listed in Table~\ref{t:LStest}, and in the remainder of this paper we adopt a circular orbit and the 68.5\%\ upper limit as error bar on the eccentricity. These new solutions are also shown in Tables~\ref{tab:sb1},~\ref{tab:sb2}, and~\ref{tab:sb2_2} and they are similar to the last ones.

\subsubsection{Systemic velocities}\label{sss:gamma}

For SB2 systems, we investigated the possibility of assigning an individual systemic velocity to each component of the system ($\gamma_1$ and $\gamma_2$). While a given binary obviously has only one true systemic velocity, several physical mechanisms -- including a line formed in a moving atmosphere or temperature structures at the surface of one or both stars -- may modify the apparent systemic velocity. In such a case, a better fit can be obtained by adopting a two-$\gamma$ orbital solution. To decide which systems would benefit from this approach, we computed one-$\gamma$ and two-$\gamma$ orbital solutions for all SB2 systems. We estimated the improvement brought by the second systemic velocity by considering the mean of an F-test and we adopt a 1\%\ significance threshold for the improvement to be significant. Four systems passed the adopted 1\%\ threshold: VFTS~197, 450, 527, and 555; their two-$\gamma$ solutions are given in Tables~\ref{tab:sb2} and ~\ref{tab:sb2_2}.

\subsection{Search for triple systems}\label{sss:triple}

To search for additional components in the confirmed SB1 and SB2 systems (see Section~\ref{ss:orbits}), we analyzed the residuals of the RV fittings following two procedures. We searched for trends in the residuals of the fit and for periodicity in the residuals by computing their Lomb-Scargle periodogram. 

In the first case, we were looking for evidence of third bodies with long orbital periods, i.e., periods longer than the baseline of our data. We did not find any positive signal as the 1 $\sigma$ uncertainty on the slope through the residuals was always larger than the slope value.   

In the second procedure, we found one system, VFTS 887 ($P_{\rm orb}\approx2.7$), with some probability of having a third component. However, the tentative orbital period for the third body candidate ( $P_{\rm orb} \sim 12.5$) seems too short for a stable configuration, unless the system is in resonance. Thus, we refrain from drawing any firm conclusions on the triple nature of this object.

To summarize, we found no clear-cut evidence of physically bound triple systems in our data set. This possibly reflects the limitation of the TMBM data in terms of $S/N$, sampling, and total baseline of the observational campaign. Indeed, the vast majority of known triple systems in the Milky Way have orbital periods of the outer components longer than 1 year \citep[e.g.,][]{Sana2014}. These periods are already challenging to characterize for binaries in TMBM (see Section~\ref{ss:orbits}), let alone for higher order multiples.

Alternatively, if further observations were to demonstrate that the frequency of triples among the massive star population in 30~Dor is lower than that in the Milky Way, this would constitute an interesting result, possibly revealing either the effect of metallicity in the formation process or that of a dense and violent environment.

\longtab{
\small
\begin{landscape}
\begin{longtable}{lccccccccc}
\caption{Orbital spectroscopic solutions for the SB1 binaries.\label{tab:sb1}}\\
\hline\hline
\vspace*{-2mm}\\
ID & Spec. Type & $\chi^2_\mathrm{red}$ & $P_\mathrm{orb}$~(days) & $T_0$~(HJD - 2400000) &  $e$ & $\omega$~(deg)  & $\gamma$~(\kms) & $K_1$~(\kms) & $a\sin i$~(R$_{\odot}$)\\
\vspace*{-2mm}\\
\hline
\endfirsthead
\caption{continued.}\\
\hline\hline
\vspace*{-2mm}\\
ID & Spec. Type & $\chi^2_\mathrm{red}$ & $P_\mathrm{orb}$~(days) & $T_0$~(HJD - 2400000) &  $e$ & $\omega$~(deg)  & $\gamma$~(\kms) & $K_1$~(\kms) & $a\sin i$~(R$_{\odot}$)\\
\vspace*{-2mm}\\
\hline
\endhead
\hline
\endfoot
064\tablefootmark{b}& O7.5\,II(f) & 4.3 &  902.9$\pm$3.9   & 56349.5$\pm$2.3   & 0.528$\pm$0.010 & 351.3$\pm$1.3 & 296.10$\pm$0.46 &  57.22$\pm$0.59 &  867.62$\pm$20.15  \\
073    & O9.5\,III        & 1.9 &    150.60$\pm$0.13       & 54939.7$\pm$4.2   & 0.203$\pm$0.031 & -12.5$\pm$9.0 & 280.67$\pm$0.54 &  26.99$\pm$0.72 &  78.68$\pm$1.60   \\
086    & O9\,III((n))     & 3.2 &     182.95$\pm$0.14      & 55026.2$\pm$1.3   & 0.514$\pm$0.030 & 344.9$\pm$4.2 & 286.31$\pm$0.26 &  12.90$\pm$0.54 &  40.04$\pm$0.82   \\
093    & O9.2\,III-IV     & 3.8 &      250.13$\pm$0.33     & 54881.4$\pm$4.4   & 0.203$\pm$0.027 & 224.8$\pm$8.5 & 264.07$\pm$0.19 &  10.89$\pm$0.25 &  52.74$\pm$0.93   \\
120    & O9.5 IV:         & 4.7 &   15.6546$\pm$0.0011     & 54802.88$\pm$0.18 & 0.280$\pm$0.015 &  47.8$\pm$4.0 & 269.78$\pm$0.31 &  24.73$\pm$0.43 &   7.347$\pm$0.093 \\
171    & O8\,II-III(f)    & 9.8 &     677.00$\pm$0.76      & 55535.2$\pm$1.8   & 0.555$\pm$0.011 &  77.4$\pm$2.1 & 269.13$\pm$0.13 &  11.77$\pm$0.14 & 131.01$\pm$0.57 \\
184    & O6.5 Vnz         & 2.1 &   32.128$\pm$0.022       & 54873.4$\pm$2.9   & 0.200$\pm$0.075 &  53.1$\pm$40.0& 273.06$\pm$0.67 &  12.05$\pm$1.09 &   7.50$\pm$0.48 \\
191    & O9.5 V           & 2.7 &      358.90$\pm$0.83     & 54971.5$\pm$19.3  &  0.22$\pm$0.07  & 236.6$\pm$18.9&  289.1$\pm$2.6  &  22.97$\pm$2.37 & 159.00$\pm$5.80\\
201    & O9.7 V + sec     & 1.3 &       15.327$\pm$0.002   & 54872.51$\pm$0.28 & 0.463$\pm$0.041 &   5.5$\pm$3.4 &  283.8$\pm$1.1  &  50.58$\pm$4.28 & 13.59$\pm$0.78 \\
225    & B0.7-1III-II     & 2.6 &    8.23371$\pm$0.00042   & 54850.25$\pm$0.34 & 0.021$\pm$0.008 & 319.7$\pm$13.4& 265.08$\pm$0.15 &  29.16$\pm$0.24 & 4.745$\pm$0.039 \\
231    & O9.7 IV:(n) + sec& 1.4 &    7.92911$\pm$0.00022   & 54837.36$\pm$0.10 & 0.406$\pm$0.037 & 200.5$\pm$5.4 &  292.5$\pm$1.2  &  50.23$\pm$2.55 & 7.20$\pm$0.22 \\
243    & O7 V(n)((f))     & 4.1 &   10.40284$\pm$0.00014   & 54859.82$\pm$0.43 & 0.016$\pm$0.008 &  30.4$\pm$15.2& 261.50$\pm$0.42 &  82.84$\pm$0.53 & 17.04$\pm$0.11 \\
243\tablefootmark{a}& O7 V(n)((f))& 4.2 &   10.40291$\pm$0.00012 & 54851.13$\pm$0.02 & 0.0       &  90.0         & 261.41$\pm$0.39 &  83.31$\pm$0.57 & 17.14$\pm$0.11 \\
256    & O7.5-8 V((n))z   & 1.6 &      246.00$\pm$0.45     & 55065.6$\pm$2.2   & 0.629$\pm$0.024 &  75.1$\pm$3.3 & 273.58$\pm$0.38 &  19.21$\pm$0.60 & 72.63$\pm$0.32 \\
277    & O9 V             & 2.7 &      240.42$\pm$0.13     & 54875.2$\pm$0.6   & 0.928$\pm$0.014 & 302.7$\pm$4.5 & 287.43$\pm$0.35 &  63.61$\pm$7.88 & 112.66$\pm$10.51\\      
314    & O9.7 IV:(n) + sec& 6.6 &    2.55091$\pm$0.00002   & 54889.93$\pm$0.04 & 0.166$\pm$0.012 & 248.9$\pm$4.7 & 273.89$\pm$0.84 & 110.80$\pm$1.26 & 5.511$\pm$0.050 \\
318    & O((n))p          & 1.0 &    14.0043$\pm$0.0029    & 54878.04$\pm$0.51 & 0.083$\pm$0.044 &  -7.5$\pm$15.7& 276.03$\pm$0.79 &  23.27$\pm$1.09 & 6.42$\pm$0.27 \\
318\tablefootmark{a}& O((n))p & 1.0 &14.0032$\pm$0.0023    & 54881.92$\pm$0.26 & 0.0             &  90.0         & 276.42$\pm$0.81 &  23.08$\pm$1.09 & 6.39$\pm$0.25\\
329    & O9.7 II-III(n)   & 5.9 &    7.04907$\pm$0.00044   & 54855.94$\pm$0.09 & 0.439$\pm$0.021 & 334.6$\pm$2.9 & 279.75$\pm$0.78 &  61.32$\pm$1.88 & 7.68$\pm$0.14 \\  
332\tablefootmark{b}& O9.2 II-III      & 6.7 &       1025.3$\pm$9.3     & 55064.0$\pm$6.8   & 0.813$\pm$0.057 & 185.4$\pm$1.4 & 268.65$\pm$0.78 &  48.17$\pm$11.37& 568.55$\pm$26.53 \\  
333\tablefootmark{b}& O8 II-III((f))   & 45.5&        980.1$\pm$1.5     & 55336.1$\pm$1.4   & 0.746$\pm$0.003 &115.55$\pm$0.49&268.246$\pm$0.079&  26.04$\pm$0.16 & 336.33$\pm$1.16 \\  
350    & O8 V             & 2.7 &    69.5695$\pm$0.0048    & 54904.26$\pm$0.20 & 0.351$\pm$0.008 &  93.4$\pm$1.5 & 266.94$\pm$0.36 &  60.23$\pm$0.42 & 77.57$\pm$0.29 \\  
386    & O9 IV(n)         & 9.5 &    20.4371$\pm$0.0026    & 54818.51$\pm$0.55 & 0.249$\pm$0.028 & 166.0$\pm$10.3& 258.83$\pm$0.31 &  14.27$\pm$0.55 & 5.58$\pm$0.17   \\ 
390    & O5-6 V(n)((fc))z & 1.8 &   21.90638$\pm$0.00088   & 54990.91$\pm$0.09 & 0.495$\pm$0.017 & 274.1$\pm$2.5 & 276.05$\pm$0.65 &  69.71$\pm$1.30 & 26.24$\pm$0.19 \\ 
404    & O3.5 V(n)((fc))  & 7.0 &    145.761$\pm$0.082     & 54993.20$\pm$0.86 & 0.718$\pm$0.016 &  99.6$\pm$2.6 & 264.06$\pm$0.35 &  27.28$\pm$0.84 & 54.72$\pm$0.49 \\ 
409    & O4 V((f))z       & 1.3 &     22.1909$\pm$0.0012   & 54876.60$\pm$0.16 & 0.294$\pm$0.012 & 105.2$\pm$3.5 & 273.21$\pm$0.41 &  43.18$\pm$0.68 & 18.11$\pm$0.21 \\ 
429    & O7.5-8 V         & 3.8 &     30.0439$\pm$0.0011   & 54874.33$\pm$0.05 & 0.559$\pm$0.005 &22.36$\pm$0.72 & 263.69$\pm$0.36 &  92.37$\pm$0.74 & 45.50$\pm$0.18 \\ 
440\tablefootmark{b}& O6-6.5 II(f)     & 6.3 &     1019.1$\pm$9.1       & 54904.12$\pm$36.6 & 0.277$\pm$0.026 & 160.7$\pm$14.2& 258.19$\pm$0.36 &  11.65$\pm$0.65 & 225.32$\pm$16.02 \\
441    & O9.5 V           & 0.9 &     6.86858$\pm$0.00022  & 54861.19$\pm$0.09 & 0.217$\pm$0.020 & 340.1$\pm$5.6 & 264.95$\pm$0.85 &  65.66$\pm$1.53 &  8.70$\pm$0.16 \\
475    & O9.7 III         & 3.0 &     4.05424$\pm$0.00012  & 54862.30$\pm$0.06 & 0.573$\pm$0.057 & -0.1$\pm$2.6  & 279.8$\pm$1.3   &  63.92$\pm$6.15 & 4.20$\pm$0.21  \\
479    & O4-5 V((fc))z    & 2.5 &   14.72542$\pm$0.00086   & 54872.85$\pm$0.13 & 0.310$\pm$0.016 & 189.4$\pm$2.2 & 269.89$\pm$0.65 &  72.95$\pm$1.09 & 20.19$\pm$0.19 \\
481    & O8.5 III         & 10.7&  141.8229$\pm$0.0091     & 54986.20$\pm$0.16 & 0.929$\pm$0.004 &37.99$\pm$0.74 & 273.89$\pm$0.24 & 128.30$\pm$5.55  & 133.14$\pm$1.93 \\ 
514    & O9.7 III         & 3.3 &    184.92$\pm$0.11       & 54842.3$\pm$1.5   & 0.411$\pm$0.019 &  41.2$\pm$2.6 & 259.73$\pm$0.35 &  22.93$\pm$0.38 & 76.42$\pm$0.51 \\ 
532    & O3 V(n)((f*))z+OB& 5.9 &     5.79608$\pm$0.00043  & 54861.49$\pm$0.07 & 0.460$\pm$0.032 & 159.0$\pm$3.1 & 267.37$\pm$0.52 &  34.41$\pm$1.19 & 3.501$\pm$0.051  \\ 
603    & O4 III(fc)       & 6.9 &    1.756777$\pm$0.000024 & 54865.06$\pm$0.12 & 0.107$\pm$0.032 & 139.2$\pm$27.1& 278.21$\pm$0.28 &  11.40$\pm$0.28 & 0.394$\pm$0.008  \\ 
613    & O8.5 Vz          & 1.7 &     69.158$\pm$0.039     & 54804.3$\pm$4.8   & 0.351$\pm$0.061 &293.9$\pm$25.7 &   275.9$\pm$2.1 &  31.97$\pm$4.86 & 40.93$\pm$4.96  \\ 
619    & O7-8 V(n)        & 1.2 &     14.5043$\pm$0.0026   & 54869.1$\pm$1.7   & 0.085$\pm$0.040 &161.6$\pm$45.0 &   274.0$\pm$1.1 &  36.84$\pm$1.36 & 10.53$\pm$0.34 \\
619\tablefootmark{a}& O7-8 V(n)        & 1.3 &    14.5025$\pm$0.0019    & 54866.39$\pm$0.19 & 0.0           &  90.0           &   273.7$\pm$1.1 &  36.84$\pm$1.42 & 10.56$\pm$0.41 \\
631    & O9.7 III(n)      & 5.5 &    5.37487$\pm$0.00018   & 54870.37$\pm$0.07 & 0.007$\pm$0.005 &  37.0$\pm$6.6 & 266.06$\pm$0.80 &  48.61$\pm$1.10 & 5.17$\pm$0.12  \\
631\tablefootmark{a}& O9.7 III(n)      & 5.5 &     5.37485$\pm$0.00015  & 54871.16$\pm$0.04 & 0.0           &  90.0           & 266.0$\pm$0.65  &  48.71$\pm$0.94 & 5.18$\pm$0.10\\
645    & O9.5 V((n))      & 1.5 &   12.5458$\pm$0.0016     & 54870.75$\pm$0.41 & 0.235$\pm$0.070 &  -2.8$\pm$10.7& 271.4$\pm$1.2   &  31.60$\pm$2.46 & 7.62$\pm$0.43  \\ 
657    & O7-8 II(f)       & 1.4 &   63.4658$\pm$0.0078     & 54858.73$\pm$0.42 & 0.480$\pm$0.021 & 312.1$\pm$3.9 & 276.54$\pm$0.80 &  44.61$\pm$1.24 & 49.11$\pm$0.68  \\ 
702    & O8 V(n)          & 3.4 &    1.981437$\pm$0.000022 & 54868.59$\pm$0.04 & 0.008$\pm$0.006 & 366.4$\pm$7.9 & 267.3$\pm$2.1   & 105.64$\pm$3.14 & 4.14$\pm$0.12 \\
702\tablefootmark{a}& O8 V(n) & 3.4 &  1.98144$\pm$0.00002 & 54869.05$\pm$0.01 & 0.0             &  90.0         & 267.1$\pm$1.9   & 105.92$\pm$3.04 & 4.15$\pm$0.12 \\
733    & O9.7p            & 10.4&    5.922121$\pm$0.000068 & 54869.75$\pm$0.16 & 0.002$\pm$0.001 & 196.4$\pm$10.2& 251.52$\pm$0.22 &  64.91$\pm$0.25 & 7.601$\pm$0.072 \\
733\tablefootmark{a}& O9.7p            & 10.4&    5.92212$\pm$0.00005   & 54868.00$\pm$0.02 & 0.0           &  90.0           & 251.49$\pm$0.22 &  65.05$\pm$0.32 & 7.617$\pm$0.037 \\
736    & O9.5 V           & 2.6 &   68.800$\pm$0.021       & 54922.1$\pm$2.6   & 0.086$\pm$0.020 &255.3$\pm$13.7 & 267.25$\pm$0.43 &  24.58$\pm$0.49 & 33.31$\pm$0.71 \\ 
743    & O9.5 V((n))      & 2.9 &   14.94732$\pm$0.00089   & 54866.91$\pm$0.31 & 0.012$\pm$0.008 & 42.4$\pm$7.9  & 257.79$\pm$0.37 &  23.43$\pm$0.56 & 6.93$\pm$0.17 \\
743\tablefootmark{a}& O9.5 V((n))      & 2.9 &   14.9475$\pm$0.0008     & 54868.87$\pm$0.07 & 0.0           &  90.0           & 257.84$\pm$0.34 &  23.44$\pm$0.50 & 6.93$\pm$0.15 \\
750\tablefootmark{b}& O9.5 IV          & 1.6 &  416.7$\pm$8.4           & 55246.8$\pm$9.6   & 0.779$\pm$0.039 &  44.2$\pm$6.1 & 255.9$\pm$1.2   &  29.49$\pm$1.55 & 152.36$\pm$19.69 \\ 
769    & O9.7 II-III      & 1.5 &    2.365644$\pm$0.000016 & 54868.34$\pm$0.03 & 0.007$\pm$0.005 & 333.9$\pm$3.5& 264.70$\pm$0.54 &  40.76$\pm$0.81 & 1.907$\pm$0.039 \\
769\tablefootmark{a}& O9.7 II-III      & 1.5 &    2.36564$\pm$0.00002   & 54869.11$\pm$0.01 & 0.0           &  90.0        & 264.67$\pm$0.50 &  40.81$\pm$0.74 & 1.956$\pm$0.013\\
779    & B1 II-Ib         & 1.7 &       59.945$\pm$0.025   & 54903.3$\pm$2.4   & 0.046$\pm$0.011 &  84.3$\pm$12.2& 260.16$\pm$0.22 &  30.81$\pm$0.21 & 36.48$\pm$0.27  \\
802    & O7.5 Vz          & 1.8 &  181.883$\pm$0.041       & 54942.98$\pm$0.25 & 0.602$\pm$0.007 & 47.18$\pm$0.78& 232.77$\pm$0.23 &  58.46$\pm$0.49 & 167.87$\pm$2.52  \\
810    & O9.7 V + B1: V:  & 9.9 &   15.6886$\pm$0.0006     & 54892.32$\pm$0.06 & 0.678$\pm$0.008 & 359.0$\pm$1.3 & 264.57$\pm$0.82 & 100.74$\pm$1.92 & 22.97$\pm$0.67 \\
812    & O4-5 V((fc))     & 2.1 &   17.28443$\pm$0.00035   & 54856.95$\pm$0.04 & 0.624$\pm$0.009 & 339.5$\pm$1.3 & 276.68$\pm$0.30 &  42.66$\pm$0.56 & 11.39$\pm$ 0.25 \\
827    & B1.5 Ib          & 2.0 &   43.221$\pm$0.017       & 54870.68$\pm$0.51 & 0.244$\pm$0.011 & 97.6$\pm$3.4  & 250.57$\pm$0.24 &  25.31$\pm$0.34 & 20.97$\pm$0.35 \\
829    & B1.5-2 II        & 2.6 &  202.93$\pm$0.88         & 55023.5$\pm$9.5   & 0.273$\pm$0.043 & 145.8$\pm$11.4& 243.55$\pm$0.44 &  12.58$\pm$0.71 & 48.57$\pm$3.33 \\ 
887    & O9.5 II-IIIn     & 17.0&    2.672807$\pm$0.000035 & 54870.39$\pm$0.03 & 0.056$\pm$0.019 &  53.5$\pm$4.8 & 251.41$\pm$0.42 &  30.07$\pm$0.66 & 1.586$\pm$0.037 \\
\end{longtable}
\flushleft
\tablefoottext{a}{Solutions obtained by keeping $e = 0$ and $\omega = 90^o$, see Section~\ref{sss:ecc}.}\\
\tablefoottext{b}{Systems with $P_{\rm orb} > 1$\,yr which need confirmation due to intrinsic limitation in our time series.}
\end{landscape}
}

\longtab{
\small
\begin{landscape}
\begin{longtable}{lllrrrrrrr}
\caption{Fitted parameters of the orbital spectroscopic solutions for the SB2 binaries.\label{tab:sb2}}\\
\hline\hline
\vspace*{-2mm}\\
ID & $\chi^2_\mathrm{red}$ & $P_\mathrm{orb}$~(days) & $T_0$~(HJD - 2400000) &
$e$ & $\omega$~(deg) & $K_1$~(\kms) &
$K_2$~(\kms) & $\gamma$~(\kms) & $\gamma_2$~(\kms) \\
\vspace*{-2mm}\\
\hline
\endfirsthead
\caption{continued.}\\
\hline\hline
\vspace*{-2mm}\\
ID & $\chi^2_\mathrm{red}$ & $P_\mathrm{orb}$~(days) & $T_0$~(HJD - 2400000) &
$e$ & $\omega$~(deg) & $K_1$~(\kms) &
$K_2$~(\kms) & $\gamma$~(\kms) & $\gamma_2$~(\kms) \\ 
\vspace*{-2mm}\\
\hline
\endhead
\hline
\endfoot
042    & 3.4  &  29.3110$\pm$0.0034     & 54899.51$\pm$0.29     & 0.188$\pm$0.011 & 134.71$\pm$4.45 &  52.15$\pm$0.87 &  62.45$\pm$0.97 &282.78$\pm$0.41 & \\
047    & 2.4  &   5.93163$\pm$0.00008   & 54893.13$\pm$0.03     & 0.047$\pm$0.007 & 235.15$\pm$2.76 & 171.16$\pm$2.63 & 174.85$\pm$2.11 &284.56$\pm$1.27 & \\ 
055    & 2.4  &   6.445026$\pm$0.000043 & 54888.99$\pm$0.04     & 0.099$\pm$0.005 &  53.21$\pm$2.00 & 147.98$\pm$0.97 & 150.86$\pm$0.92 &268.45$\pm$0.50 & \\ 
061    & 8.8  &   2.333440$\pm$0.000007 & 54892.84$\pm$0.02     & 0.028$\pm$0.004 & 162.71$\pm$2.12 & 167.64$\pm$1.73 & 278.98$\pm$0.90 &280.37$\pm$0.69 & \\ 
063    & 8.3  &  85.6950$\pm$0.0059     & 54928.23$\pm$0.16     & 0.608$\pm$0.012 & 357.66$\pm$1.25 &  83.40$\pm$2.72 & 150.93$\pm$4.47 &268.08$\pm$0.56 & \\
066    & 21.6 &   1.141160$\pm$0.000005 & 54893.0883$\pm$0.0009 & 0.008$\pm$0.005 & 196.35$\pm$2.79 &  58.89$\pm$1.06 & 104.94$\pm$0.89 &268.86$\pm$0.57 & \\
066\tablefootmark{a}& 20.3 &   1.141161$\pm$0.000004 & 54893.3200$\pm$0.0048 & 0.0             &  90.0           &  58.87$\pm$0.99 & 104.90$\pm$0.86 &268.72$\pm$0.55 & \\
094    & 25.2 &   2.256394$\pm$0.000006 & 54892.36$\pm$0.02     & 0.084$\pm$0.006 & 160.93$\pm$3.12 & 126.02$\pm$1.59 & 130.02$\pm$0.78 &263.81$\pm$0.64 & \\
114    & 13.5 &    27.7767$\pm$0.0012   & 54909.19$\pm$0.08     & 0.496$\pm$0.013 & 161.11$\pm$1.68 &  92.57$\pm$3.41 & 112.35$\pm$2.06 &279.85$\pm$0.81 & \\
116    & 3.3  &      23.9204$\pm$0.0016 & 54890.71$\pm$0.19     & 0.239$\pm$0.014 & 343.29$\pm$3.51 &  79.37$\pm$3.39 & 109.12$\pm$1.55 &274.69$\pm$0.89 & \\
140    & 7.0  &   1.611655$\pm$0.000007 & 54893.11$\pm$0.02     & 0.002$\pm$0.002 &  31.02$\pm$4.06 &  97.96$\pm$1.93 &  98.98$\pm$1.99 &271.49$\pm$0.89 & \\
140\tablefootmark{a}& 6.6  &   1.611654$\pm$0.000006 & 54893.3747$\pm$0.0046 & 0.0             &  90.0           &  97.81$\pm$1.66 &  98.81$\pm$1.79 &271.48$\pm$0.82 & \\
174    & 3.9  &   4.755980$\pm$0.000020 & 54894.38$\pm$0.02     & 0.265$\pm$0.004 &  28.01$\pm$1.44 & 162.23$\pm$0.91 & 238.34$\pm$1.54 &286.26$\pm$0.53 & \\
176    & 21.3 &   1.777593$\pm$0.000003 & 54867.1452$\pm$0.0055 & 0.032$\pm$0.004 &  18.69$\pm$1.22 & 244.60$\pm$0.98 & 391.38$\pm$2.29 &269.34$\pm$0.68 & \\
187    & 12.9 &   3.542881$\pm$0.000025 & 54890.67$\pm$0.04     & 0.183$\pm$0.013 &  96.35$\pm$3.41 & 136.46$\pm$1.82 & 178.40$\pm$3.19 &282.51$\pm$0.90 & \\
197    & 31.9 &      69.7306$\pm$0.0039 & 54898.21$\pm$0.32     & 0.109$\pm$0.004 &  40.69$\pm$1.56 &  27.42$\pm$0.58 &  79.16$\pm$0.30 &282.40$\pm$0.20 &\\
197\tablefootmark{b}& 27.1 &    69.7315$\pm$0.0042   & 54898.13$\pm$0.11     & 0.106$\pm$0.005 &  40.34$\pm$0.46 &  27.81$\pm$0.67 &  79.38$\pm$0.41 &283.76$\pm$0.29 & 277.49$\pm$0.56 \\
217    & 16.7 &   1.855341$\pm$0.000002 & 54893.77$\pm$0.01     & 0.001$\pm$0.001 & 326.15$\pm$1.86 & 222.83$\pm$0.53 & 268.31$\pm$0.62 &281.67$\pm$0.32 & \\
217\tablefootmark{a} & 16.7 & 1.855341$\pm$0.000002 & 54894.408$\pm$0.001 & 0.0   &  90.0      &   222.74$\pm$0.48 &  268.22$\pm$0.58  & 281.67$\pm$0.30 \\
327    & 8.9  &   2.955204$\pm$0.000022 & 54867.39$\pm$0.02     & 0.225$\pm$0.010 & 106.60$\pm$2.73 &  86.31$\pm$1.79 & 145.09$\pm$1.68 &267.70$\pm$0.85 & \\
352    & 3.7  &   1.124143$\pm$0.000002 & 54894.3736$\pm$0.0069 & 0.012$\pm$0.005 & 178.67$\pm$2.25 & 318.76$\pm$1.74 & 314.77$\pm$1.87 & 269.84$\pm$0.84 & \\
352\tablefootmark{a}& 6.9  &   1.124142$\pm$0.000001 & 54894.0976$\pm$0.0015 & 0.0             &  90.0           & 318.15$\pm$1.65 & 314.09$\pm$1.57 & 273.35$\pm$0.81 & \\
450    & 106.7&   6.892352$\pm$0.000015 & 54762.1710$\pm$0.0057 & 0.067$\pm$0.002 &  21.84$\pm$0.29 & 190.74$\pm$1.79 & 206.05$\pm$0.20 &249.60$\pm$0.19 & \\
450\tablefootmark{b}& 85.9 &   6.892336$\pm$0.000027 & 54762.1605$\pm$0.0027 & 0.070$\pm$0.002 &  20.97$\pm$0.63 & 206.01$\pm$0.33 & 201.11$\pm$2.86 &248.78$\pm$0.29 & 289.28$\pm$2.54 \\
487    & 11.0 &   4.121166$\pm$0.000017 & 54894.74$\pm$0.01     & 0.089$\pm$0.003 & 275.97$\pm$0.75 & 166.13$\pm$0.97 & 191.40$\pm$0.95 &266.14$\pm$0.57 & \\
500    & 9.2  &   2.875370$\pm$0.000004 & 54893.54$\pm$0.06     & 0.001$\pm$0.001 & 167.05$\pm$7.12 & 237.47$\pm$0.65 & 240.15$\pm$0.80 &270.62$\pm$0.40 & \\
500\tablefootmark{a}& 8.9  &   2.875371$\pm$0.000004 & 54892.9234$\pm$0.0016 & 0.0             &  90.0           & 237.52$\pm$0.72 & 240.23$\pm$0.84 &270.64$\pm$0.42 & \\
508    & 14.7 & 128.586$\pm$0.025       & 54941.00$\pm$0.59     & 0.396$\pm$0.012 & 254.67$\pm$0.72 &  80.14$\pm$1.34 & 115.18$\pm$2.07 &254.82$\pm$2.03 & \\
527    & 244.6&   153.9572$\pm$0.0034   & 54808.52$\pm$0.06     & 0.462$\pm$0.001 & 126.95$\pm$0.23 &  95.94$\pm$0.19 & 121.59$\pm$0.25 &262.37$\pm$0.086 & \\
527\tablefootmark{b}& 145.2&     153.9416$\pm$0.0092 & 54809.5529$\pm$0.0086 & 0.422$\pm$0.005 & 130.63$\pm$0.34 &  94.37$\pm$0.47 & 119.73$\pm$0.56 &256.88$\pm$0.33 & 270.06$\pm$0.40 \\
538    & 17.4 &   4.159758$\pm$0.000023 & 54897.6500$\pm$0.0003 & 0.001$\pm$0.001 &  67.81$\pm$0.91 & 145.62$\pm$3.07 & 226.04$\pm$1.25 &264.23$\pm$0.82 & \\
538\tablefootmark{a}& 16.6 &   4.159757$\pm$0.000018 & 54897.91$\pm$0.01     & 0.0             &  90.0           & 145.46$\pm$2.67 & 225.88$\pm$1.06 &264.30$\pm$0.75 &  \\
543    & 6.4  &   1.383987$\pm$0.000003 & 54893.89$\pm$0.01     & 0.016$\pm$0.007 & 127.70$\pm$3.91 & 211.87$\pm$2.35 & 189.89$\pm$2.15 &266.37$\pm$1.17 & \\
543\tablefootmark{a}& 6.2  &   1.383989$\pm$0.000003 & 54894.4364$\pm$0.0029 & 0.0             & 90.0            & 189.87$\pm$1.96 & 211.84$\pm$2.32 &266.35$\pm$1.12 & \\
555    & 6.1  &    66.1008$\pm$0.0085   & 54902.53$\pm$0.19     & 0.802$\pm$0.016 & 102.01$\pm$2.04 &  59.26$\pm$2.91 &  86.65$\pm$4.12 &269.10$\pm$0.48 & \\
555\tablefootmark{b}& 5.5  &      66.0995$\pm$0.0015 & 54902.5436$\pm$0.0061 & 0.829$\pm$0.017 & 103.36$\pm$1.89 &  65.07$\pm$1.55 &  93.96$\pm$5.90 &263.45$\pm$1.44 & 270.27$\pm$0.63 \\
563    & 3.4  &   1.217341$\pm$0.000007 & 54894.08$\pm$0.04     & 0.015$\pm$0.010 & 230.78$\pm$8.61 & 162.09$\pm$3.56 & 202.70$\pm$4.03 &268.74$\pm$1.86 & \\
563\tablefootmark{a}&  3.2 &   1.217342$\pm$0.000006 & 54893.6004$\pm$0.0068 &           0.0   &  90.0           & 160.35$\pm$3.21 & 200.47$\pm$3.65 &268.65$\pm$1.80 & \\
642    & 2.8  &   1.726822$\pm$0.000011 & 54894.74$\pm$0.01     & 0.008$\pm$0.005 & 268.865$\pm$4.53& 141.01$\pm$1.25 & 219.94$\pm$1.69 &271.82$\pm$0.77 & \\
642\tablefootmark{a}& 2.6  &   1.726824$\pm$0.000008 & 54893.8800$\pm$0.0074 & 0.0             &  90.0           & 141.06$\pm$1.22 & 220.37$\pm$1.74 &271.89$\pm$0.79 & \\
652    & 39.2 &   8.58909$\pm$0.00015   & 54894.4914$\pm$0.0074 & 0.000$\pm$0.000 &  48.04$\pm$1.18 &  64.06$\pm$3.43 & 202.33$\pm$0.42 &256.42$\pm$0.31 & \\
661    & 20.6 &   1.266430$\pm$0.000004 & 54892.82$\pm$0.03     & 0.024$\pm$0.005 &  54.33$\pm$10.85& 265.19$\pm$0.69 & 374.27$\pm$1.57 &286.39$\pm$0.92 & \\
771    & 3.9  &      29.8688$\pm$0.0015 & 54867.49$\pm$0.11     & 0.509$\pm$0.013 & 168.54$\pm$2.40 &  66.10$\pm$1.93 &  67.70$\pm$1.47 &270.70$\pm$0.70 & \\
806    & 15.7 &    2.584883$\pm$0.000004& 54889.0752$\pm$0.0068 & 0.005$\pm$0.002 & 233.38$\pm$1.01 & 159.37$\pm$0.44 & 194.19$\pm$0.54 &239.53$\pm$0.23 & \\
\end{longtable}
\flushleft
\tablefoottext{a}{Solutions obtained by keeping $e = 0$ and $\omega = 90\degr$, see Section~\ref{sss:ecc}.}\\
\tablefoottext{b}{Solutions obtained by using two systemic velocities, see Section~\ref{sss:gamma}.}
\end{landscape}
}

\longtab{
\small
\begin{landscape}
\begin{longtable}{lcccccccccc}
\caption{Derived parameters of the orbital spectroscopic solutions for the SB2 binaries.\label{tab:sb2_2}}\\
\hline\hline
\vspace*{-2mm}\\
ID & Spec. Type & $m_1\sin^3 i$~(M$_{\odot}$) & $m_2\sin^3 i$~(M$_{\odot}$)\\
\vspace*{-2mm}\\
\hline
\endfirsthead
\caption{continued.}\\
\hline\hline
\vspace*{-2mm}\\
ID & Spec. Type & $m_1\sin^3 i$~(M$_{\odot}$) & $m_2\sin^3 i$~(M$_{\odot}$)\\
\vspace*{-2mm}\\
\hline
\endhead
\hline
\endfoot
042    & O9.5 III((n))                           & 1.970$\pm$0.066 &  2.359$\pm$0.076  \\
047    & O9 V + O9.5 V                           & 12.82$\pm$0.35 & 12.549$\pm$0.350 \\ 
055    & O8.5 V + O9.5 IV                        & 8.86$\pm$0.13 &  8.693$\pm$0.124\\ 
061    & ON8.5 III: + O9.7: V:                   & 13.44$\pm$0.14 &  8.073$\pm$0.149 \\ 
063    & O5 III(n)(fc) + sec                     & 36.82$\pm$2.04 & 20.345$\pm$1.057 \\
066    & O9.5 III(n)                             & 0.333$\pm$0.008 &  0.187$\pm$0.006 \\
066\tablefootmark{a}& O9.5 III(n)                             & 0.332$\pm$0.007 &  0.186$\pm$0.005 \\
094    & O3.5 Inf*p + sec?                       & 1.969$\pm$0.033 &  1.969$\pm$0.034 \\
114    & O8.5 IV + sec                           & 8.91$\pm$0.65 &  7.34$\pm$0.72 \\
116    & O9.7: V: + B0: V:                       & 8.81$\pm$0.42 &  6.41$\pm$0.55 \\
140    & O8.5 Vz                                 & 0.638$\pm$0.027 &  0.632$\pm$0.025 \\
140\tablefootmark{a}& O8.5 Vz                                 & 0.638$\pm$0.026 &  0.632$\pm$0.025 \\
174    & O8 V + B0: V:                           & 16.98$\pm$0.25 & 11.55$\pm$0.15 \\
176    & O6 V:((f)) + O9.5: V:                   & 29.08$\pm$0.37 & 18.18$\pm$0.18 \\
187    & O9 IV: + B0: V:                         & 6.01$\pm$0.19 &  4.67$\pm$0.12 \\
197    & O9 III                                  & 6.374$\pm$0.089 &  2.205$\pm$0.071 \\
197\tablefootmark{b}& O9 III                                  & 6.47$\pm$0.12 &  2.267$\pm$0.087 \\
217    & O4 V((fc)): + O5 V((fc)):               & 12.429$\pm$0.062 & 10.322$\pm$0.049 \\
217\tablefootmark{a} & O4 V((fc)): + O5 V((fc)): & 12.425$\pm$0.059 & 10.318$\pm$0.047 \\
327    & O8.5 V(n) + sec                         & 2.192$\pm$0.065 &  1.303$\pm$0.053 \\
352    & O4.5 V(n)((fc)):z: + O5.5 V(n)((fc)):z: & 14.71$\pm$0.15 & 14.90$\pm$0.16 \\
352\tablefootmark{a}& O4.5 V(n)((fc)):z: + O5.5 V(n)((fc)):z: & 14.62$\pm$0.16 & 14.81$\pm$0.16 \\
450    & O9.7 III: + O7::                        & 23.01$\pm$3.12 & 21.30$\pm$5.89 \\
450\tablefootmark{b}& O9.7 III: + O7::                        & 23.63$\pm$0.67 & 24.20$\pm$0.37\\
487    & O6.5: IV:((f)): + O6.5: IV:((f)):       & 10.32$\pm$0.27 &  8.92$\pm$0.17 \\
500    & O6.5 IV((fc)) + O6.5 V((fc))            & 16.33$\pm$0.12 & 16.15$\pm$0.11 \\
500\tablefootmark{a}& O6.5 IV((fc)) + O6.5 V((fc))            & 16.33$\pm$0.12 & 16.15$\pm$0.11\\
508    & O9.5 V                                  & 45.31$\pm$3.59 & 31.48$\pm$2.37  \\
527    & O6.5 Iafc + O6 Iaf                      & 63.97$\pm$0.32 & 50.48$\pm$0.24 \\
527\tablefootmark{b}& O6.5 Iafc + O6 Iaf                      & 65.21$\pm$0.80 & 51.40$\pm$0.65 \\
538    & ON9 Ia: + O7.5: I:(f):                  & 13.43$\pm$0.26 &  8.65$\pm$0.32 \\
538\tablefootmark{a}& ON9 Ia: + O7.5: I:(f):                  & 13.43$\pm$0.18 &  8.65$\pm$0.19 \\
543    & O9 IV + O9.7: V                         & 4.90$\pm$0.11 &  4.39$\pm$0.11 \\
543\tablefootmark{a}& O9 IV + O9.7: V                         & 4.91$\pm$0.12 &  4.400$\pm$0.096 \\
555    & O9.5 Vz                                 & 2.69$\pm$0.36 &  1.85$\pm$0.20 \\
555\tablefootmark{b}& O9.5 Vz                                 & 2.85$\pm$0.13 &  1.972$\pm$0.077 \\
563    & O9.7 III: + B0: V:                      & 3.34$\pm$0.18 &  2.67$\pm$0.18 \\
563\tablefootmark{a}& O9.7 III: + B0: V:                      & 3.30$\pm$0.14 &  2.64$\pm$0.12 \\
642    & O5 Vz: + O8 Vz:                         & 5.138$\pm$0.095 &  3.293$\pm$0.061 \\
642\tablefootmark{a}& O5 Vz: + O8 Vz:                         & 5.14$\pm$0.11 &  3.294$\pm$0.062 \\
652    & B2 Ip + O9 III:                         & 12.72$\pm$0.23 &  3.983$\pm$0.214 \\
661    & O6.5 V(n) + O9.7: V:                    & 20.25$\pm$0.25 & 14.30$\pm$0.15 \\
771    & O9.7 III:(n)                            & 2.39$\pm$0.12 &  2.33$\pm$0.13 \\
806    & O5.5 V((fc)):z + O7 Vz:                 & 6.505$\pm$0.041 &  5.338$\pm$0.032 \\
\end{longtable}
\flushleft
\tablefoottext{a}{Solutions obtained by keeping $e = 0$ and $\omega = 90^o$, see Section~\ref{sss:ecc}.}\\
\tablefoottext{b}{Solutions obtained by using two systemic velocities, see Section~\ref{sss:gamma}.}
\end{landscape}
}

\section{Discussion}\label{s:discuss}
\subsection{Orbital distributions}\label{ss:sb_discuss}
This section presents the observed distributions of orbital parameters for the binaries in our sample. It also describes a first attempt to correct for observational biases using the VFTS binary detection probabilities computed in \citet{SdKdM13}.

\begin{figure}
\centering
\includegraphics[width=\hsize]{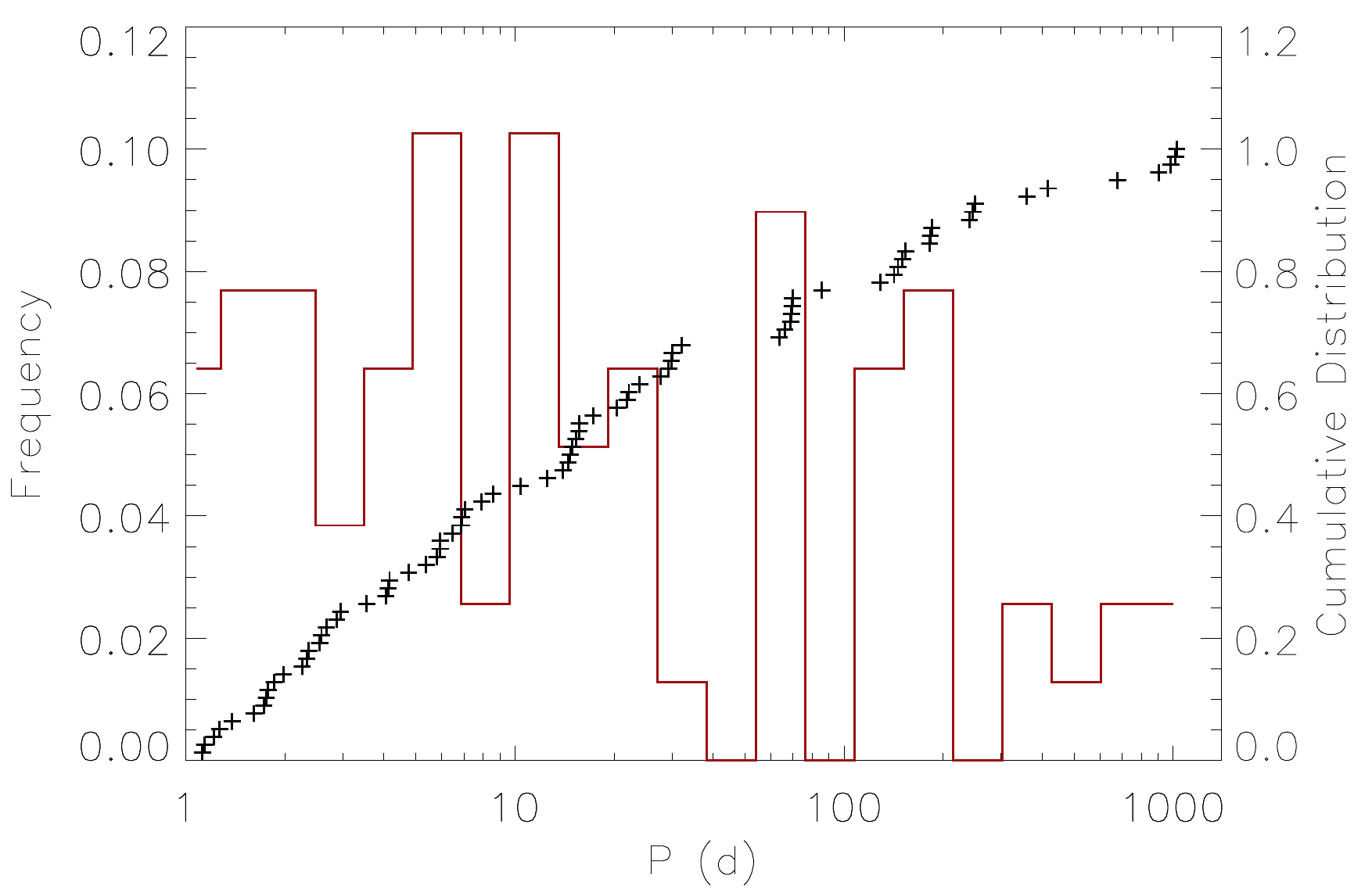}
\includegraphics[angle=-90,width=\hsize]{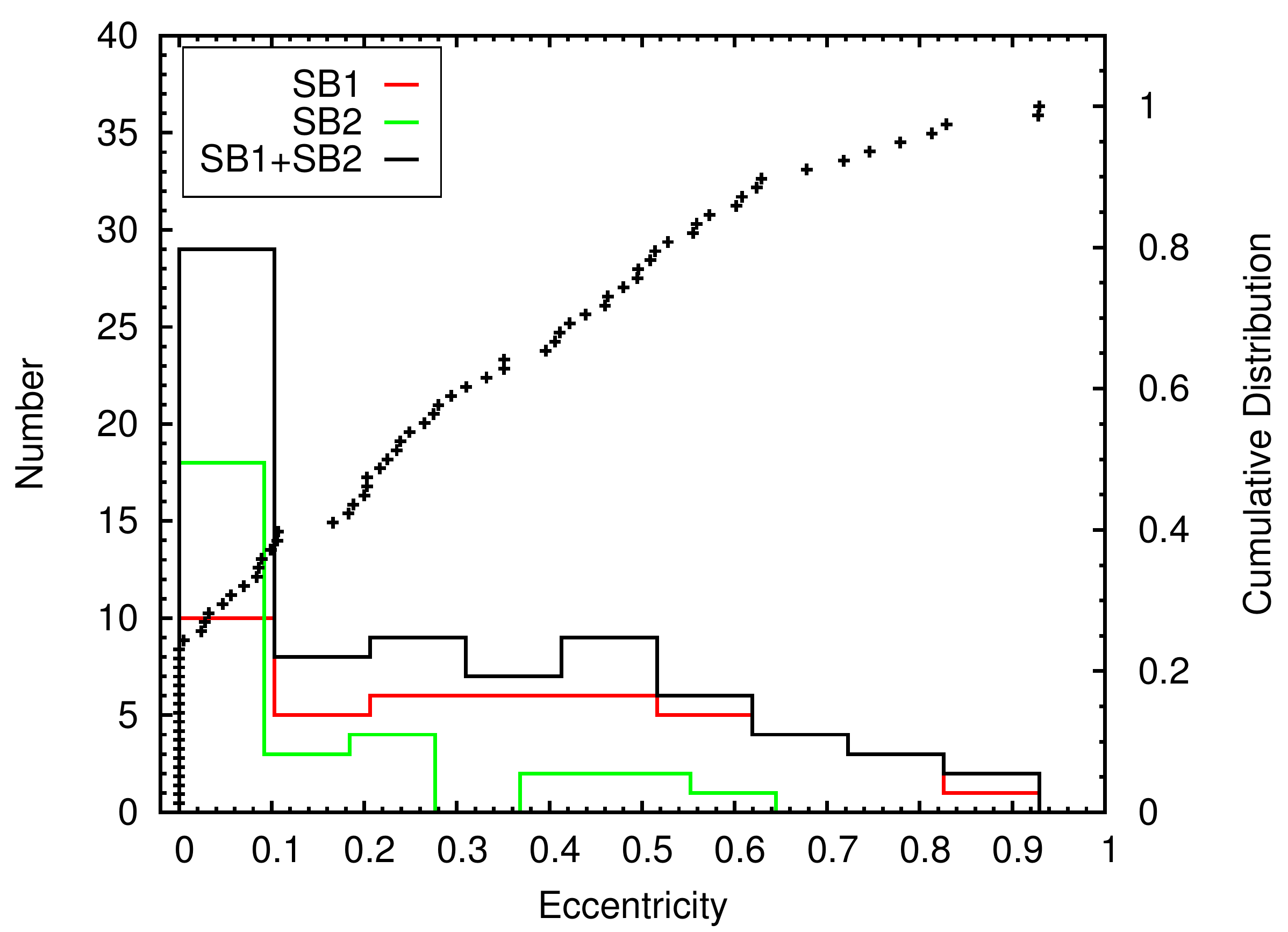}
\includegraphics[angle=-90,width=\hsize]{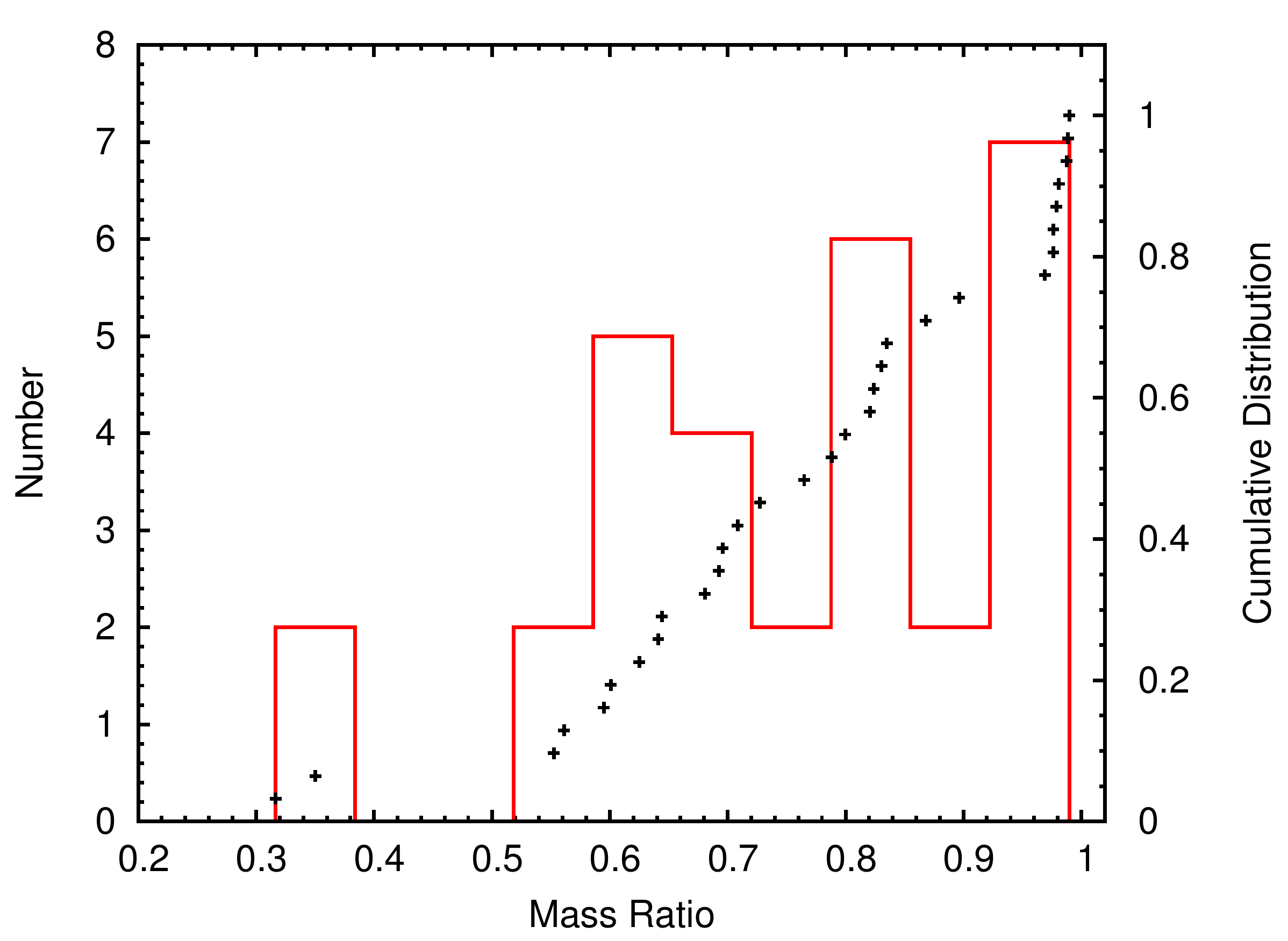}
\caption{Observed histograms (left-hand scale) and cumulative distributions (right-hand scale) of orbital periods ($P_\mathrm{orb}$), eccentricities ($e$), and mass ratios ($q$) in the TMBM sample. Periods and eccentricity distributions include both SB1 and SB2 systems, while mass ratios are restricted to SB2 systems.}
\label{f:obs_dist}
\end{figure}

\subsubsection{Orbital periods}\label{ss:p_discuss}

The observed orbital period distribution of the TMBM O-type binaries covers the whole range from about 1 to 1000~days (Fig.~\ref{f:obs_dist}). About 40\%\ of our systems have an orbital period of less than a week, 70\%\, less than a month, and 90\%, less than a year. Some structures are visible throughout the distribution, with the most important ones at about 15 and 60~days. The significance of these wiggles is hard to assess. A Kuiper test indeed indicates that a smooth power-law distribution with an index of $-0.3$ to $-0.5$, depending on the minimum period of the fit (see discussion in Sect.~\ref{ss:bias}), adequately reproduces the data  (Kuiper's probability: $P_\mathrm{Kuiper} > 0.18$).

\begin{figure}
\centering
\includegraphics[angle=-90,width=\hsize]{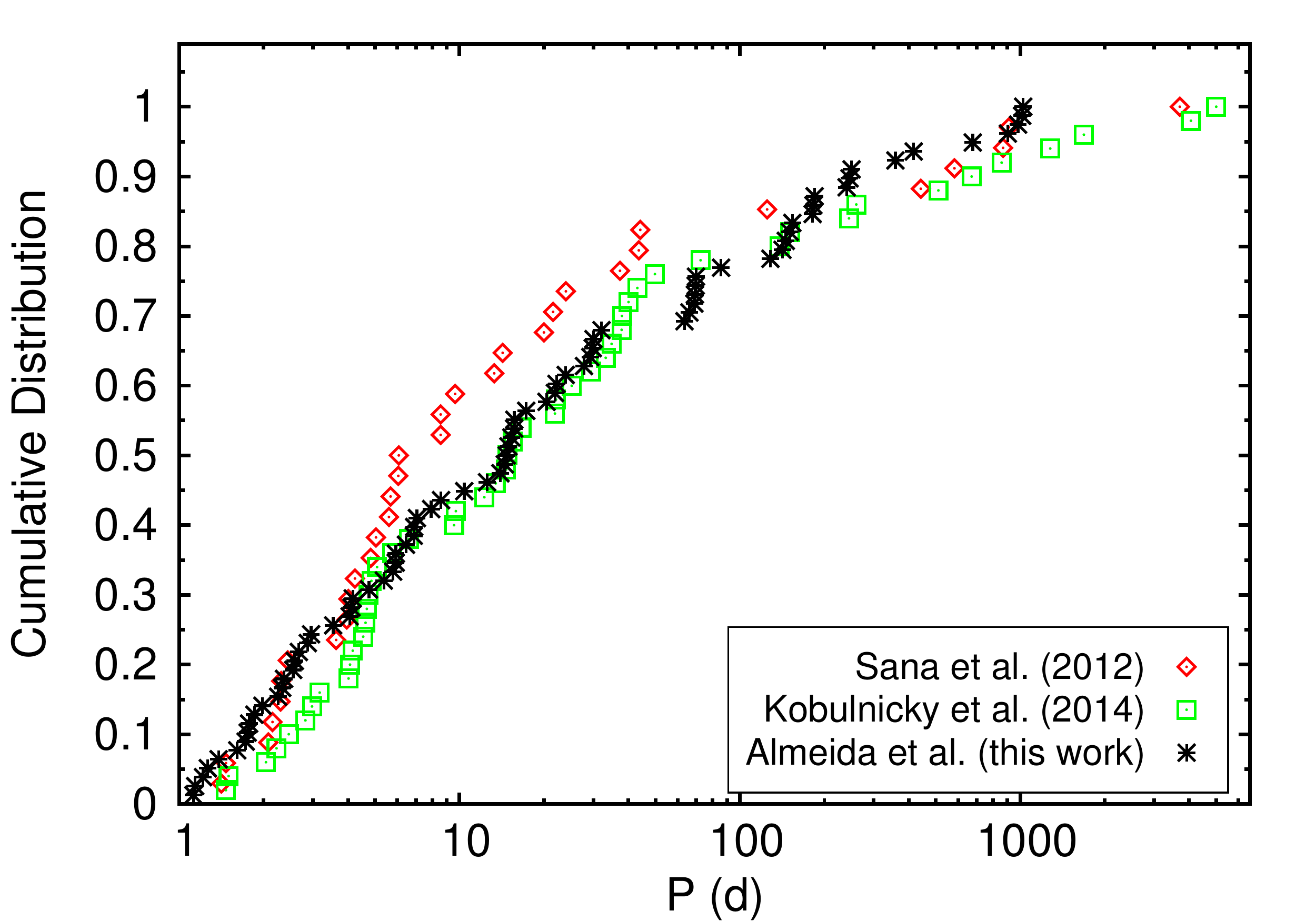}
\caption{Cumulative distributions of measured orbital periods in different samples. Red diamonds, green squares, and black asterisks are orbital periods from \citet[][]{SdMdK12}, \citet[][]{KK+2014}, and this work, respectively.}
\label{f:obs_comp}
\end{figure}

\begin{figure}
\centering
\includegraphics[width=\hsize]{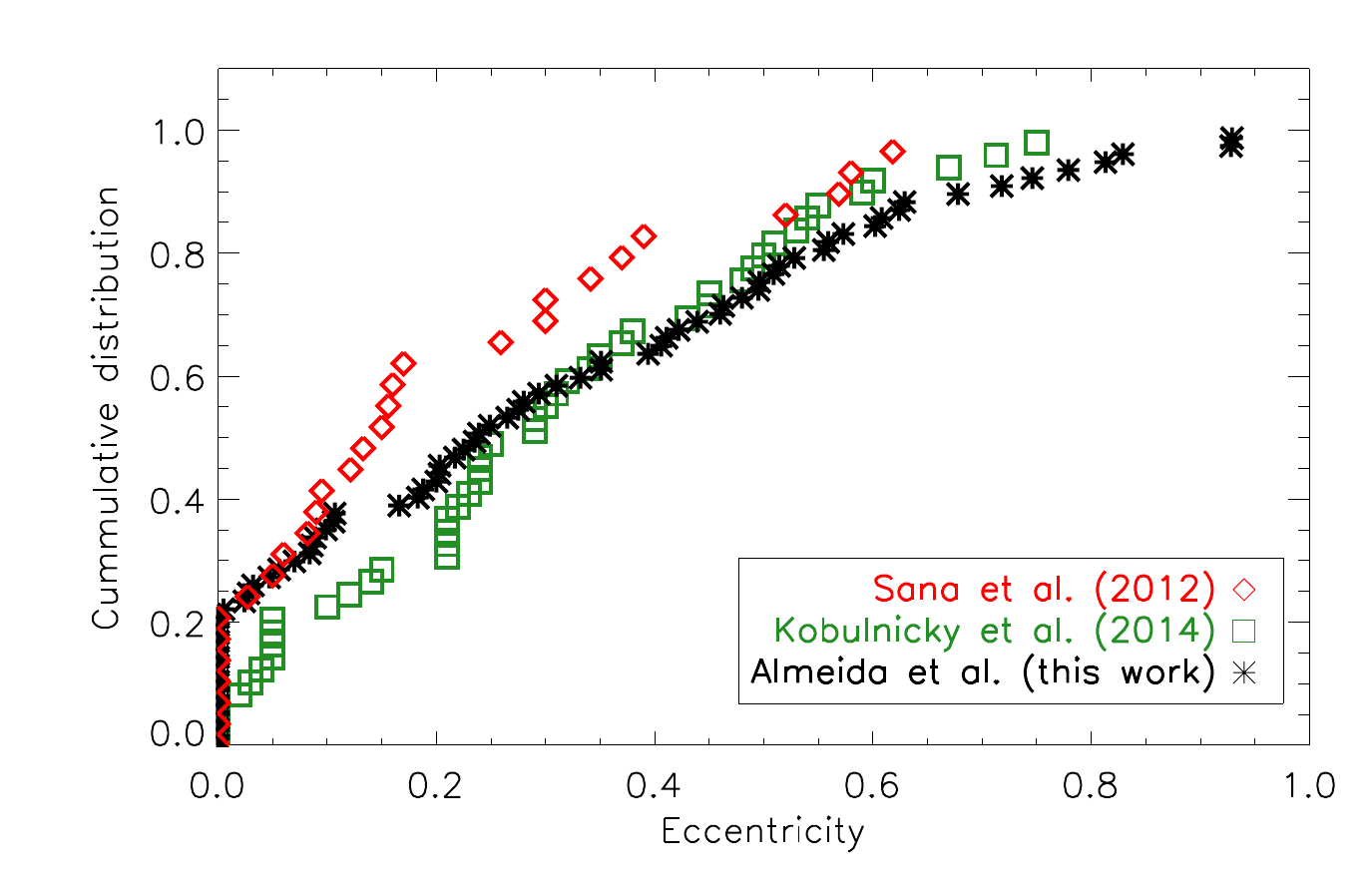}
\caption{Cumulative distributions of measured eccentricities in different samples. Red diamonds, green squares, and black asterisks are eccentricities from \citet[][]{SdMdK12,KK+2014}, and this work, respectively.}
\label{f:E_compar}
\end{figure}

Figure~\ref{f:obs_comp} compares the observed orbital period distribution of the TMBM sample to that of two recent works, \citet[][\citetalias{SdMdK12}]{SdMdK12} and \citet[][\citetalias{KK+2014}]{KK+2014}. \citetalias{SdMdK12} analyzed O-type binaries from six Galactic open clusters with an average age of less than 4 Myr. \citetalias{KK+2014} investigated OB-type binaries in the somewhat older Cygnus OB2 association \citep[up to 7~Myr-old,][]{Wright2015}. These three distributions are not bias-corrected, and this simple exercise should be taken with care. However, the observational campaigns and methodology behind these campaigns similar, which means that, to the first order, they should share similar detection biases. Interestingly, TMBM has five objects with a period shorter than any Galactic object in \citetalias{SdMdK12} and \citetalias{KK+2014}\footnote{The shortest period system in these two Galactic studies is FO15 ($P_\mathrm{orb}=1.41$~d); however, shorter period Galactic systems are known. 
HD~64315 (
V402 Pup; $P_\mathrm{orb}=1.02$~d) is one such case.}. Up to about four days, TMBM and \citetalias{SdMdK12} are overabundant by about 10\%\ compared to \citetalias{KK+2014}. Between one week and three months, \citetalias{SdMdK12} has more binaries than TMBM and \citetalias{KK+2014}, while all three distributions catch up at longer periods. The TMBM is obviously limited to orbital period $\lesssim1000$~days owing to the more limited campaign baseline. Importantly, however, Kuiper tests in between any of these three distributions fail to prove that the differences spotted by eye are statistically significant ($P_\mathrm{Kuiper}> 0.27$).

\subsubsection{Eccentricities}\label{ss:e_discuss}
The cumulative distribution of the eccentricities (Fig.~\ref{f:obs_dist}) is characterized by an overabundance of systems with circular and low eccentricity orbits. About 40\%\ of the systems exhibit quasi-circular orbits $e < 0.1$. This characteristic can be qualitatively explained by the large number of short-period systems for which tidal effects and/or mass transfer will tend to circularize the orbit \citep[][]{Zahn1977}. Between $e=0.1$ and $e=0.6$ the distribution of eccentricities is almost constant and flattens somewhat afterwards. 

Figure~\ref{f:E_compar} displays the observed eccentricity distributions of O-type binaries in young Galactic clusters (\citetalias{SdMdK12}), from OB-type binaries in Cyg-OB2  (\citetalias{KK+2014}) and from TMBM O-type binaries. The fraction of circularized orbits in the S+2102 and TMBM sample are in remarkable agreement while that of \citetalias{KK+2014} seems lower by about a factor of 2. However, applying the Lucey-Sweeney test to investigate the significance of the eccentricities (see Sect.~\ref{sss:ecc}), the fraction of systems with insignificant eccentricities in the \citetalias{KK+2014} sample rises to 0.35. The Cyg OB2 and the TMBM samples both show a more populated tail towards high eccentricities than the Galactic clusters sample. This probably reflects a limitation of the latter observational campaign, which was less suited to detect high-eccentricity systems. This is confirmed by further results below, after detection biases were taken into account. As for the period distribution, no clear-cut 
statistically significant difference can be observed among the three observational samples discussed here ($P_\mathrm{Kuiper}> 0.12$).

\subsubsection{Mass ratios}\label{ss:q_discuss}

The distribution of mass ratios $q$ for the 30 O-type SB2 binaries in our sample is displayed in Fig.~\ref{f:obs_dist}. It shows that about 20\%\ of the SB2 systems have a mass ratio larger than 0.95. The distribution is mostly flat below that and down to a mass ratio of 0.55. There is only one O-type system with $q<0.55$. This lack of low mass-ratio systems can be explained by observational biases as the detection of the secondary signature for binaries with low mass-ratios and large flux contrasts requires high signal-to-noise data which is not always available. As a consequence, 61\%\ of our objects are SB1 systems, many of which are likely to populate the $q<0.55$ region.

 The possible overabundance of (near) equal-mass systems is barely consistent with statistical fluctuations despite our limited sample size. Whether this represents a genuine twin population as proposed by \citet{PS2006} for the eclipsing binaries in the Small Magellanic Cloud \citep[see also][]{Luc06, CaD14} or results from evolutionary effects (see Sects.~\ref{ss:space} and \ref{ss:evol}) remains to be quantified.
 
\begin{figure}
\centering
\includegraphics[angle=-90,width=\hsize]{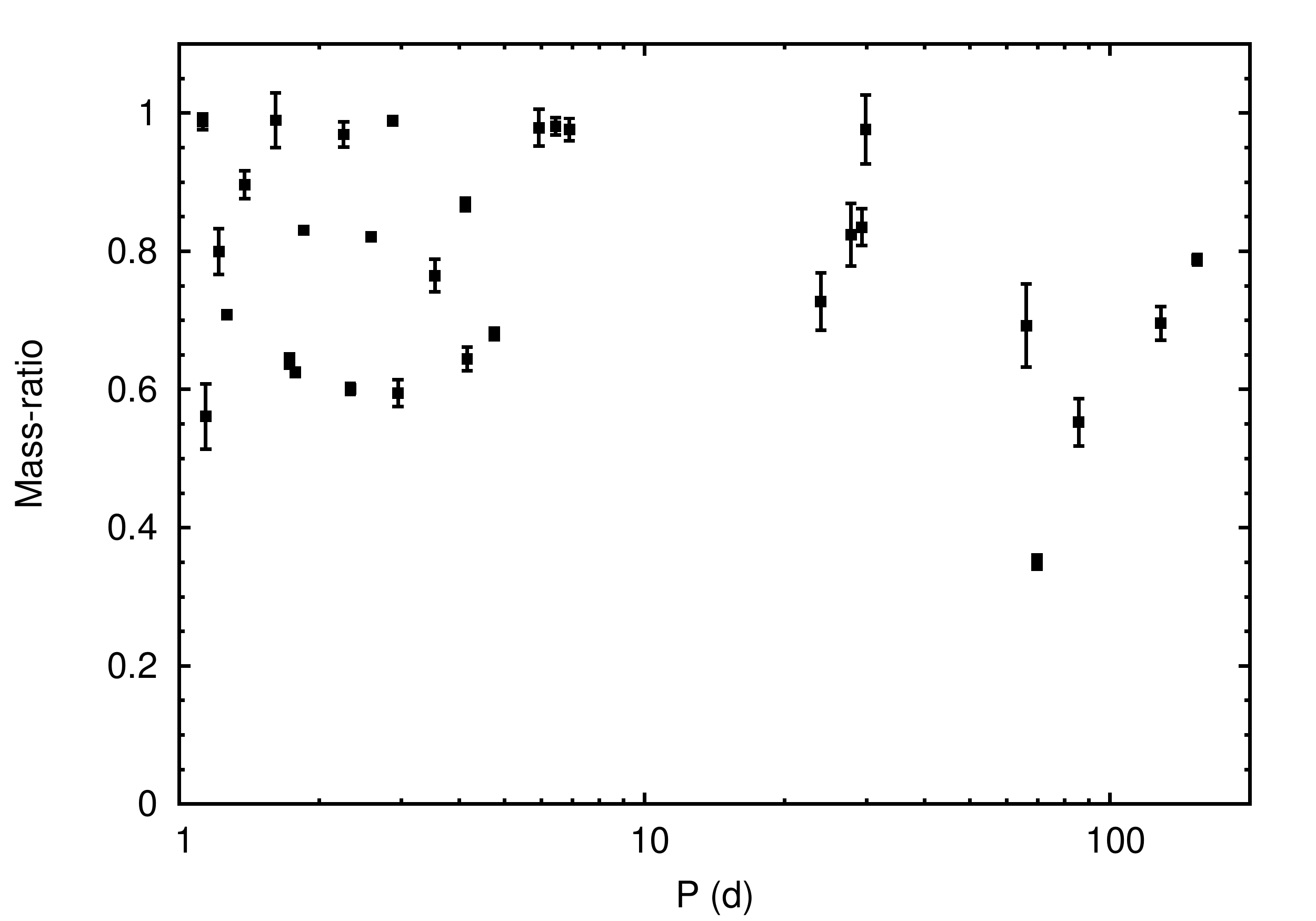}
\caption{Mass ratio vs. orbital period for the SB2 systems.}
\label{f:q_vs_P}
\end{figure}

 Interestingly, no obvious correlation of mass ratio with the orbital period -- hence the orbital separation -- can be identified in the present sample (see Fig.~\ref{f:q_vs_P}). The ratio of SB1 ($f_\mathrm{SB1}$) to the total number of binaries below and above orbital periods of 20 days, however, changes by a factor of two (from 0.25 to 0.51). This could either reflect the difficulty of separating the two components in longer period systems (as the RV separation decreases proportionally to $P_\mathrm{orb}^{1/3}$), or it could be related to lower mass ratios in the long-period regime as suggested by \citet{Moe2016}. Only in-depth modeling of the observational biases will allow us to investigate this question, which will be addressed in a separate study of the TMBM project.

\begin{figure}
\centering
\includegraphics[width=\hsize]{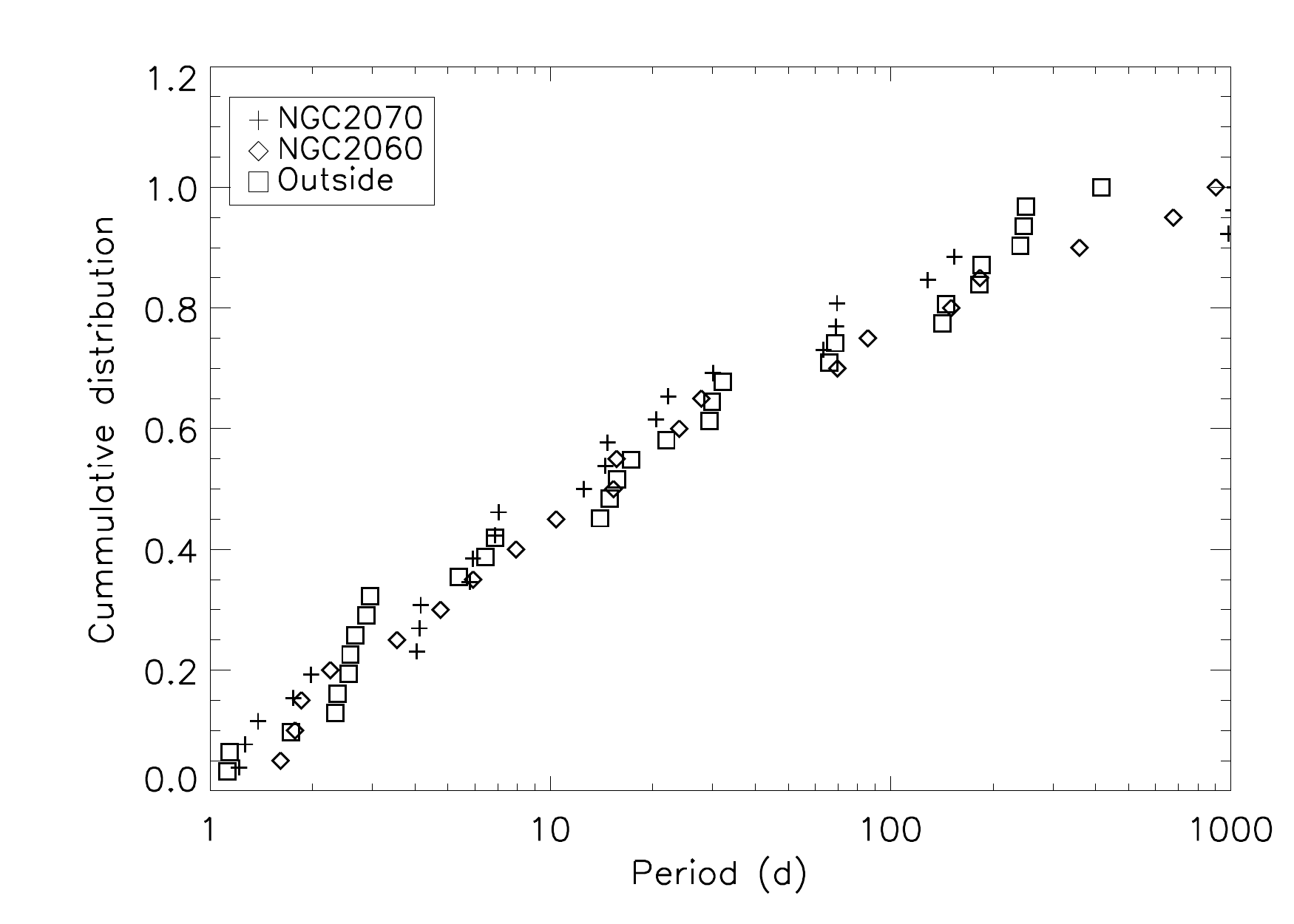}
\includegraphics[width=\hsize]{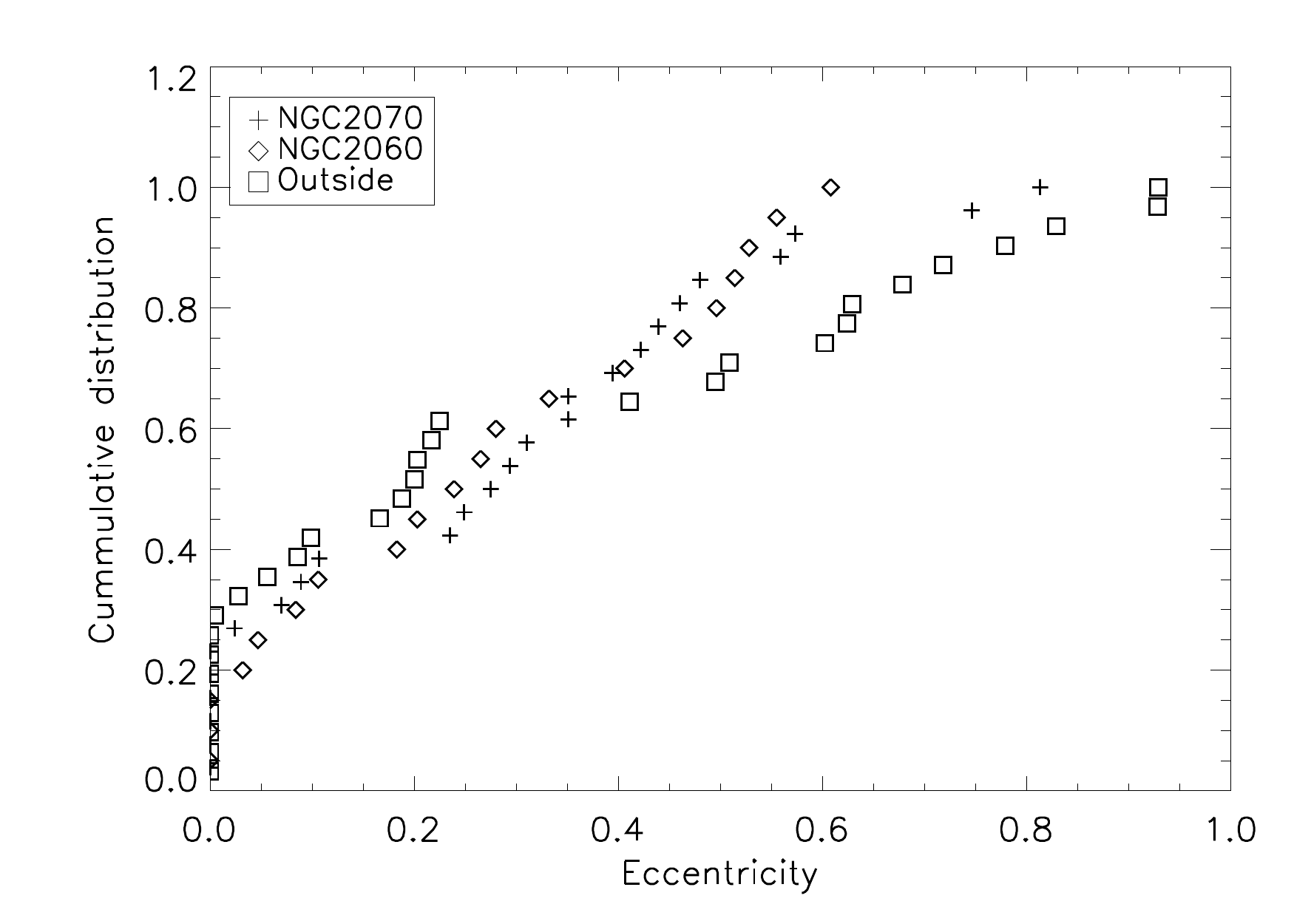}
\includegraphics[width=\hsize]{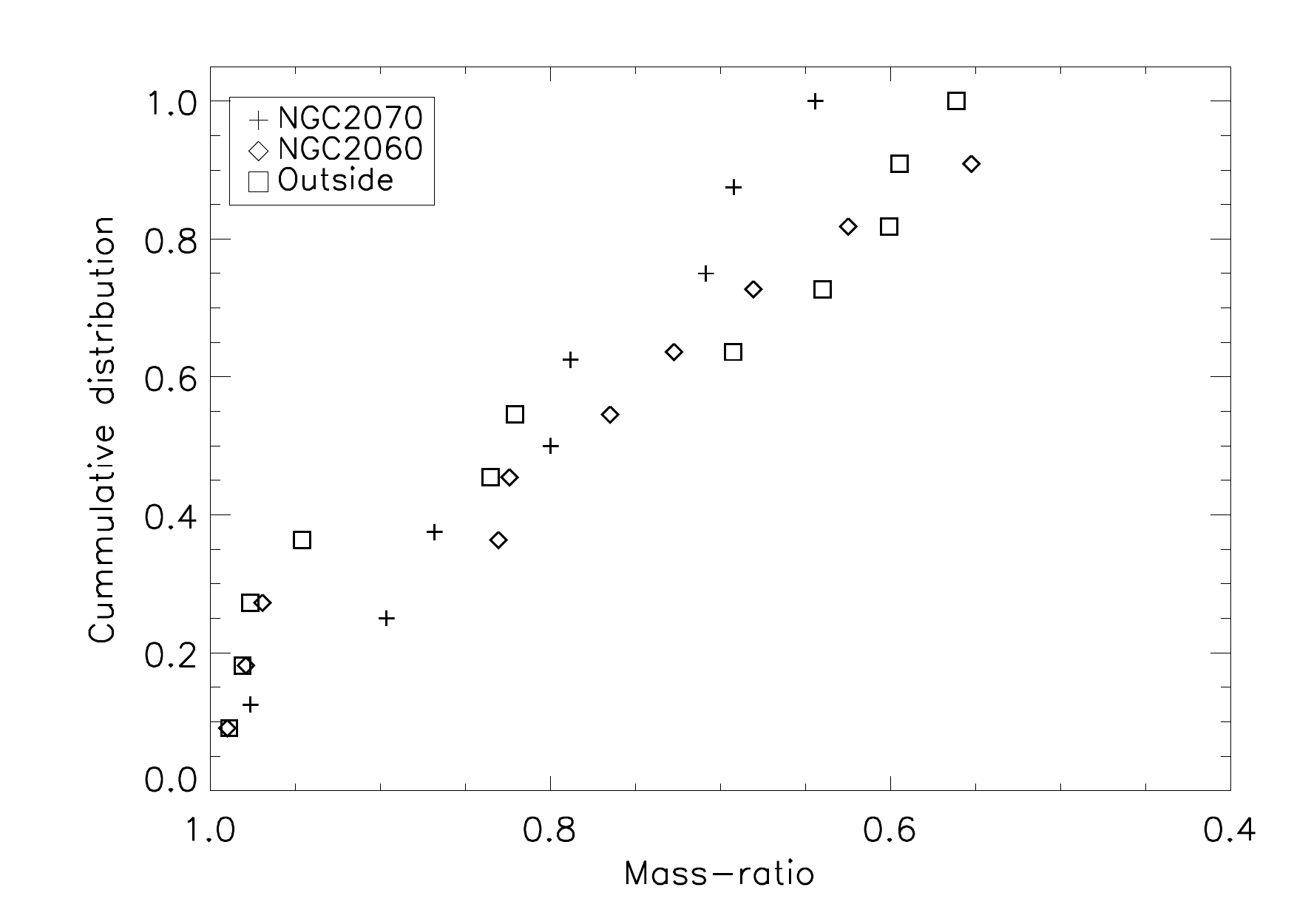}
\caption{Cumulative distribution function of orbital periods, eccentricities, and mass ratios for spatially selected subsamples of the 30 Dor O-type binary population (see legend). }
\label{f:spat_var_cdf}
\end{figure}
\begin{figure}
\centering
\includegraphics[width=6.9cm]{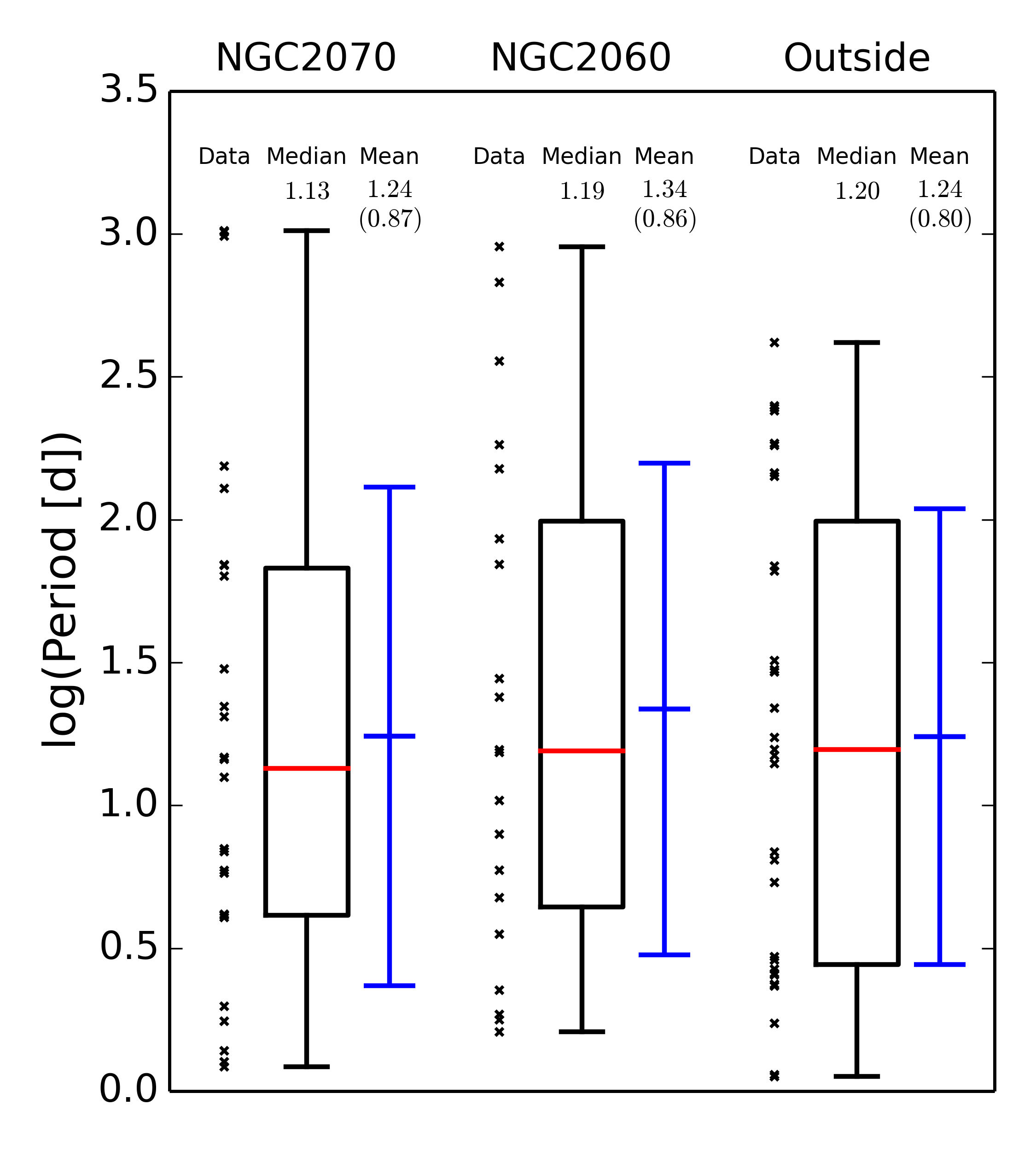}
\includegraphics[width=6.9cm]{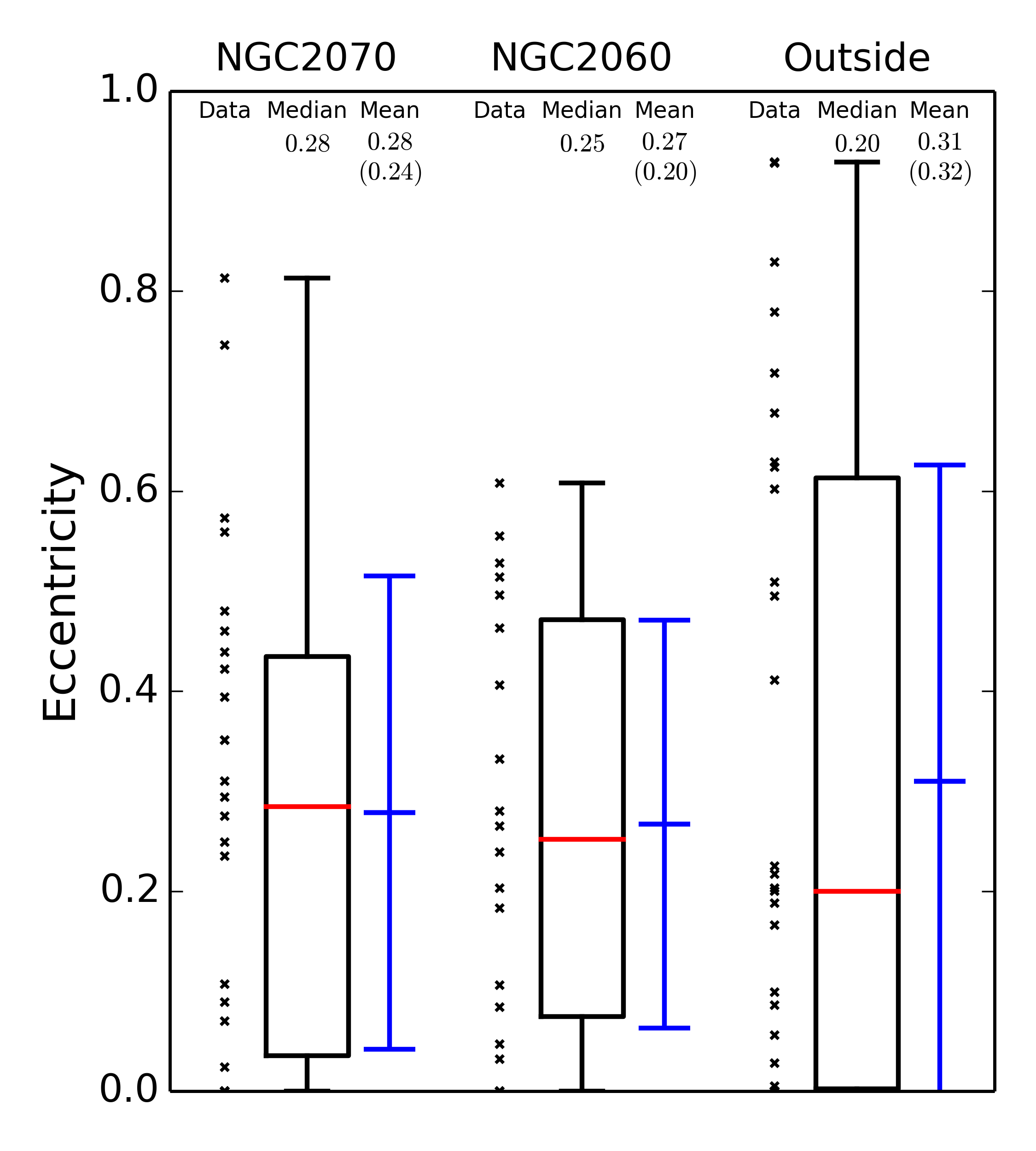}
\includegraphics[width=6.9cm]{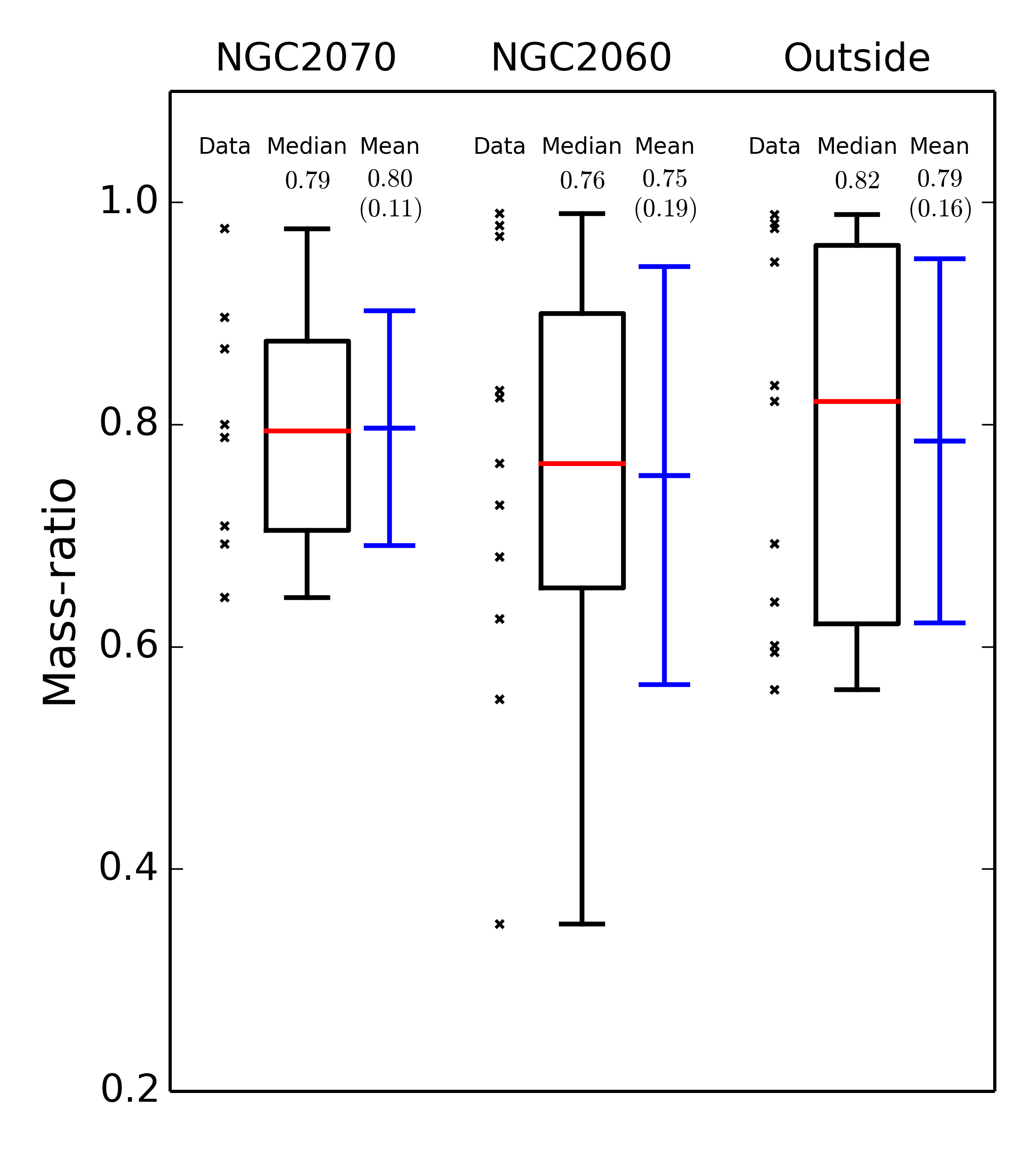}
\caption{Comparisons of the  orbital periods, eccentricities, and mass ratios for our three spatially selected subsamples. Each figure gives the data points (black crosses), quartiles (boxes), and standard deviation (blue whiskers) for each of the three subsamples.}
\label{f:spat_var_boxplot}
\end{figure}

\subsection{Spatial variations}\label{ss:space}
The previous section showed the lack of significant differences between the orbital period distributions obtained from two Galactic samples and that from 30~Dor. Here we take advantage of the large TMBM binary database to search for further  variations across three different, spatially selected populations in 30~Dor. Following \citet{SdKdM13}, we split the sample according to three regions: NGC~2070, NGC~2060 and Outside Clusters (see also Table~\ref{t:alloc}). In \citet{SdKdM13}, we could not find a significant difference in the observed binary fraction between these three regions. Here we further investigate the ratio of SB2 and SB1 systems, their orbital periods, eccentricities, and the mass ratio of SB2 systems. 

\begin{figure}
\centering
\includegraphics[width=\columnwidth]{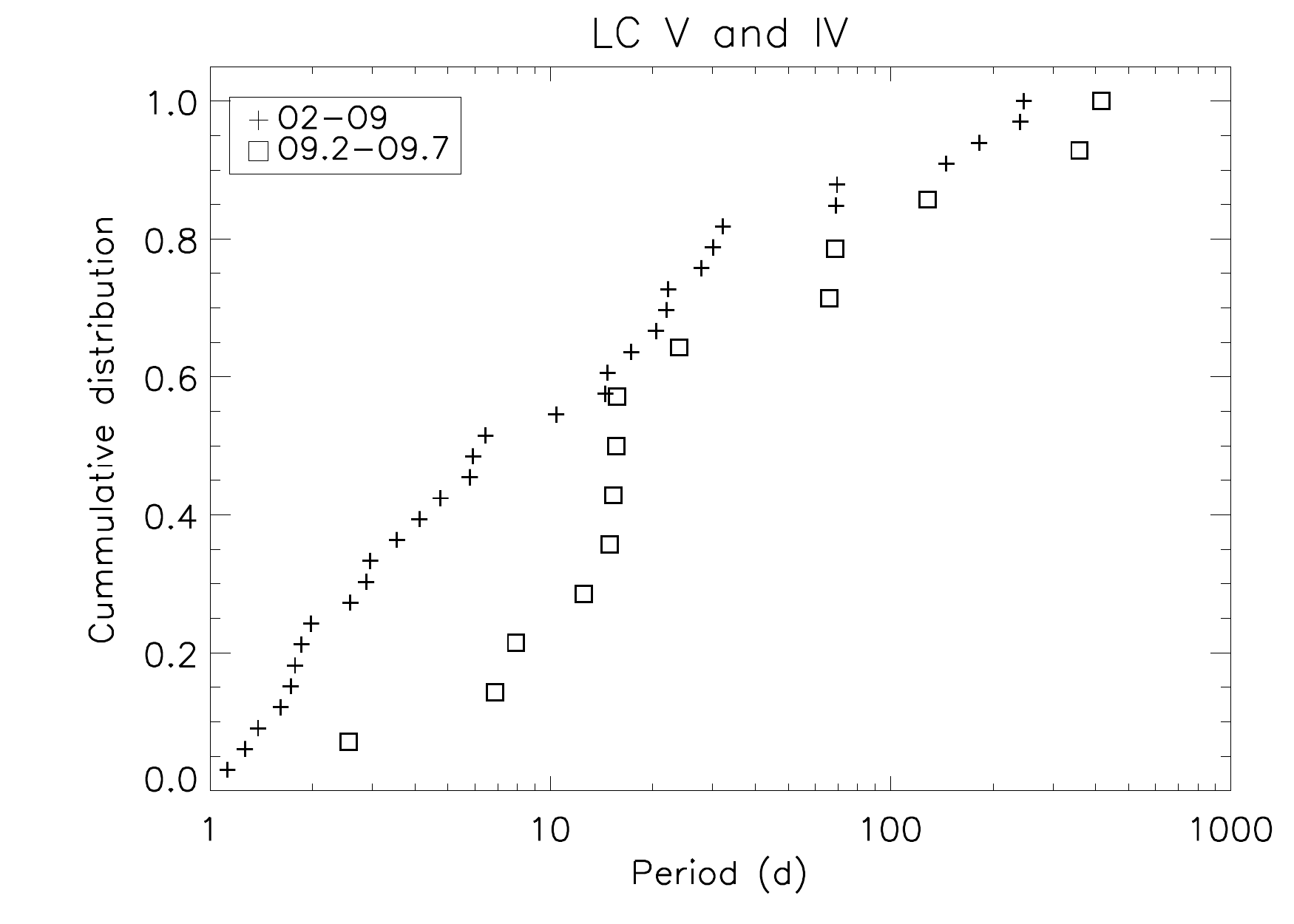}
\includegraphics[width=\columnwidth]{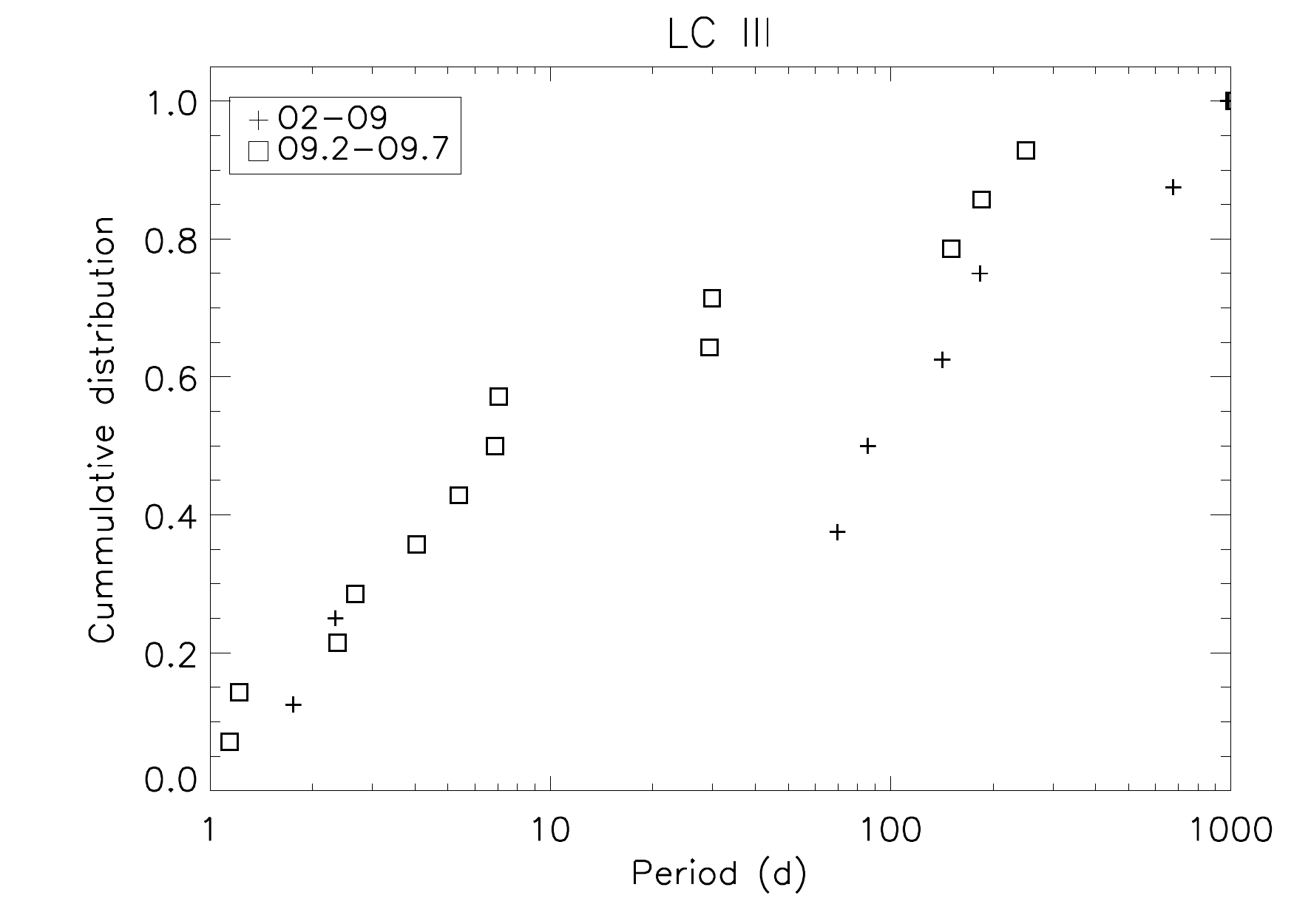}
\caption{Cumulative distribution function of orbital periods of O stars earlier and later than spectral subtype O9 (see legend). The upper figure is for LC V and IV; the bottom figure for LC III.}
\label{f:spt_var_P_LC345}
\end{figure}
\begin{figure}
\centering
\includegraphics[width=\columnwidth]{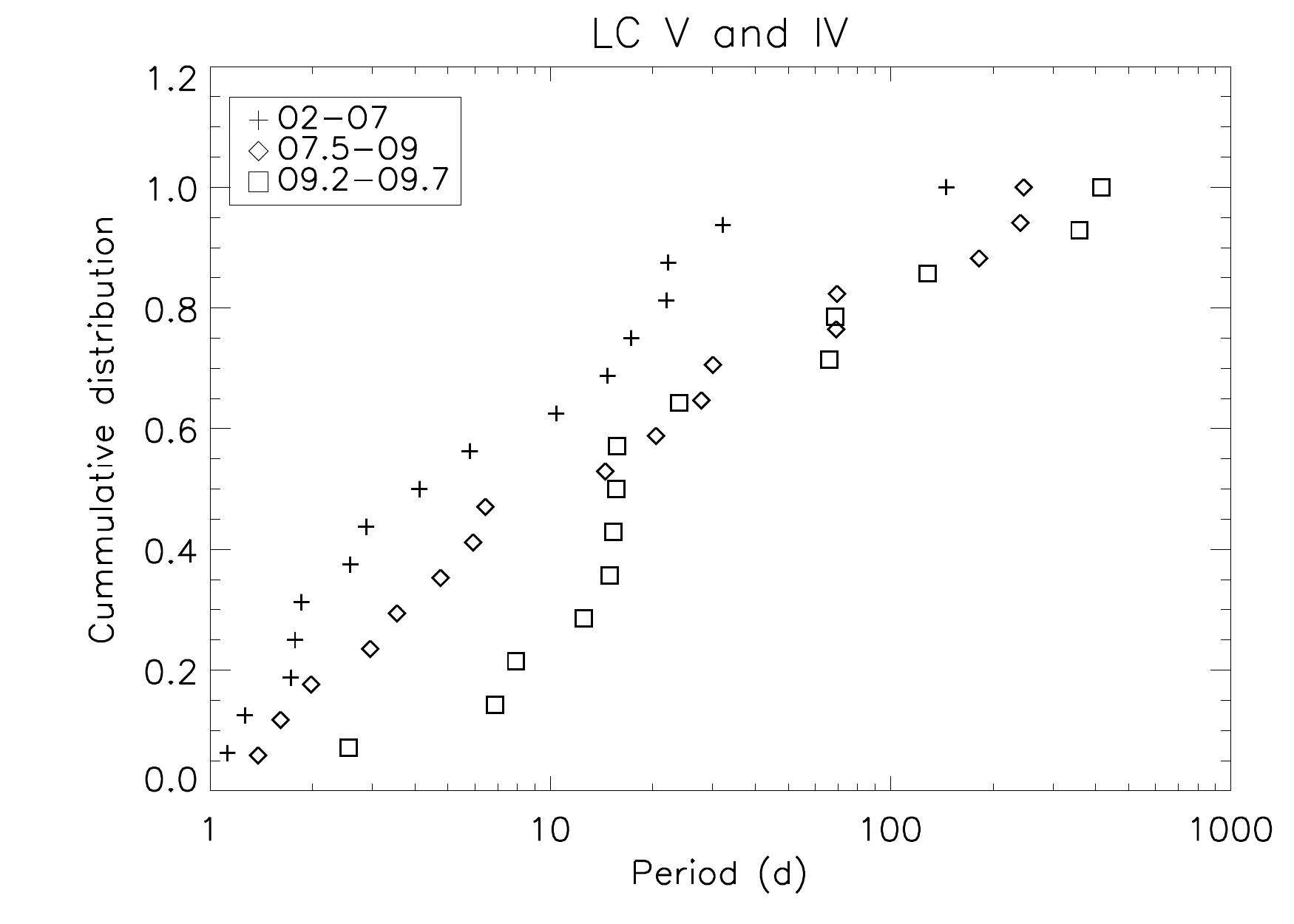}
\includegraphics[width=7.5cm]{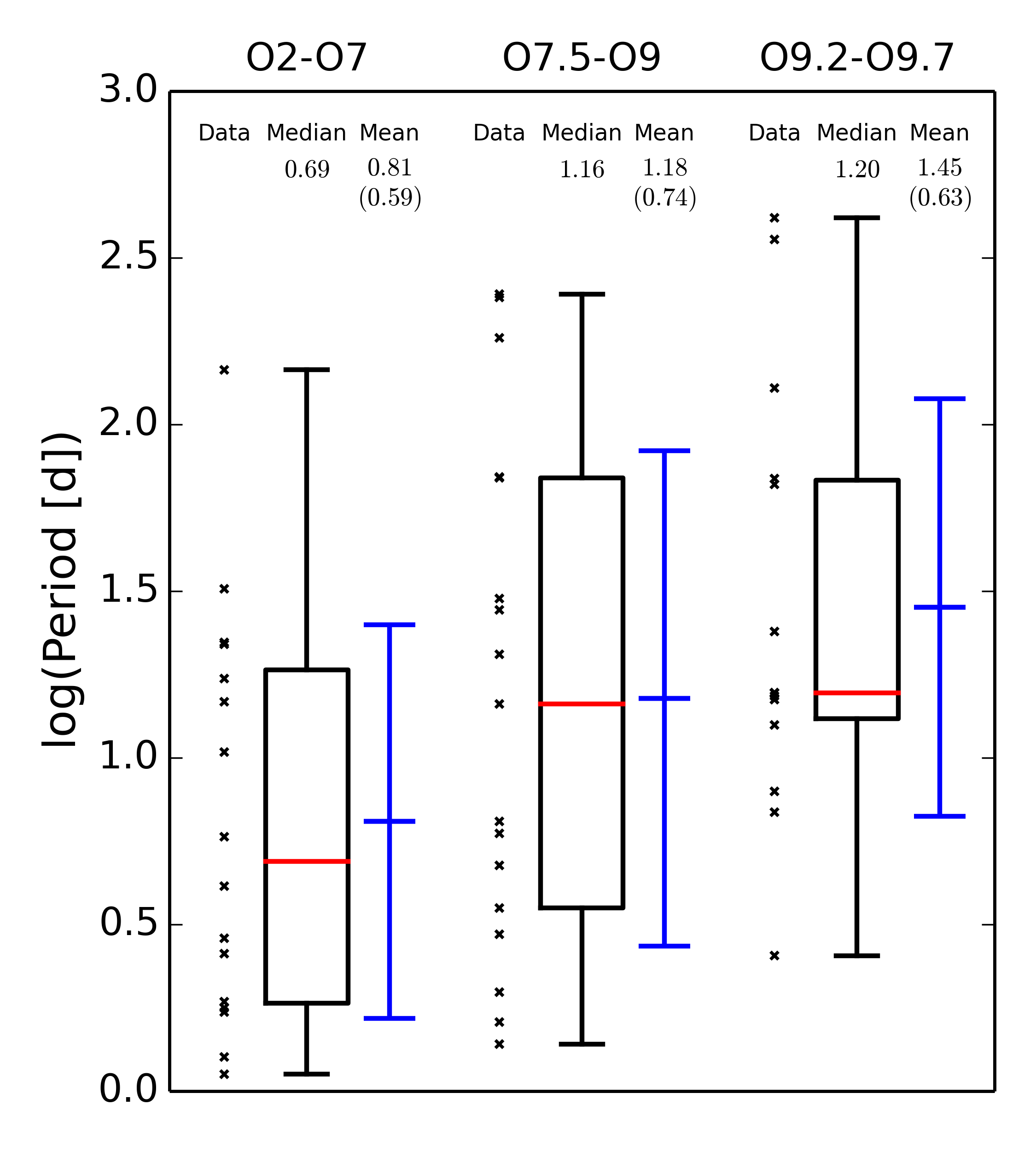}
\caption{Cumulative distribution functions and box diagram of orbital periods of LC V and IV stars for three ranges of spectral subtypes (see labels).}
\label{f:box_spt}
\end{figure}
The fraction of SB2, i.e.\ $f_\mathrm{SB2}=N_\mathrm{SB2}/(N_\mathrm{SB1}+N_\mathrm{SB2}$) in the three regions  are $0.67\pm 0.09$ (NGC~2070), $0.45\pm 0.11$ (NGC~2060) and $0.69\pm 0.08$ (Outside), where binomial statistics have been used to compute the 1$\sigma$ uncertainties. While NGC~2070 and the field yield identical values, NGC~2060 tends to display a smaller fraction of SB2 systems. This may indicate  overall smaller mass ratios, and a larger fraction of optically faint companions.

The observed period distributions in the three regions are, however, remarkably similar (Fig.~\ref{f:spat_var_cdf}, top panel), as confirmed by Kuiper tests ($P_\mathrm{Kuiper}>0.8$). The means and medians of the three samples are all situated in a narrow range, between 10 and 20 days (Fig.~\ref{f:spat_var_boxplot}). There is a tendency for slightly shorter periods outside the clusters and in NGC 2070, but this trend is unlikely to be significant.

Similarly, the eccentricity and, for SB2s, the mass ratio distributions show no statistically significant differences ($P_\mathrm{Kuiper}>0.15$), with the notable exception of the eccentricity distributions between NGC~2070 and Outside the clusters ($P_\mathrm{Kuiper}=0.047$). The O-star binary population outside NGC~2060 and NGC~2070 indeed contains a larger abundance of systems with $e\sim0.2$ and with large eccentricities ($e>0.6$). The latter group is three times more frequent in the field than in the associations. Some interesting trends can also be observed.  Most (quasi) equal-mass systems ($q>0.95$) are found outside NGC~2070, which is believed to be the youngest population (but see Schneider et al., in prep.). NGC 2060 itself tends to show a smaller number of both low ($e<0.05$) and high ($e>0.6$) eccentricity systems. 

Despite these small differences, the significance of which is hard to assess, the overall period, eccentricity and mass ratio properties of the three samples show a large degree of coherence, while the evolutionary stages of  these regions are likely different \citep{WaB97}, as are their dynamical properties. Spatial and evolutionary variations are thus  hard to identify even with the large binary sample at hand. This suggests that they only have a moderate impact on the distribution of orbital parameters in the spectroscopic regime considered here (i.e., $P_\mathrm{orb} \lesssim 1$~yr). As a corollary, our findings also point to the fact that that the orbital properties of the stars in our sample may be set by local conditions, possibly during the star formation process, rather than by overall properties of their respective parent clusters or associations, such as stellar density and total mass.

\subsection{Correlation with spectral properties}\label{ss:spt}

In this section we split the sample according to the O spectral subtypes and luminosity class (LC). As a first example, Fig.~\ref{f:spt_var_P_LC345} compares the period distribution of LC V and IV (upper panel) with the LC~III (lower panel) of late O stars (O9.2-O9.7) and earlier subtypes. Strikingly, there is a significant depletion of short-period systems ($P_\mathrm{orb}<15$~d) among late dwarfs and a si\-mi\-lar depletion among earlier giants. It is hard to imagine a physical explanation. Alternatively, it is possible to imagine that dilution and line blending due to the presence of the companion affect the accuracy of luminosity classification (see a similar case for single O stars in Ramirez-Agudelo et al., submitted).

To investigate further, we split the sample of LC V and IV in early (O2-O7), intermediate (O7.5-O9) and late (O9.2-O9.7) subtypes. The separations were chosen to split the sample into roughly equal sizes. Figure~\ref{f:box_spt} reveals interesting patterns. It shows that the earlier type binaries tend to have shorter periods than later type ones. While the sample is small and a self-consistent bias analysis would be required, the signal seems genuine. The physical mechanism that led to shorter periods among the earliest binaries is yet to be identified.

\begin{figure}
\centering
\includegraphics[width=\hsize]{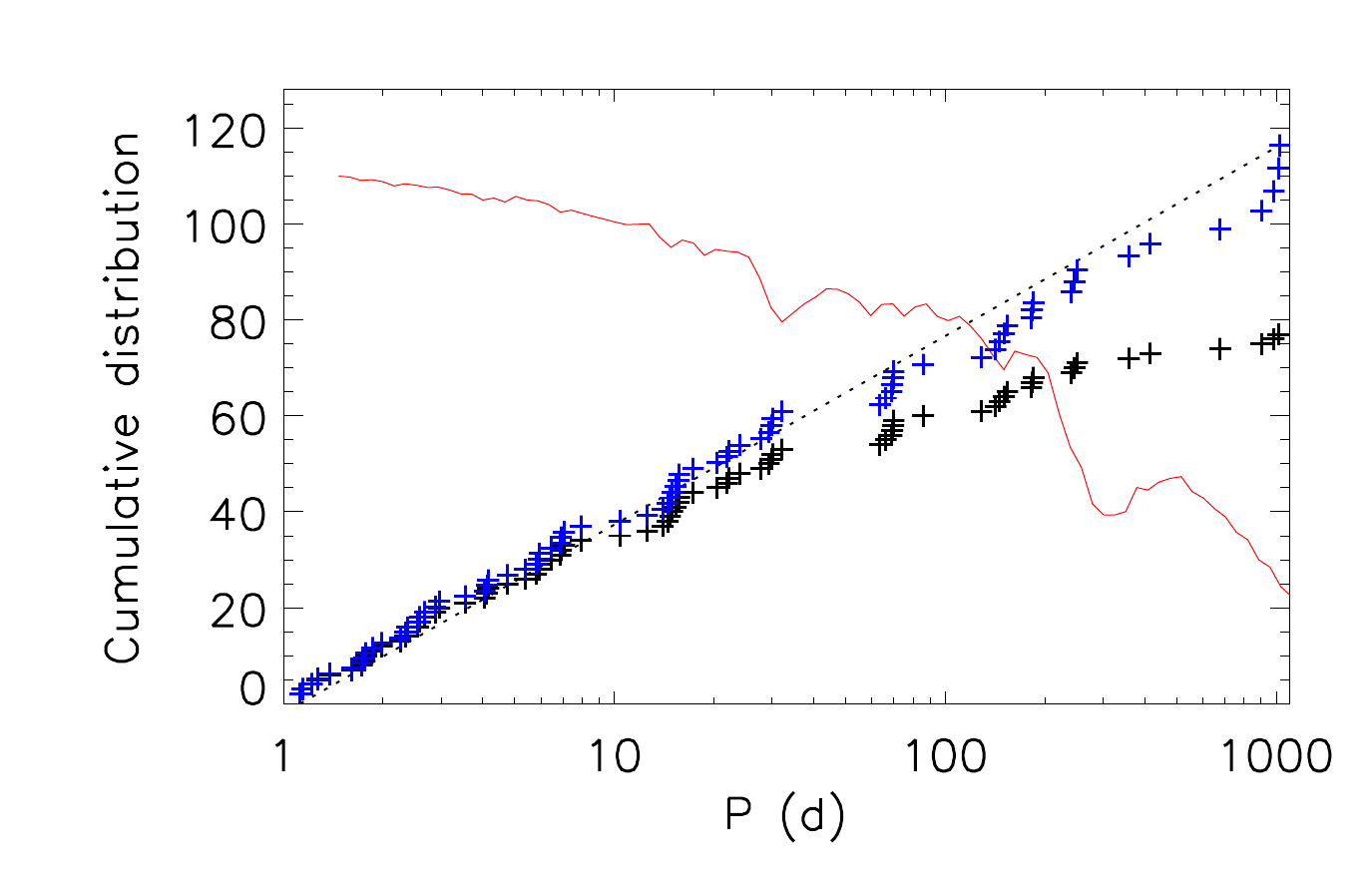}
\includegraphics[width=\columnwidth]{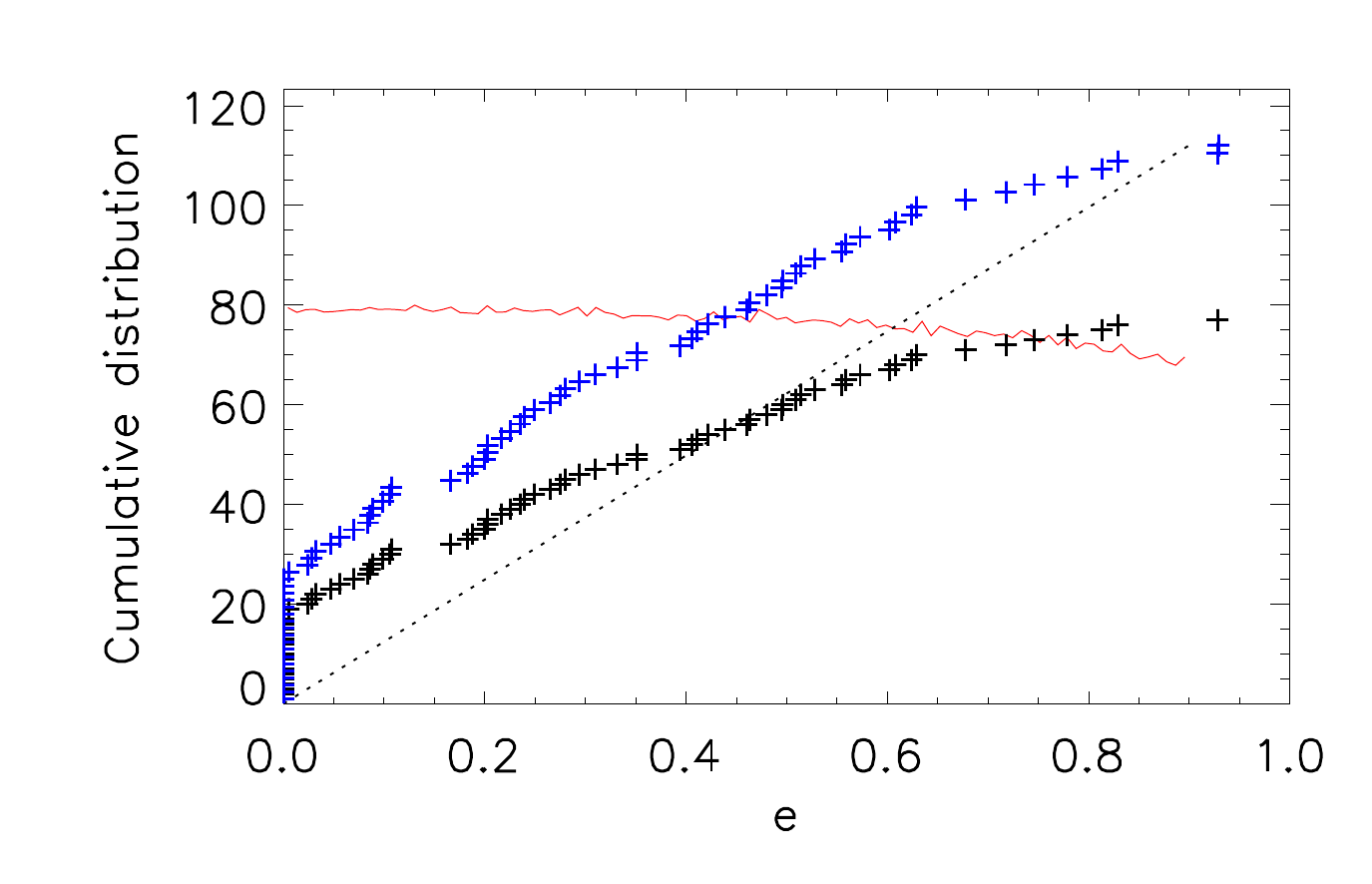}
\caption{Observed (black) and bias-corrected (blue) distribution of periods ($P_\mathrm{orb}$) and eccentricities ($e$)  for the TMBM sample. The red curves indicate the VFTS binary detection probability curves computed by \citet{SdKdM13}. Dotted lines show a uniform distribution.}
\label{f:bias_dist1}
\end{figure}

\subsection{Observational biases}\label{ss:bias}
As discussed in other works \citep{SdMdK12, SdKdM13}, the impact of the observational biases on the observed distributions depends on the value of the parent distribution, which are unknown quantities. As a consequence, a self-consistent modeling of the entire VFTS and TMBM data set is required to constrain the parent distributions. This will be addressed in a dedicated paper in the TMBM series. Here, we used the VFTS detection probability curves \citep[Fig.~8 in][]{SdKdM13} to apply a first order of magnitude correction to the observed distributions. We limit ourselves to the orbital period and eccentricity distributions because the mass ratio distribution is heavily affected by incompleteness (61\%\ of our O-type binary sample are SB1 systems). 
The results are shown in Fig.~\ref{f:bias_dist1} together with  the adopted VFTS detection probability distributions. The accuracy of the results depends on how realistic are the parent period, mass ratio, and eccentricity distributions derived based on the modeling of the handful multi-epoch observations of the VFTS.

\begin{figure}
\centering
\includegraphics[width=\hsize]{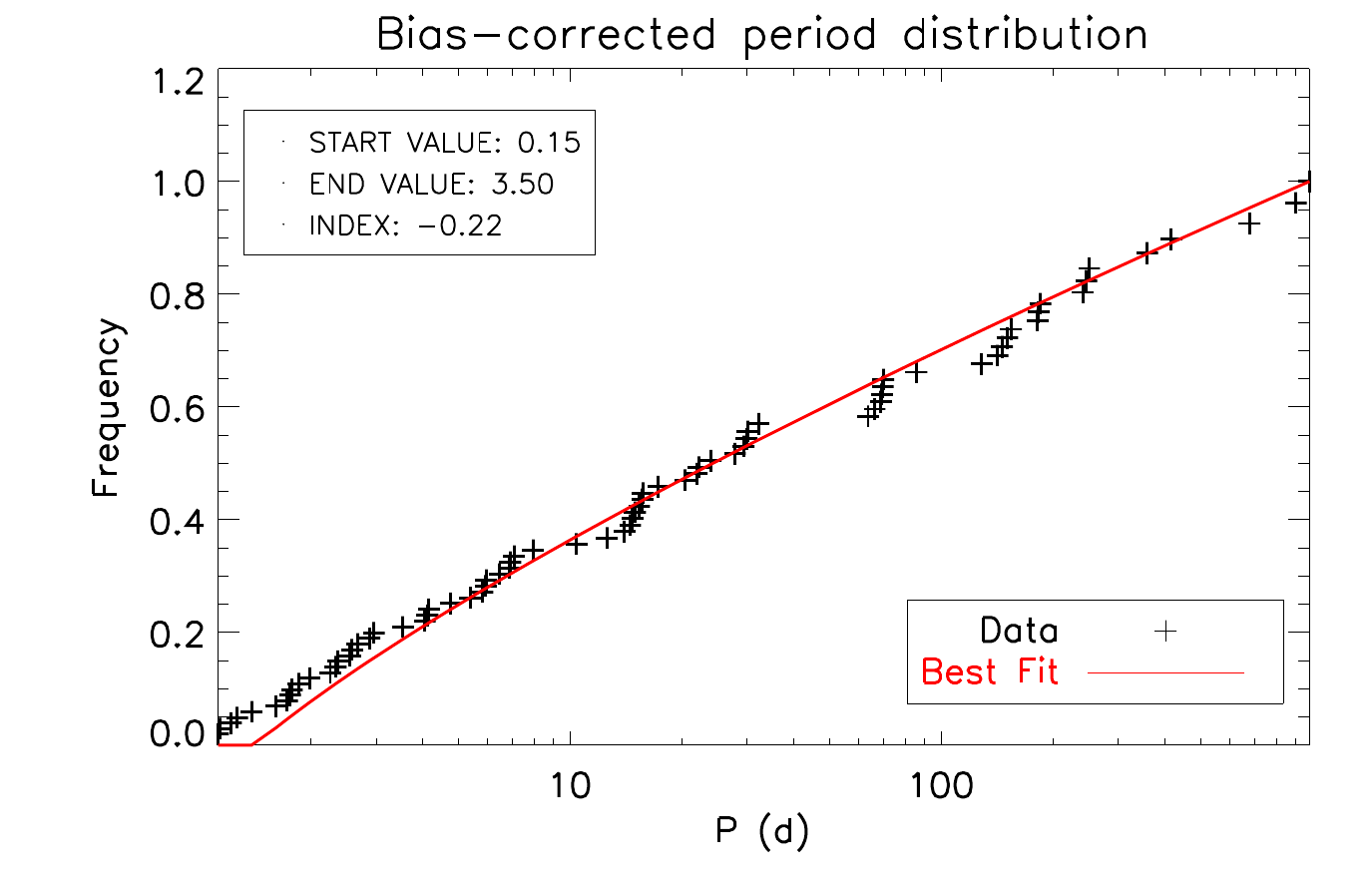}
\includegraphics[width=\hsize]{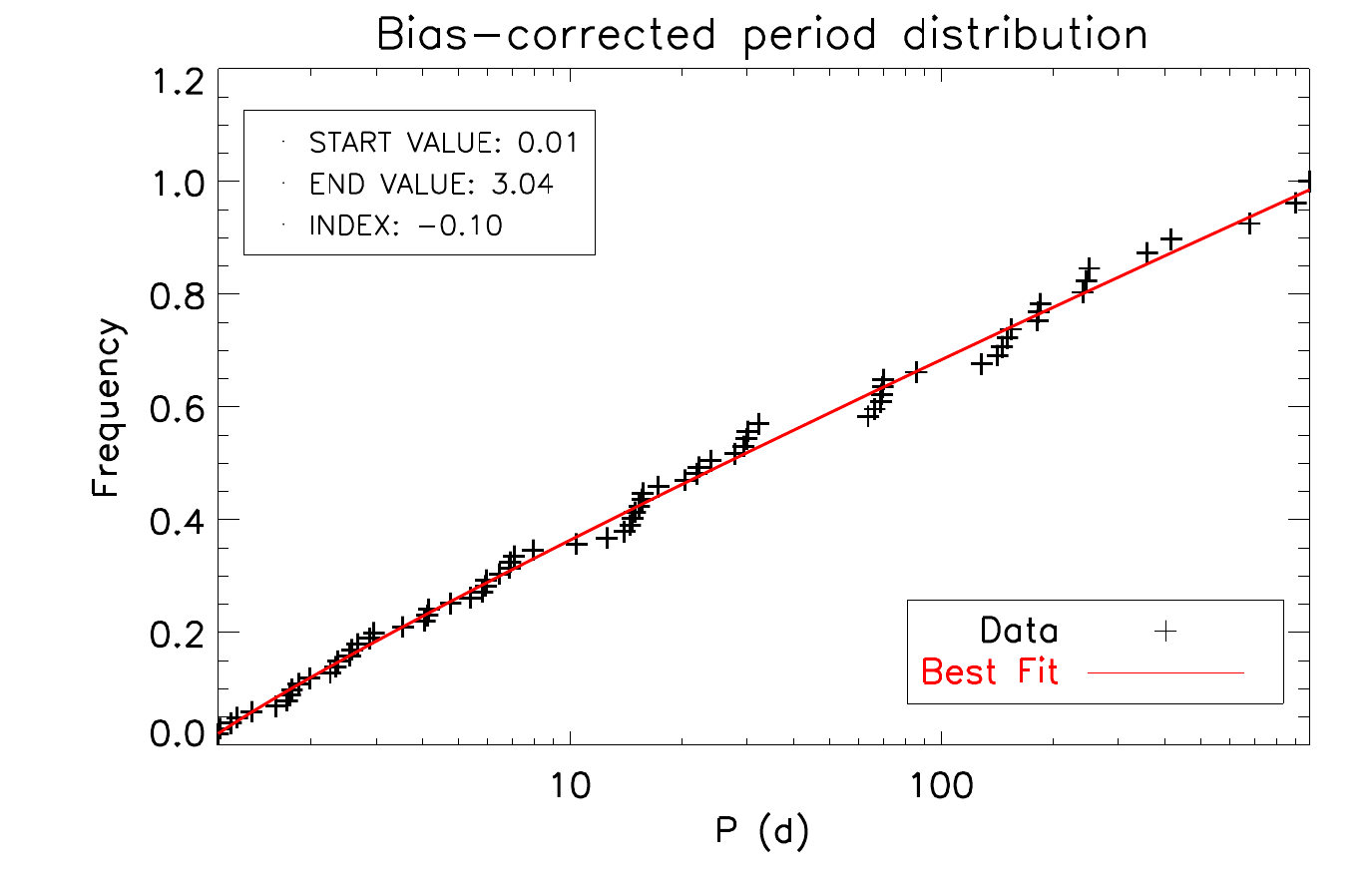}
\caption{Best-fit power laws to the bias-corrected period distribution. The upper panel has fixed period boundaries while these boundaries are adjusted together with the power-law index in the lower panel. Formal errors on the index values are on the order of 0.02, but likely underestimate the true uncertainty as the formal errors do not take into account the uncertainties on the period measurements.}
\label{f:P_pl}
\end{figure}

\subsubsection*{Power-law fitting}
To compare the present results with those of previous studies, we adjust the bias-corrected distributions with a power law
\begin{equation}
f_x=x^\alpha \hspace*{4mm}\mathrm{for}\hspace*{4mm} x\in [x_\mathrm{min}:x_\mathrm{max}],
\end{equation}
where $x$ is either the orbital period ($x=\log P$) or the eccentricity ($x=e$).

When performing such a fit, it is important to realize that the adopted fitting range ($x_\mathrm{min}$ and $x_\mathrm{max}$) has an impact on the obtained best-fit index ($\alpha$), so that it is not possible to directly compare power-law indexes obtained in different fitting ranges. This is illustrated in Fig.~\ref{f:P_pl}, where the upper panel shows the result of a fit with a lower limit of $P_\mathrm{min}=10^{0.15}\approx1.4$~d, i.e.\ the same values as adopted in \citet{SdMdK12} for the Galactic clusters. The choice of the minimum period value was then guided by the shortest period in the sample. In the absence of period measurements,  \citet{SdKdM13} adopted the same lower boundary for the VFTS analysis of the 30~Dor region to allow for a direct comparison with our previous works. 

The TMBM has now revealed that orbital periods as short as 1.1~d are present in the 30~Dor region, which suggests that the lower fitting boundary needs to be adapted. In the lower panel of Fig.~\ref{f:P_pl}, we have allowed the fitting routine to adjust both the lower and upper boundaries of the period range. The period index goes from $-0.2$ to $-0.1$ and the best-fit relation now better represents the short-period end of the distribution. The former case lays within the errors of $-0.45\pm0.30$ obtained for the VFTS sample based on a modeling of the RV variations \citep{SdKdM13}. In the latter case, the presence of extremely short-period systems, not accounted for in \citet[][]{SdKdM13}, results in a flatter power-law index as explained above.

\begin{figure}
\centering
\includegraphics[width=\hsize]{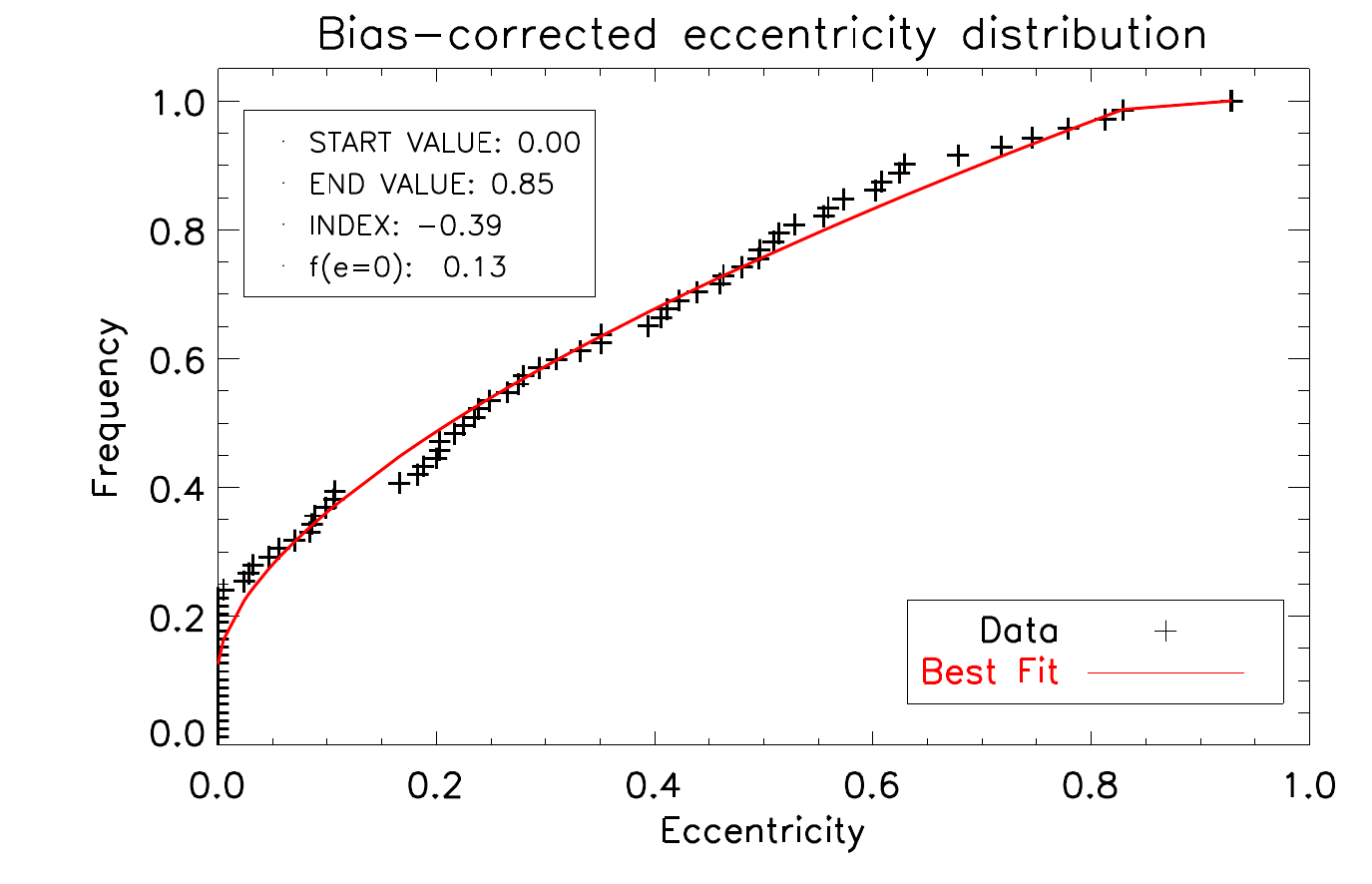}
\caption{Best-fit distribution to the bias-corrected eccentricity distribution. The fitted distribution combined a threshold $f(e=0)$ of circular orbits and a power law for $e>0$.}
\label{f:E_pl}
\end{figure}

The eccentricity distribution is adequately reproduced by a power law with an index of $-0.5$ (Fig.~\ref{f:E_pl}). A finer modeling that includes an adjustable contribution of circularized systems yields a slightly better fit (Fig.~\ref{f:E_pl}). Best-fit parameters in this case are $-0.4$ for the power-law index and $f(e=0)=0.13$ for the threshold of circularized systems.  Models with and without threshold are compatible with findings from  Galactic samples although the TMBM data contains higher eccentricity systems than those seen in the Galactic samples.

\subsection{Evolutionary impact}\label{ss:evol}

In the above sections, we point out a number of possible differences between the orbital parameter distributions in 30~Doradus compared to previous Galactic studies. While these differences may not all be significant, it is interesting to discuss them in regard to expectations from binary evolution theory.

 \citet{SdKdM13} found an O-type star binary fraction in 30~Dor ($f_\mathrm{bin}=0.54\pm0.04$) that is lower than that in Galactic open clusters ($f_\mathrm{bin}=0.69\pm0.09$). Interestingly, the index of the period distribution that we measured here is flatter than that found in \citet{SdKdM13}. As a consequence, based on the two-dimensional projections of the merit function in \citet[][their Fig. 6]{SdKdM13}, the true binary fraction in 30~Dor may actually be closer to 60\%\, and the mass ratio distribution significantly flatter. A self-consistent bias correction is needed to obtain accurate numbers. While the binary fraction may not be as different as previously thought, binary interaction is expected to decrease the apparent binary fraction as most post-interaction products would not be detected as such by RV surveys \citep{dMSL14}. 

The power-law index that we obtain for the period distribution also seems flatter than that obtained in Galactic regions (\citetalias{SdMdK12}: $-0.5$, \citetalias{KK+2014}: $-0.2$; this work: $-0.1$), although admittedly the figures are within 2$\sigma$ of one another. This again follows expectations of binary evolution. The shortest period binary systems will be affected first and, following a similar reasoning to that for the binary fraction, will be harder or impossible to detect after the interaction and will disappear from our sample. This will decrease the presence of short- vs.\ long-period systems and will lead to a flattening of the orbital period distribution.

Furthermore, conservative long-duration case-A interactions will tend to shorten the orbital period, and to equalize and then inverse the mass ratio. They may be responsible for the presence of shorter period systems in 30~Dor compared to the Galactic samples. Similarly, the formation of a contact phase in the shortest period systems, as observed in VFTS~352 \citep{Almeida+2015}, will quickly equalize the mass ratio and is expected to contribute to the peak (or part of it) close to unity in the mass ratio distribution. In this respect, it is interesting to note that the equal-mass binaries tend to be found away from NGC~2070, i.e., not in the youngest part of the 30~Dor region. 

Further work is definitely needed, on the one hand to assess the significance of the trends in the TMBM results and on the other hand to perform detailed binary evolution simulations taking into account the star formation history in 30~Dor. However, the above discussion suggests that the overall trends within the 30 Dor sample are in qualitative agreement with expectations from binary evolution.

\section{Summary}\label{s:ccl}

We introduced the Tarantula Massive Binary Monitoring (TMBM), a multi-epoch spectroscopic campaign targeting 100 massive binary candidates discovered by the VLT FLAMES-Tarantula Survey (VFTS). Combining the VFTS data with 32 new epochs collected by using the FLAMES/GIRAFFE spectrograph between 2013 and 2014, TMBM characterizes the  orbits of 82 systems: 51 SB1s and 31 SB2 binaries. For the remaining 18 systems, 14 do not show any periodicity, while the other 4 do, but we have not been able to find a satisfactory orbital solution.

The observed distribution of orbital periods for the TMBM sample is remarkably similar to equivalent distributions measured in Galactic regions \citep{SdMdK12,KK+2014}. After a first-order correction for detection biases, the obtained distribution seems slightly flatter with a power-law index of $-0.1$ in $\log$ space in the range between 1 and 1000 days. Such a difference is small and possibly not significant. We thus conclude that the metallicity difference between the LMC and the Milky Way only has a small effect, if any, on the period distribution of massive stars. 

We further search for spatial variations within the 30~Dor field of view. To the first order, the orbital distributions are remarkably similar across the entire region, indicating that environmental and evolutionary effects are of second-order at best. Small features are identified, however: (i) the eccentricity distribution in NGC~2070 is slightly different from that of the field; (ii) the somewhat older cluster NGC~2060 seems to have a larger fraction of SB1 systems; and (iii) the equal-mass binaries ($q>0.95$) are all found outside NGC~2070, the central association that surrounds R136, the very young and massive cluster at 30~Dor's core. While the significance of these results is hard to assess, items (ii) and (iii) are in qualitative agreement with expectations from binary evolution. 

Intriguing differences in the period distribution of dwarfs and giants are found. We suspect that multiplicity has impacted the luminosity class criteria used in their classification. Differences are also found among dwarfs of different spectral subtypes: earlier types -- likely more massive at birth -- indeed tend to display somewhat shorter periods than later types -- likely less massive at birth. In the future, spectral disentangling will allow atmosphere analysis of the individual stars to be performed. This will help to investigate these trends further. Spectral disentangling will also help in finding lower mass companions whose signature is currently lost in the noise of the individual spectra. This will allow us to complement the mass ratio distribution which is currently limited to only 31 SB2 systems. 

Finally, self-consistent modeling of the observational biases and comparison with detailed binary population synthesis will allow further investigation into the evolutionary impact on the present-day distribution. This is important to estimate the initial distribution of the orbital parameters (and the initial binary fraction). These quantities can indeed be used to constrain the outcome of the still poorly understood process of massive star formation and the origin of the large fraction of close massive binaries. They are also needed as input distributions for population synthesis models that aim to predict the outcome of massive young starburst regions in the near to far universe.

\begin{acknowledgements}
This work is based on observations collected at the European Southern Observatory under program IDs 090.D-0323 and 092.D-0136. The authors are grateful to the ESO staff for their support in the preparation and execution of the observations. L.A.A. acknowledges support from the Fundac\~ao de Amparo \`a Pesquisa do Estado de S\~{a}o Paulo - FAPESP (2013/18245-0 and 2012/09716-6). J.M.A. acknowledges support from the Spanish Government Ministerio de Econom{\'\i}a y Competitividad (MINECO) through grant AYA2013-40\,611-P. N.D.R. is grateful for postdoctoral support from the University of Toledo and the Helen Brooks endowment. S.d.M. acknowledges support by a Marie Sklodowska-Curie Action (H2020 MSCA-IF-2014, project ID 661502) and National Science Foundation under Grant No. NSF PHY11-25915. AFJM is grateful for financial aid from NSERC (Canada) and FQRNT (Quebec). R.B. acknowledges support from the Project FONDECYT Regular 1140076. MG acknowledges the Royal Society (University Research Fellowship) and the European Research Council (Starting Grant, grant agreement n. 335936) for support. O.H.R.A acknowledges funding from the European Union's Horizon 2020 research and innovation programme under the Marie Sklodowska-Curie grant agreement No. 665593 awarded to the Science and Technology Facilities Council. N.J.G. is part of the International Max Planck Research School (IMPRS) for Astronomy and Astrophysics at the Universities of Bonn
and Cologne.
\end{acknowledgements}

\appendix 
\section{Notes on individual objects}\label{app:notes}

\subsubsection*{VFTS~042}
With a combined spectral type of O9~III((n)), VFTS~042 is a clear SB2 system in our data, but the line profiles never fully separate. We fitted the line profiles using a single-Gaussian and a double-Gaussian profile to measured the RVs. The period obtained, 29.31\,d, is identical in both cases, but the SB2 approach provides a better representation of the line profiles throughout the epochs. The SB2 RV curve is also convincing and we adopt the SB2 orbital solution for this system. Obviously, the determination of accurate orbital parameters would benefit from spectral disentangling.   

\subsubsection*{VFTS~064}
This a long-period system for which we only observe one periastron passage. Within the range of our periodogram computation, the 900~d period has the highest peak and yields a plausible RV solution. However, adopting a period twice as long ($P = 1811~\rm d$) significantly improves the $\chi^2$ of the fit and brings the systemic velocity to $\gamma=279.1\pm1.4$~\kms, i.e. in perfect agreement with the average value of the 30~Dor region.

\subsubsection*{VFTS~087}
This object shows two peaks larger than 1\%\ FAP in the periodogram, the first one corresponding to a periodicity of 1 day; and the second to $P_\mathrm{orb} > 1000$ days. Both periods are possible observational aliases and yield unlikely RV curves and orbital solution.

\subsubsection*{VFTS~113}
The VFTS~\#113 time series has two peaks that pass the 0.1\%\ FAP threshold: 1~d, 450~d, 650~d, and 1000~d. Both result from the time sampling of our time series and, indeed, do not yield any realistic RV curves. No other significant periodicity was identified in the data.

\subsubsection*{VFTS~114}
VFTS~114 with $P_{\rm orb} \sim 27.28$\,d and $e \sim 0.5$ shows clear SB2 profiles with one of the components (the primary according to the line strength) being hotter and displaying a larger spin rate. Classified by \citep{WSSD14} as O8.5~IV + sec, the almost complete absence of \heb\l4541 in the secondary spectrum points towards a very late-O or an early-B spectral type for the secondary companion. We attempted to fit both an SB1 (based on single-Gaussian RV measurements)  and an SB2 solution but the RVs obtained in the former case are not correct in this very clear SB2 system. While the lines do not deblend well at all phases, the fact that the \heb\ lines are only present in one of the components helps to bootstrap the Gaussian fitting. The SB2 solution indicates an almost equal-mass system. Given the likely difference of rotation rate, VFTS~\#114 is a candidate post-/current-interaction binary where the component with the faster spin has been spun up by mass and angular momentum transfer in a past or 
ongoing RLOF event.

\subsubsection*{VFTS~116}
VFTS~116 ($P_{\rm orb} \sim 23.9$\,d, $e \sim 0.24$) is a clear SB2 system (O9.7: V: + B0: V:). The spectrum is relatively noisy, but the signature of the two components is clearly seen in the \hea\ lines. However, it is unclear whether the two Gaussian profiles towards which our code converges is the best representation of the two line profiles or, even, that the solution is unique. Nevertheless, the SB2 solution gives better results than the SB1 solution and we adopted the former. We caution, however, that in addition to the orbital period, the other orbital parameters should be viewed as preliminary only.

\subsubsection*{VFTS~184}
The periodogram displays one clear peak at $P_\mathrm{orb}\approx 32$~d with a significance better than 0.1\%. The RV curve obtained when adopting the 32 d period is plausible albeit noisy.

\subsubsection*{VFTS~259}
This target classified as O6\,Iaf shows peaks in its periodogram that pass the 1\%\ significance cutoff at 1 and 3.7 d, but these do not yield convincing orbital solutions. Dedicated  study of the line profile variability is needed to shed more light on the nature of these variations.

\subsubsection*{VFTS~267}
This object has a spectral type O3\,III-I(n)f* and shows two peaks just larger than 1\%\ FAP at periods close to 1000~d, however, they are likely the results of the sampling and lead to a clustering of the data in the phase diagram. Inspection of the spectra seems to indicate line profile variability, including changing asymmetry. Further work is needed to decide whether this is the result of a strong blend in a SB2 system or the signature of atmospheric activities.

\subsubsection*{VFTS~352}
VFTS~352 is a SB2 system (O4.5 V(n)((fc)):z:\,$+$\,O5.5 V(n)((fc)):z:) with $P_{\rm orb}\,\sim\,1.12$~d and $e\,\sim\,0.0$. Recently, \citet[][]{Almeida+2015} showed a detailed study of this binary and concluded that it is the most massive and hottest overcontact binary known so far. Although our analysis was only with spectroscopic data and \citet[][]{Almeida+2015} additionally used photometry, our solutions are in good agreement.

\subsubsection*{VFTS~404}
VFTS~404 \heb\ lines show clear line profile variations indicative of a double-lined binary. However, the components never separate well enough to allow us to fit two Gaussians. The periodogram reveals a clear peak at 146~d, whose significance is better than 1\%\, but does not reach the 0.1\%\ FAP cutoff. The limited significance despite the clarity of the binary signal likely results from the fact that we have been unable to separate the line profiles of the two components. While the SB1 RV curve is well behaved, parameters such as the amplitude of the RV curve and the eccentricity should be considered preliminary.

\subsubsection*{VFTS~432}
VFTS~432 is a O3.5 V((f)), a quite rapid rotator and one of the poorest orbital solutions. The RV curve seems to indicate some clustering of the RV measurement either above or below the curve. This may result from an undetected companion or from line profile variability. We include this system among our targets with periodicity but no coherent RV curve.

\subsubsection*{VFTS~440}
This O6-6.5 II(f) star shows a shift of about 15~\kms{} between the VFTS and TMBM campaigns, possibly indicating a long-period system. The periodogram reveals a strong peak at 1012~d which does not coincide with the 1600~d peak in the power spectrum of the object time series. Adopting this period yields a reasonable RV curve and an orbital solution with $e \sim 0.28$, although we caution that our data do not cover a full cycle.

\subsubsection*{VFTS~445}
VFTS~445 shows strong SB2 line profiles in the \heb\ll4200, 4541 lines at three separate epochs in our TMBM time series and once in the VFTS data. The separations between the components reach 275~\kms. Unfortunately, no clear period is identified in the periodogram: peaks at $\approx$2.1~d, 0.7~d and 2.9~d pass the 1\%~FAP threshold. However, none of these lead to realistic RV curves. This may result from the limited quality of the RV measurements resulting from the fact that the lines never fully deblend, hence the shape of the Gaussian profiles, and more importantly their relative width and intensities, are poorly constrained. More work possibly including disentangling is required, but is beyond the scope of this paper.

\subsubsection*{VFTS~450}
VFTS~450 is a clear SB2 binary (O9.7 III: + O7:) with $P_{\rm orb} \sim 6.89$\,d, $e \sim 0.06$ and showing very large RV variation ($\drv > 400$~\kms). The O9.7~III companion shows a \hea\ line about 5 to 6 times stronger than the O7 companion, indicating a large luminosity ratio. While the secondary RV measurements are relatively imprecise ($\sigma_\mathrm{RV2}\approx 10$~\kms), the best-fit orbital solution indicates that the companions have very similar masses ($M_1/M_2=1.01\pm0.03$). This system was recently studied by \citet{Howat+2015}. Further work beyond Gaussian fitting is needed to improve the secondary RVs.

\subsubsection*{VFTS~526}
The periodogram displays one clear peak at $P_\mathrm{orb}\approx 10$~d with a significance better than 1\%; however, no coherent RV curve was found.

\subsubsection*{VFTS~527}
VFTS~527 shows clear double-line profiles and it was spectroscopically classified as O6.5 Iafc $+$ O6 Iaf by \citet[]{WSSD14}. This system was studied in detail by \citet[][]{TES11} and has $P_{\rm orb} \sim 153.9$\,d and $e \sim 0.46$. Our fitted orbital period and eccentricity are in good agreement with the less precise values from the latter; however, our estimates of the component masses ($m\,sin^3\,i$) are lower than those done in \citet[][]{TES11} by $\sim18\%$ and $24\%$ for the primary and secondary, respectively.

\subsubsection*{VFTS~588}
The VFTS~588 periodogram shows a peak at 1.46~d, but the peak does not pass the 1\%\ FAP threshold. A RV curve folded with this period indicates a highly eccentric system, a very unlikely configuration given the short orbital period. A visual inspection of the spectra seems to indicate the presence of line profile variations but the robustness of this conclusion is hard to assess given the relatively limited S/N of the data. Given that the period does not pass the significance threshold and yields an improbable orbital solution, we list VFTS~588 among our targets with no periodicity found.

\subsubsection*{VFTS~739}
VFTS 739 seems to present a composite spectrum resulting from two single objects in the same line of sight. The brightest one would be in the LMC ($\gamma \sim 256$~\kms) and the second, fainter one, in the foreground ($\gamma \approx 65$~\kms).

\subsubsection*{VFTS~750}
VFTS~750 is a late O-type star with clear RV variations with peak-to-peak amplitude of about 60~\kms{} during the VFTS campaign. We did not observe such a large RV variation during the TMBM campaign even though the object is clearly variable. In the periodogram the peak at about 421~d passes the 1\%\ significance and such a period allows us to derive a reasonable though quite eccentric ($e=0.74$) orbital solution. New observations covering multiple epochs of maximum RV variation are needed to confirm the orbital period and eccentricity.

\subsubsection*{VFTS~764}
VFTS~764 is a O9.7 Ia Nstr star. Its periodogram presents a peak at 1.2 d, but it yields no coherent RV curve.

\subsubsection*{VFTS~774}
VFTS~774 was classified as O7.5 IVp + O8.5: V: on the basis of a double-lined \heb\l4686 by the VFTS.  Unfortunately, our new data do not cover that line. The individual epoch data are noisy. Combined with the broadness of the lines, the RV measurement accuracy is on average 25~\kms, i.e. one of the worst of the campaign. The only peak in the periodogram that passes the 1\%\ significance threshold corresponds to the one-day alias, but does not yield a coherent RV curve.

\subsubsection*{VFTS~810}
VFTS 810 is an O9.7 V + B1: V: binary with $P_{\rm orb} \sim 15.7$\,d, $e \sim 0.68$, and clear SB2 signatures in the \hea\ lines. The lines are difficult to separate given the relatively limited S/N of this object. The B1 component seems to have a larger projected rotational velocity than the primary, but the Gaussian fitting does  model all epochs  well. The periodogram reveals one clear peak at 15.69~d and another one at 0.92~d. However, none of these pass the 1\%\ significance level. This possibly results from the large number of epochs where the spectra are blended. Given the uncertainties, we limited the orbital solution fitting to the narrower primary component; however, the obtained SB1 solution is realistic and the period is likely robust. Spectral disentangling may improve the RV fitting and the orbital solution in future work.

\section{Periodograms and radial velocity curves}\label{app:B}

\begin{figure*}
\centering
\includegraphics[width=4.4cm,angle=90]{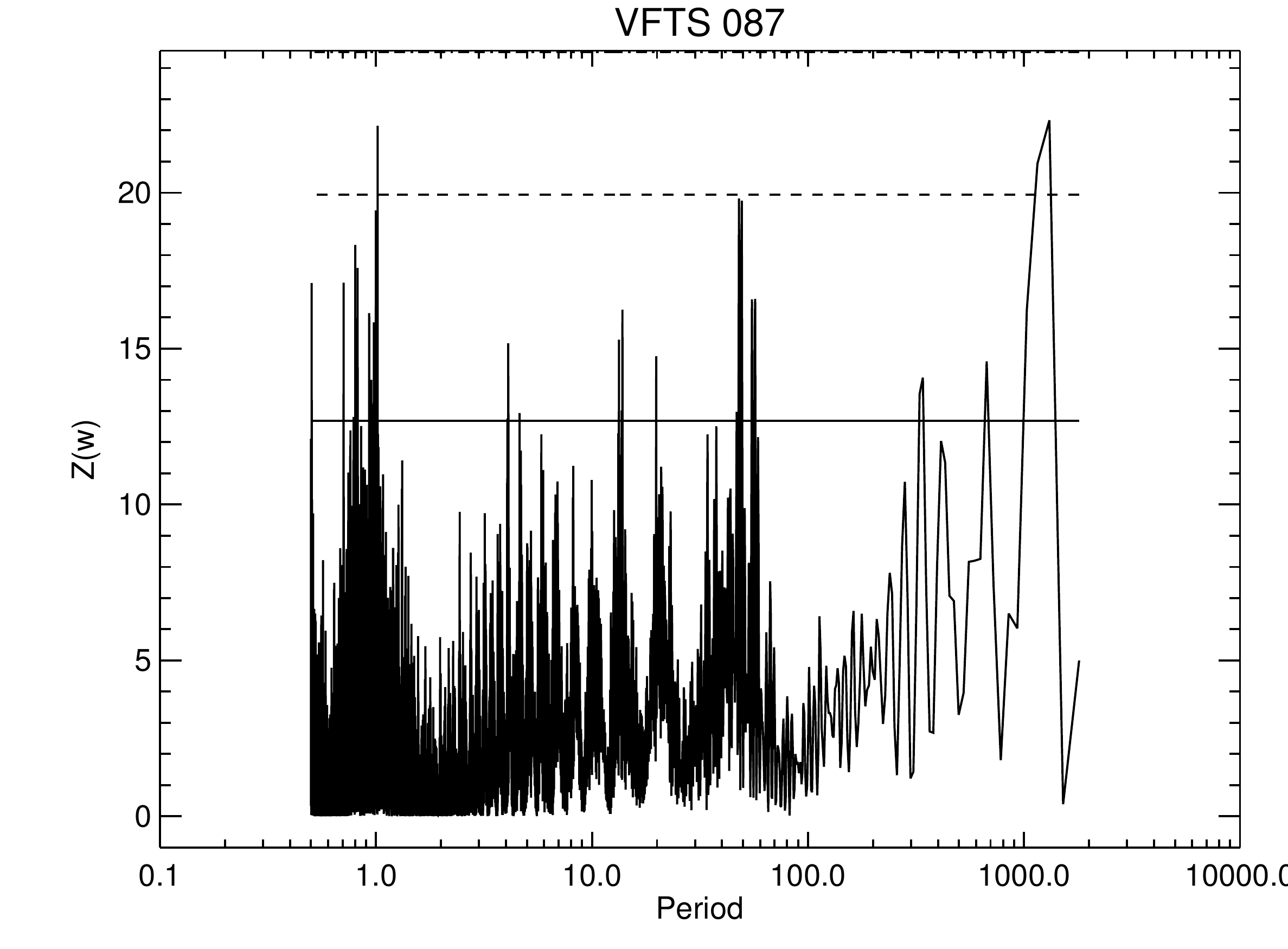}
\includegraphics[width=4.4cm,angle=90]{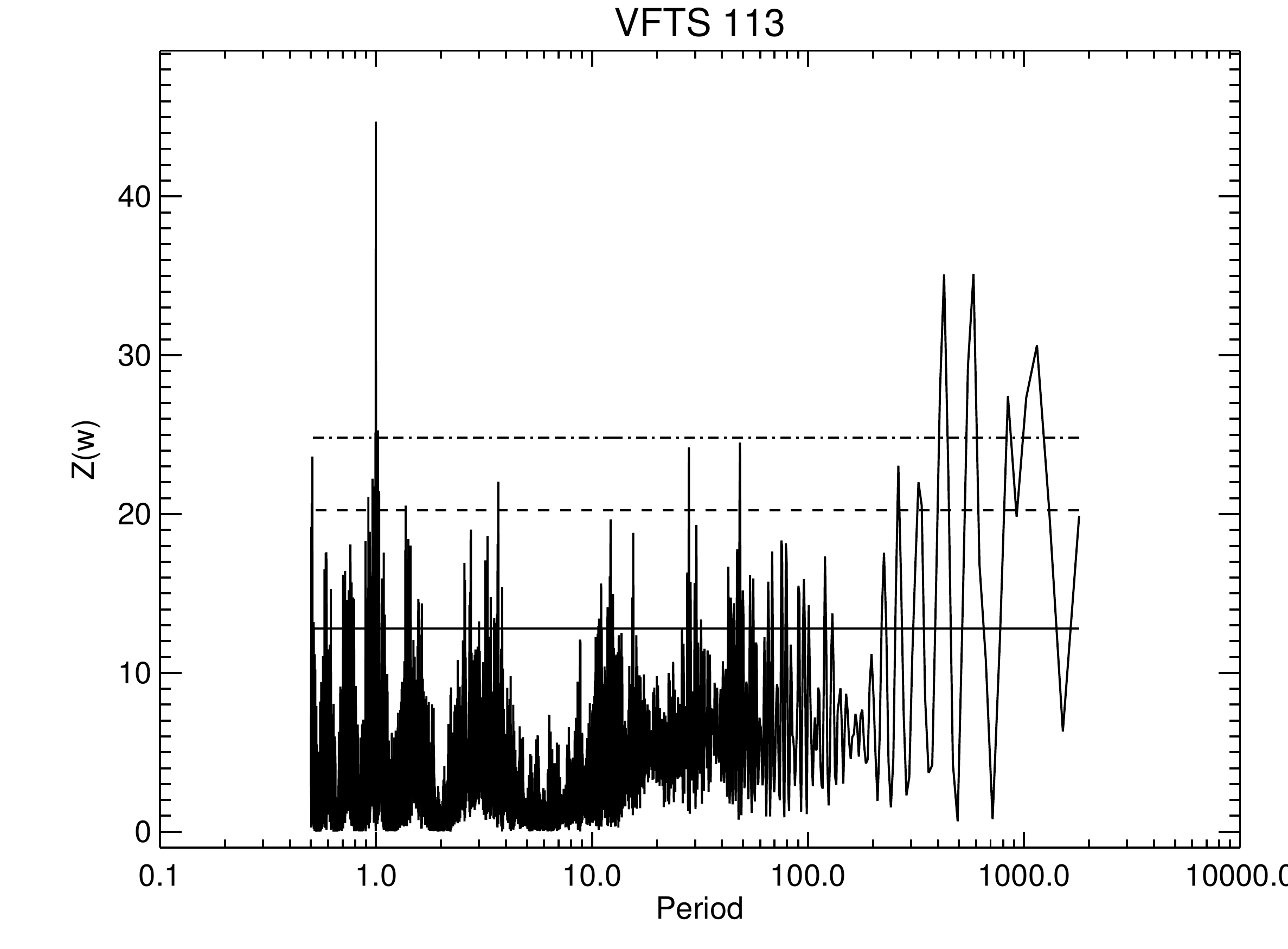}
\includegraphics[width=4.4cm,angle=90]{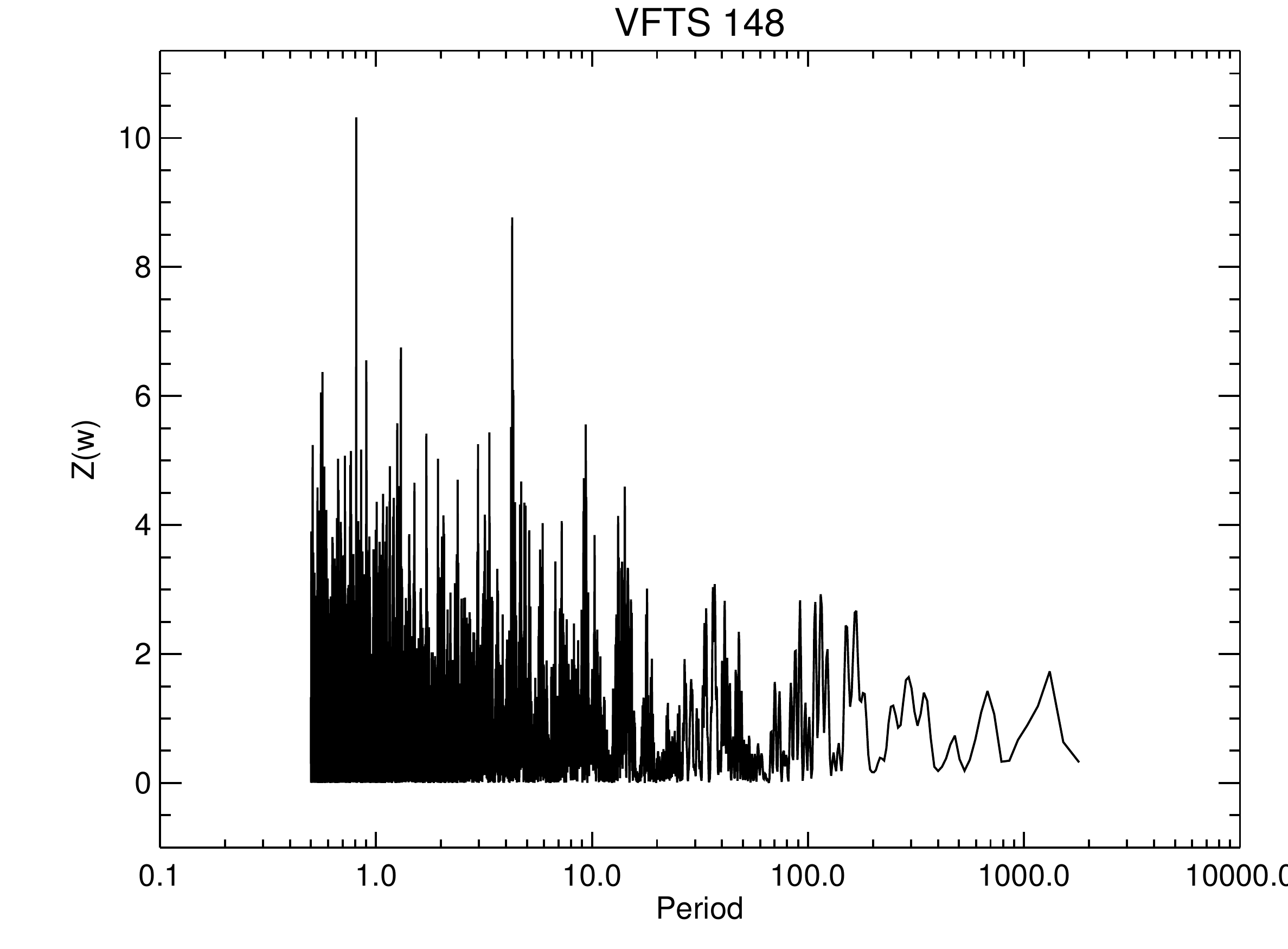}\\
\includegraphics[width=4.4cm,angle=90]{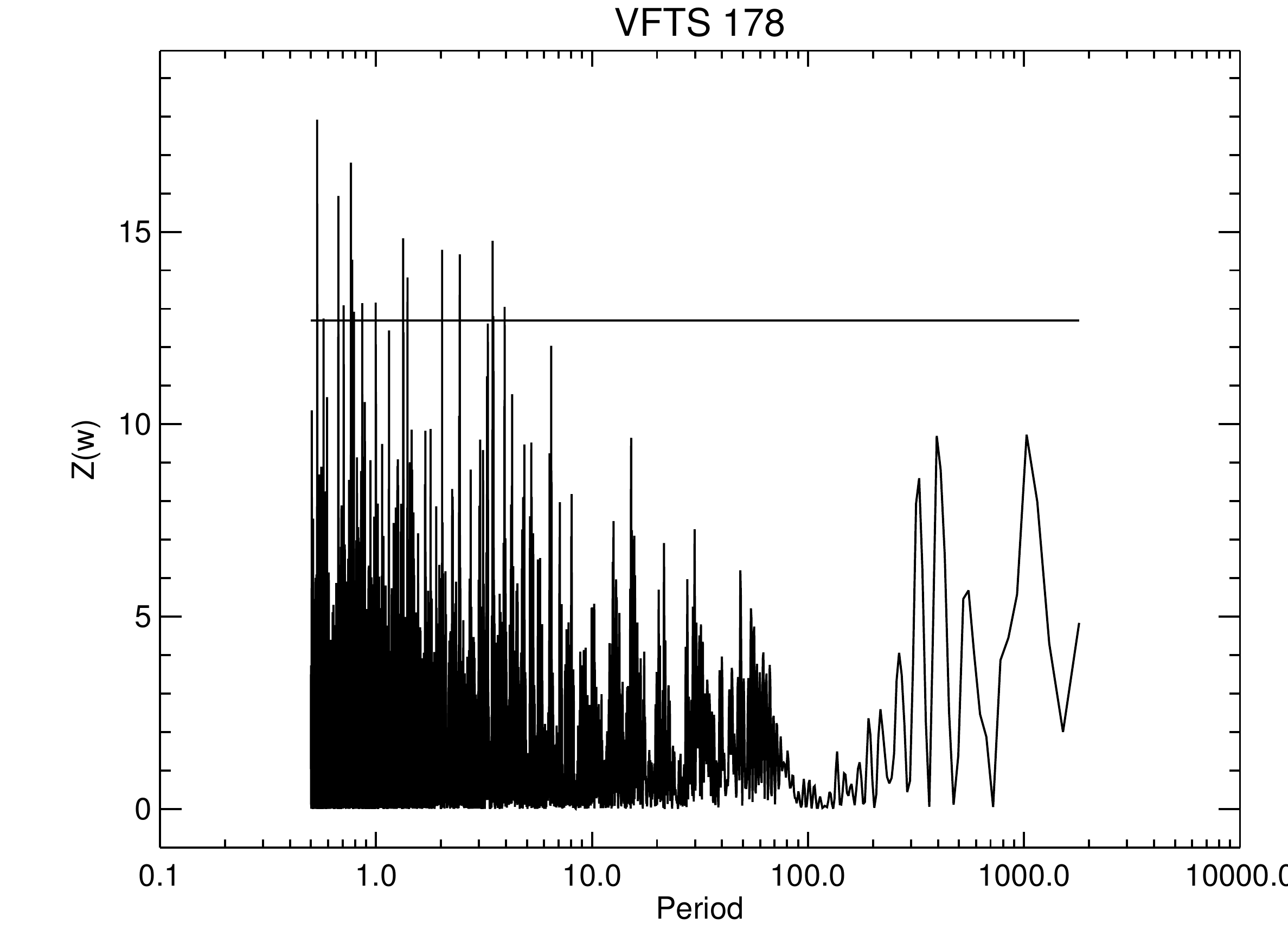}
\includegraphics[width=4.4cm,angle=90]{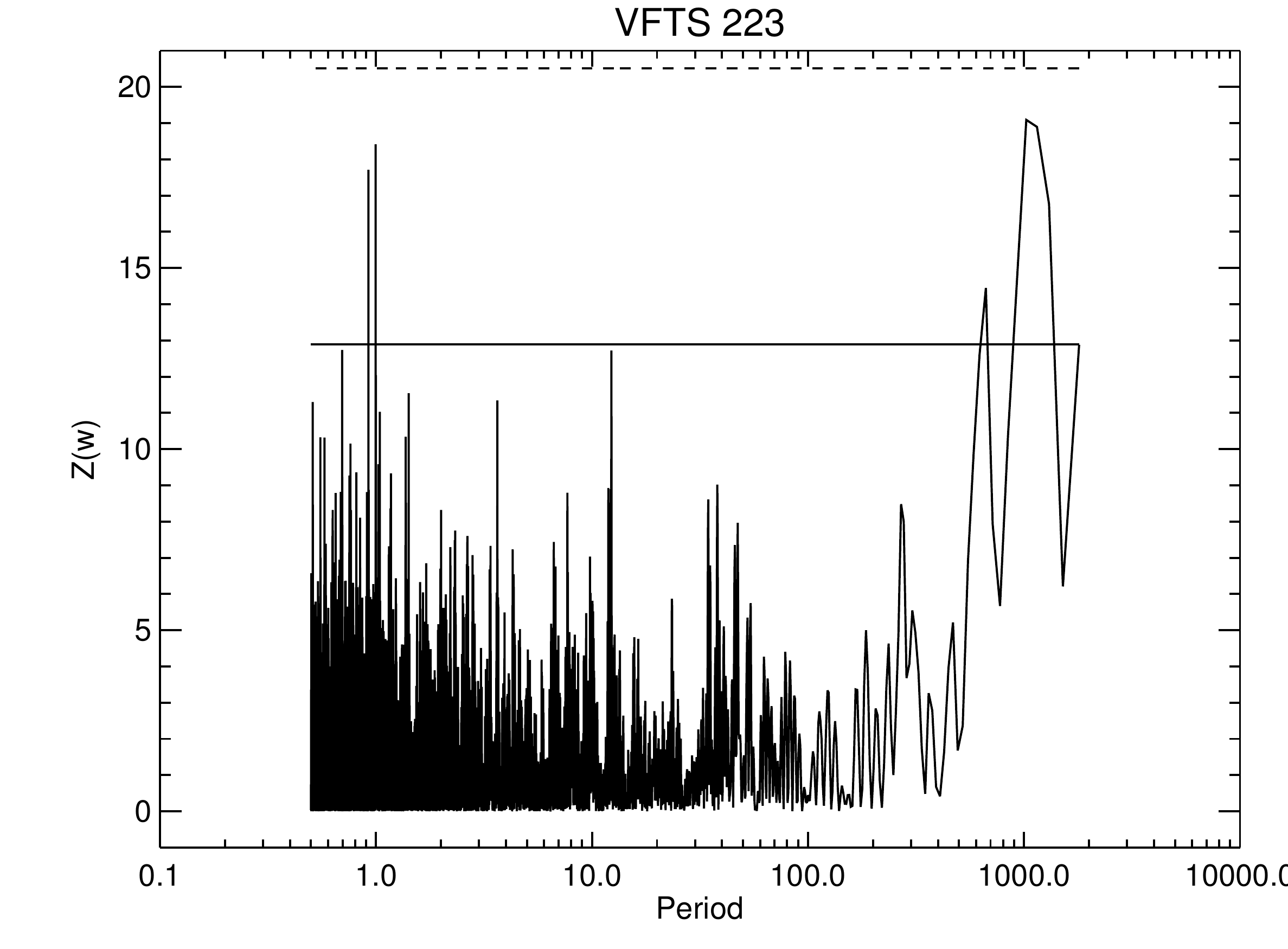}
\includegraphics[width=4.4cm,angle=90]{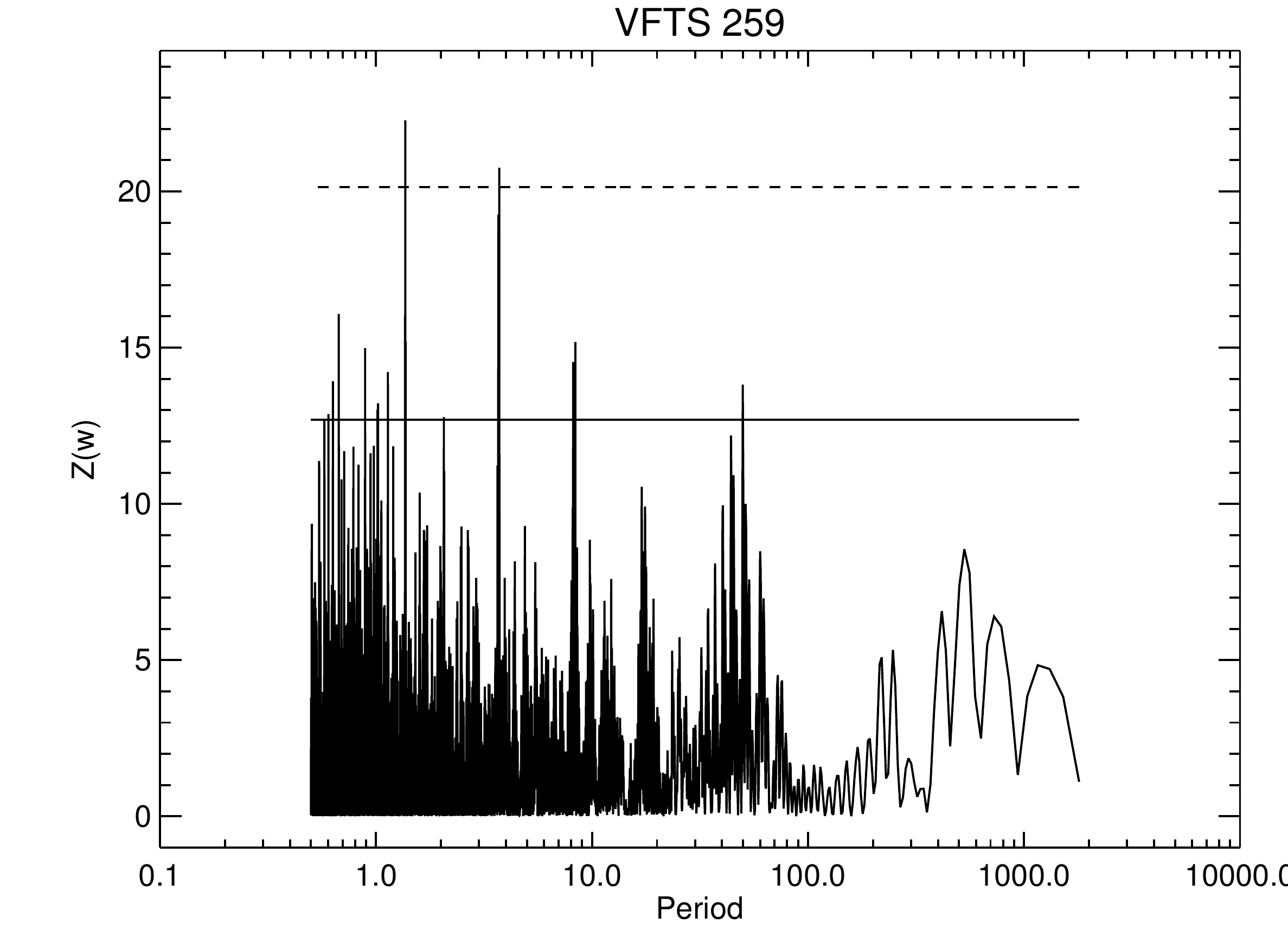}\\
\includegraphics[width=4.4cm,angle=90]{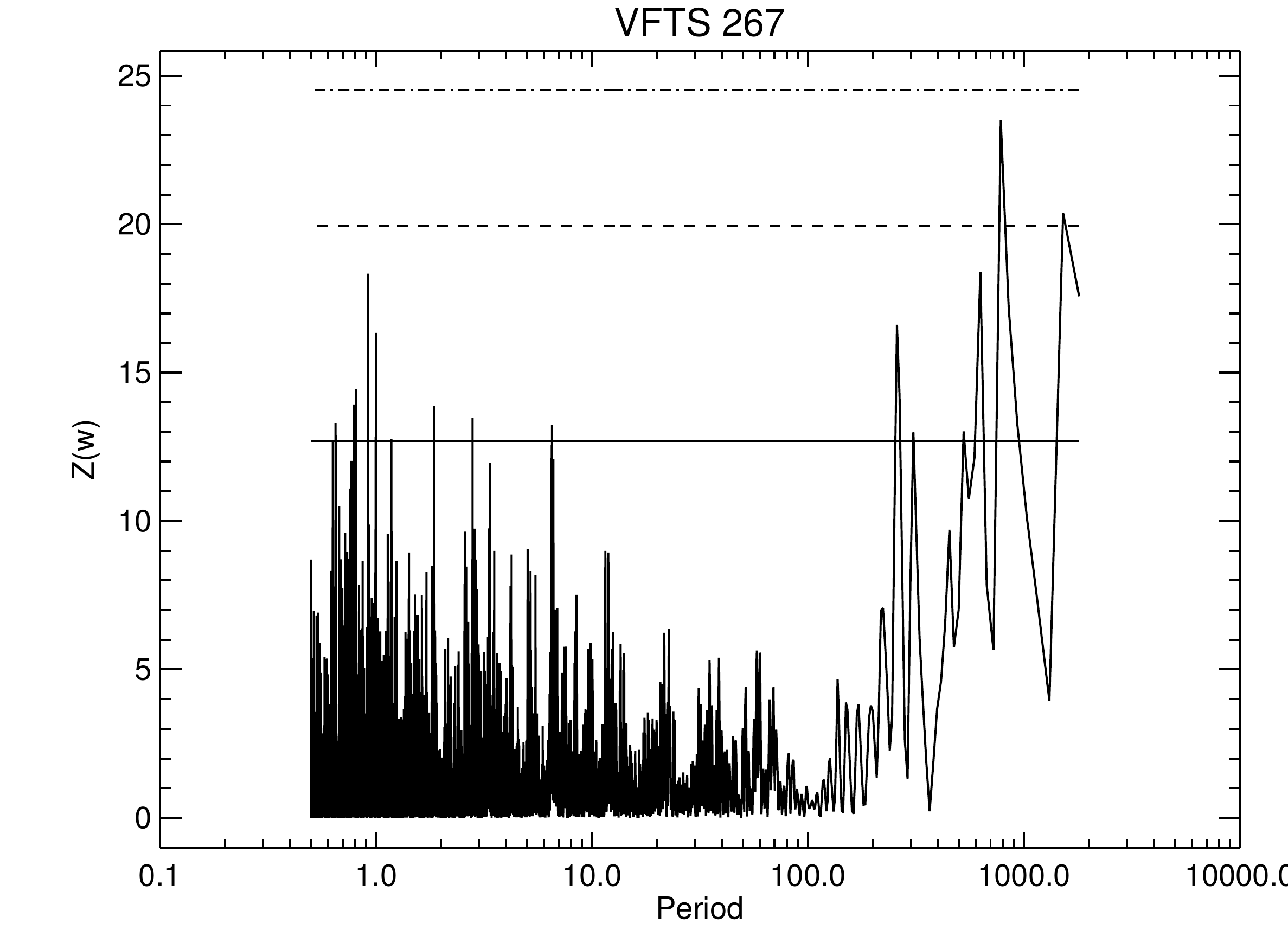}
\includegraphics[width=4.4cm,angle=90]{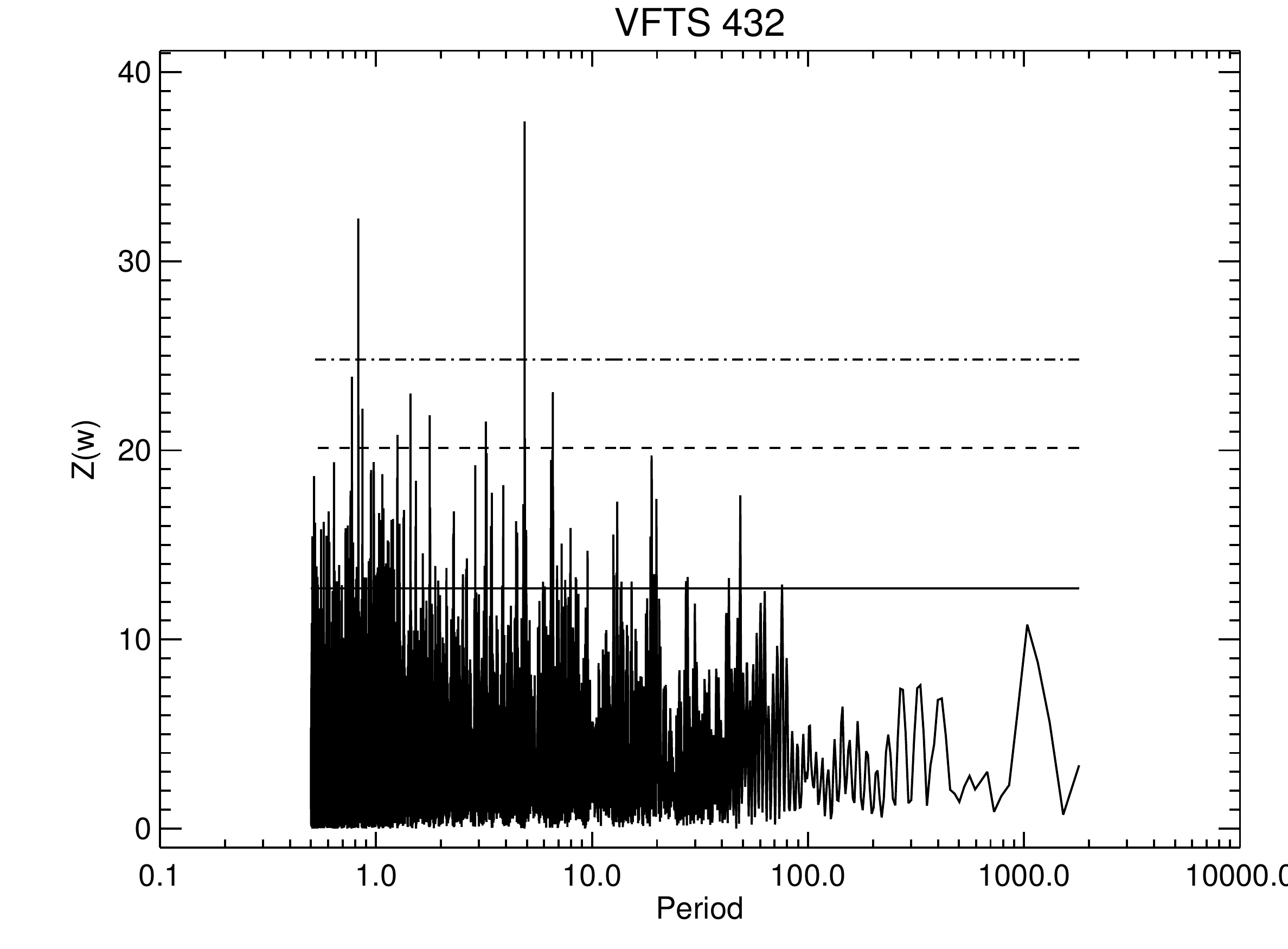}
\includegraphics[width=4.4cm,angle=90]{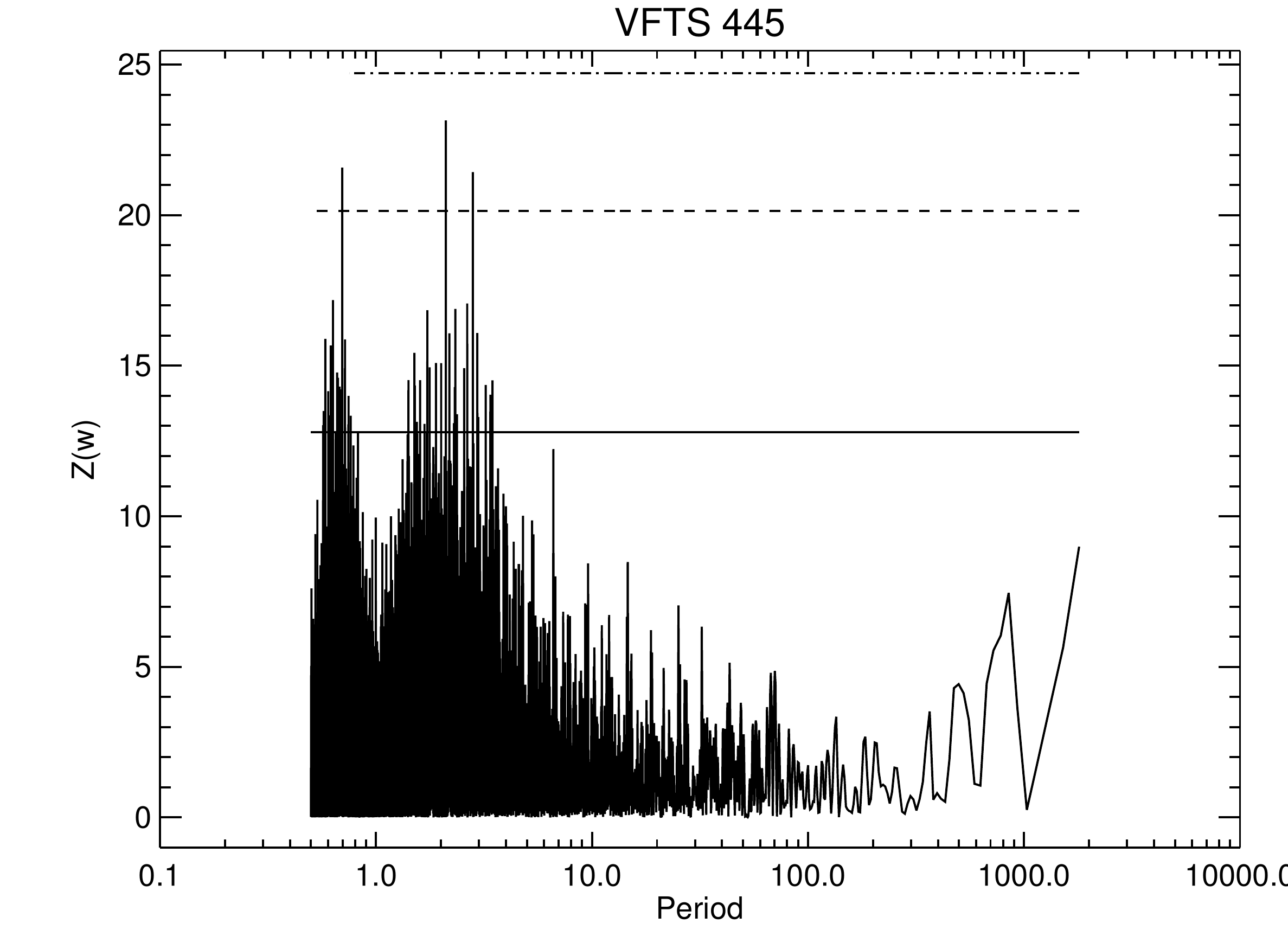}\\
\includegraphics[width=4.4cm,angle=90]{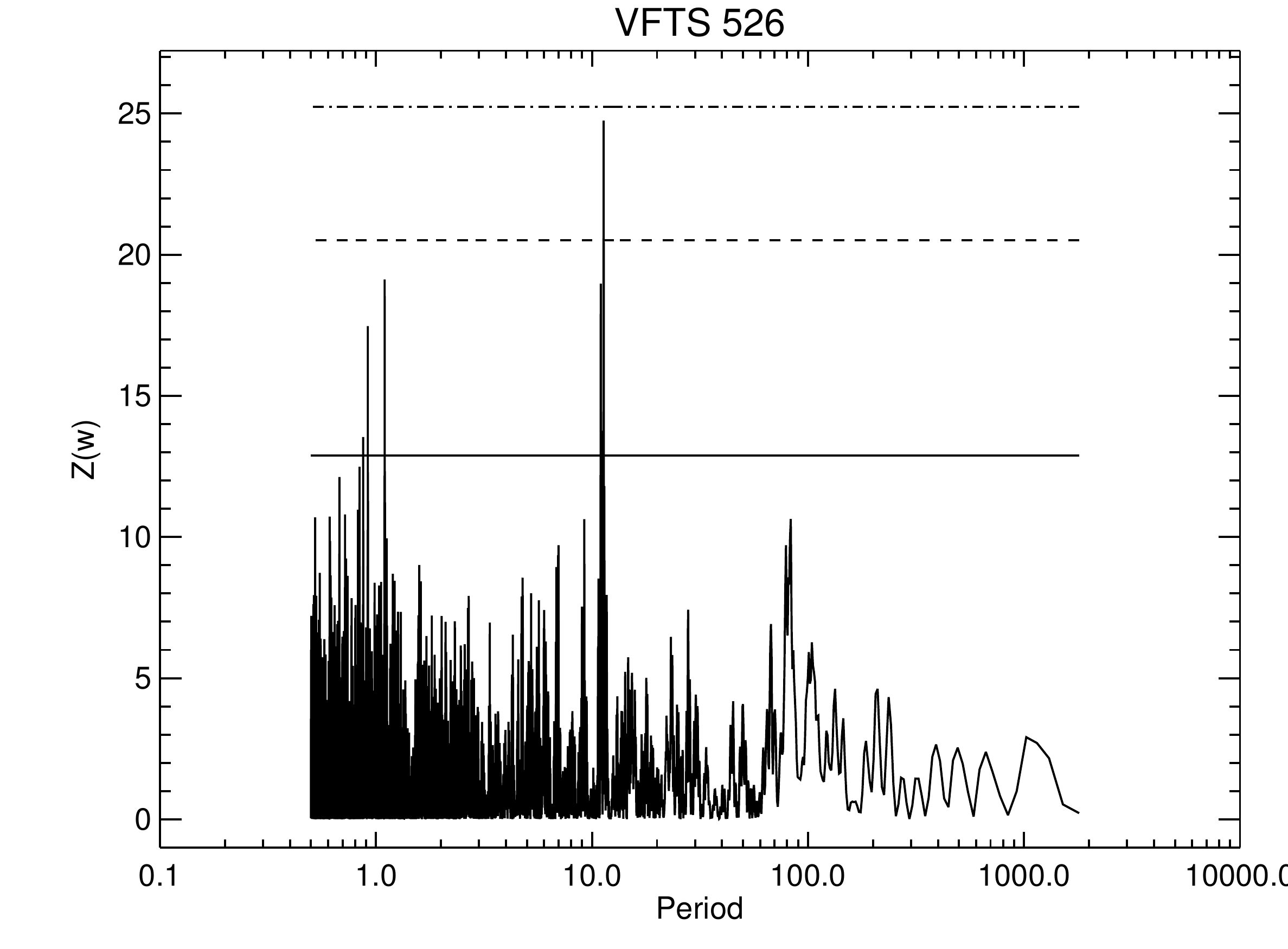}
\includegraphics[width=4.4cm,angle=90]{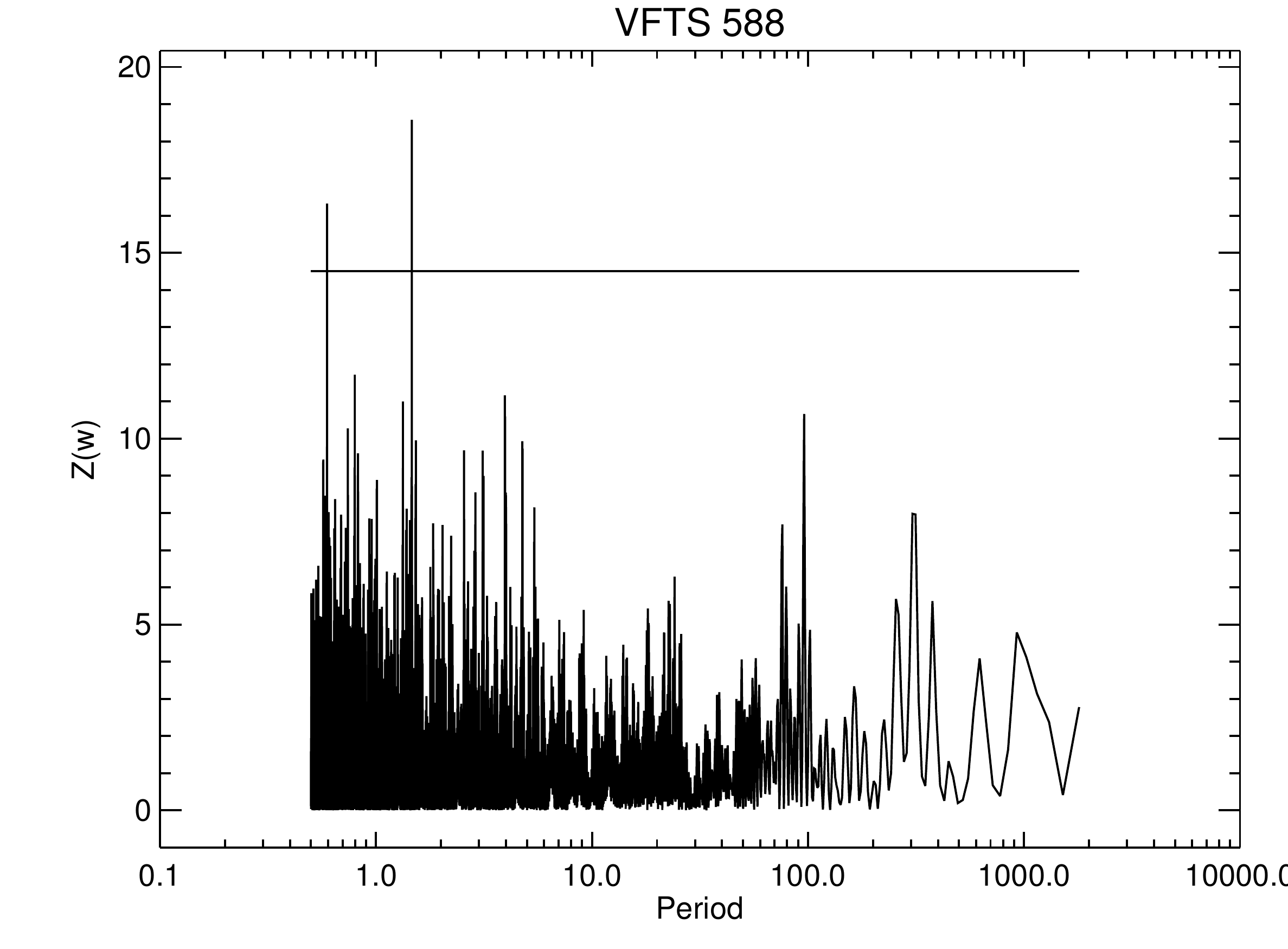}
\includegraphics[width=4.4cm,angle=90]{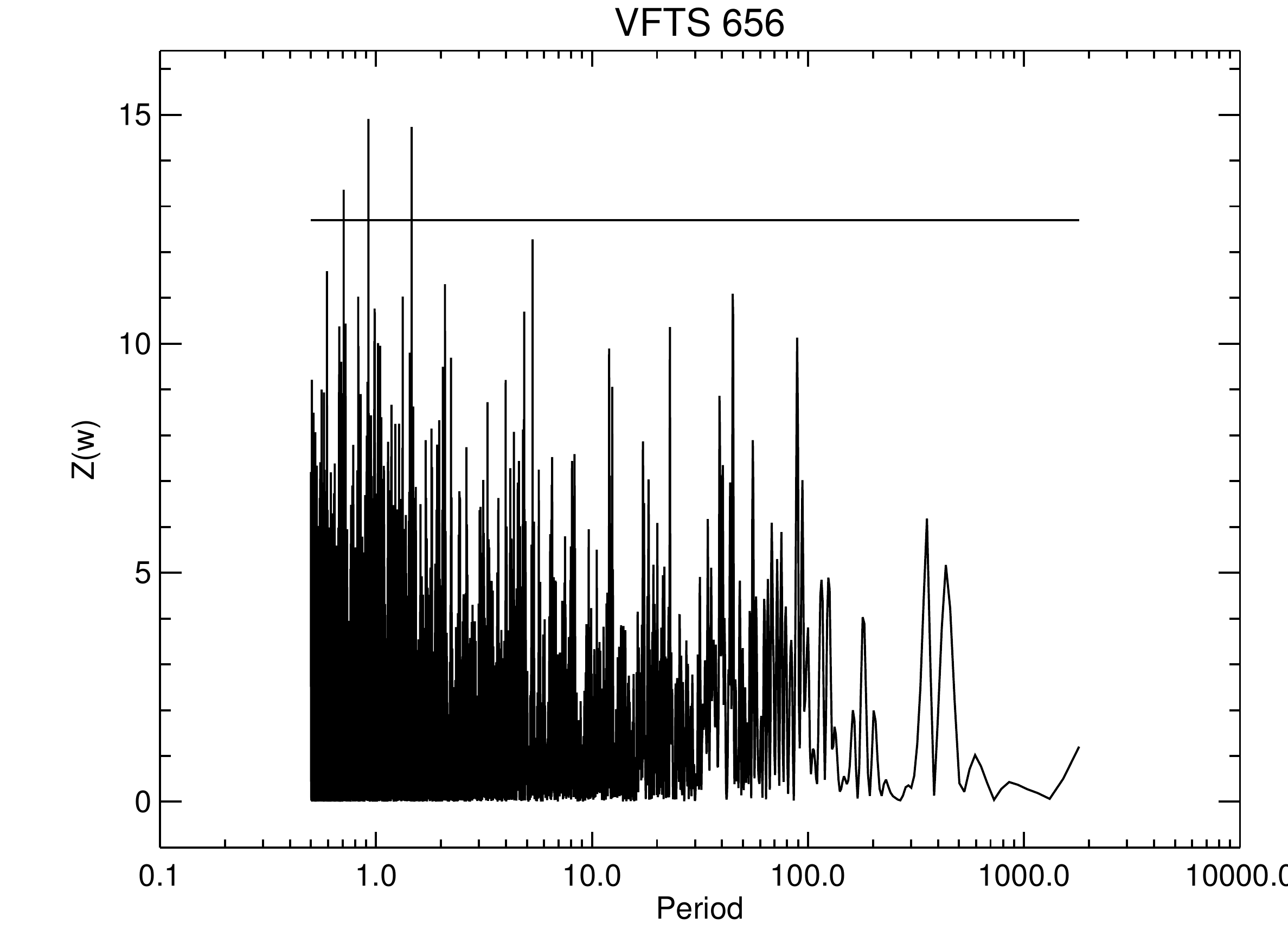}\\
\includegraphics[width=4.4cm,angle=90]{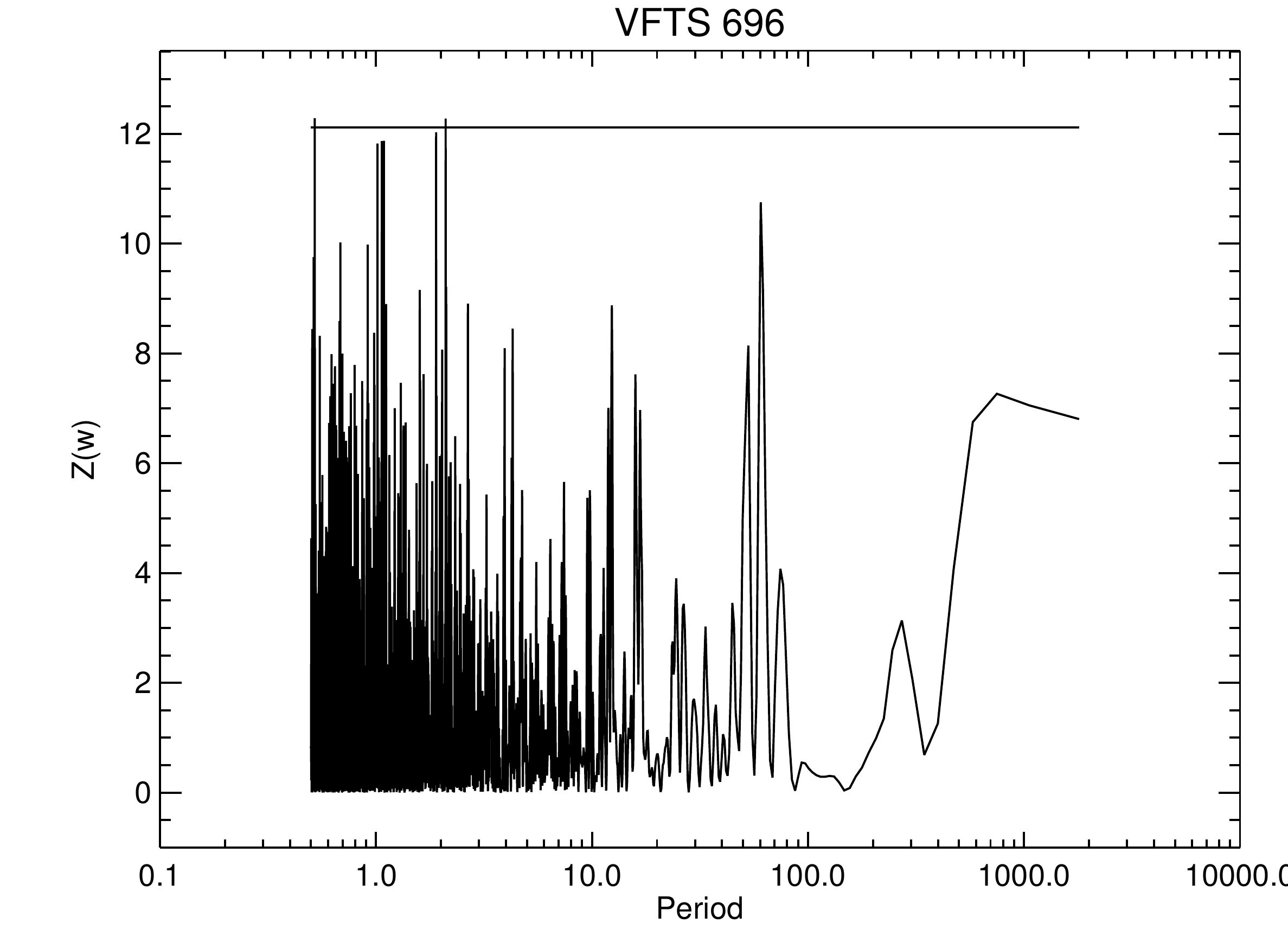}
\includegraphics[width=4.4cm,angle=90]{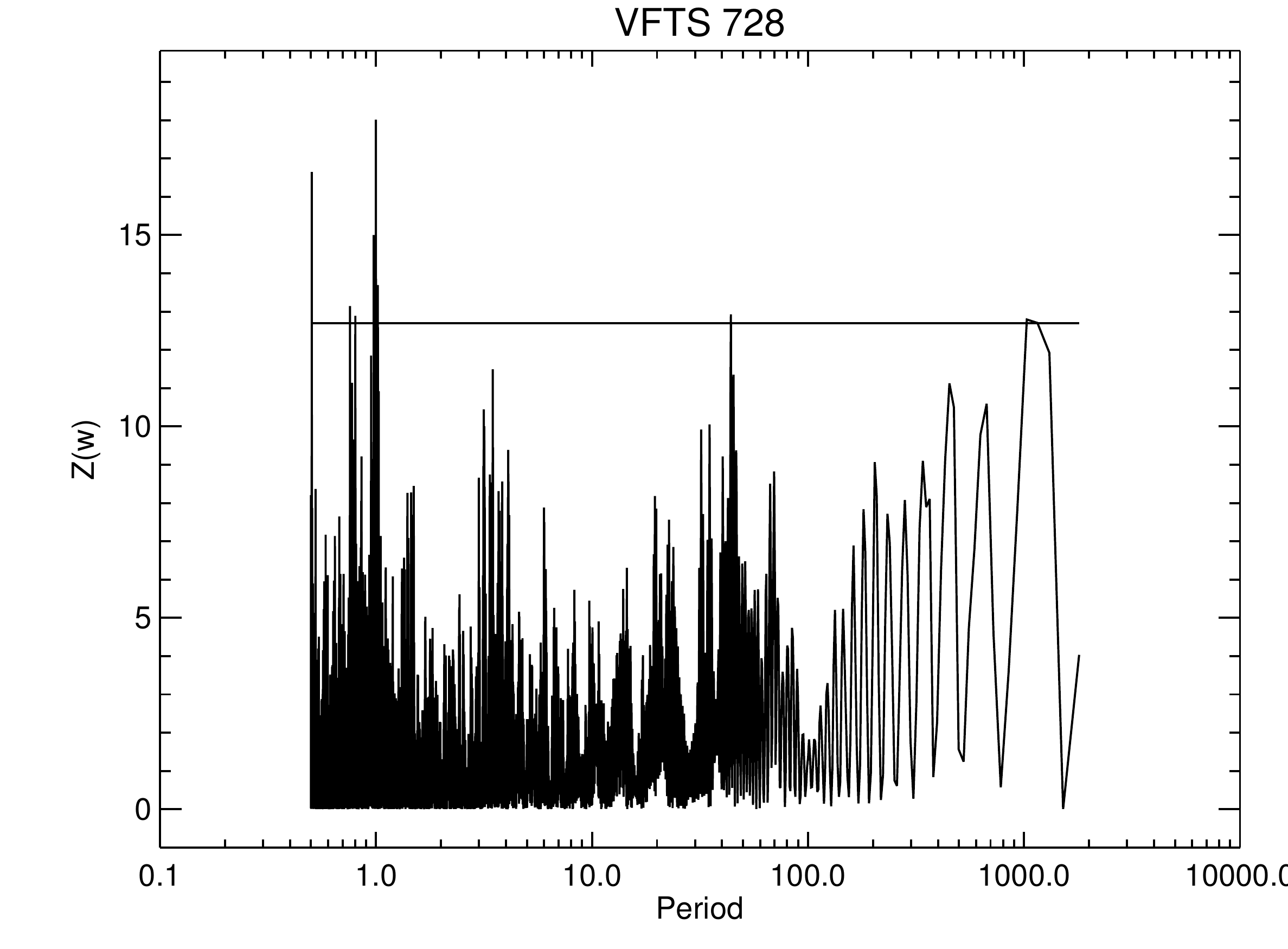}
\includegraphics[width=4.4cm,angle=90]{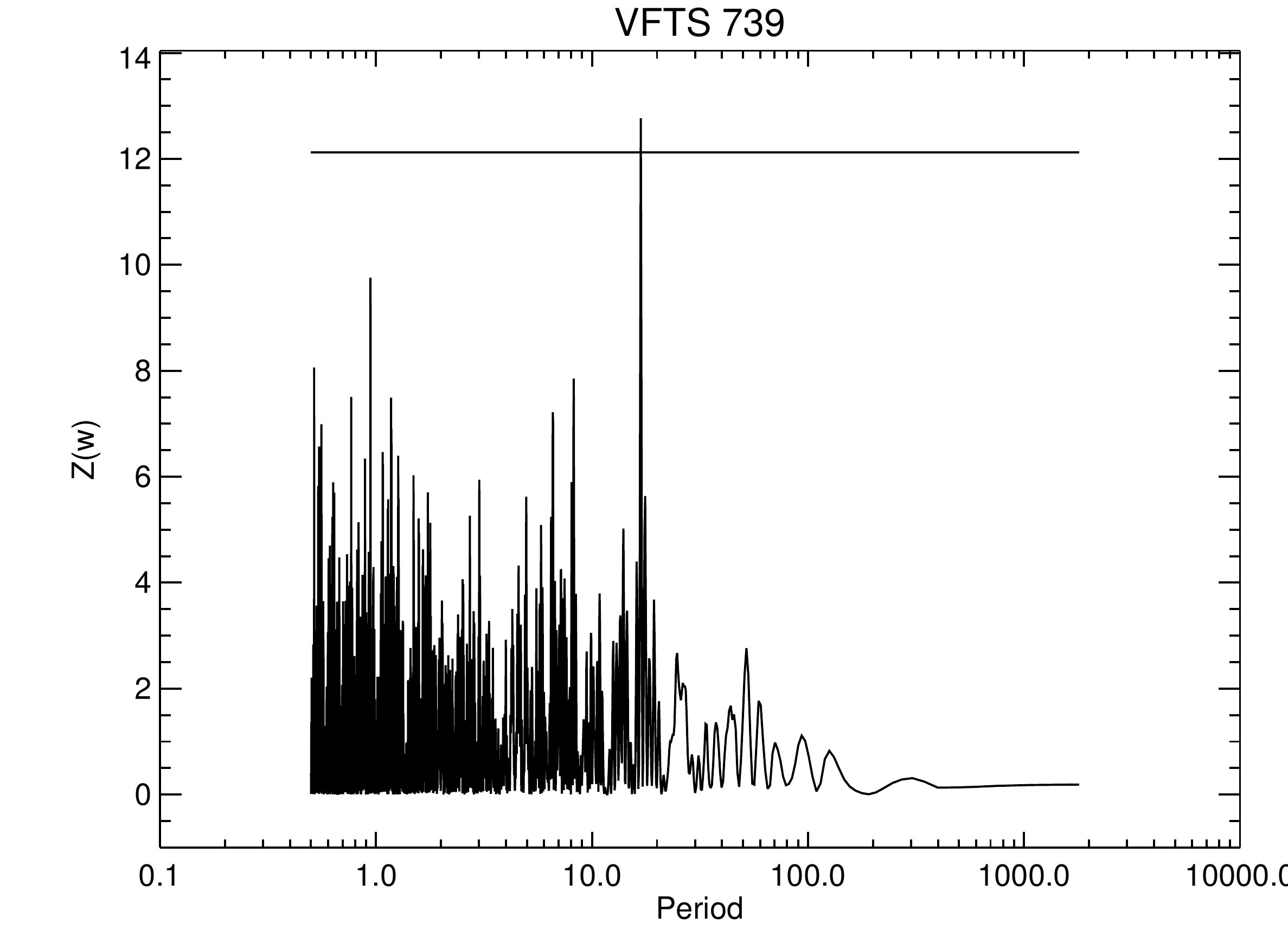}
\caption{Lomb-Scargle periodograms for the systems which do not show periodicities. The solid, dashed, and dot-dashed lines represent the 50\%, 1\%, and 0.1\% false alarm probabilities, respectively.}
\label{no_period:periodogram}
\end{figure*}

\begin{figure*}
\centering
\ContinuedFloat
\includegraphics[width=4.4cm,angle=90]{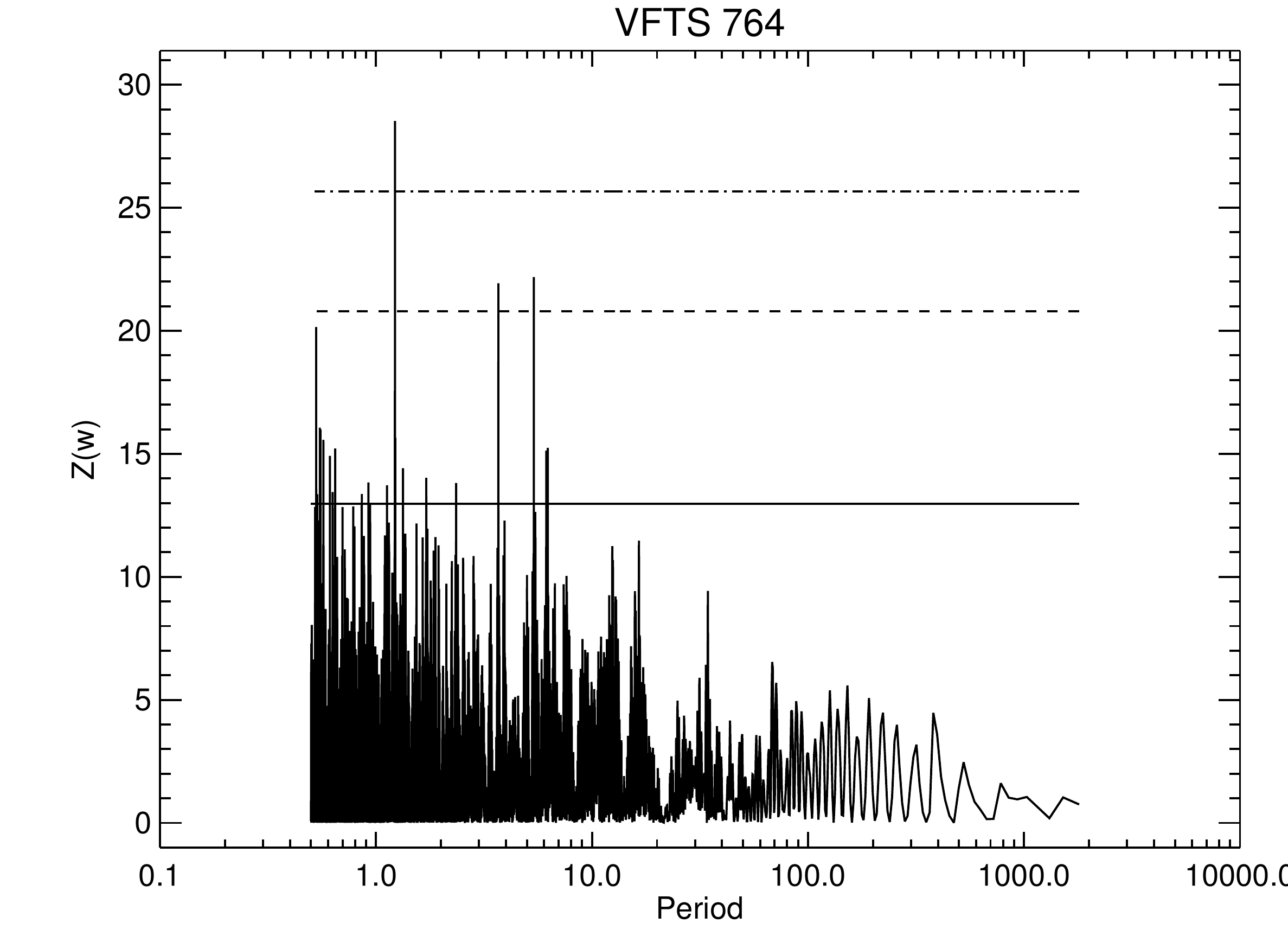}
\includegraphics[width=4.4cm,angle=90]{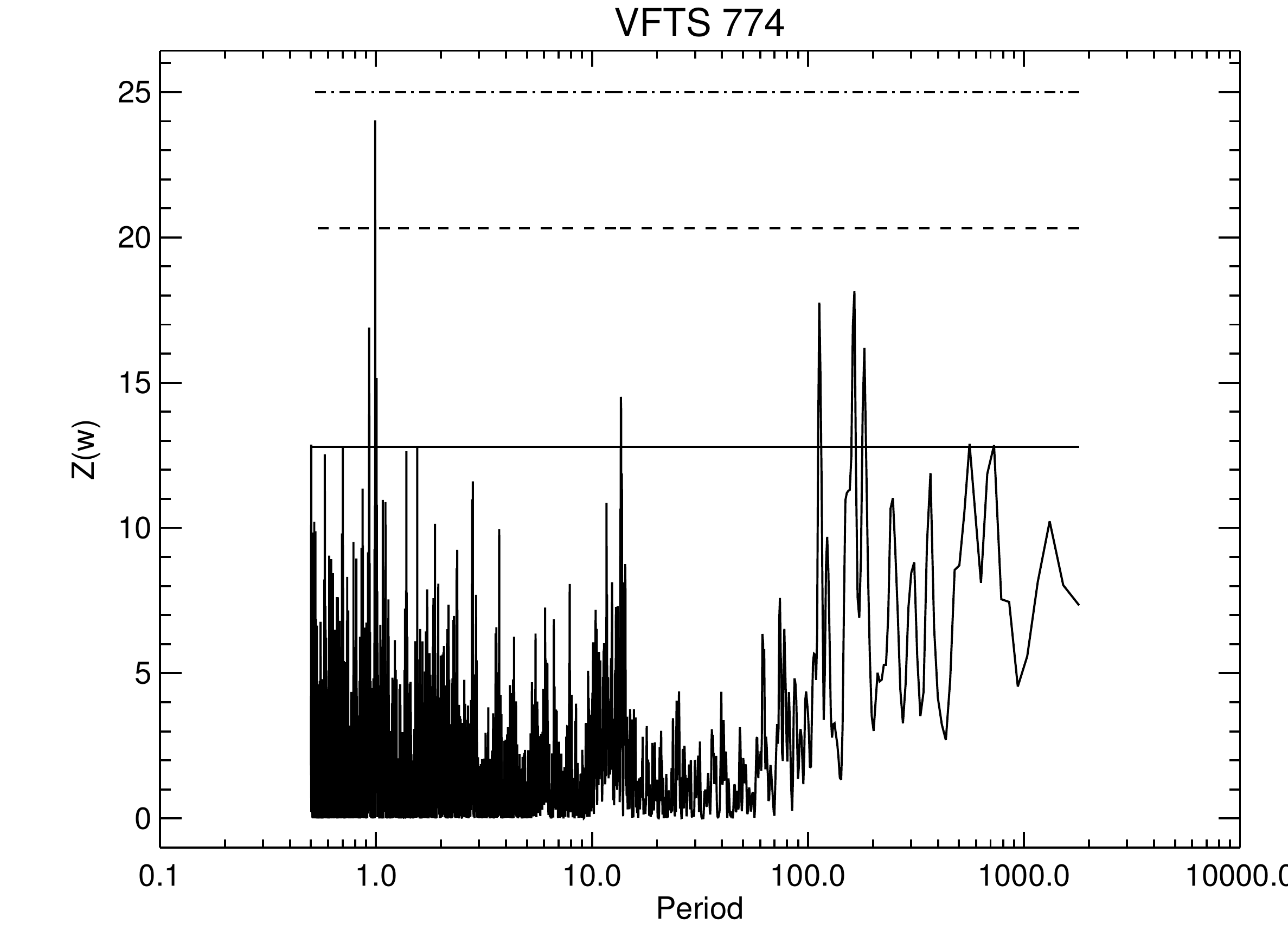}
\includegraphics[width=4.4cm,angle=90]{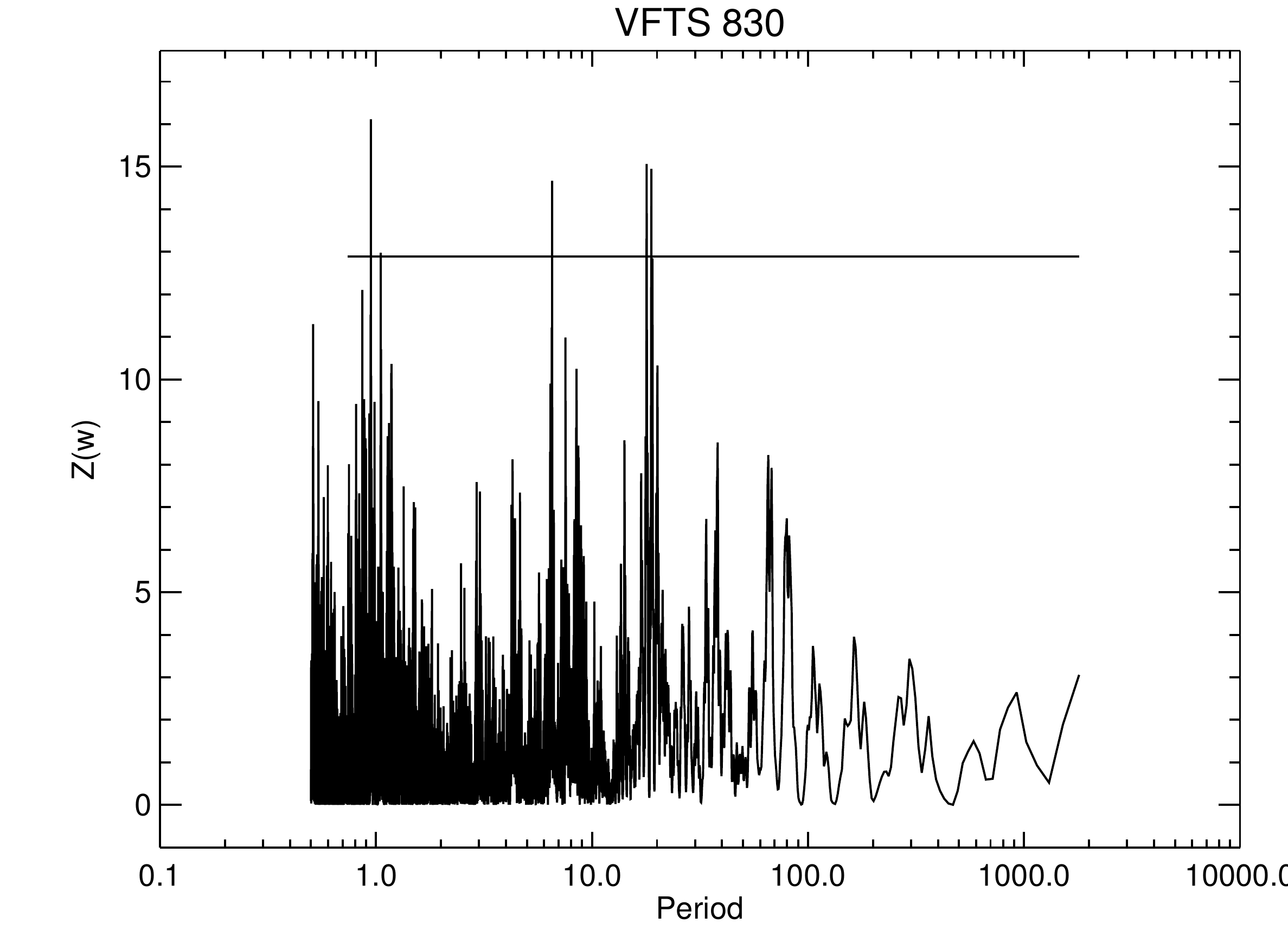}
\caption{{\it Continued...}}
\end{figure*}

\begin{figure*}
\centering
\includegraphics[width=4.4cm,angle=-90]{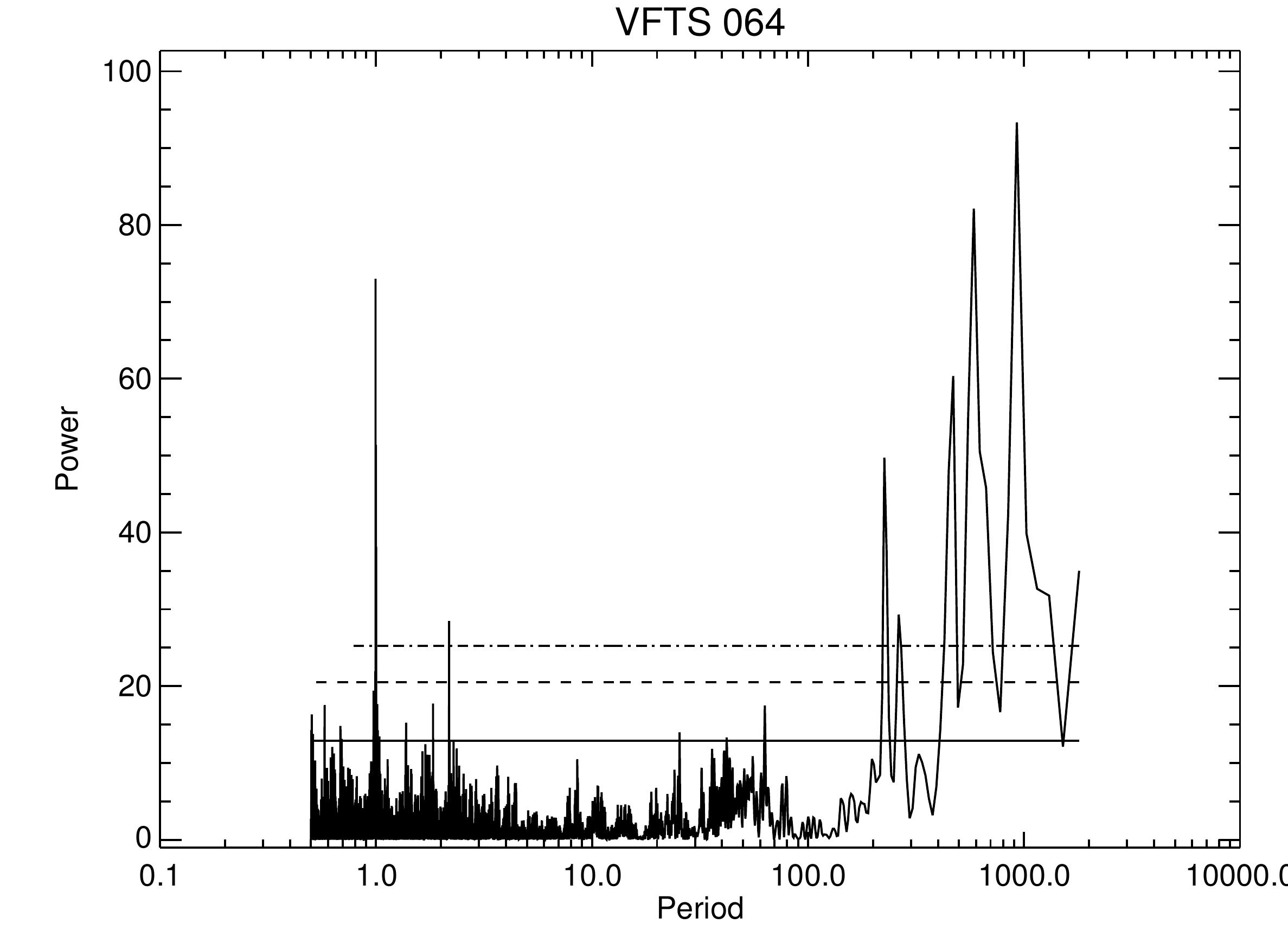}
\includegraphics[width=4.4cm,angle=-90]{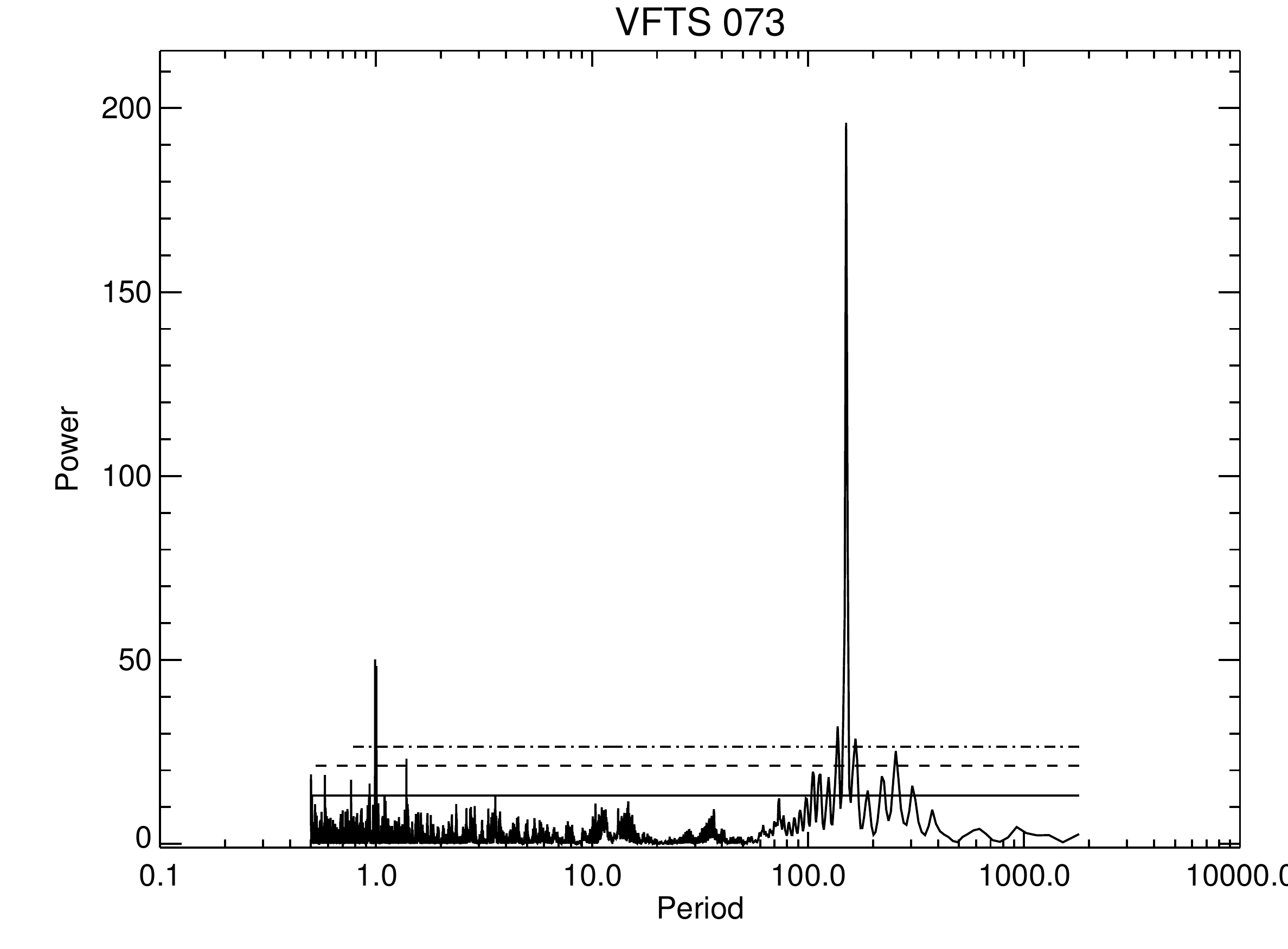}
\includegraphics[width=4.4cm,angle=-90]{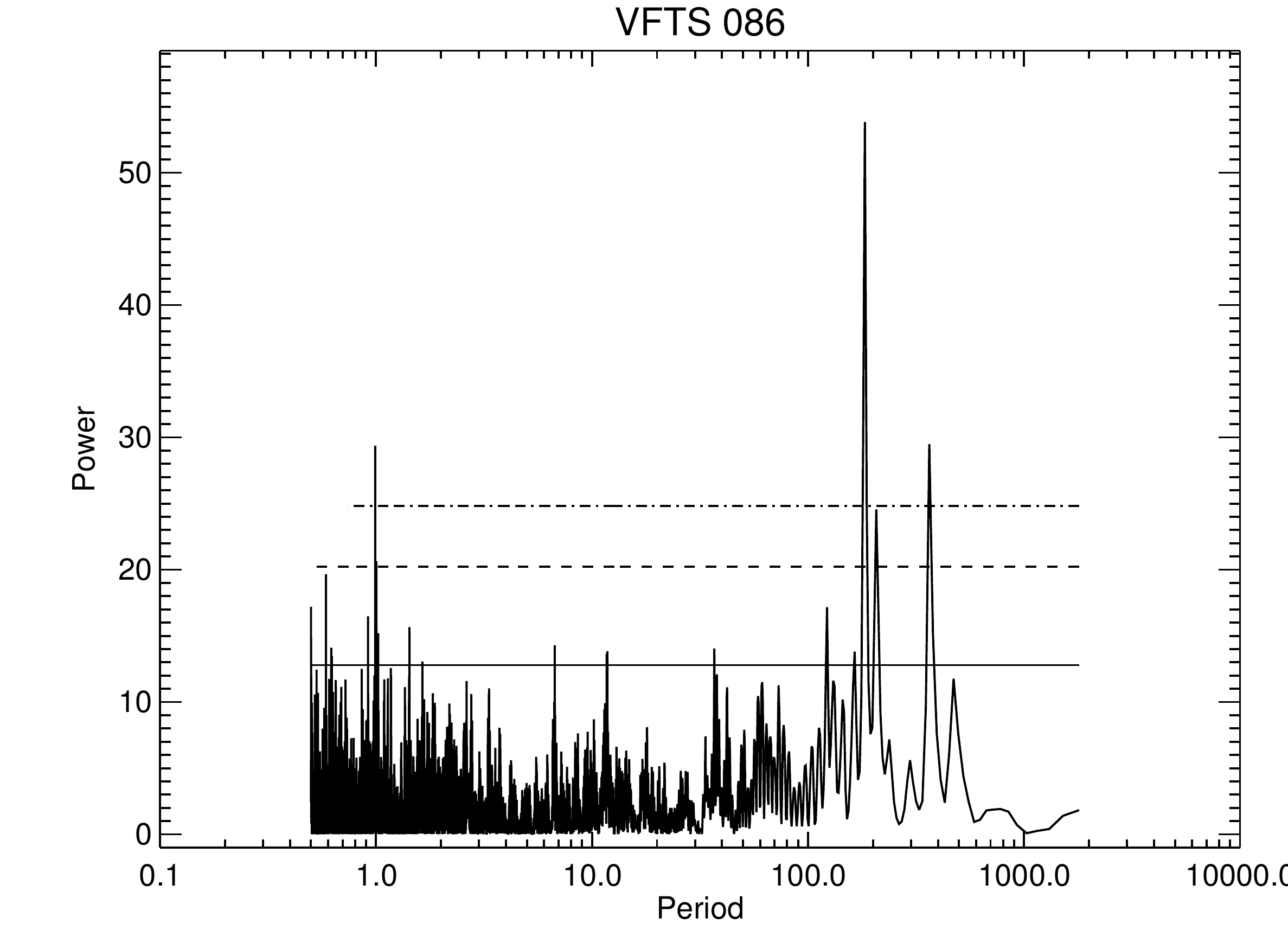}
\includegraphics[width=4.4cm,angle=-90]{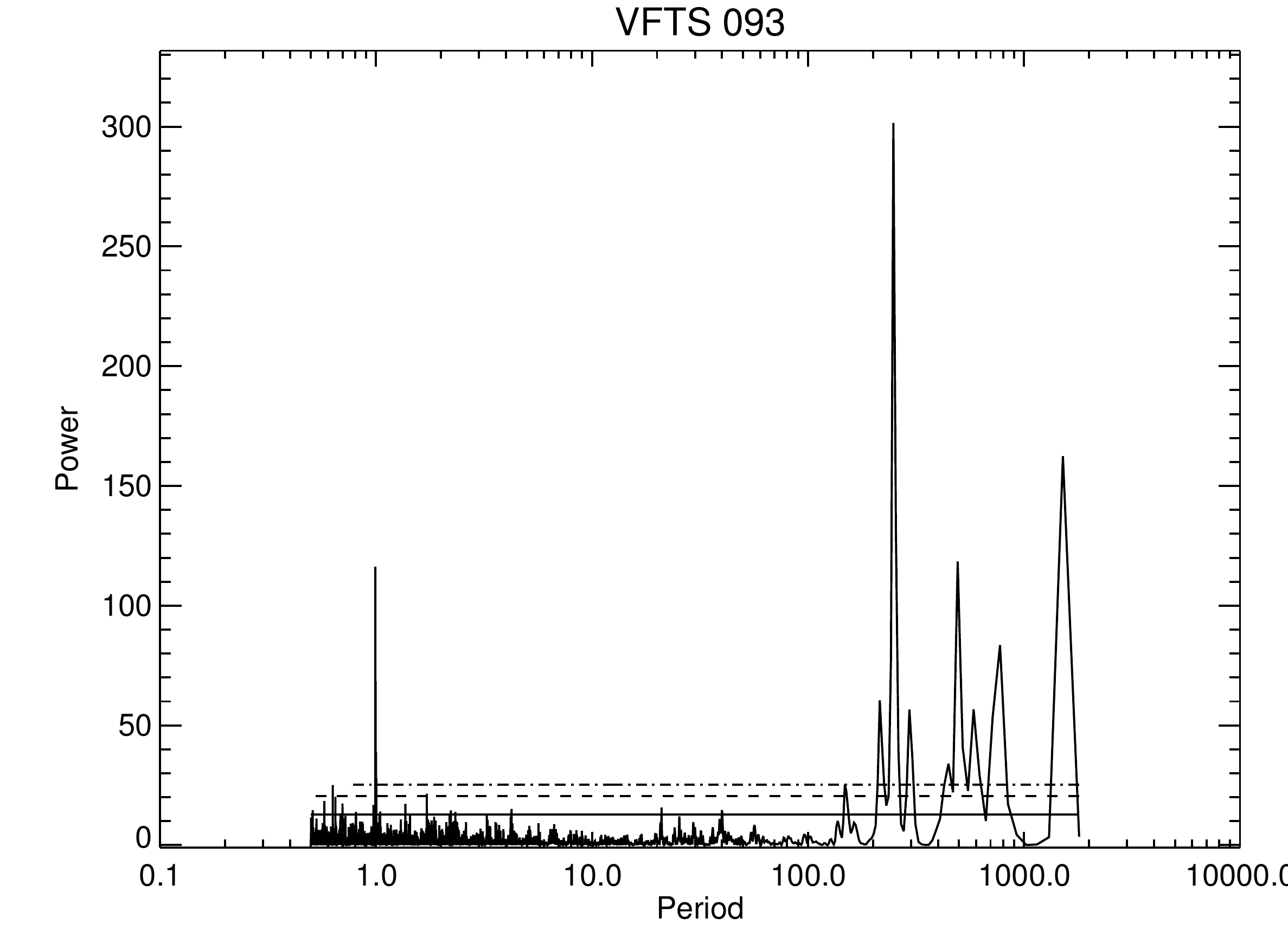}
\includegraphics[width=4.4cm,angle=-90]{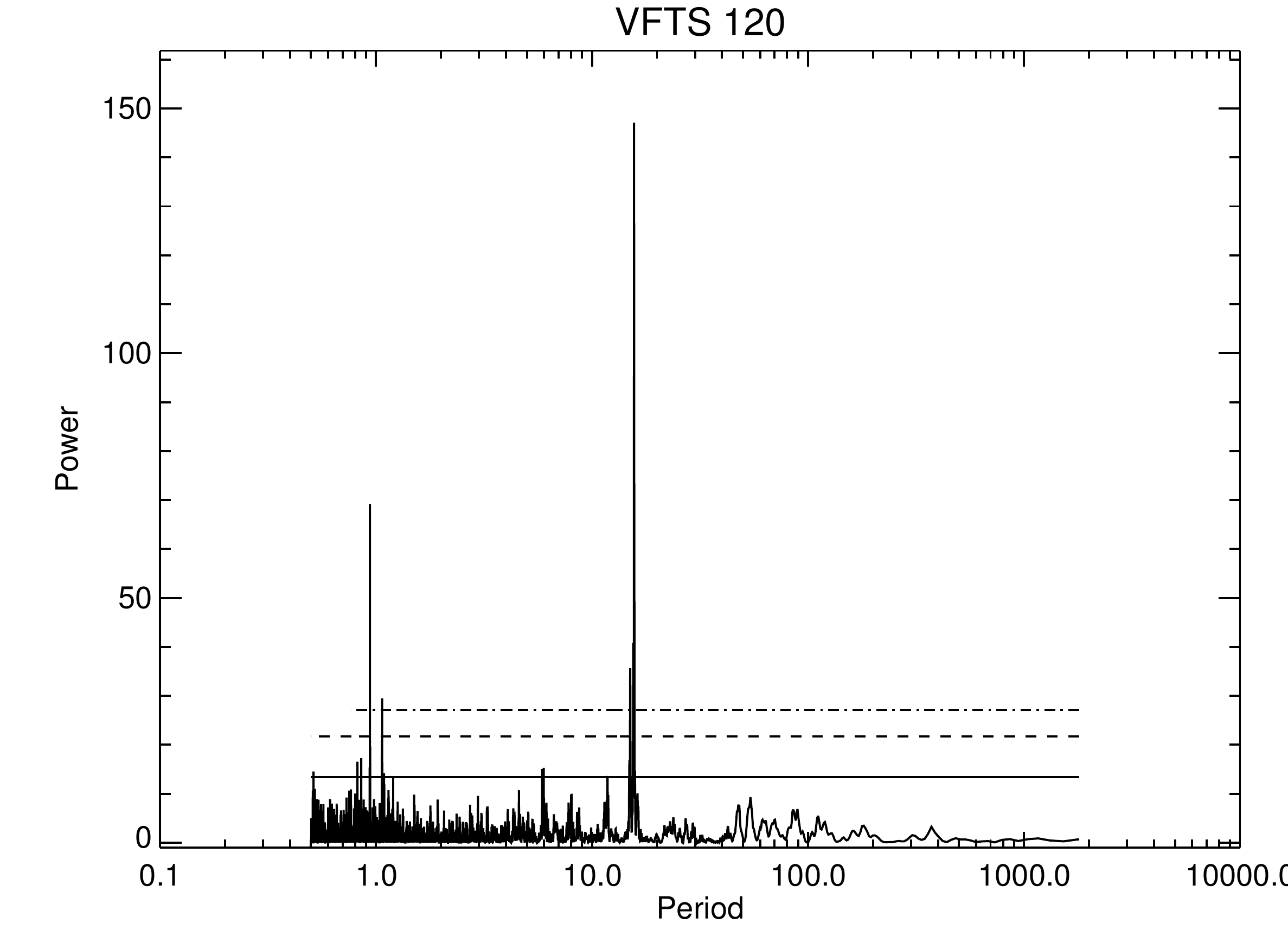}
\includegraphics[width=4.4cm,angle=-90]{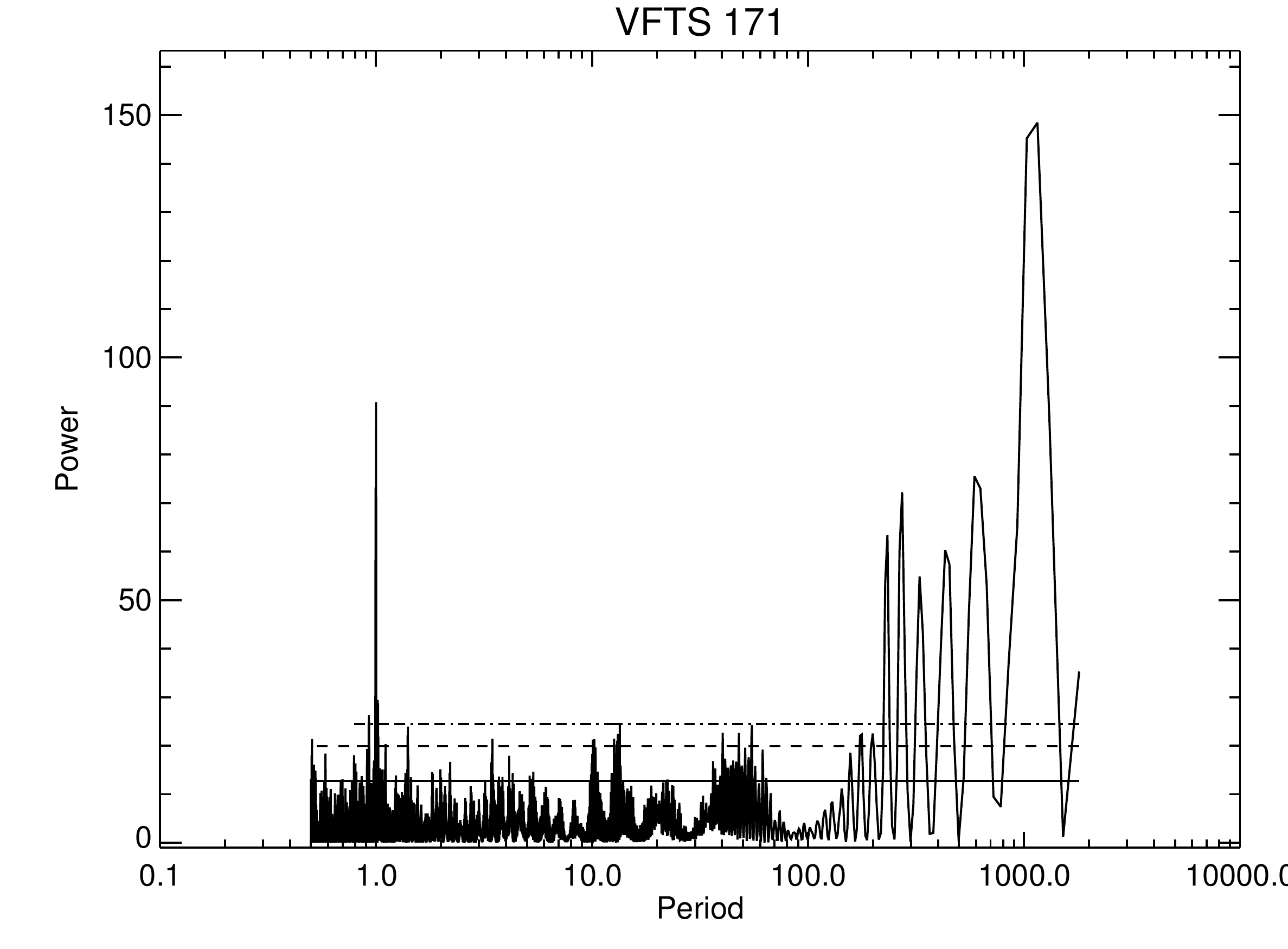}
\includegraphics[width=4.4cm,angle=-90]{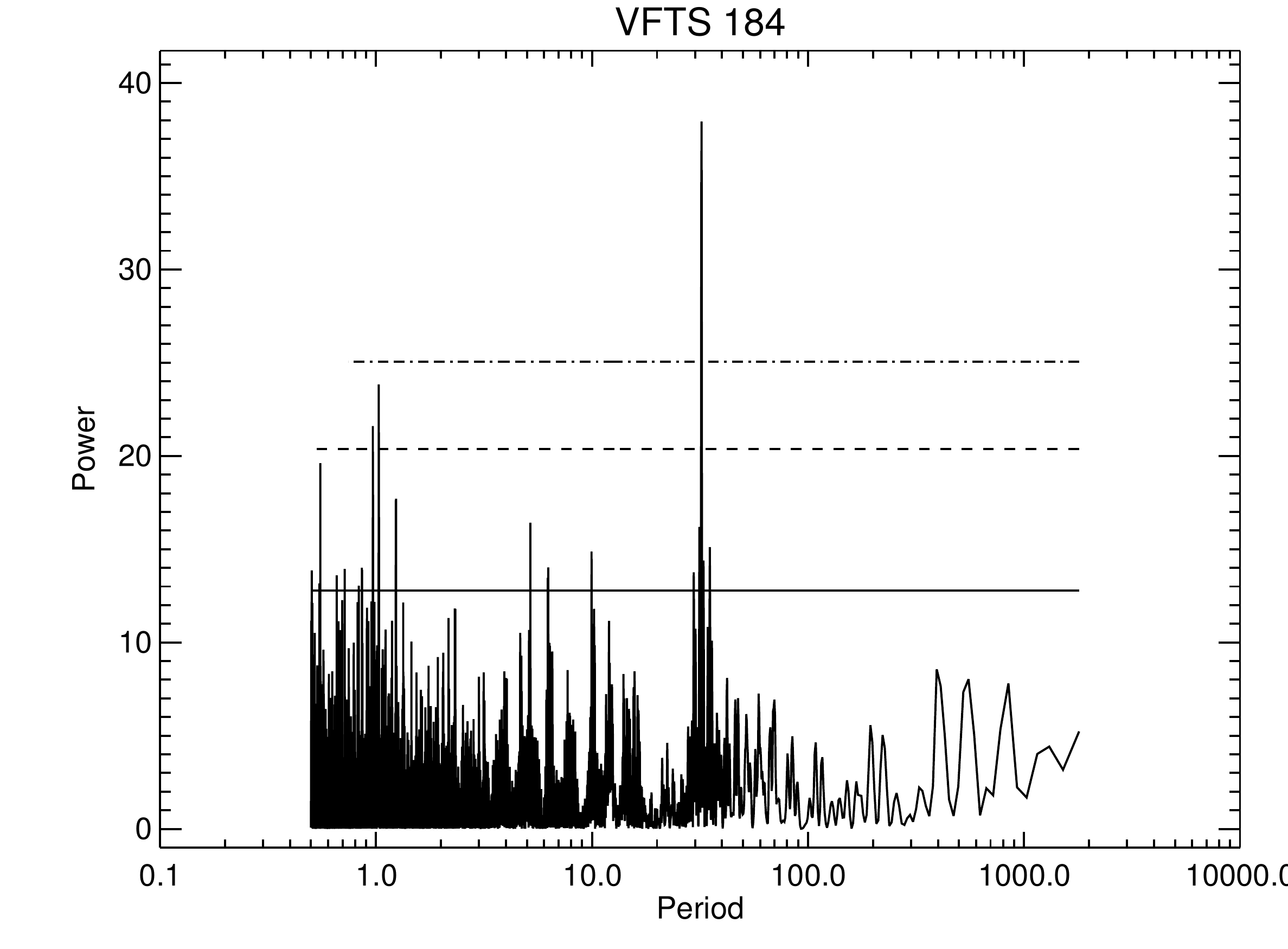}
\includegraphics[width=4.4cm,angle=-90]{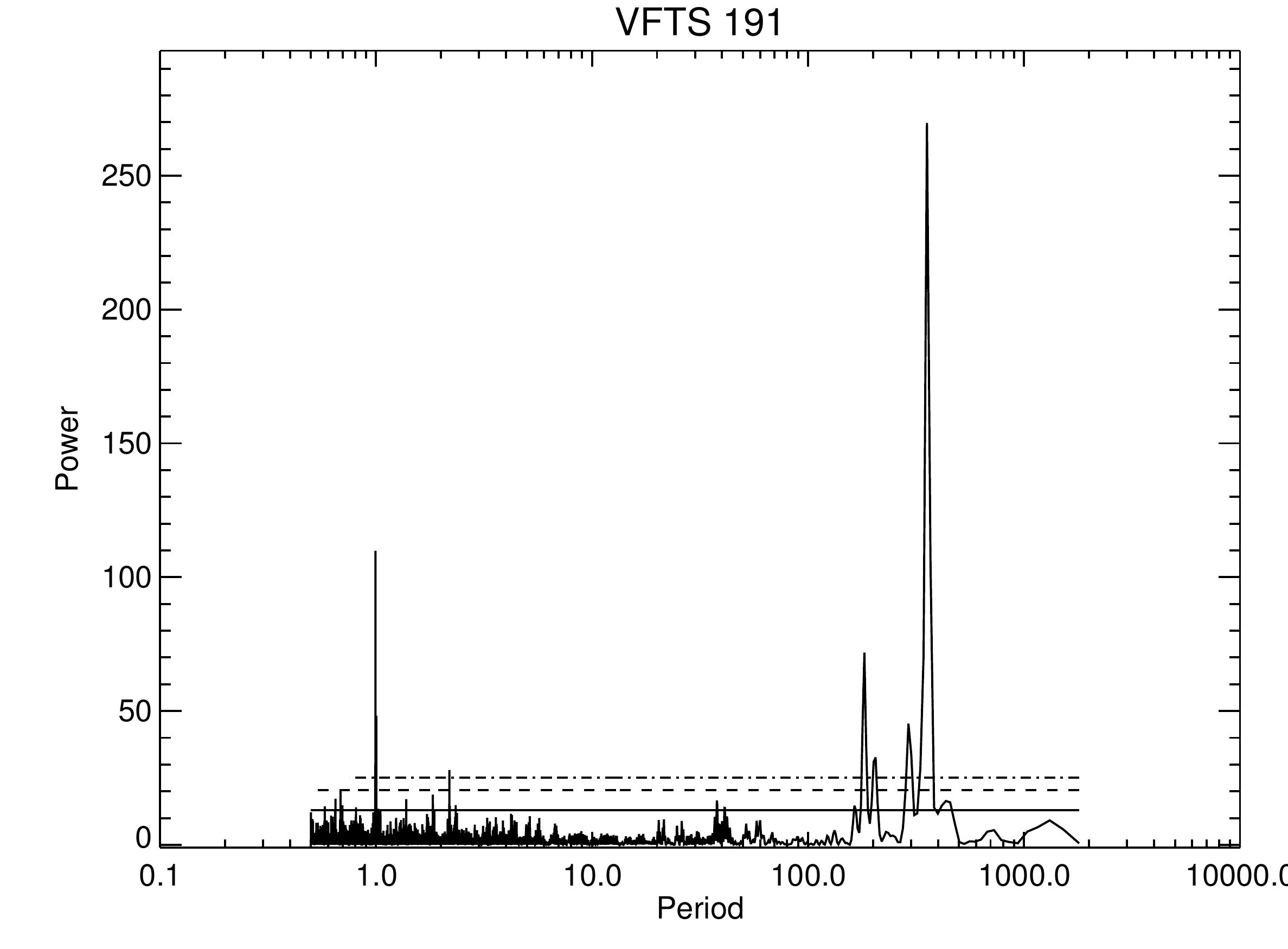}
\includegraphics[width=4.4cm,angle=-90]{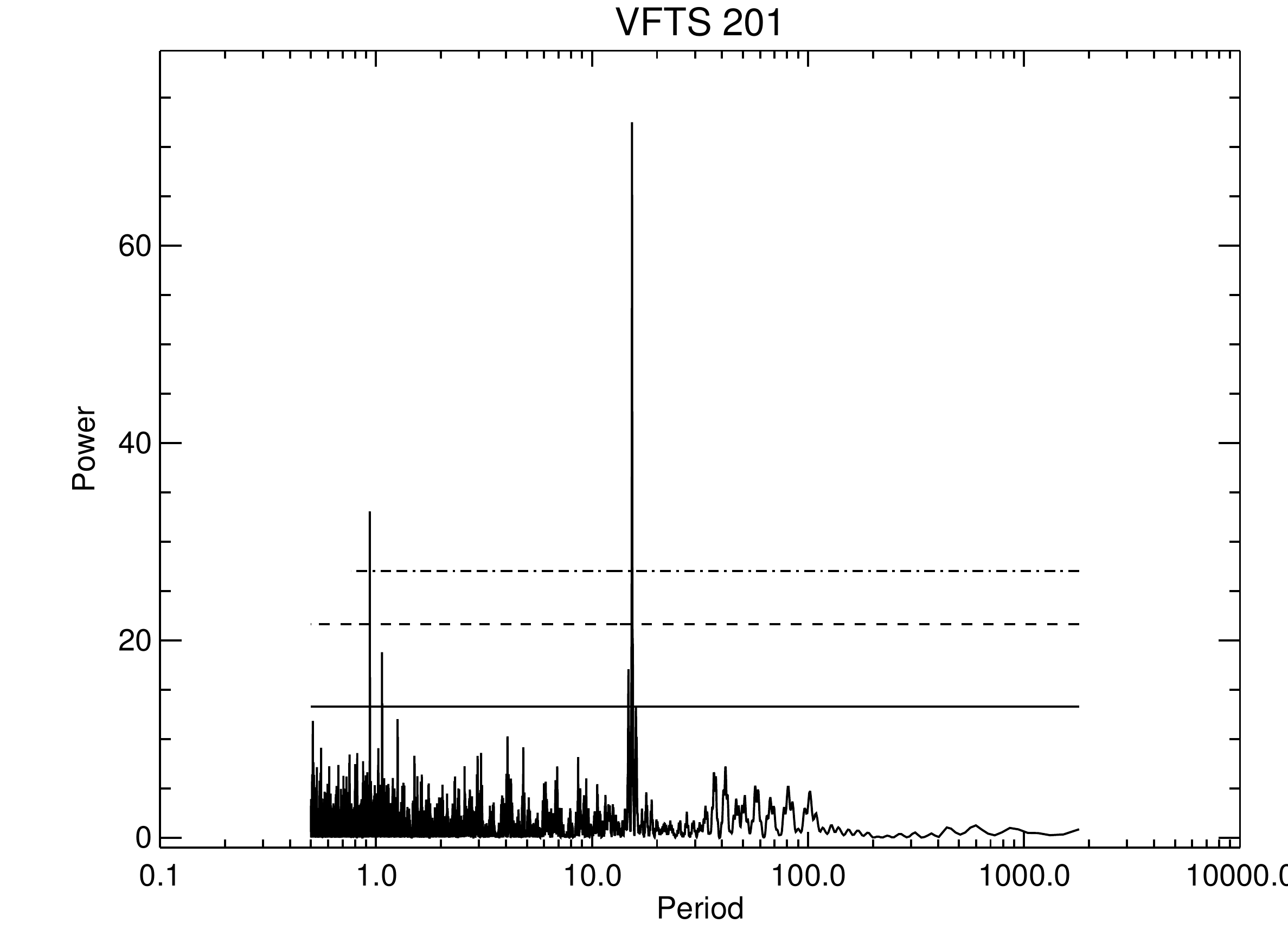}
\includegraphics[width=4.4cm,angle=-90]{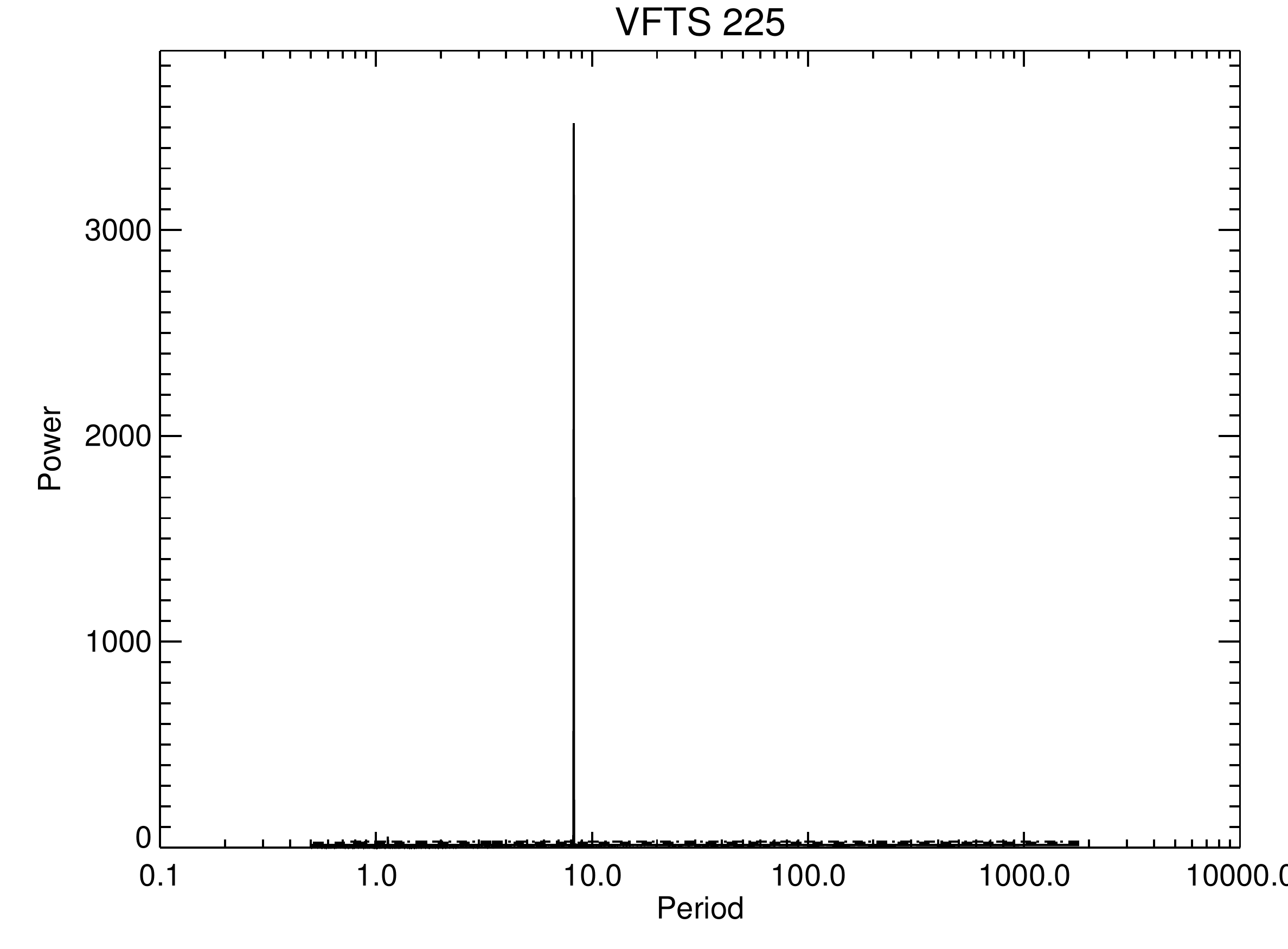}
\includegraphics[width=4.4cm,angle=-90]{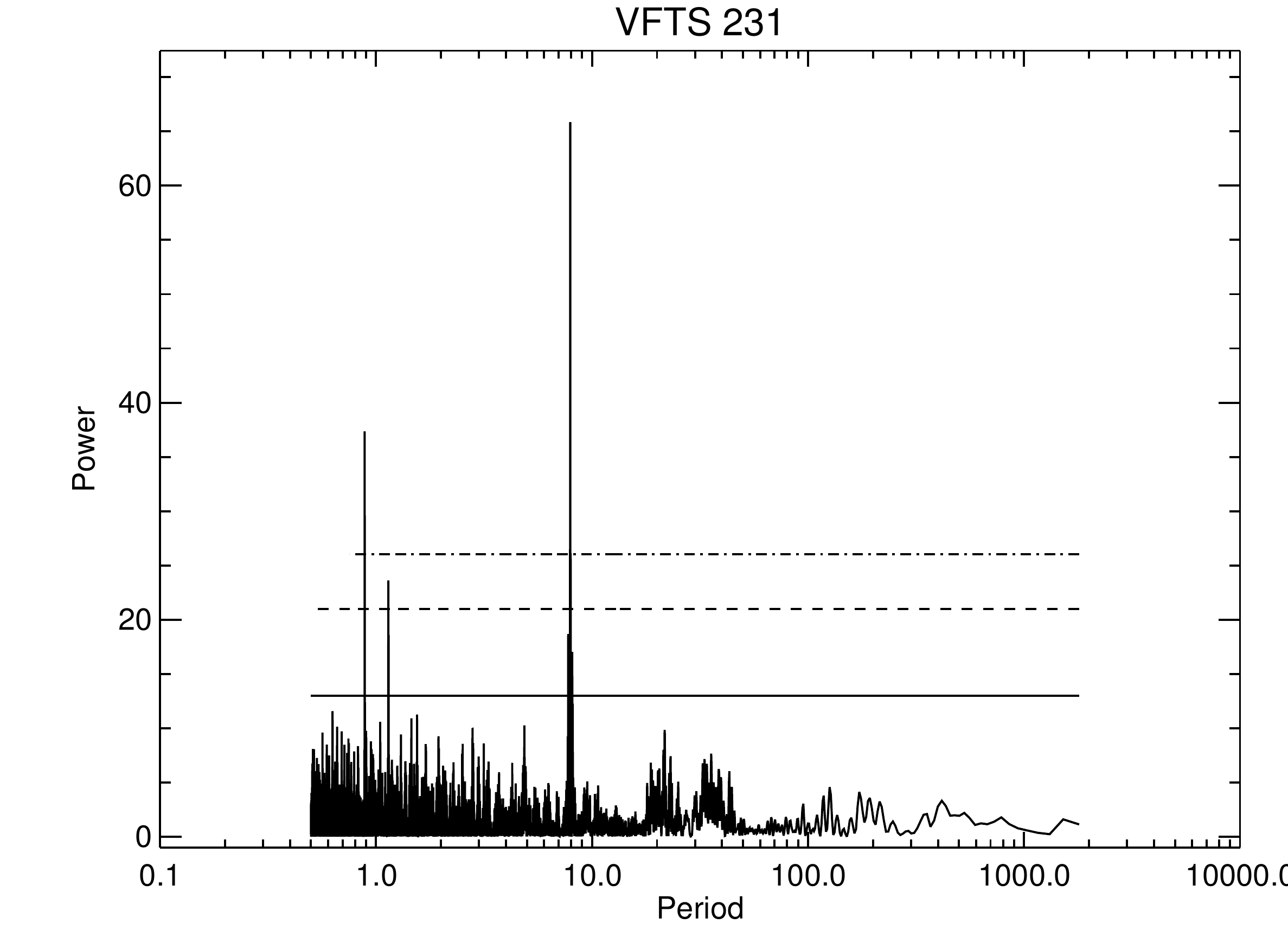}
\includegraphics[width=4.4cm,angle=-90]{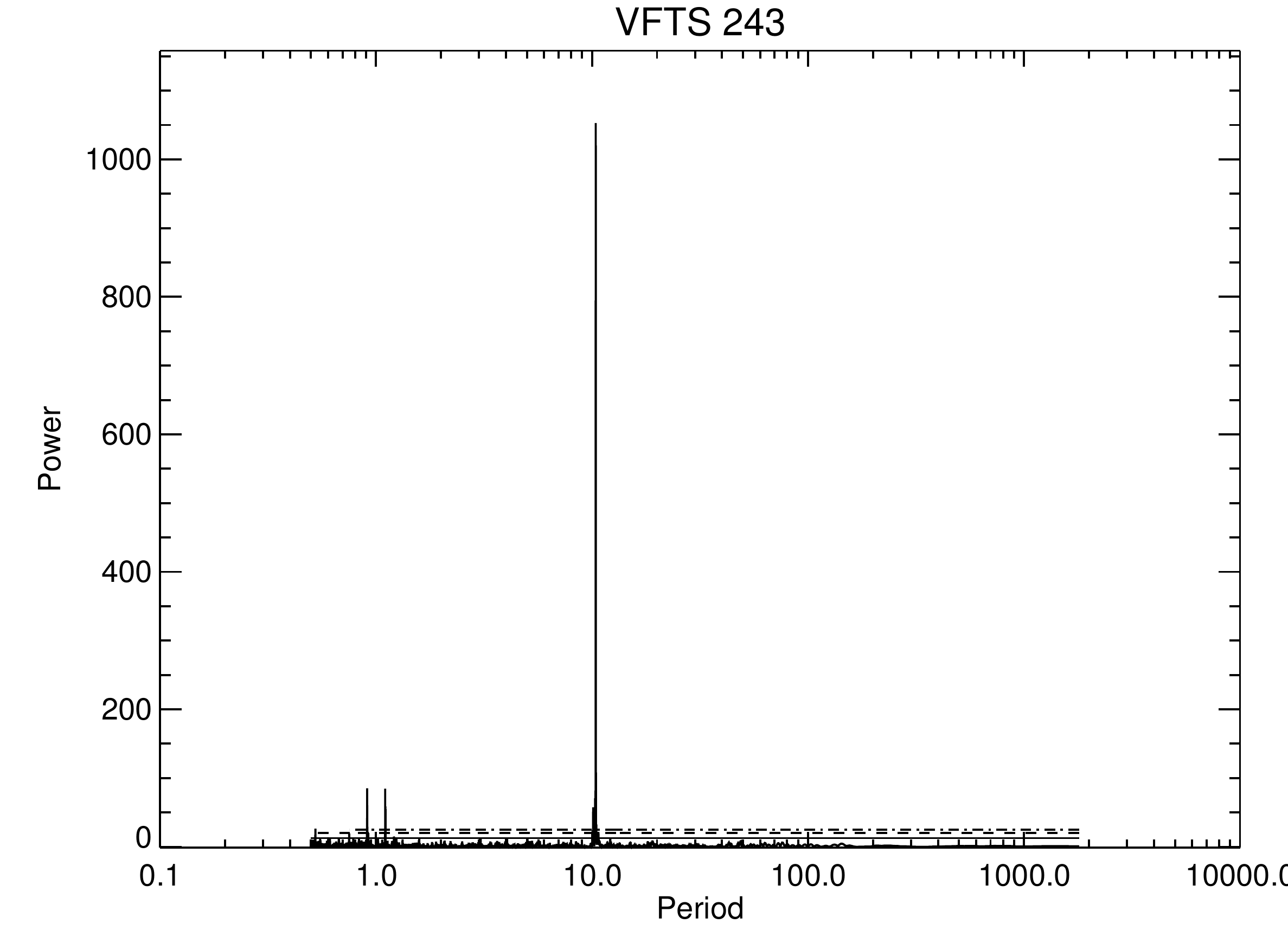}
\includegraphics[width=4.4cm,angle=-90]{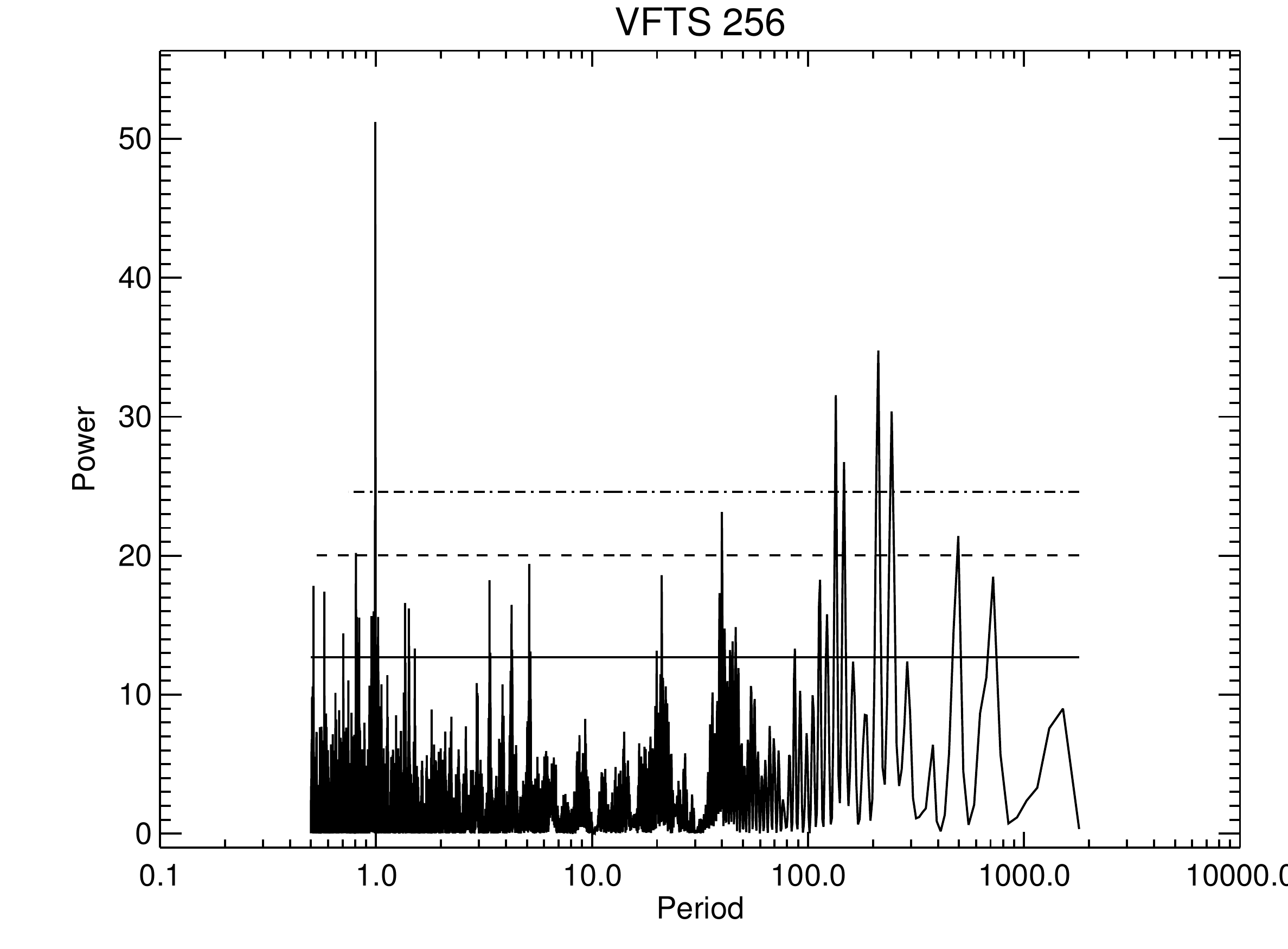}
\includegraphics[width=4.4cm,angle=-90]{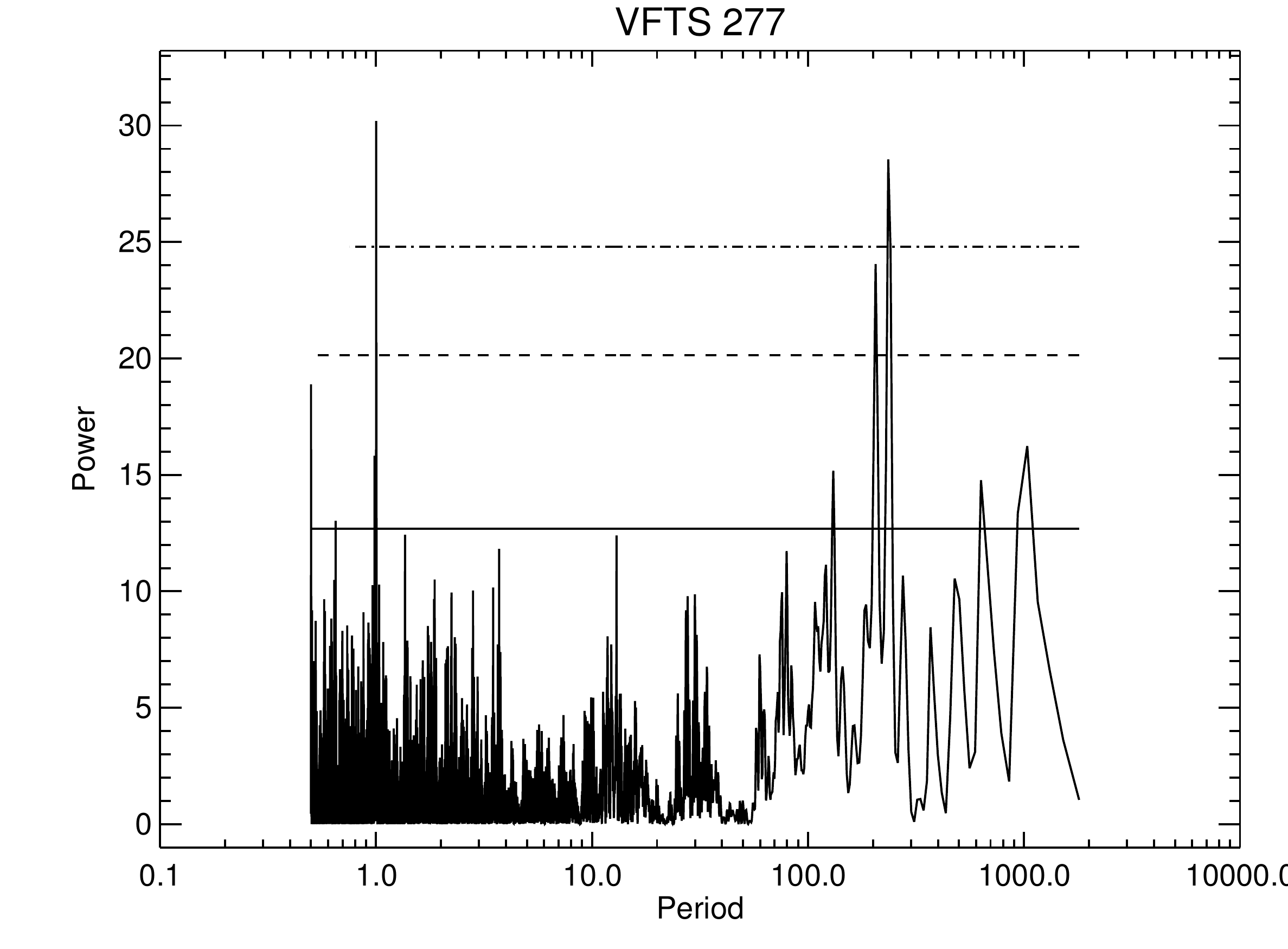}
\includegraphics[width=4.4cm,angle=-90]{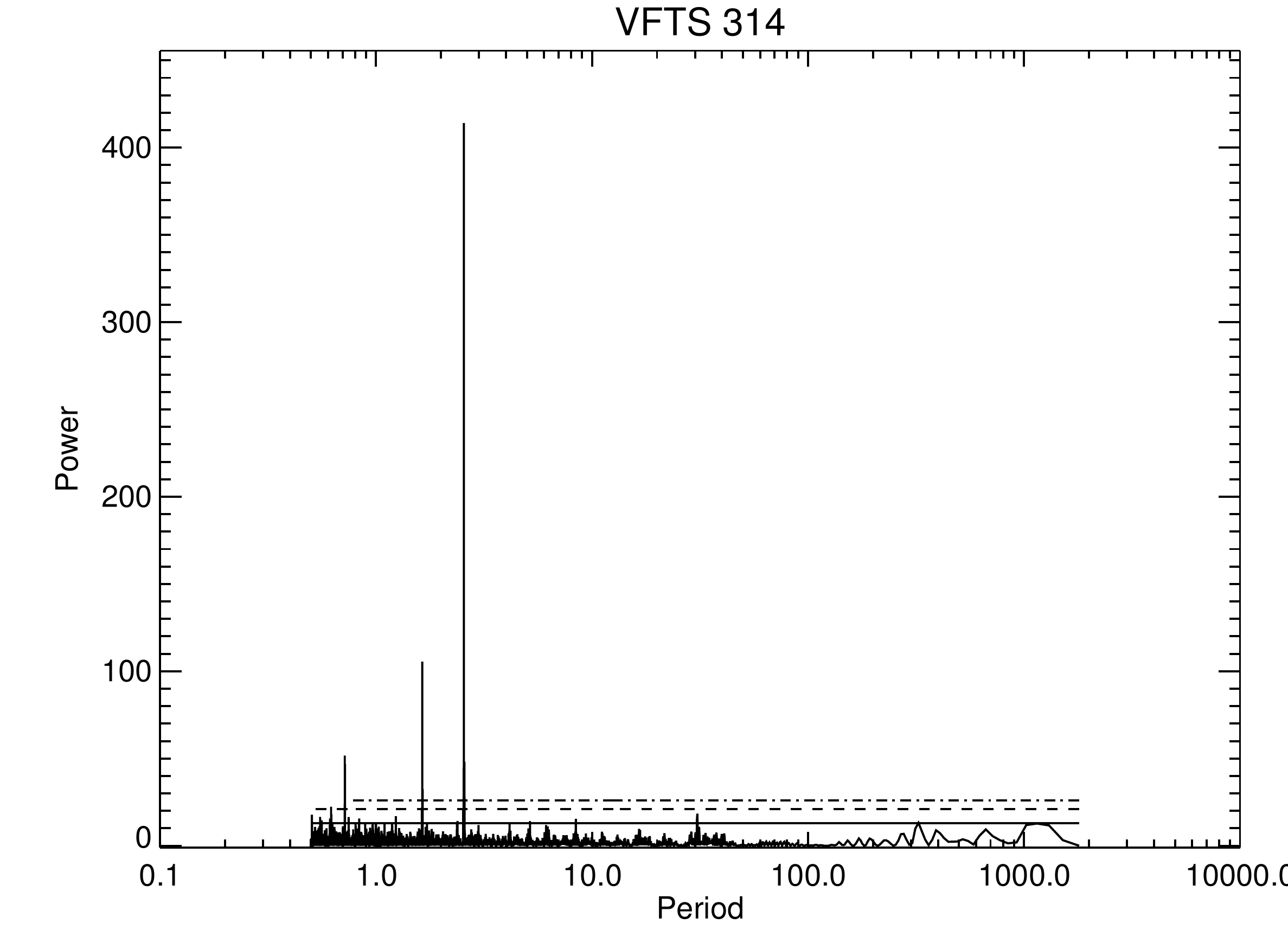}
\caption{Lomb-Scargle periodograms for the SB1 systems. The solid, dashed, and dot-dashed lines represent the 50\%, 1\%, and 0.1\% false alarm probabilities, respectively.}
\label{sb1:periodogram}
\end{figure*}

\begin{figure*}
\centering
\ContinuedFloat
\includegraphics[width=4.4cm,angle=-90]{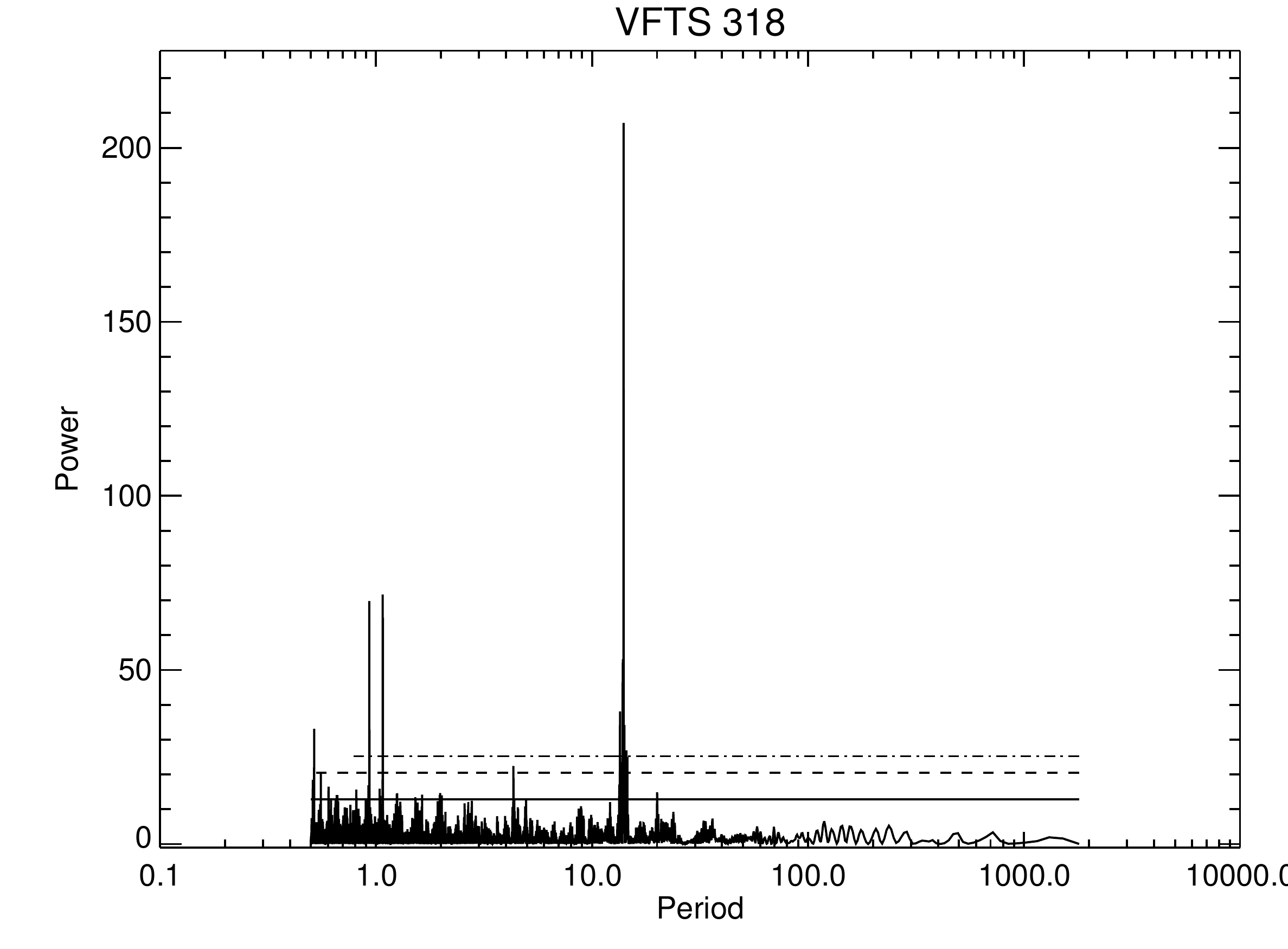}
\includegraphics[width=4.4cm,angle=-90]{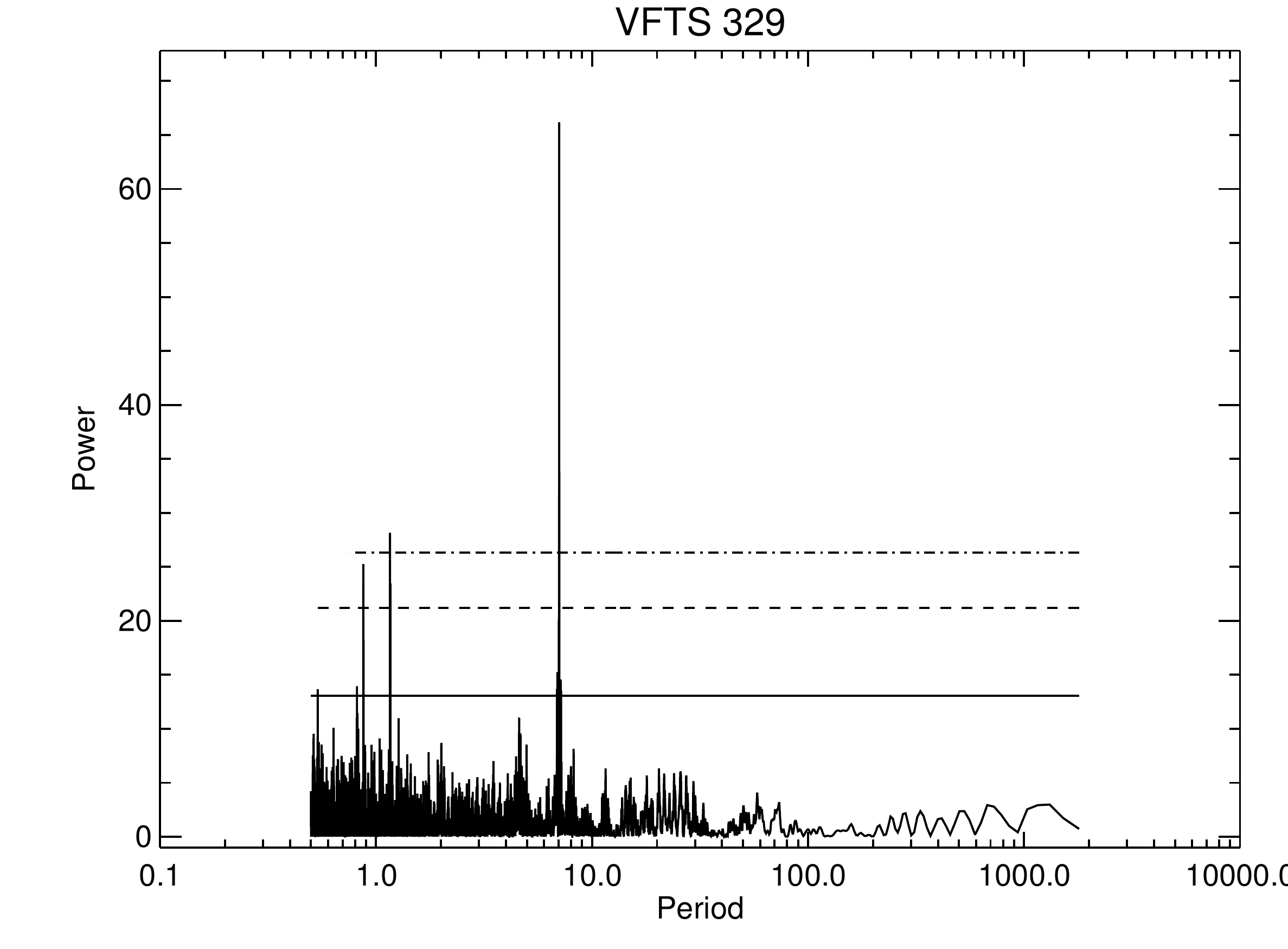}
\includegraphics[width=4.4cm,angle=-90]{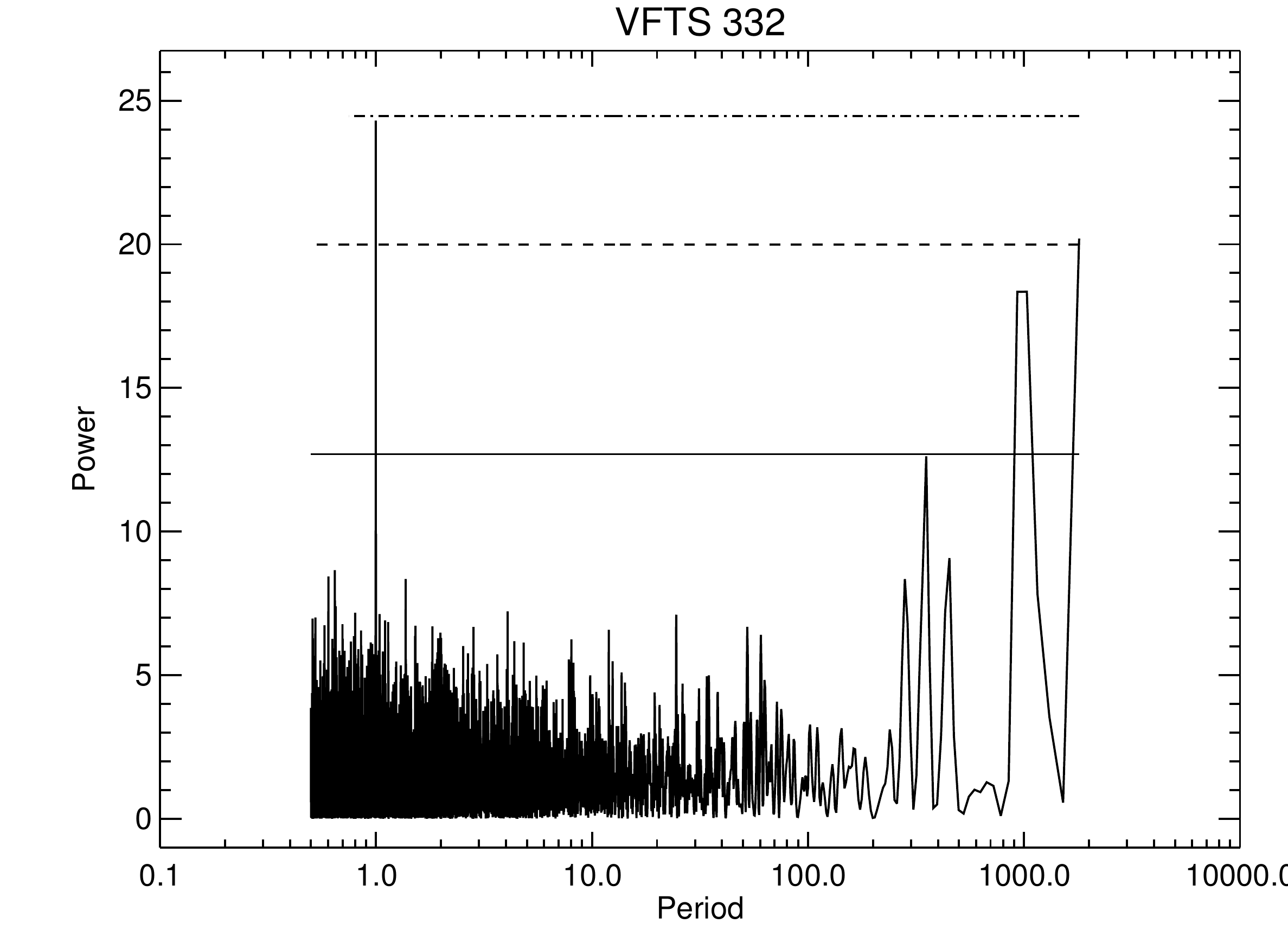}
\includegraphics[width=4.4cm,angle=-90]{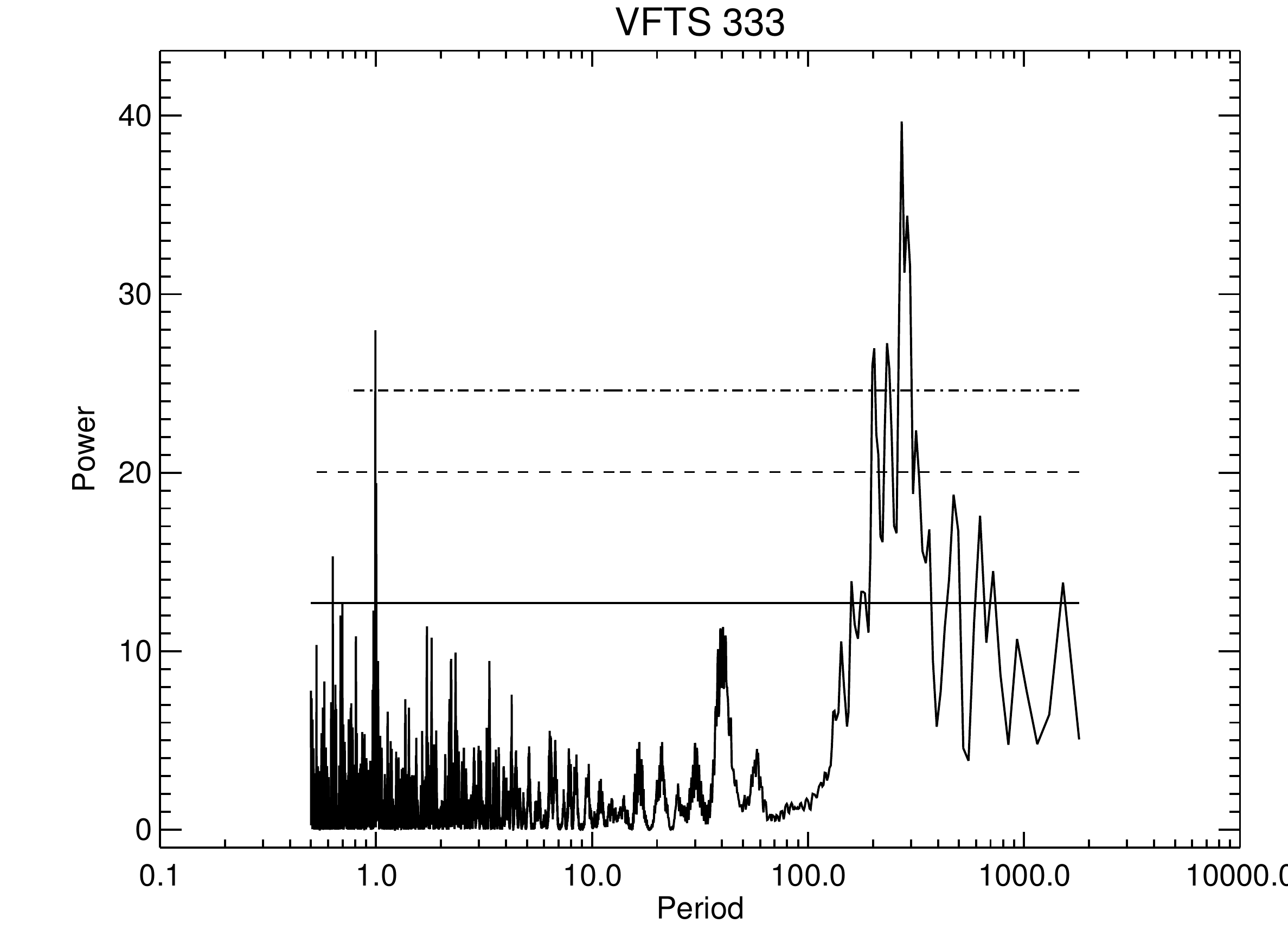}
\includegraphics[width=4.4cm,angle=-90]{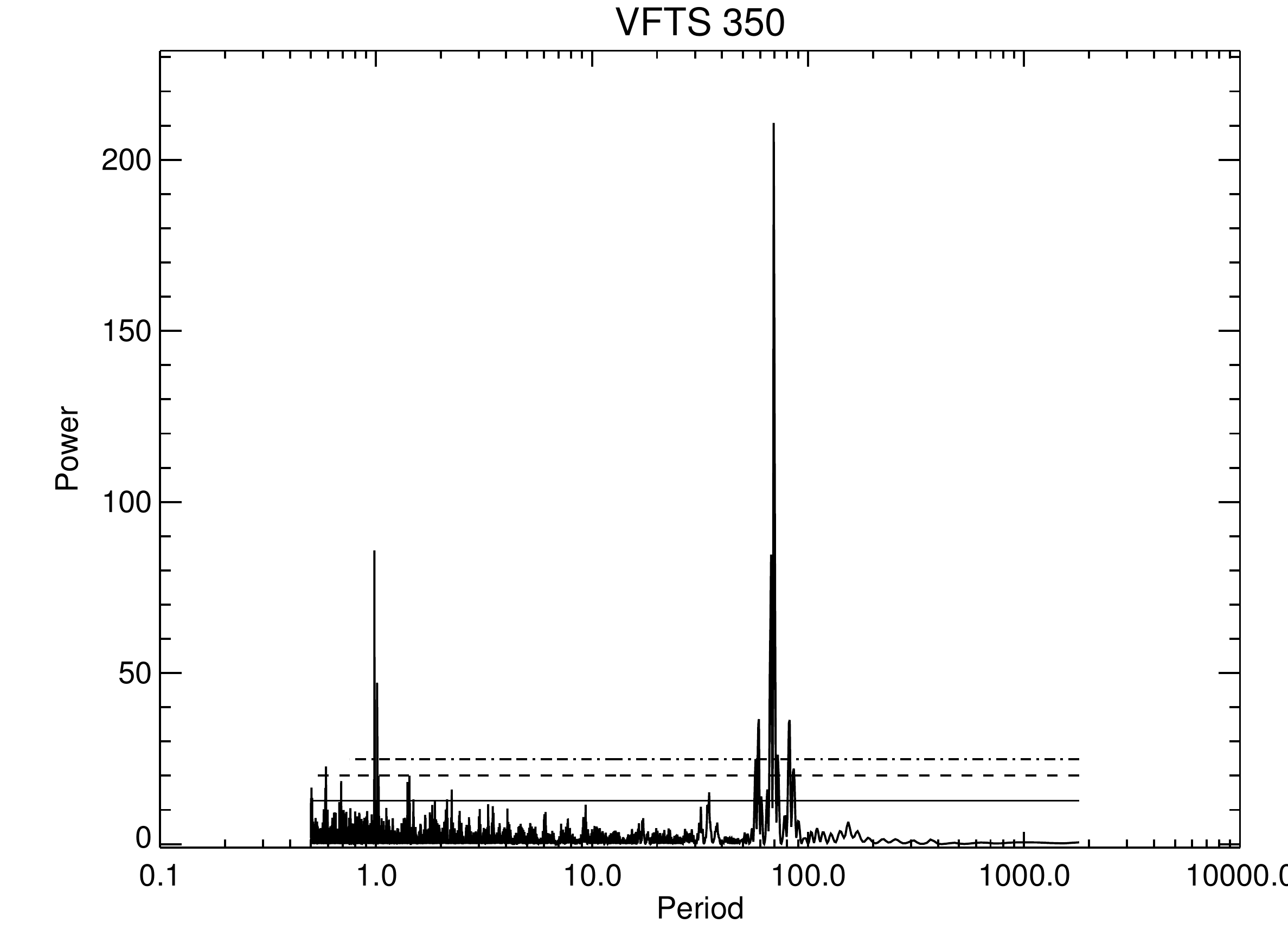}
\includegraphics[width=4.4cm,angle=-90]{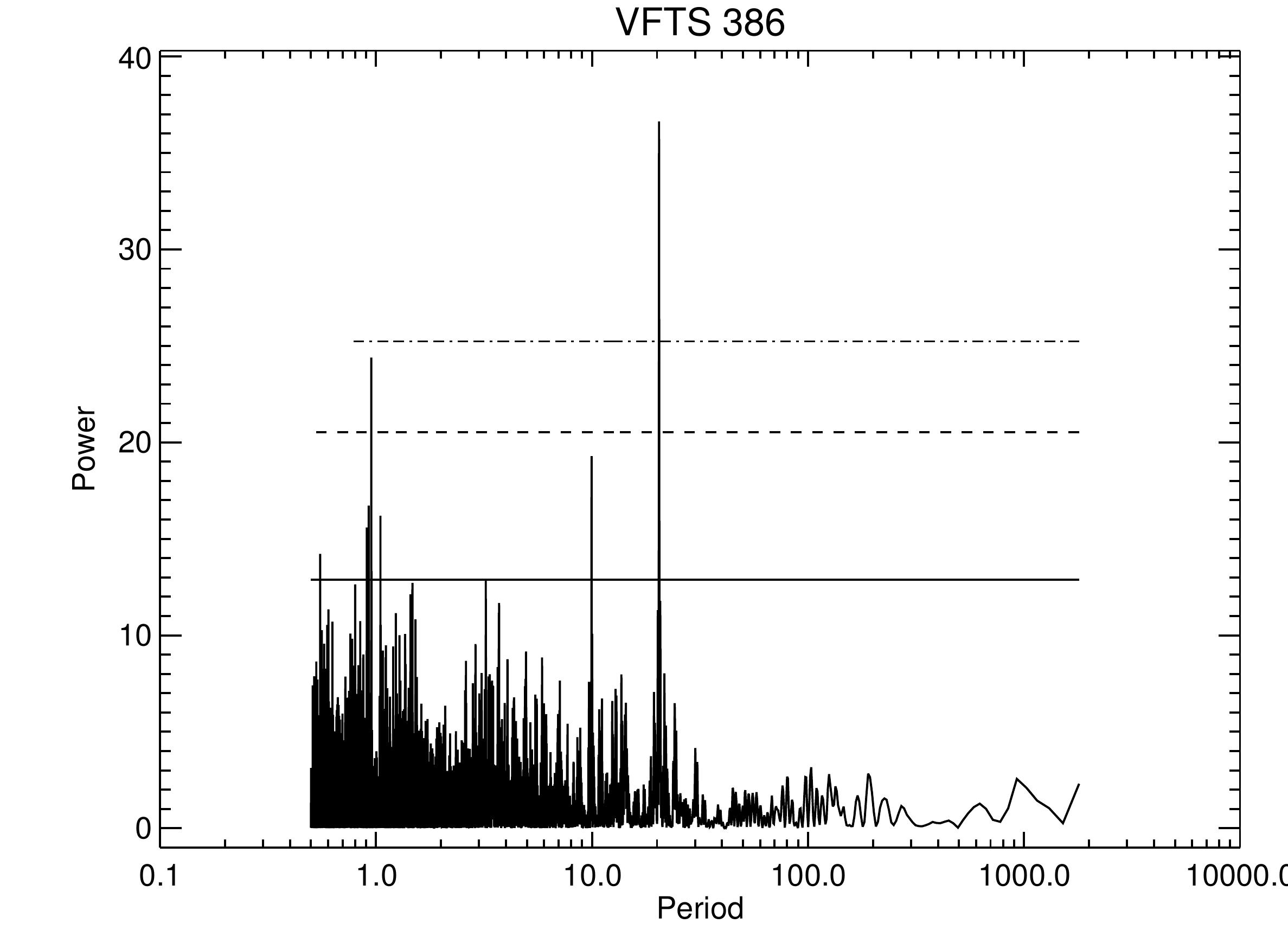}
\includegraphics[width=4.4cm,angle=-90]{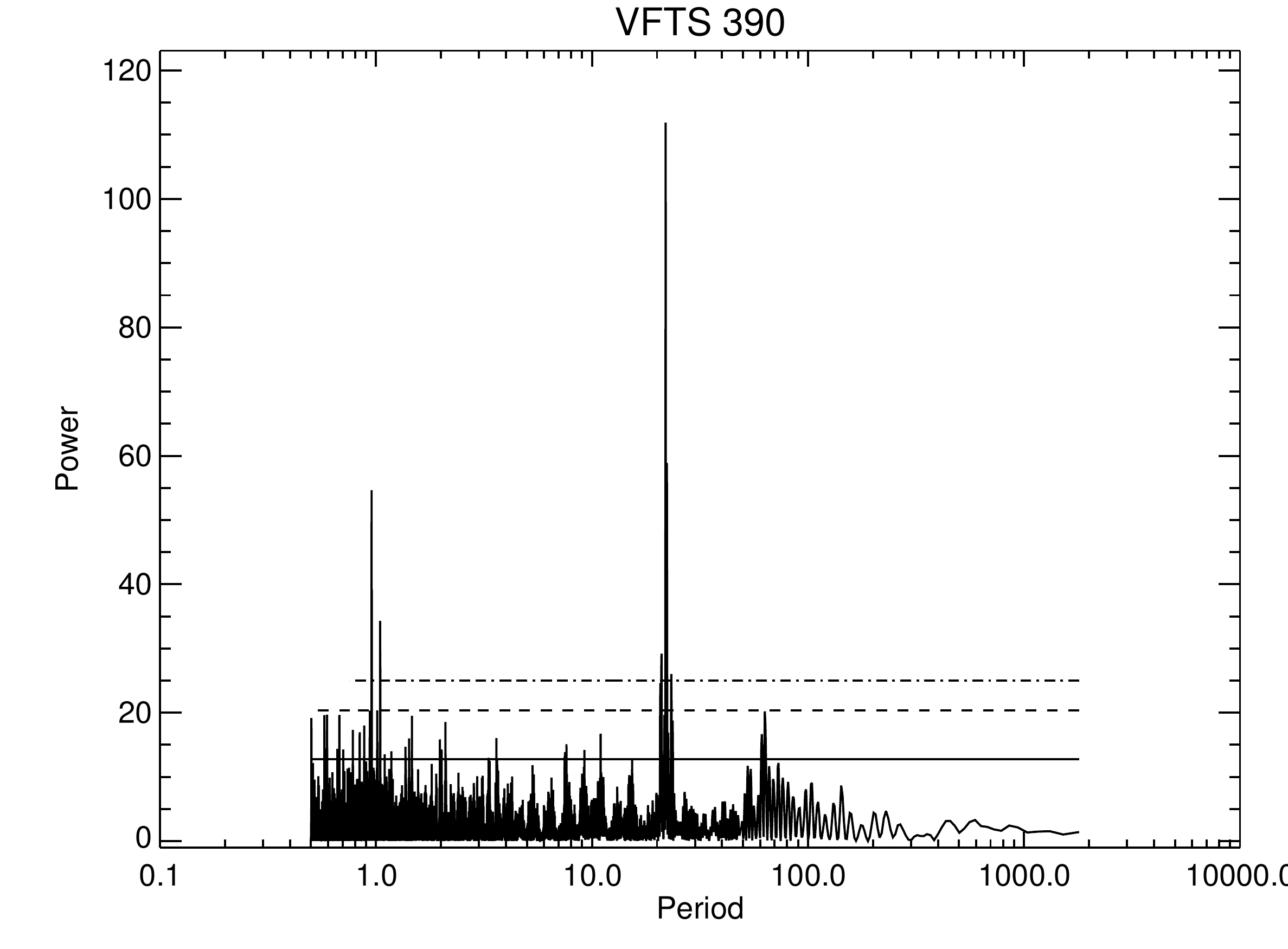}
\includegraphics[width=4.4cm,angle=-90]{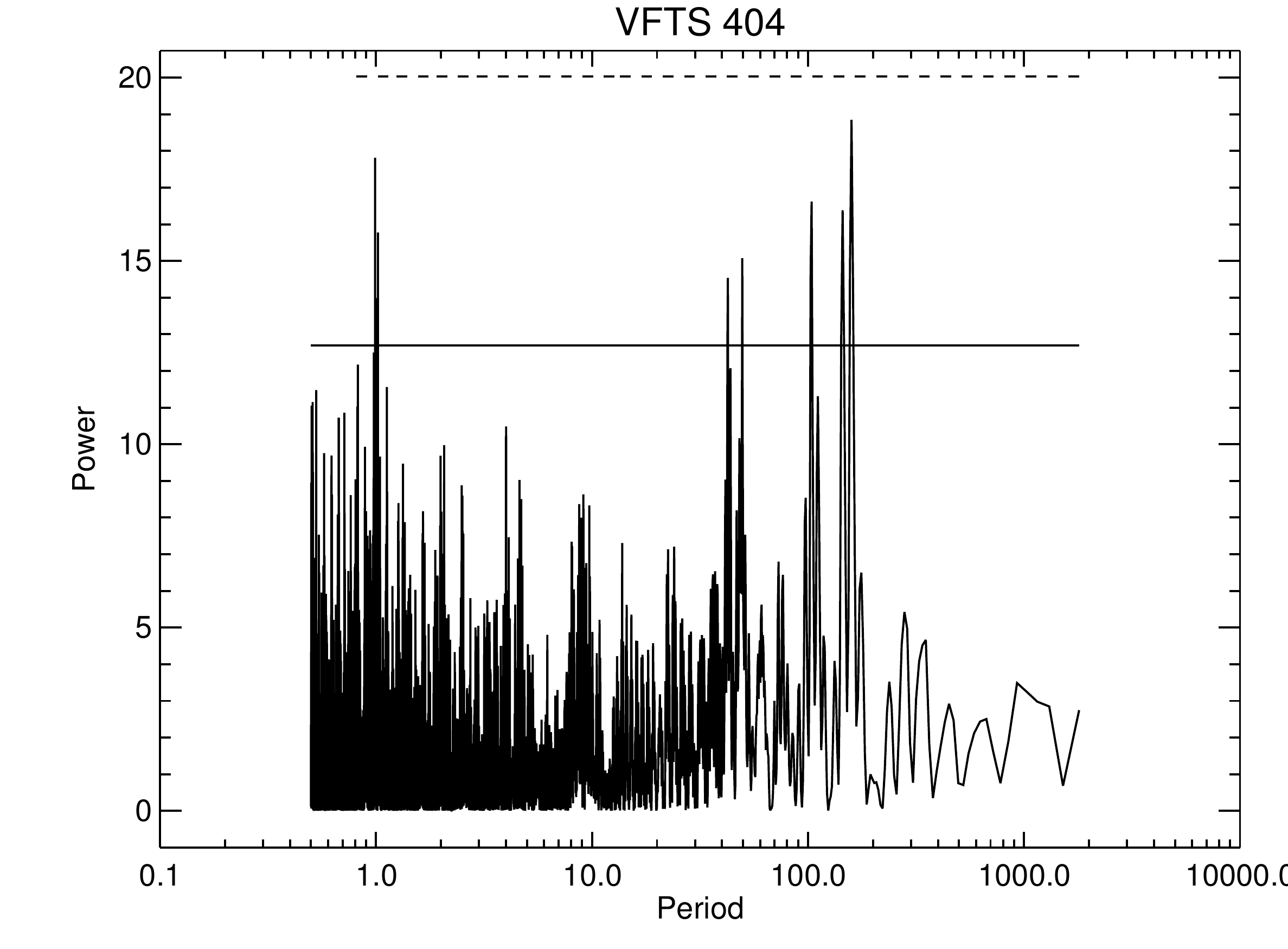}
\includegraphics[width=4.4cm,angle=-90]{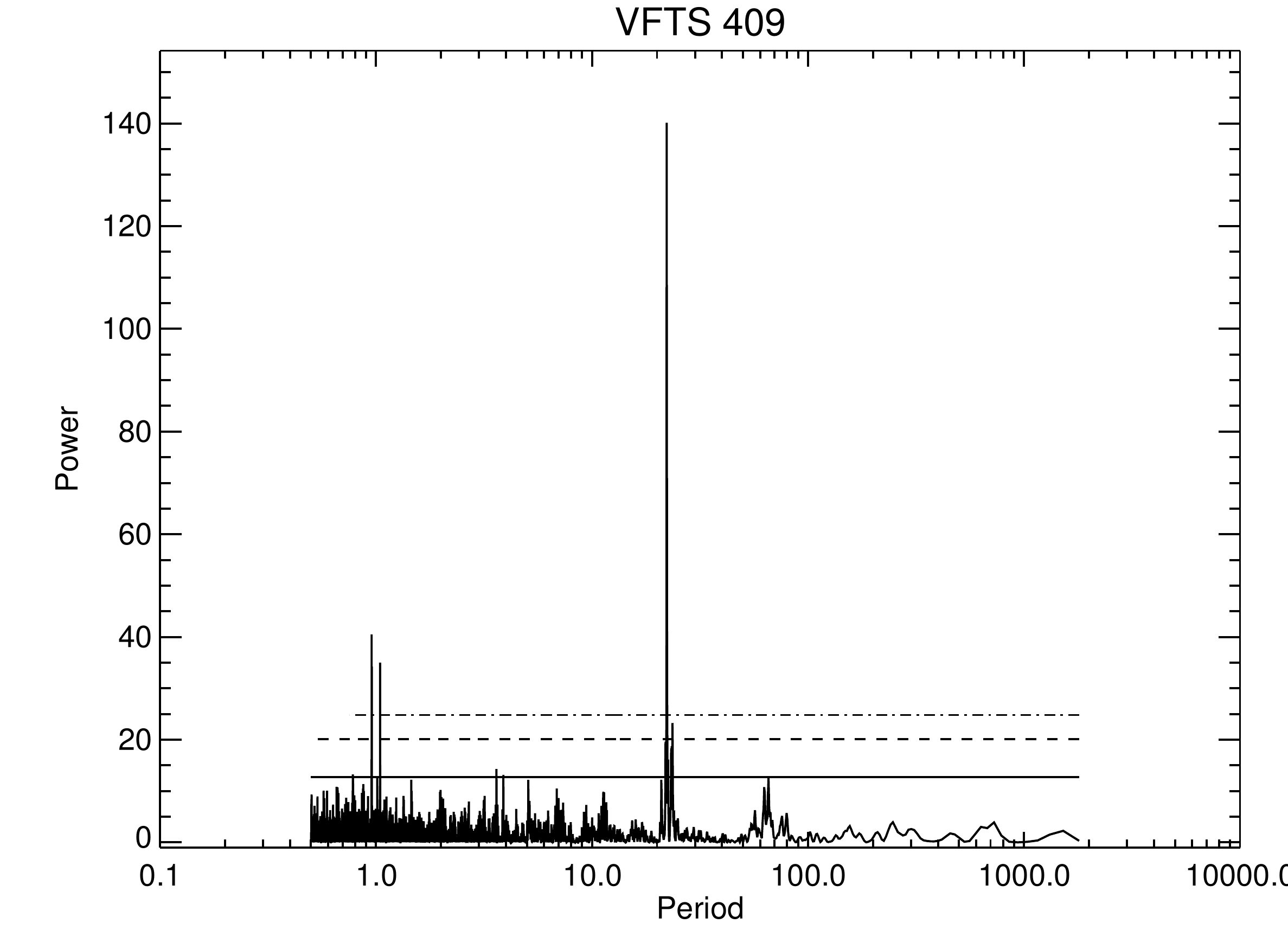}
\includegraphics[width=4.4cm,angle=-90]{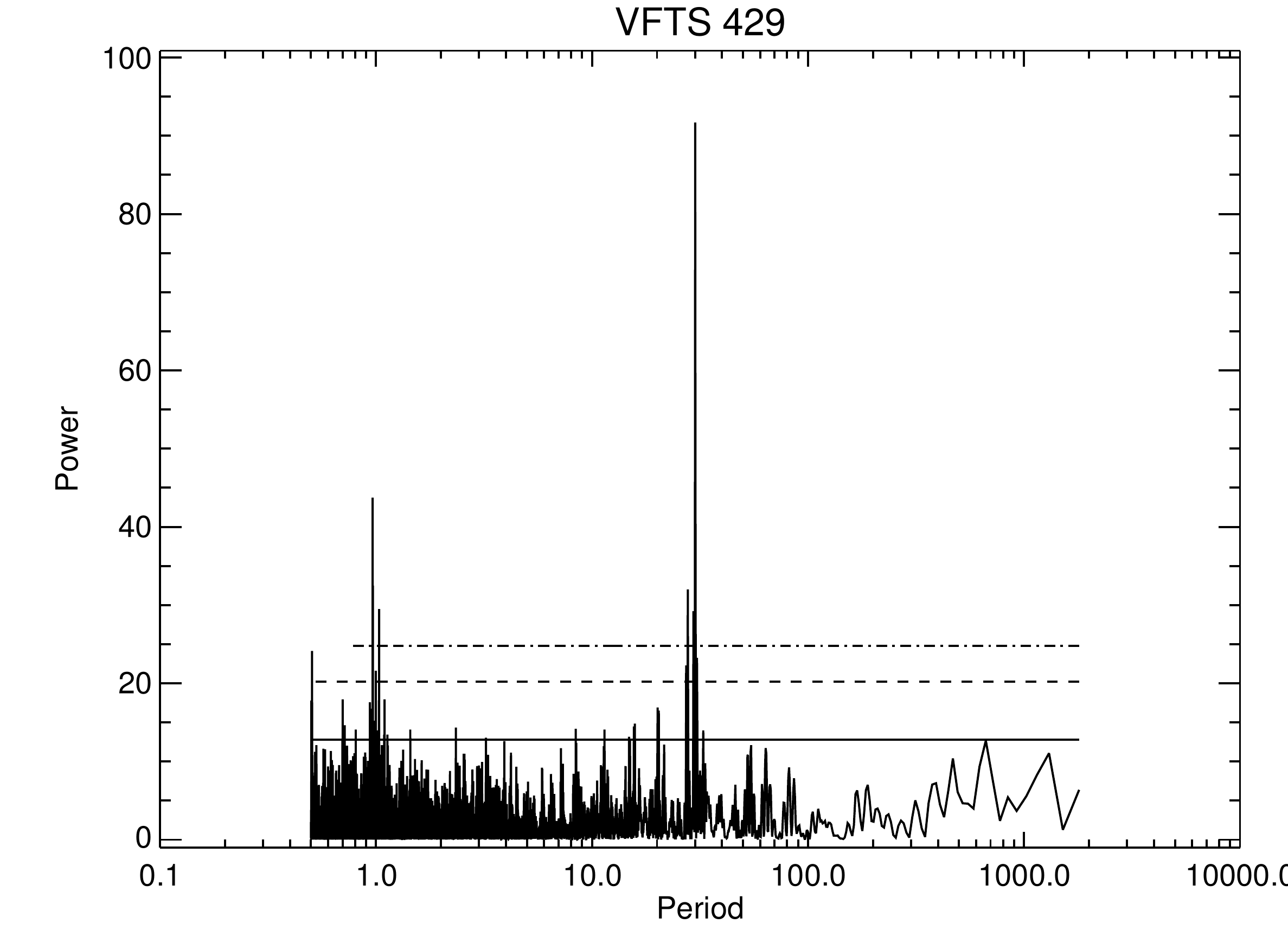}
\includegraphics[width=4.4cm,angle=-90]{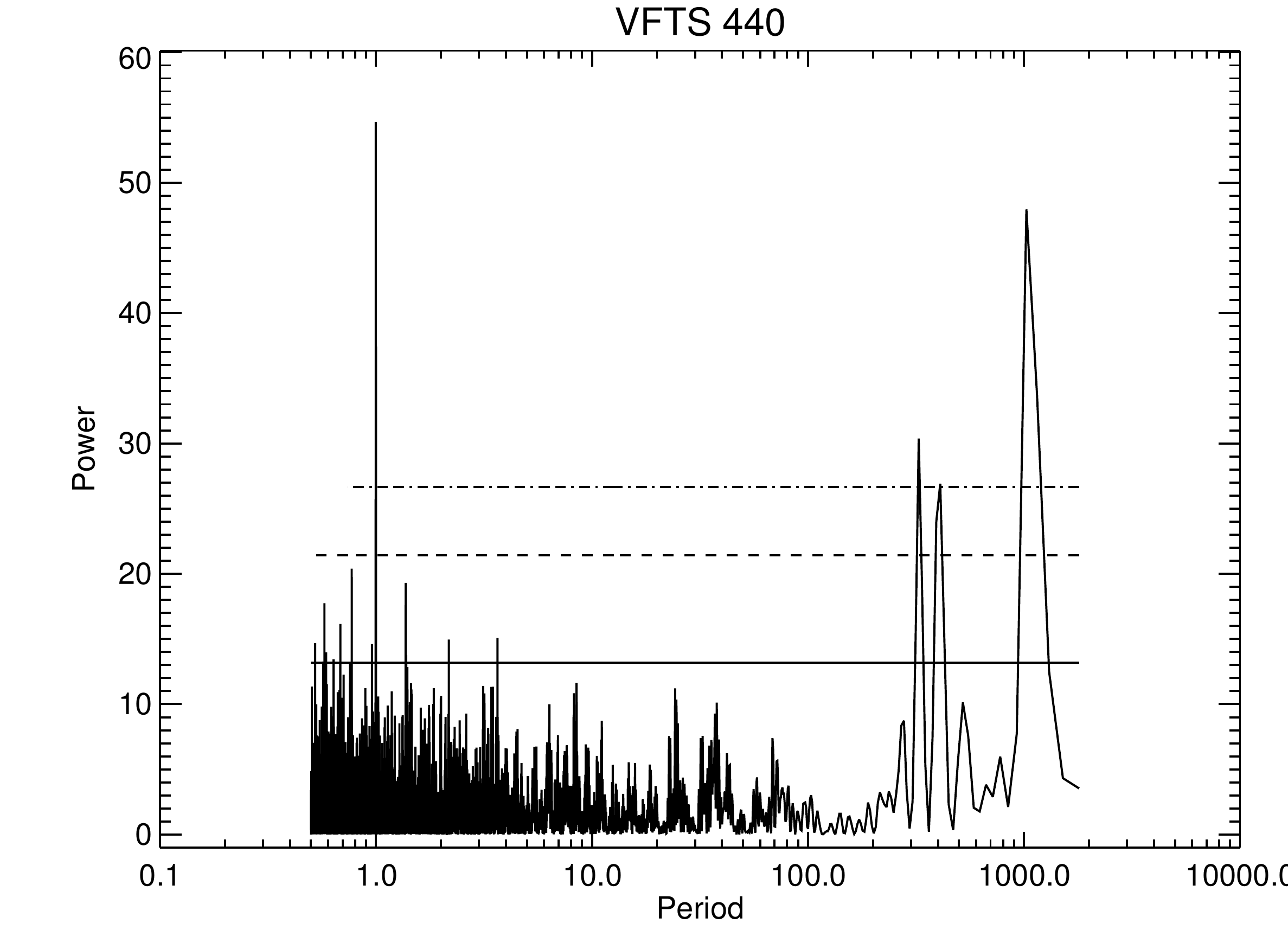}
\includegraphics[width=4.4cm,angle=-90]{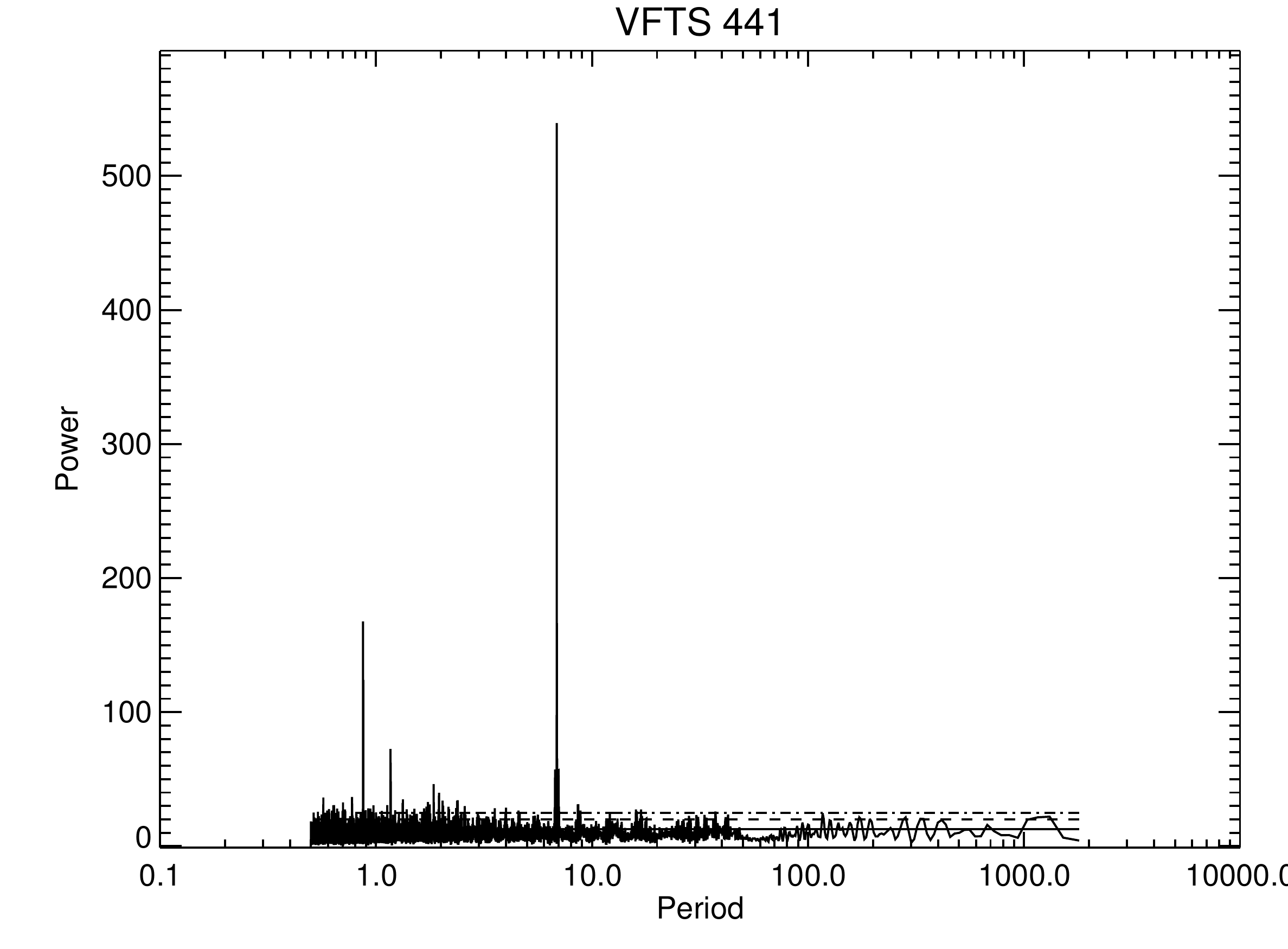}
\includegraphics[width=4.4cm,angle=-90]{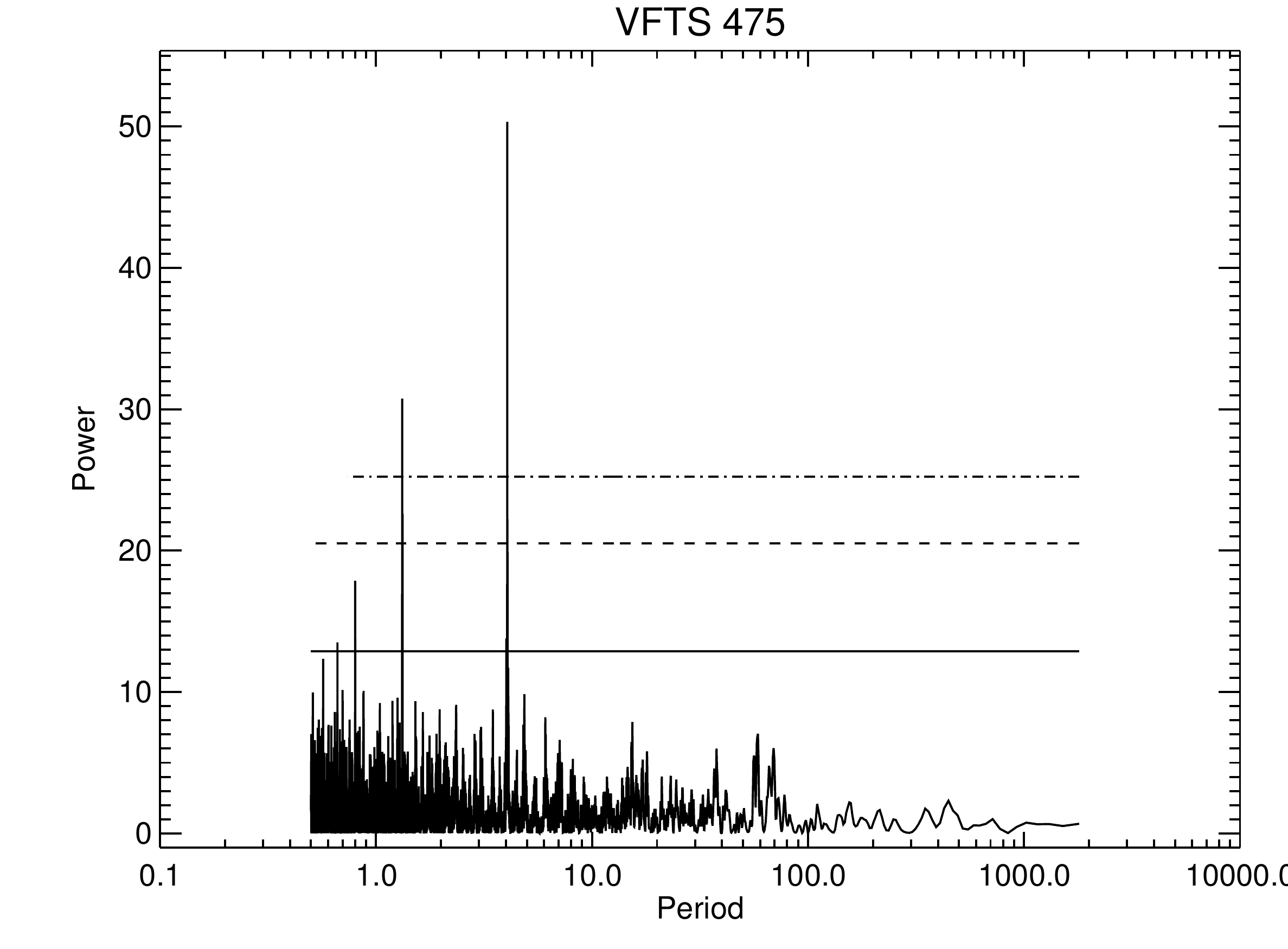}
\includegraphics[width=4.4cm,angle=-90]{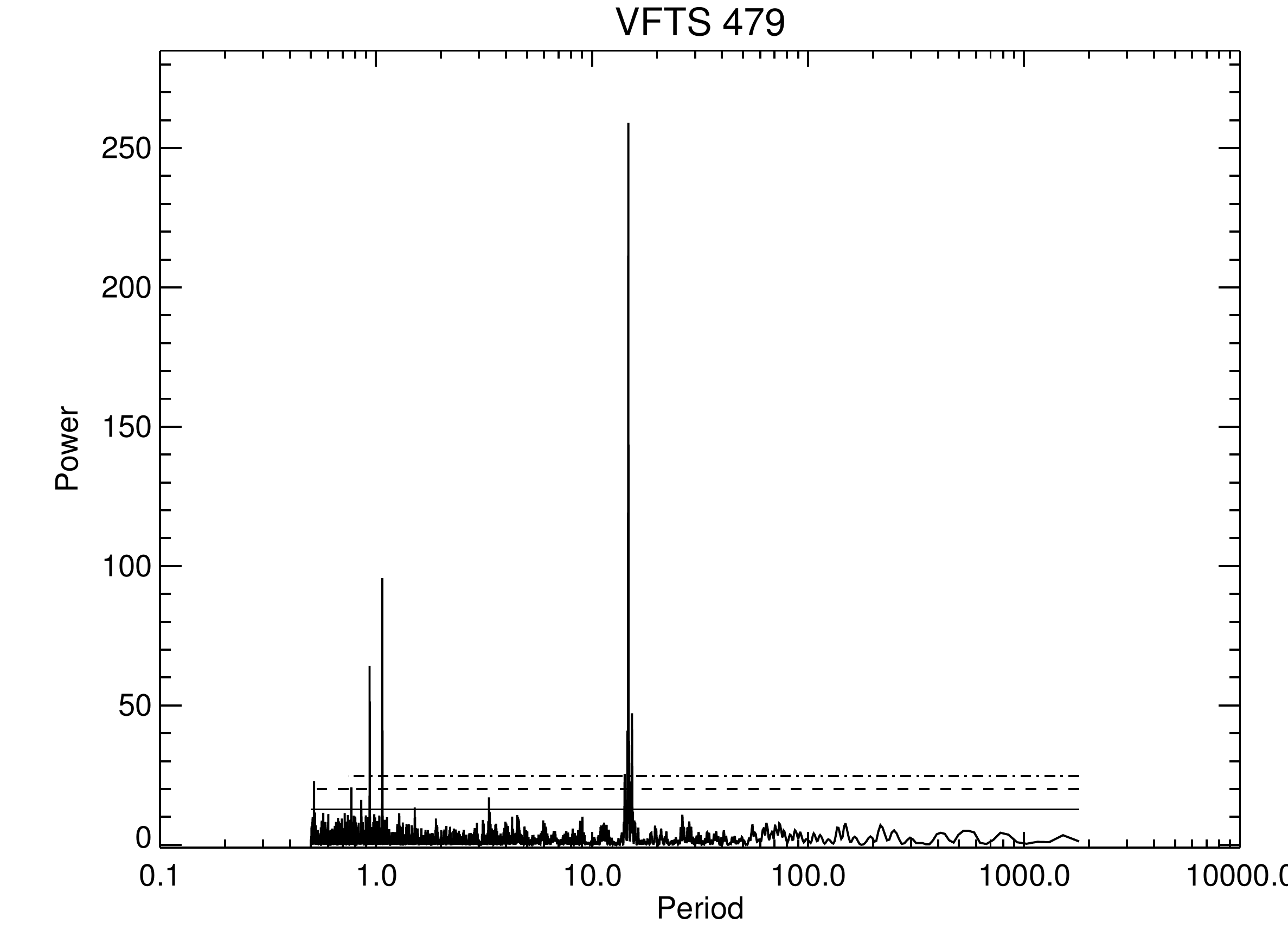}
\includegraphics[width=4.4cm,angle=-90]{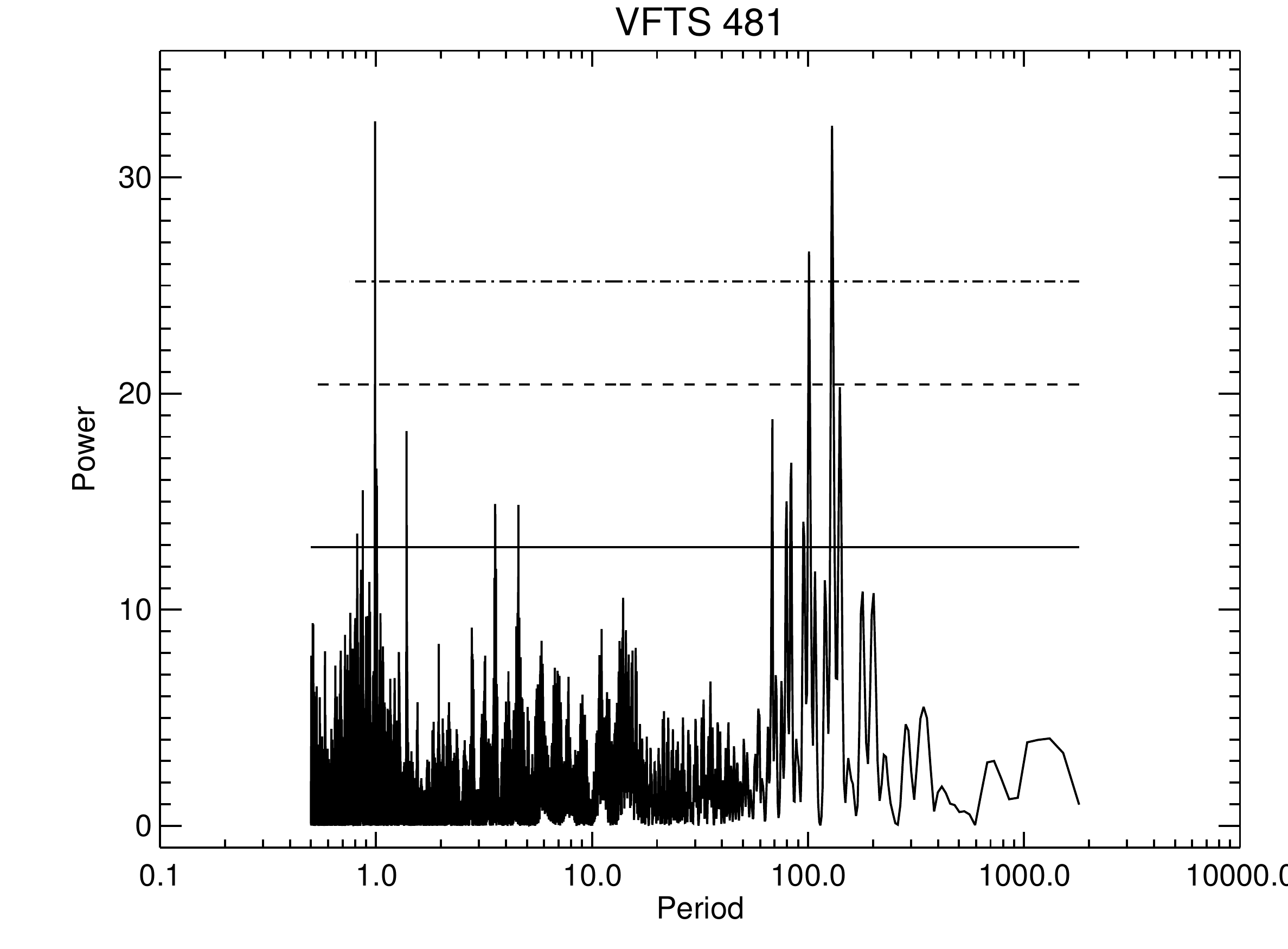}
\caption{{\it Continued...}}
\label{sb1:periodogram2}
\end{figure*}

\begin{figure*}
\centering
\ContinuedFloat
\includegraphics[width=4.4cm,angle=-90]{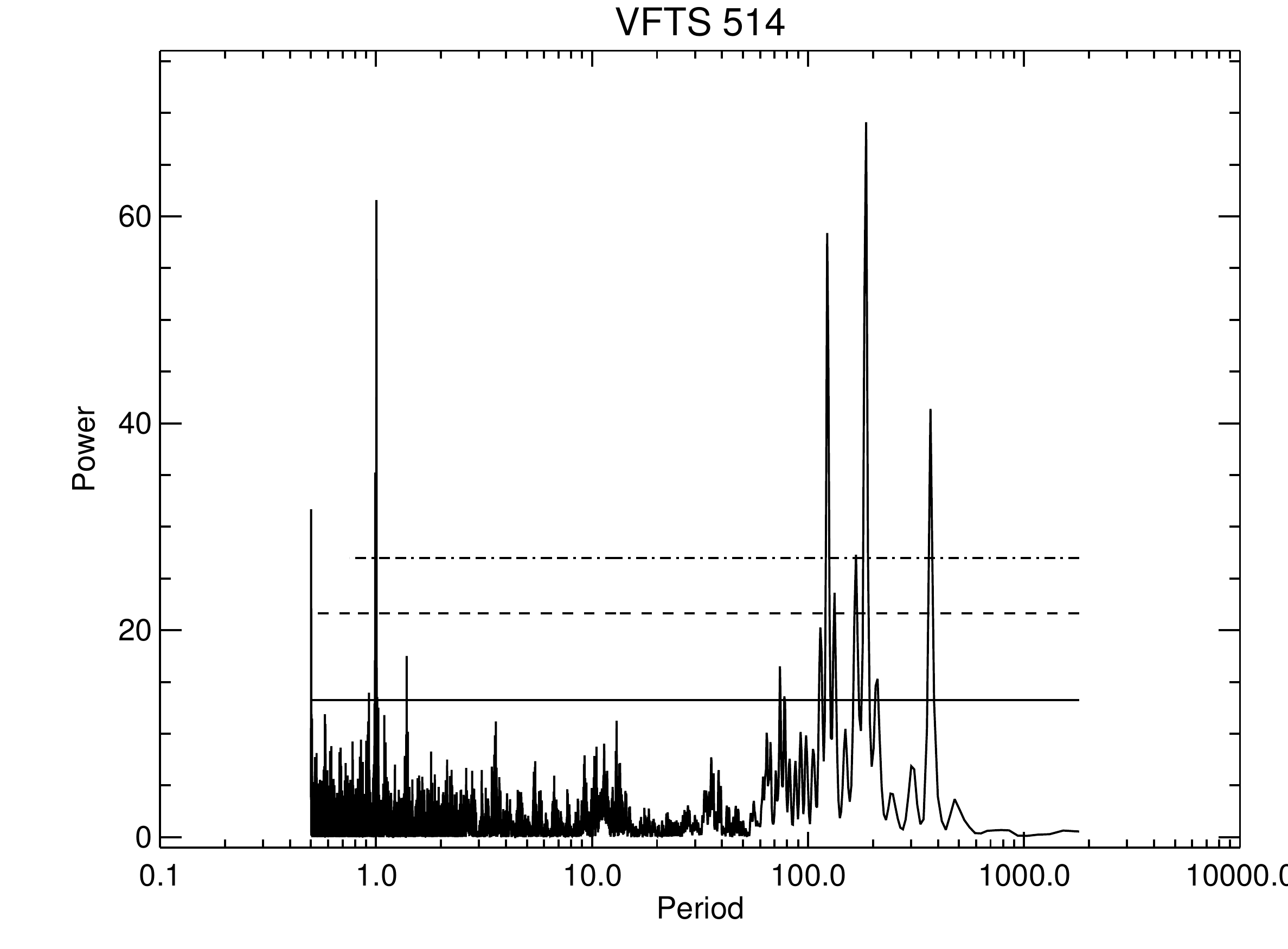}
\includegraphics[width=4.4cm,angle=-90]{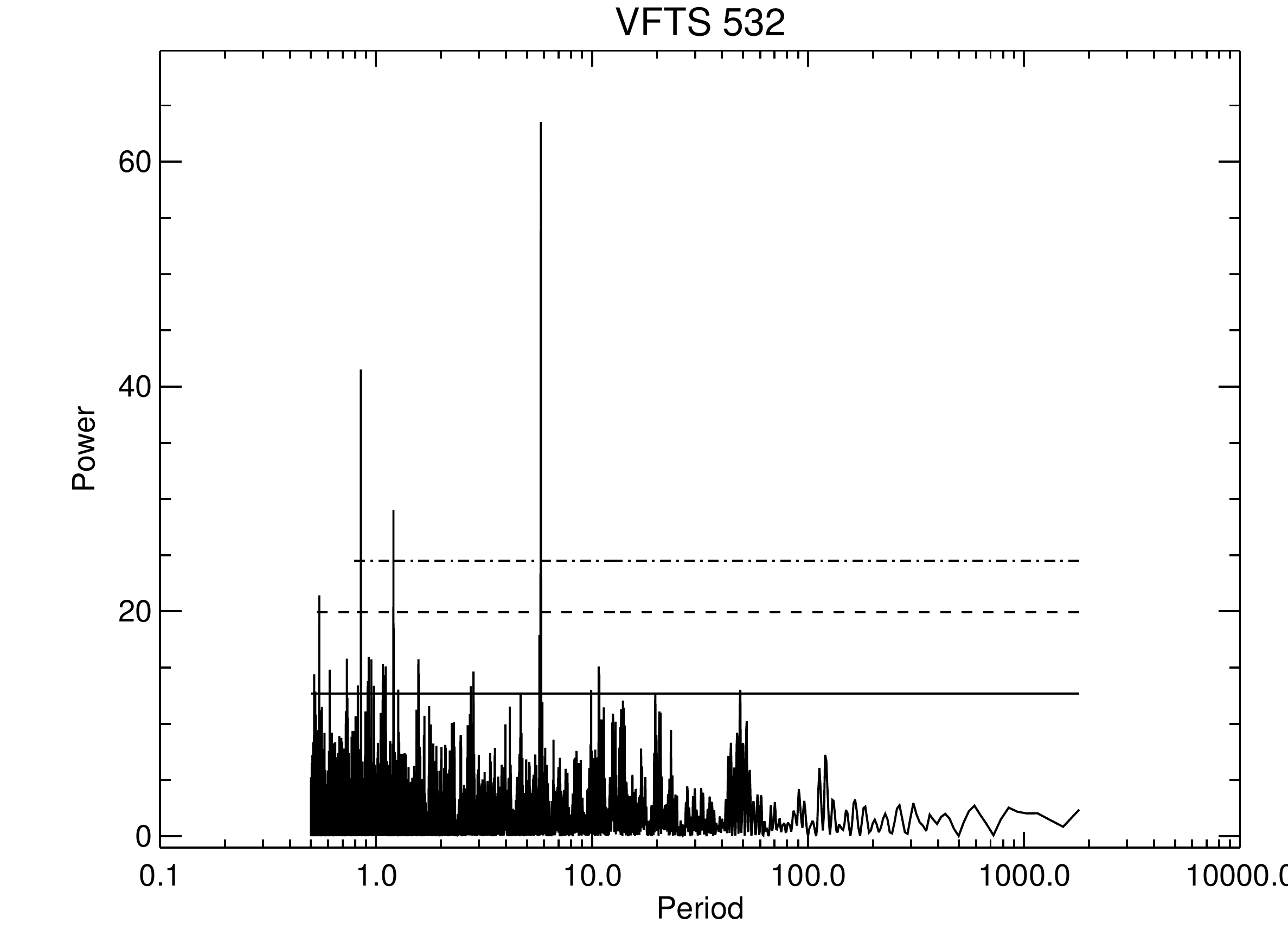}
\includegraphics[width=4.4cm,angle=-90]{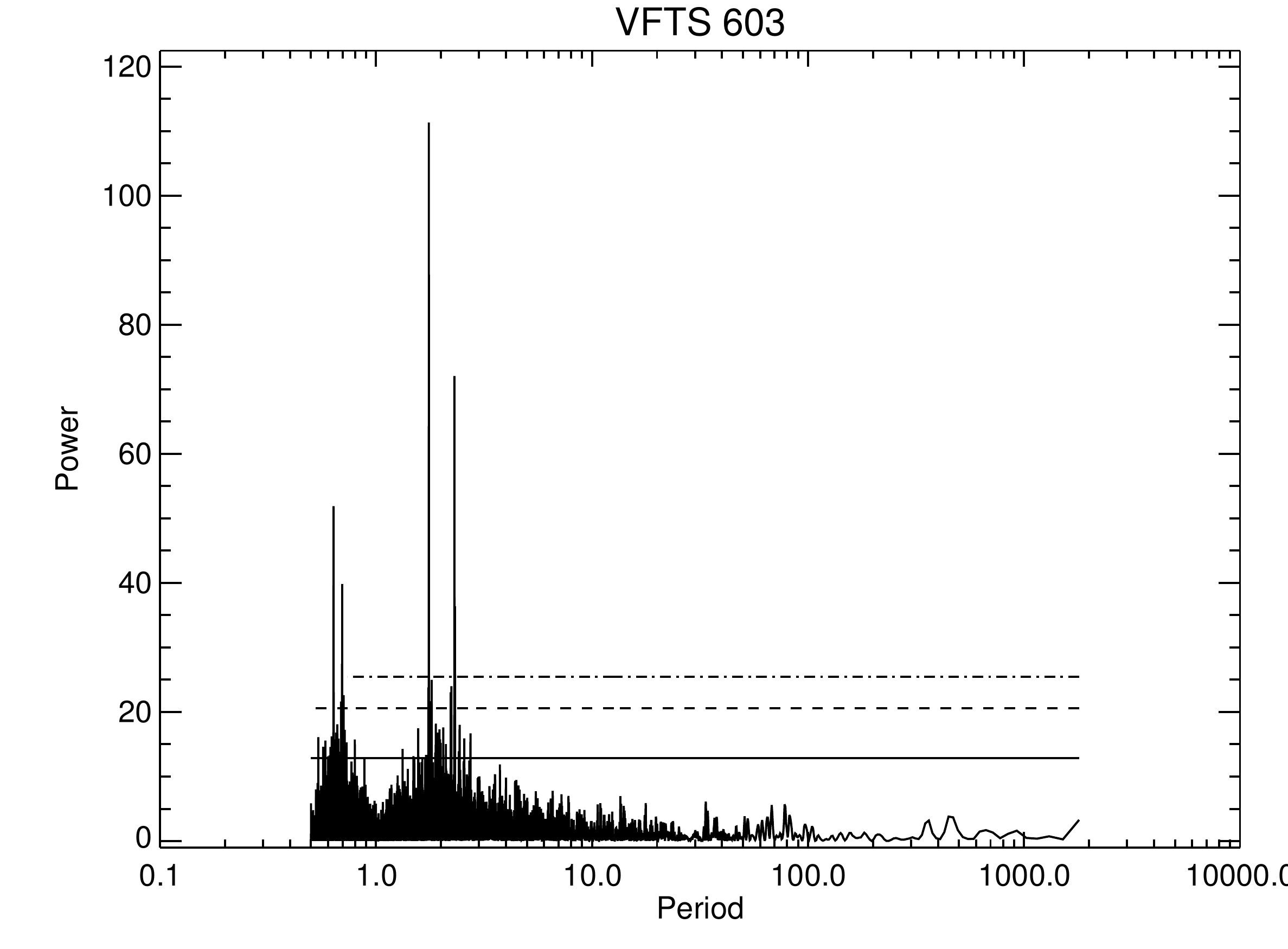}
\includegraphics[width=4.4cm,angle=-90]{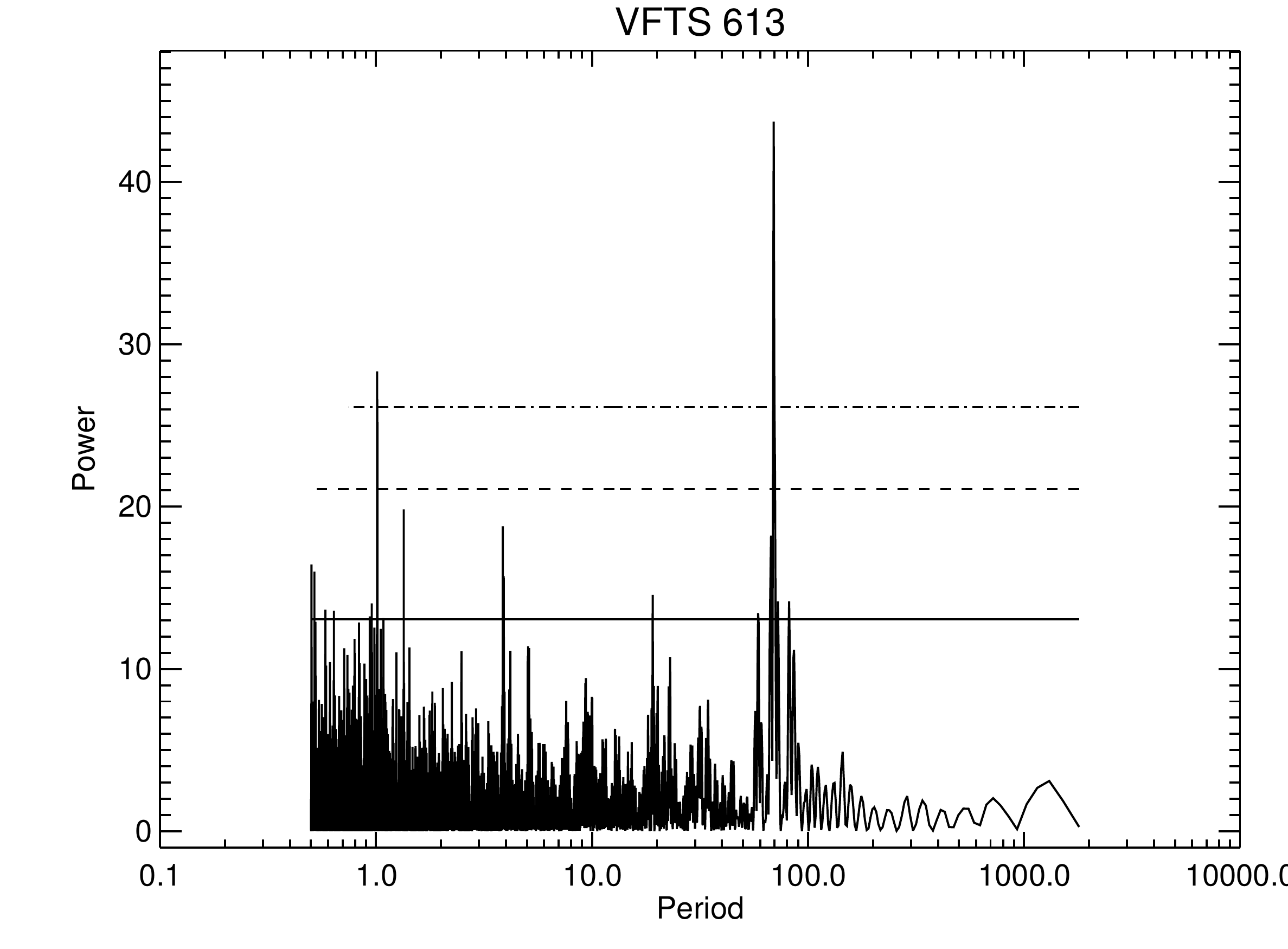}
\includegraphics[width=4.4cm,angle=-90]{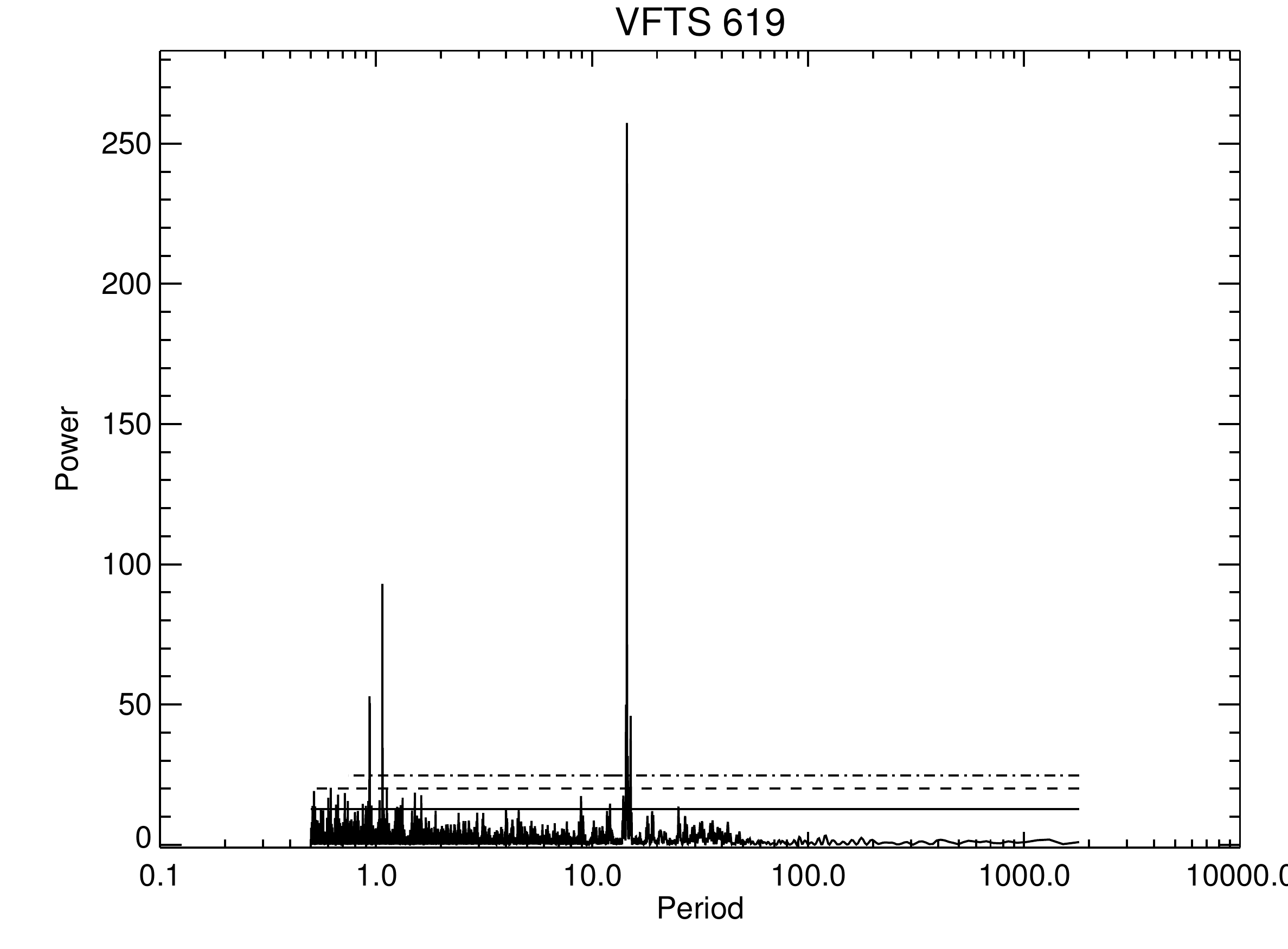}
\includegraphics[width=4.4cm,angle=-90]{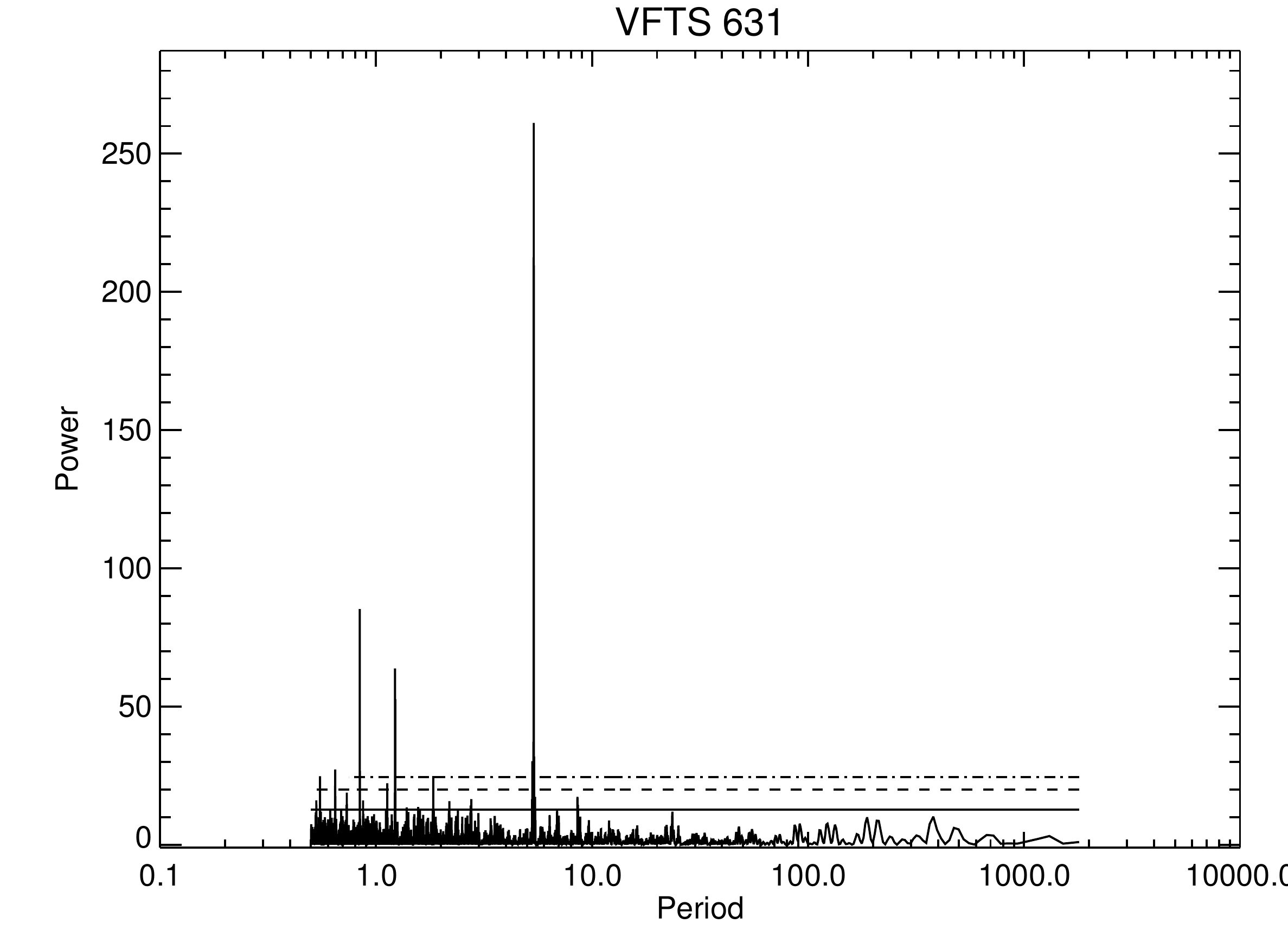}
\includegraphics[width=4.4cm,angle=-90]{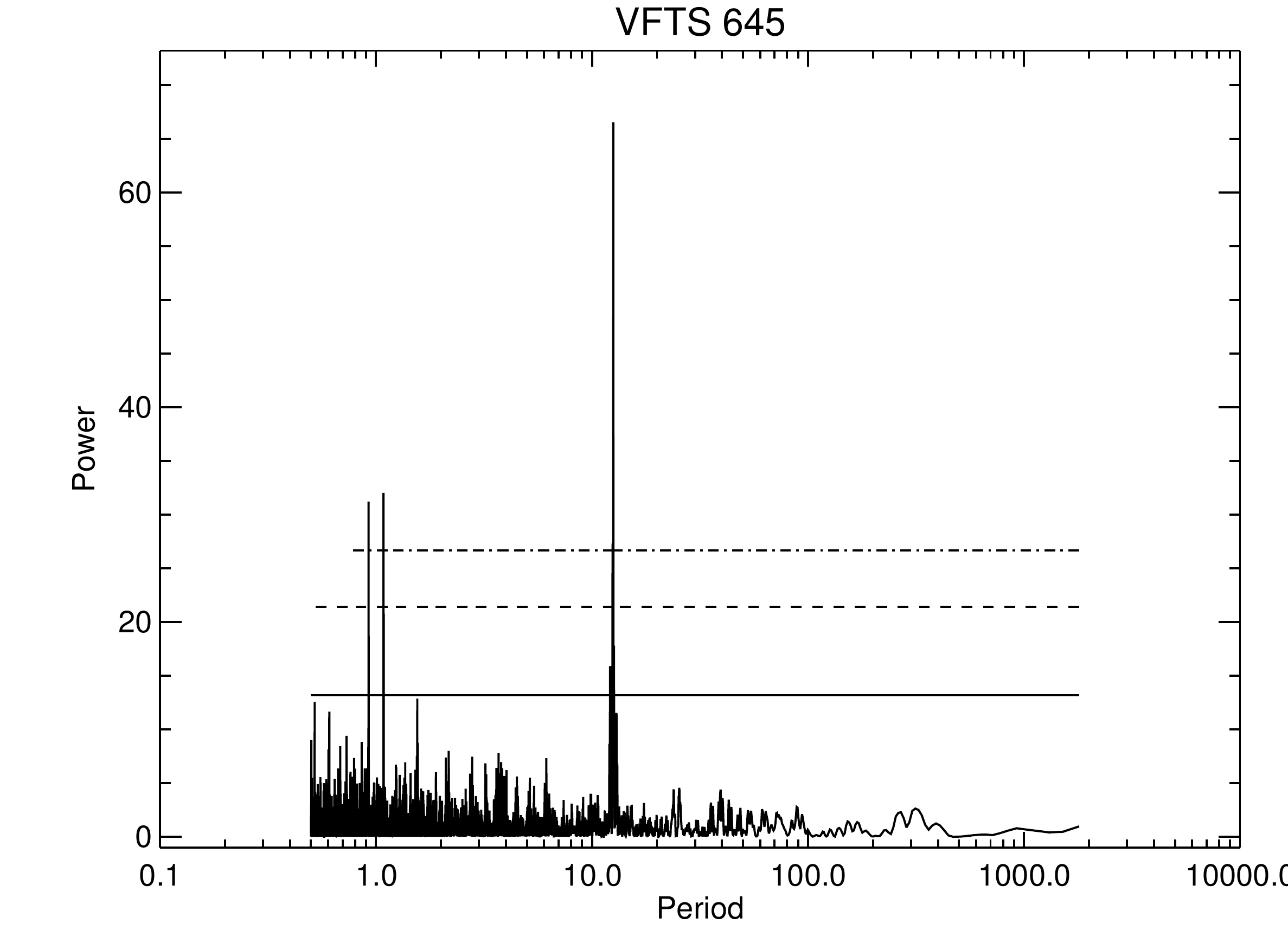}
\includegraphics[width=4.4cm,angle=-90]{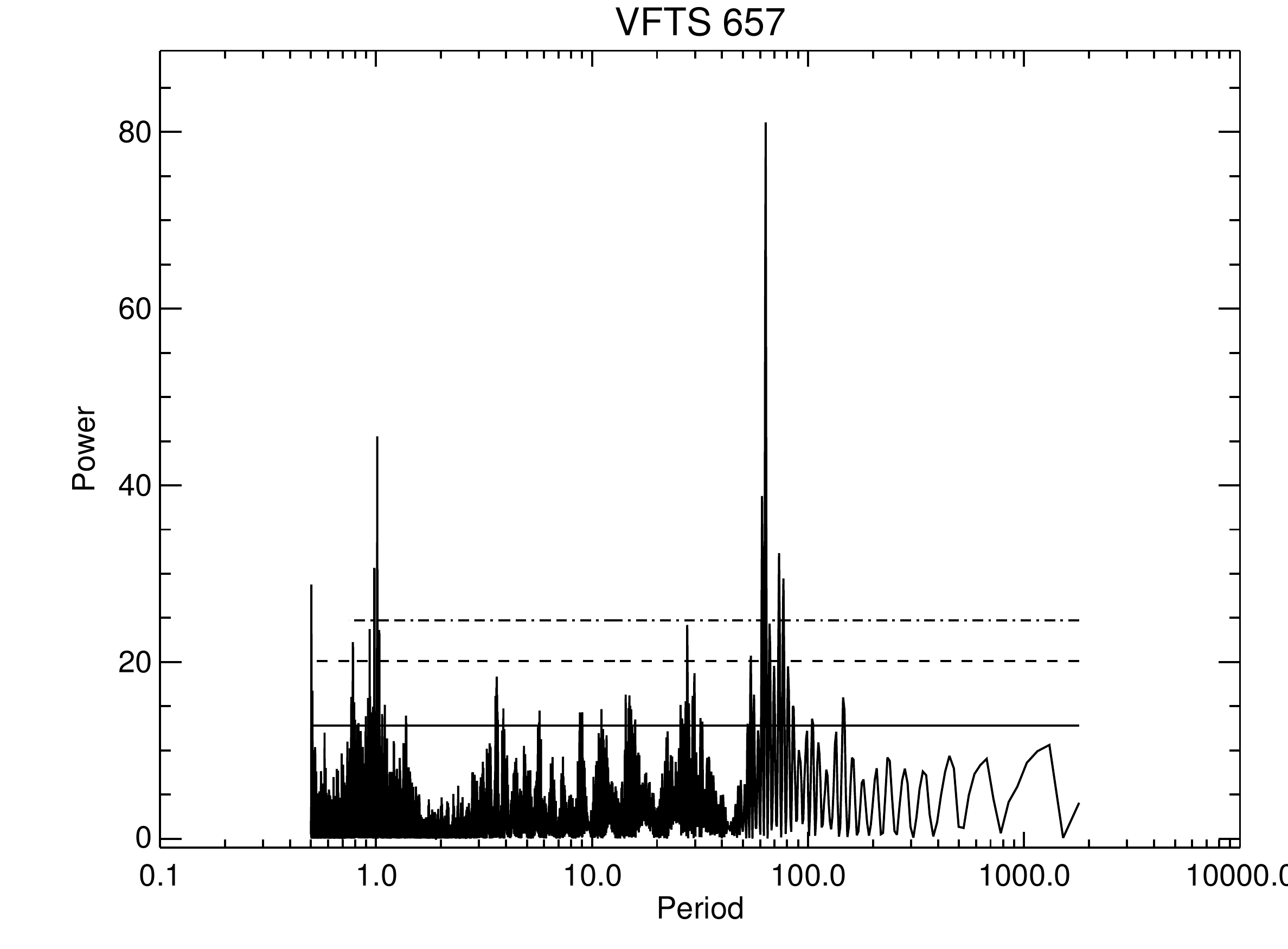}
\includegraphics[width=4.4cm,angle=-90]{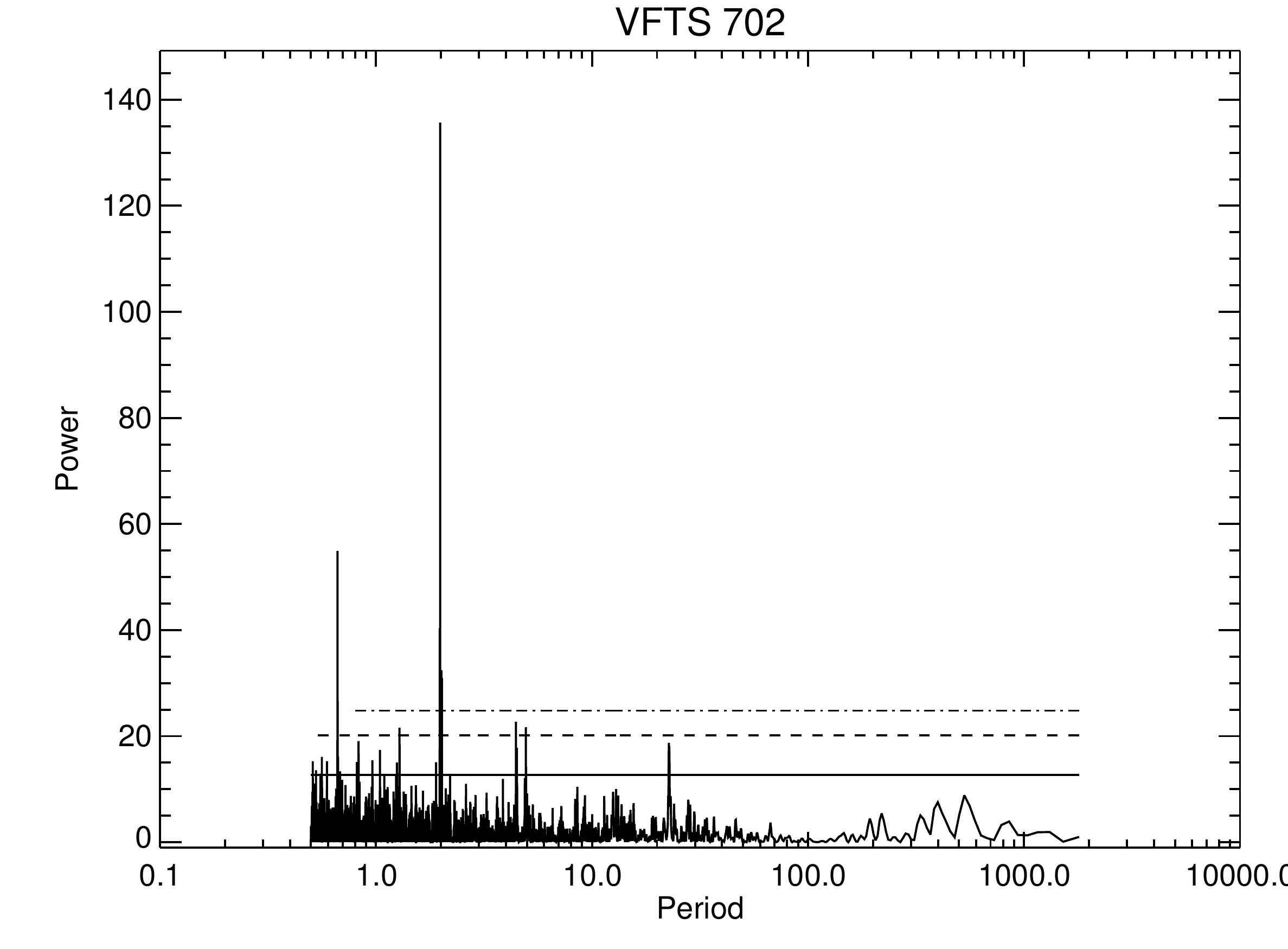}
\includegraphics[width=4.4cm,angle=-90]{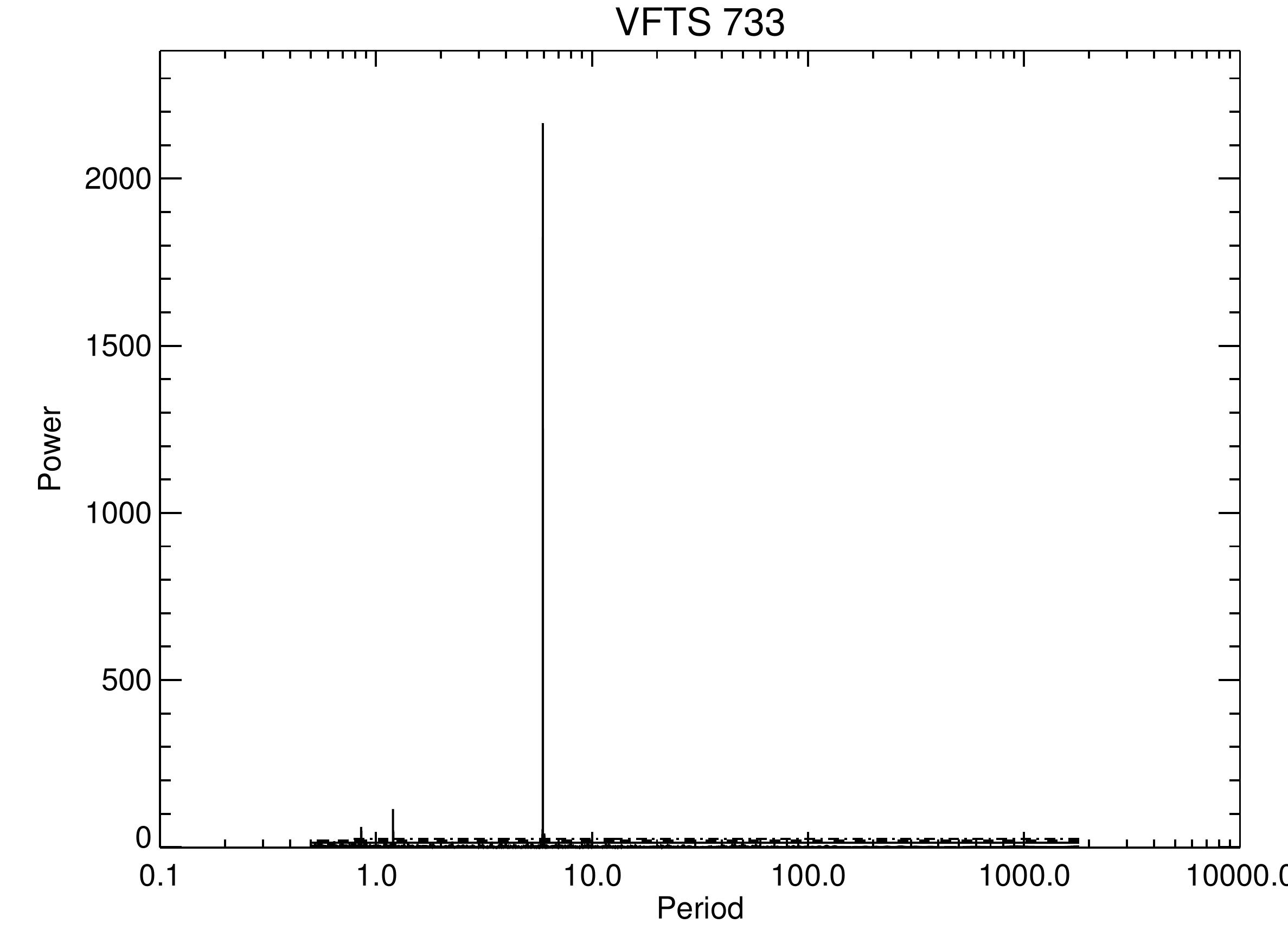}
\includegraphics[width=4.4cm,angle=-90]{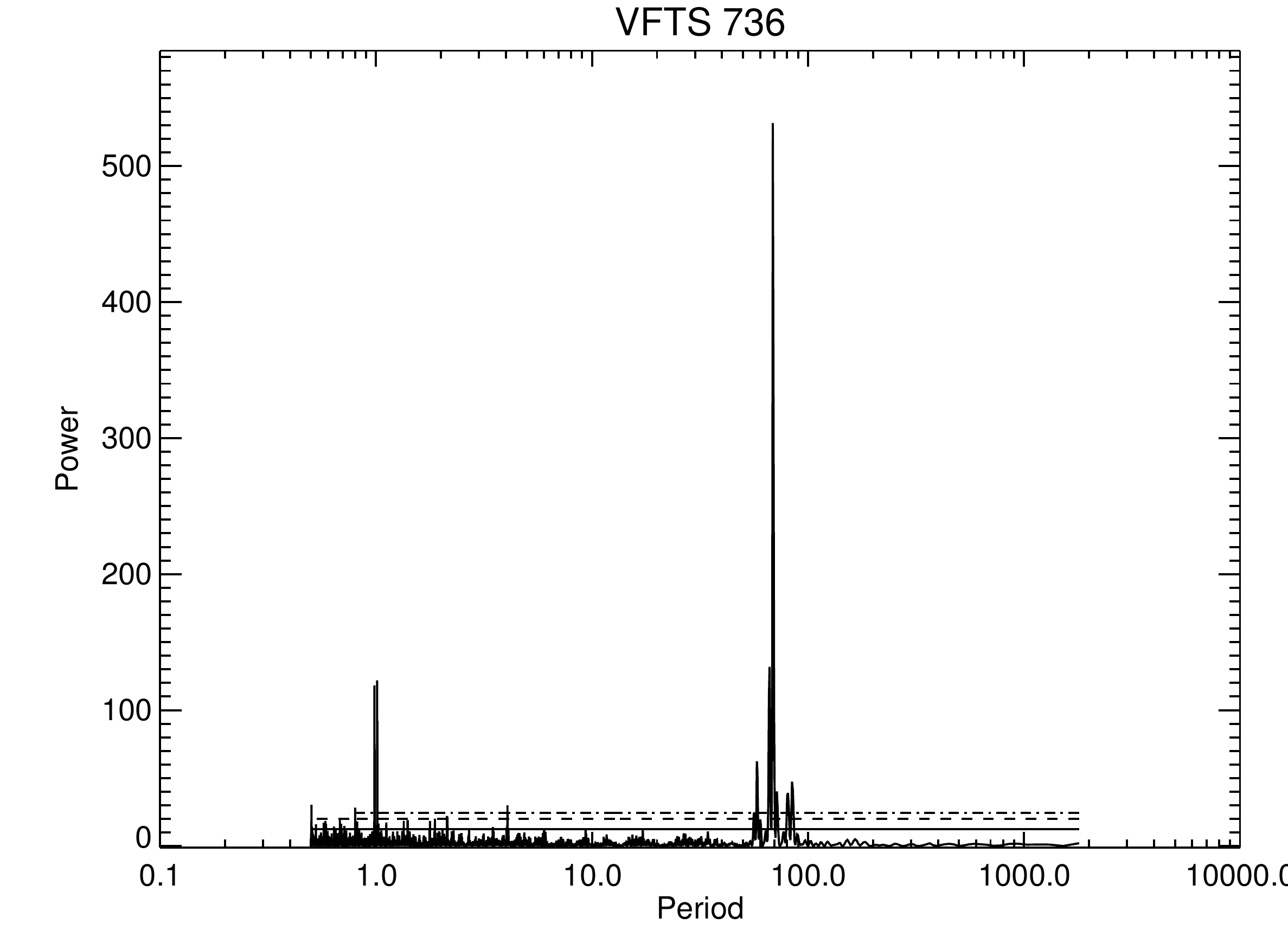}
\includegraphics[width=4.4cm,angle=-90]{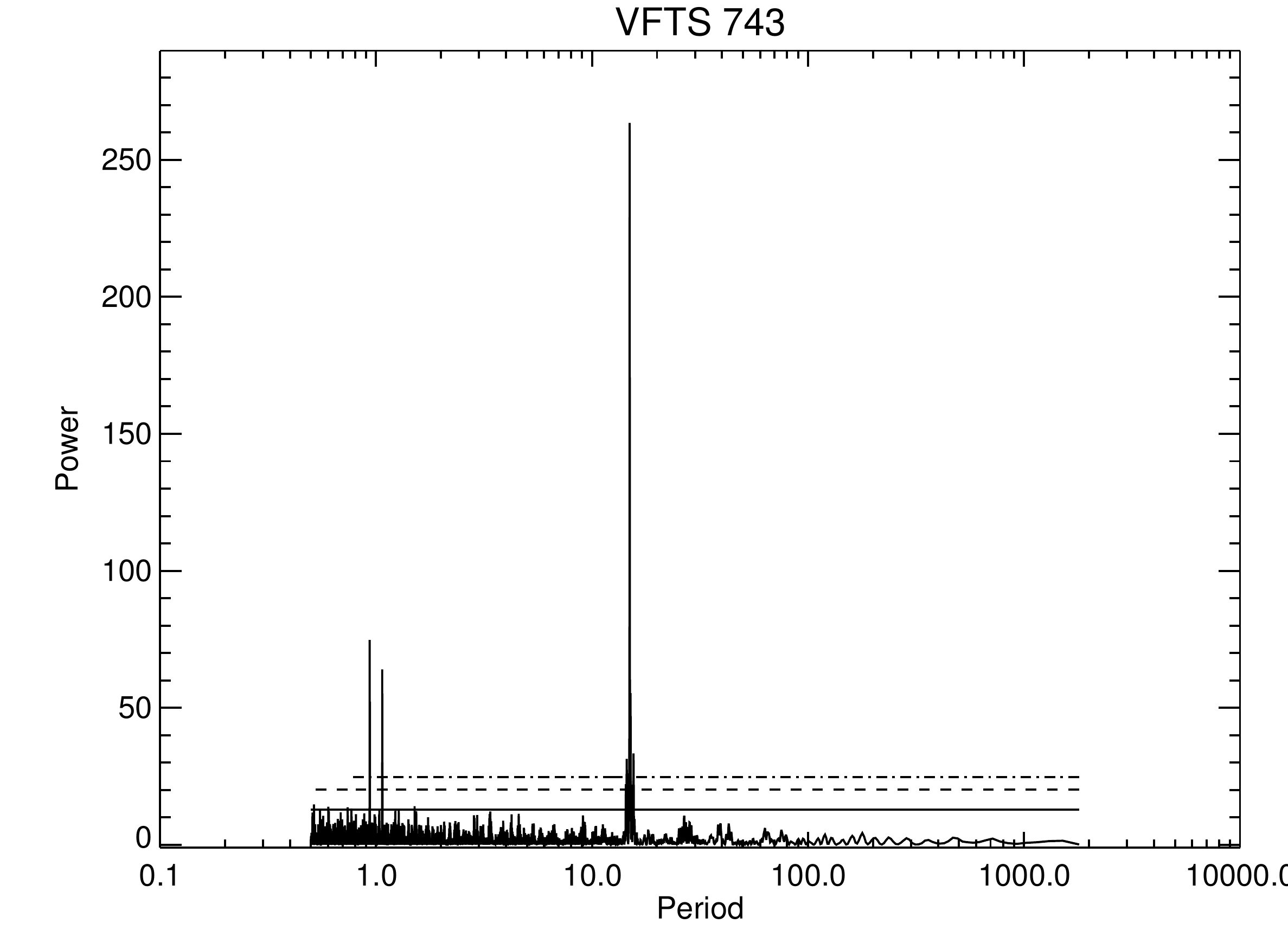}
\includegraphics[width=4.4cm,angle=-90]{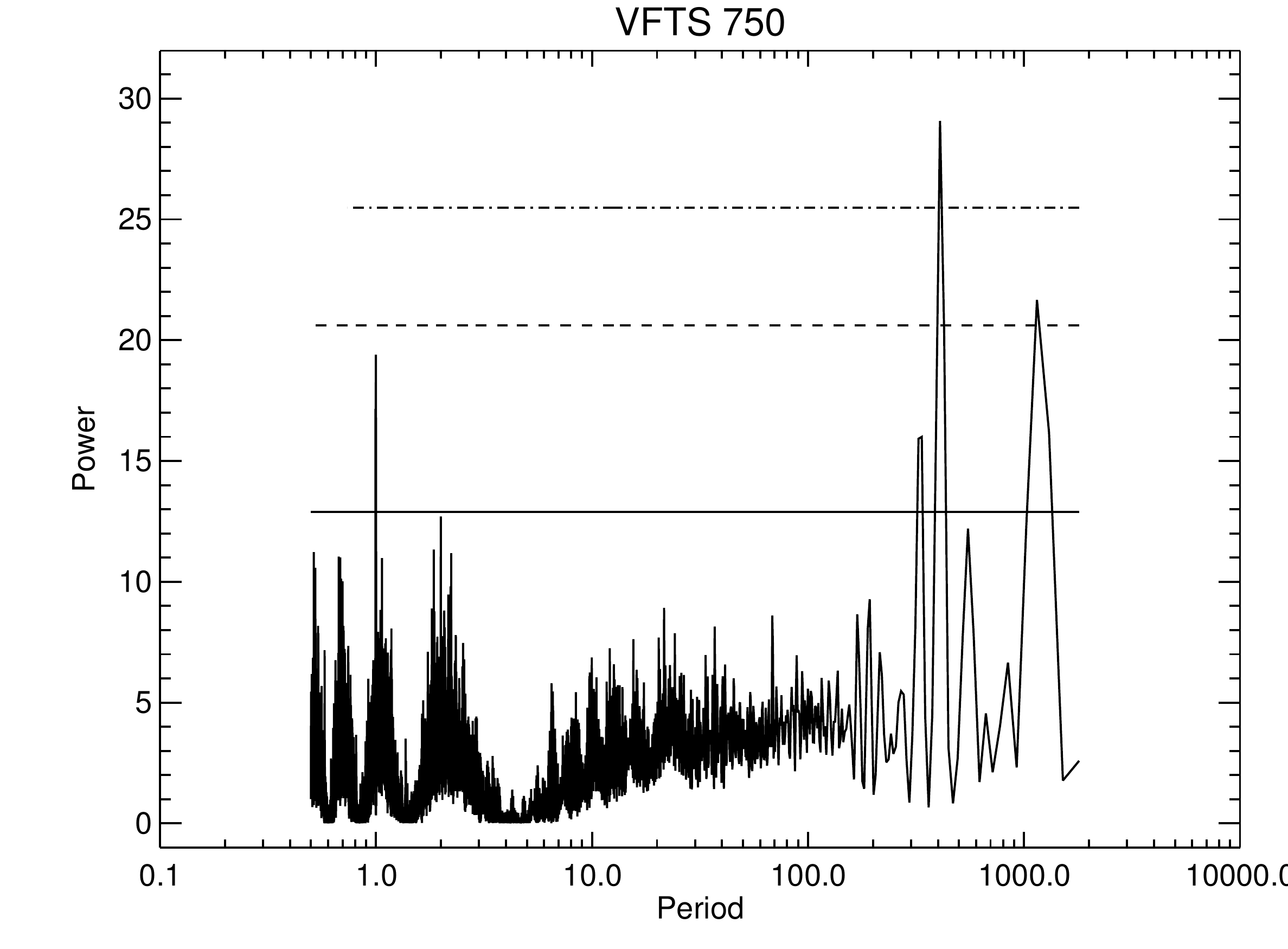}
\includegraphics[width=4.4cm,angle=-90]{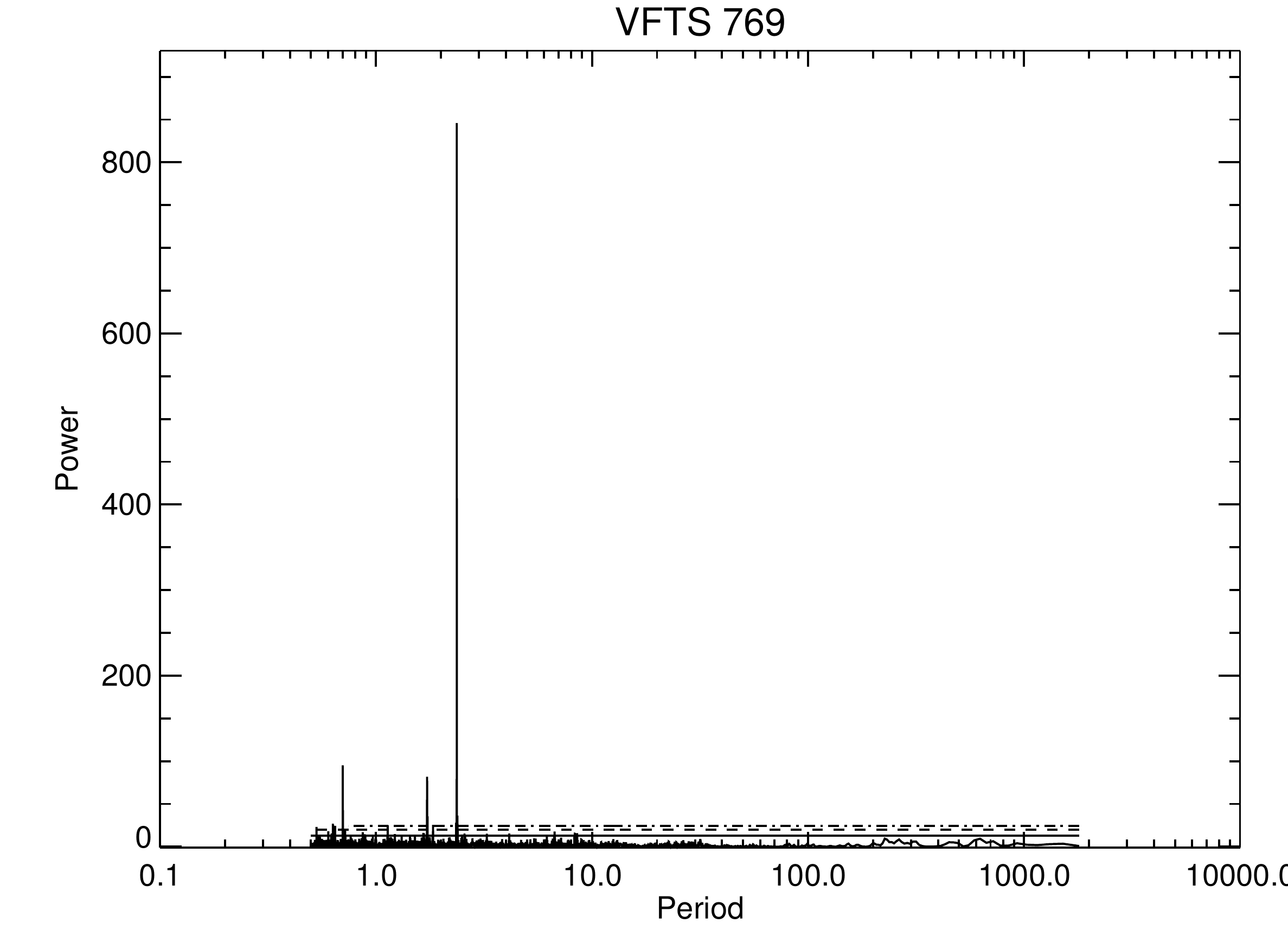}
\includegraphics[width=4.4cm,angle=-90]{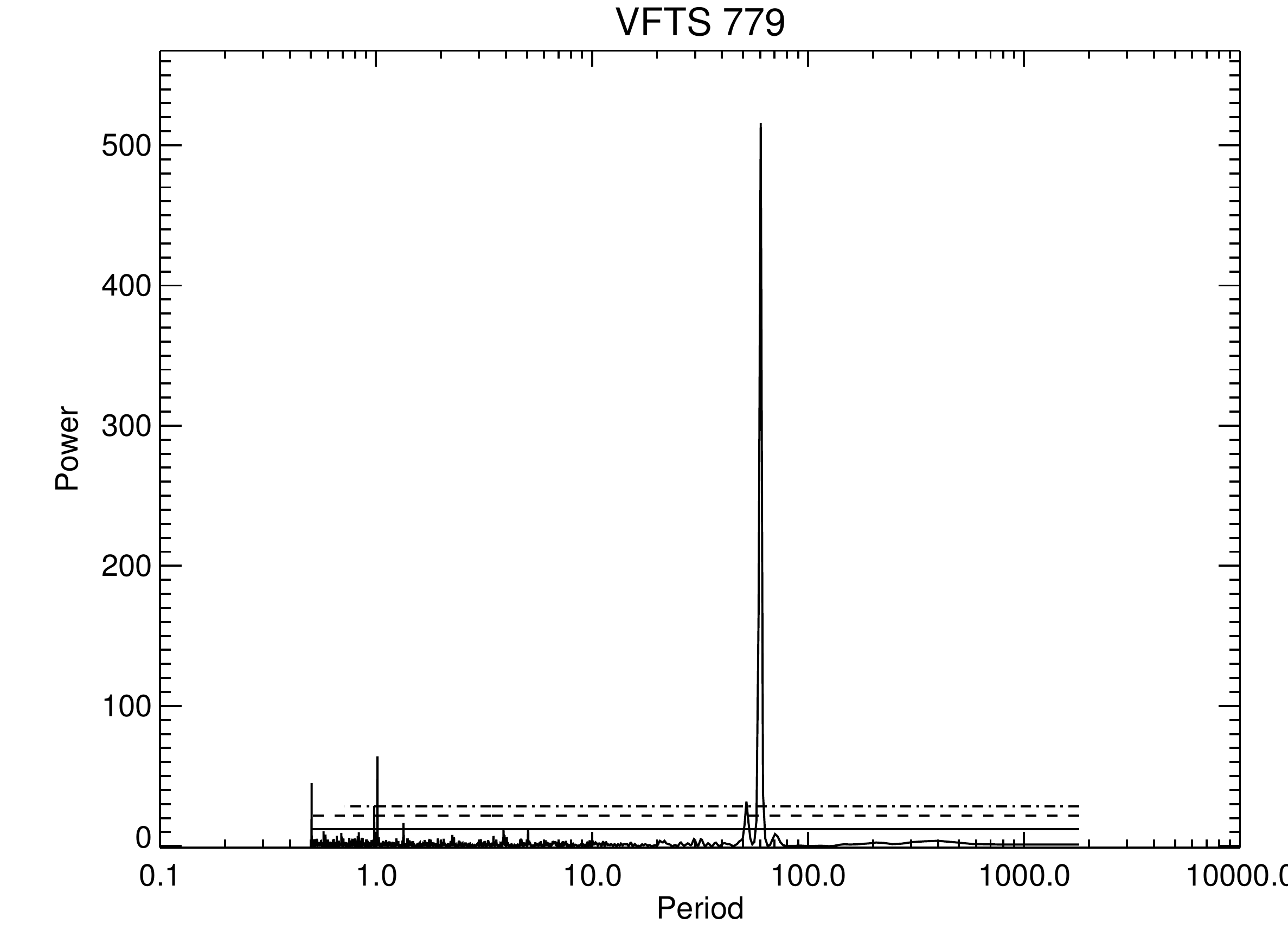}
\caption{{\it Continued...}}
\label{sb1:periodogram3}
\end{figure*}

\begin{figure*}
\centering
\ContinuedFloat
\includegraphics[width=4.4cm,angle=-90]{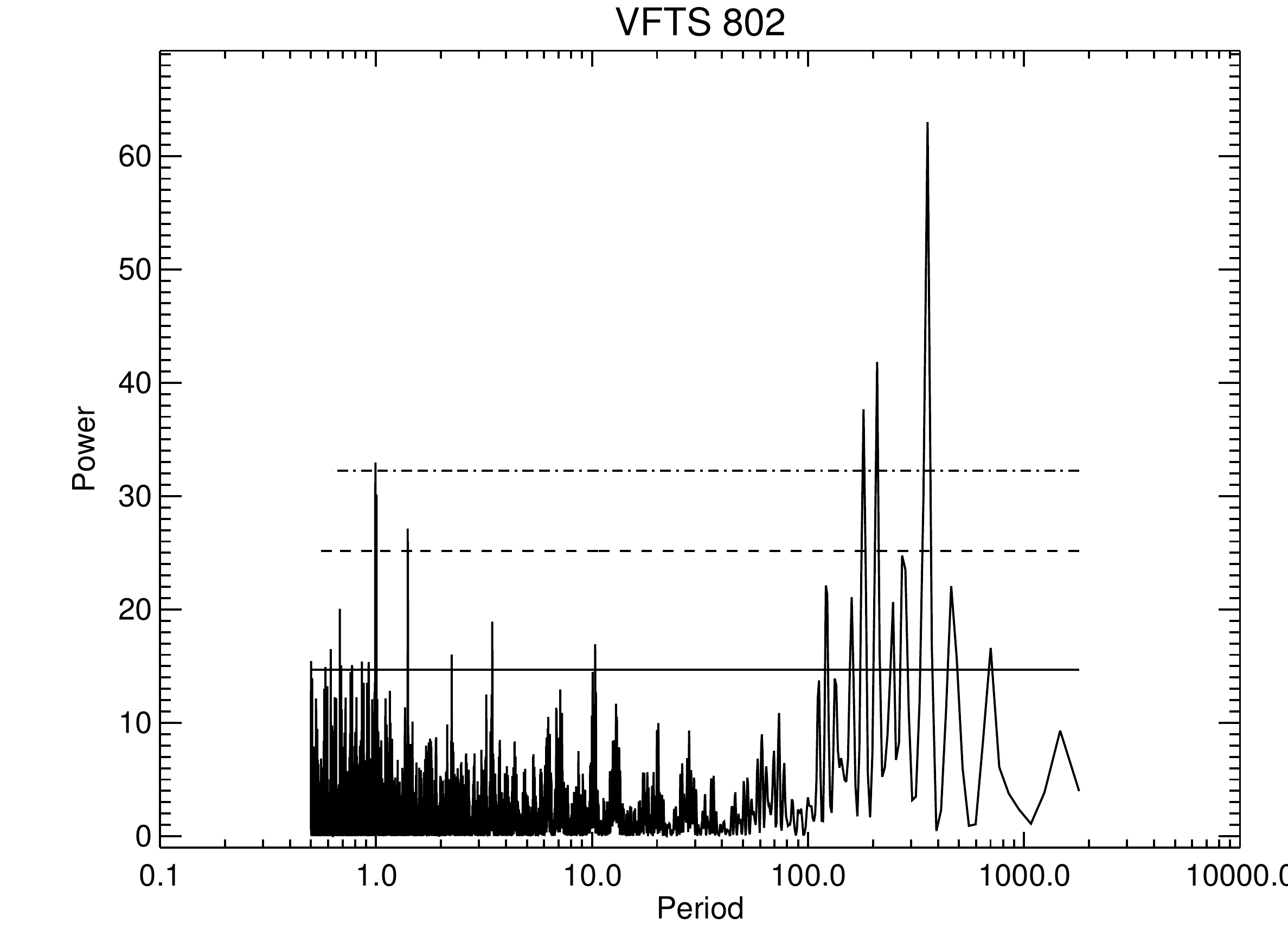}
\includegraphics[width=4.4cm,angle=-90]{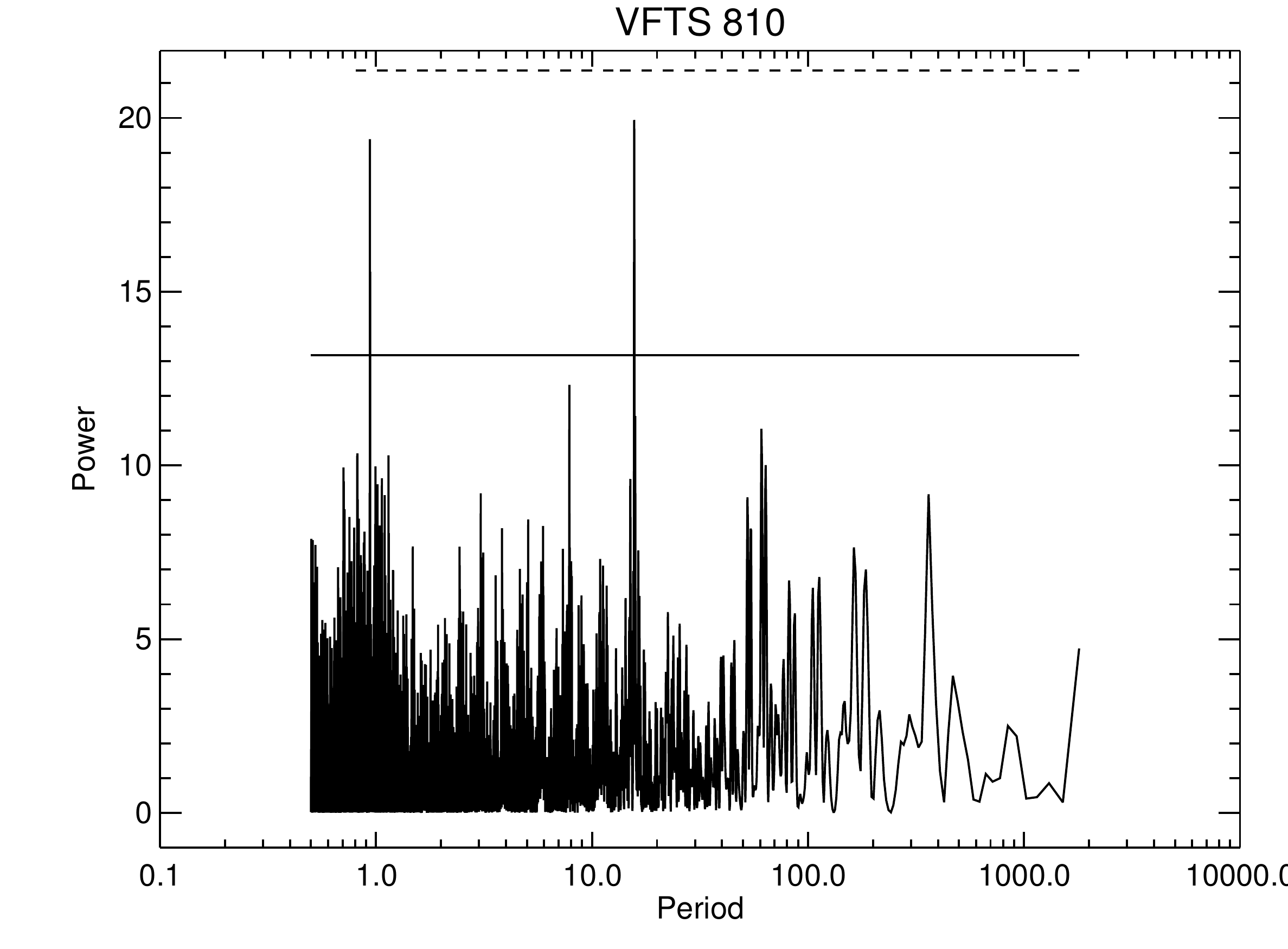}
\includegraphics[width=4.4cm,angle=-90]{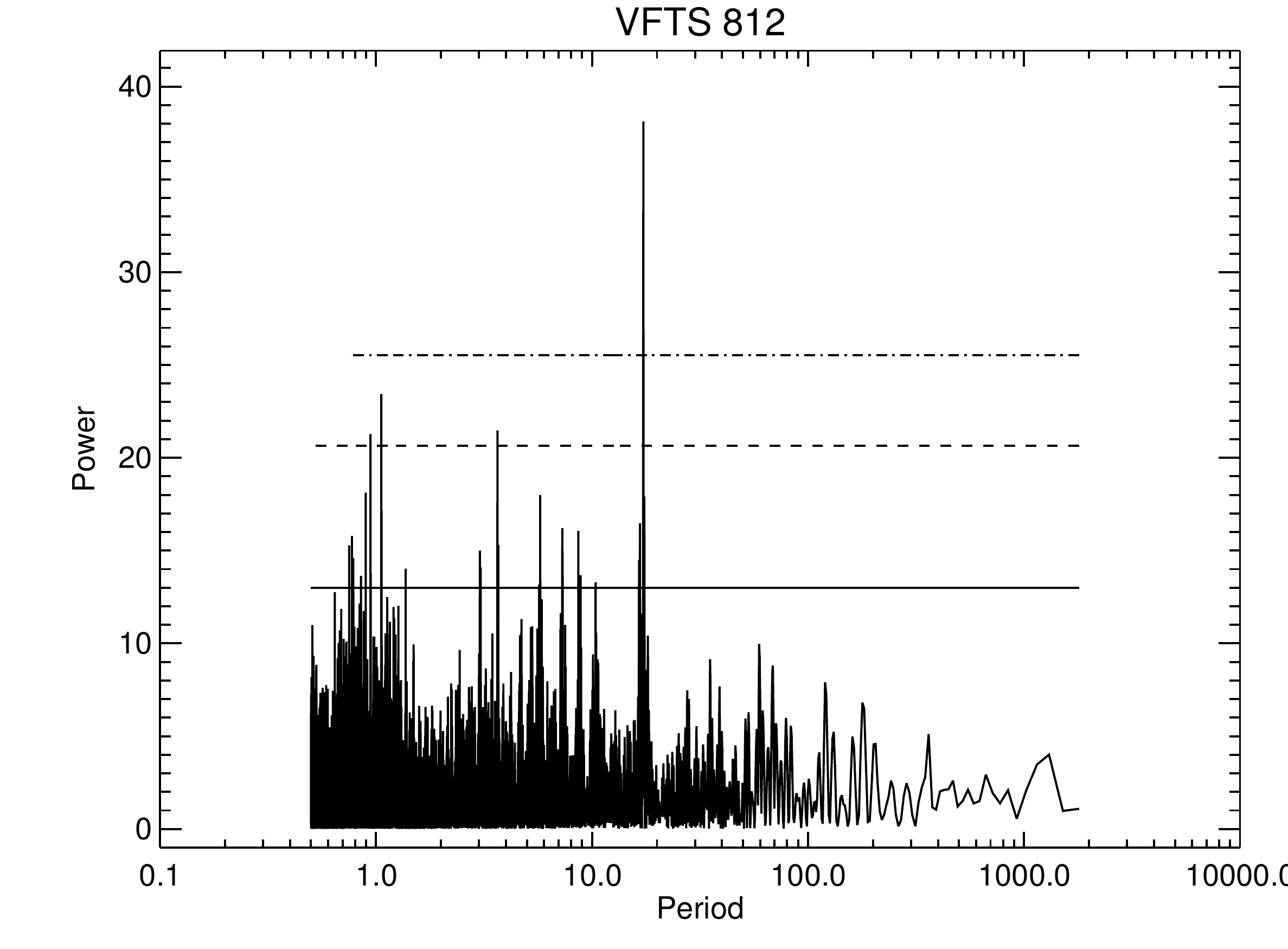}
\includegraphics[width=4.4cm,angle=-90]{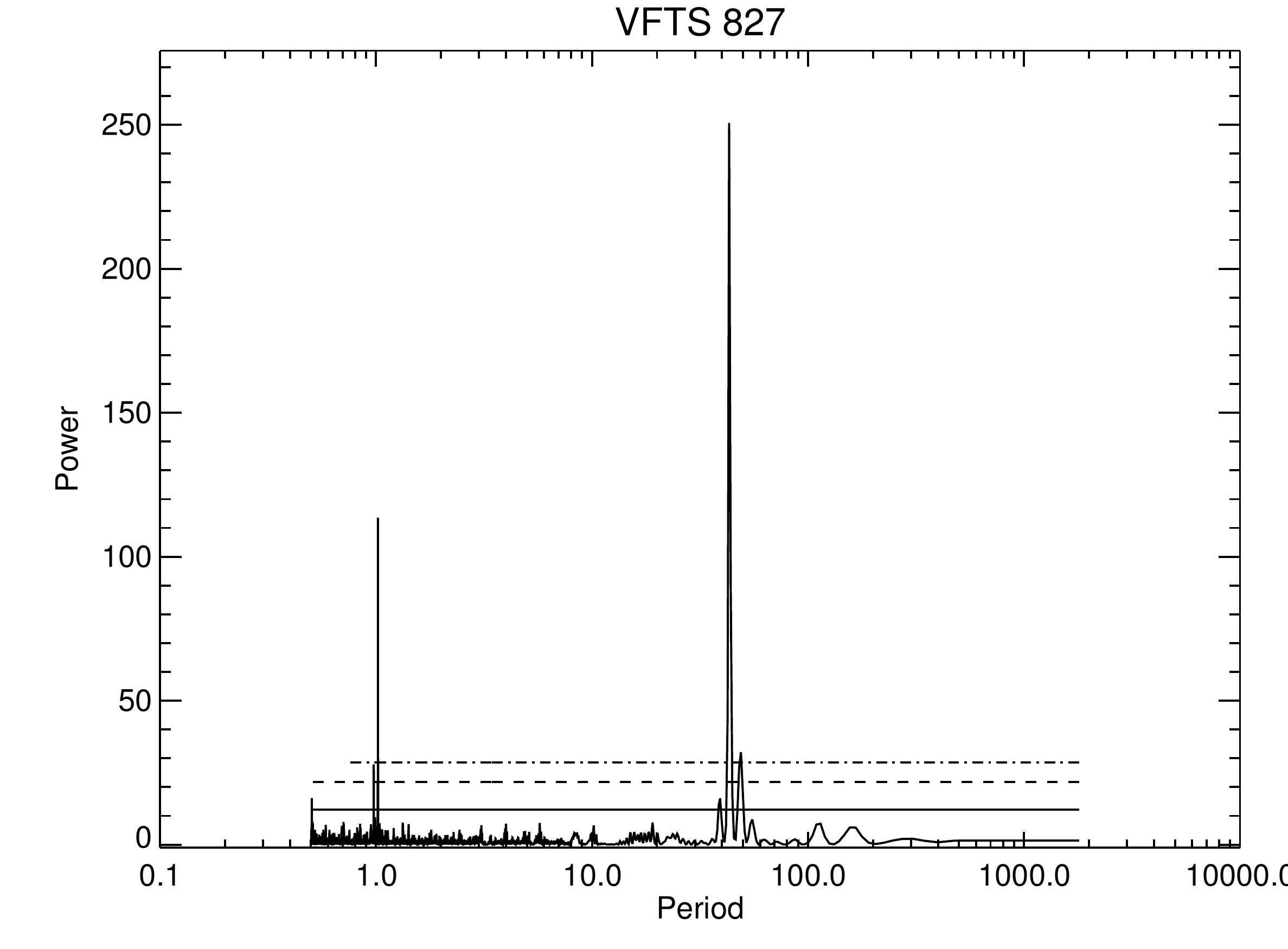}
\includegraphics[width=4.4cm,angle=-90]{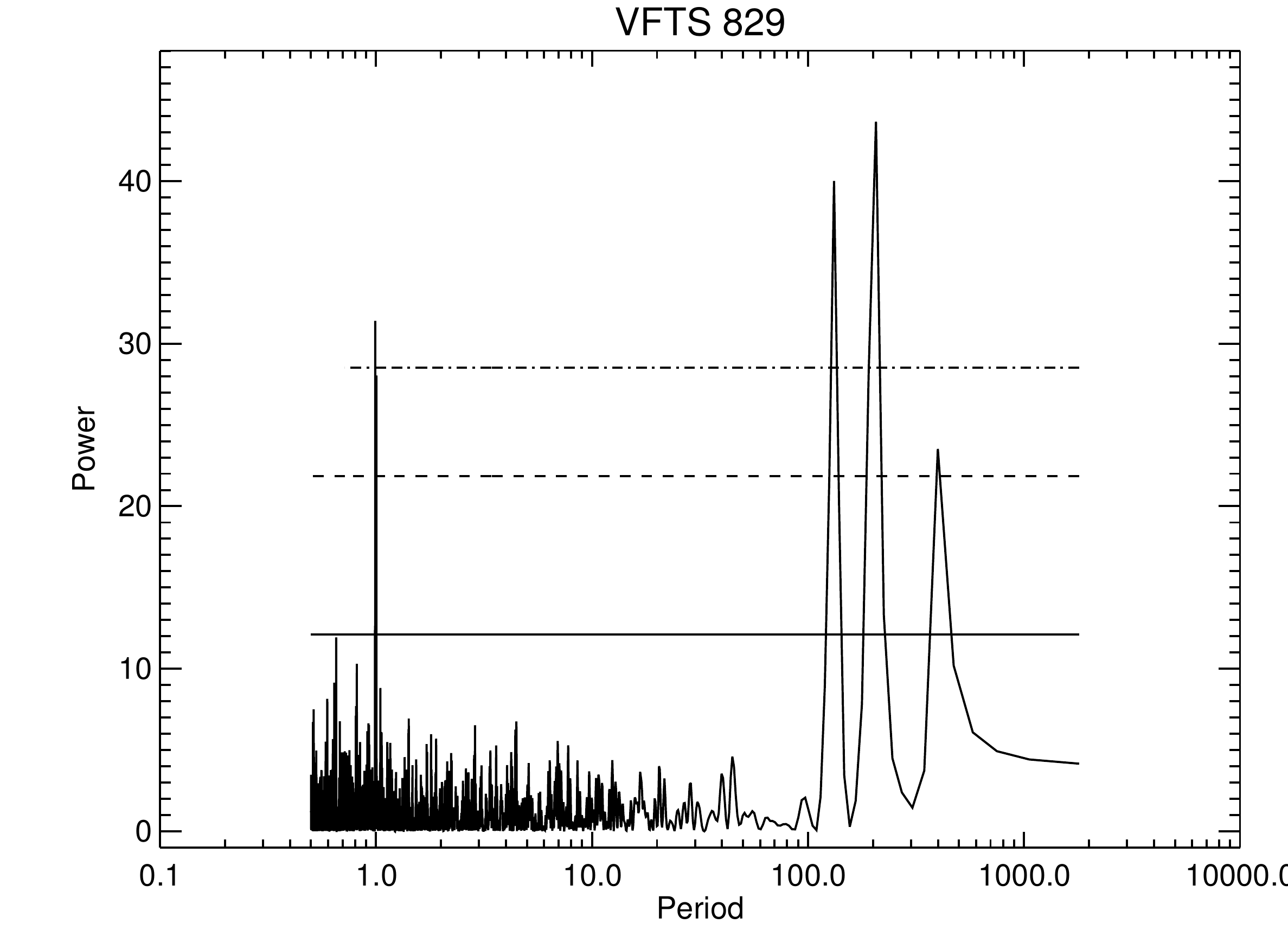}
\includegraphics[width=4.4cm,angle=-90]{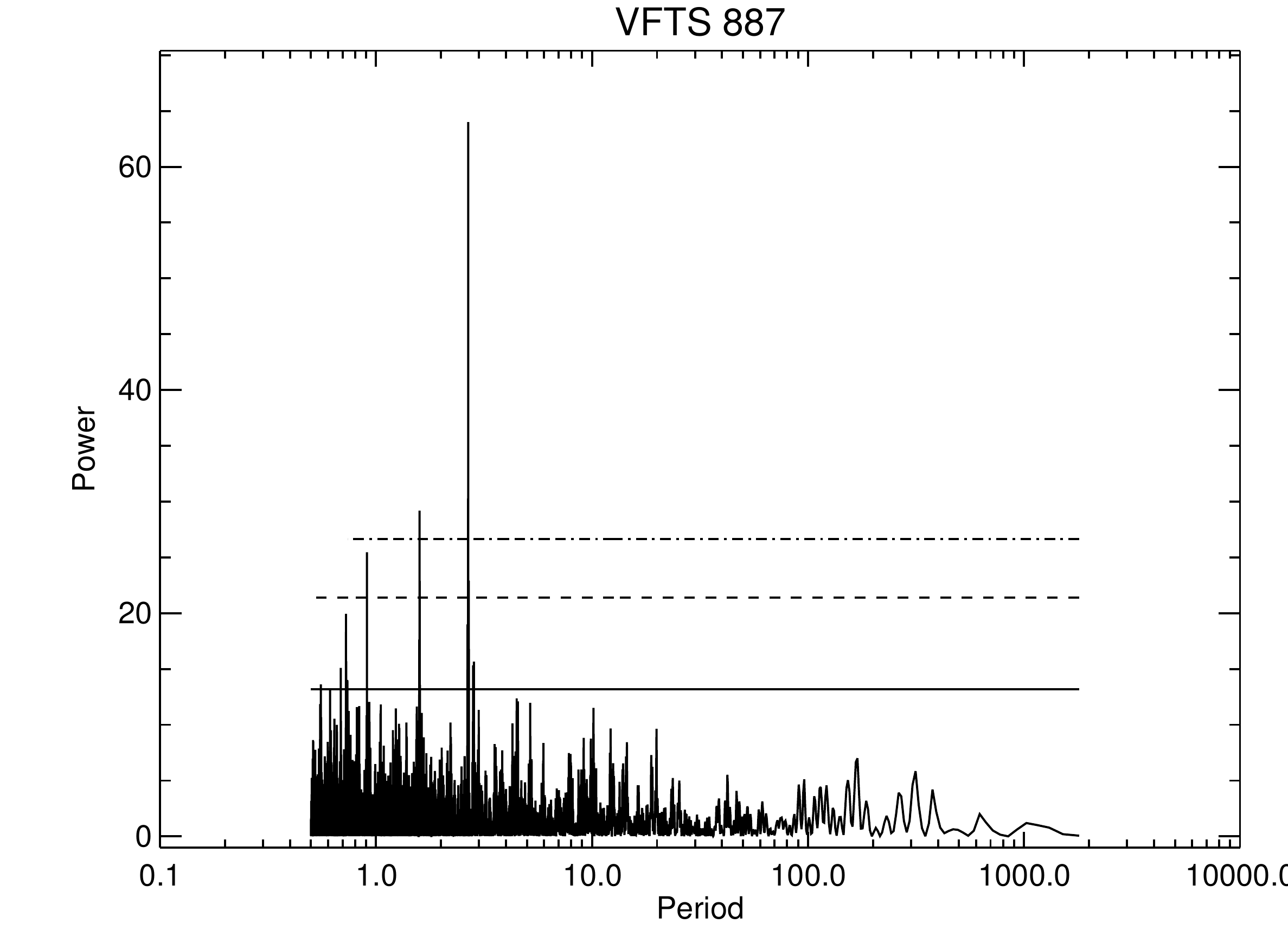}
\caption{{\it Continued...}}
\label{sb1:periodogram4}
\end{figure*}

\begin{figure*}
\centering
\includegraphics[width=4.4cm,angle=-90]{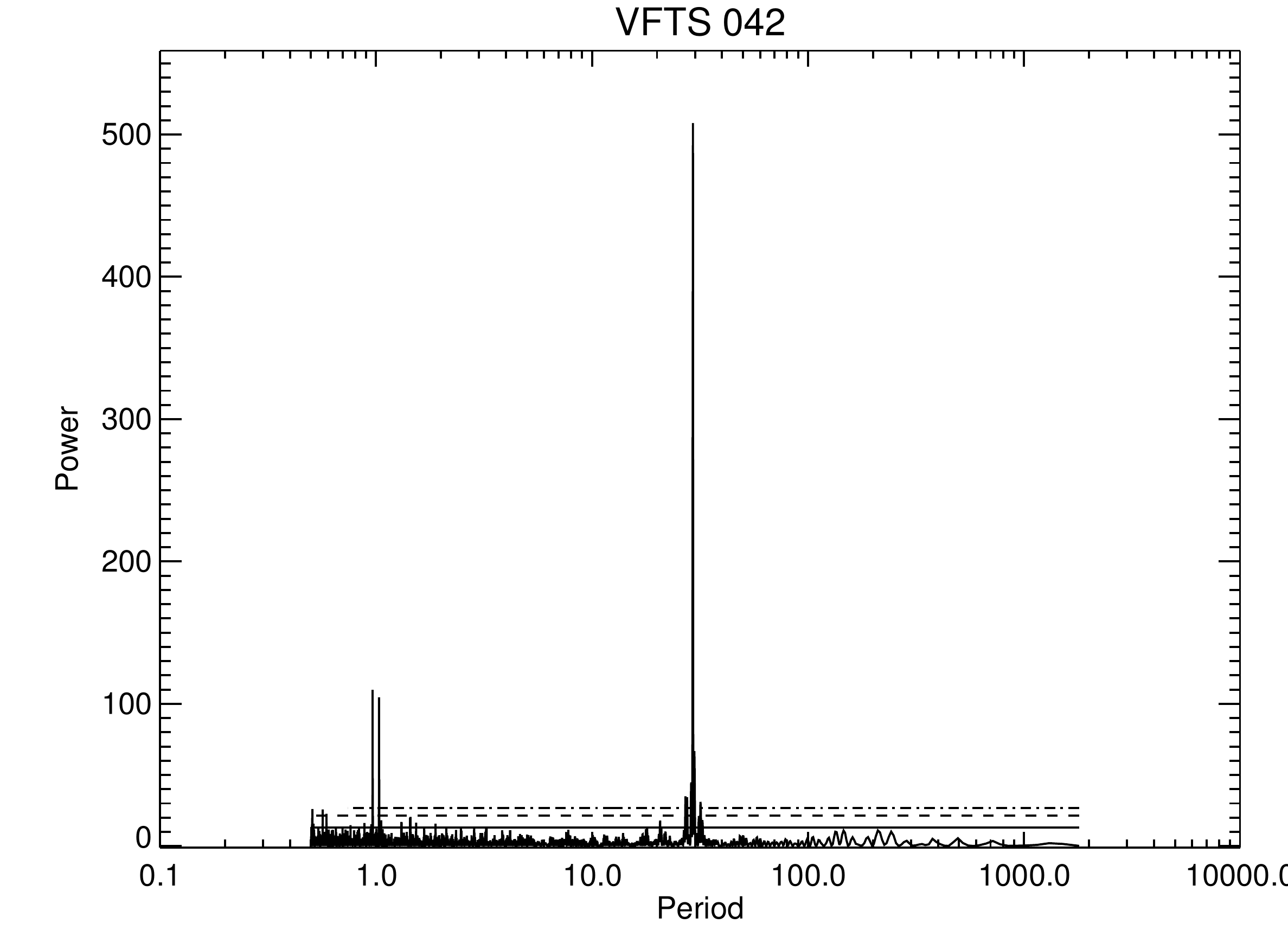}
\includegraphics[width=4.4cm,angle=-90]{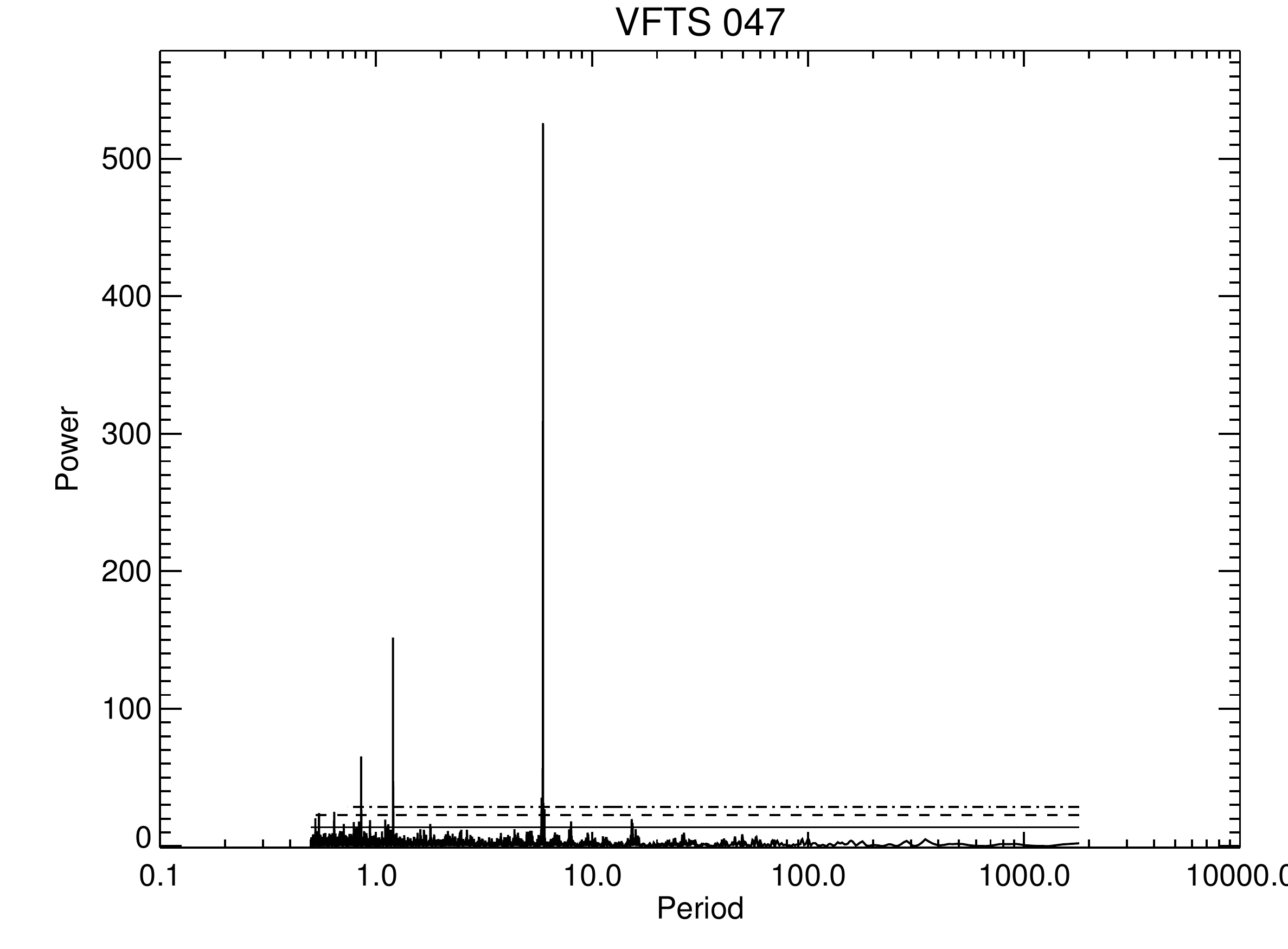}
\includegraphics[width=4.4cm,angle=-90]{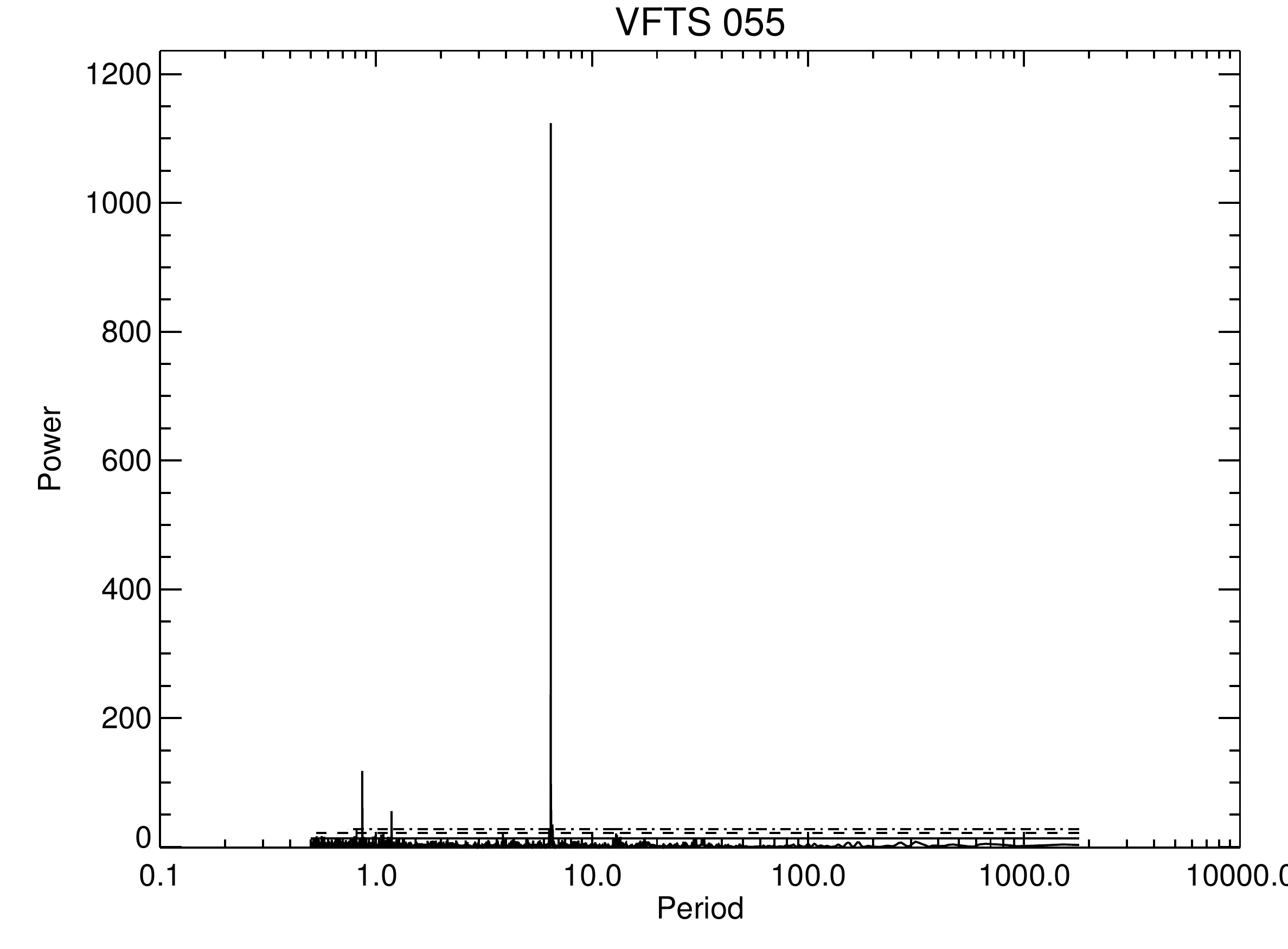}
\includegraphics[width=4.4cm,angle=-90]{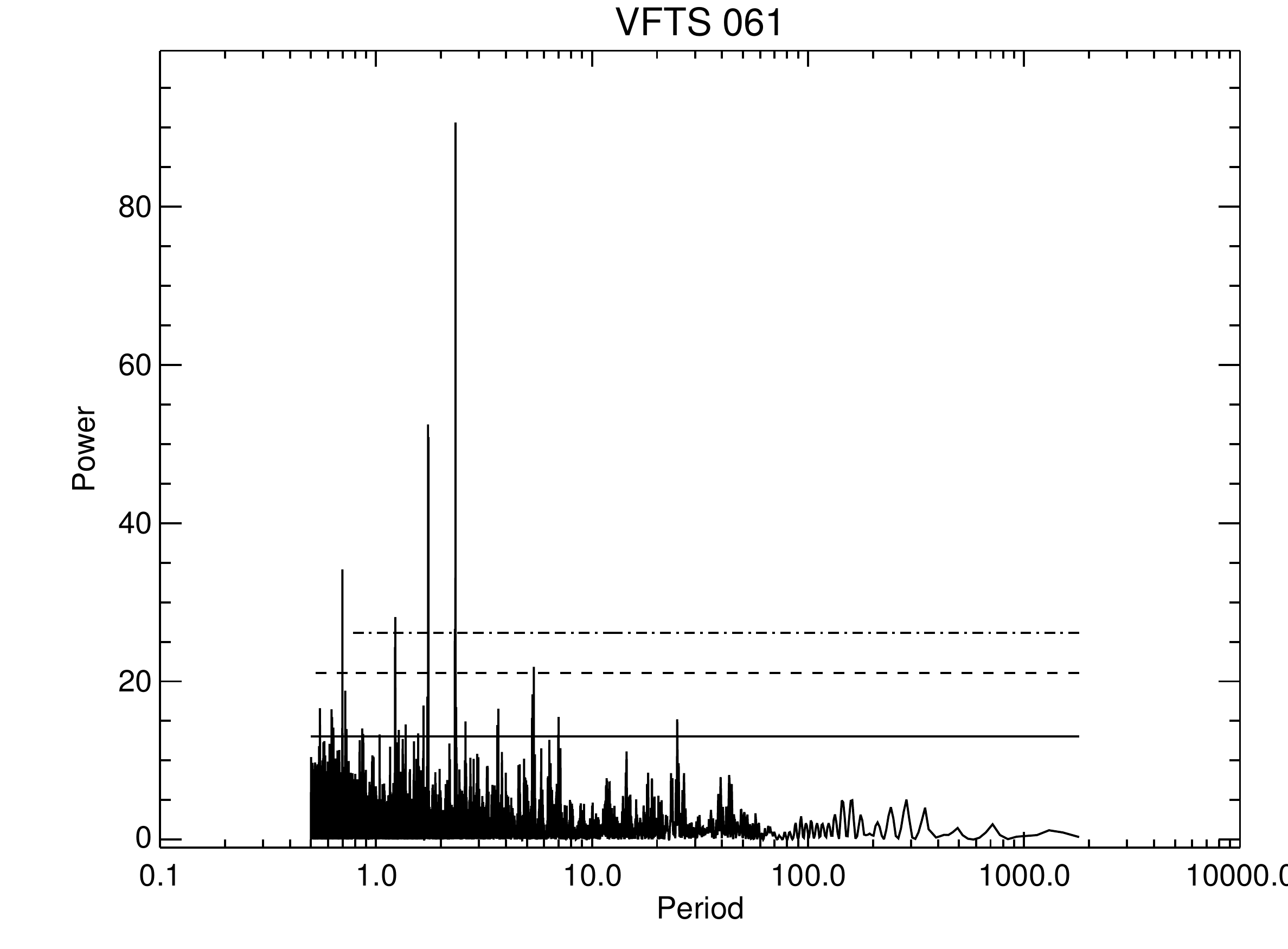}
\includegraphics[width=4.4cm,angle=-90]{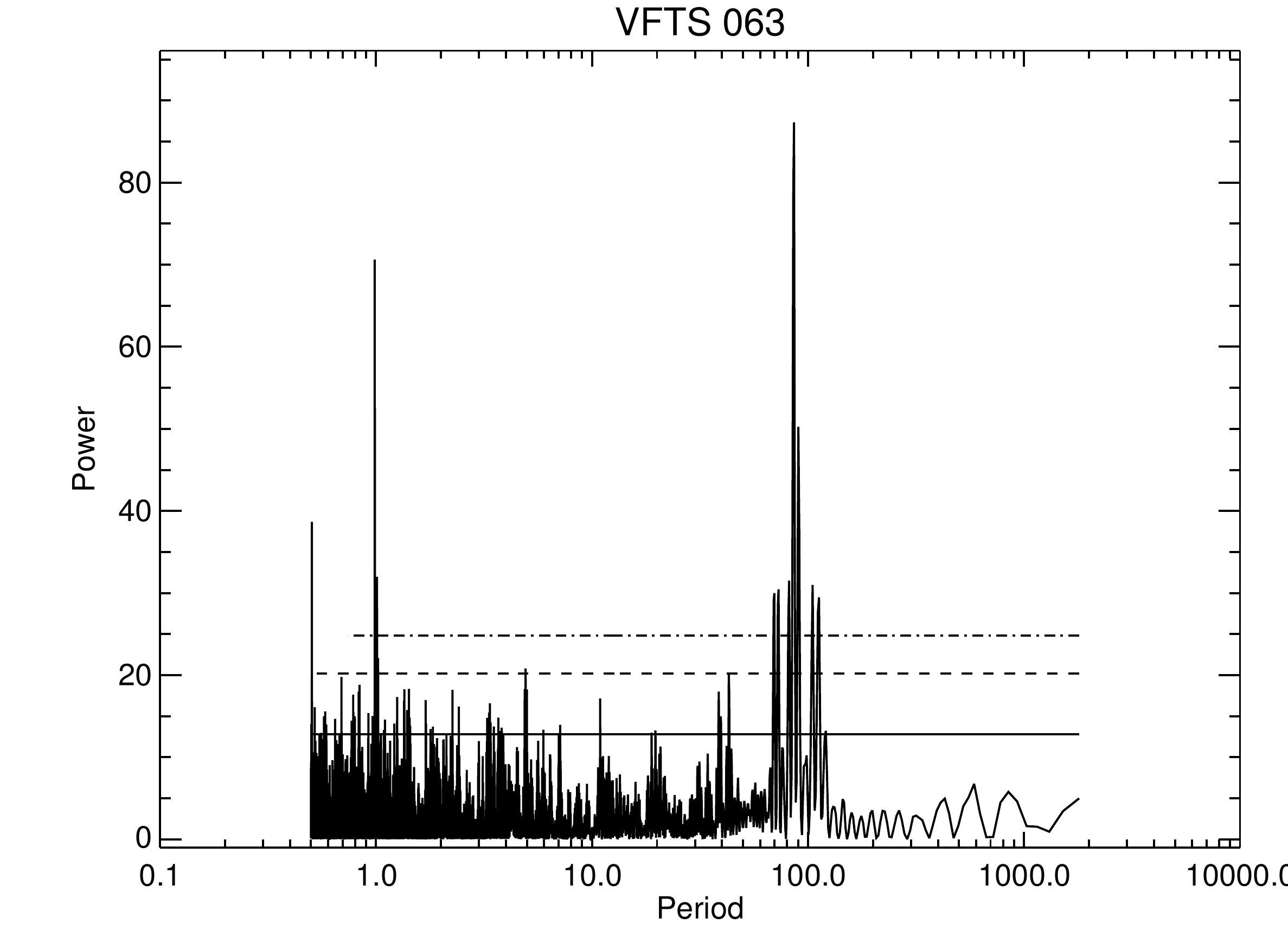}
\includegraphics[width=4.4cm,angle=-90]{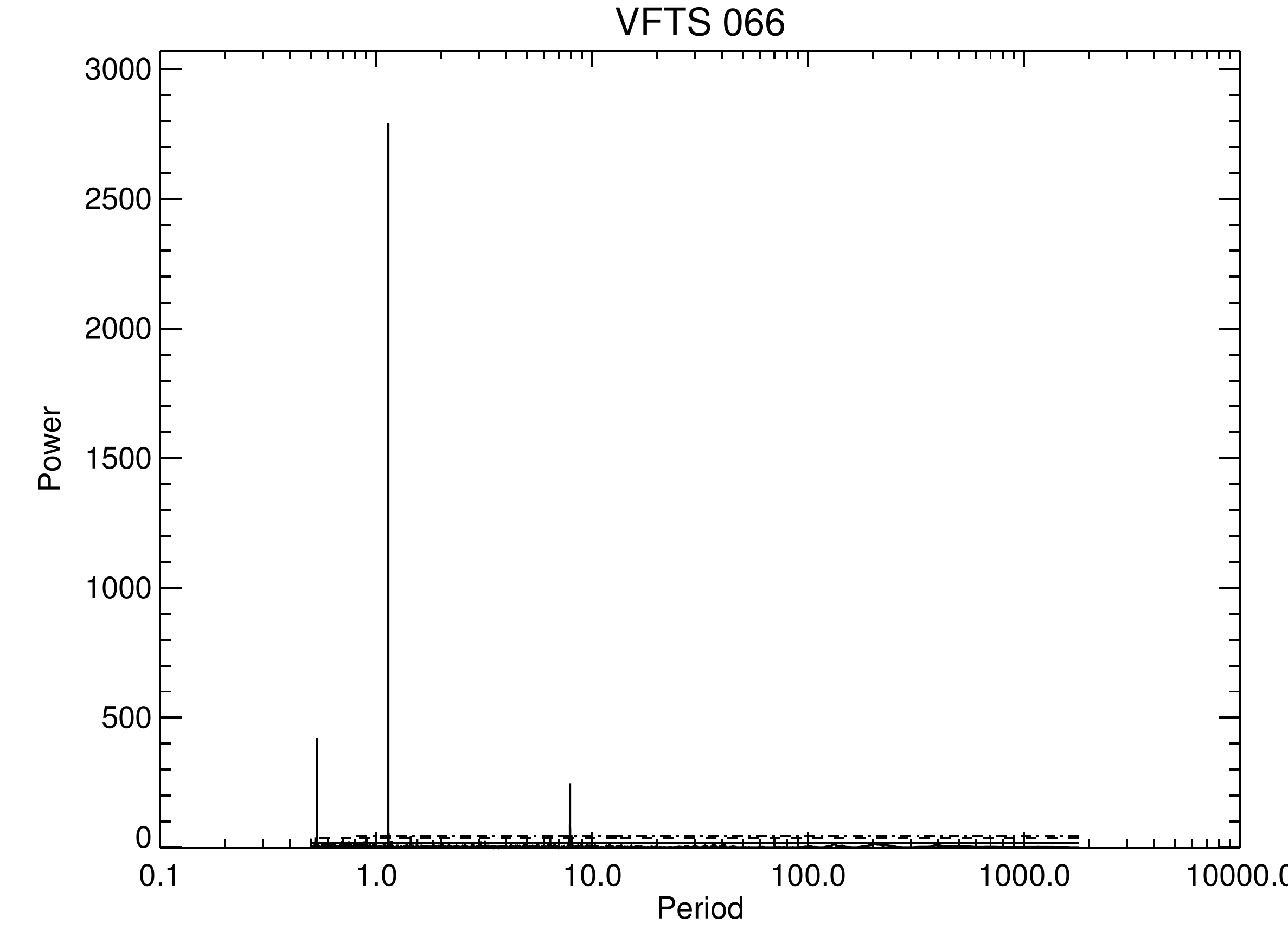}
\includegraphics[width=4.4cm,angle=-90]{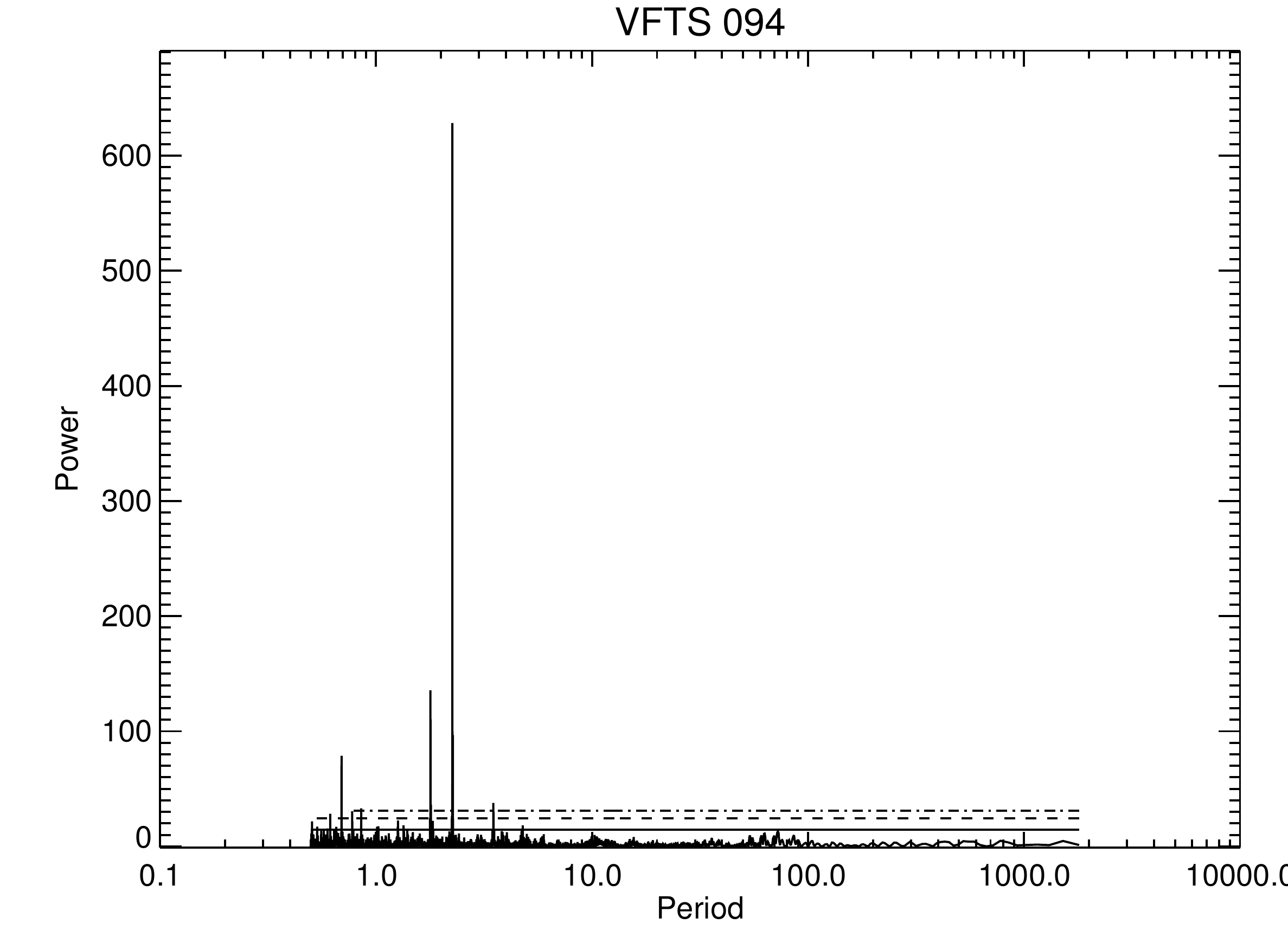}
\includegraphics[width=4.4cm,angle=-90]{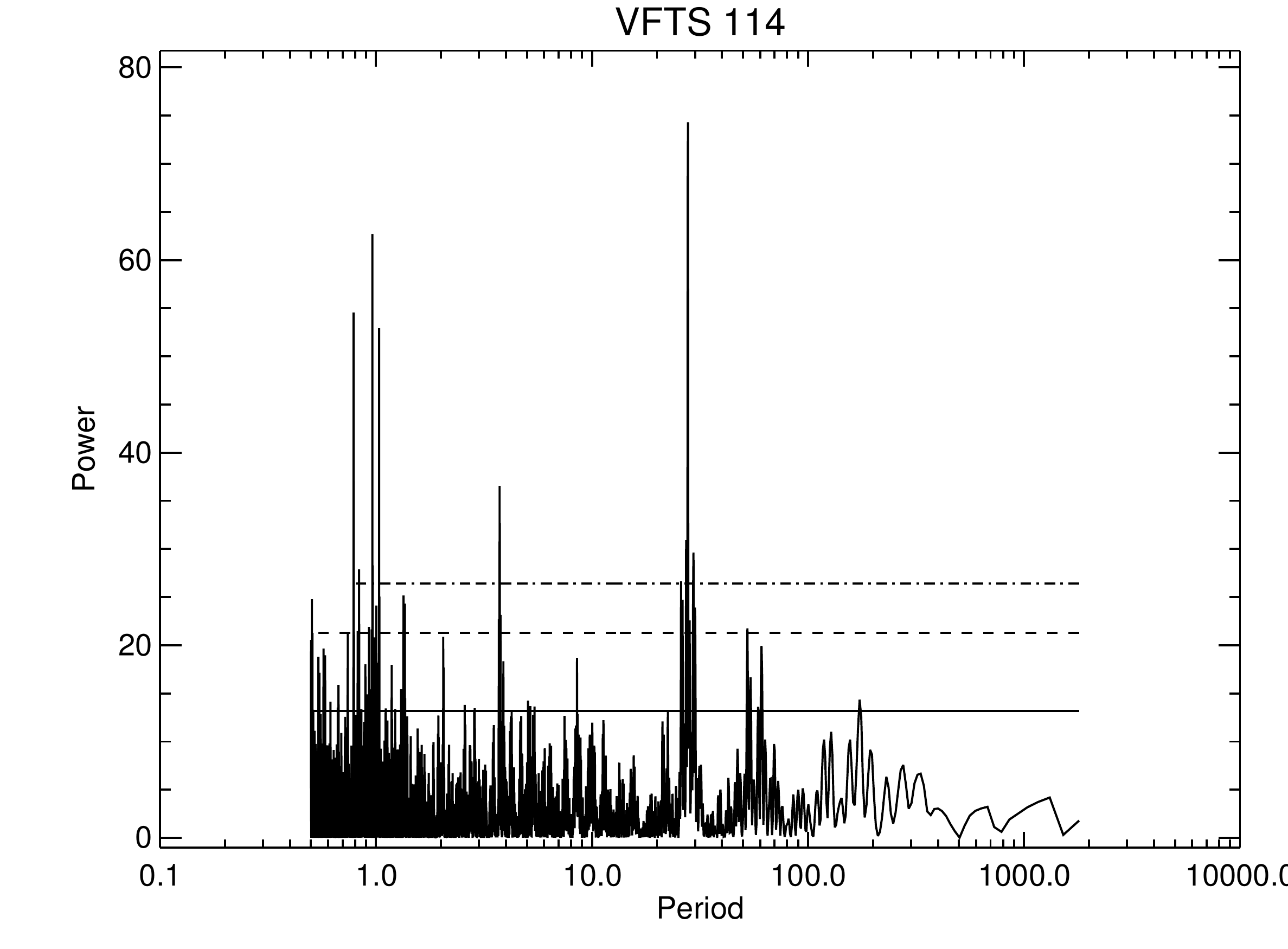}
\includegraphics[width=4.4cm,angle=-90]{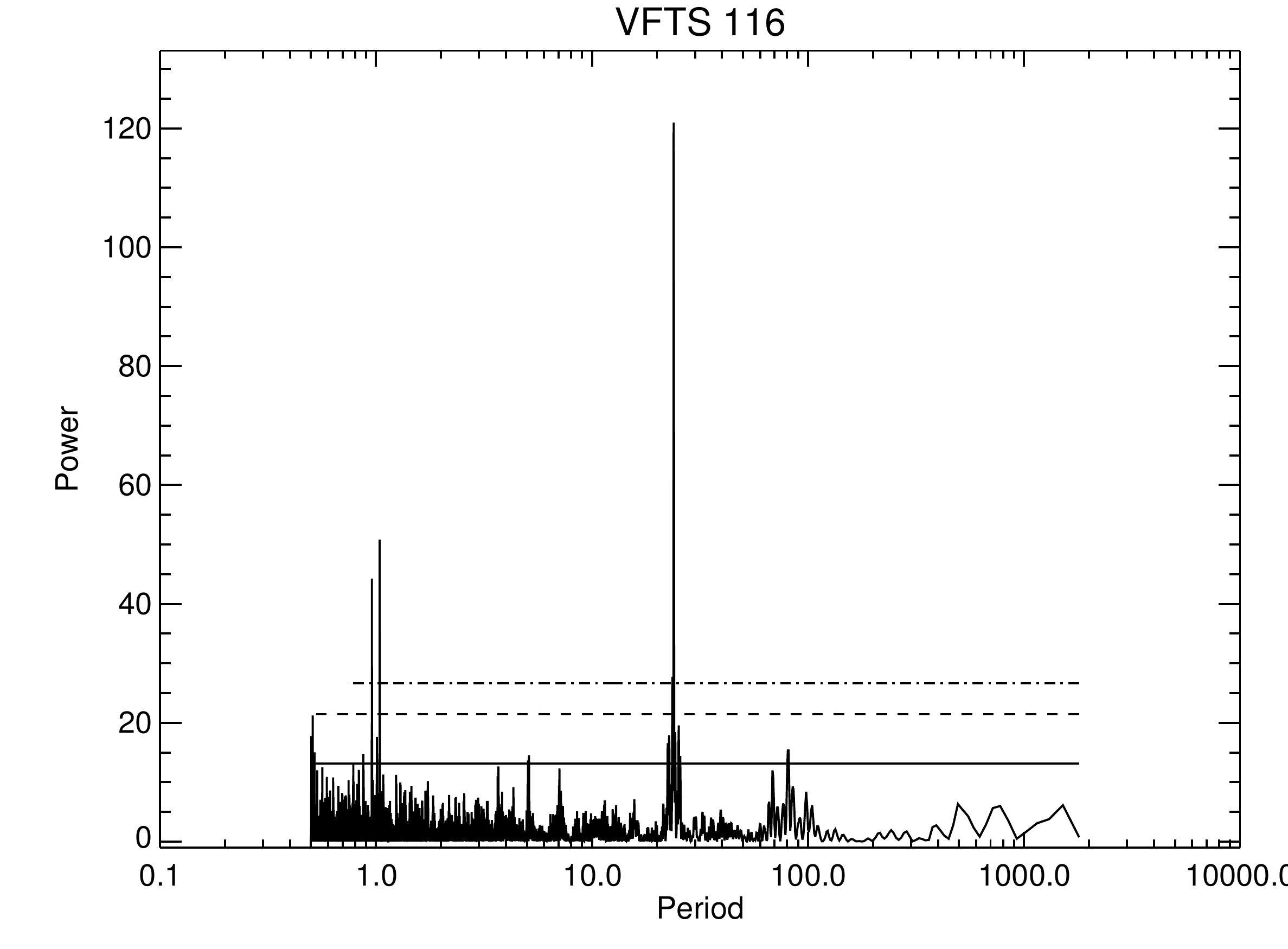}
\includegraphics[width=4.4cm,angle=-90]{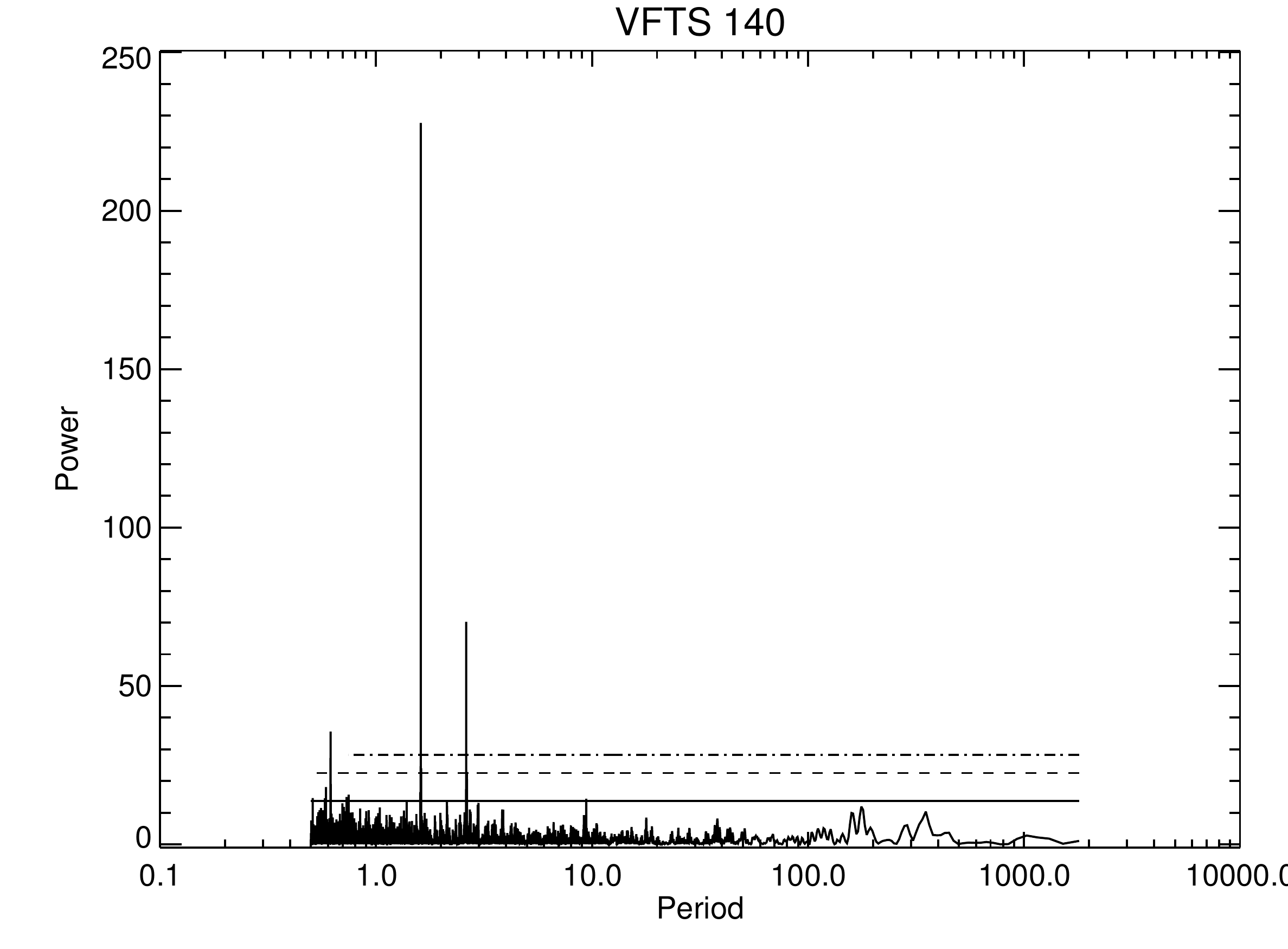}
\includegraphics[width=4.4cm,angle=-90]{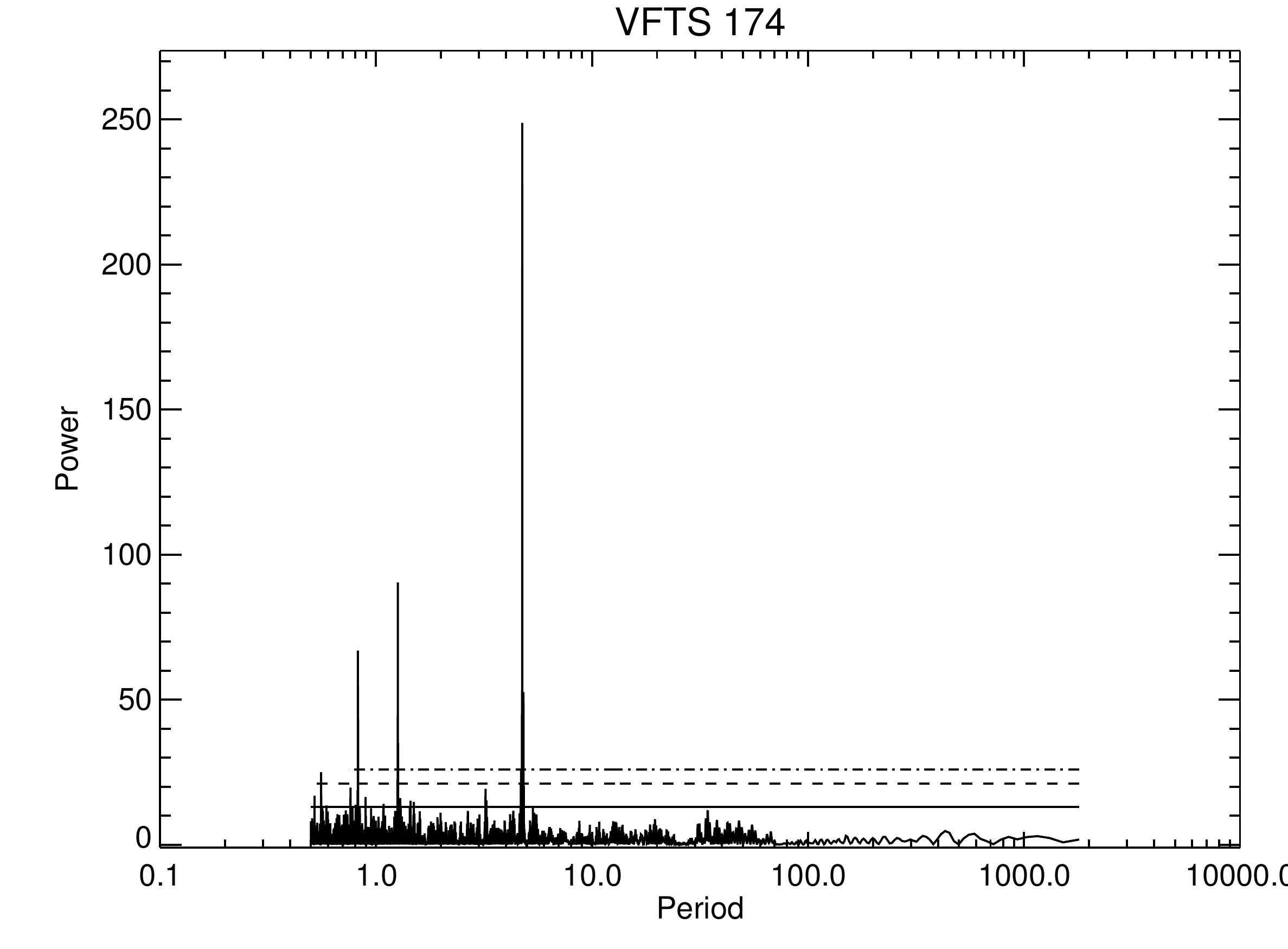}
\includegraphics[width=4.4cm,angle=-90]{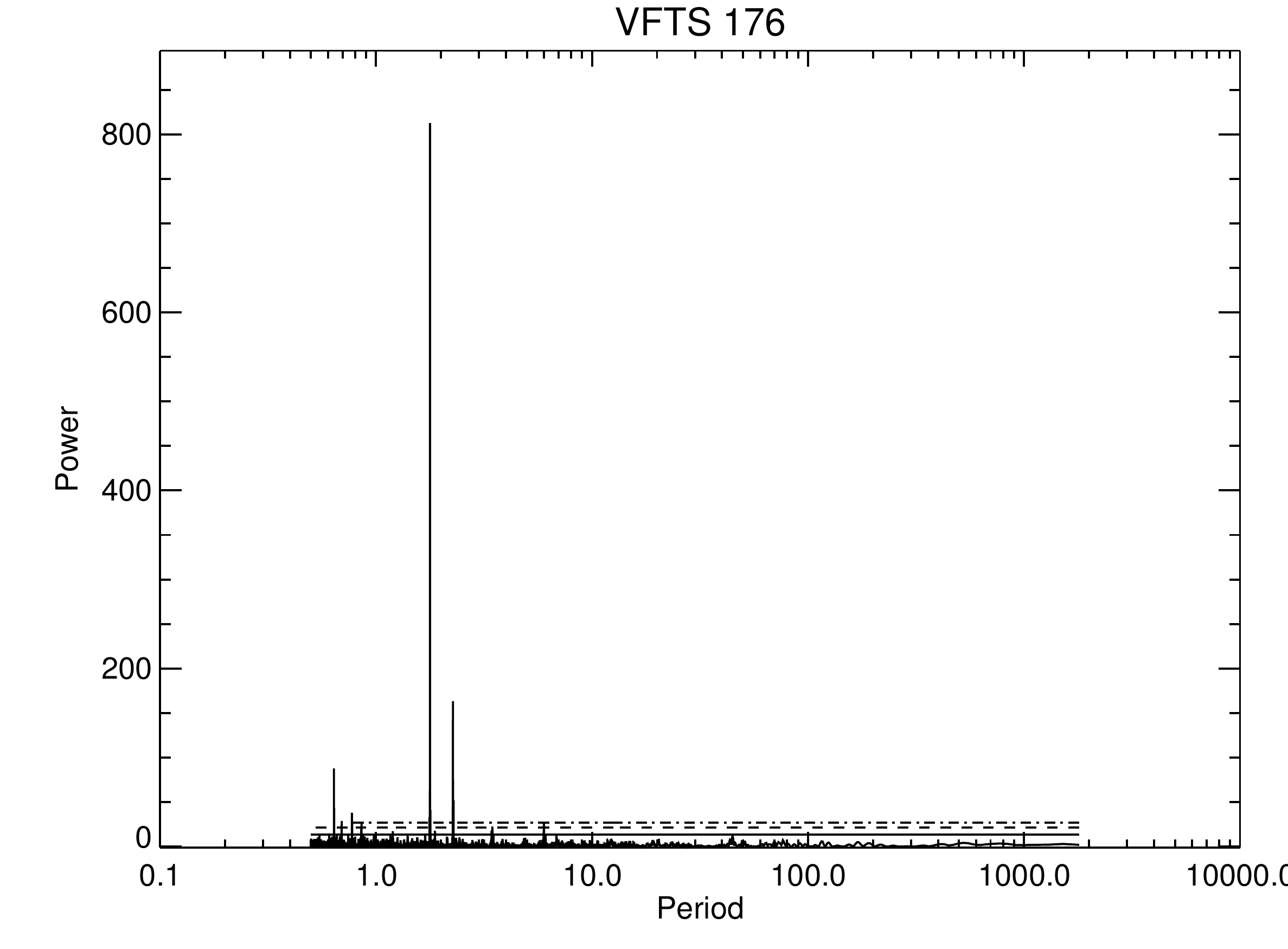}
\includegraphics[width=4.4cm,angle=-90]{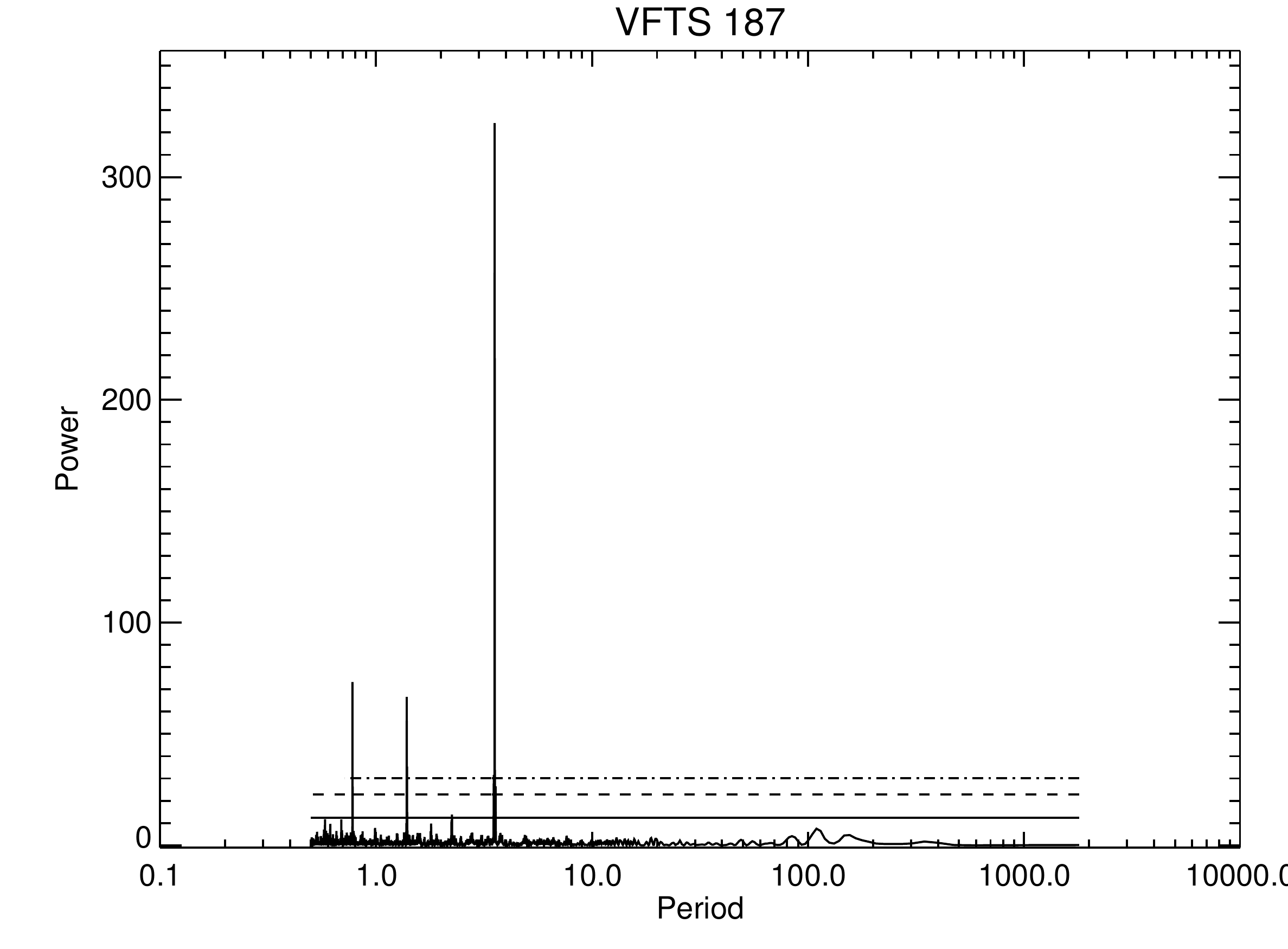}
\includegraphics[width=4.4cm,angle=-90]{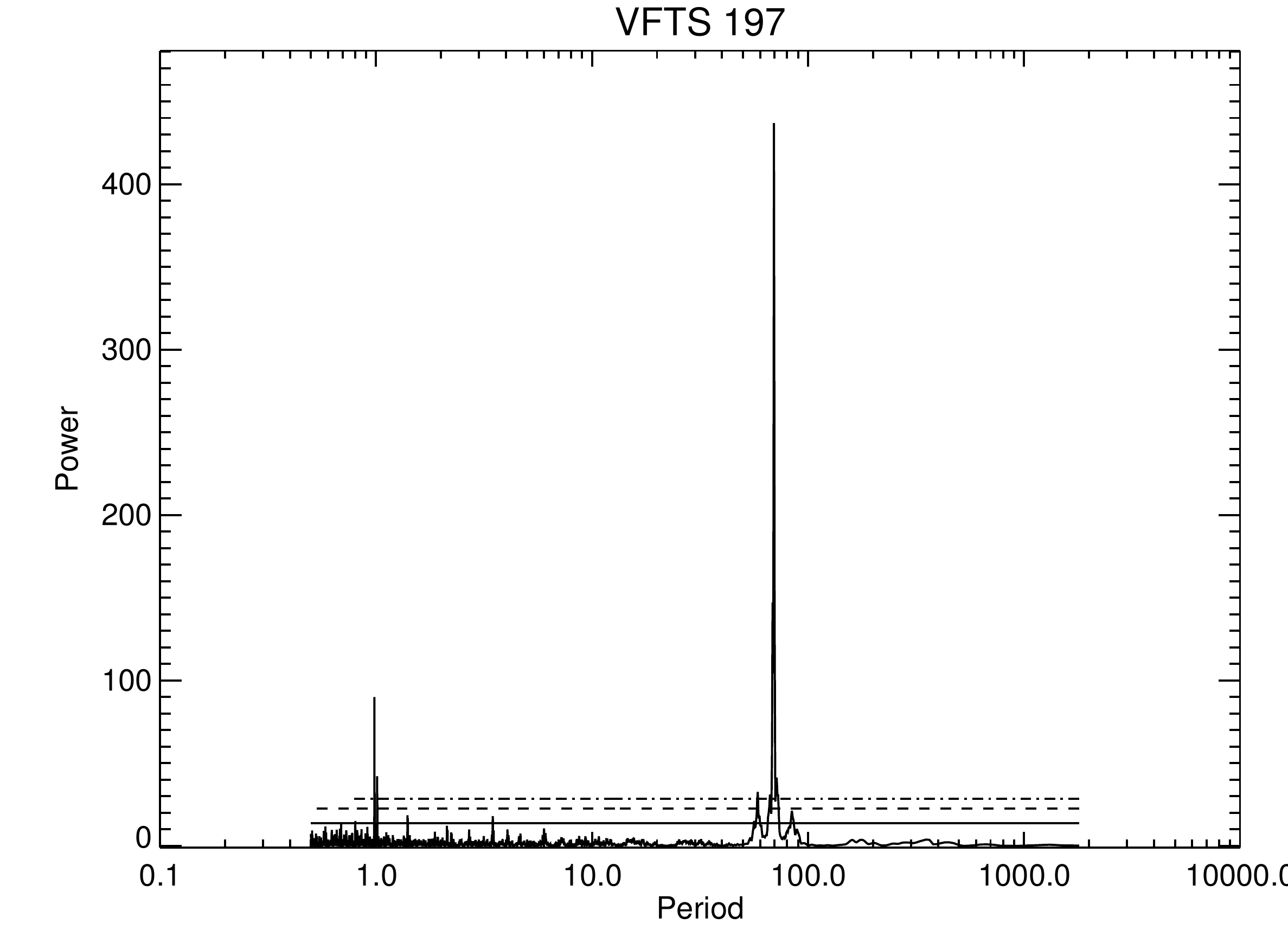}
\includegraphics[width=4.4cm,angle=-90]{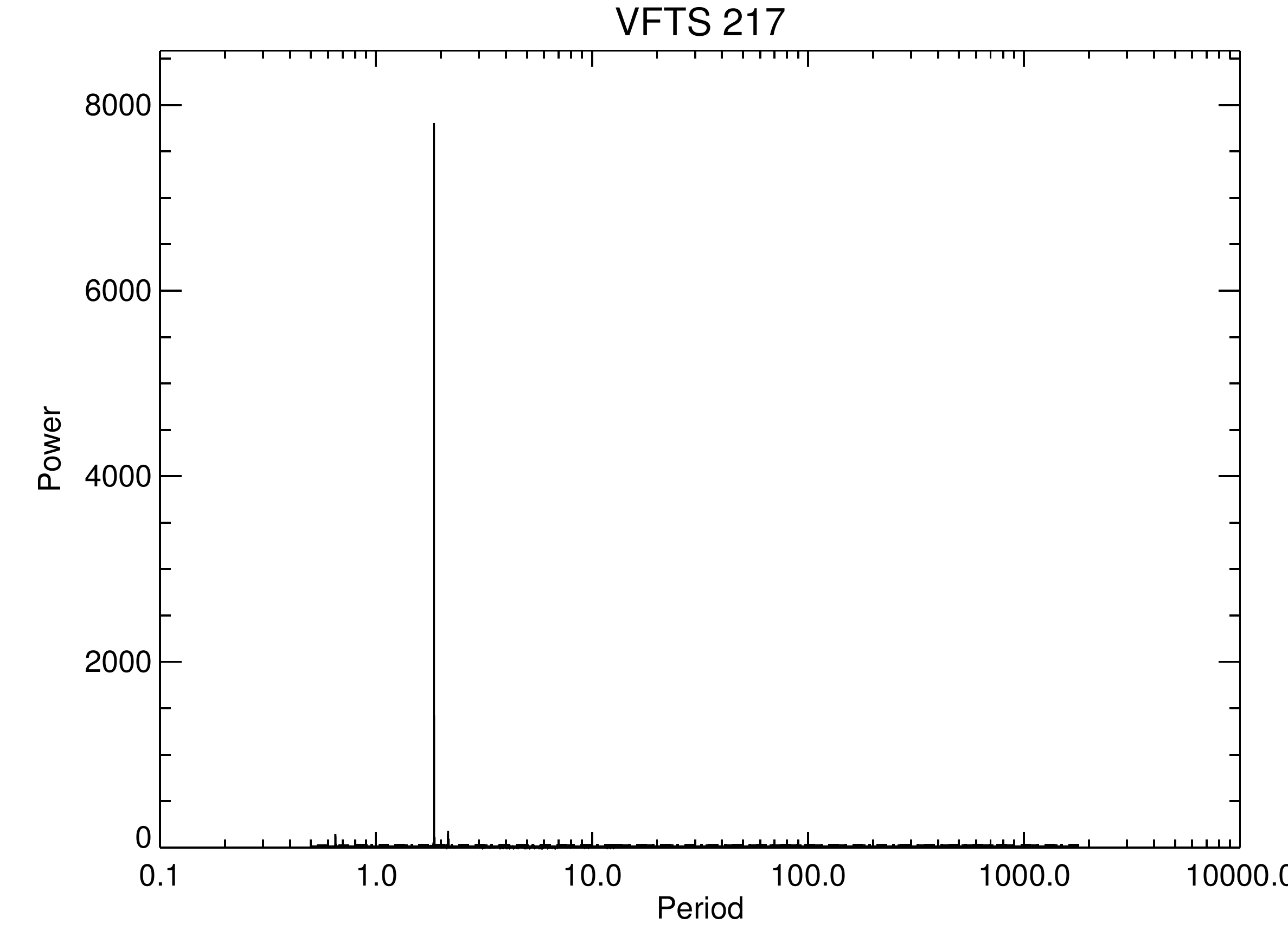}
\caption{Lomb-Scargle periodograms for the SB2 systems. The solid, dashed, and dot-dashed lines represent the 50\%, 1\%, and 0.1\% false alarm probabilities, respectively.}
\label{sb2:periodogram1}
\end{figure*}

\begin{figure*}
\centering
\ContinuedFloat
\includegraphics[width=4.4cm,angle=-90]{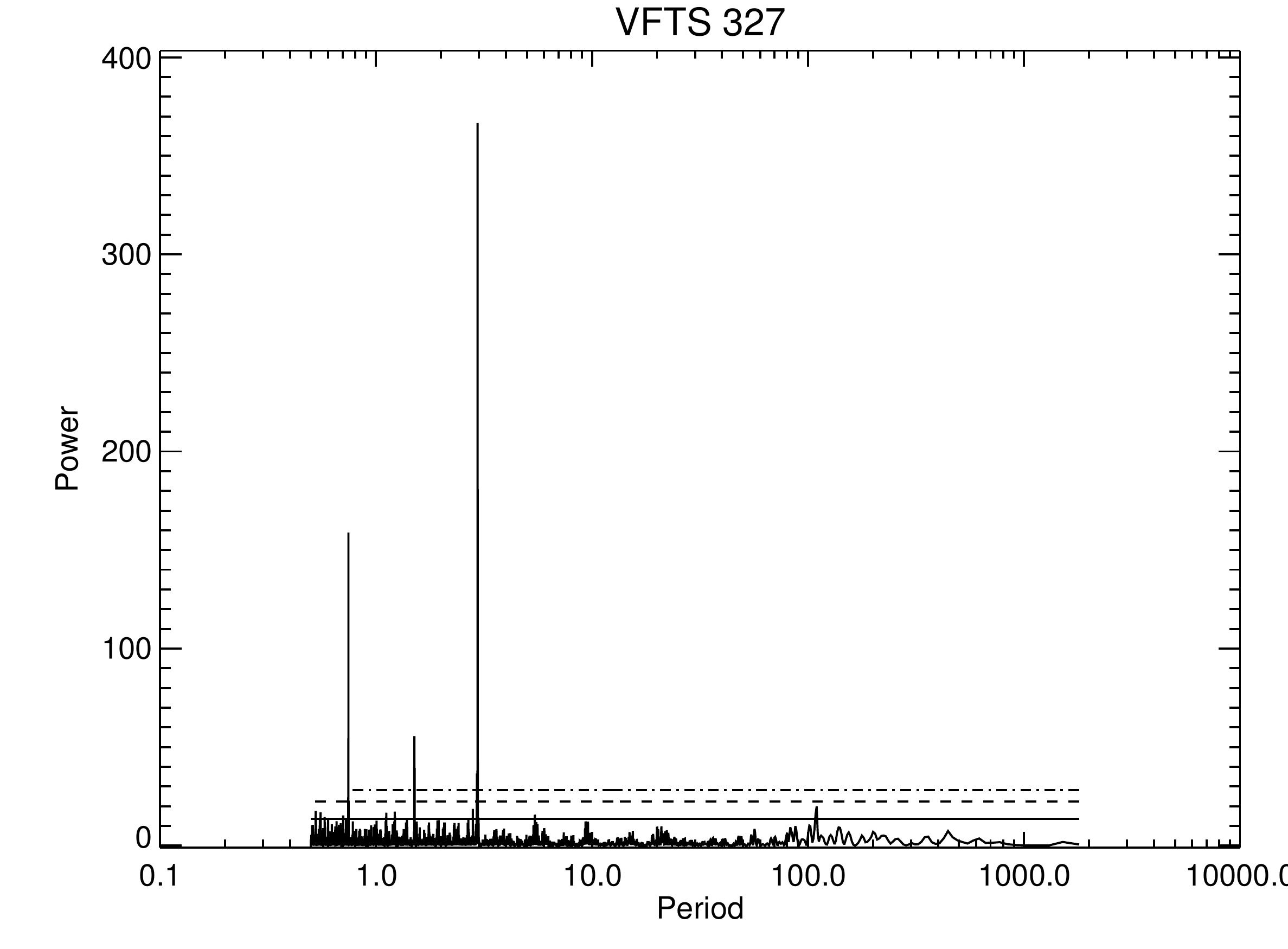}
\includegraphics[width=4.4cm,angle=-90]{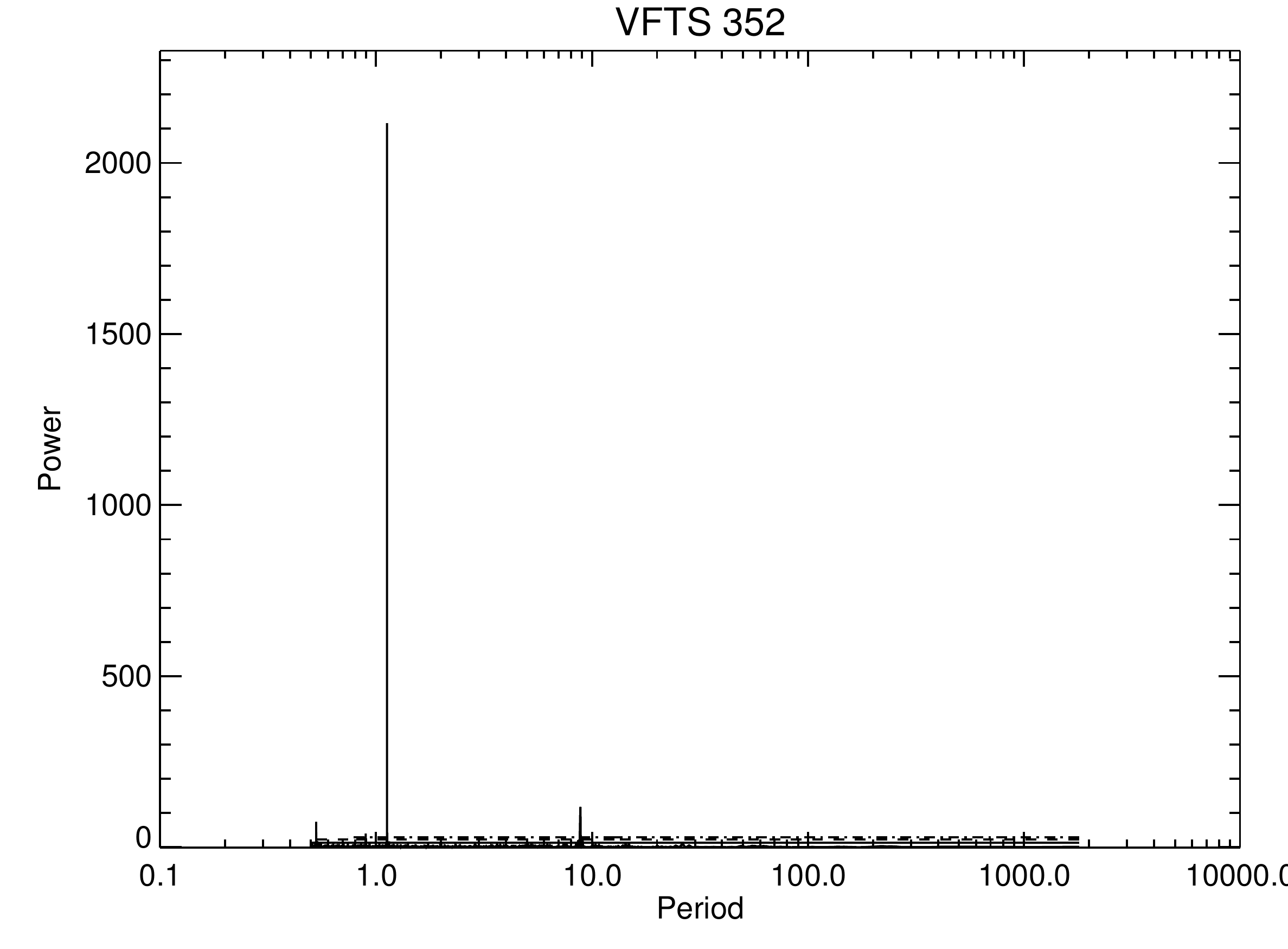}
\includegraphics[width=4.4cm,angle=-90]{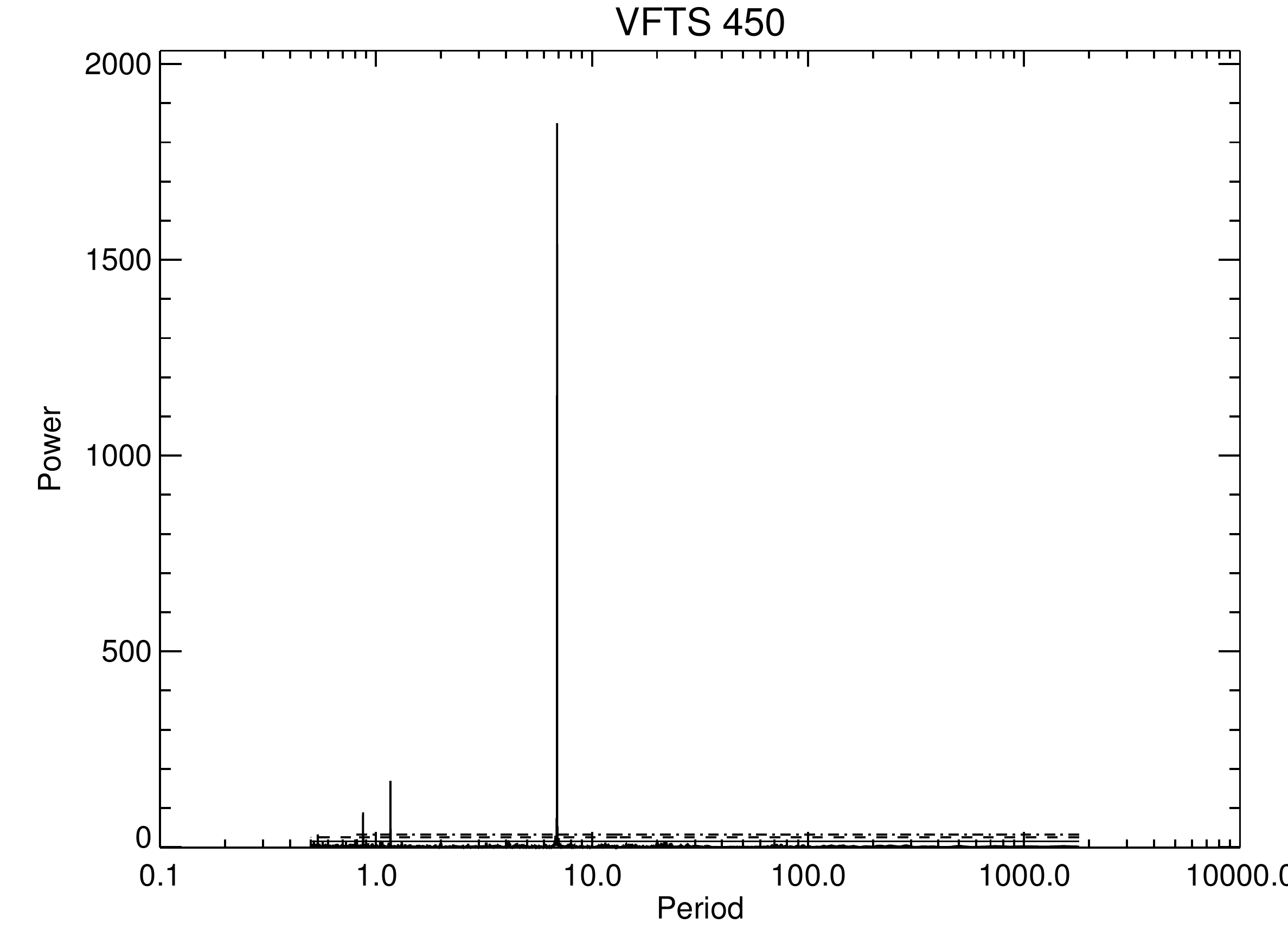}
\includegraphics[width=4.4cm,angle=-90]{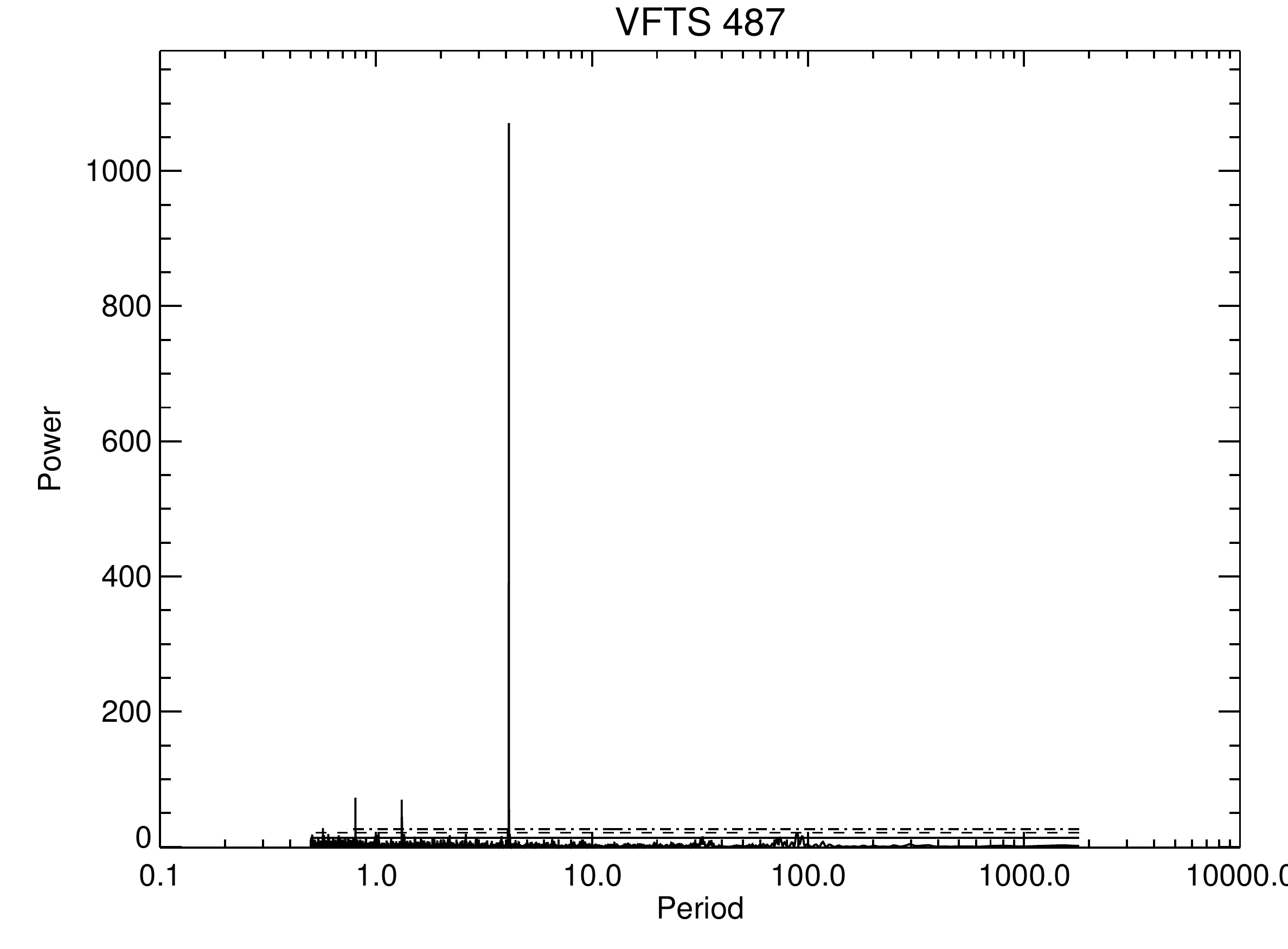}
\includegraphics[width=4.4cm,angle=-90]{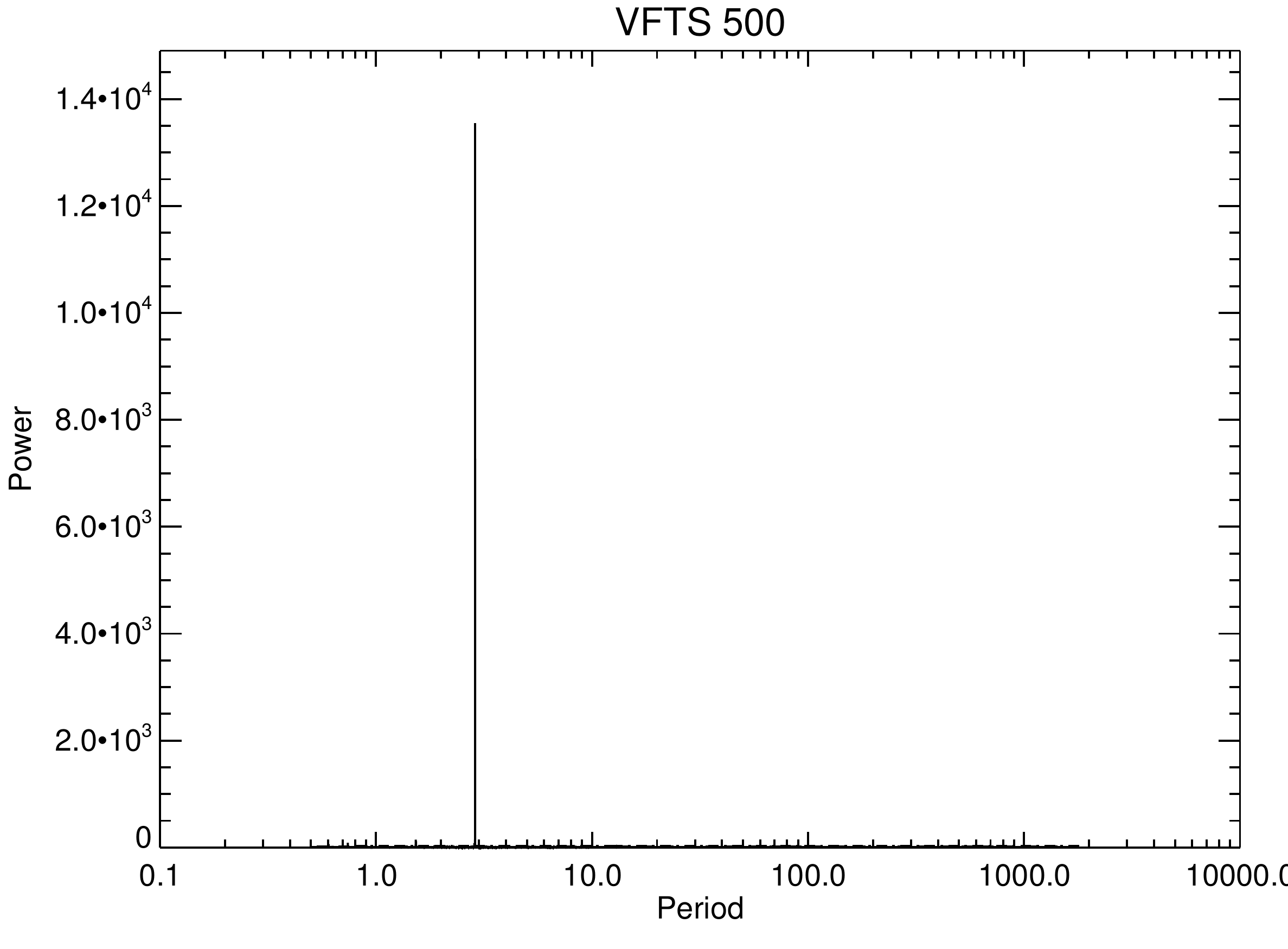}
\includegraphics[width=4.4cm,angle=-90]{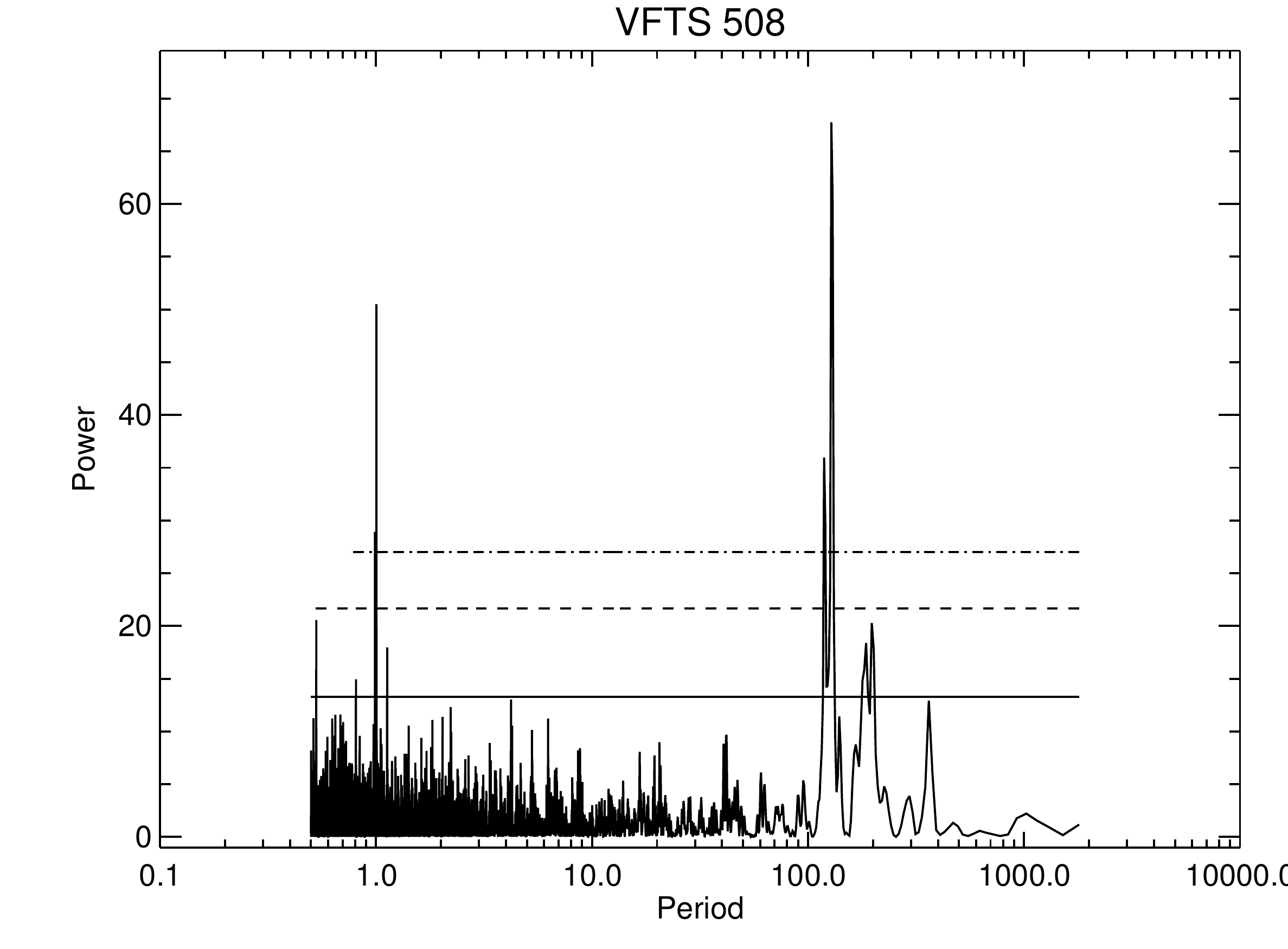}
\includegraphics[width=4.4cm,angle=-90]{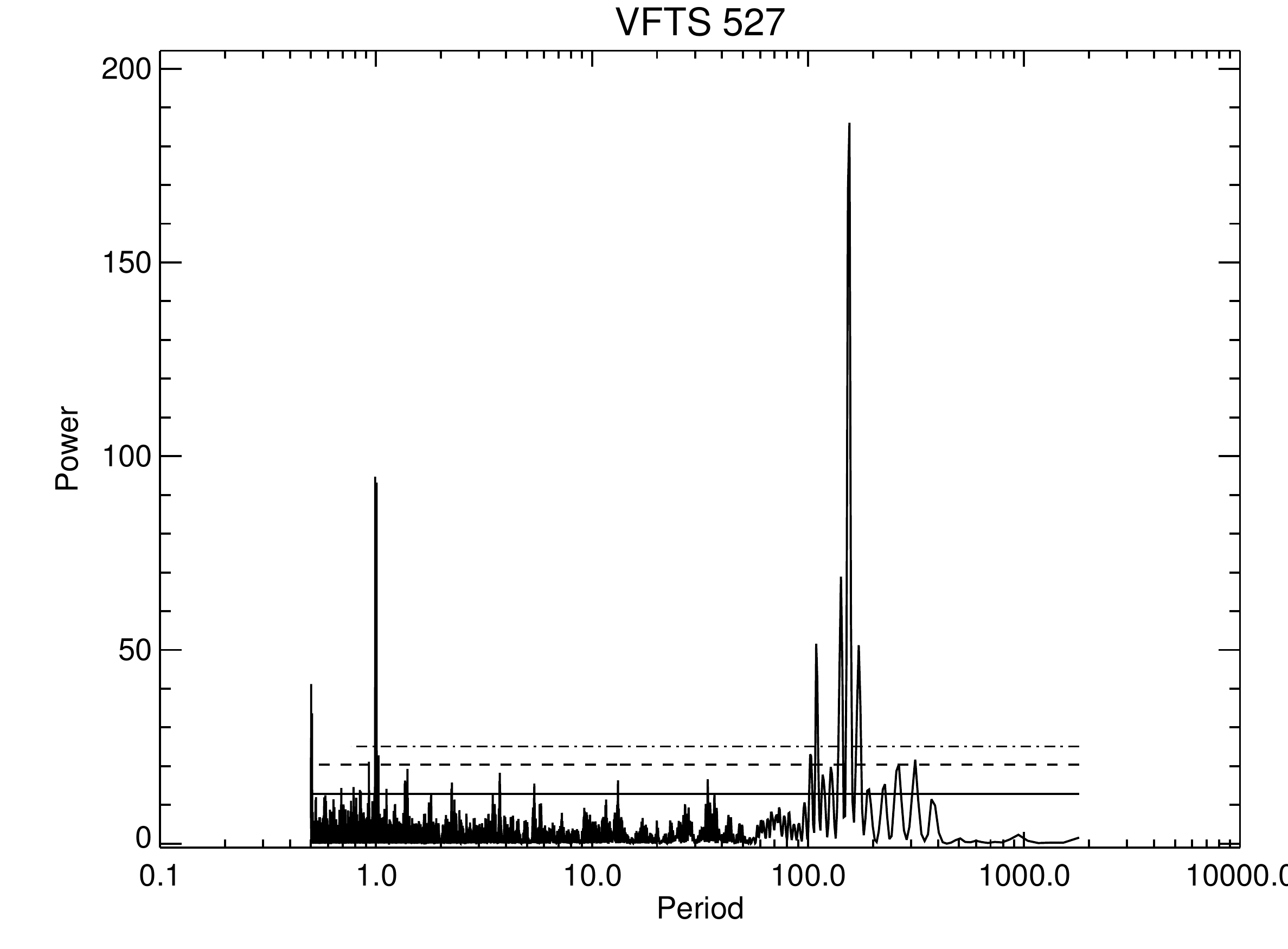}
\includegraphics[width=4.4cm,angle=-90]{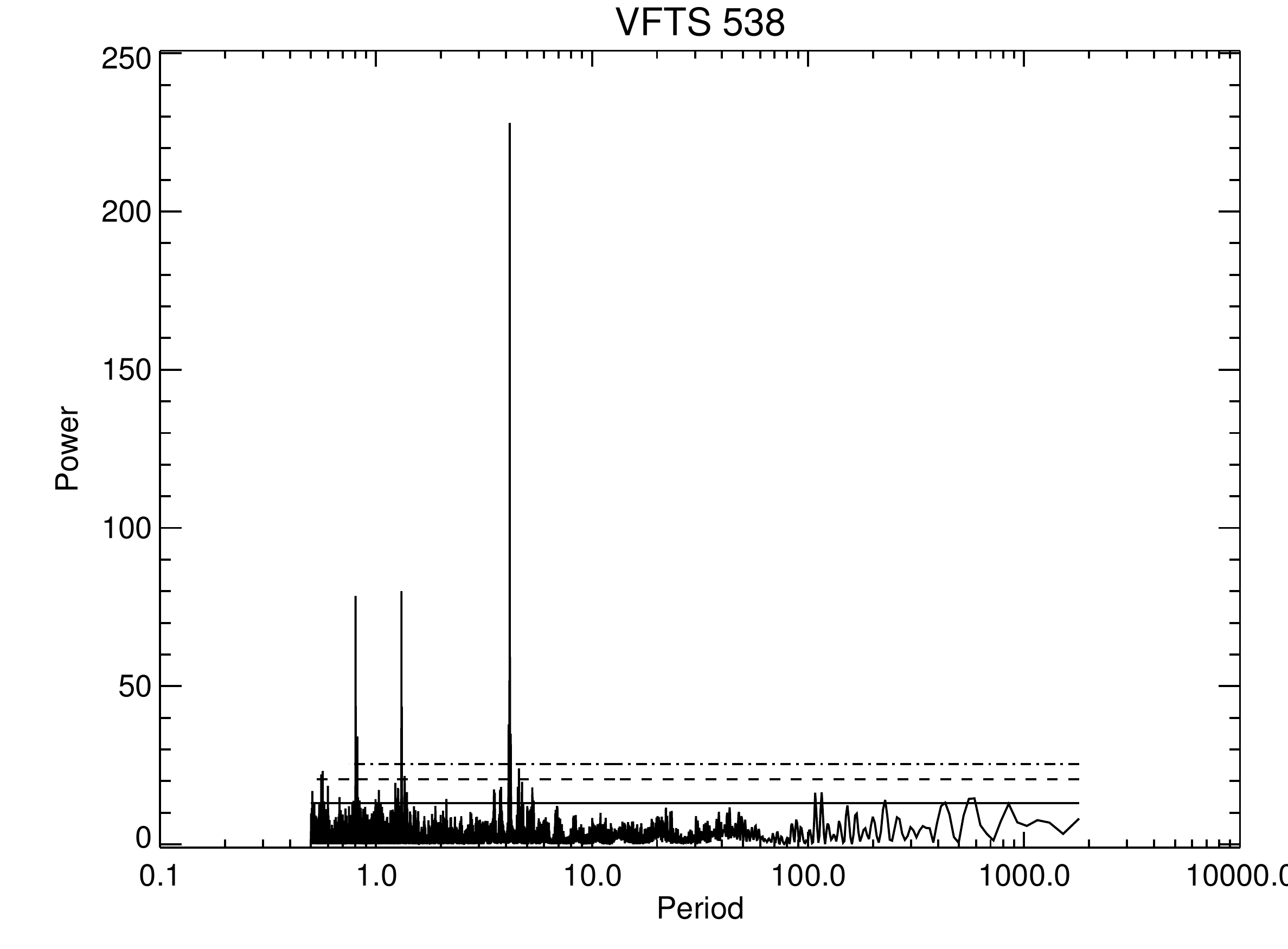}
\includegraphics[width=4.4cm,angle=-90]{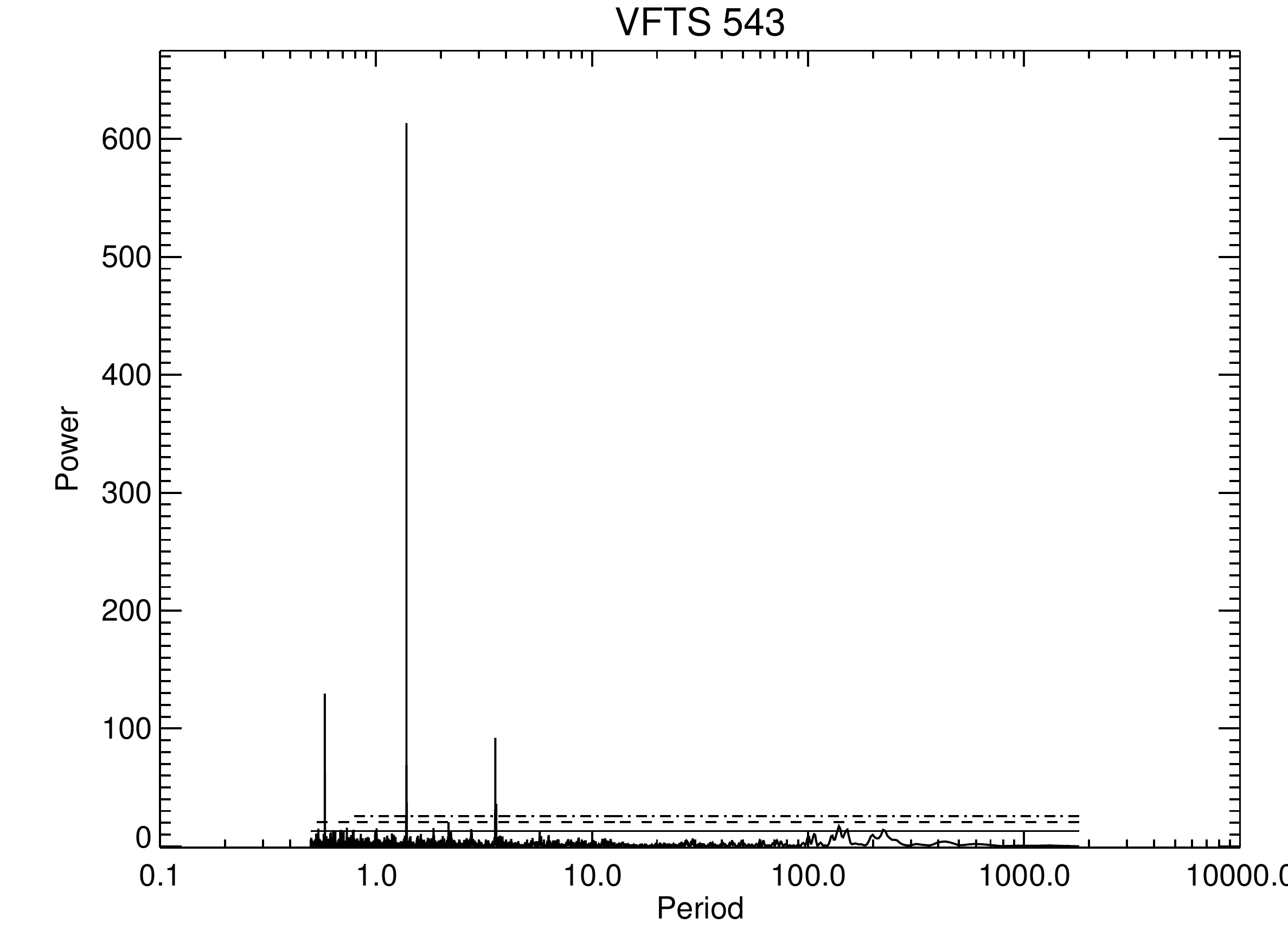}
\includegraphics[width=4.4cm,angle=-90]{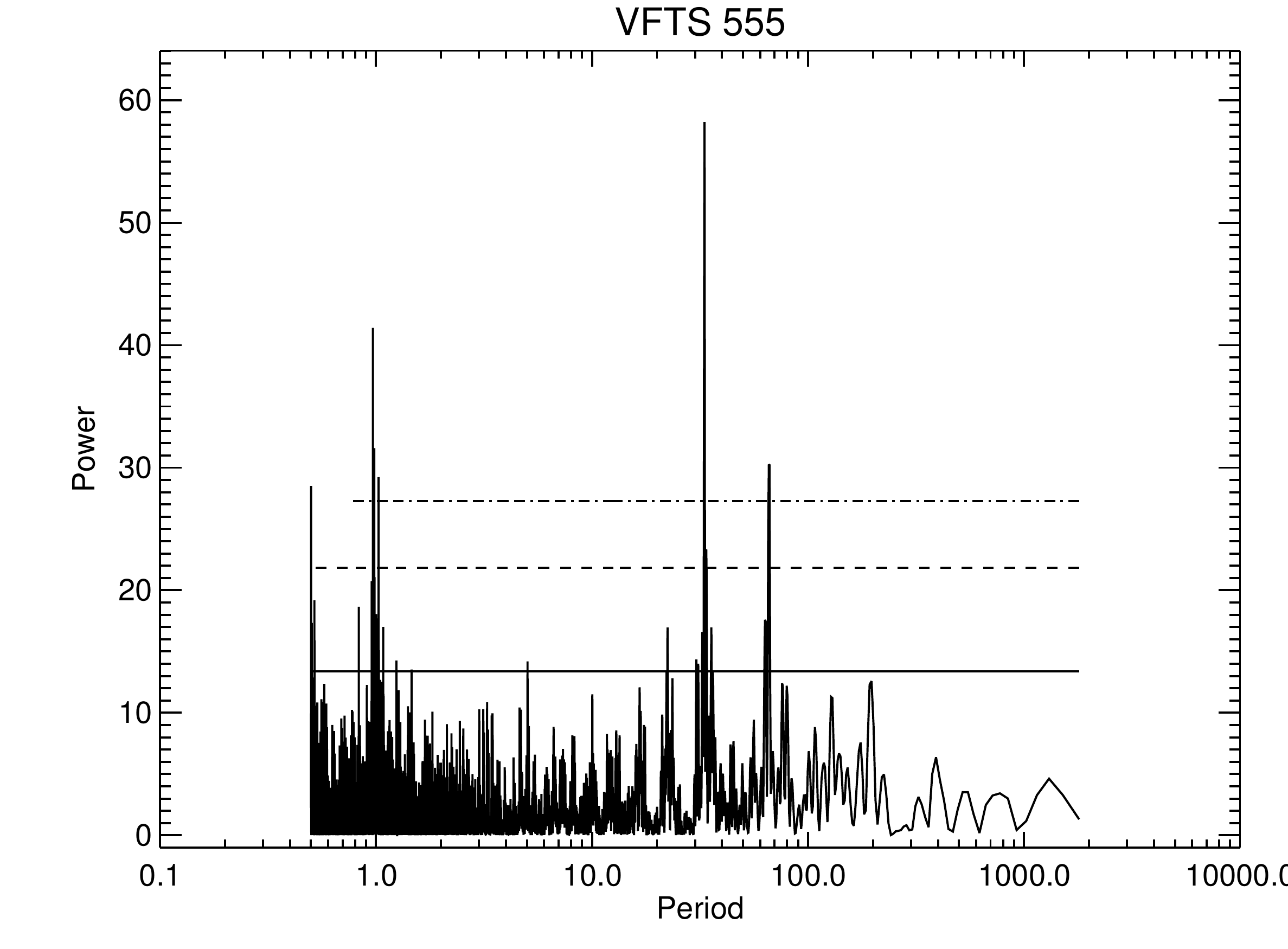}
\includegraphics[width=4.4cm,angle=-90]{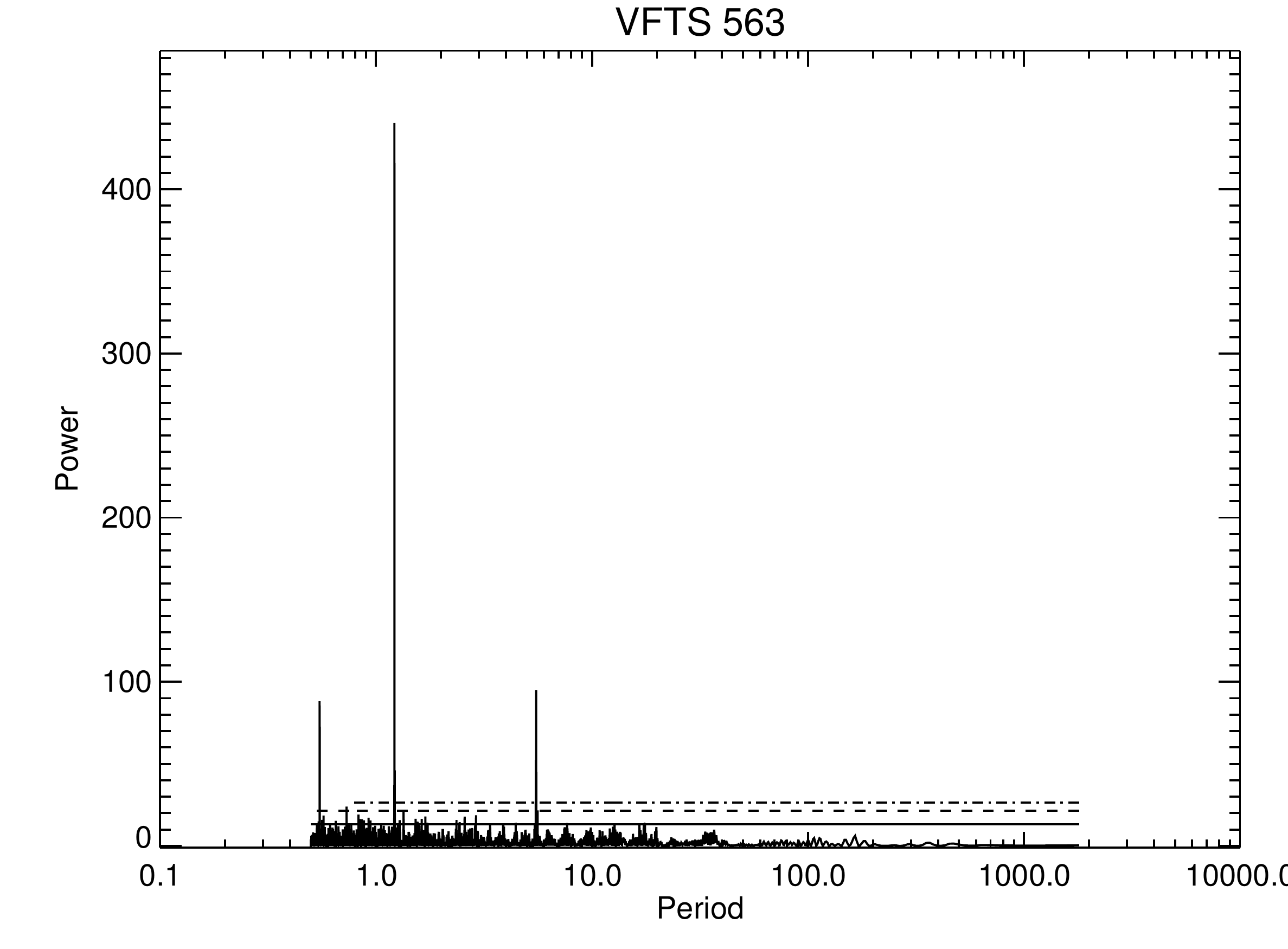}
\includegraphics[width=4.4cm,angle=-90]{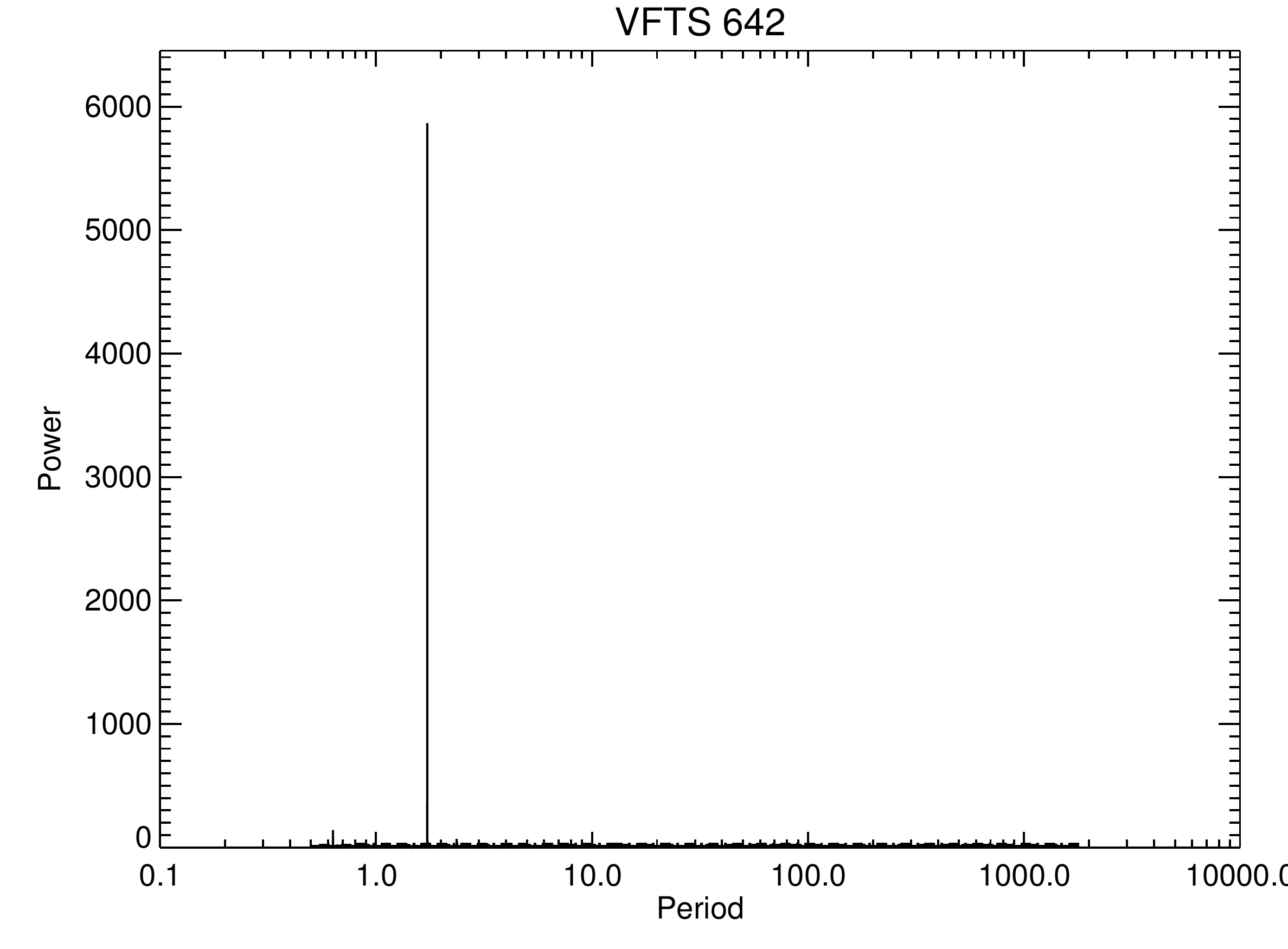}
\includegraphics[width=4.4cm,angle=-90]{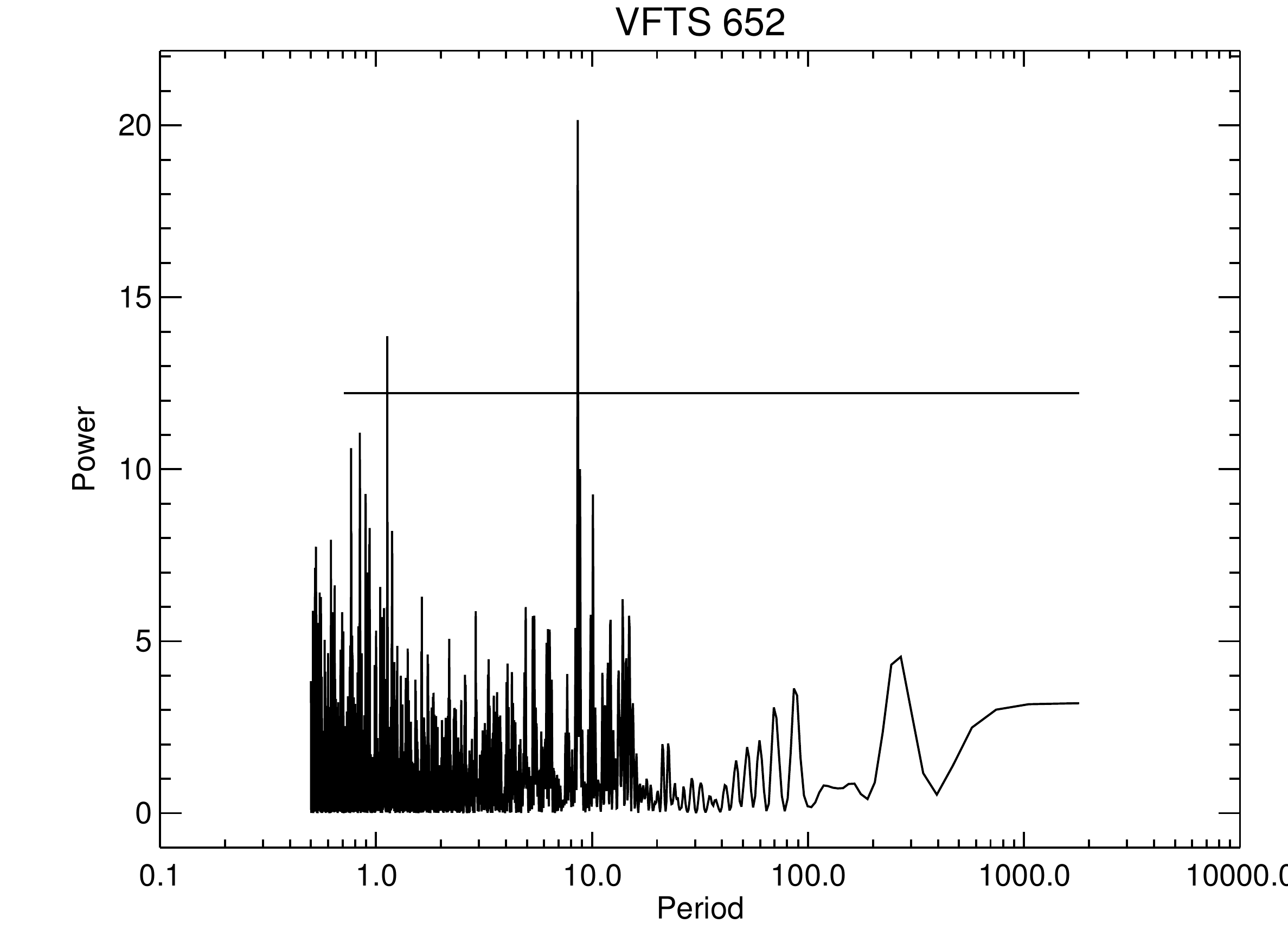}
\includegraphics[width=4.4cm,angle=-90]{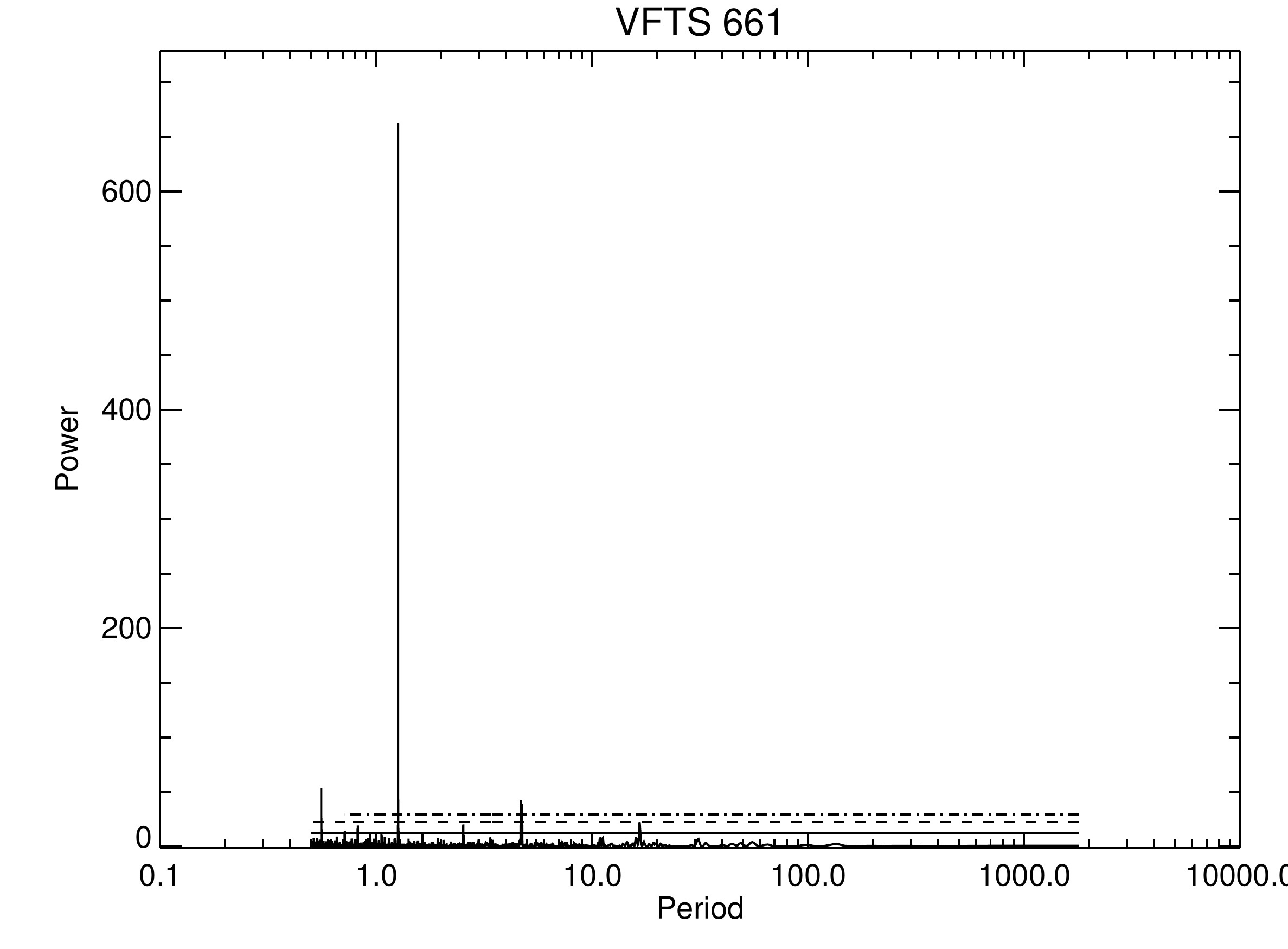}
\includegraphics[width=4.4cm,angle=-90]{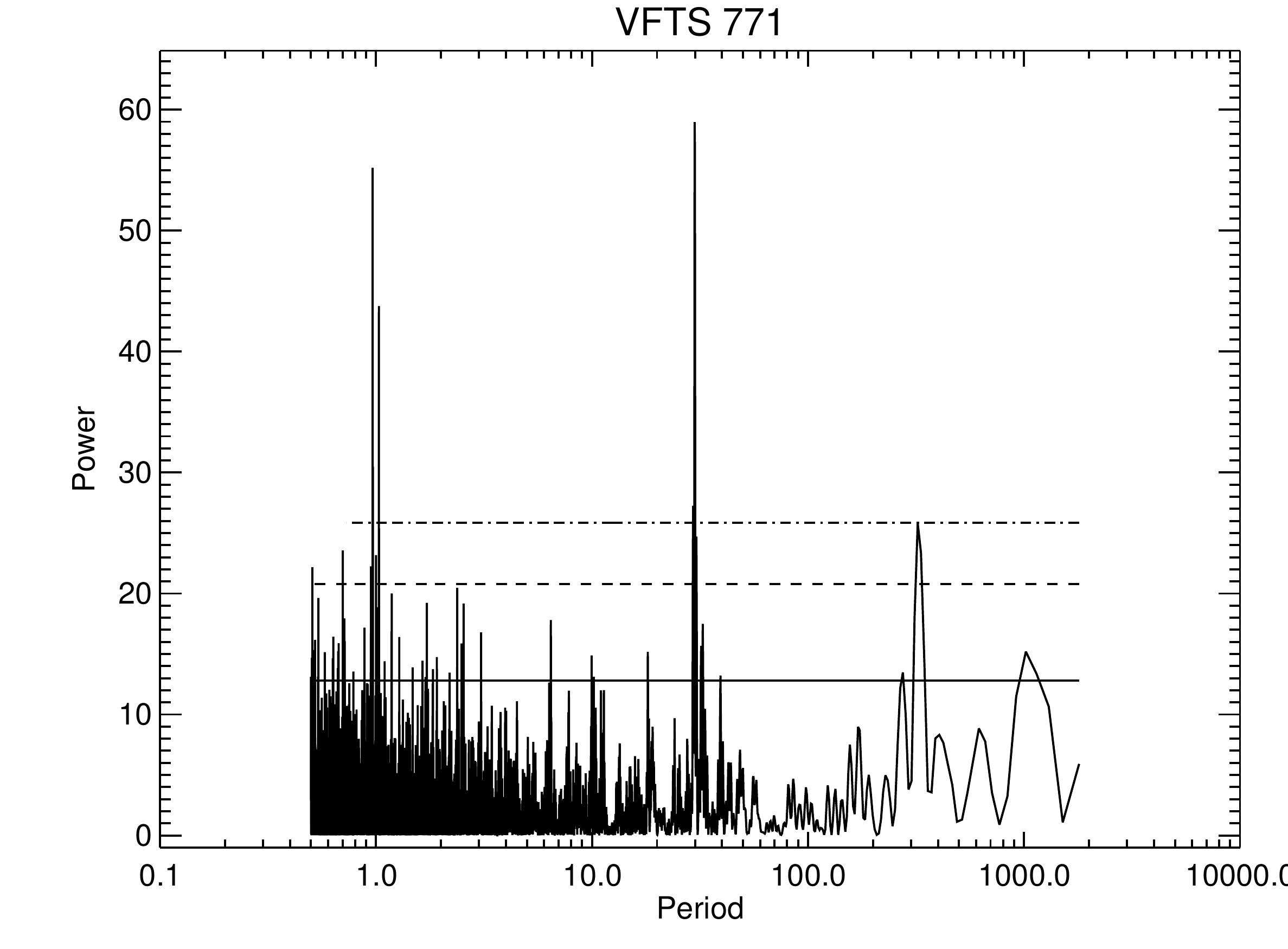}
\caption{{\it Continued...}}
\label{sb2:periodogram2}
\end{figure*}

\begin{figure*}
\centering
\ContinuedFloat
\includegraphics[width=4.4cm,angle=-90]{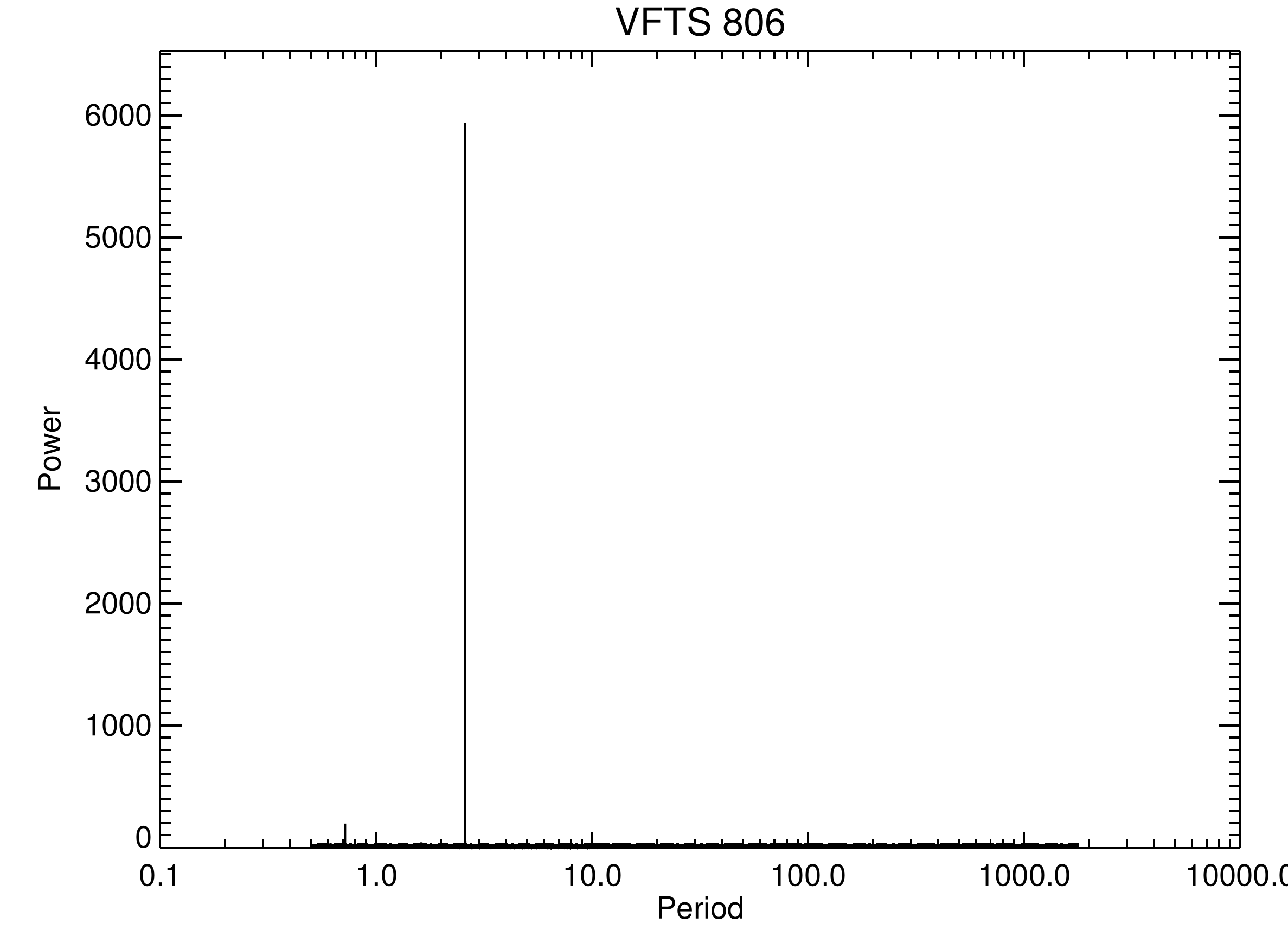}
\caption{{\it Continued...}}
\label{sb2:periodogram3}
\end{figure*}

\begin{figure*}
\centering
\includegraphics[width=4.7cm,angle=-90]{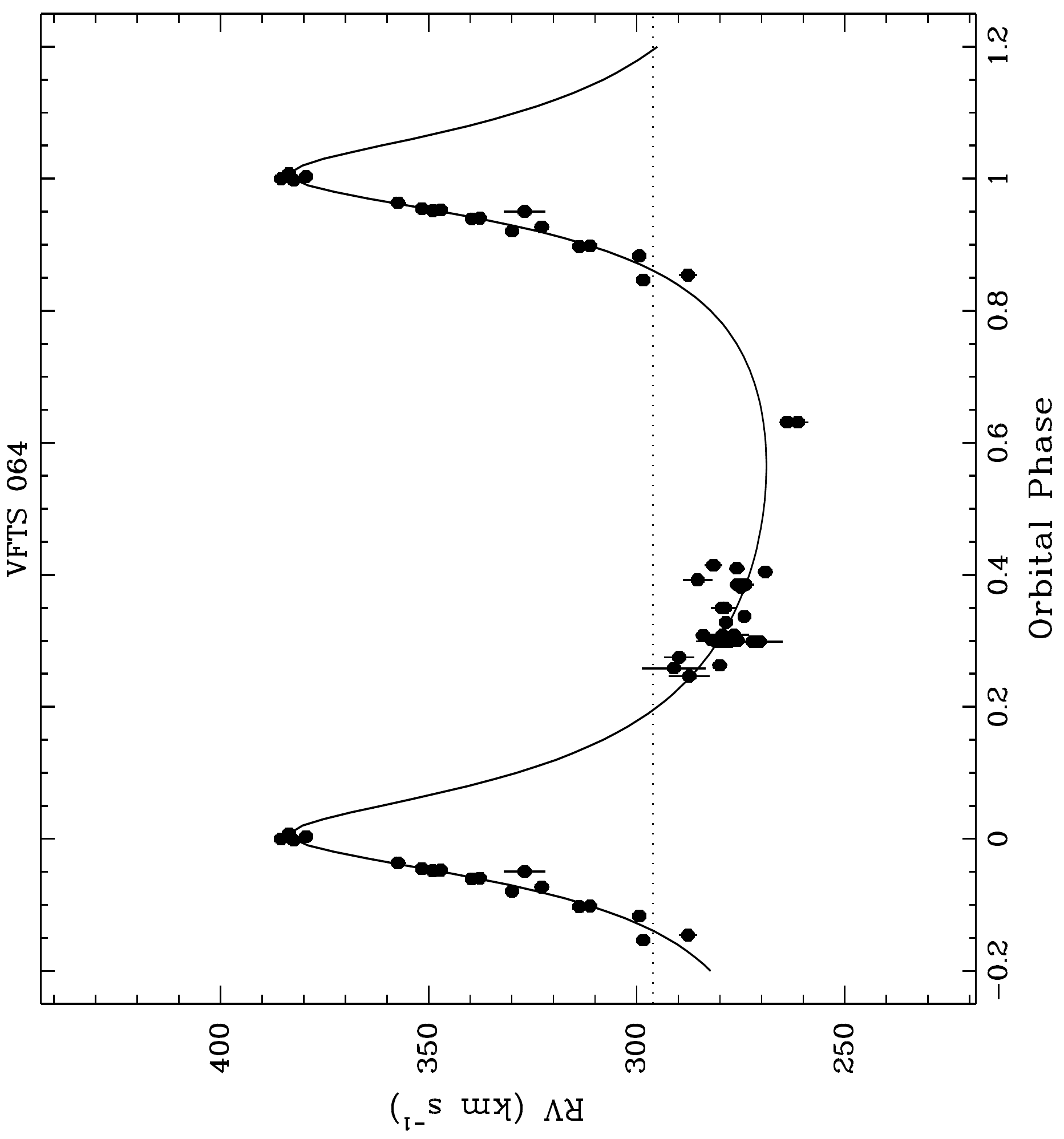}
\includegraphics[width=4.7cm,angle=-90]{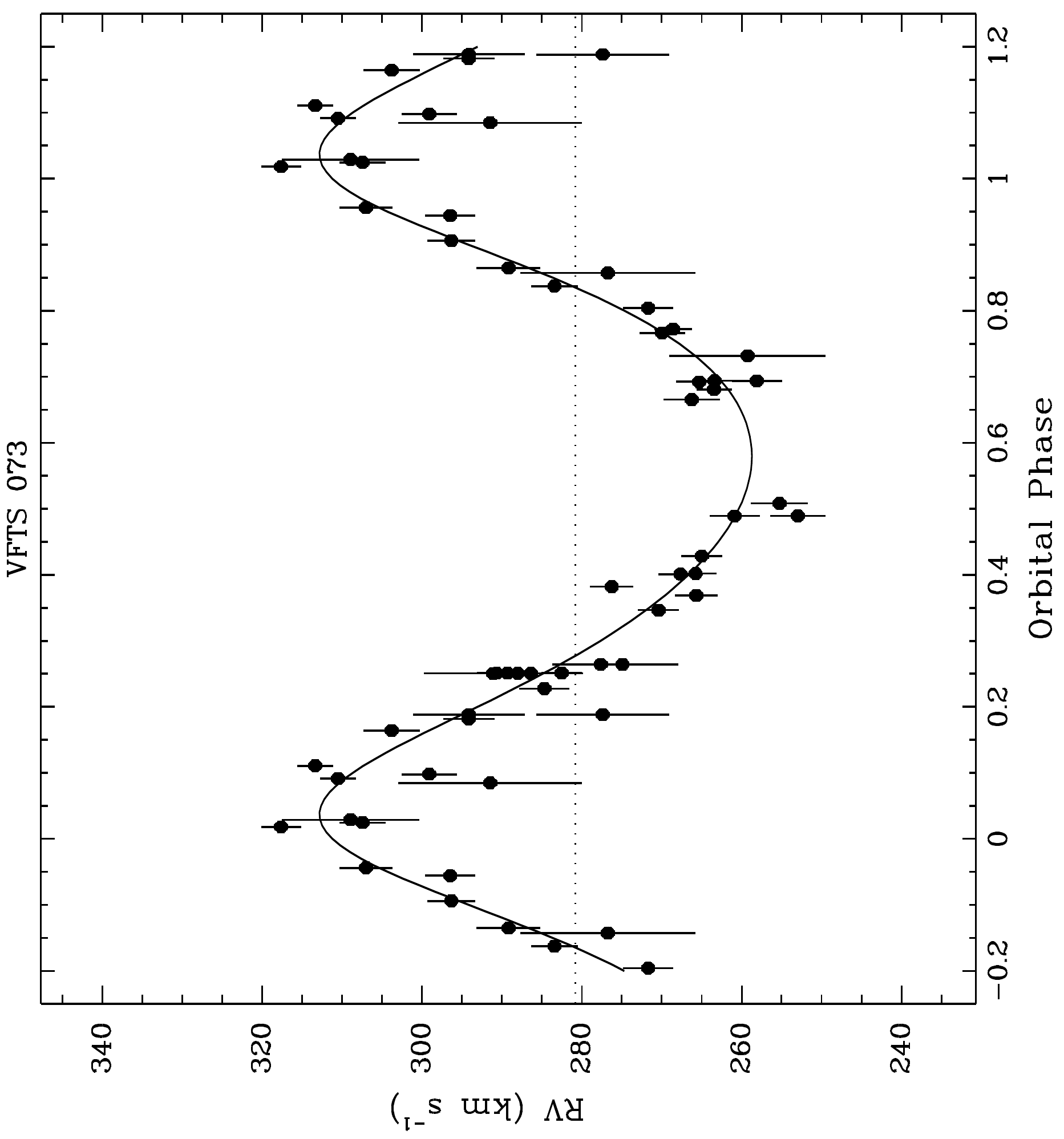}
\includegraphics[width=4.7cm,angle=-90]{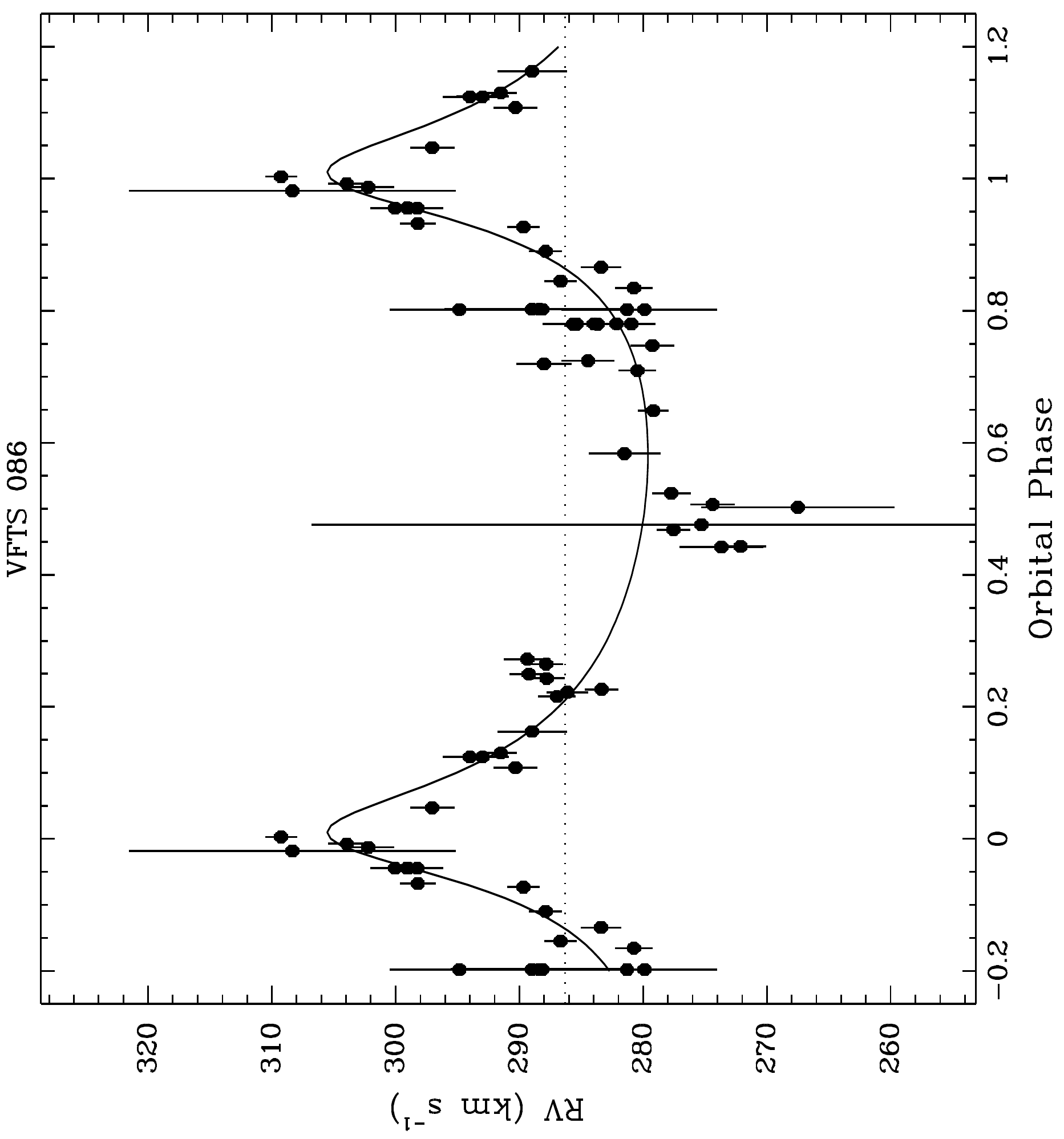}
\includegraphics[width=4.7cm,angle=-90]{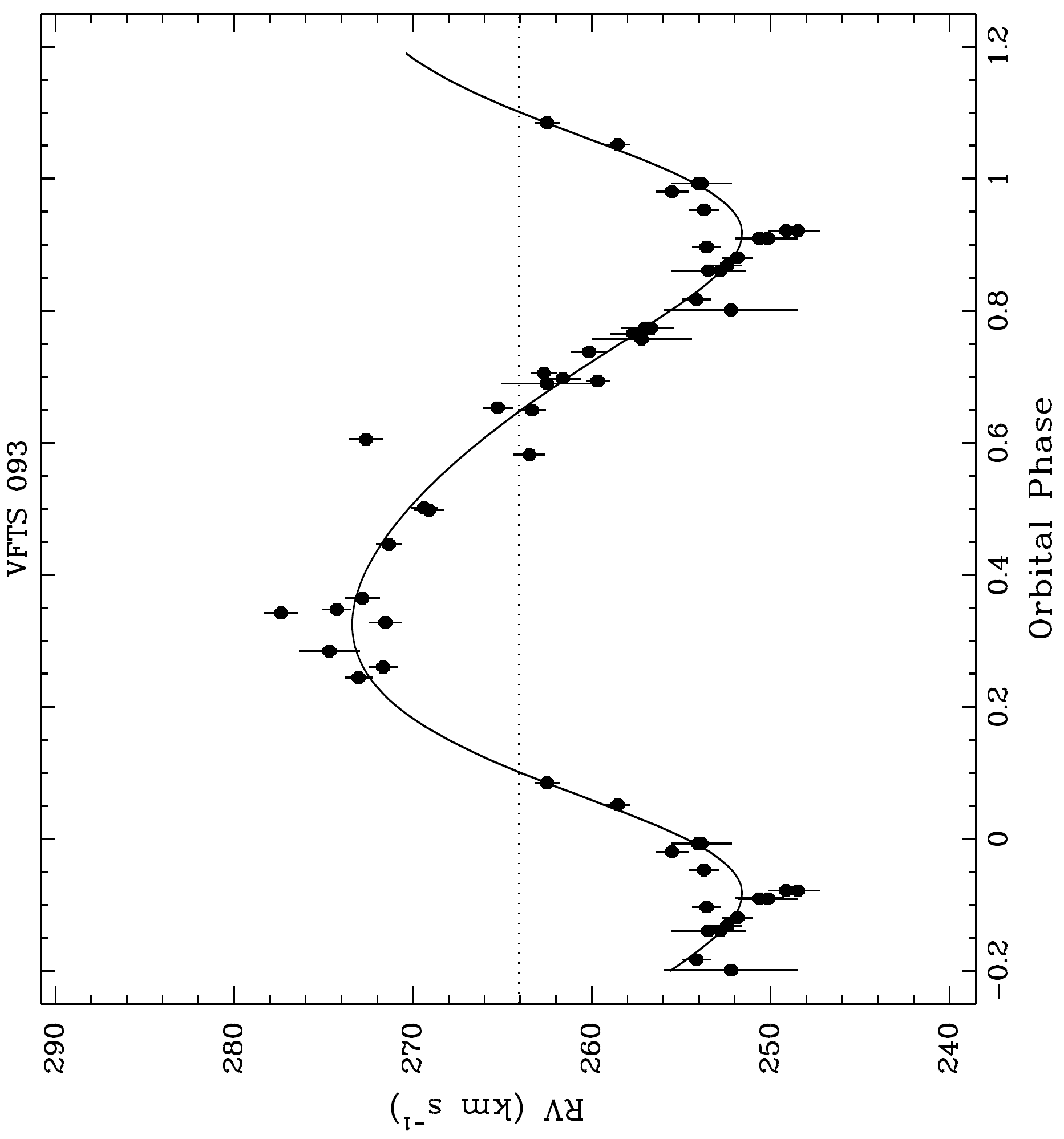}
\includegraphics[width=4.7cm,angle=-90]{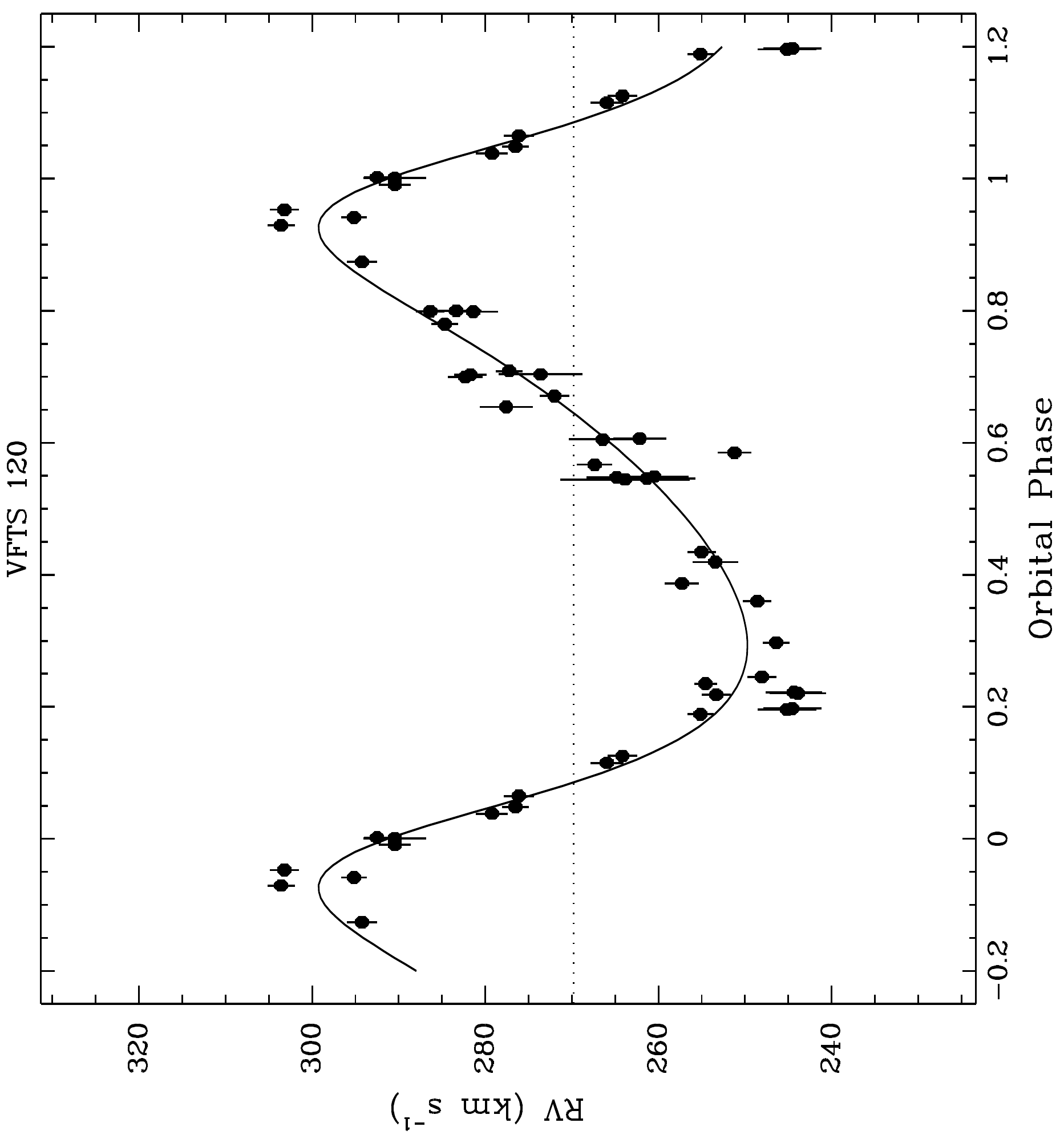}
\includegraphics[width=4.7cm,angle=-90]{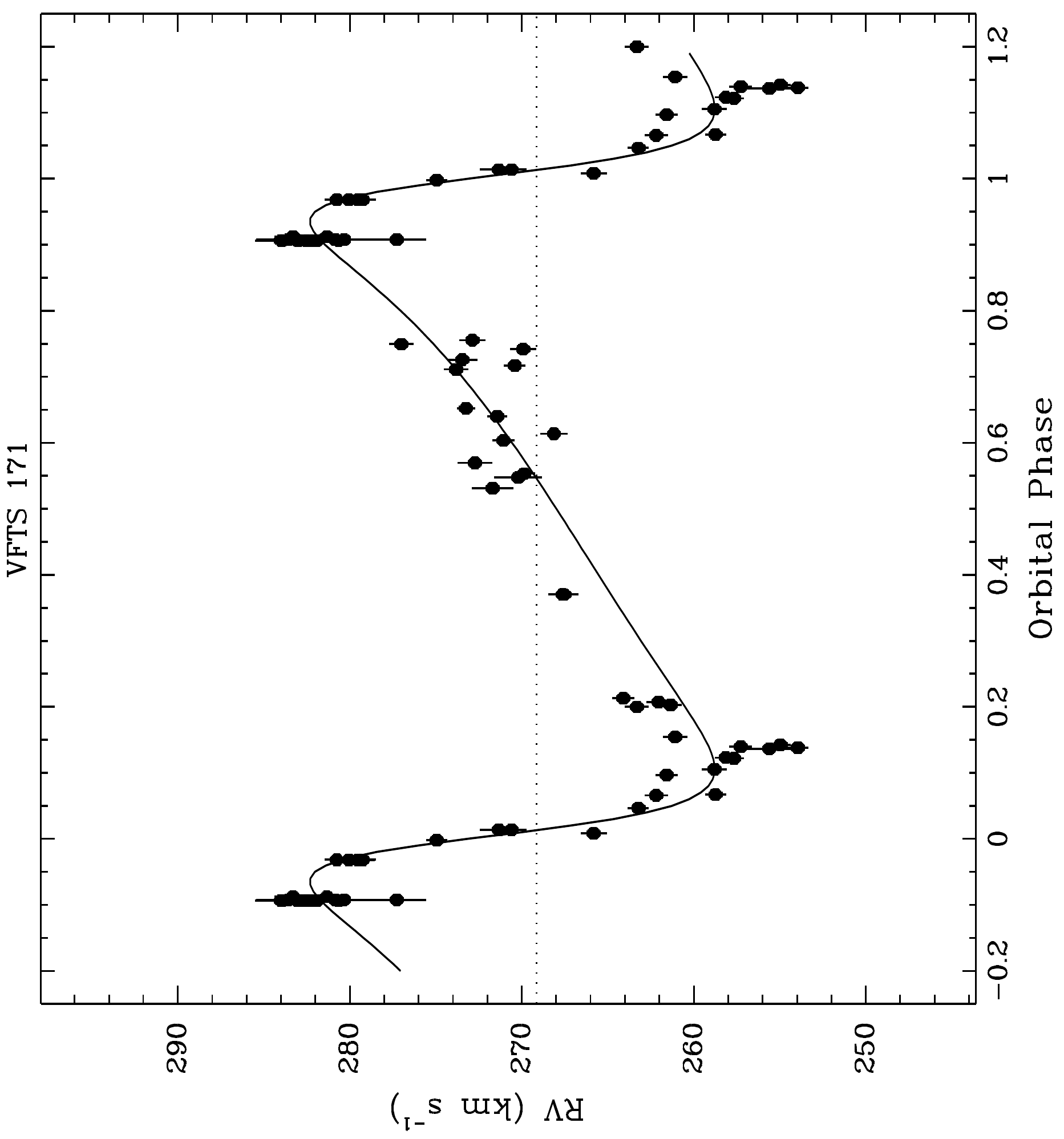}
\includegraphics[width=4.7cm,angle=-90]{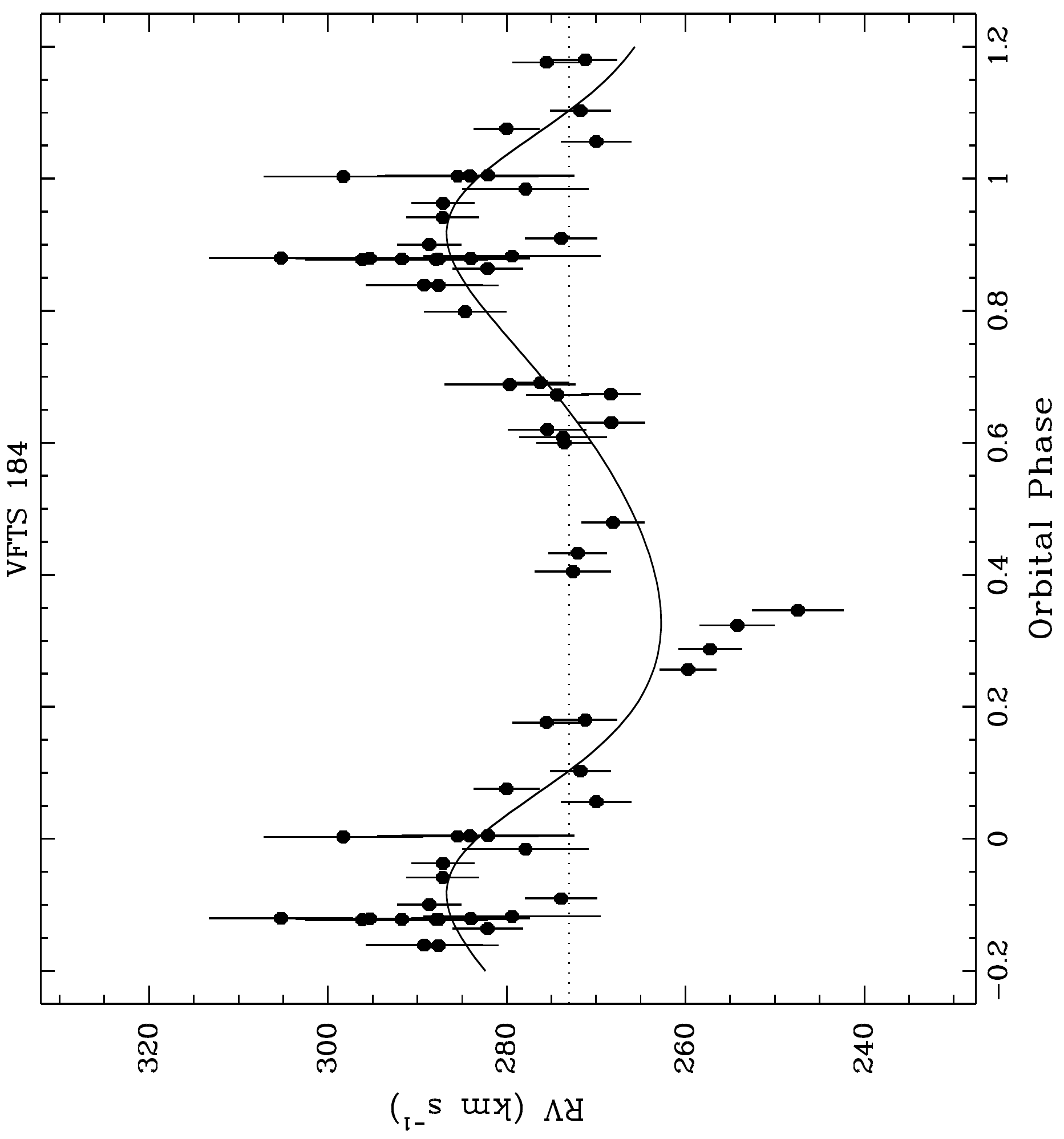}
\includegraphics[width=4.7cm,angle=-90]{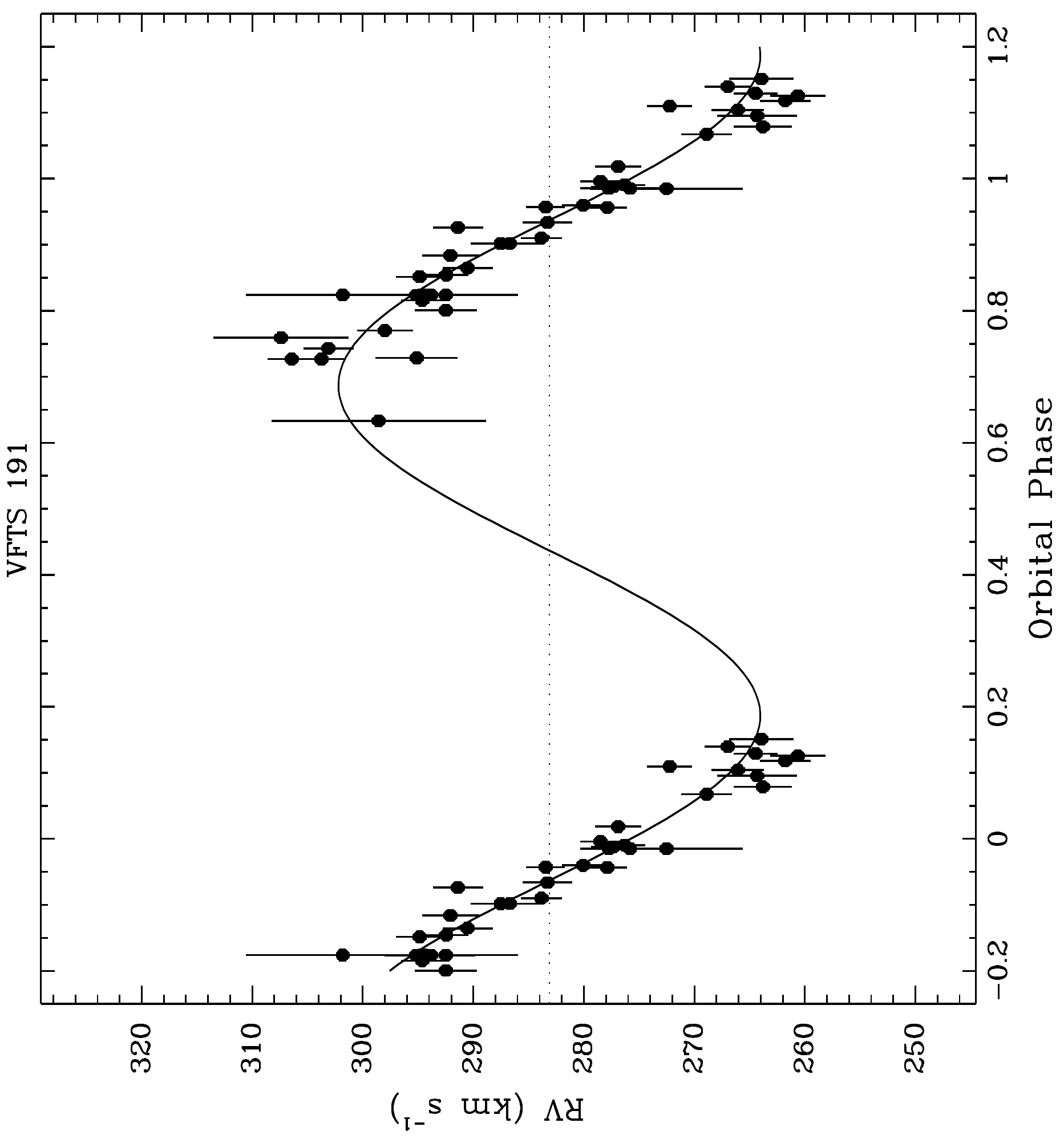}
\includegraphics[width=4.7cm,angle=-90]{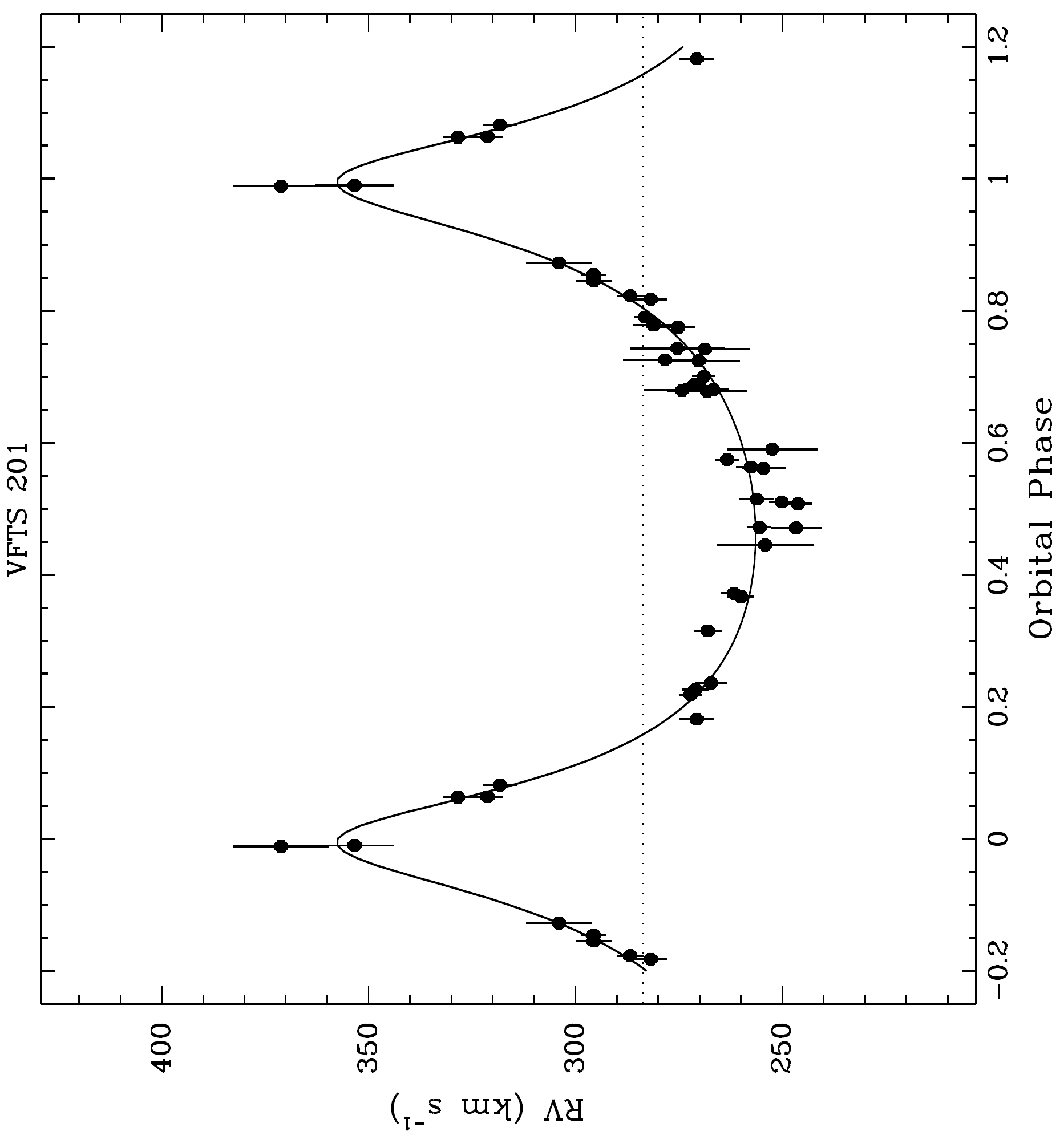}
\includegraphics[width=4.7cm,angle=-90]{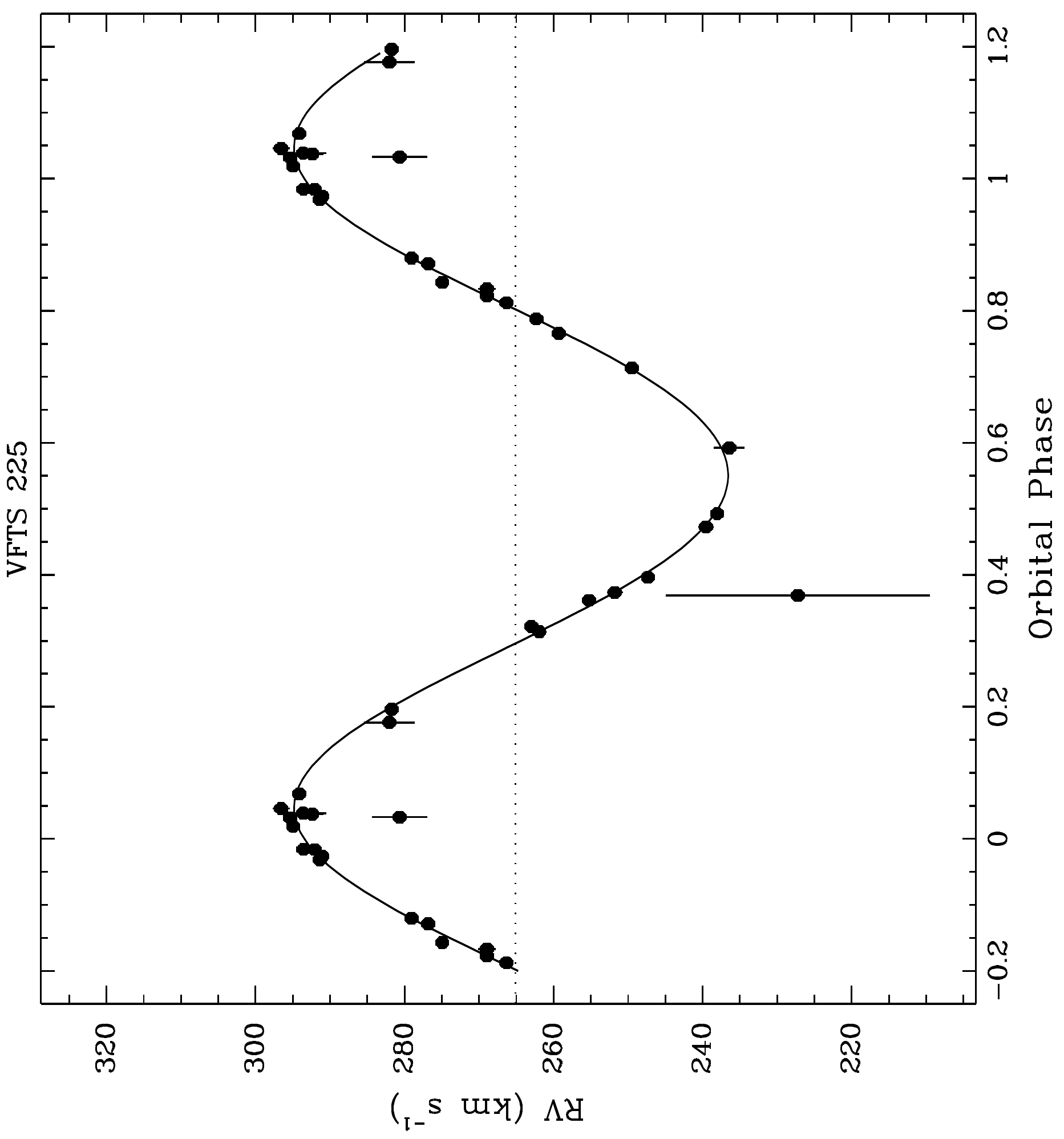}
\includegraphics[width=4.7cm,angle=-90]{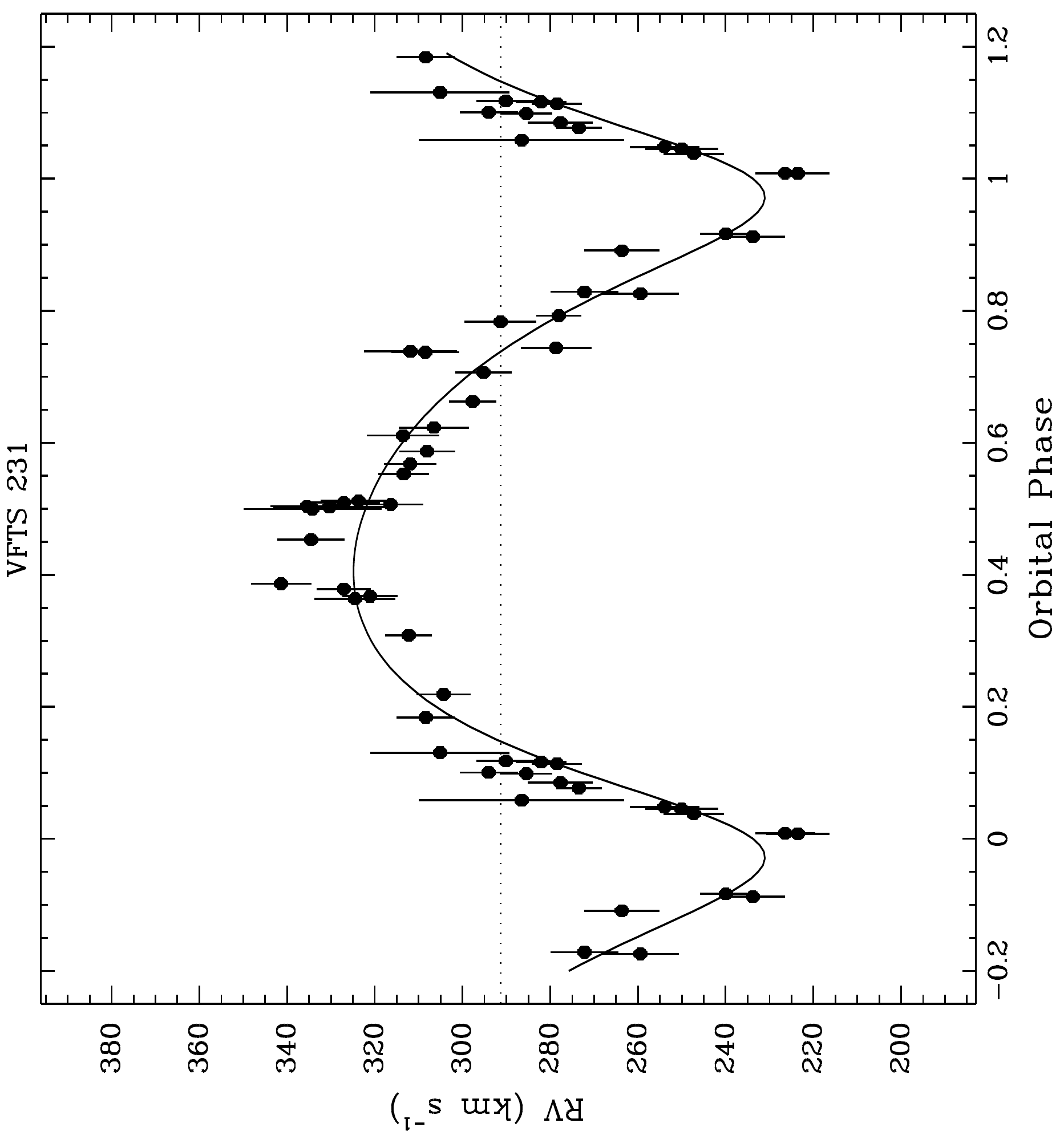}
\includegraphics[width=4.7cm,angle=-90]{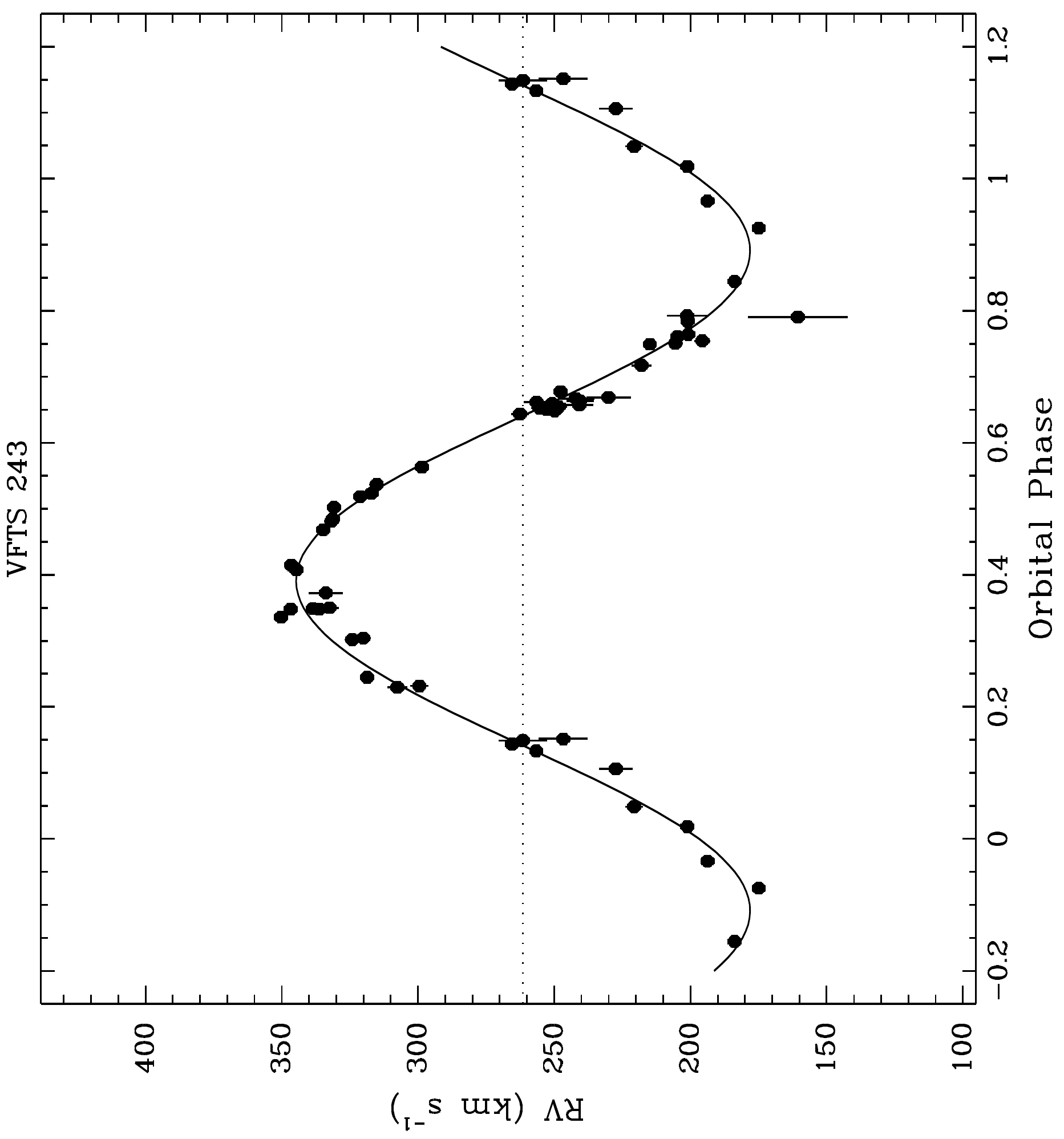}
\includegraphics[width=4.7cm,angle=-90]{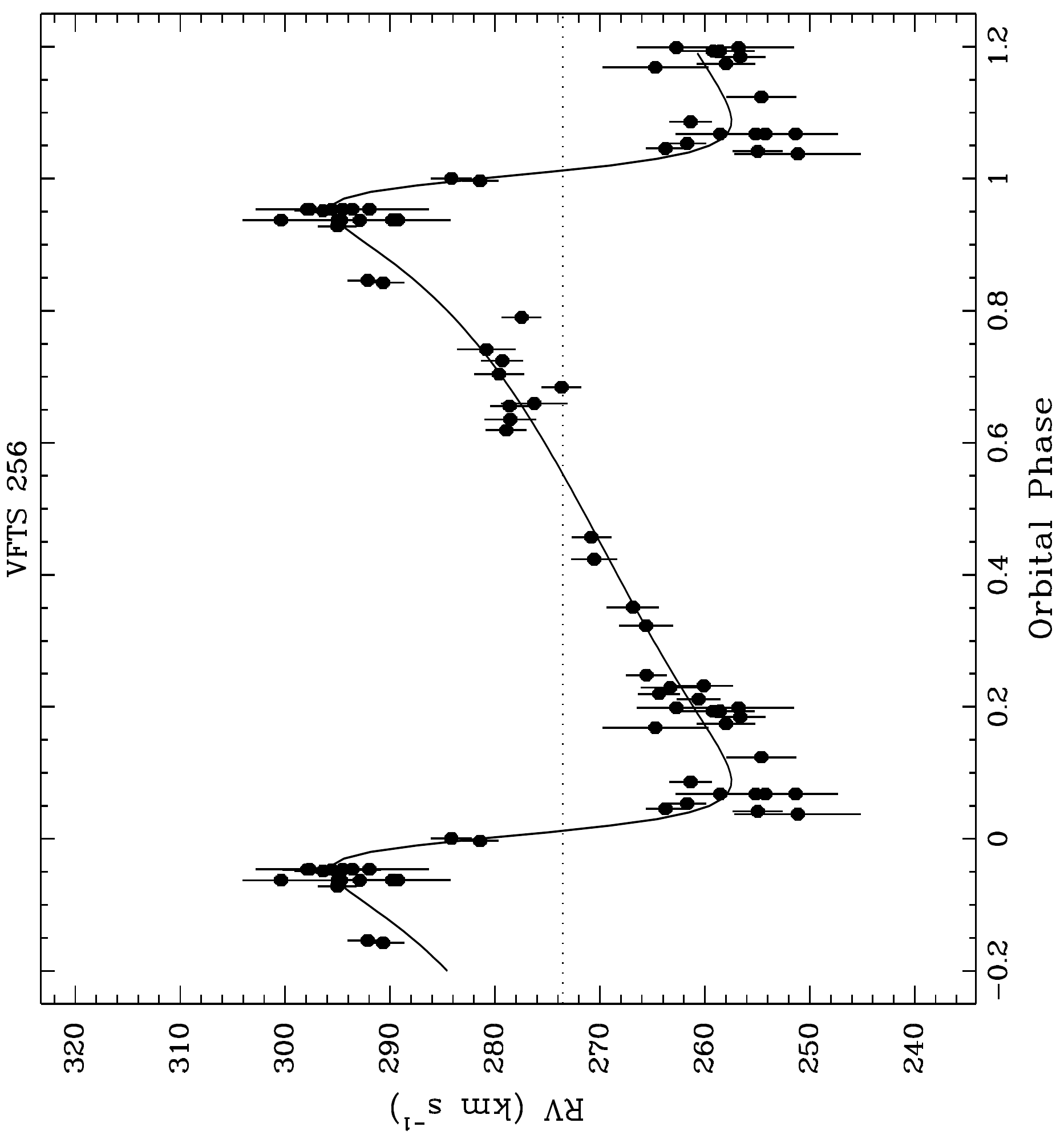}
\includegraphics[width=4.7cm,angle=-90]{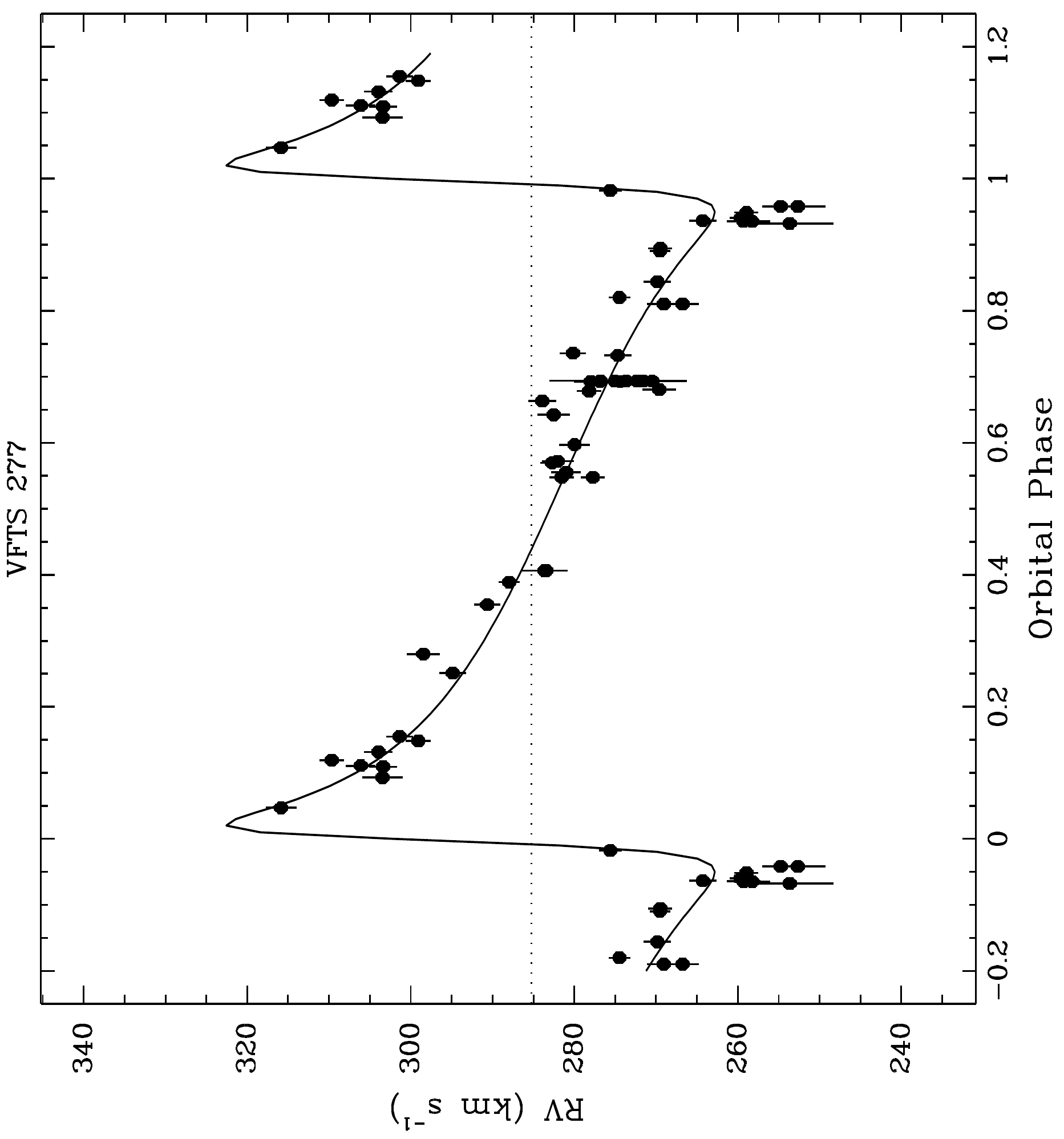}
\includegraphics[width=4.7cm,angle=-90]{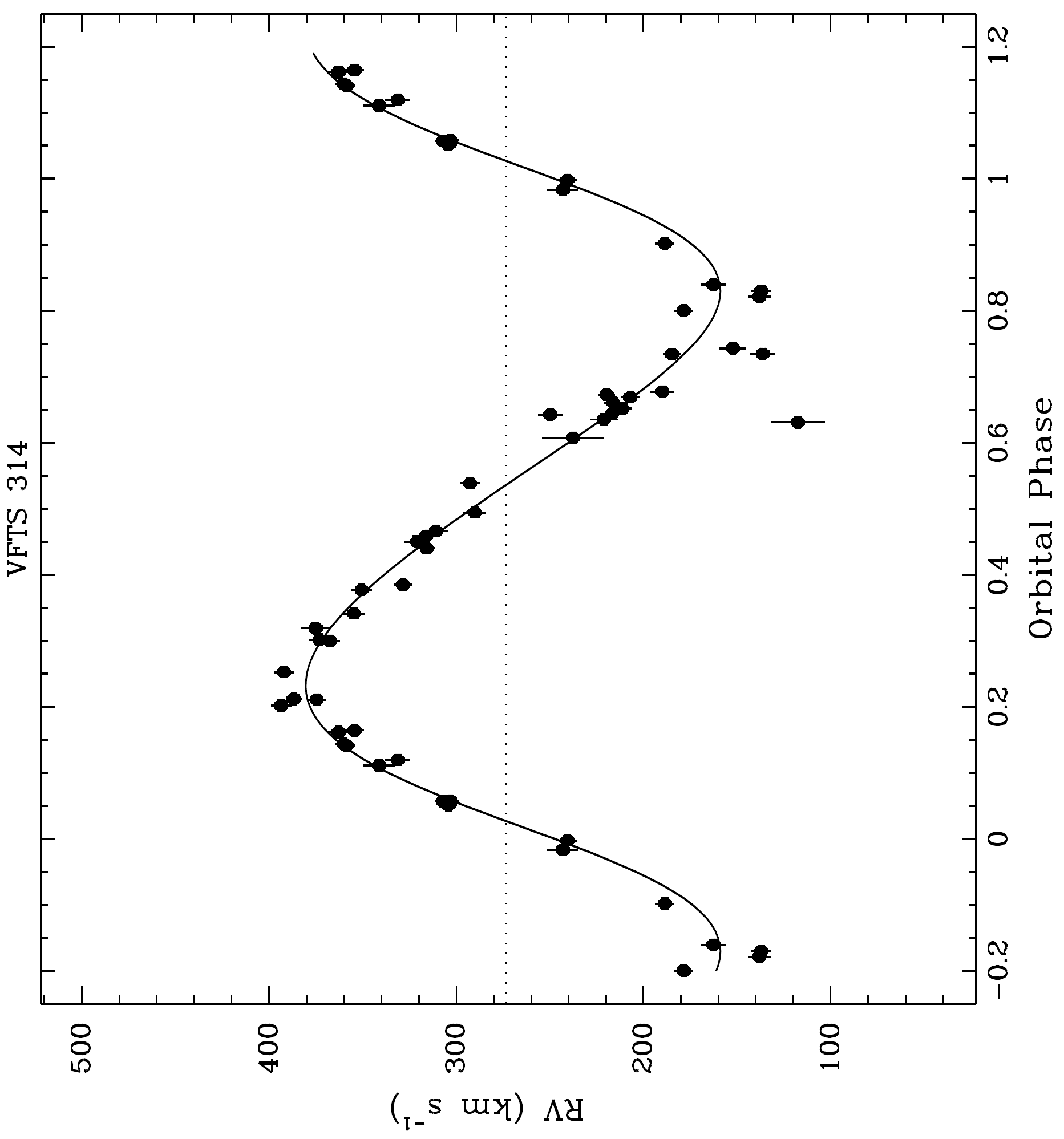}
\caption{Radial velocity (RV) curves for the SB1 systems. The filled hexagons are the RV measurements, while the dotted and solid lines are the systemic velocity and the best solution obtained by using the procedure shown in Section~\ref{ss:orbits}.}
\label{sb1:orb_solution}
\end{figure*}

\begin{figure*}
\centering
\ContinuedFloat
\includegraphics[width=4.7cm,angle=-90]{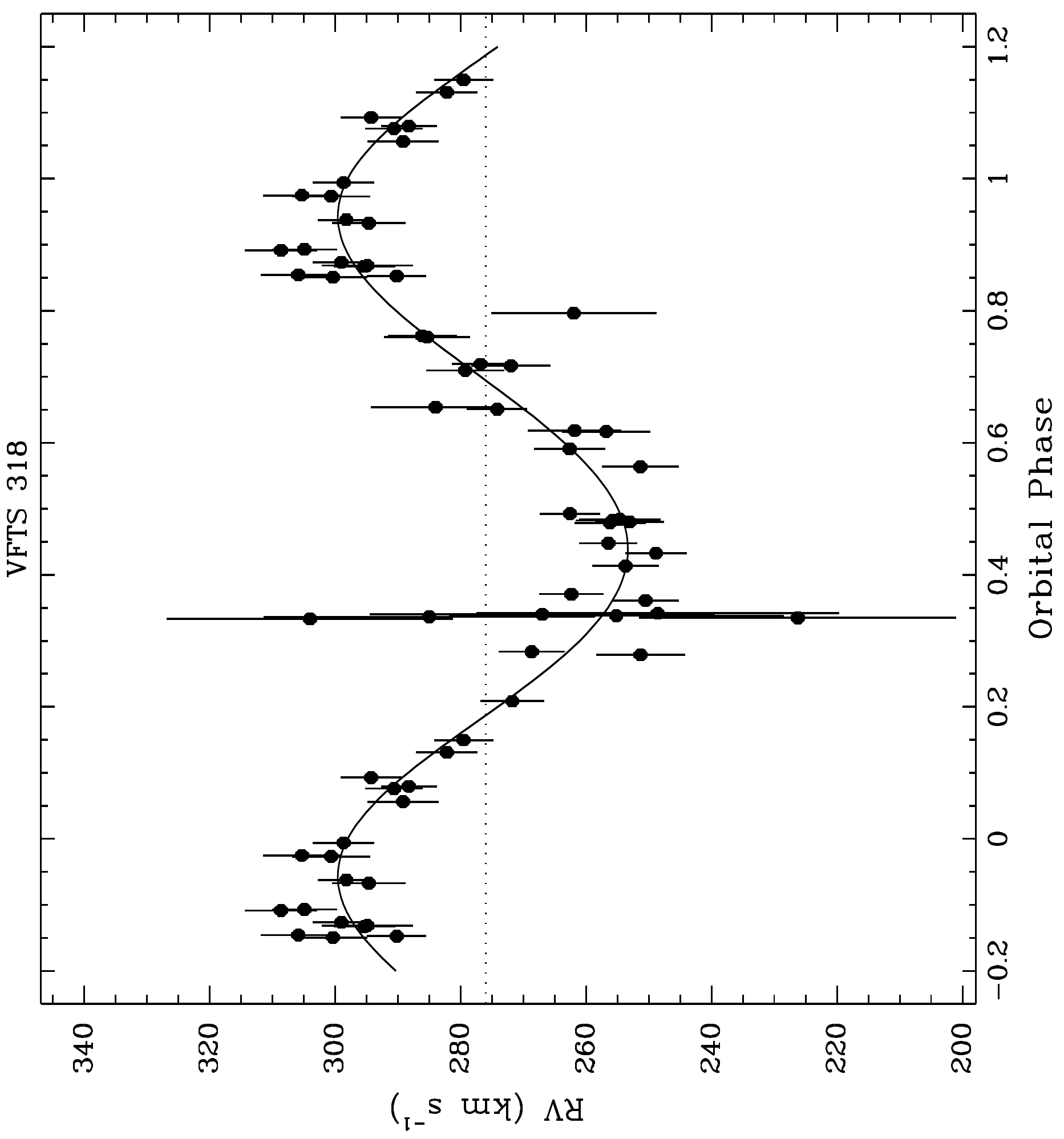}
\includegraphics[width=4.7cm,angle=-90]{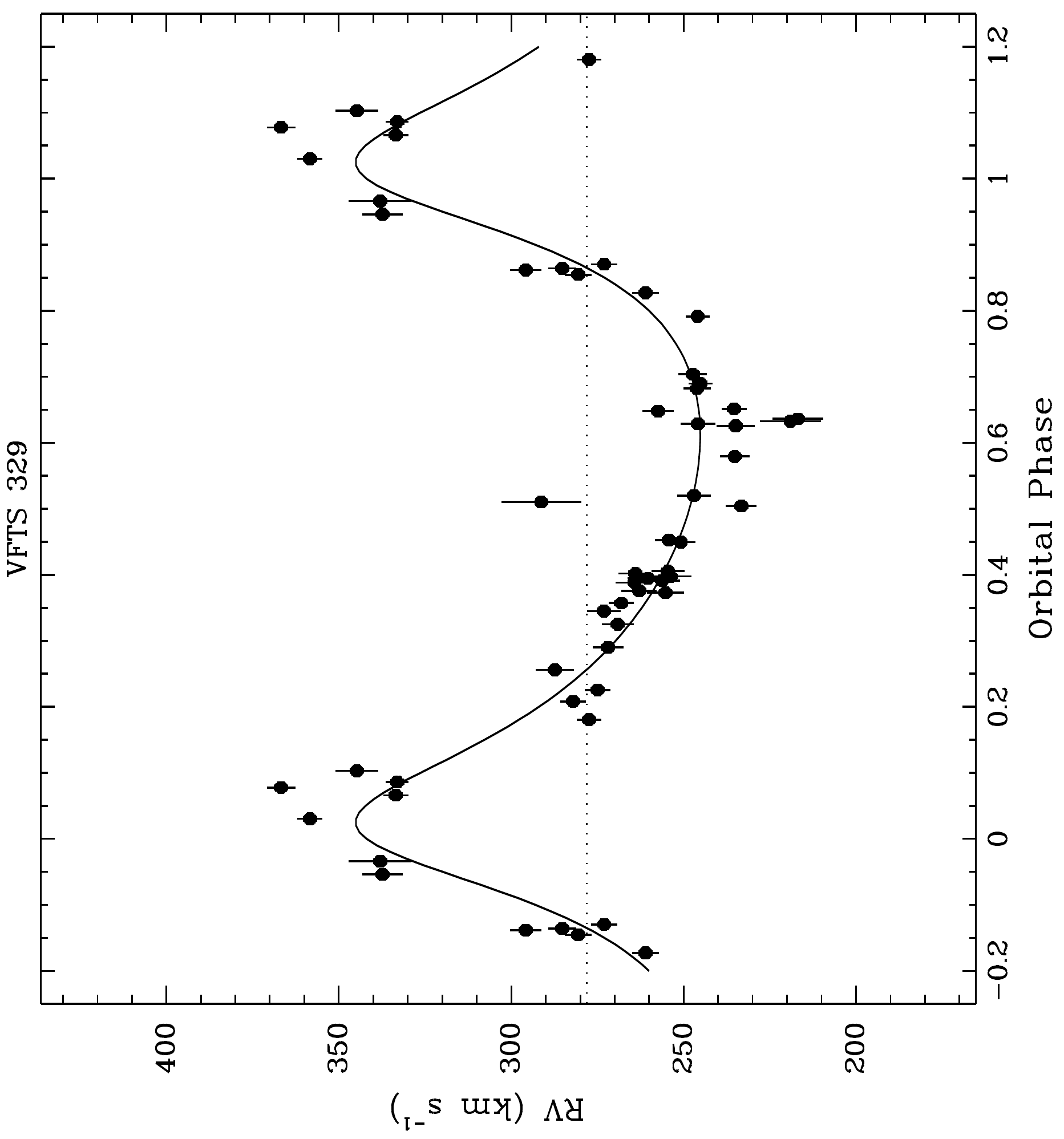}
\includegraphics[width=4.7cm,angle=-90]{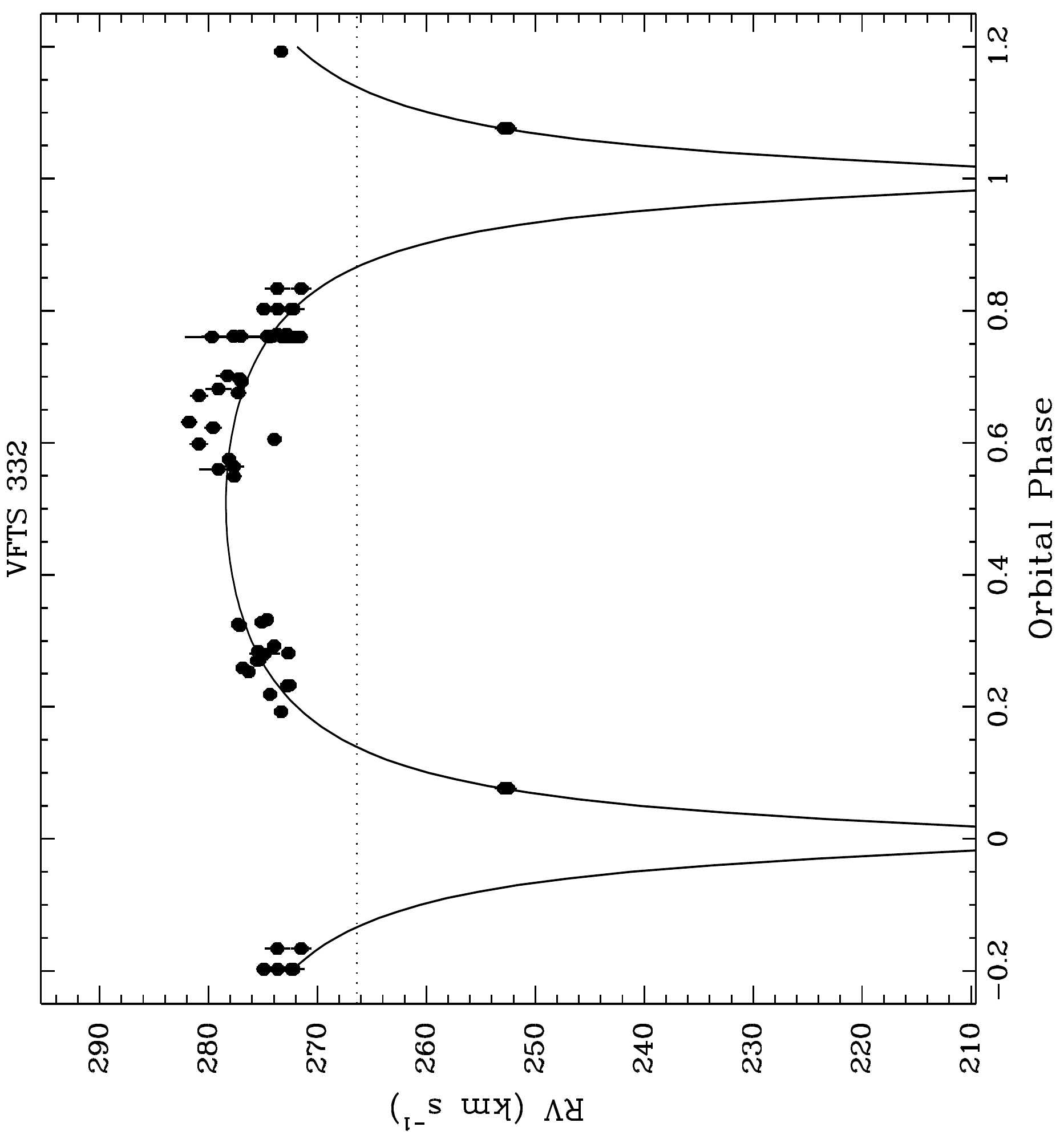}
\includegraphics[width=4.7cm,angle=-90]{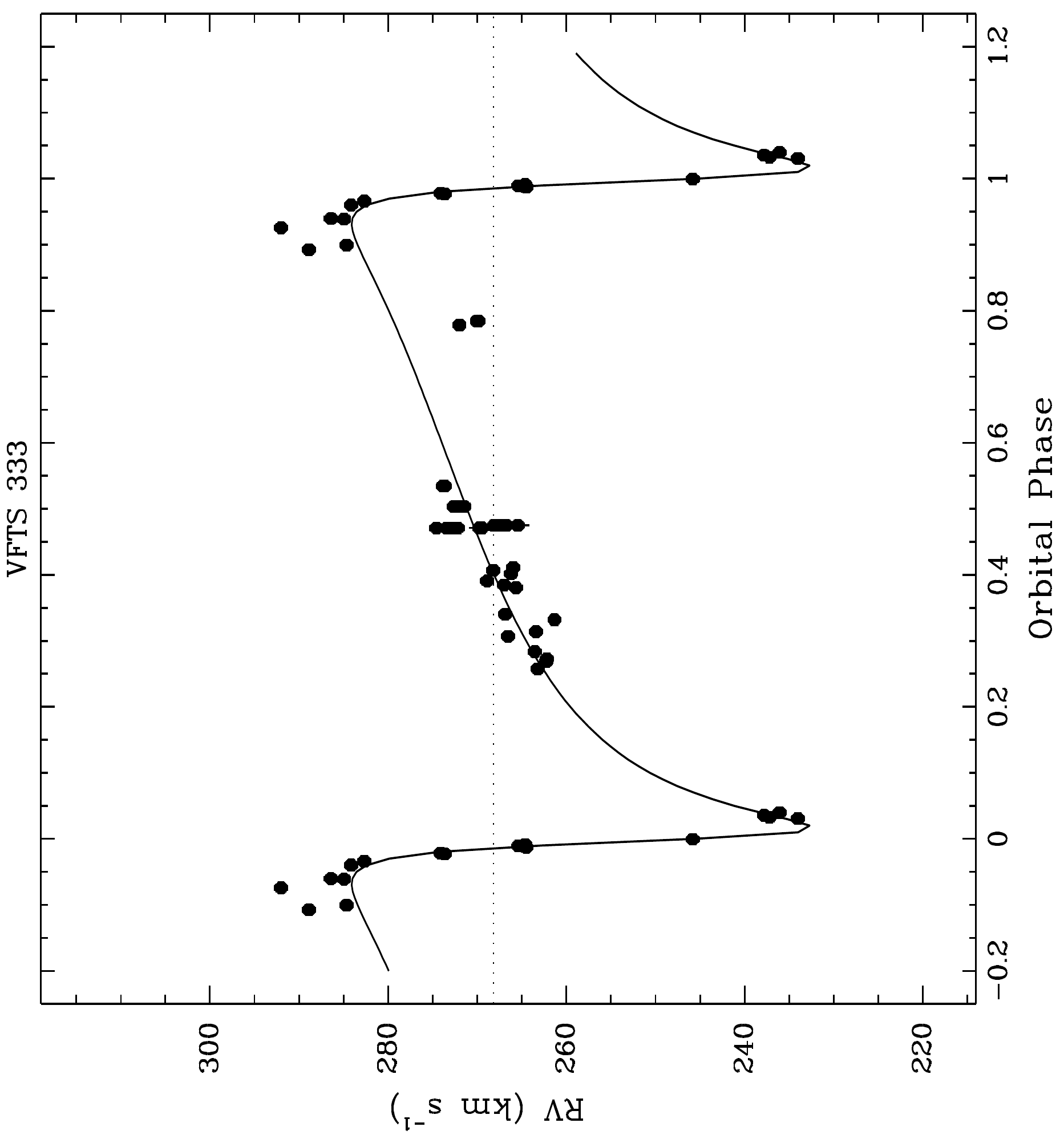}
\includegraphics[width=4.7cm,angle=-90]{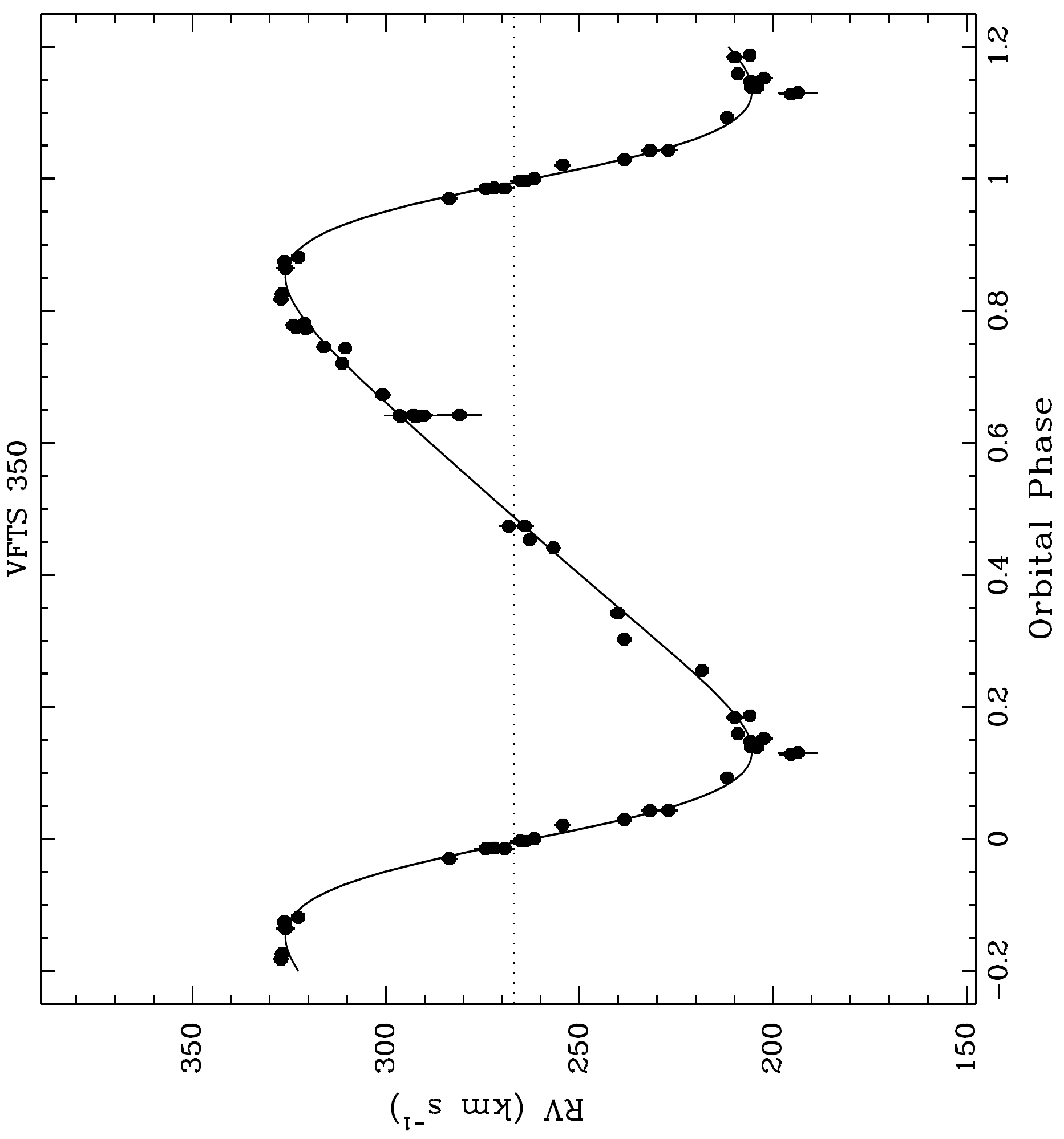}
\includegraphics[width=4.7cm,angle=-90]{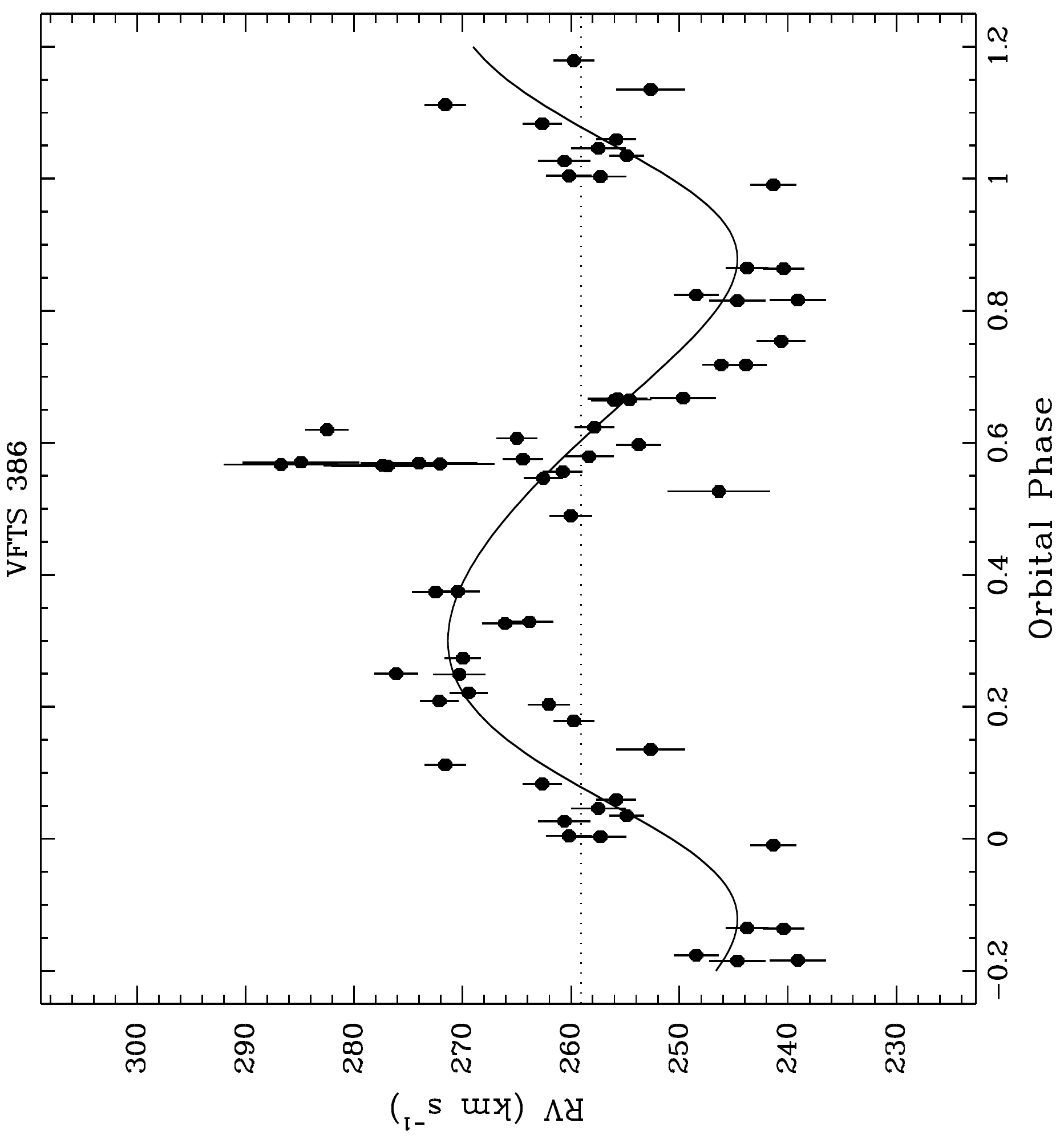}
\includegraphics[width=4.7cm,angle=-90]{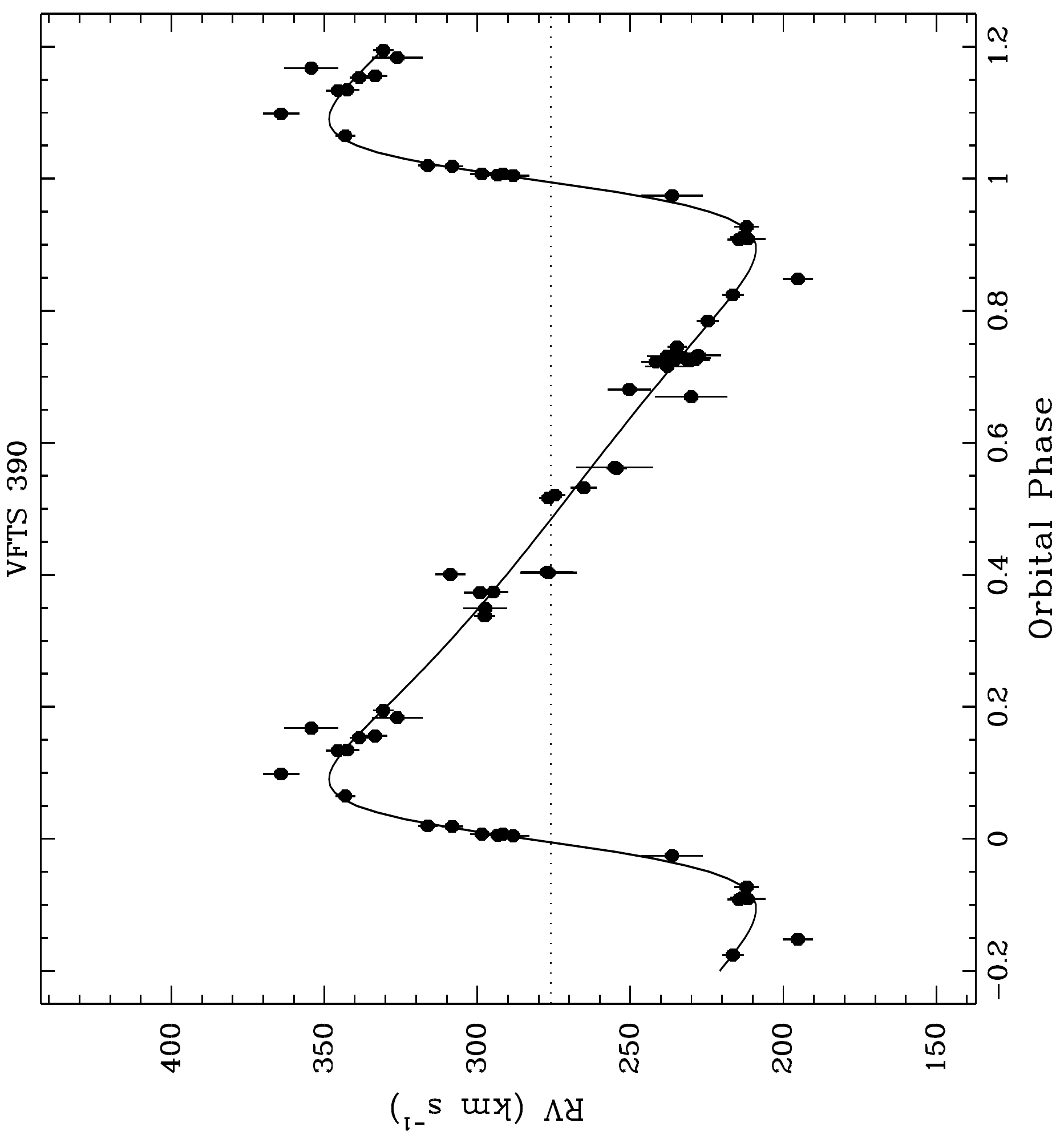}
\includegraphics[width=4.7cm,angle=-90]{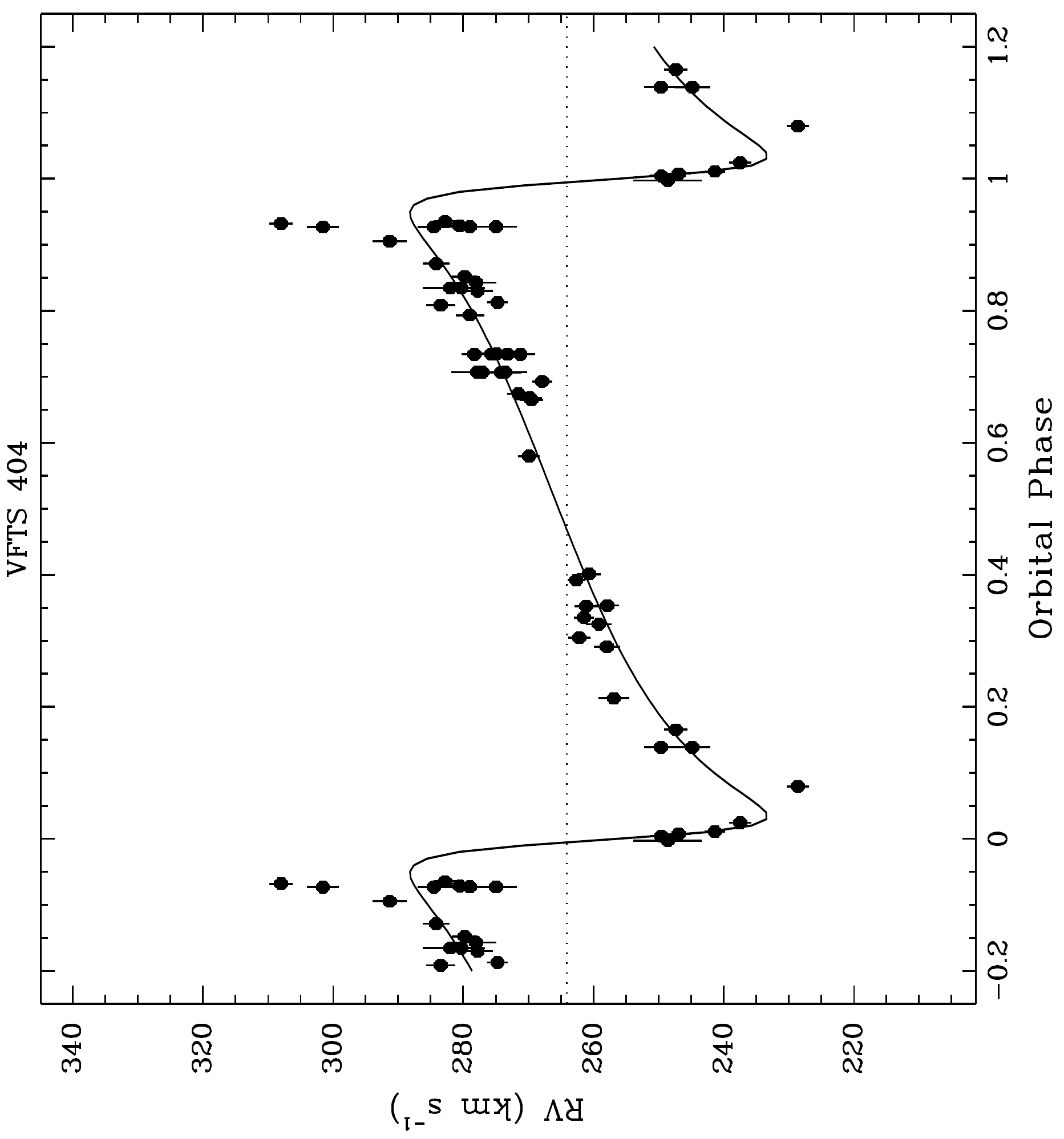}
\includegraphics[width=4.7cm,angle=-90]{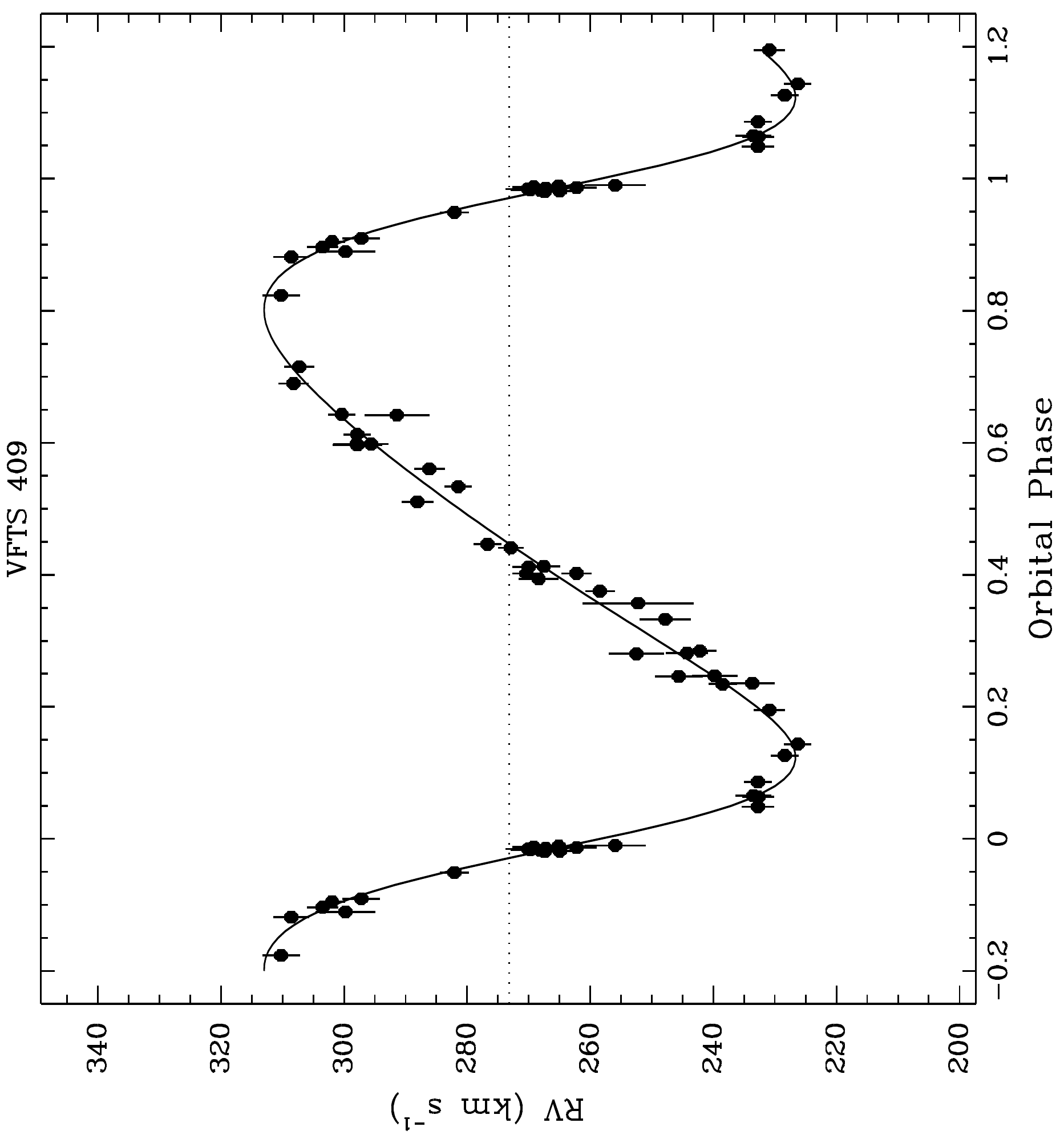}
\includegraphics[width=4.7cm,angle=-90]{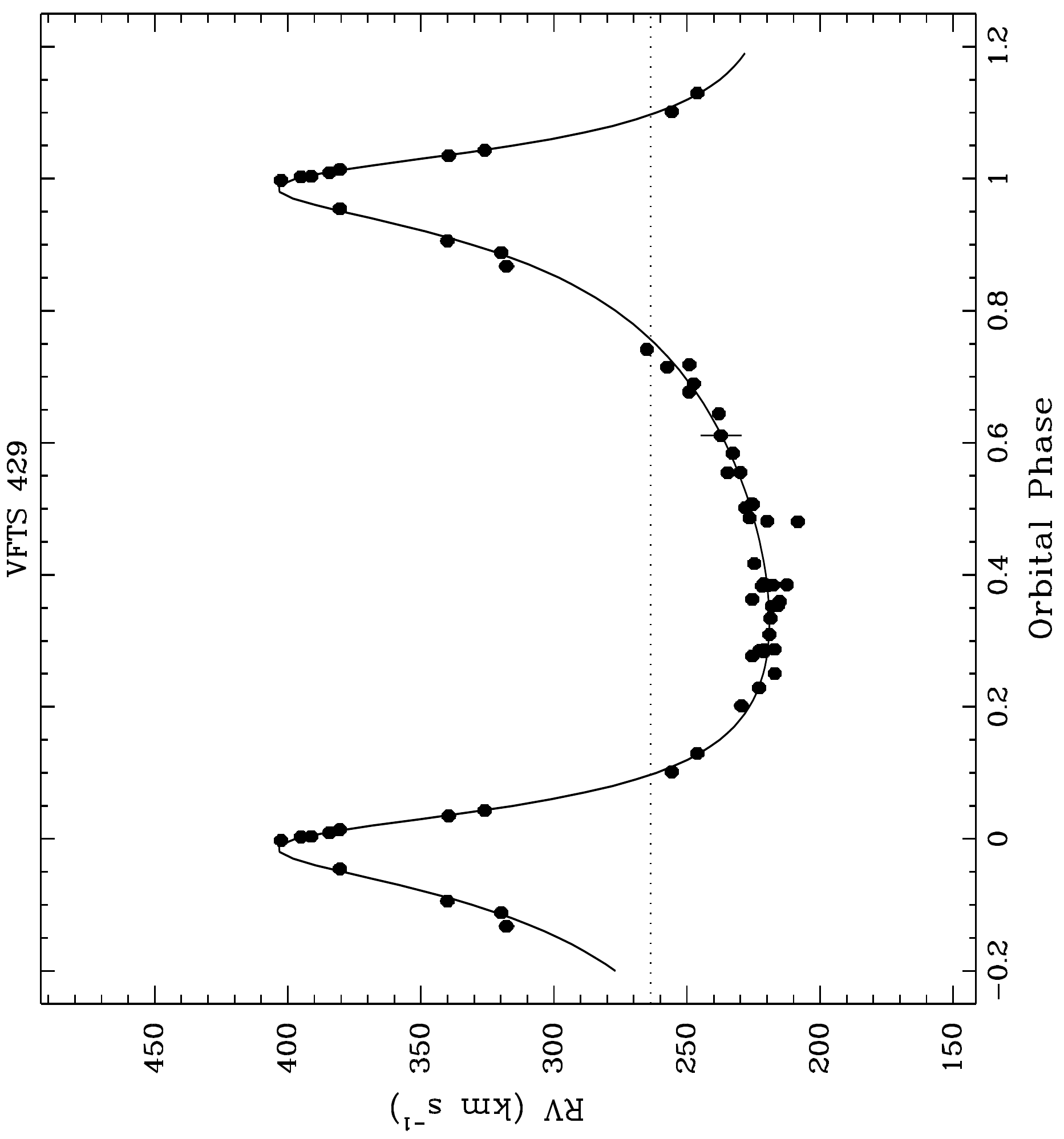}
\includegraphics[width=4.7cm,angle=-90]{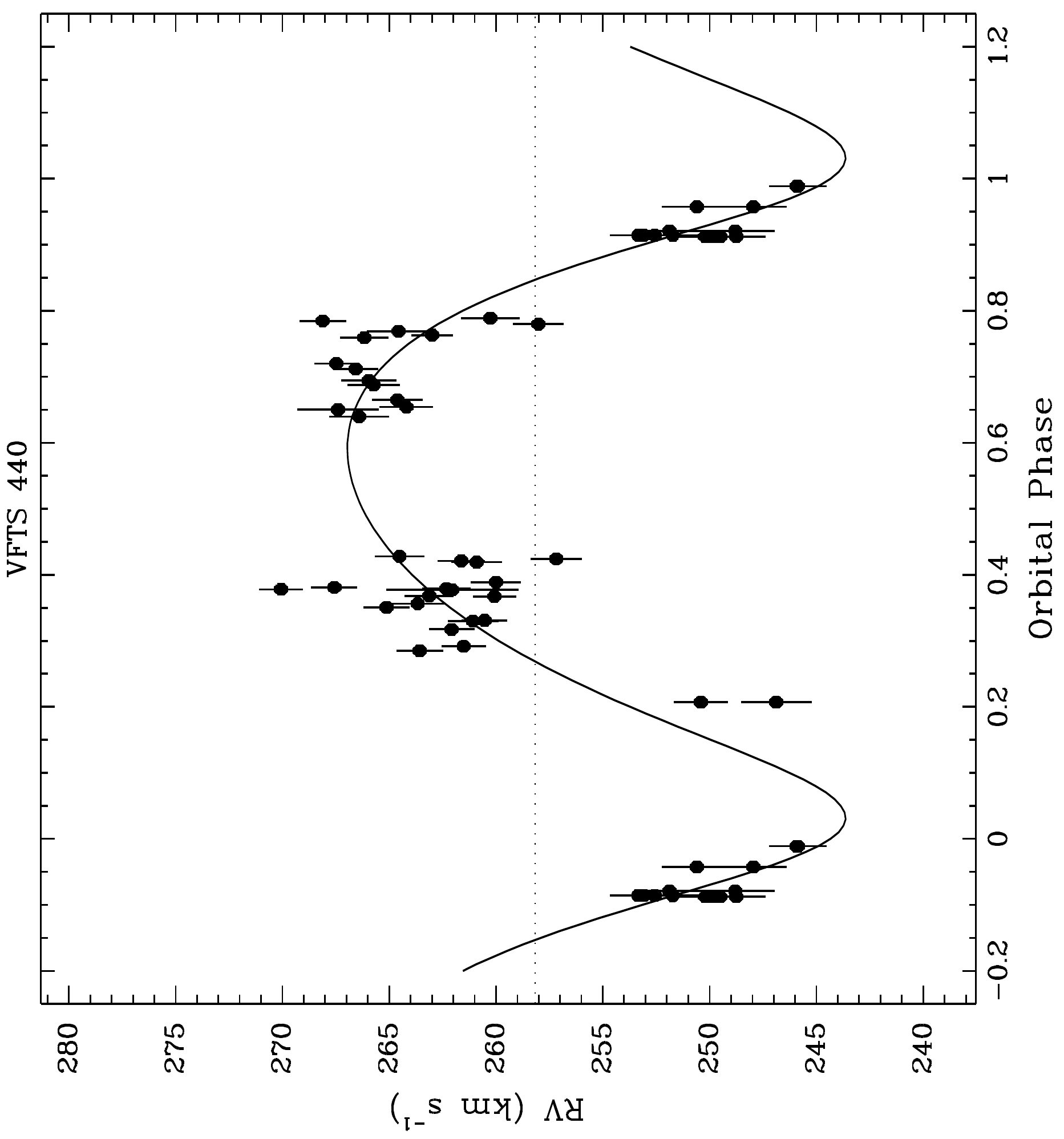}
\includegraphics[width=4.7cm,angle=-90]{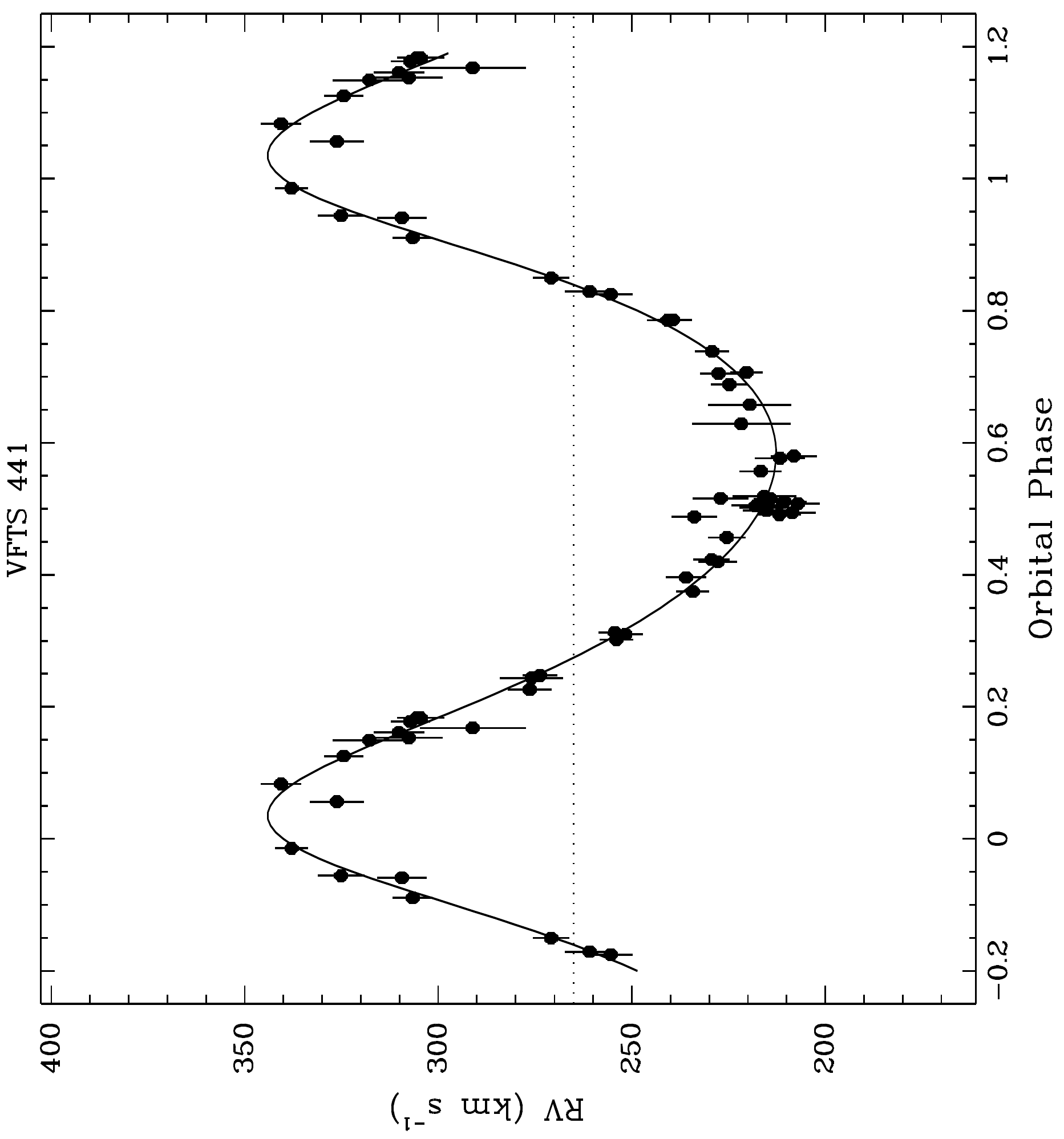}
\includegraphics[width=4.7cm,angle=-90]{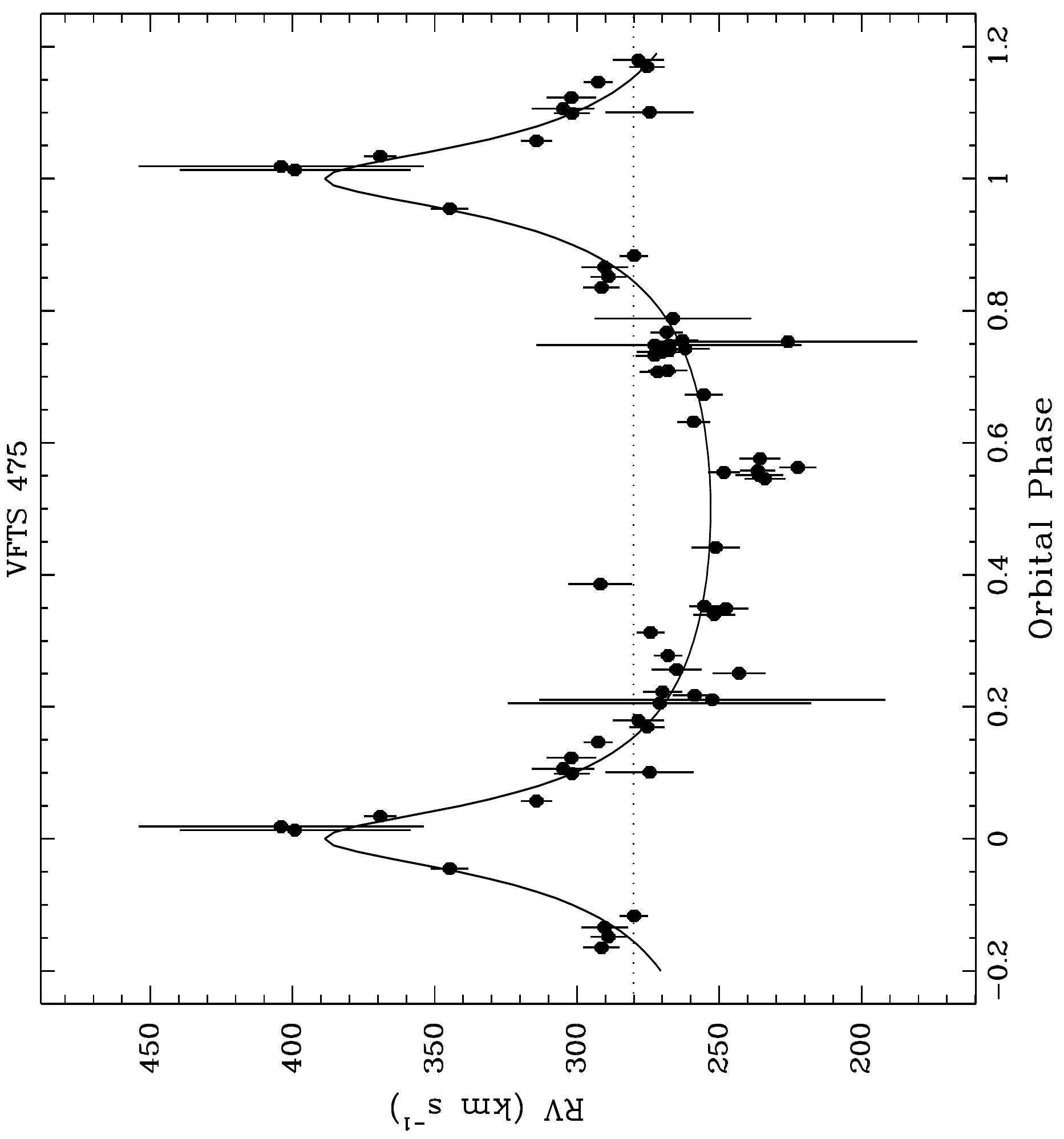}
\includegraphics[width=4.7cm,angle=-90]{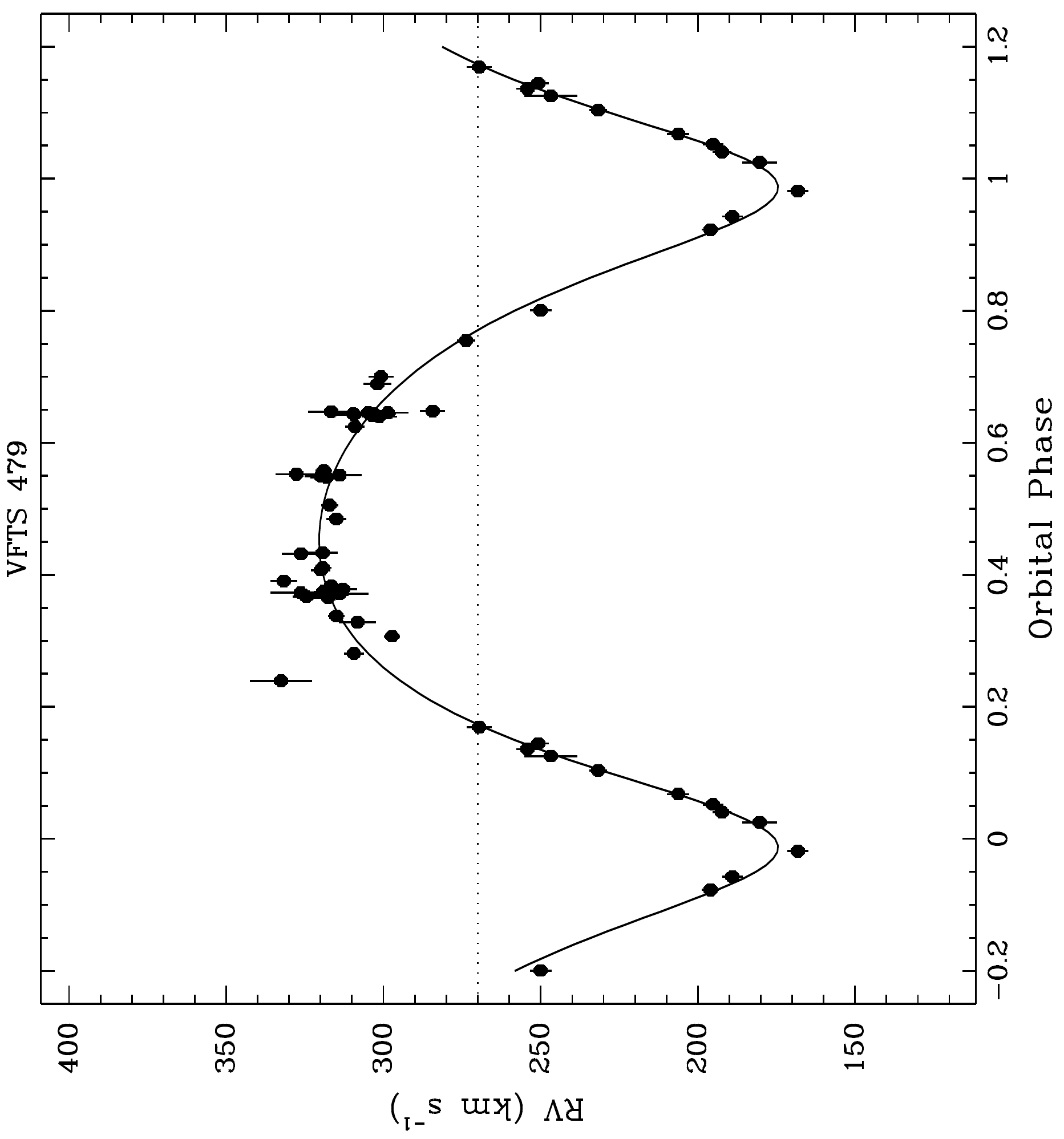}
\includegraphics[width=4.7cm,angle=-90]{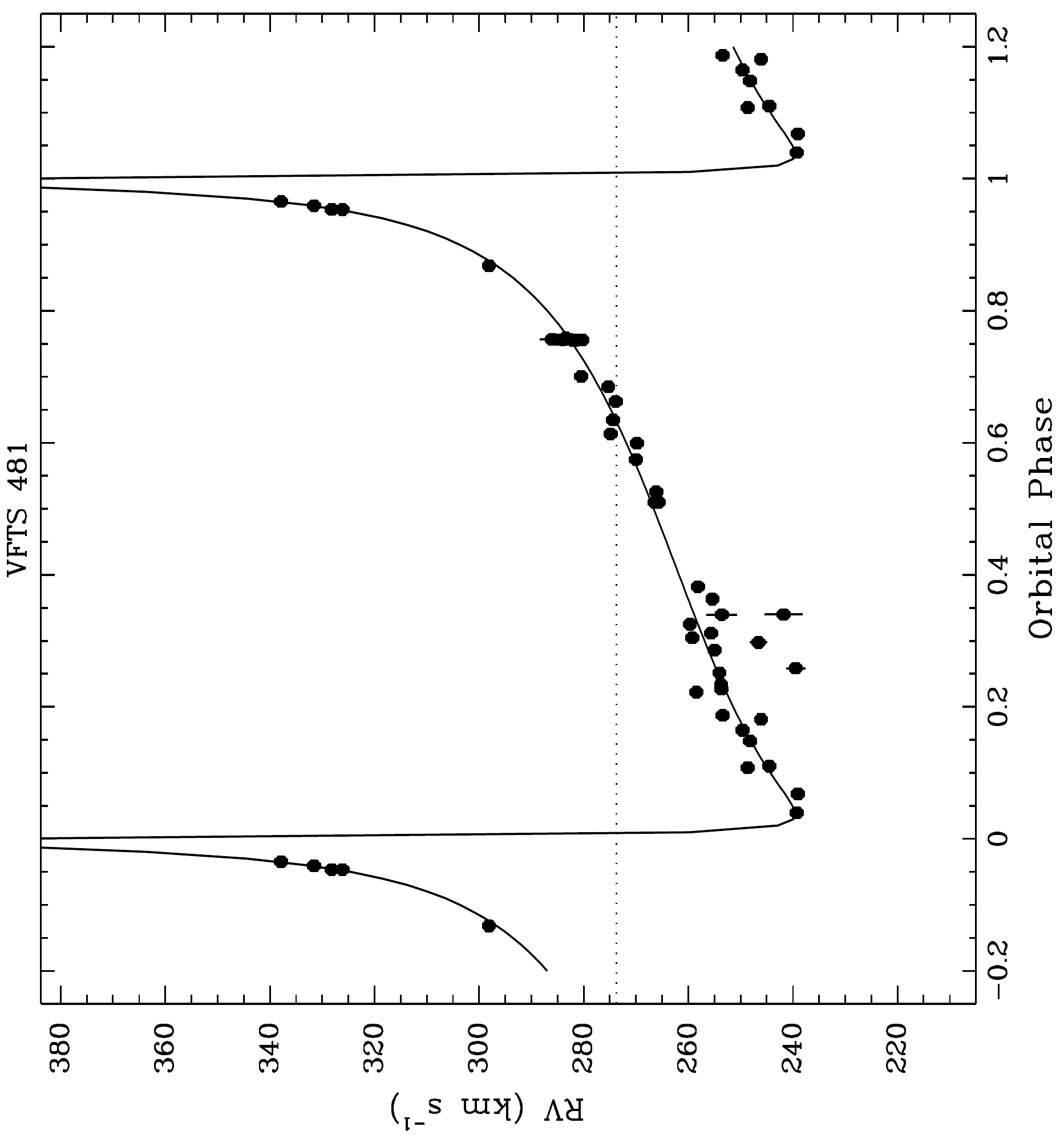}
\caption{{\it Continued...}}
\end{figure*}

\begin{figure*}
\centering
\ContinuedFloat
\includegraphics[width=4.7cm,angle=-90]{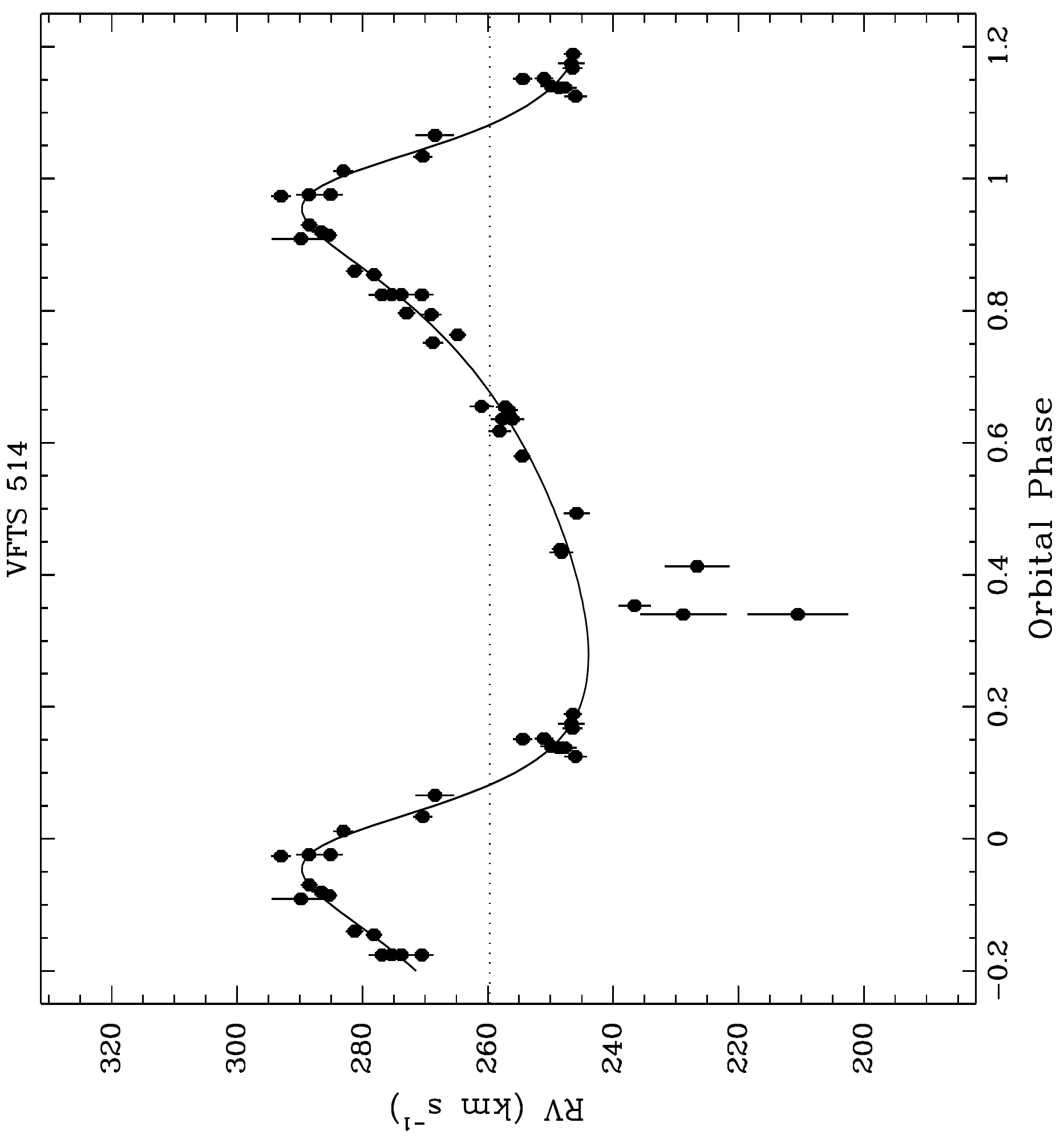}
\includegraphics[width=4.7cm,angle=-90]{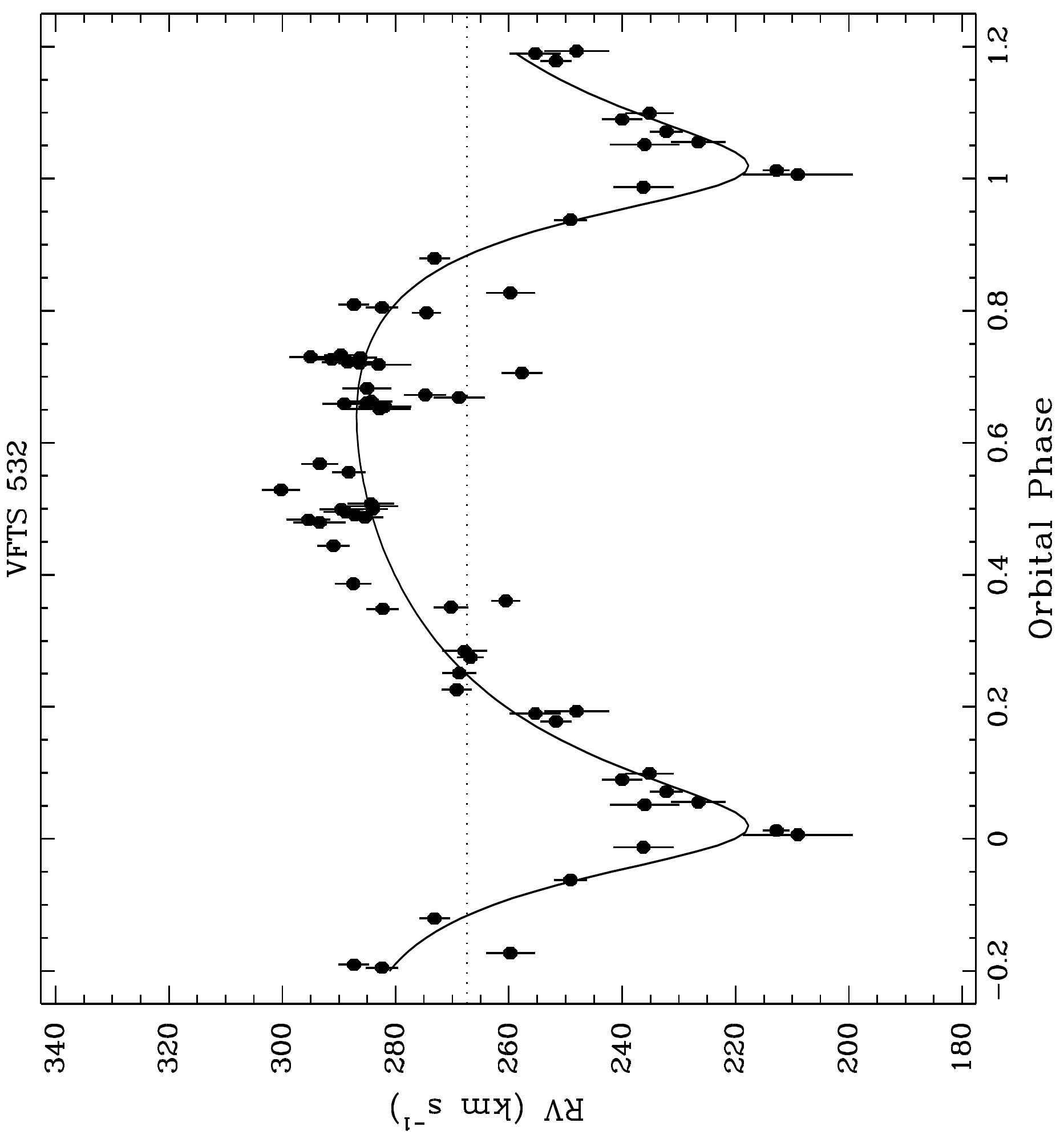}
\includegraphics[width=4.7cm,angle=-90]{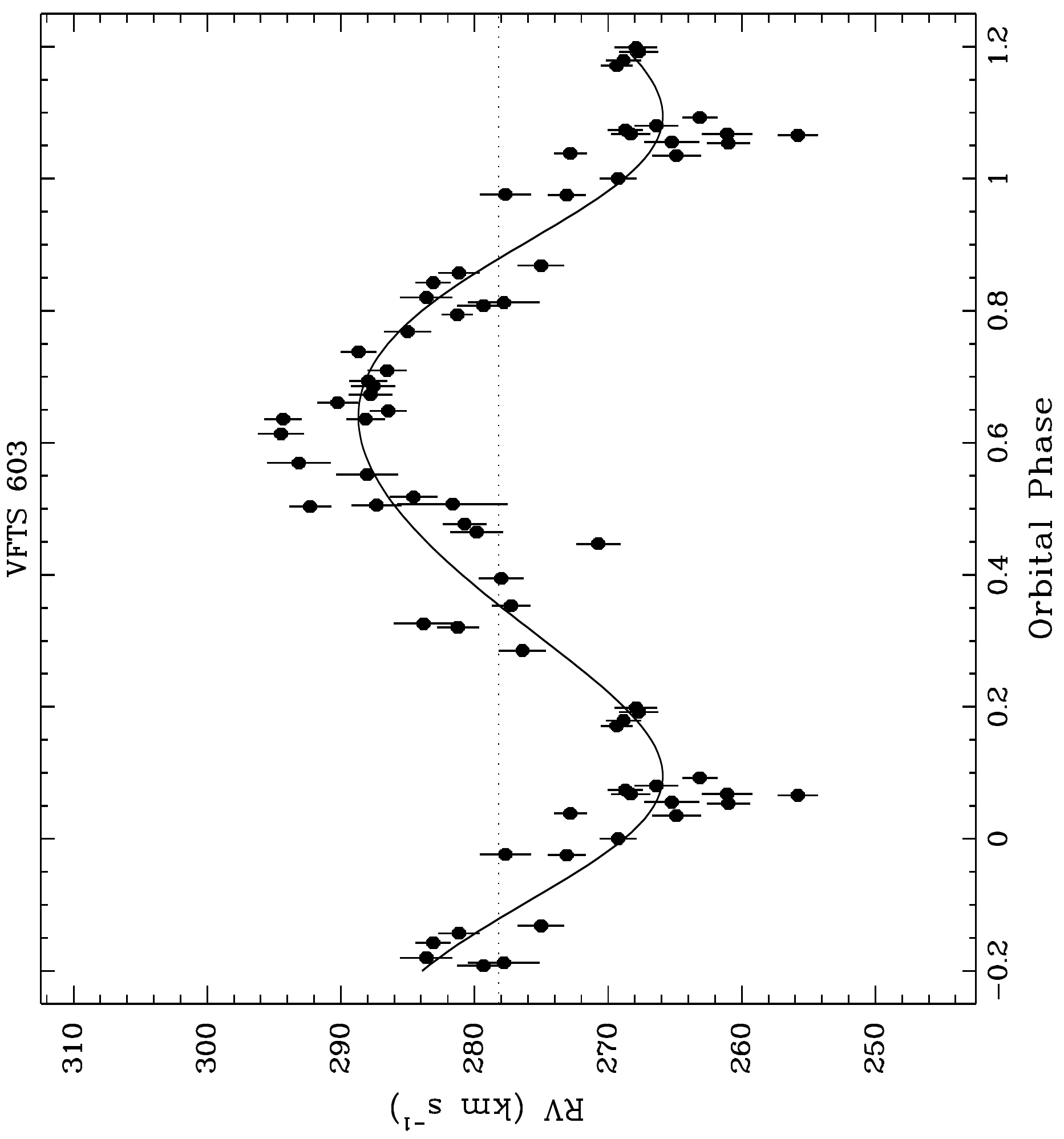}
\includegraphics[width=4.7cm,angle=-90]{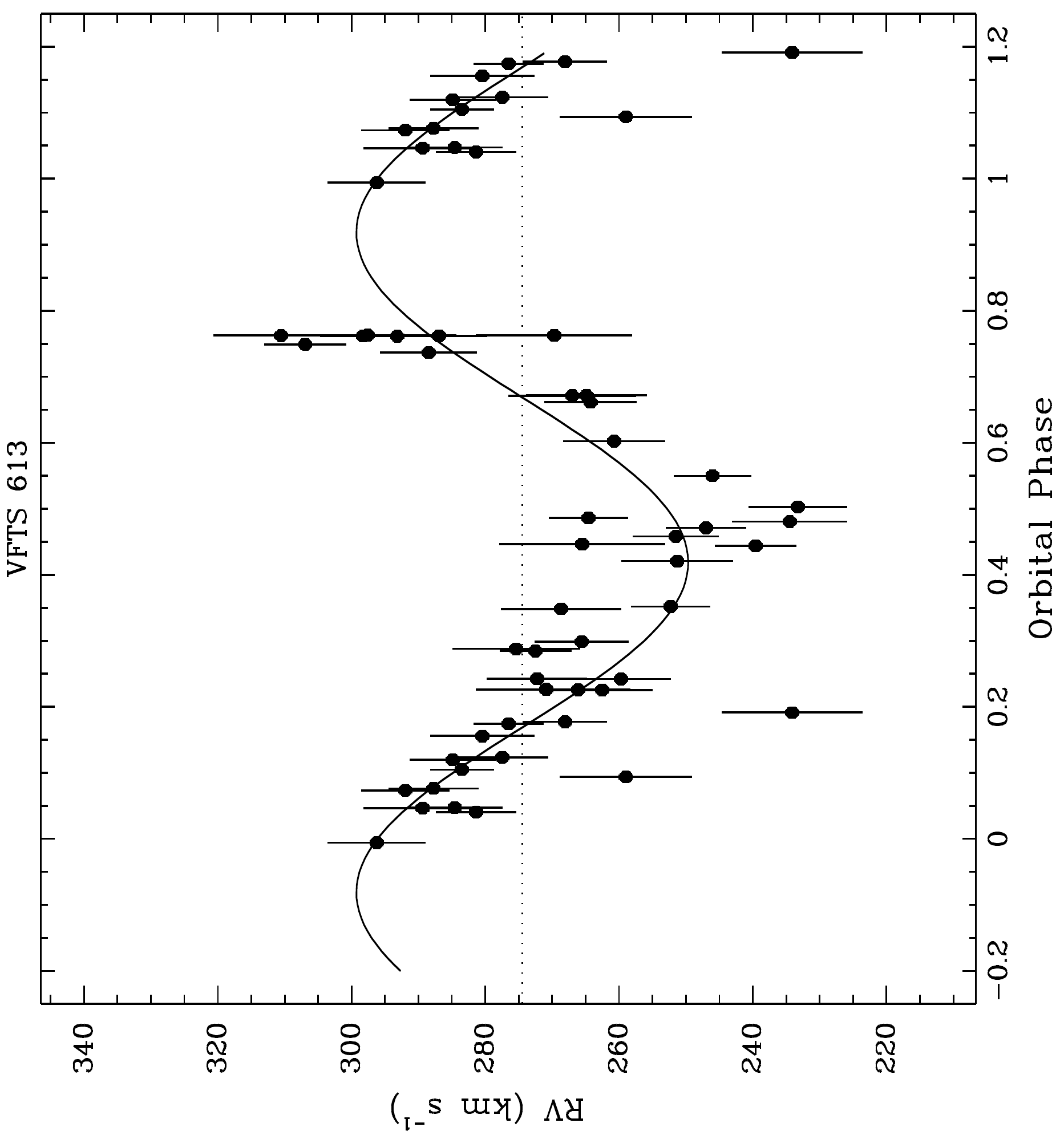}
\includegraphics[width=4.7cm,angle=-90]{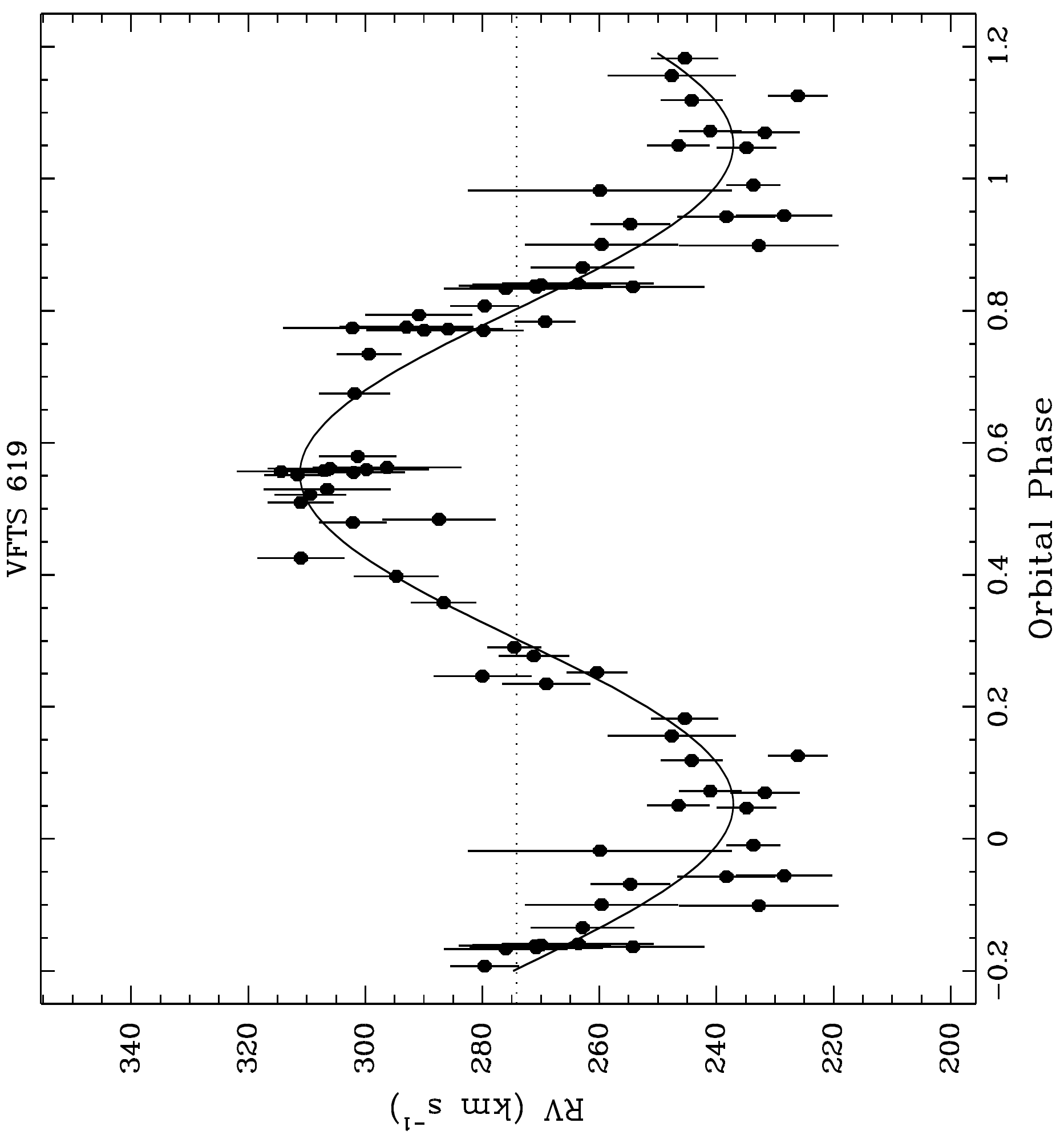}
\includegraphics[width=4.7cm,angle=-90]{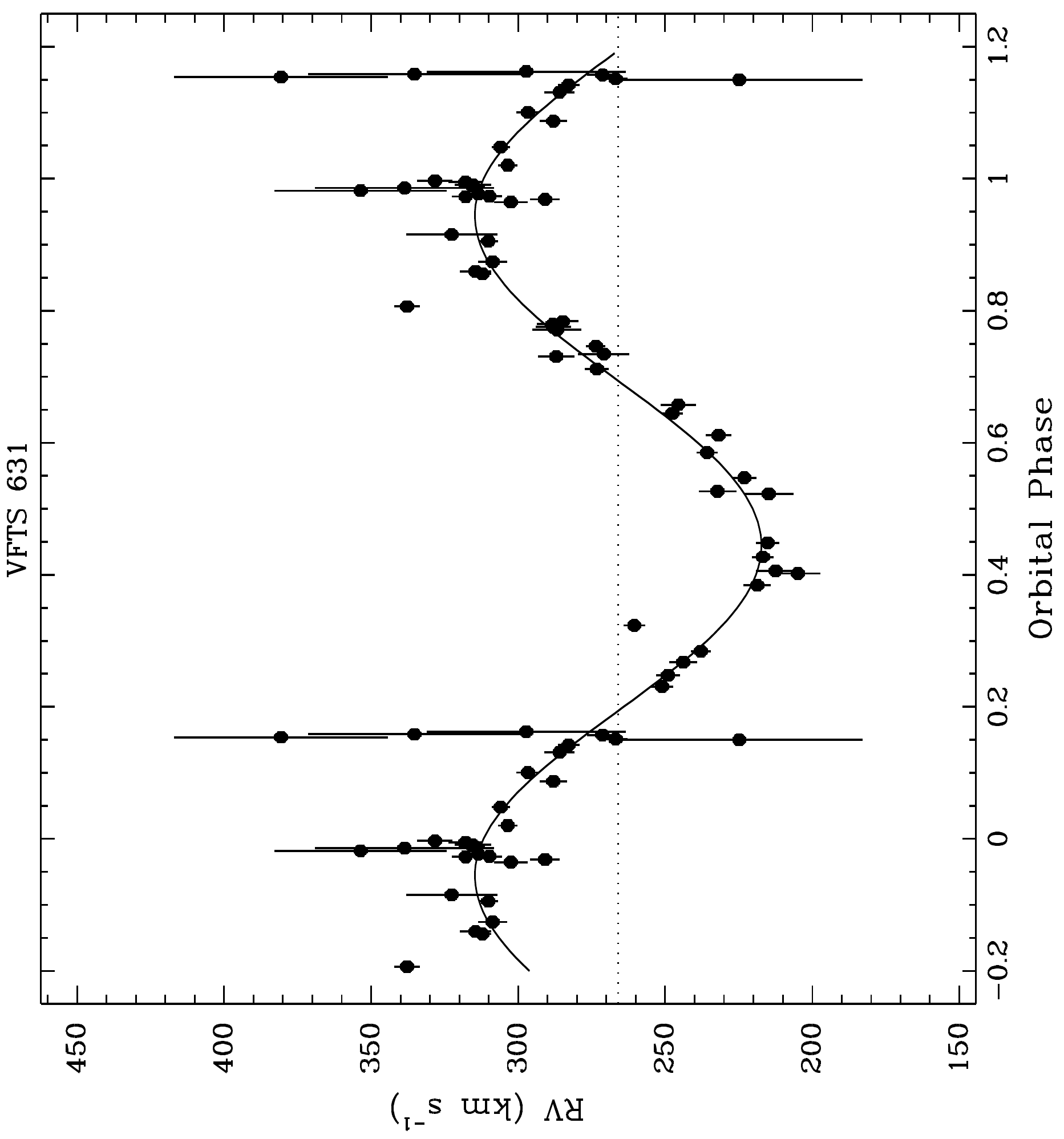}
\includegraphics[width=4.7cm,angle=-90]{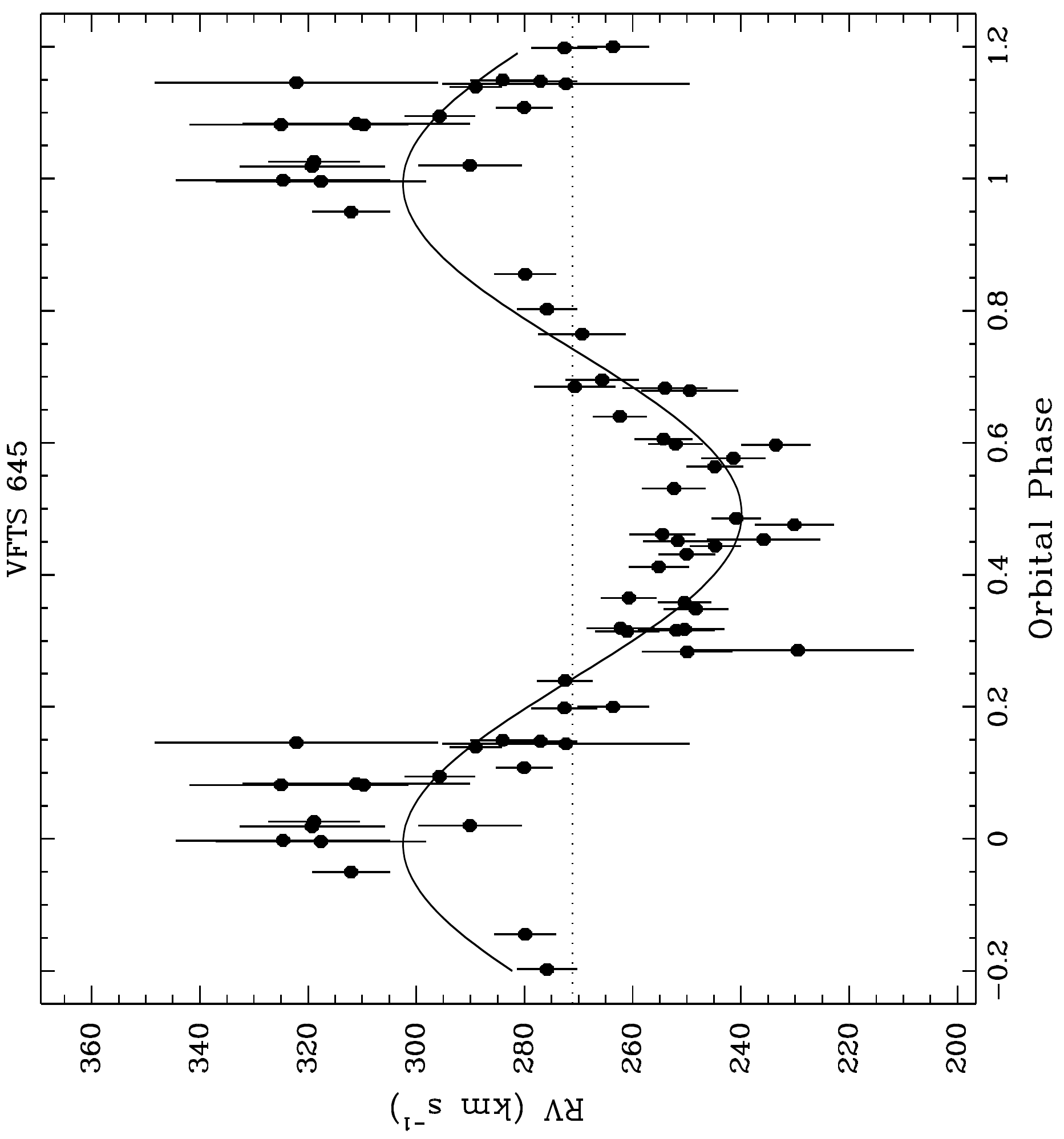}
\includegraphics[width=4.7cm,angle=-90]{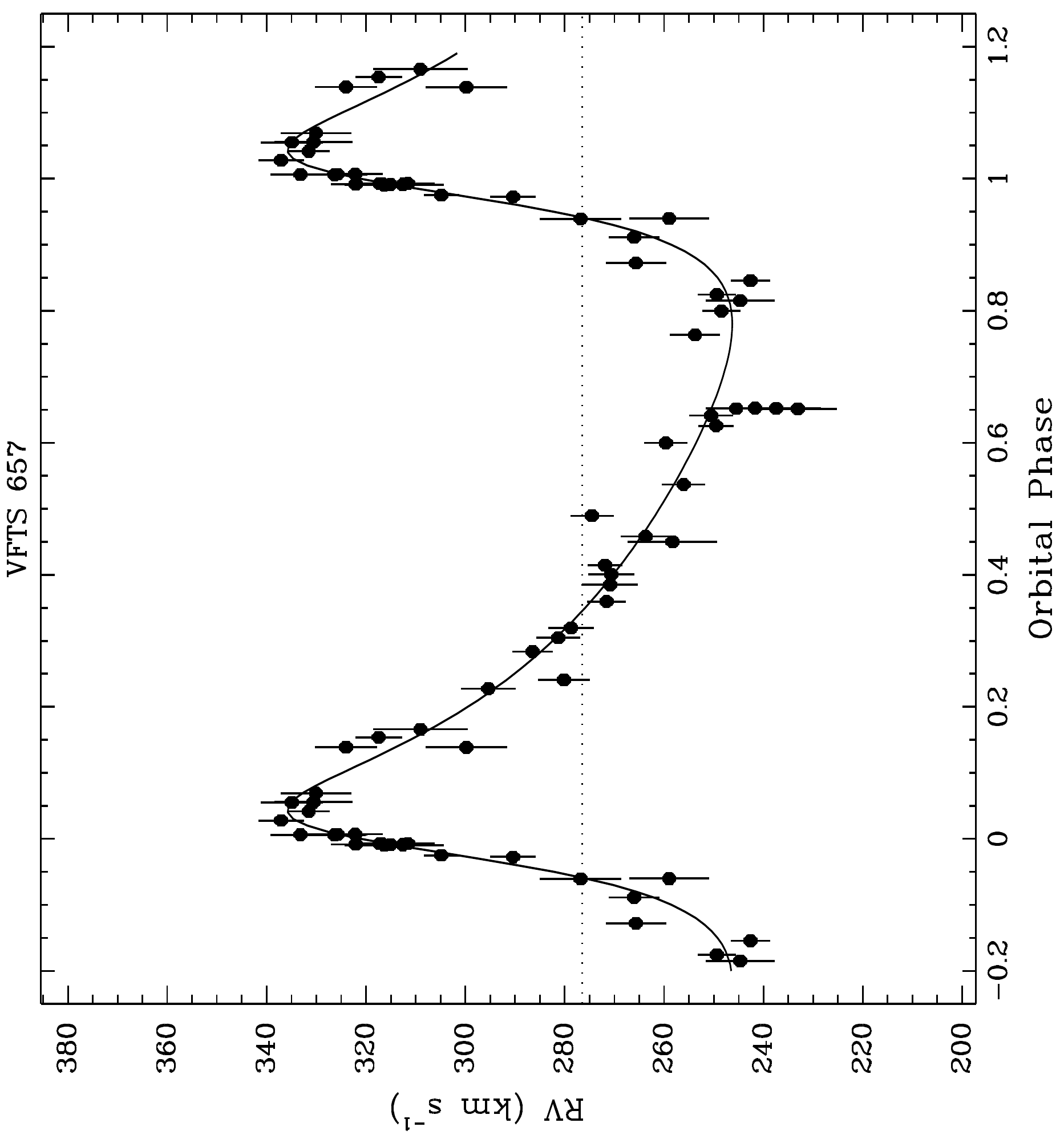}
\includegraphics[width=4.7cm,angle=-90]{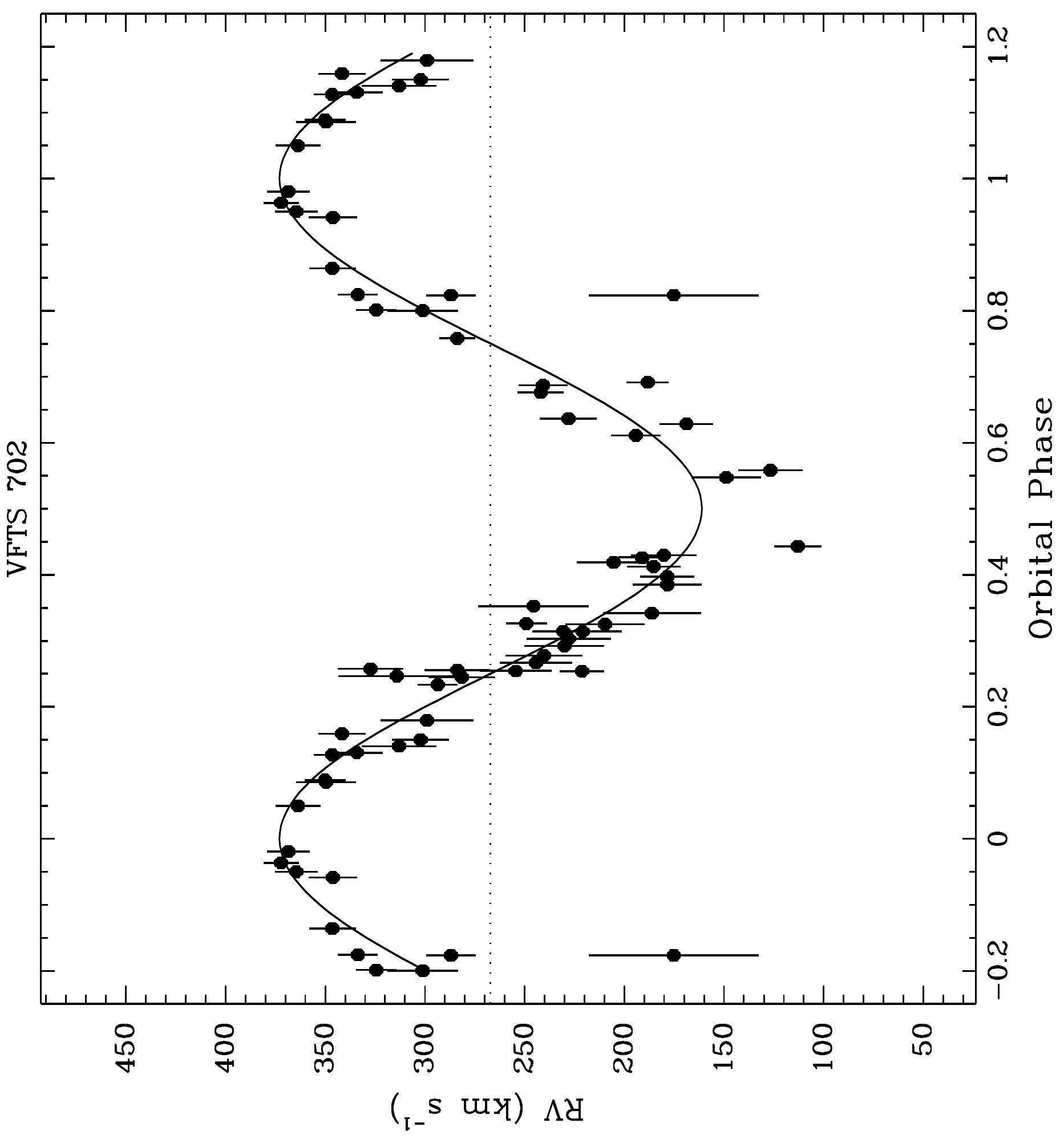}
\includegraphics[width=4.7cm,angle=-90]{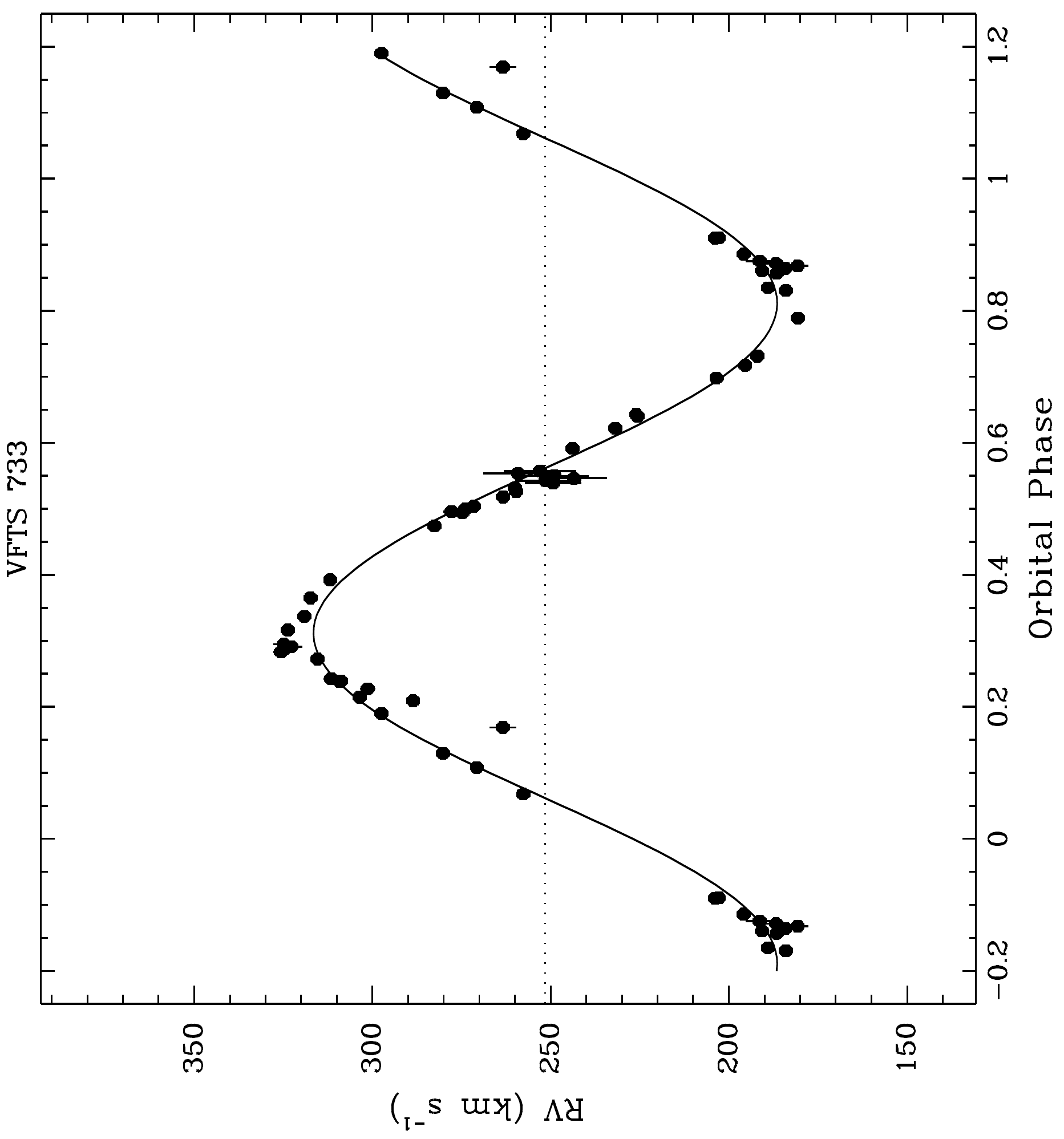}
\includegraphics[width=4.7cm,angle=-90]{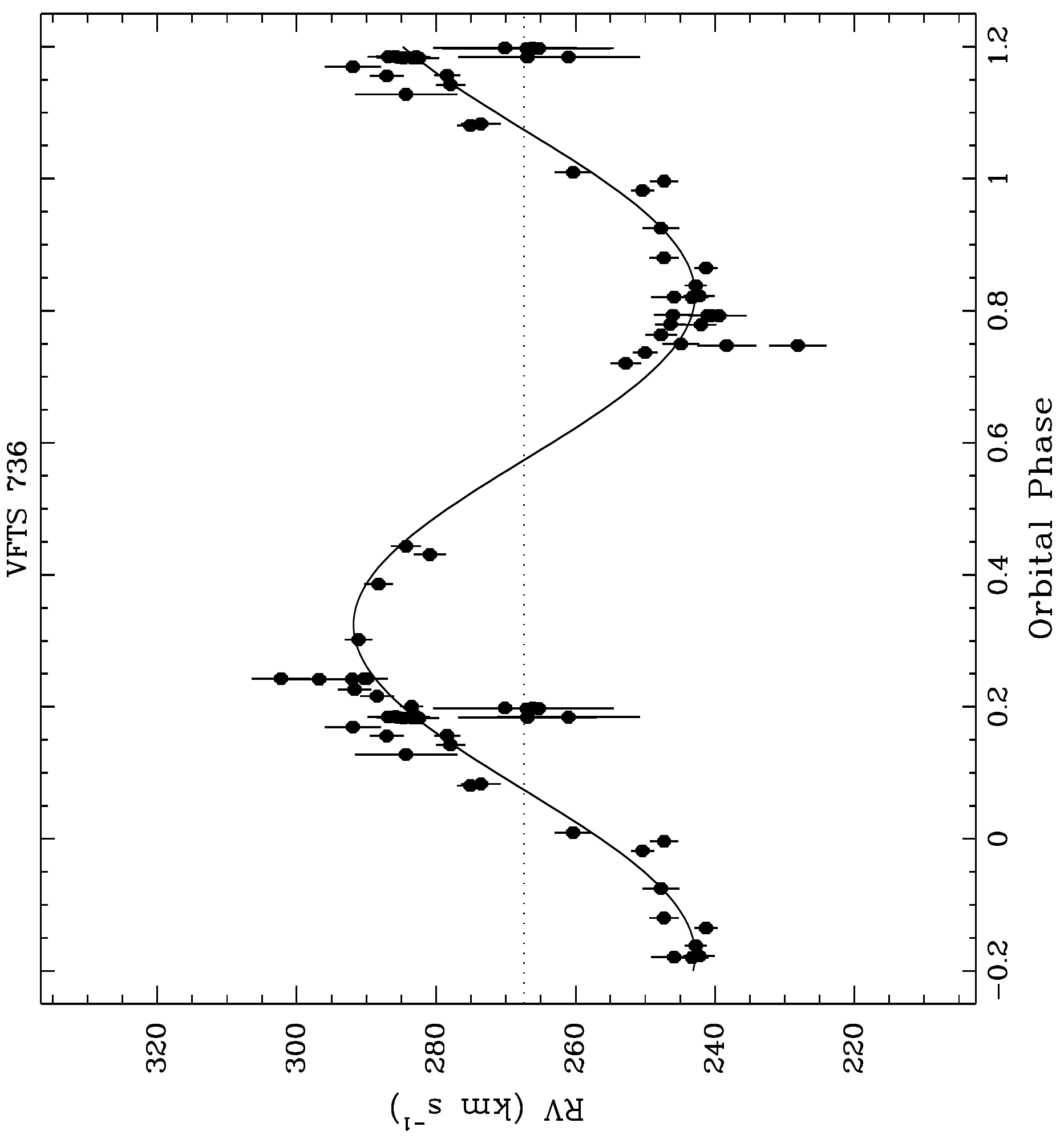}
\includegraphics[width=4.7cm,angle=-90]{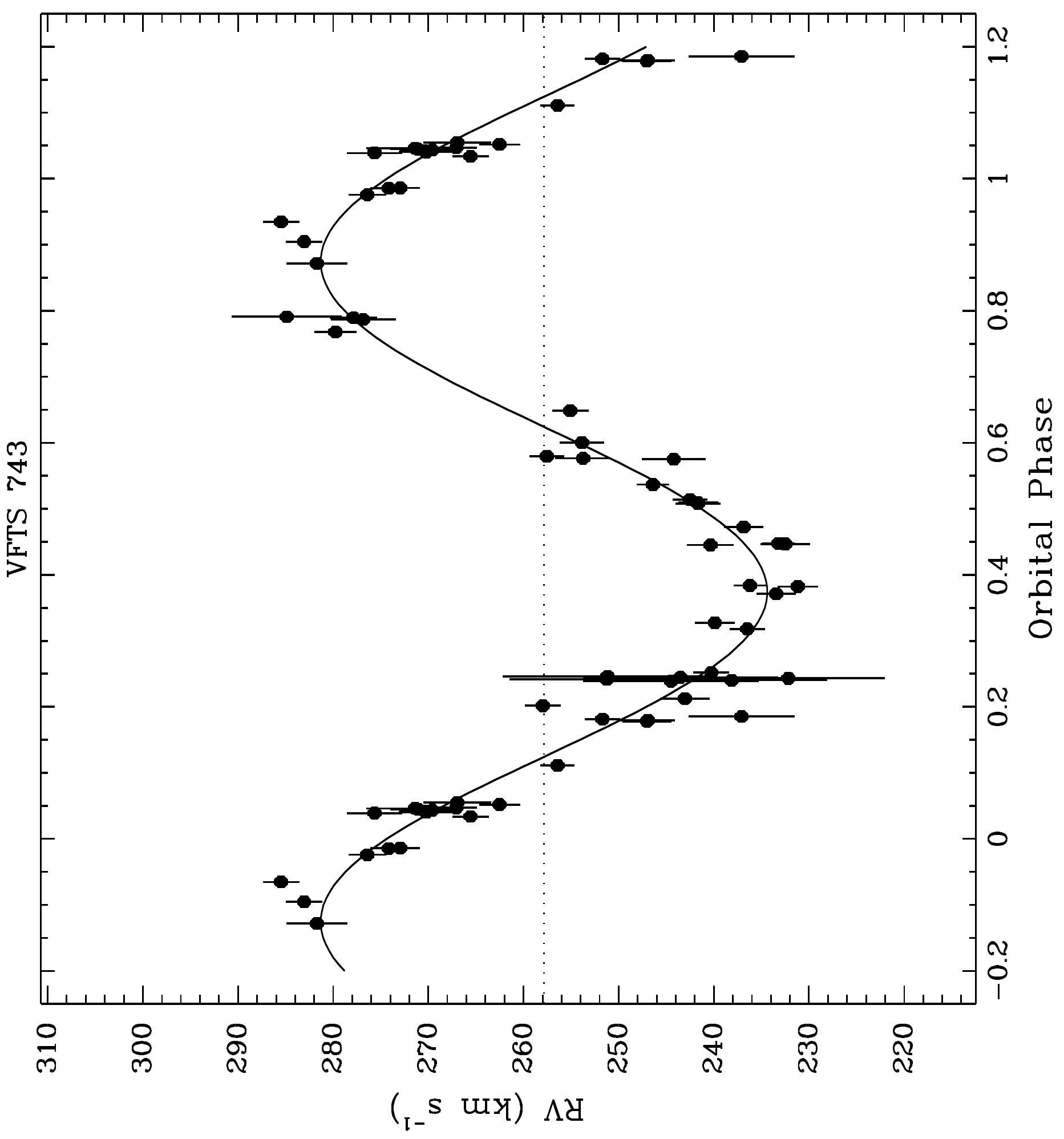}
\includegraphics[width=4.7cm,angle=-90]{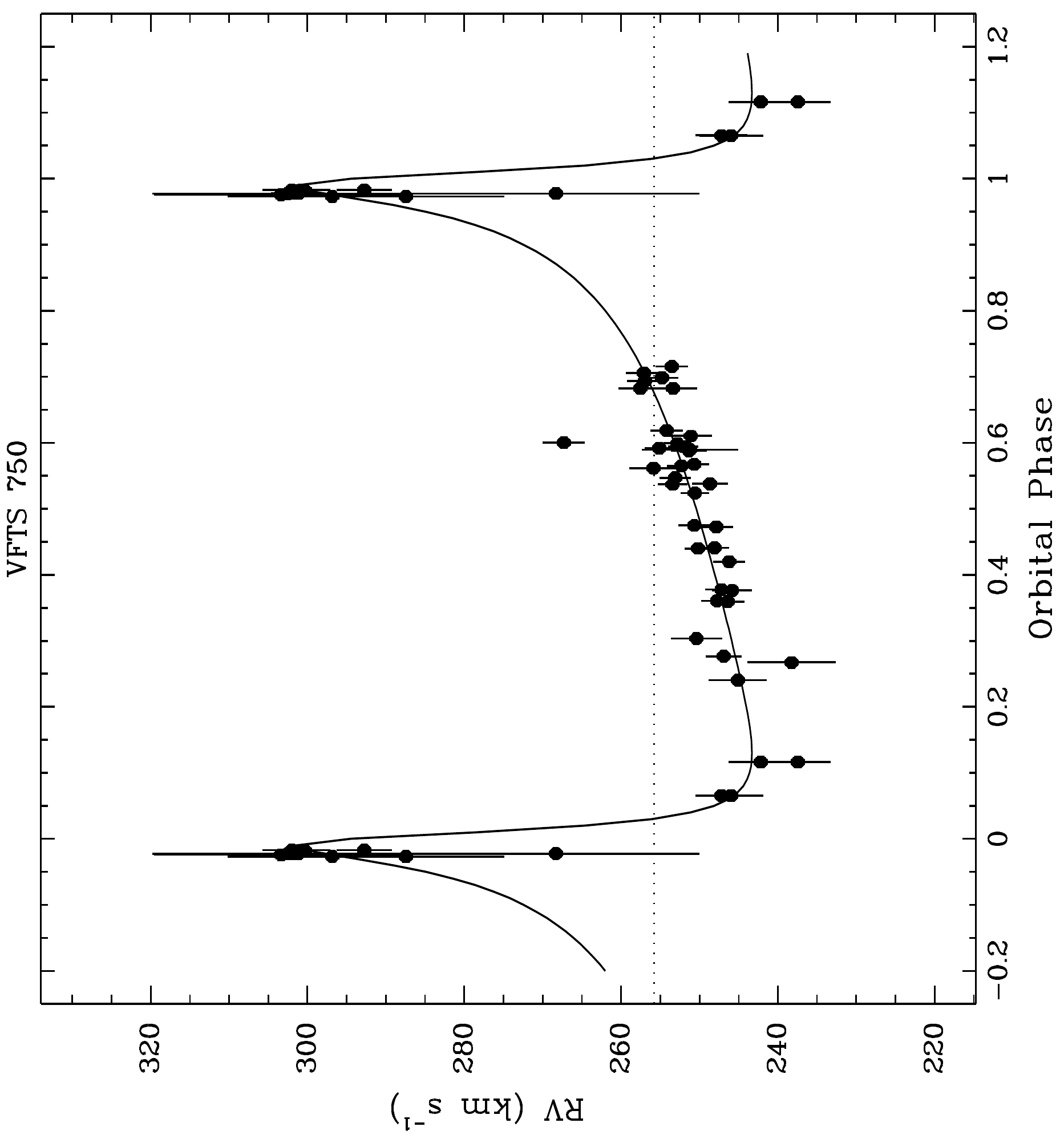}
\includegraphics[width=4.7cm,angle=-90]{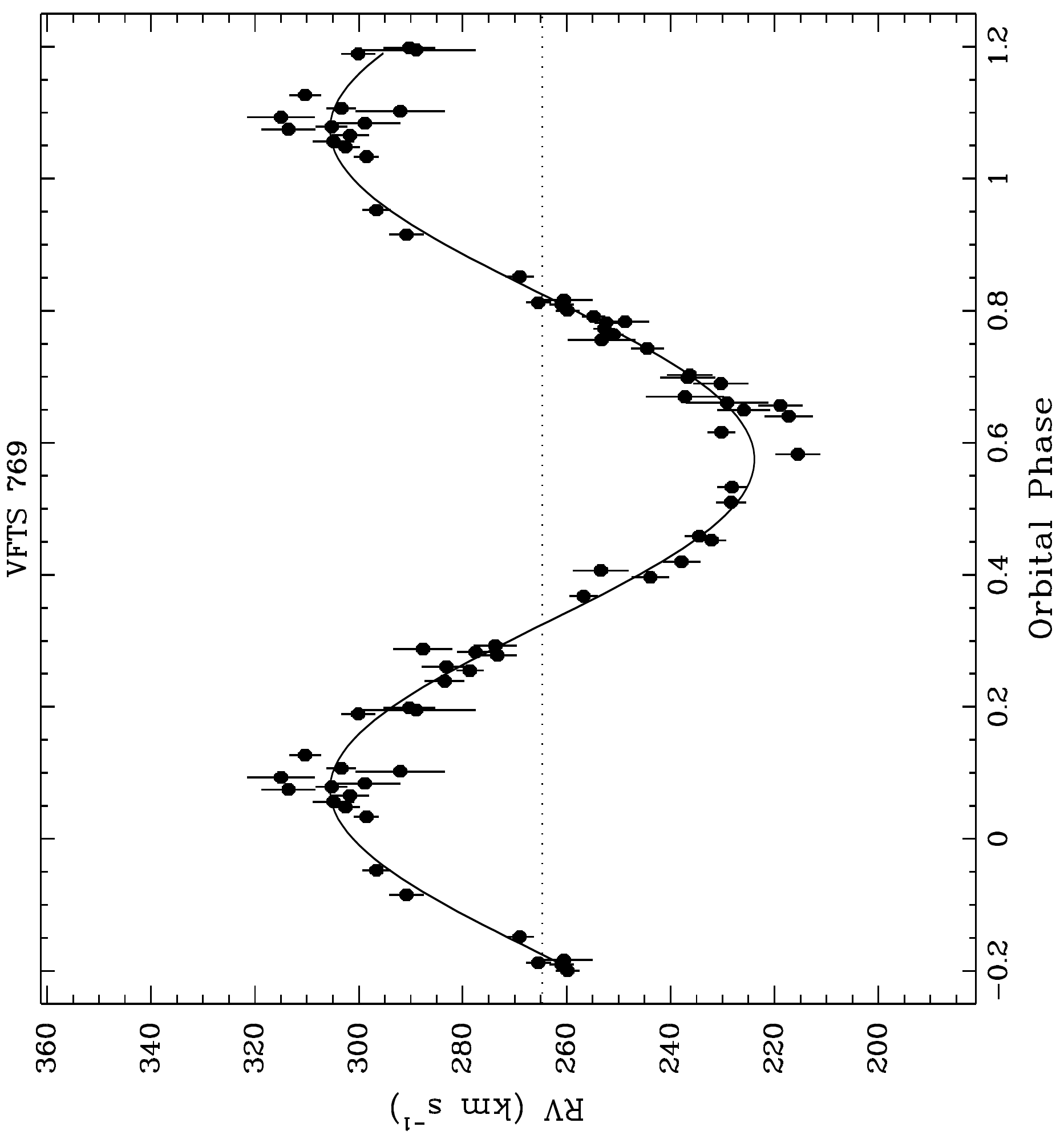}
\includegraphics[width=4.7cm,angle=-90]{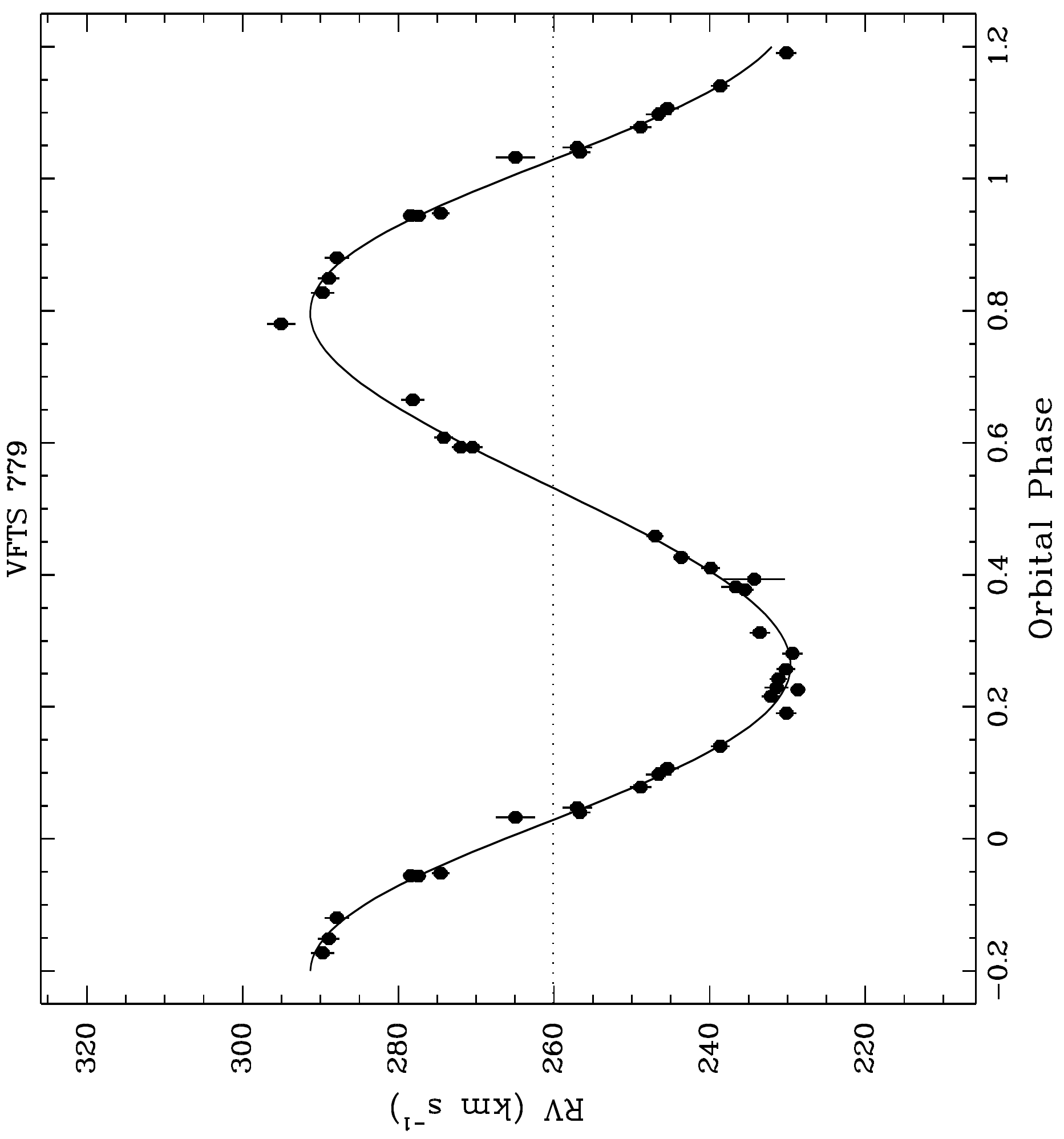}
\caption{{\it Continued...}}
\label{sb1:orb_solution3}
\end{figure*}

\begin{figure*}
\centering
\ContinuedFloat
\includegraphics[width=4.7cm,angle=-90]{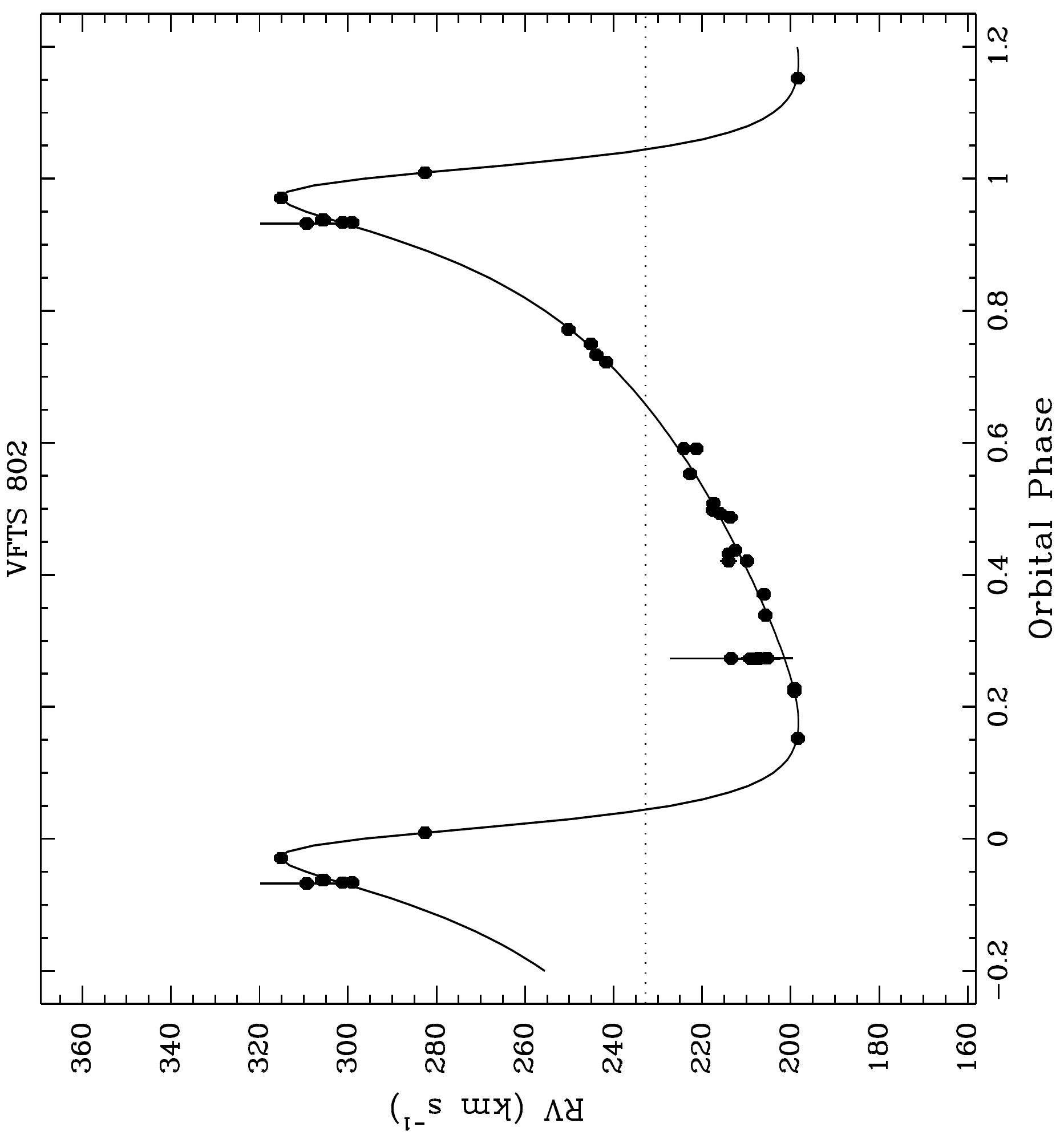}
\includegraphics[width=4.7cm,angle=-90]{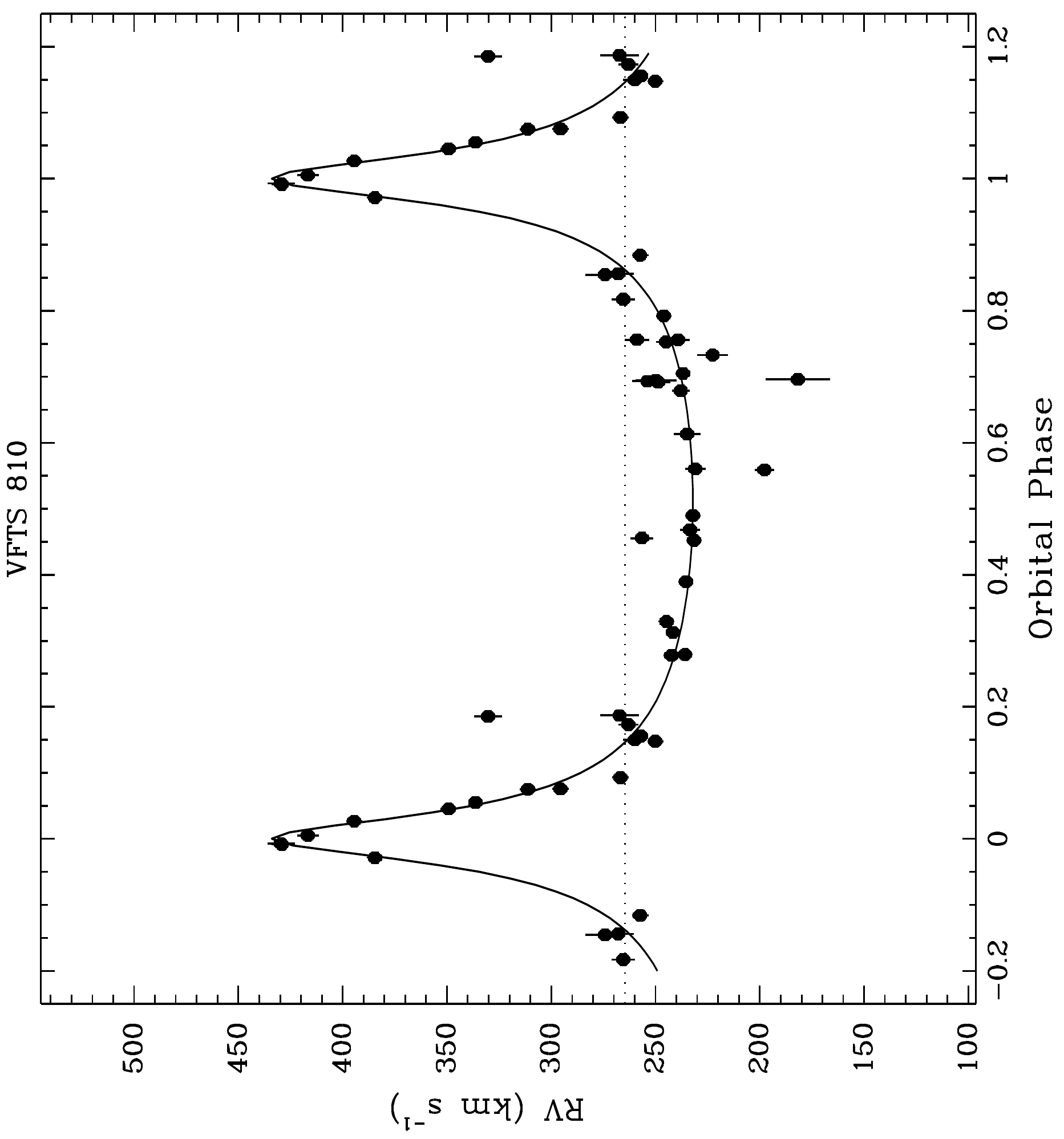}
\includegraphics[width=4.7cm,angle=-90]{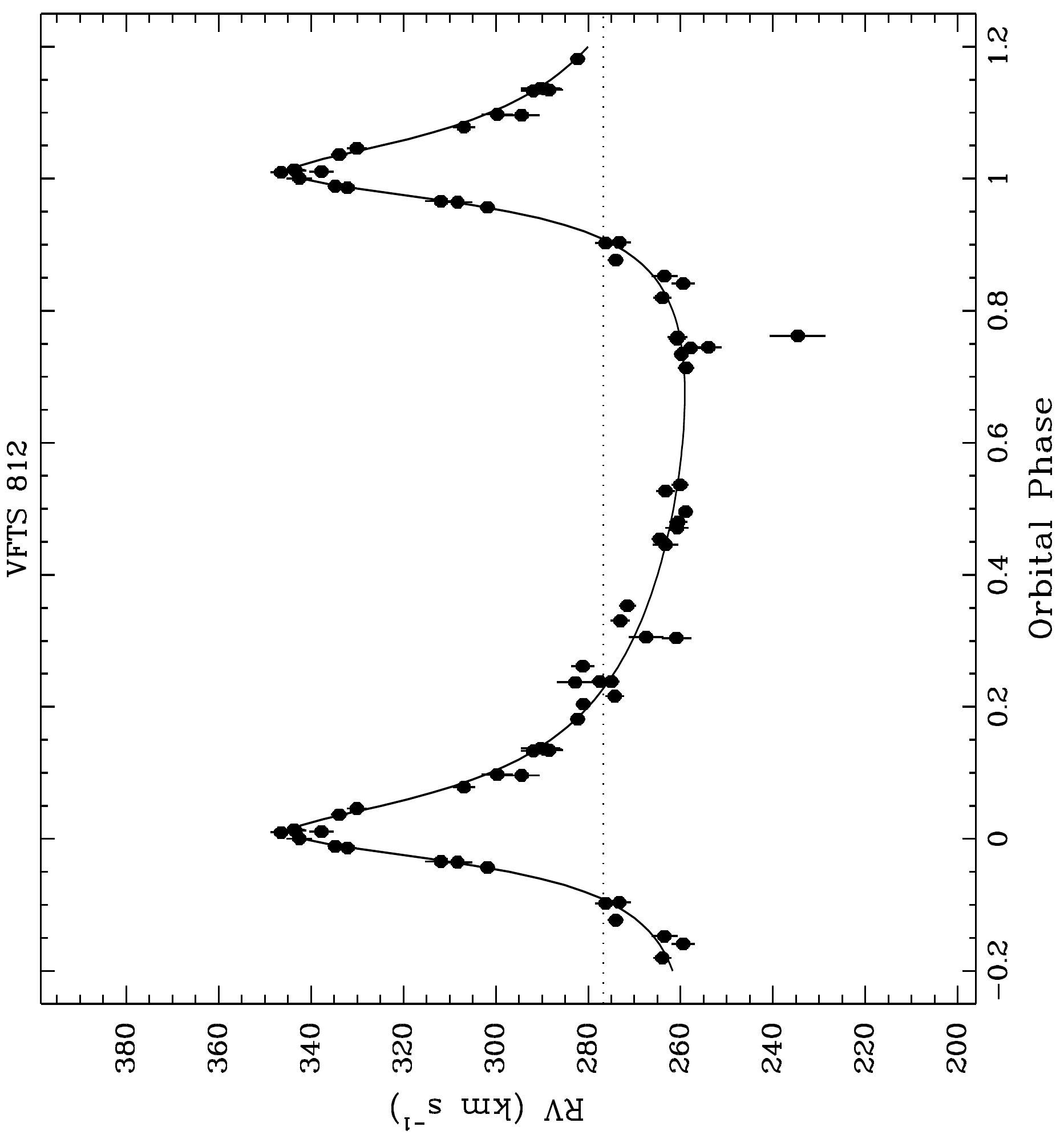}
\includegraphics[width=4.7cm,angle=-90]{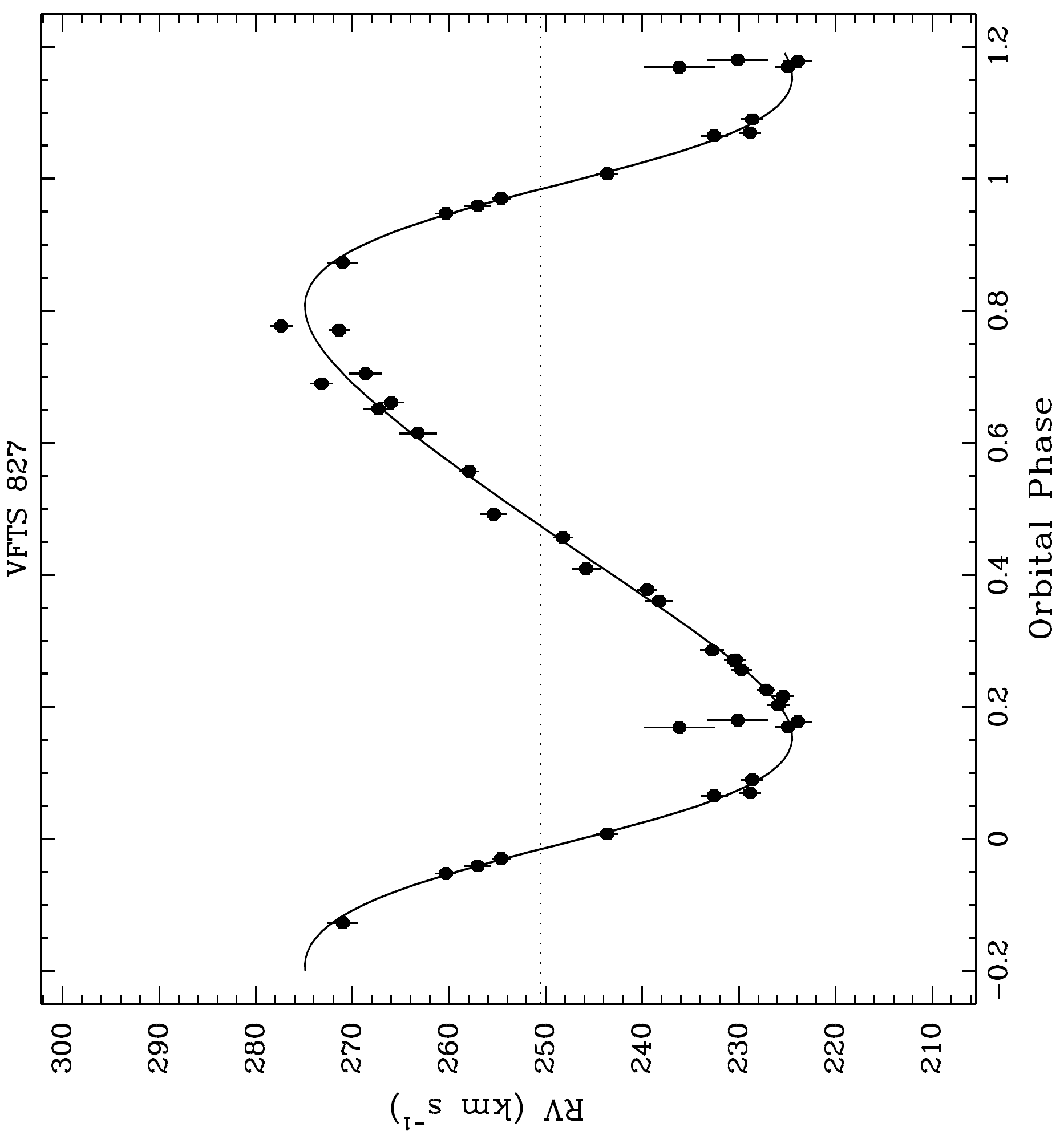}
\includegraphics[width=4.7cm,angle=-90]{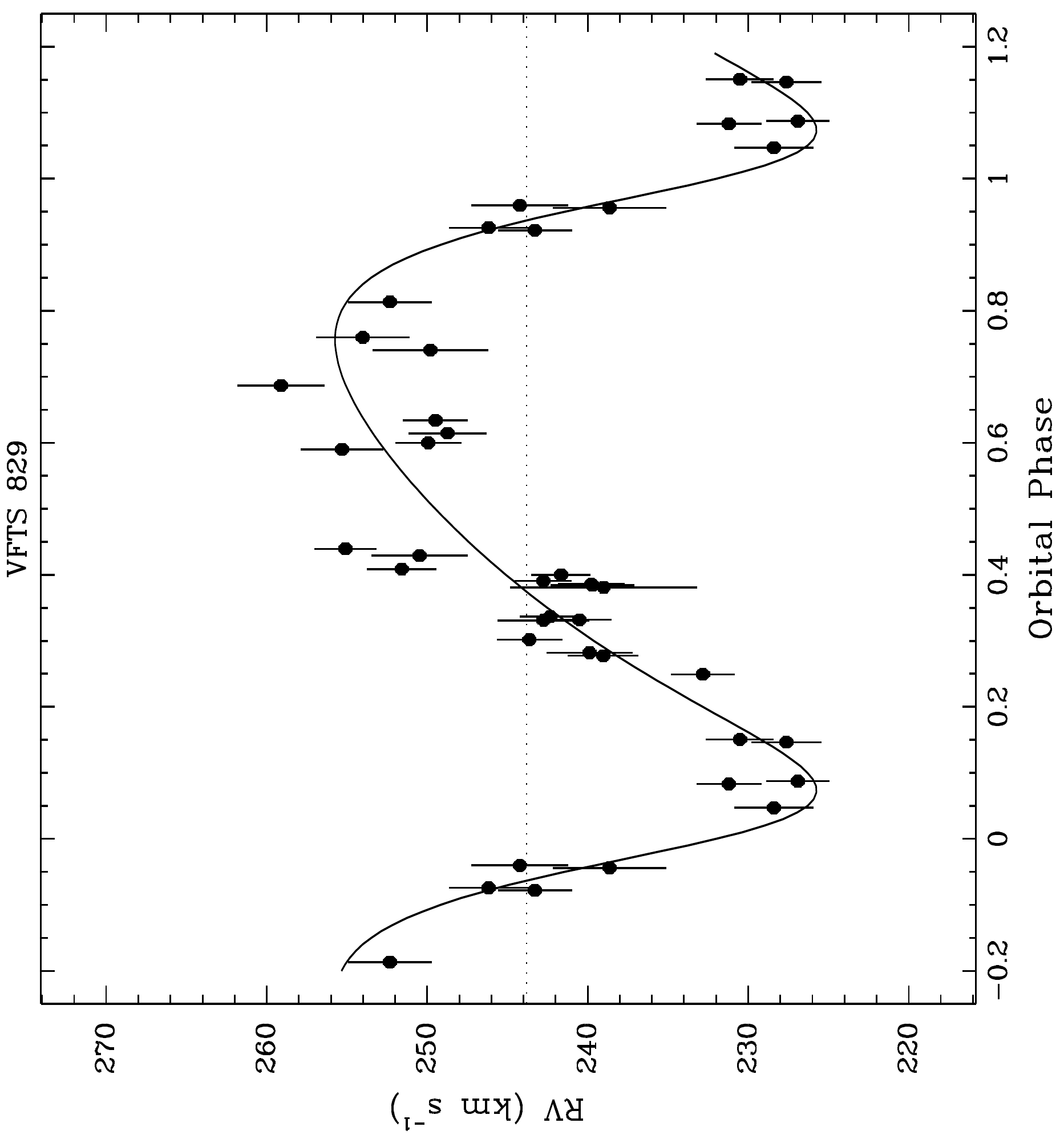}
\includegraphics[width=4.7cm,angle=-90]{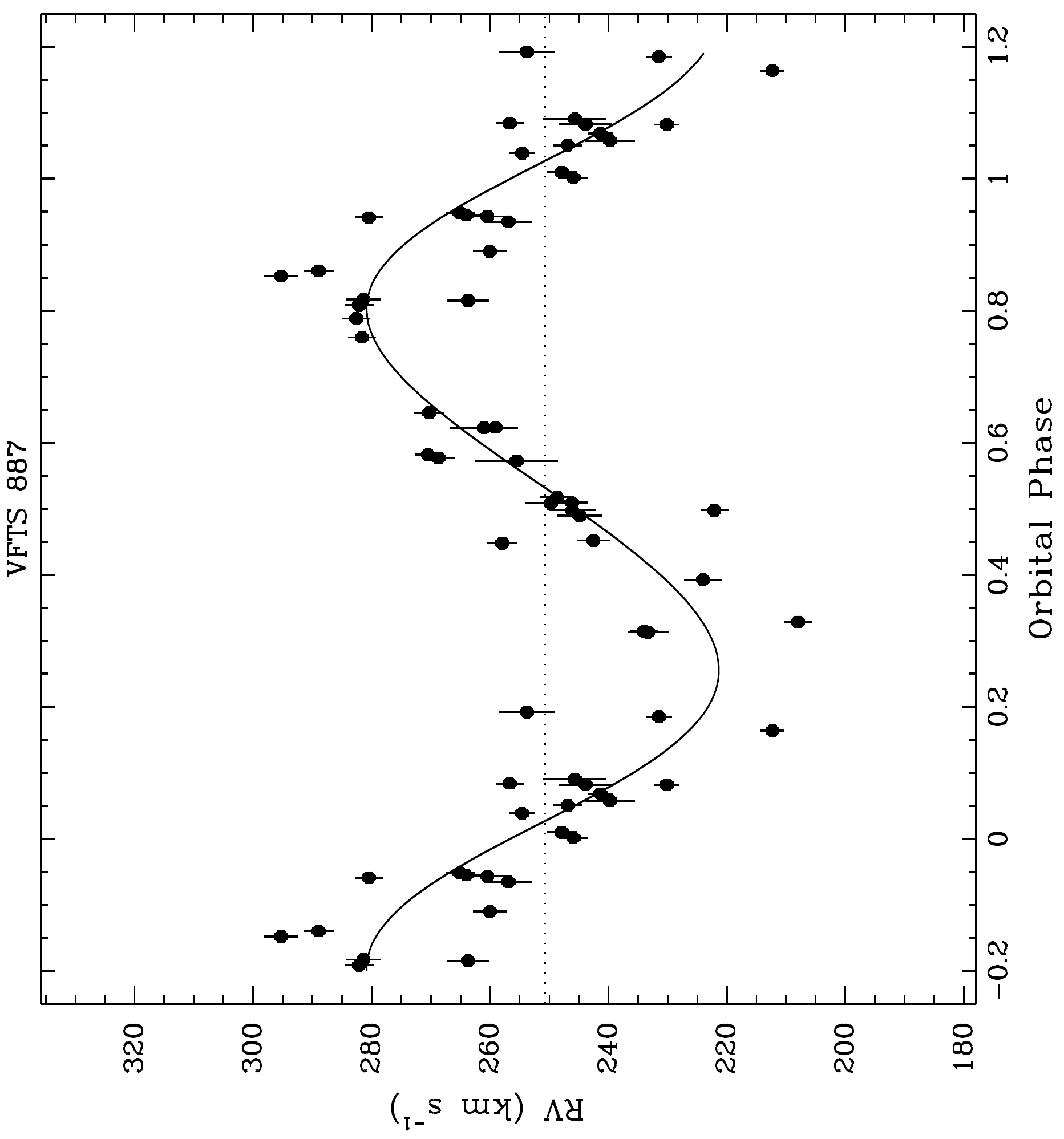}
\caption{{\it Continued...}}
\label{sb1:orb_solution4}
\end{figure*}

\begin{figure*}
\centering
\includegraphics[width=4.7cm,angle=-90]{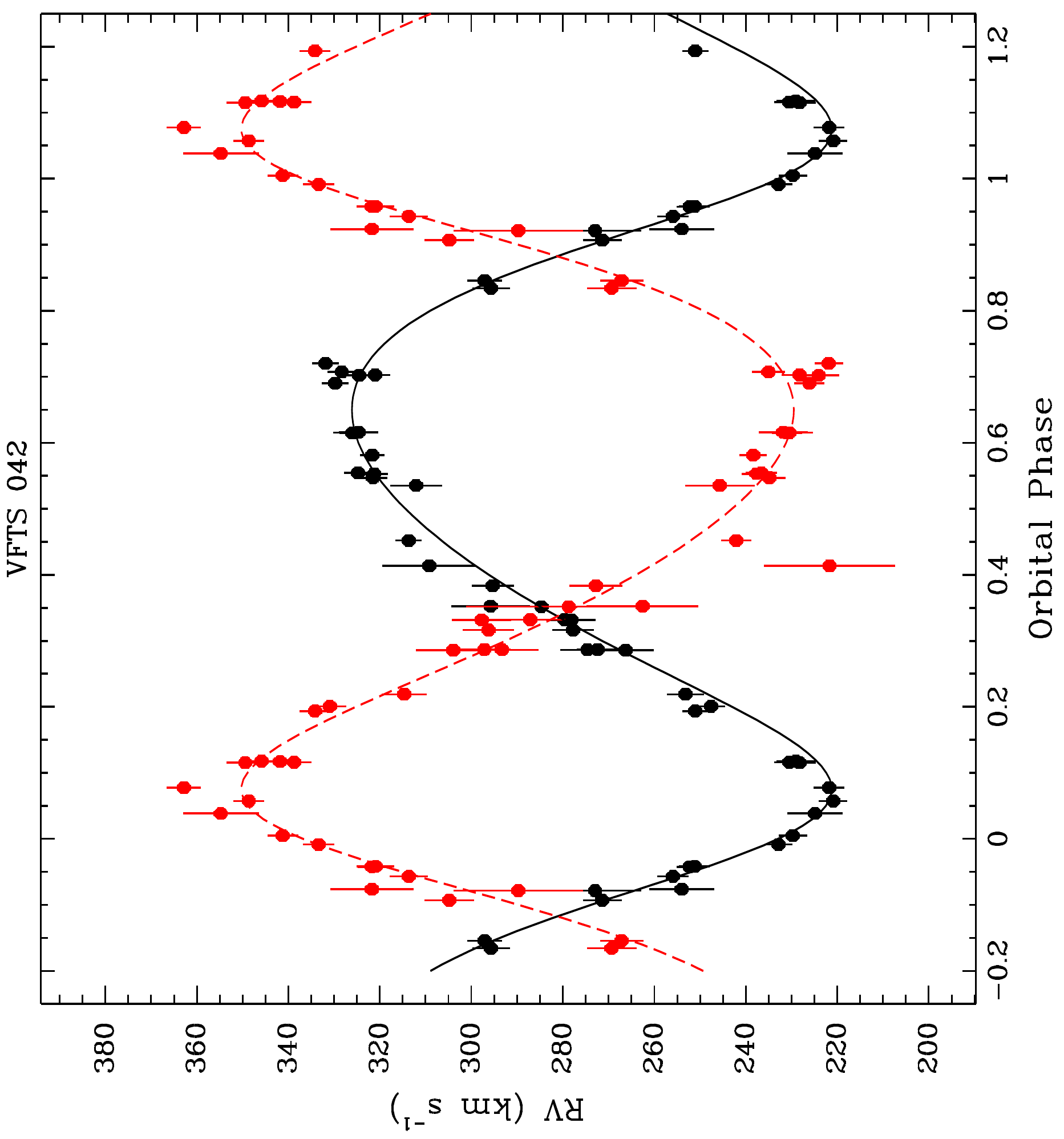}
\includegraphics[width=4.7cm,angle=-90]{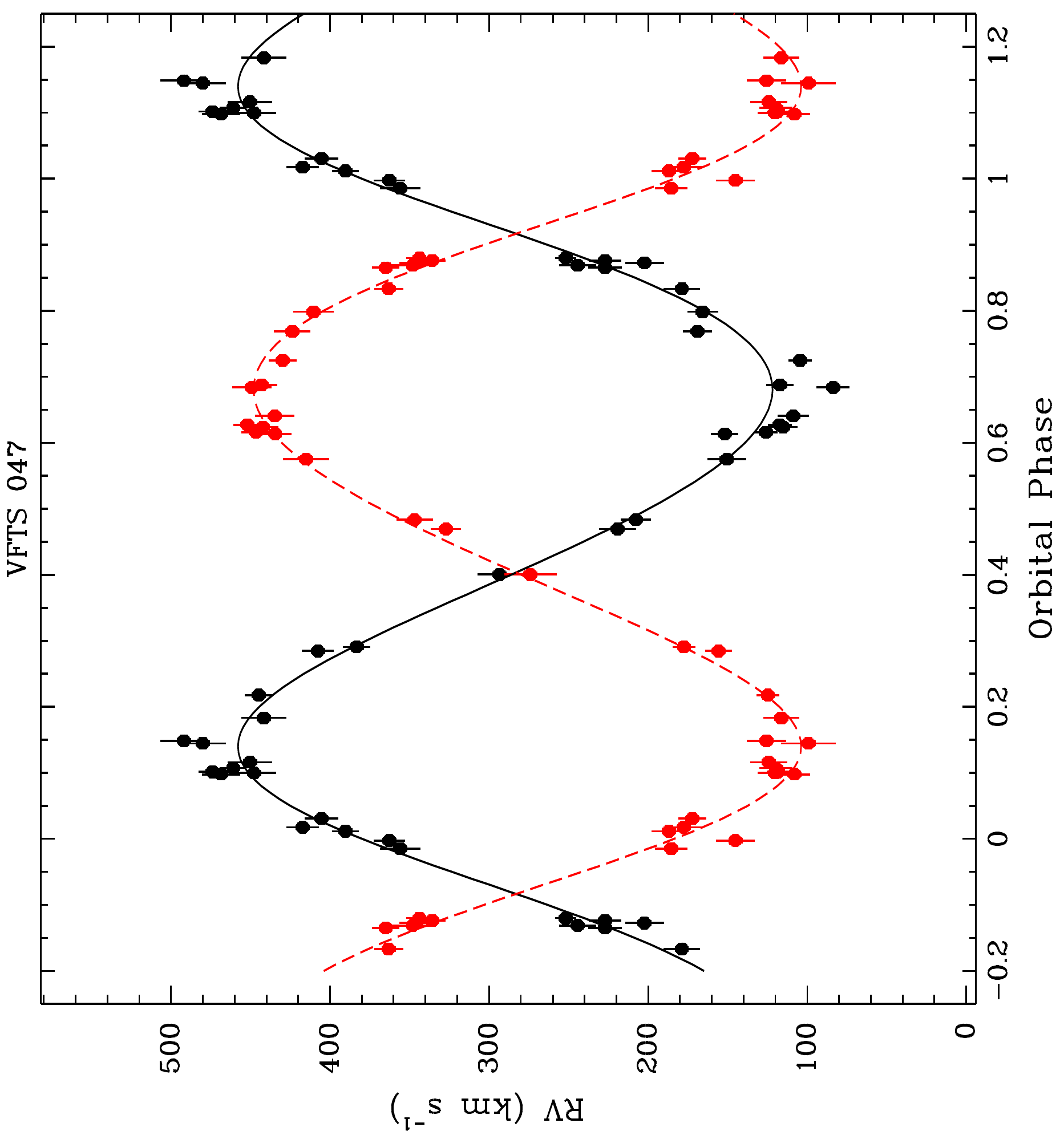}
\includegraphics[width=4.7cm,angle=-90]{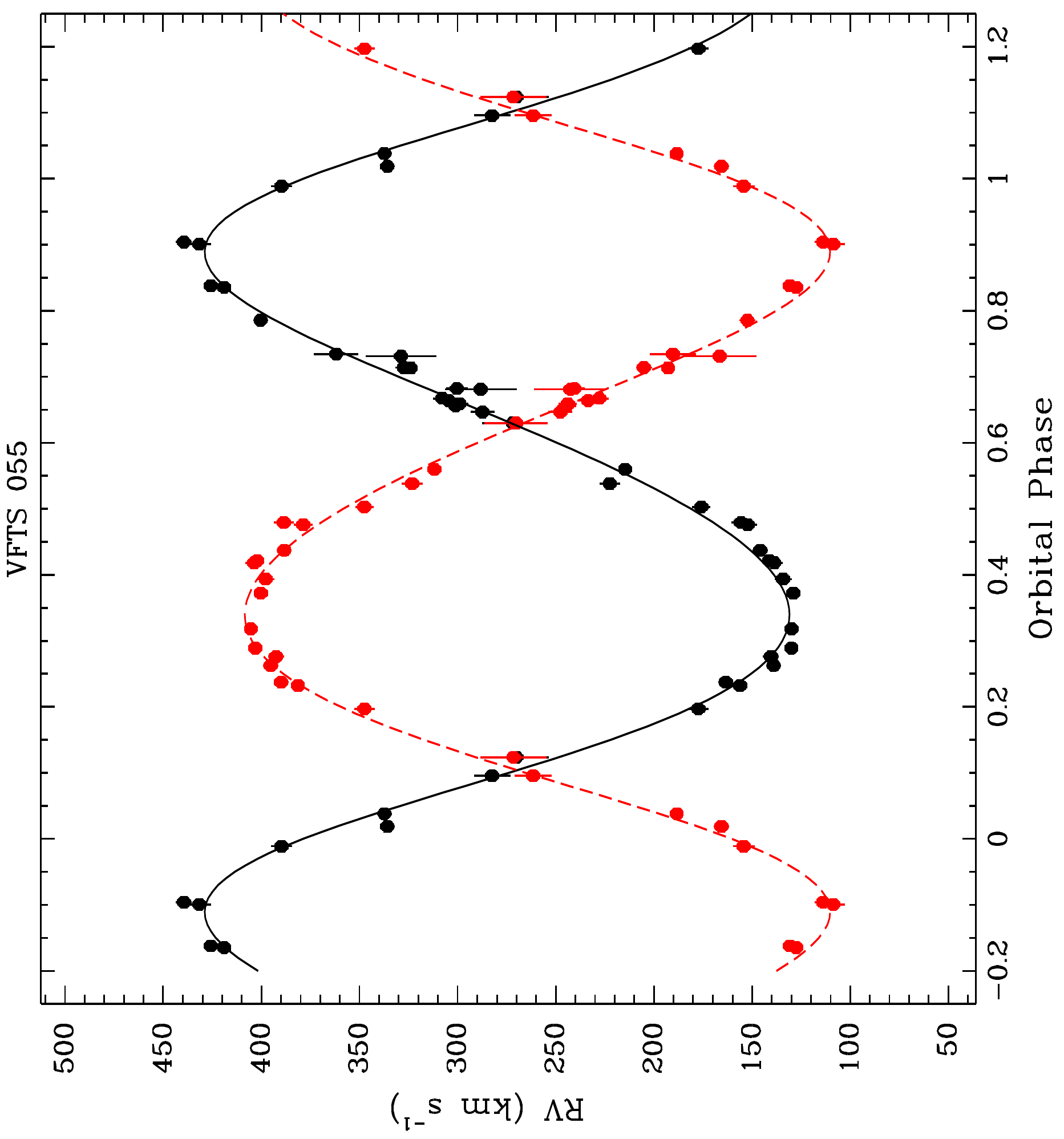}
\includegraphics[width=4.7cm,angle=-90]{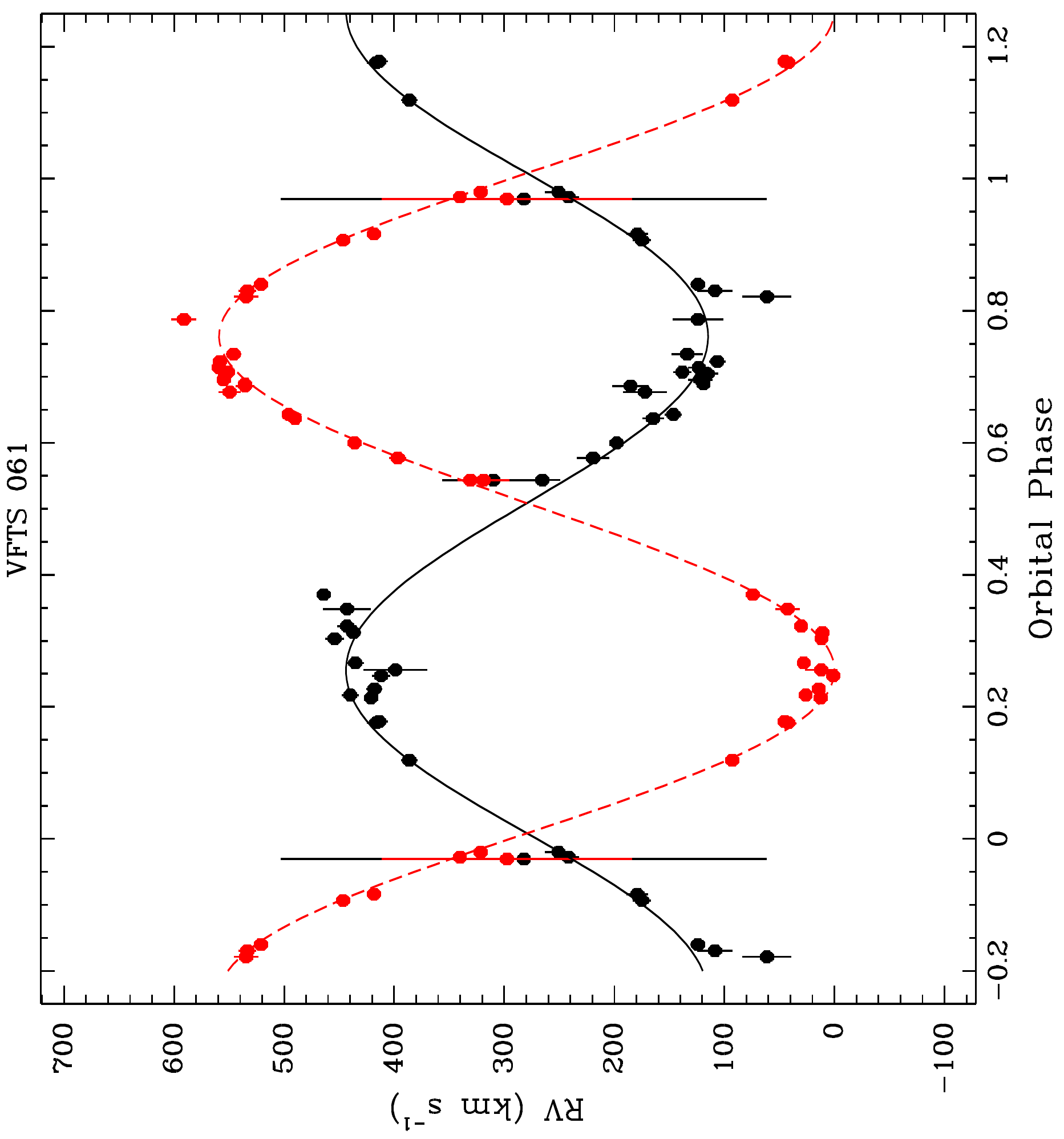}
\includegraphics[width=4.7cm,angle=-90]{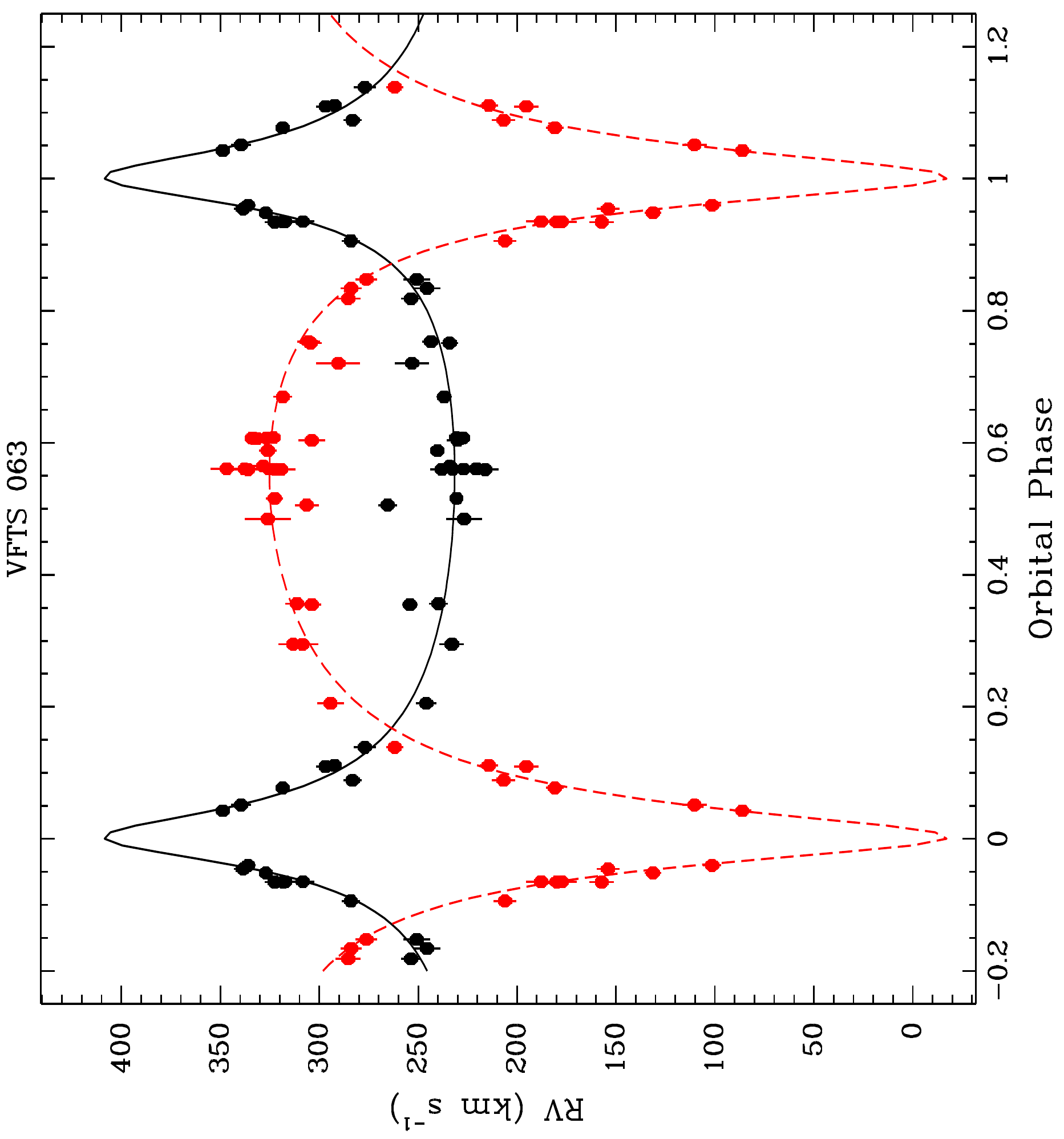}
\includegraphics[width=4.7cm,angle=-90]{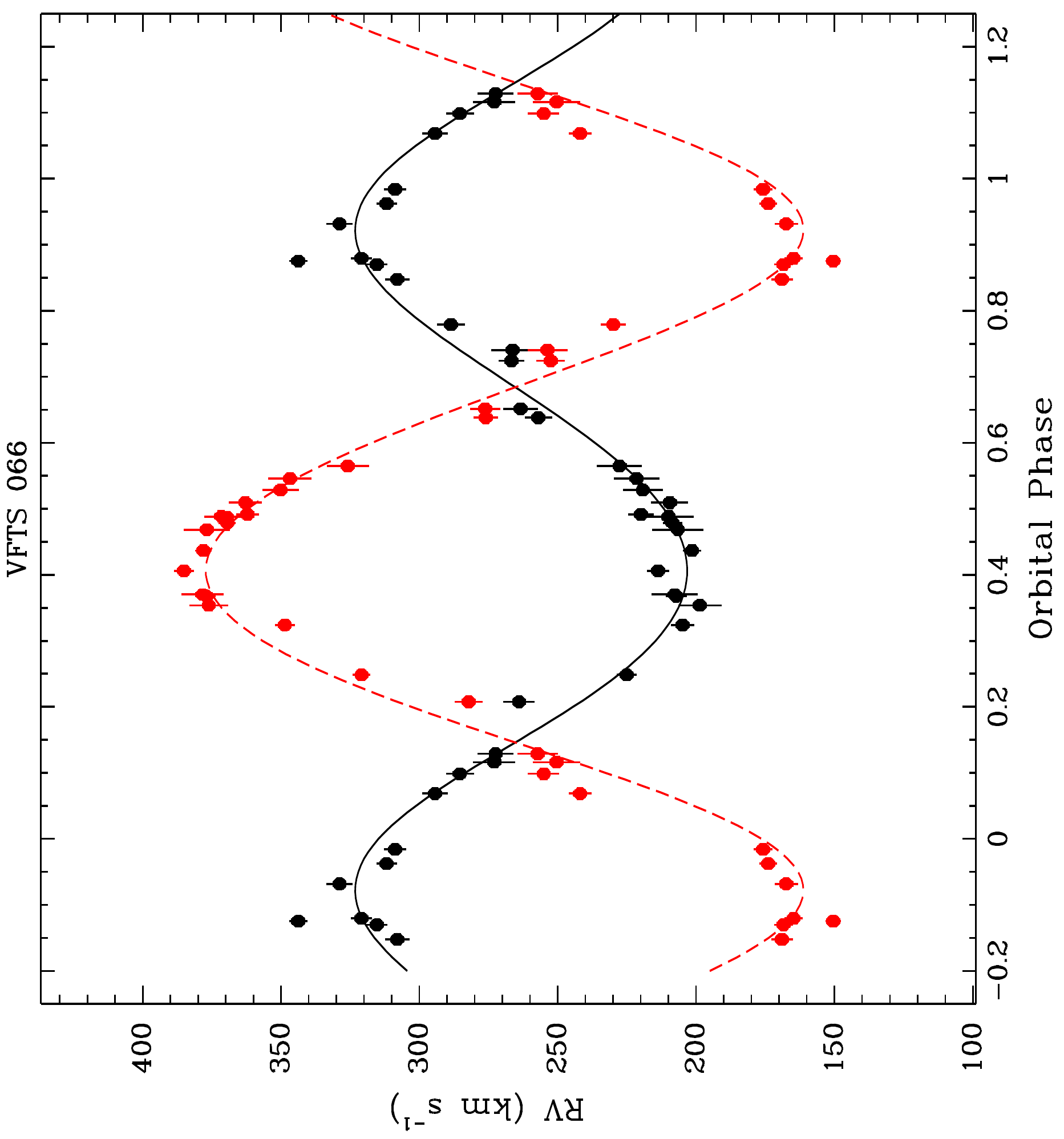}
\includegraphics[width=4.7cm,angle=-90]{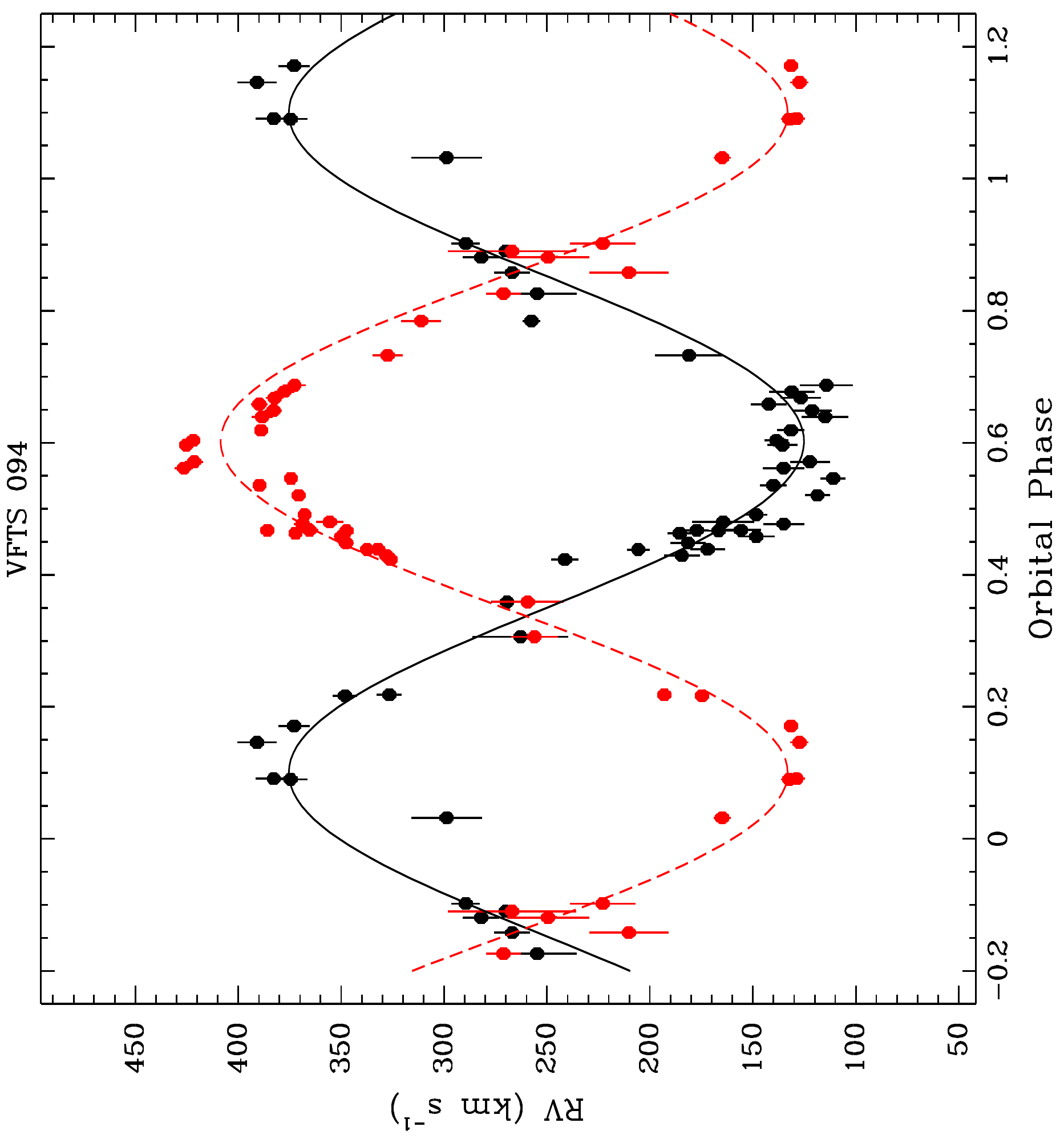}
\includegraphics[width=4.7cm,angle=-90]{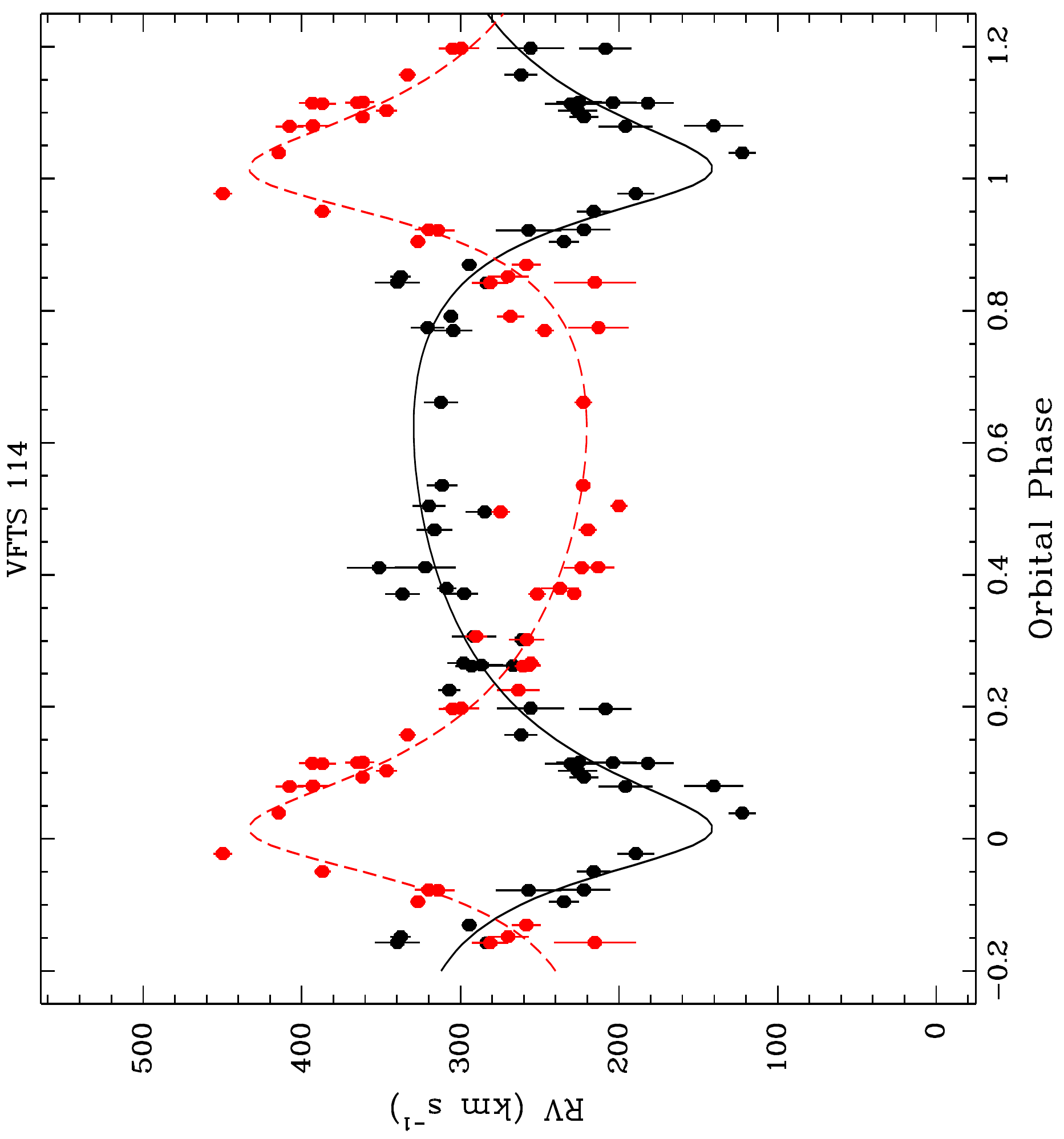}
\includegraphics[width=4.7cm,angle=-90]{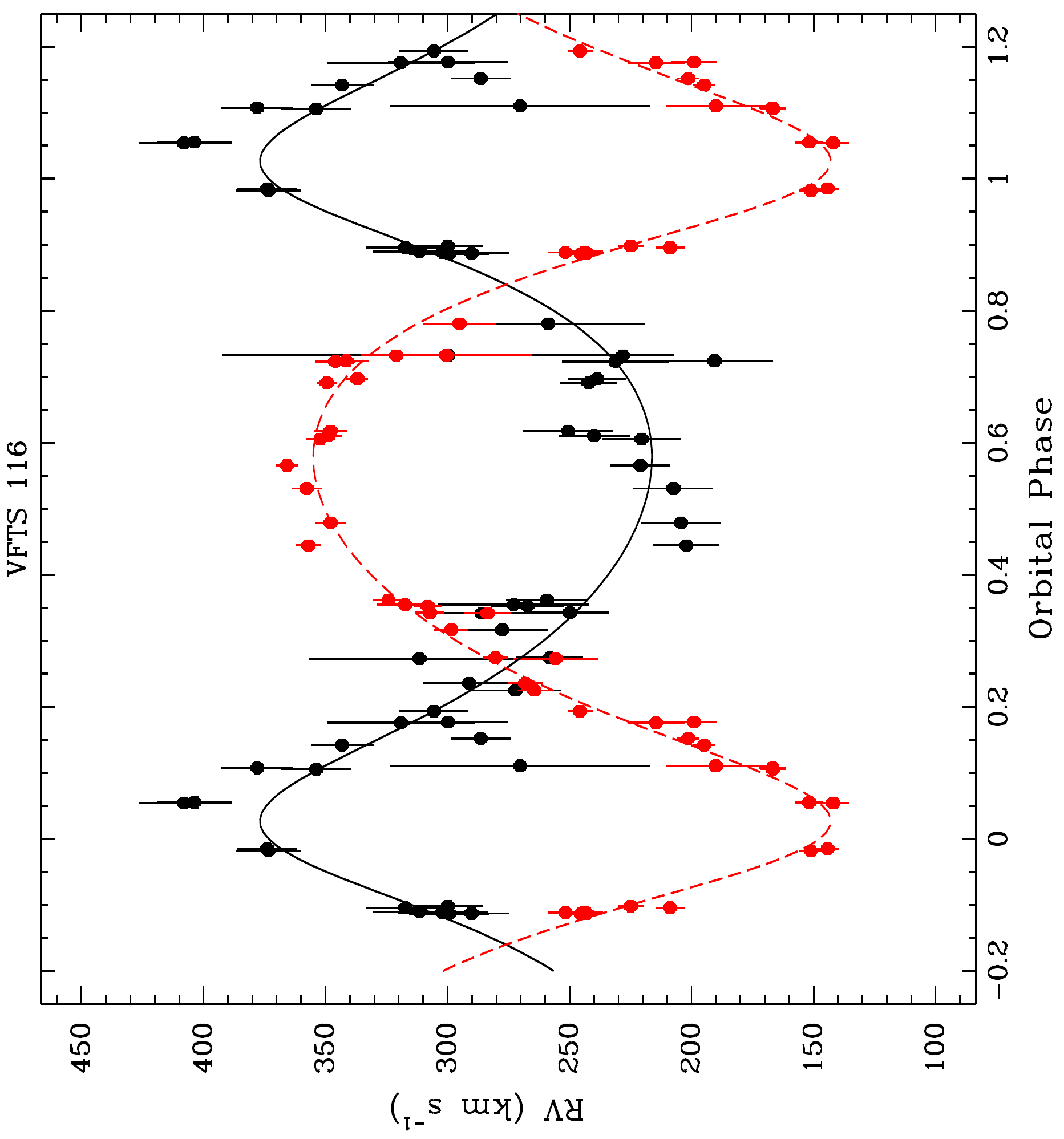}
\includegraphics[width=4.7cm,angle=-90]{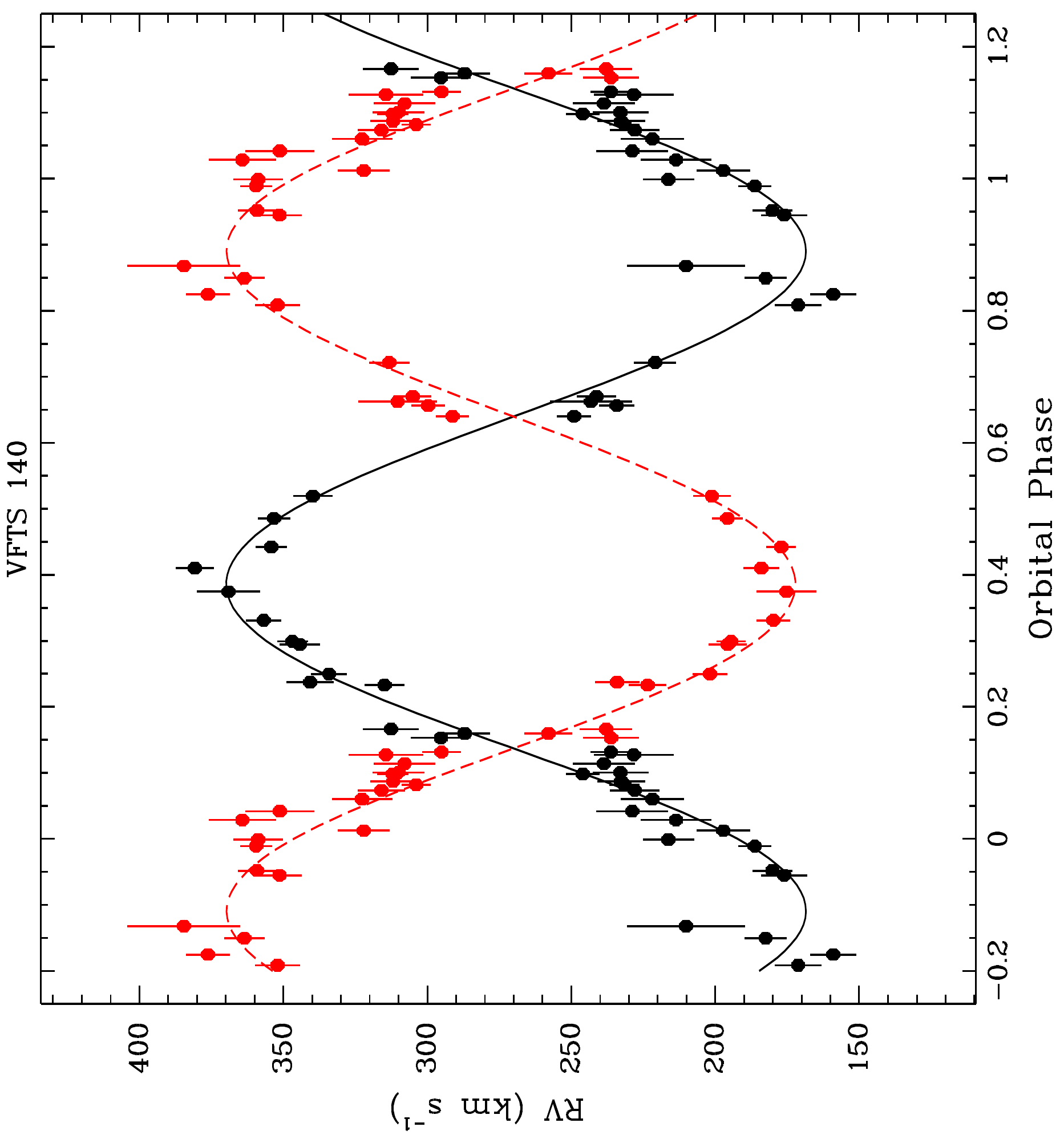}
\includegraphics[width=4.7cm,angle=-90]{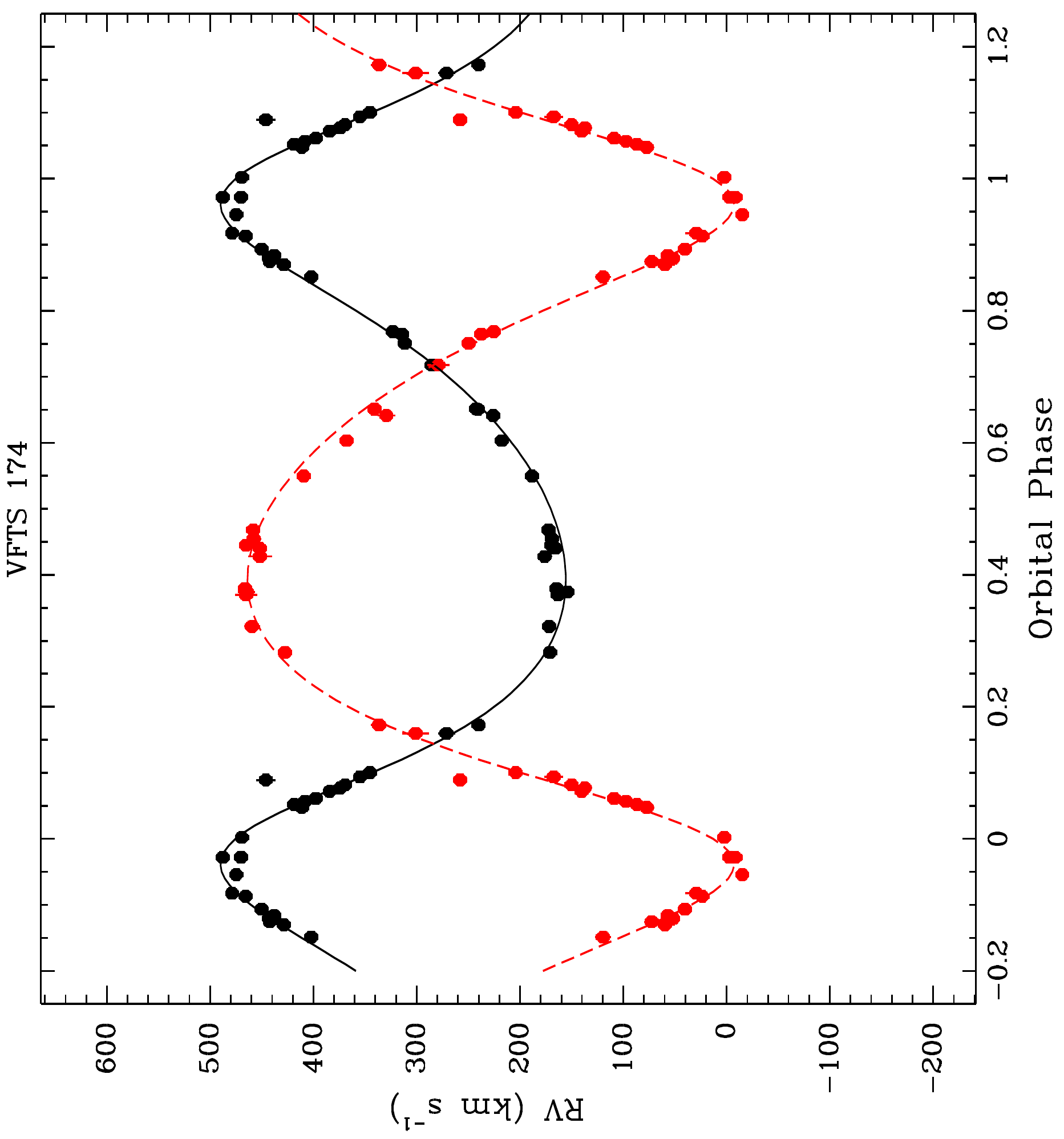}
\includegraphics[width=4.7cm,angle=-90]{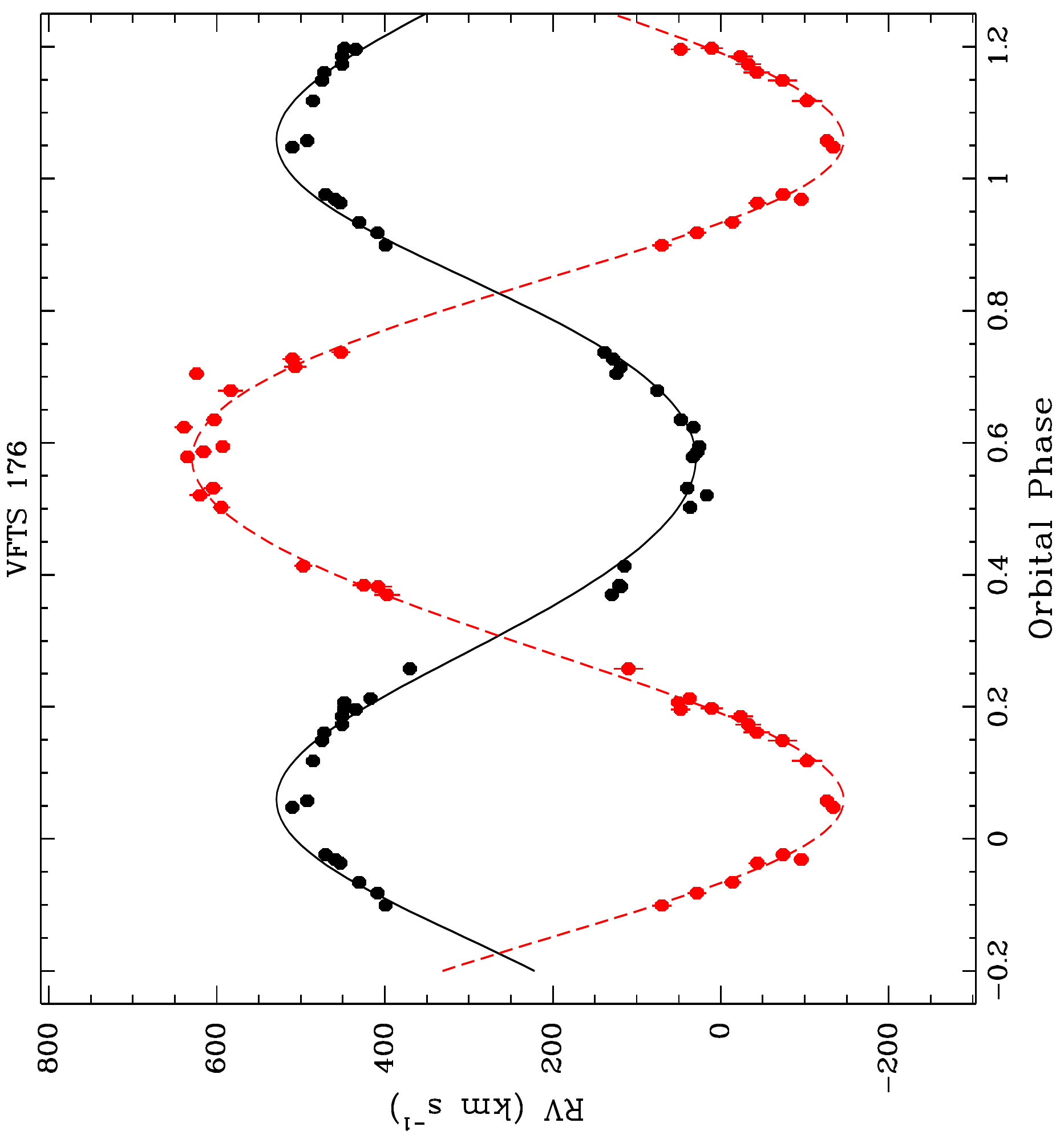}
\includegraphics[width=4.7cm,angle=-90]{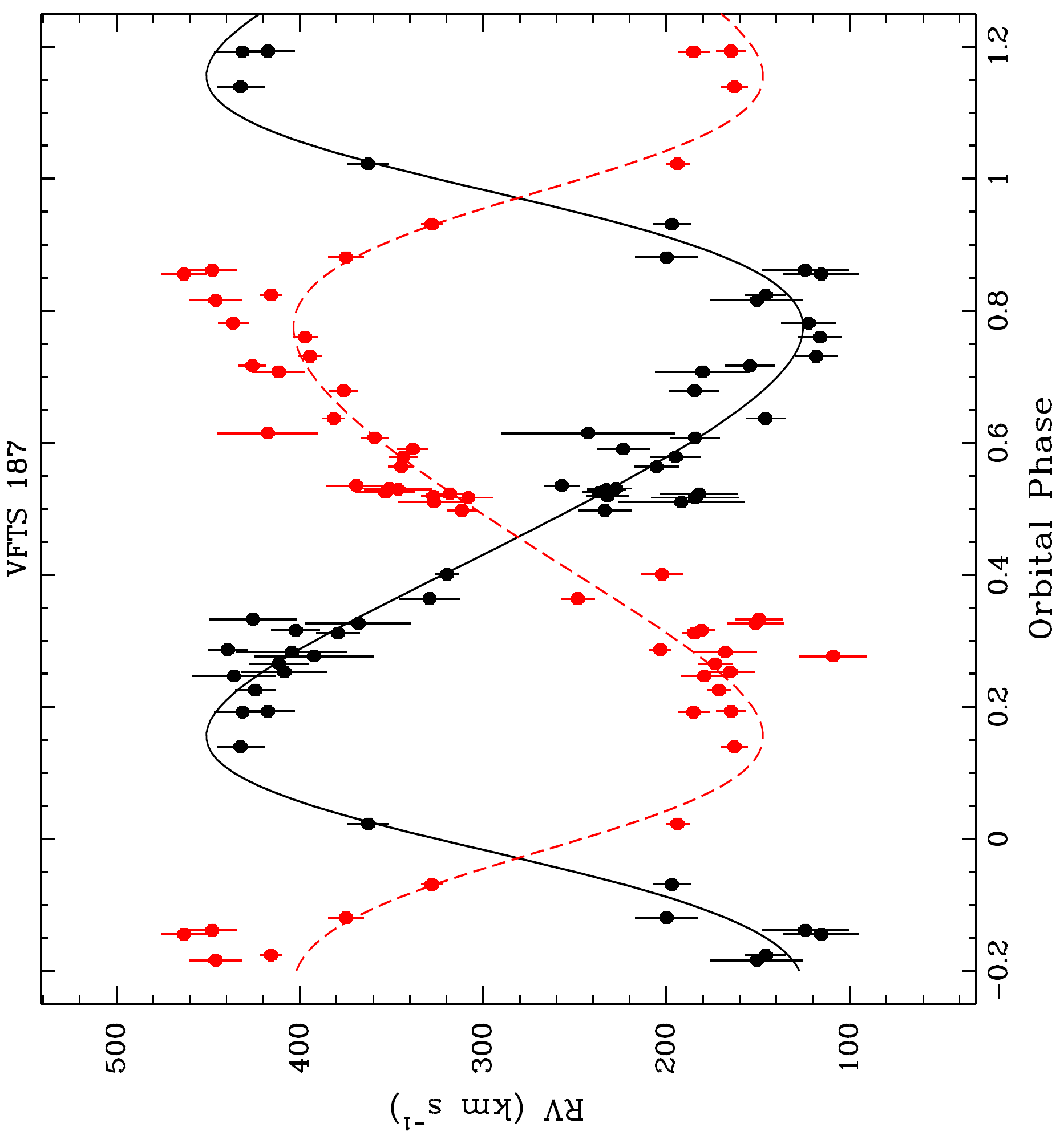}
\includegraphics[width=4.7cm,angle=-90]{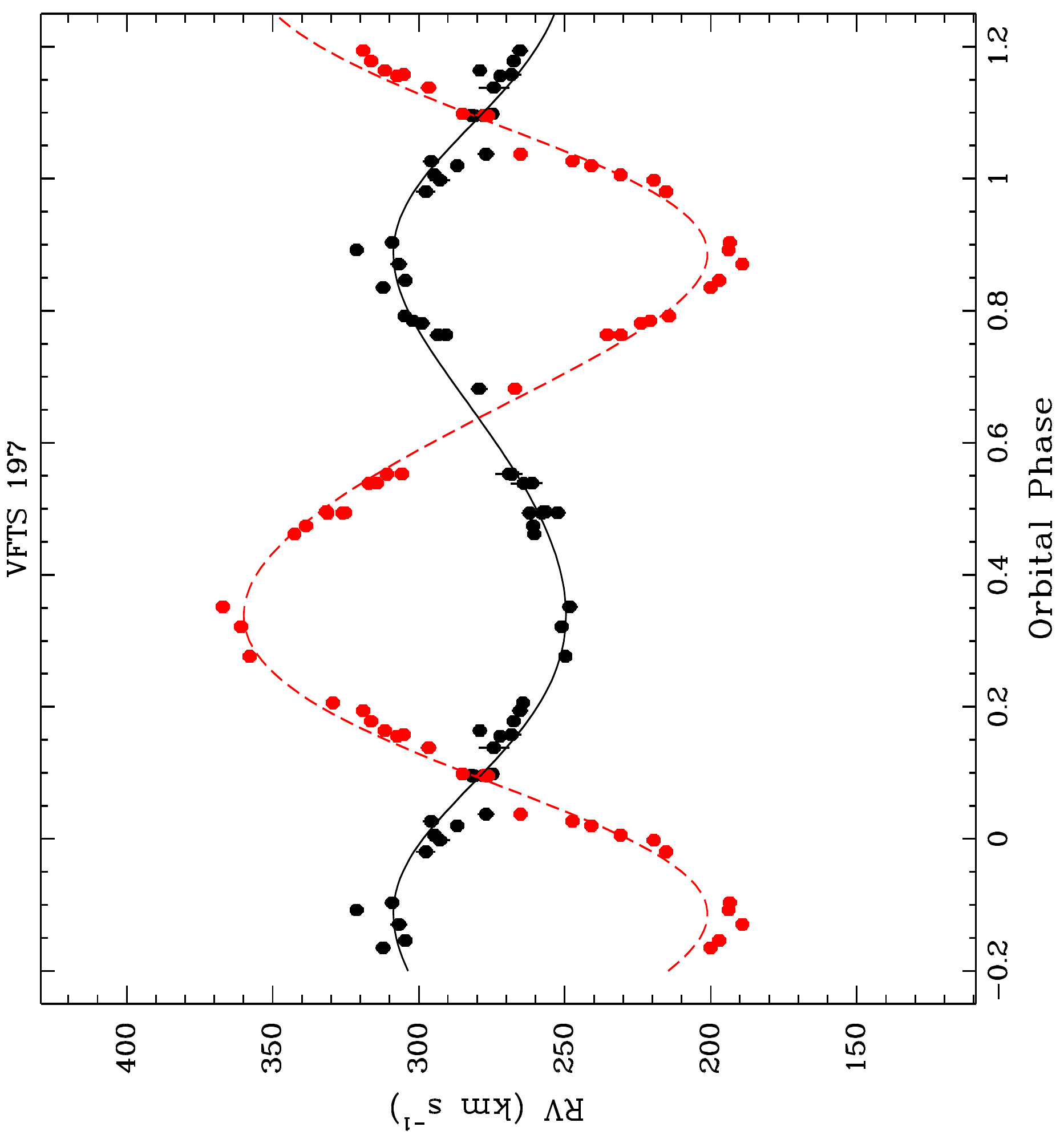}
\includegraphics[width=4.7cm,angle=-90]{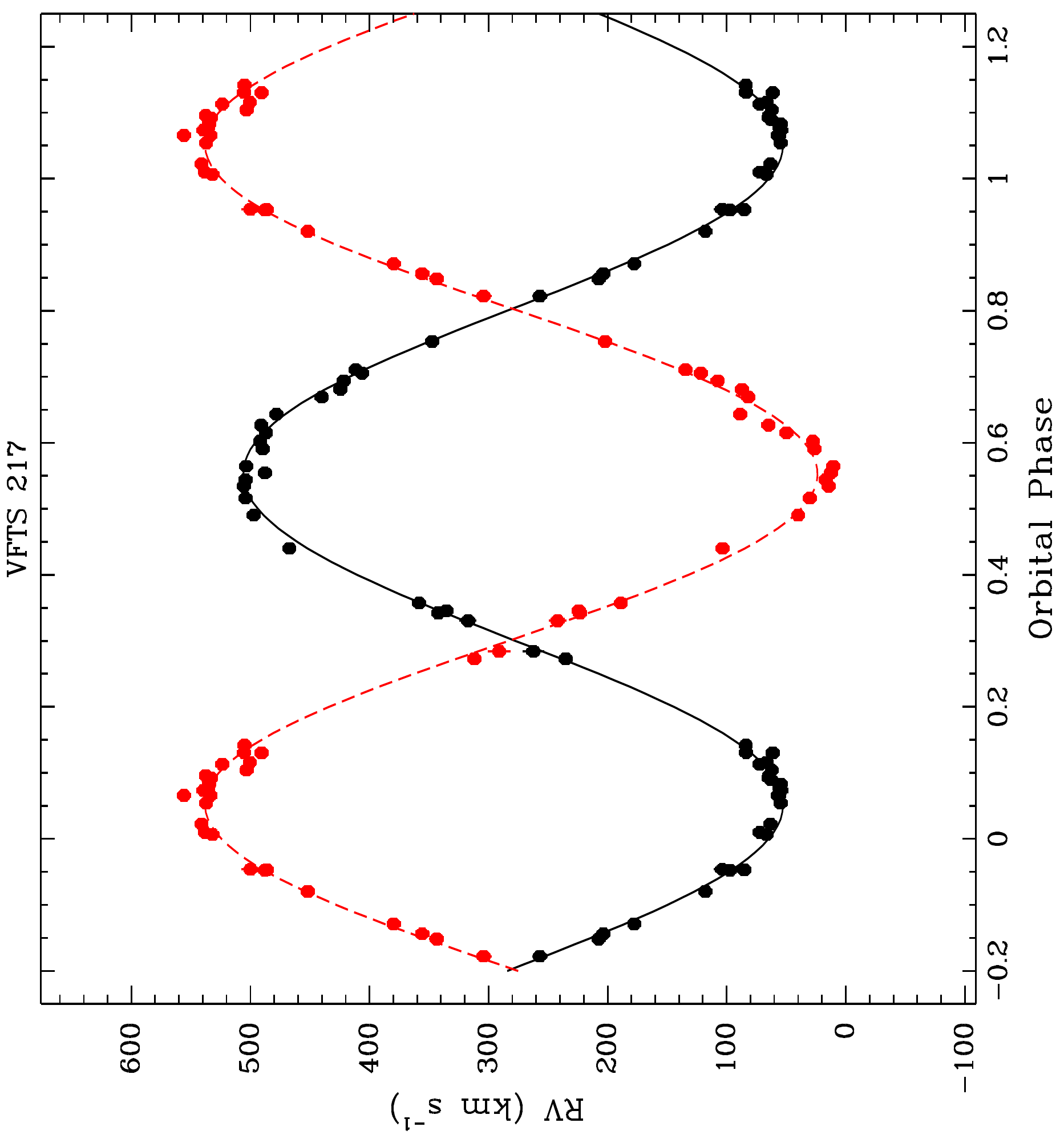}
\caption{Radial velocity (RV) curves for the SB2 systems. The filled black and red hexagons are the RV measurements for the primary and secondary components of the system and the black solid and red dashed lines are the best-fit orbital solutions  (see Section~\ref{ss:orbits}).}
\label{sb2:orb_solution1}
\end{figure*}

\begin{figure*}
\centering
\ContinuedFloat
\includegraphics[width=4.7cm,angle=-90]{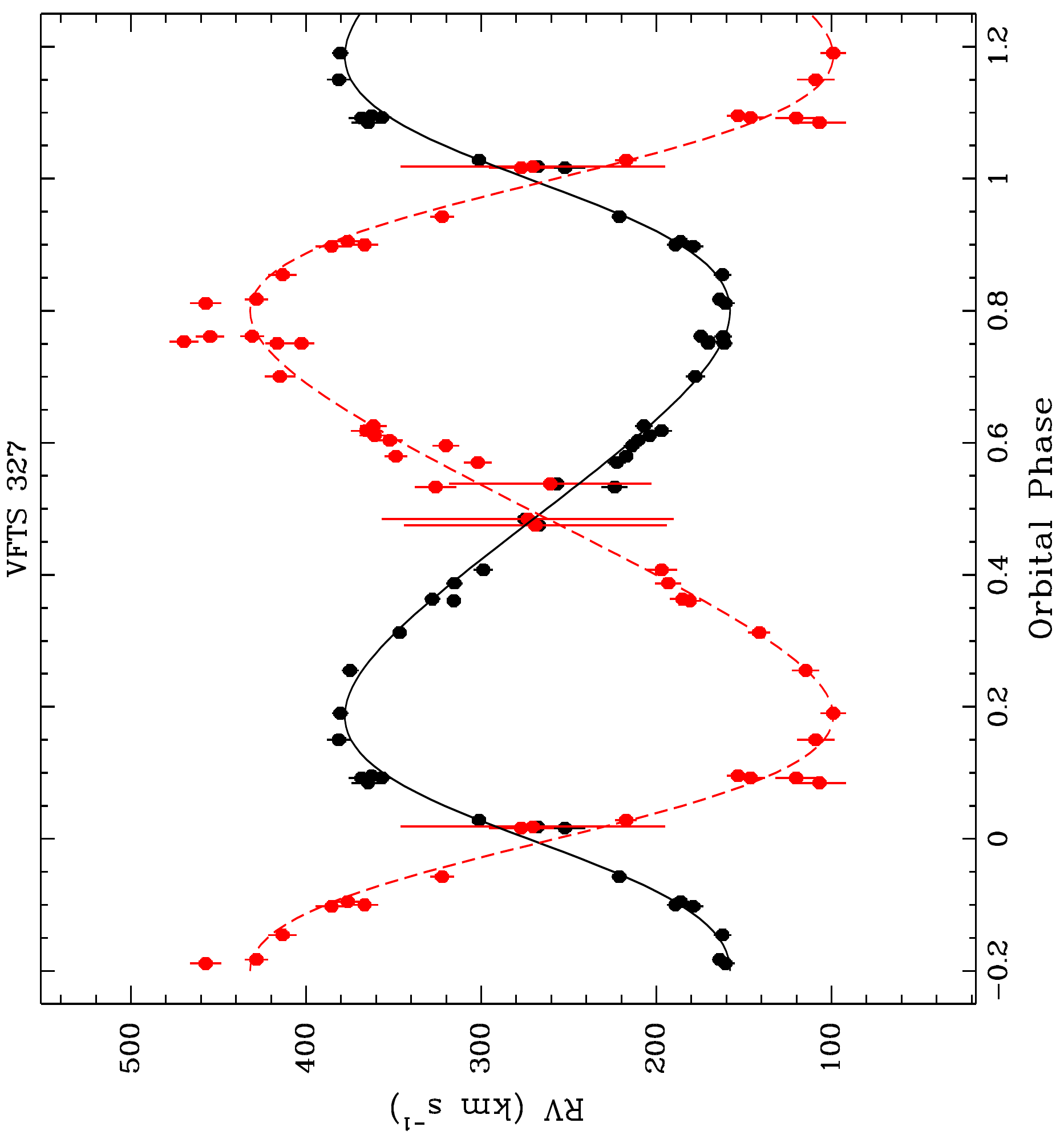}
\includegraphics[width=4.7cm,angle=-90]{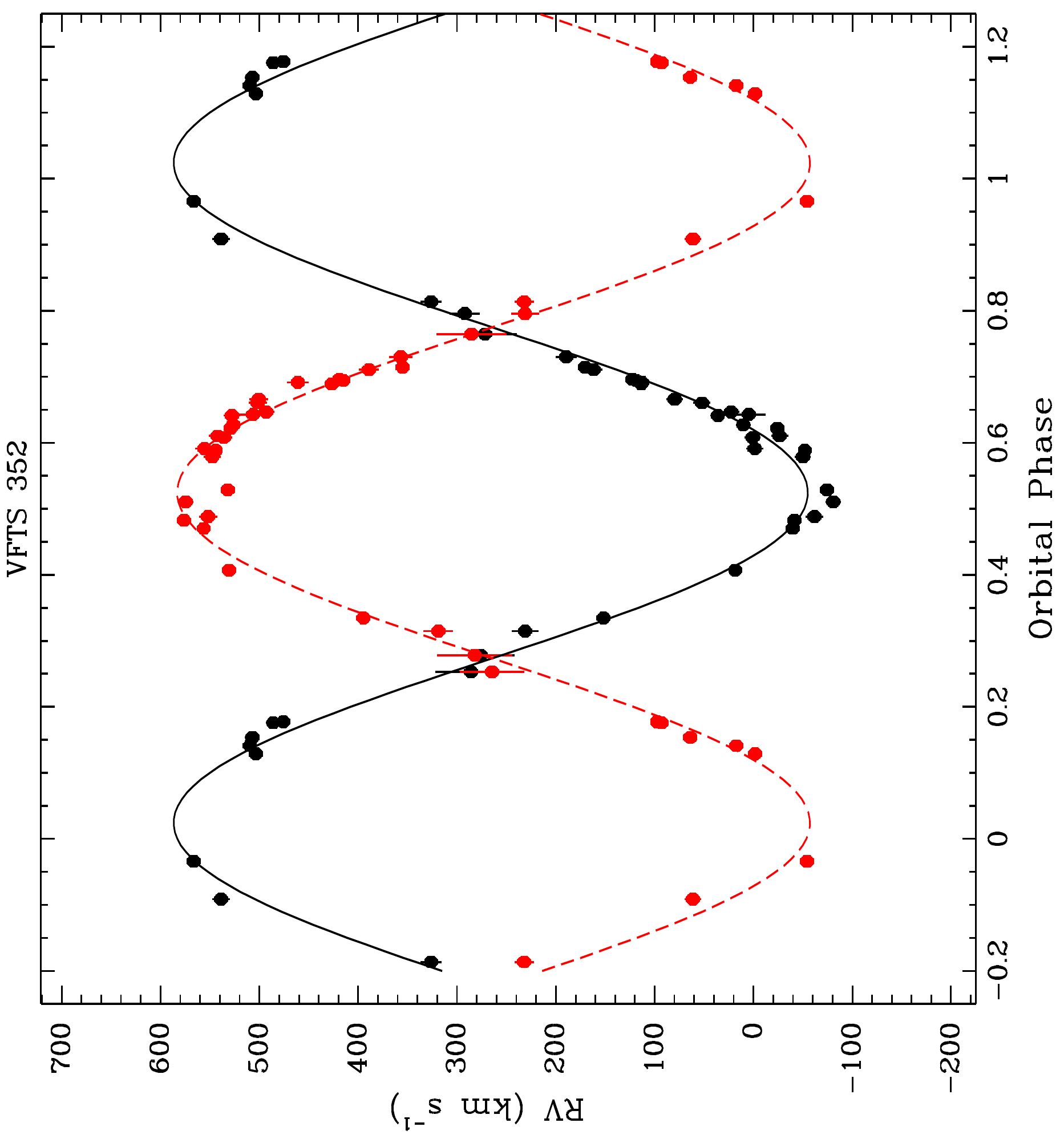}
\includegraphics[width=4.7cm,angle=-90]{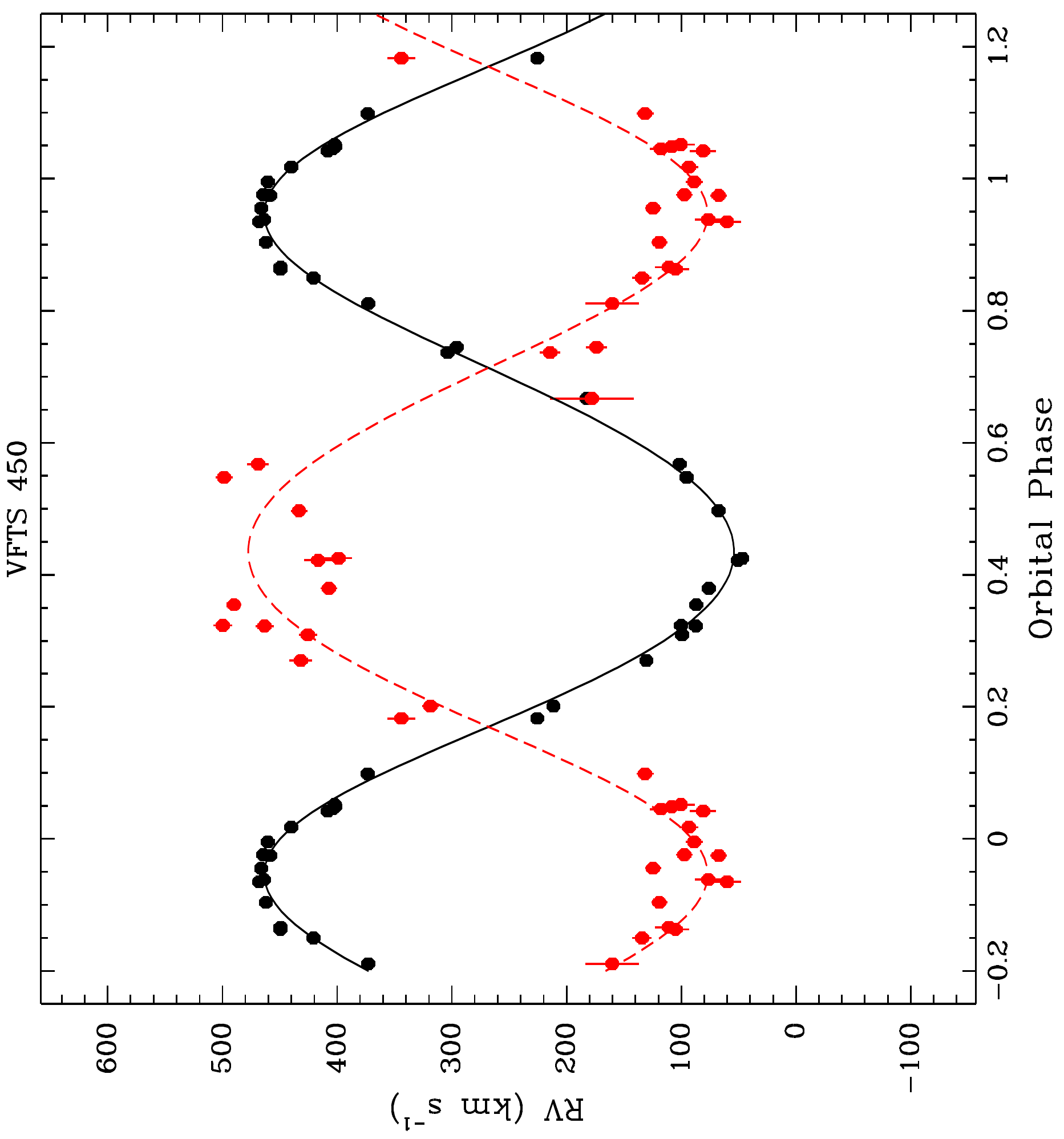}
\includegraphics[width=4.7cm,angle=-90]{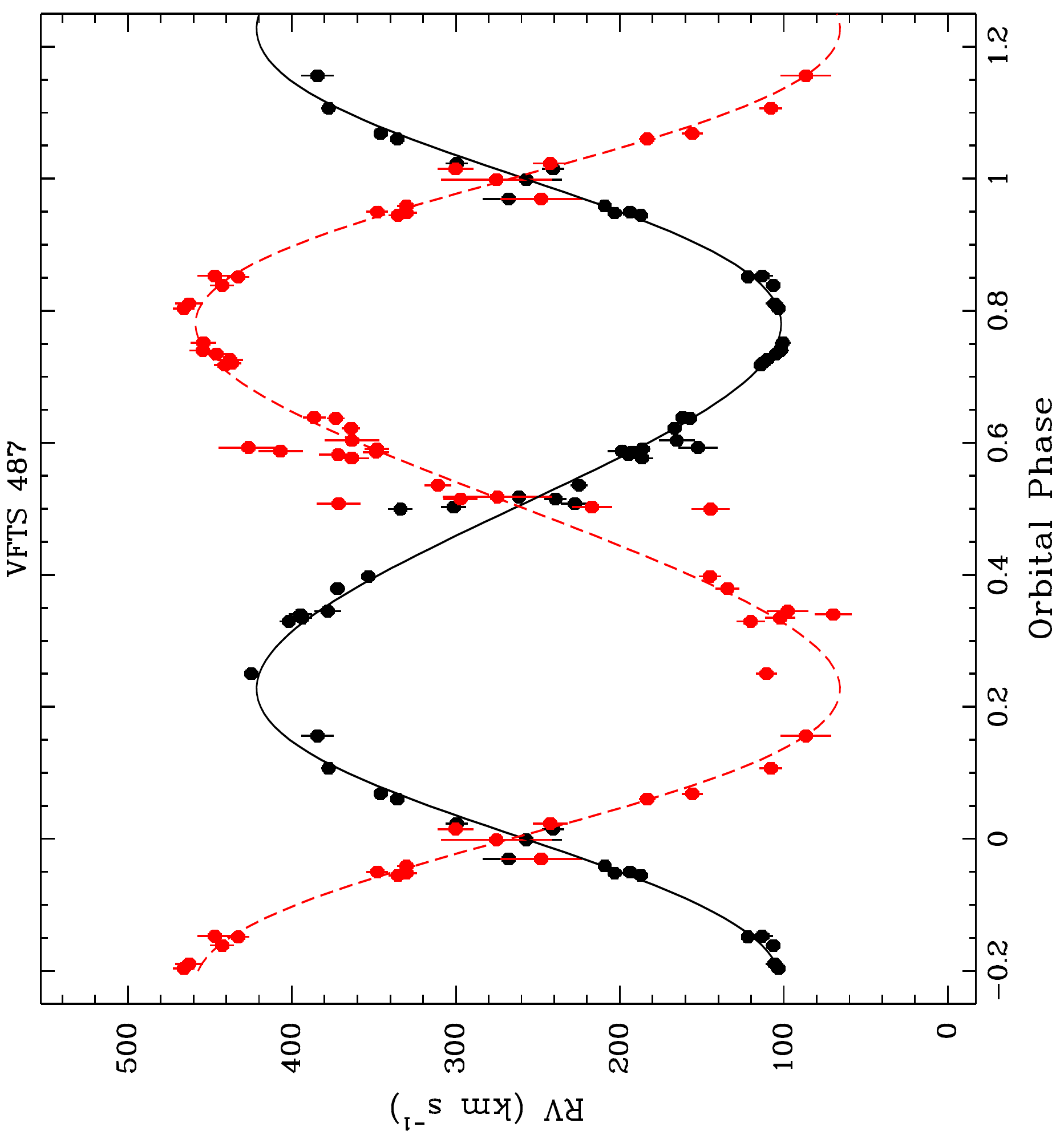}
\includegraphics[width=4.7cm,angle=-90]{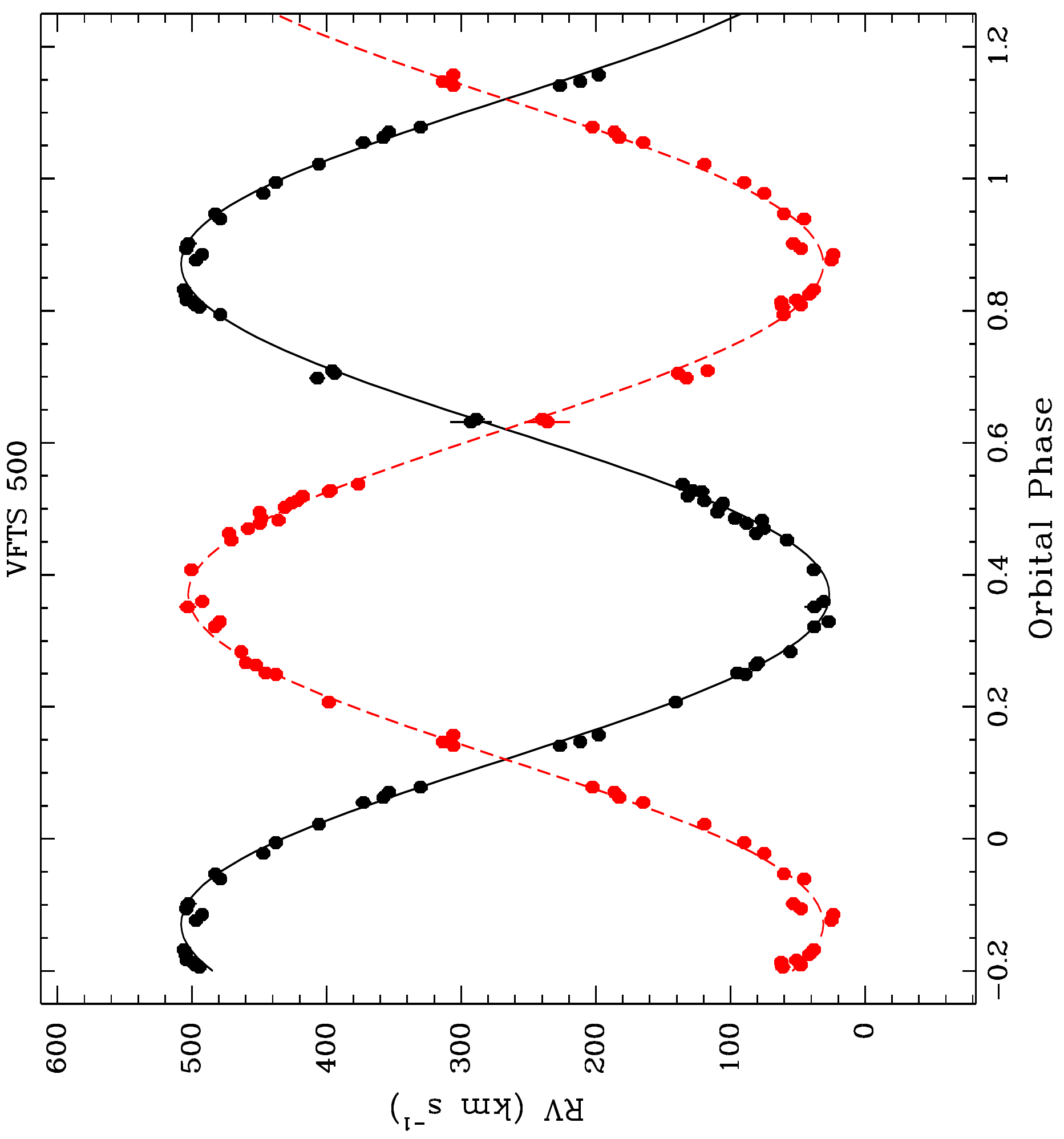}
\includegraphics[width=4.7cm,angle=-90]{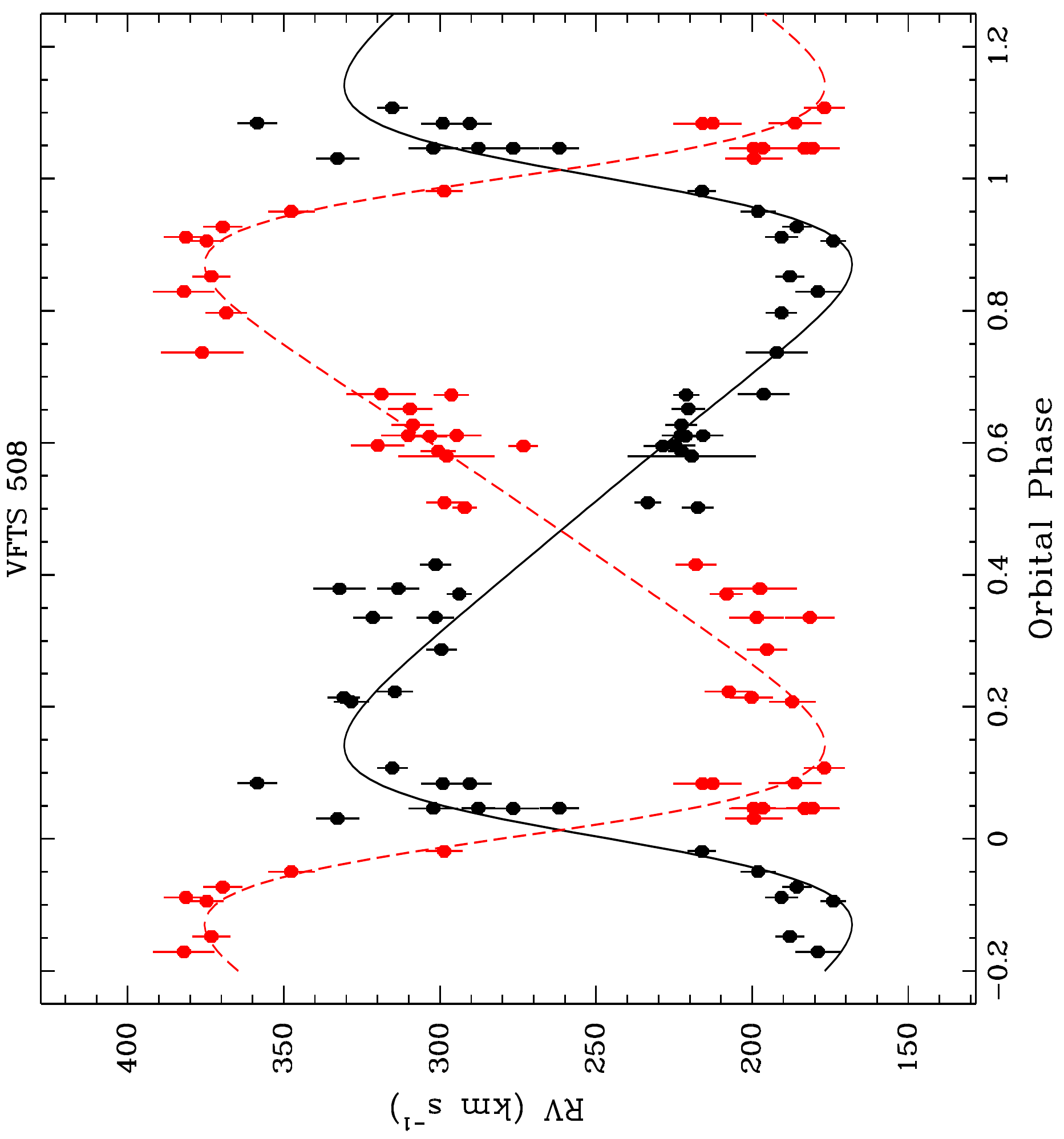}
\includegraphics[width=4.7cm,angle=-90]{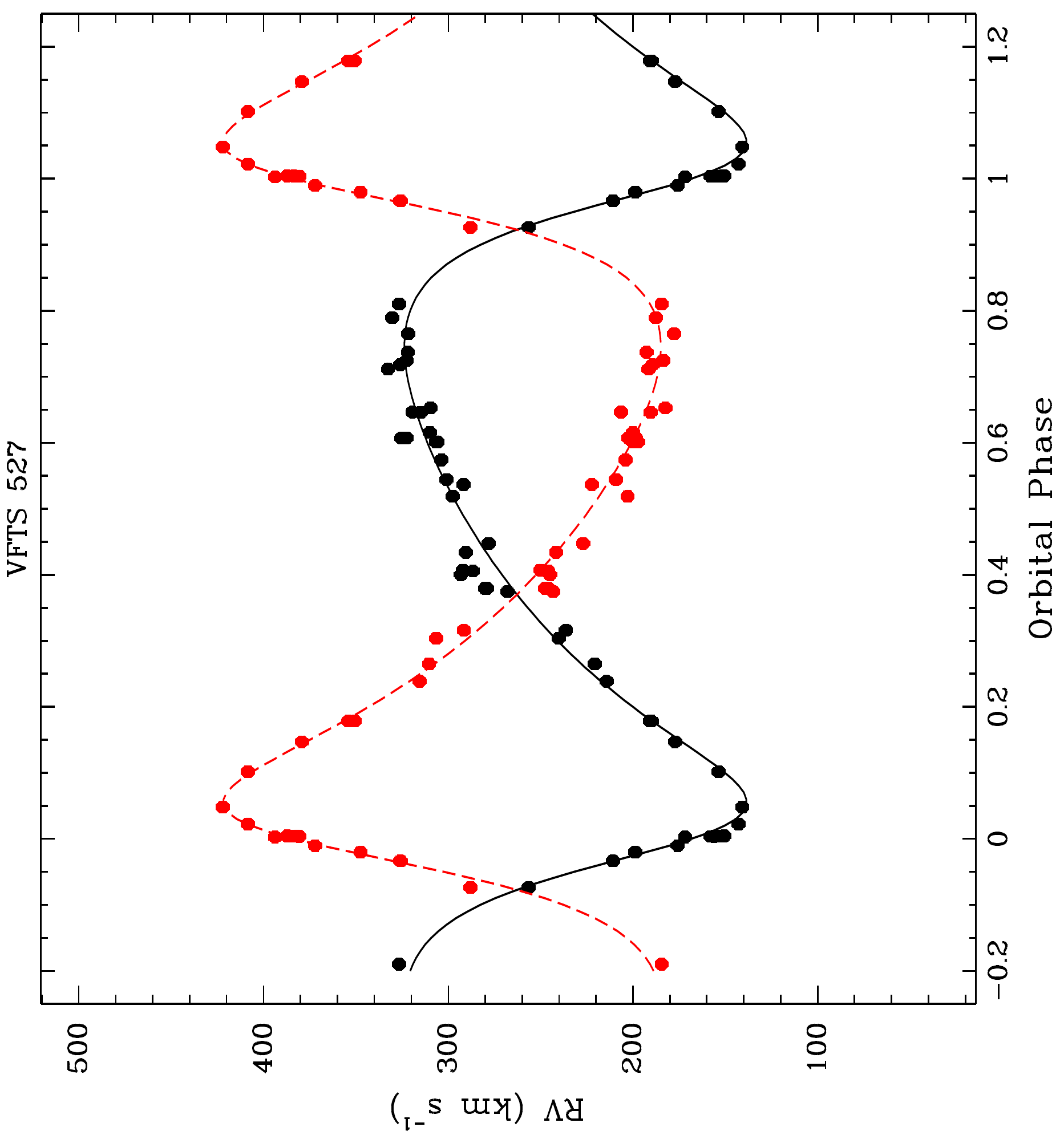}
\includegraphics[width=4.7cm,angle=-90]{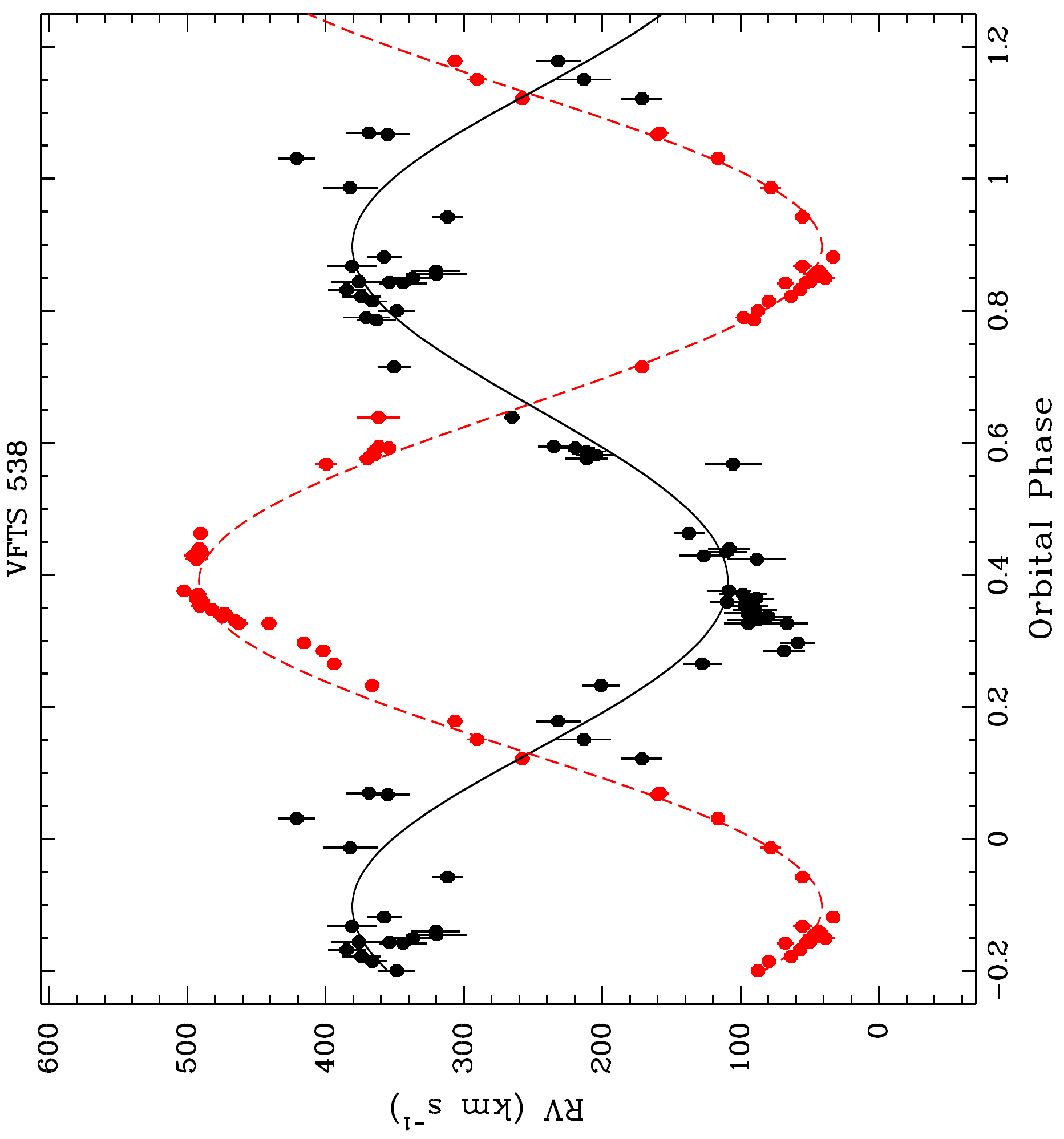}
\includegraphics[width=4.7cm,angle=-90]{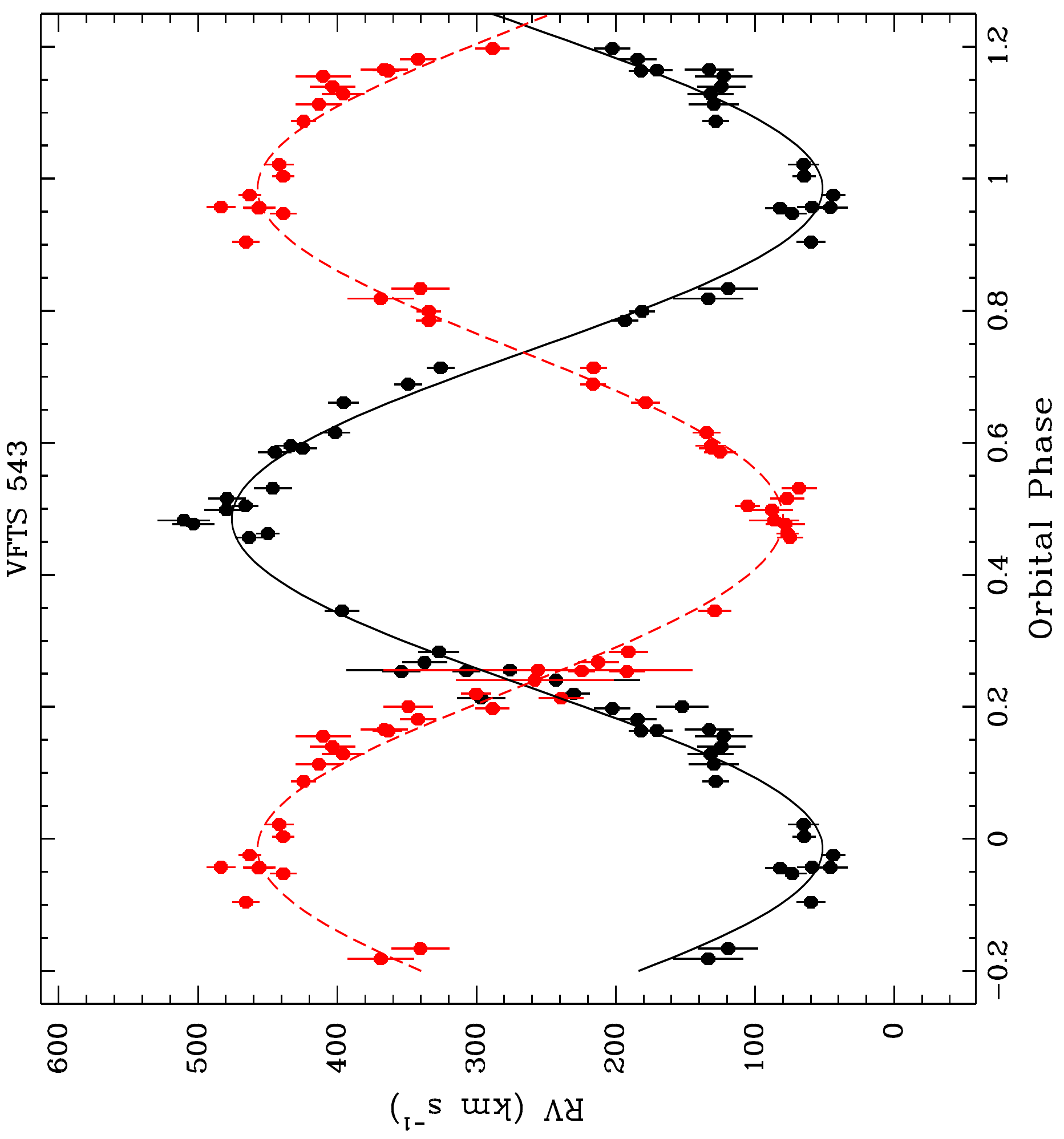}
\includegraphics[width=4.7cm,angle=-90]{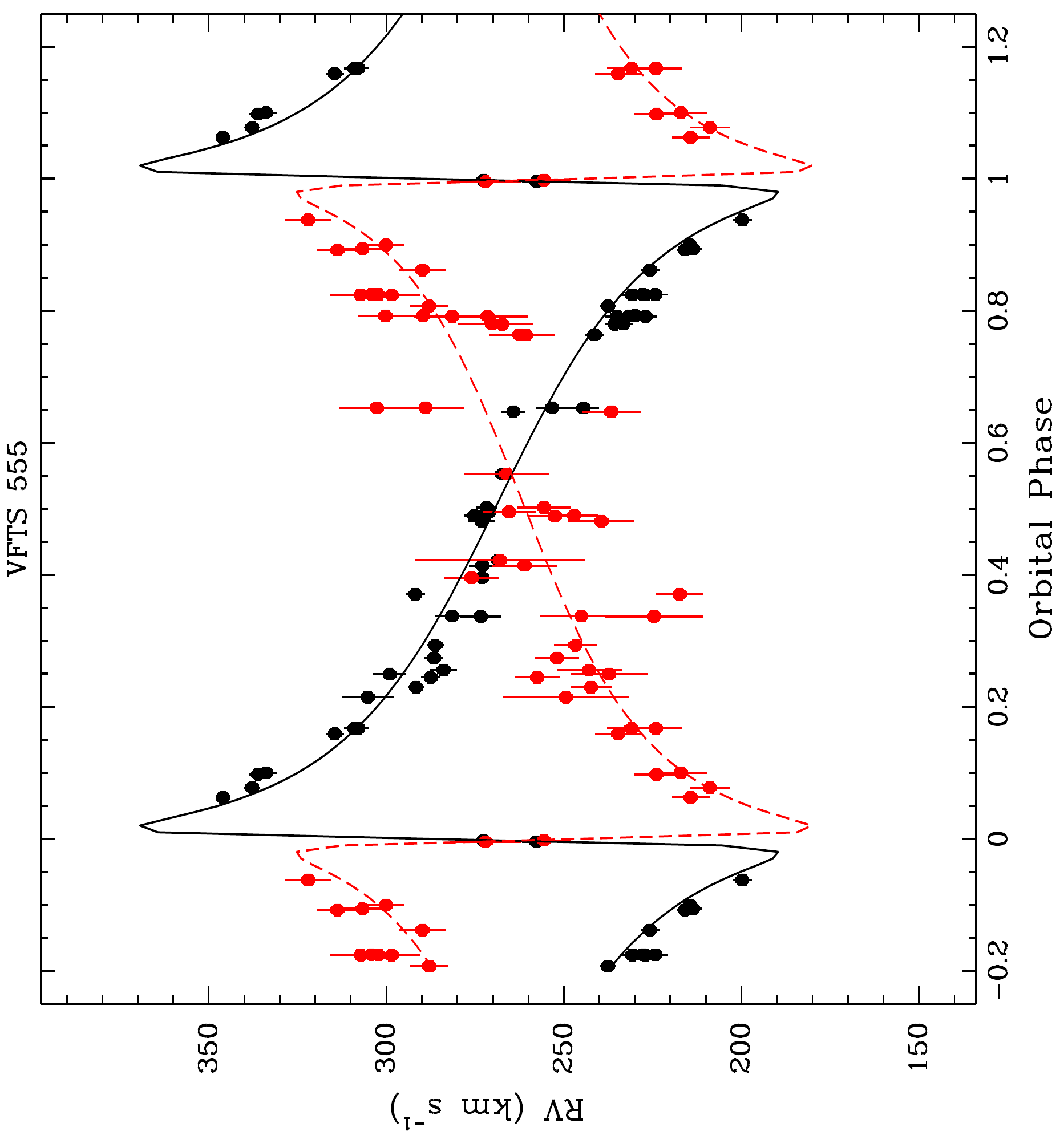}
\includegraphics[width=4.7cm,angle=-90]{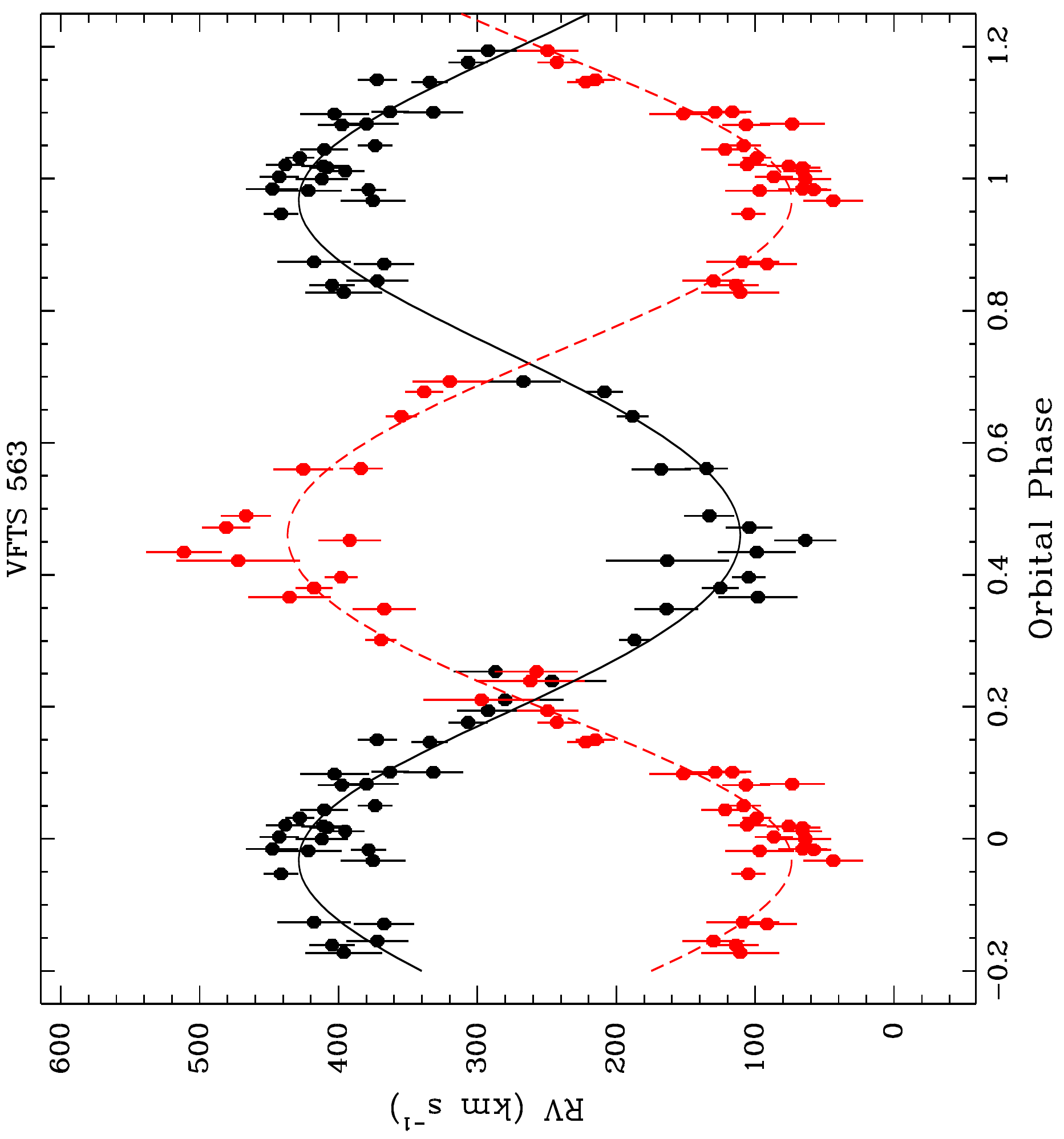}
\includegraphics[width=4.7cm,angle=-90]{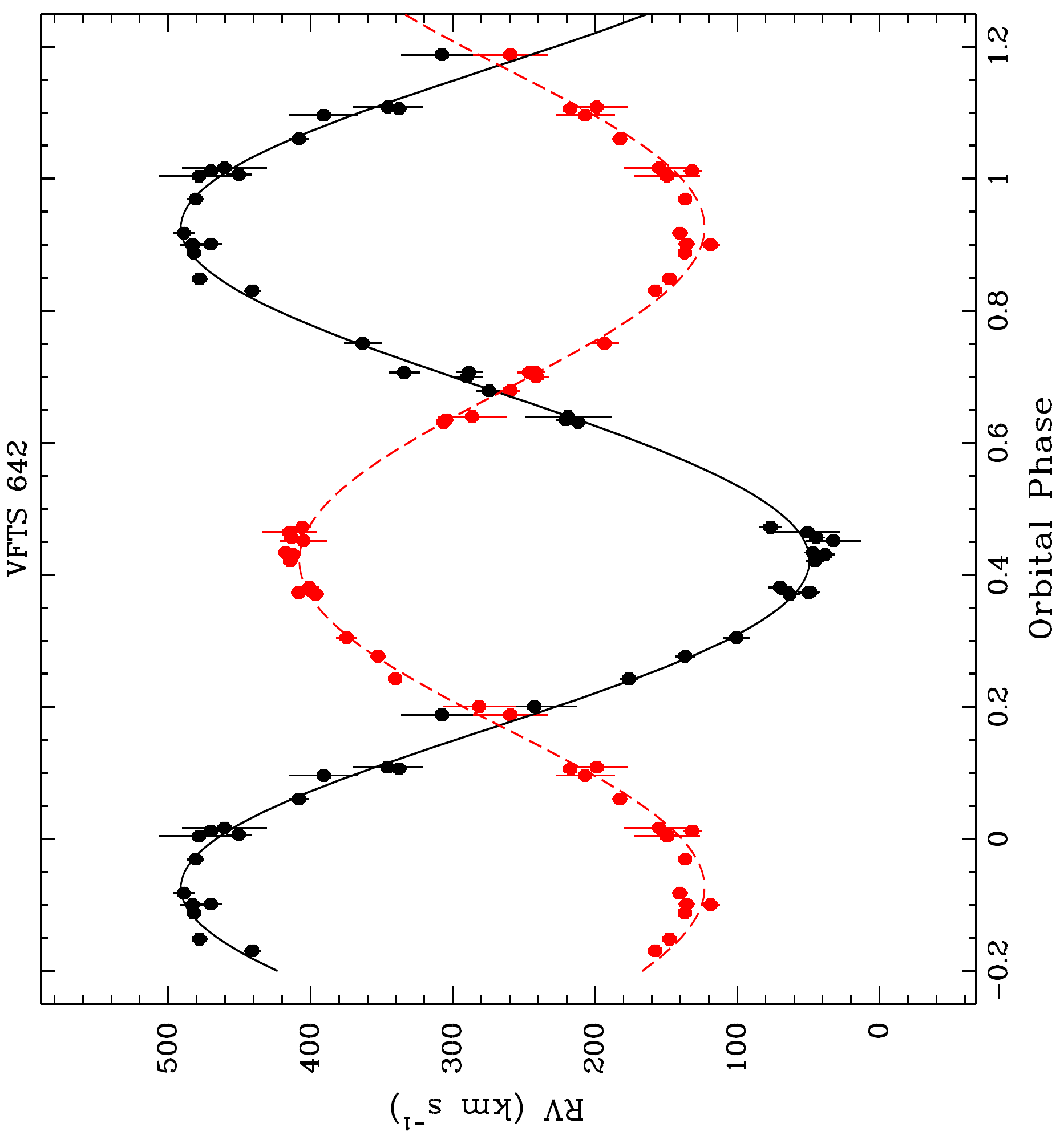}
\includegraphics[width=4.7cm,angle=-90]{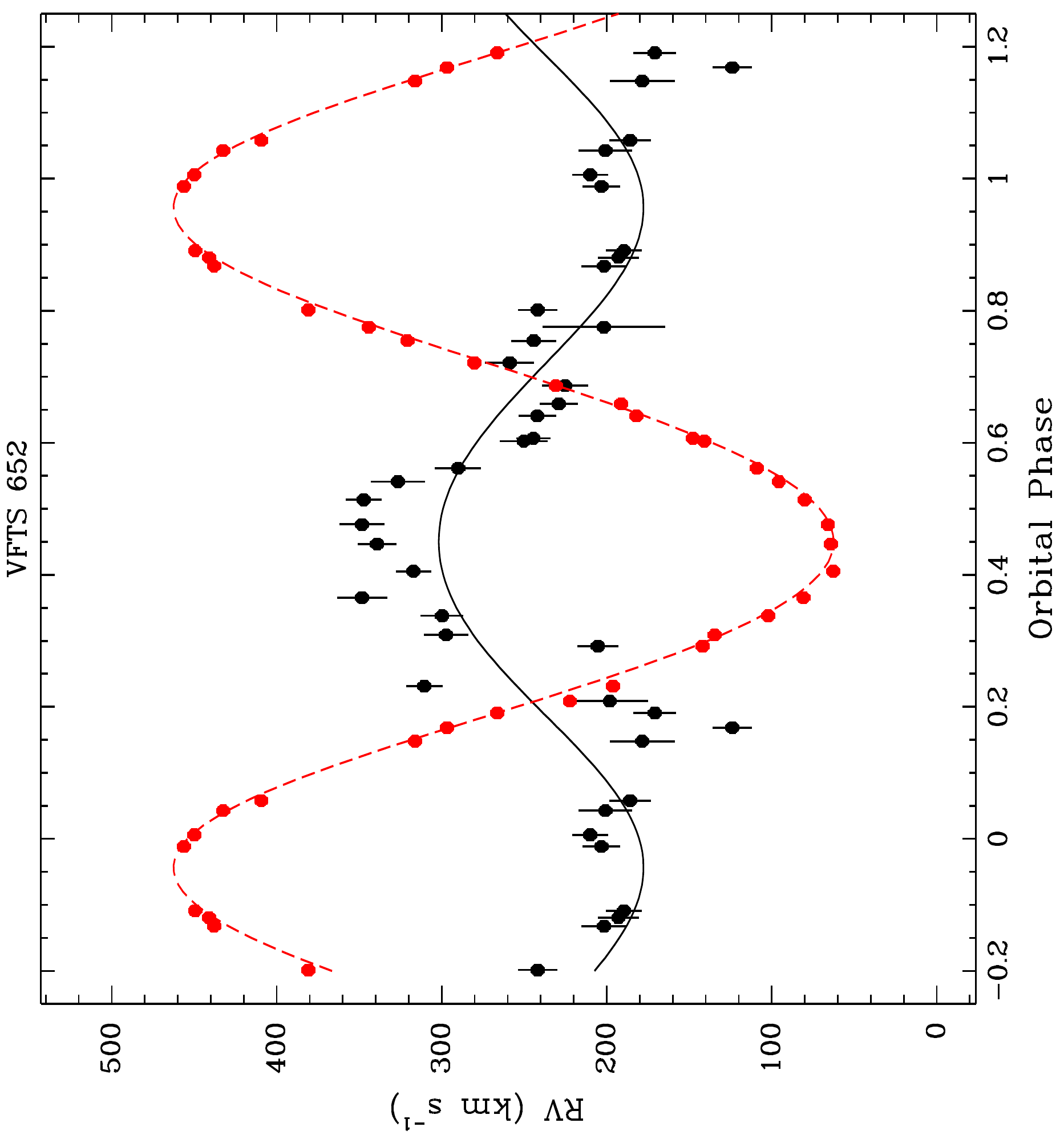}
\includegraphics[width=4.7cm,angle=-90]{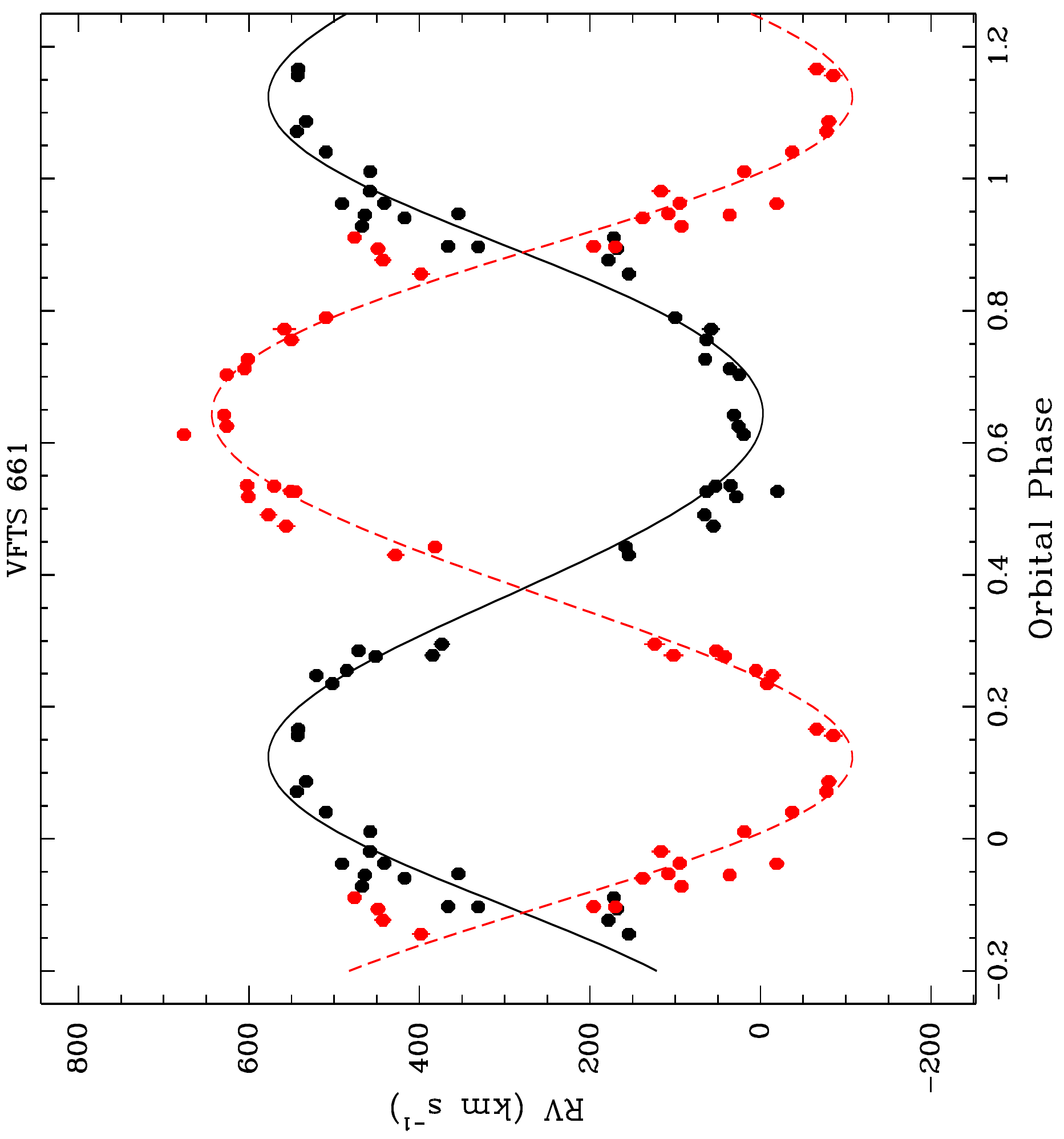}
\includegraphics[width=4.7cm,angle=-90]{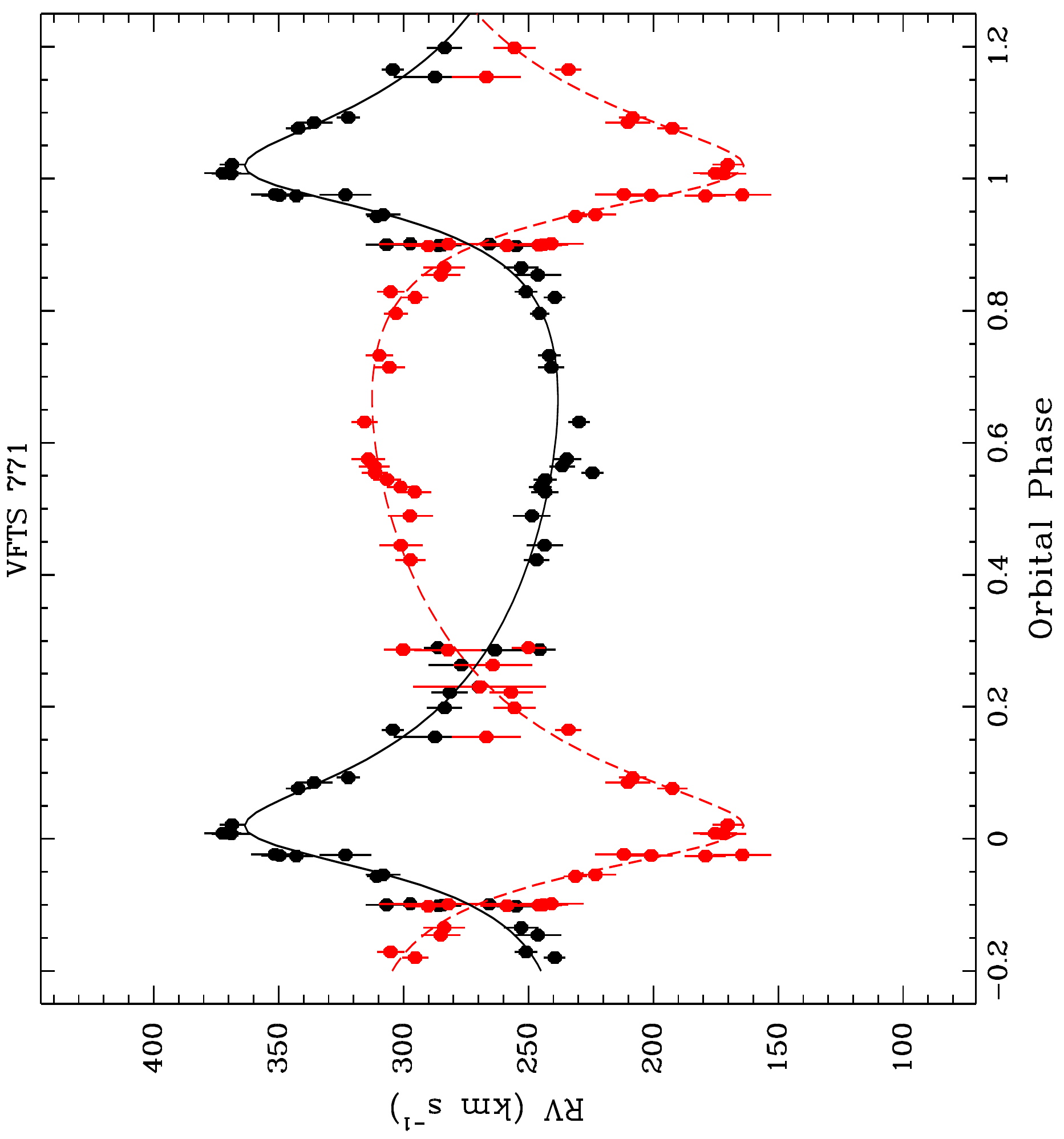}
\caption{{\it Continued...}}
\end{figure*}

\begin{figure*}
\centering
\ContinuedFloat
\includegraphics[width=4.7cm,angle=-90]{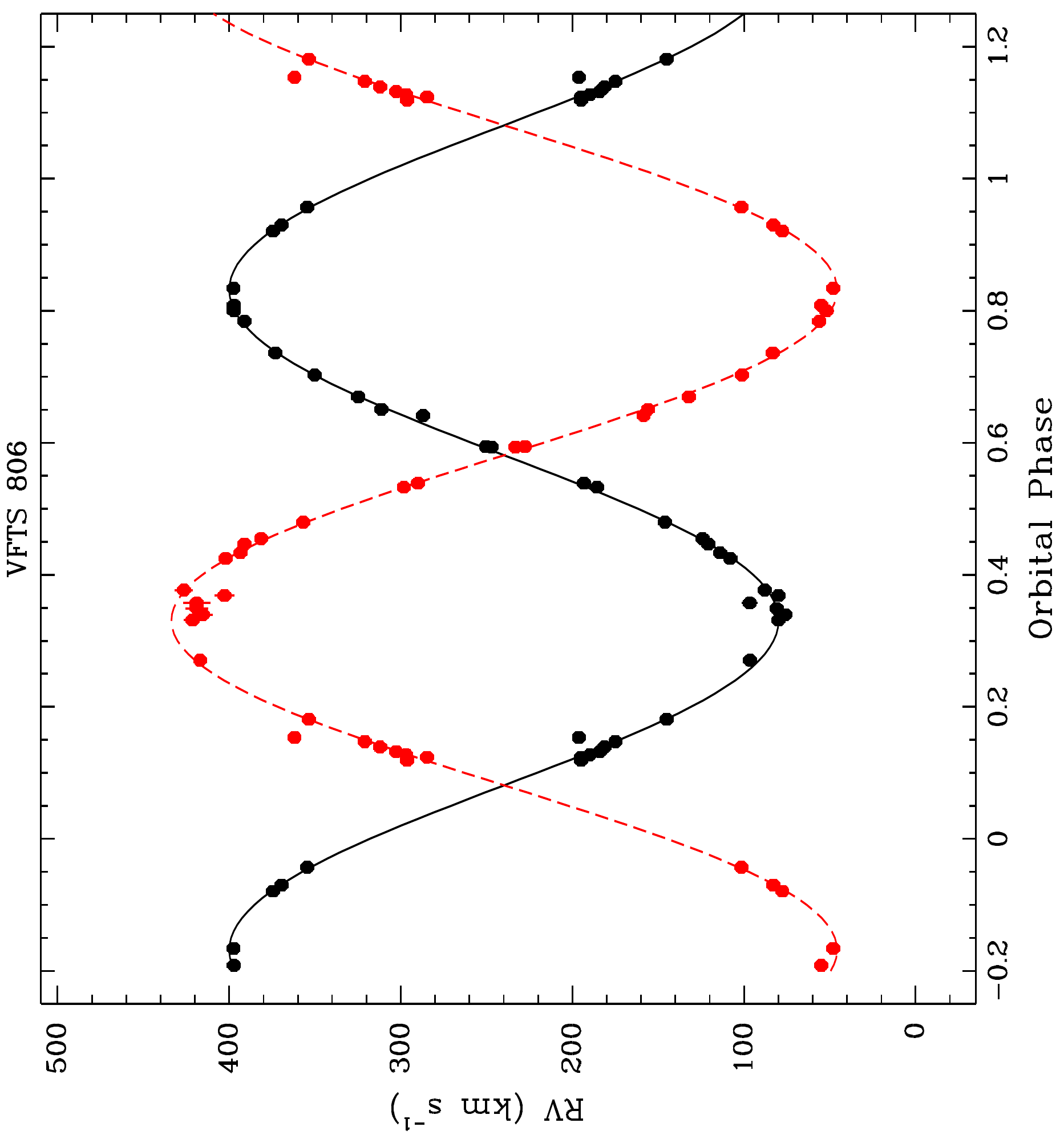}
\caption{{\it Continued...}}
\label{sb2:orb_solution3}
\end{figure*}


\begin{thebibliography}{}

\bibitem[Almeida et al.(2015)]{Almeida+2015} Almeida, L.~A., Sana, H., de Mink, S.~E., et al.\ 2015, \apj, 812, 102 

\bibitem[Barb{\'a} et al.(2010)]{barba+2010} Barb{\'a}, R.~H., Gamen, R., Arias, J.~I., et al.\ 2010, Revista Mexicana de Astronomia y Astrofisica Conference Series, 38, 30 

\bibitem[Belczynski et al.(2008)]{Belczynski2008} Belczynski, K., Kalogera, V., Rasio, F.~A., et al.\ 2008, \apjs, 174, 223-260 

\bibitem[{{Bresolin} {et~al.}(2008){Bresolin}, {Crowther}, {Puls}, \& {F.~Bresolin, P.~A.~Crowther, \& J.~Puls}}]{BCP08}{Bresolin}, F., {Crowther}, P.~A., {Puls}, J., \& {F.~Bresolin, P.~A.~Crowther, \& J.~Puls}, eds. 2008, IAU Symposium, Vol. 250, {Massive Stars as Cosmic   Engines}

\bibitem[Cantrell \& Dougan(2014)]{CaD14} Cantrell, A.~G., \& Dougan, T.~J.\ 2014, \mnras, 445, 2028 

\bibitem[de Mink \& Belczynski(2015)]{dMB15} de Mink, S.~E., \& Belczynski, K.\ 2015, \apj, 814, 58 

\bibitem[{{de Mink} {et~al.}(2009){de Mink}, {Cantiello}, {Langer}, {Pols},
  {Brott}, \& {Yoon}}]{dMCL09}{de Mink}, S.~E., {Cantiello}, M., {Langer}, N., {et~al.} 2009, \aap, 497, 243

\bibitem[de Mink et al.(2014)]{dMSL14} de Mink, S.~E., Sana, H., Langer, N., Izzard, R.~G., \& Schneider, F.~R.~N.\ 2014, \apj, 782, 7 

\bibitem[De Lee et al.(2013)]{DeLee2013} De Lee, N., Ge, J., Crepp, J.~R., et al.\ 2013, \aj, 145, 155 

\bibitem[Dunstall et al.(2015)]{DDS15} Dunstall, P.~R., Dufton, P.~L., Sana, H., et al.\ 2015, \aap, 580, A93 


\bibitem[Dworetsky(1983)]{Dworetsky1983} Dworetsky, M.~M.\ 1983, \mnras, 203, 917

\bibitem[Eldridge et al.(2008)]{Eldridge+2008} Eldridge, J.~J., Izzard, R.~G., \& Tout, C.~A.\ 2008, \mnras, 384, 1109 

\bibitem[Eldridge et al.(2013)]{Eldridge+2013} Eldridge, J.~J., Fraser, M., Smartt, S.~J., Maund, J.~R., \& Crockett, R.~M.\ 2013, \mnras, 436, 774 

\bibitem[{{Evans} {et~al.}(2011){Evans}, {Taylor}, {H{\'e}nault-Brunet},  {Sana}, {de Koter}, {Sim{\'o}n-D{\'{\i}}az}, {Carraro}, {Bagnoli}, {Bastian}, {Bestenlehner}, {Bonanos}, {Bressert}, {Brott}, {Campbell}, {Cantiello},
{Clark}, {Costa}, {Crowther}, {de Mink}, {Doran}, {Dufton}, {Dunstall},
{Friedrich}, {Garcia}, {Gieles}, {Gr{\"a}fener}, {Herrero}, {Howarth},
{Izzard}, {Langer}, {Lennon}, {Ma{\'{\i}}z Apell{\'a}niz}, {Markova},
{Najarro}, {Puls}, {Ramirez}, {Sab{\'{\i}}n-Sanjuli{\'a}n}, {Smartt},
{Stroud}, {van Loon}, {Vink}, \& {Walborn}}]{ETHB11}{Evans}, C.~J., {Taylor}, W.~D., {H{\'e}nault-Brunet}, V., {et~al.} 2011, \aap, 530, A108

\bibitem[Evans et al.(2010)]{Evans2010} Evans, C.~J., Walborn, N.~R., Crowther, P.~A., et al.\ 2010, \apjl, 715, L74 

\bibitem[Garmany et al.(1980)]{GCM80} Garmany, C.~D., Conti, P.~S., \& Massey, P.\ 1980, \apj, 242, 1063 

\bibitem[Gvaramadze(2009)]{Gvaramadze2009} Gvaramadze, V.~V.\ 2009, \mnras, 395, L85 

\bibitem[{{Howarth} {et~al.}(2015){Howarth}, {Dufton}, {Dunstall}, {Evans},
{Almeida}, {Bonanos}, {Clark}, {Langer}, {Sana}, {Sim{\'o}n-D{\'{\i}}az},
{Soszy{\'n}ski}, \& {Taylor}}]{Howat+2015} {Howarth}, I.~D., {Dufton}, P.~L., {Dunstall}, P.~R., {et~al.} 2015, \aap, 582, A73

\bibitem[Kato et al.(2007)]{kato+2007} Kato, D., Nagashima, C., Nagayama, T., et al.\ 2007, \pasj, 59, 615 

\bibitem[Kobulnicky \& Fryer(2007)]{Kobulnicky+2007} Kobulnicky, H.~A., \& Fryer, C.~L.\ 2007, \apj, 670, 747 

\bibitem[Kiminki \& Kobulnicky(2012)]{KK2012} Kiminki, D.~C., \& Kobulnicky, H.~A.\ 2012, \apj, 751, 4


\bibitem[Kobulnicky et al.(2014)]{KK+2014} Kobulnicky, H.~A., Kiminki, D.~C., Lundquist, M.~J., et al.\ 2014, \apjs, 213, 34 

\bibitem[{{Langer} {et~al.}(2008){Langer}, {Cantiello}, {Yoon}, {Hunter}, {Brott}, {Lennon}, {de Mink}, \& {Verheijdt}}]{LCY08}{Langer}, N., {Cantiello}, M., {Yoon}, S.-C., {et~al.} 2008, in {Massive Stars as Cosmic Engines}, Vol. 250, IAU Symposium, ed. F.~{Bresolin}, P.~A. {Crowther}, \& J.~{Puls}, 167--178

\bibitem[Lomb(1976)]{Lomb1976} Lomb, N.~R.\ 1976, \apss, 39, 447 

\bibitem[Lucy(2006)]{Luc06} Lucy, L.~B.\ 2006, \aap, 457, 629 


\bibitem[{{Lucy}(2013)}]{Luc13}{Lucy}, L.~B. 2013, \aap, 551, A47

\bibitem[{{Lucy} \& {Sweeney}(1971)}]{LuS71}{Lucy}, L.~B., \& {Sweeney}, M.~A. 1971, \aj, 76, 544

\bibitem[Mason et al.(2009)]{MHG09} Mason, B.~D., Hartkopf, W.~I., Gies, D.~R., Henry, T.~J., \& Helsel, J.~W.\ 2009, \aj, 137, 3358  

\bibitem[McEvoy et al.(2015)]{MEDE15} McEvoy, C.~M., Dufton, P.~L., Evans, C.~J., et al.\ 2015, \aap, 575, A70 

\bibitem[Moe \& Di Stefano(2016)]{Moe2016} Moe, M., \& Di Stefano, R.\ 2016, arXiv:1606.05347 

\bibitem[Pinsonneault \& Stanek(2006)]{PS2006} Pinsonneault, M.~H., \& Stanek, K.~Z.\ 2006, \apjl, 639, L67 

\bibitem[{{Podsiadlowski} {et~al.}(1992){Podsiadlowski}, {Joss}, \& {Hsu}}]{PJH92}{Podsiadlowski}, P., {Joss}, P.~C., \& {Hsu}, J.~J.~L. 1992, \apj, 391, 246

\bibitem[Podsiadlowski et al.(2010)]{Podsiadlowski+2010} Podsiadlowski, P., Ivanova, N., Justham, S., \& Rappaport, S.\ 2010, \mnras, 406, 840 

\bibitem[Ram{\'{\i}}rez-Agudelo et al.(2015)]{RASdK15} Ram{\'{\i}}rez-Agudelo, O.~H., Sana, H., de Mink, S.~E., et al.\ 2015, \aap, 580, A92 


\bibitem[Ram{\'{\i}}rez-Agudelo et al.(2013)]{RASDS13} Ram{\'{\i}}rez-Agudelo, O.~H., Sim{\'o}n-D{\'{\i}}az, S., Sana, H., et al.\ 2013, \aap, 560, A29 

\bibitem[Sadowski et al.(2008)]{Sadowski+2008} Sadowski, A., Belczynski, K., Bulik, T., et al.\ 2008, \apj, 676, 1162 

\bibitem[{{Sana} \& {Evans}(2011)}]{SaE11}{Sana}, H., \& {Evans}, C.~J. 2011, in IAU Symposium, ed. {C.~Neiner, G.~Wade, G.~Meynet, \& G.~Peters}, Vol. 272, 474

\bibitem[Sana et al.(2011)]{SLBDB11} Sana, H., Le Bouquin, J.-B., De Becker, M., et al.\ 2011, \apjl, 740, L43 

\bibitem[{{Sana} {et~al.}(2012){Sana}, {de Mink}, {de Koter}, {Langer},
  {Evans}, {Gieles}, {Gosset}, {Izzard}, {Le Bouquin}, \&
  {Schneider}}]{SdMdK12}{Sana}, H., {de Mink}, S.~E., {de Koter}, A., {et~al.} 2012, Science, 337, 444

\bibitem[{{Sana} {et~al.}(2013){Sana}, {de Koter}, {de Mink}, {Dunstall},
  {Evans}, {H{\'e}nault-Brunet}, {Ma{\'{\i}}z Apell{\'a}niz},
  {Ram{\'{\i}}rez-Agudelo}, {Taylor}, {Walborn}, {Clark}, {Crowther},
  {Herrero}, {Gieles}, {Langer}, {Lennon}, \& {Vink}}]{SdKdM13}
{Sana}, H., {de Koter}, A., {de Mink}, S.~E., {et~al.} 2013, \aap, 550, A107

\bibitem[Sana et al.(2013)]{SvBT13} Sana, H., van Boeckel, T., Tramper, F., et al.\ 2013, \mnras, 432, 26 

\bibitem[Sana et al.(2014)]{Sana2014} Sana, H., Le Bouquin, J.-B., Lacour, S., et al.\ 2014, \apjs, 215, 15 

\bibitem[Scargle(1982)]{Scargle1982} Scargle, J.~D.\ 1982, \apj, 263, 835

\bibitem[Schnurr et al.(2009)]{SMVS09} Schnurr, O., Moffat, A.~F.~J., Villar-Sbaffi, A., St-Louis, N., \& Morrell, N.~I.\ 2009, \mnras, 395, 823 


\bibitem[Shenar et al.(2016)]{Shenar} Shenar, T., Richardson, N.~D., Sablowski, D.~P., et al.\ 2016, \aap, accepted, arXiv:1610.07614

\bibitem[Sota et al.(2014)]{sota+2014} Sota, A., Ma{\'{\i}}z Apell{\'a}niz, J., Morrell, N.~I., et al.\ 2014, \apjs, 211, 10 

\bibitem[Taylor et al.(2011)]{TES11} Taylor, W.~D., Evans, C.~J., Sana, H., et al.\ 2011, \aap, 530, L10 


\bibitem[Tramper et al.(2016)]{TSF16} Tramper, F., Sana, H., Fitzsimons, N.~E., et al.\ 2016, \mnras, 455, 1275 

\bibitem[Udalski et al.(2008)]{Udalski+2008} Udalski, A., Szymanski, M.~K., Soszynski, I., \& Poleski, R.\ 2008, \actaa, 58, 69

\bibitem[Udalski et al.(2015)]{Udalski+2015} Udalski, A., Szyma{\'n}ski, M.~K., \& Szyma{\'n}ski, G.\ 2015, \actaa, 65, 1 

\bibitem[{{Walborn} \& {Blades}(1997)}]{WaB97}{Walborn}, N.~R., \& {Blades}, J.~C. 1997, \apjs, 112, 457

\bibitem[{{Walborn} {et~al.}(2014){WSSD14}, {Sana}, {Sim{\'o}n-D{\'{\i}}az},
  {Ma{\'{\i}}z Apell{\'a}niz}, {Taylor}, {Evans}, {Markova}, {Lennon}, \& {de
  Koter}}]{WSSD14}{Walborn}, N.~R., {Sana}, H., {Sim{\'o}n-D{\'{\i}}az}, S., {et~al.} 2014, \aap, 564, A40
  
\bibitem[{Wright} {et al.}(2015)]{Wright2015} Wright, N.~J., Drew, J.~E., \& Mohr-Smith, M.\ 2015, \mnras, 449, 741 

\bibitem[{{Yoon} {et~al.}(2010){Yoon}, {Woosley}, \& {Langer}}]{YWL10}
{Yoon}, S.-C., {Woosley}, S.~E., \& {Langer}, N. 2010, \apj, 725, 940

\bibitem[Zahn(1977)]{Zahn1977} Zahn, J.-P.\ 1977, \aap, 57, 383 

\bibitem[{{Zinnecker} \& {Yorke}(2007)}]{ZiY07}{Zinnecker}, H., \& {Yorke}, H.~W. 2007, \araa, 45, 481



\end{thebibliography}
\end{document}